\documentclass[a4paper,11pt]{report}
\usepackage{graphicx}
\usepackage{epsfig}
\usepackage[english]{babel}
\usepackage{amssymb}
\usepackage{amsmath}
\usepackage{indentfirst}
\usepackage{multirow}
\def\td{{\rm d}}
\def\D{{\rm D}}
\def\lp{\left. }
\def\rp{\right. }
\def\lr{\left( }
\def\rr{\right) }
\def\le{\left[ }
\def\re{\right] }

\def\beq{\begin{equation}}
\def\eeq{\end{equation}}
\def\bea{ \begin{eqnarray}}
\def\eea{\end{eqnarray}}
\def\as{\alpha_s}
\def\nn{\nonumber}
\def\stau{\tilde{\tau}}
\def\ms{m_{\tilde{q}}}
\def\hs{\theta_{\tilde{q}}}
\def\hp{\theta_{\tilde{t}}}
\def\hm{\theta_{\tilde{b}}}
\def\mg{m_{\tilde{g}}}
\def\mi{m_{\tilde{\chi}^0_i}}
\def\mj{m_{\tilde{\chi}^0_j}}
\def\tg{t_{\tilde{g}}}
\def\ti{t_{\tilde{\chi}^0_i}}
\def\tj{t_{\tilde{\chi}^0_j}}
\def\nbar{\bar{N}}
\def\bbar{\bar{b}}

\RequirePackage[pdfpagemode=none,
pdftoolbar=true,pdffitwindow=true,pdfcenterwindow=true]{hyperref}

\graphicspath{{/}} 
\begin{document}

\hspace{-1cm}\begin{minipage}[t]{\textwidth} \thispagestyle{empty}
\vspace{-2cm} \hspace{-2cm} \small LPSC 07-70\\
\vspace{2cm} \normalsize
\begin{center}

Universit\'e Joseph Fourier - Grenoble 1 \\
Ecole doctorale de Physique\\

\vspace{2cm}

\large
\textbf{Th\`ese de doctorat}\\
Sp\'ecialit\'e: Physique des particules\\
\vspace{1cm}
pr\'esent\'ee par\\
\textbf{Benjamin Fuks}\\ \vspace{1cm}
en vue de l'obtention du grade de \\
\textbf{Docteur en Sciences de l'Universit\'e Joseph Fourier}\\

\vspace{1.5cm}

\LARGE \textbf{QCD-resummation and non-minimal flavour-violation
for supersymmetric particle production at hadron colliders}\\

\vspace{4cm} \normalsize

Soutenue le 26 juin 2007 devant le jury compos\'e de:\\
\vspace{1cm}

\renewcommand{\arraystretch}{.9}
\begin{tabular}{l c l}
Prof. Aldo Deandrea & \hspace{1cm} & Rapporteur\\
Dr. Jonathan Ellis & & Rapporteur\\
Prof. Wolfgang Hollik & & Examinateur\\
Prof. Michael Klasen & & Directeur de th\`ese\\
Dr. Serge Kox & & Examinateur\\
Prof. Fran\c cois Le Diberder & & Examinateur\\
Prof. G\'erard Sajot & & Pr\'esident du jury\\
\end{tabular}\end{center}
\end{minipage}

\newpage  \thispagestyle{empty} $~$\\  \newpage
\pagenumbering{roman}

\chapter*{Remerciements}

Par cette premi\`ere page, je d\'edicace ce manuscrit \`a toutes
les personnes qui m'ont aid\'e et soutenu pendant ces trois
derni\`eres ann\'ees.\\

Je remercie sinc\`erement Johann Collot et Serge Kox, directeurs
du LPSC de Grenoble, de m'avoir accueilli au sein du laboratoire
durant ma th\`ese.\\

Mes remerciements s'adressent \'egalement \`a mes rapporteurs,
Aldo Deandrea et John Ellis, ainsi qu'aux membres de mon jury,
Wolfgang Hollik, Serge Kox, Fran\c cois Le Diberder et G\'erard
Sajot, pour leur lecture attentive de ce manuscrit, leurs
commentaires et suggestions.\\

Entamer une th\`ese consiste \`a d\'ebuter un long travail, et ce
travail ne peut \^etre effectu\'e sans guide. J'aimerais remercier
Michael, mon directeur de th\`ese, qui m'a accord\'e sa confiance
tout au long de ces trois ann\'ees. En f\'evrier 2004, tu m'as
propos\'e un sujet de th\`ese en physique th\'eorique des
particules alors que je n'avais jamais effectu\'e le moindre
travail de recherche dans ce domaine. Depuis, je pense m'\^etre
rattrap\'e, les projets s'\'etant succ\'ed\'es les uns aux autres.
J'aimerais \'egalement te t\'emoigner ma plus sinc\`ere
reconnaissance pour la patience et la gentillesse dont tu as fait
preuve face \`a mes innombrables questions, et pour nos
discussions, sans quoi ce travail ne serait pas ce qu'il est
aujourd'hui. J'esp\`ere de tout c\oe ur que notre collaboration
durera encore de nombreuses ann\'ees. Encore merci.\\

D\`es le d\'ebut de ma th\`ese, le jeune novice que j'\'etais put
b\'en\'eficier des conseils et de l'exp\'erience d'un post-doc
avis\'e, Giuseppe. J'ai essay\'e de t'apprendre quelques
expressions du parler bruxellois. En \'echange, tu m'as enseign\'e
que foncer t\^ete baiss\'ee dans un tas de probl\`emes afin d'y
extirper les solutions n'est pas toujours la m\'ethode la plus
efficace, et qu'il vaut parfois mieux prendre du recul et attaquer
les probl\`emes un par un. De plus, m'avoir comme voisin de bureau
n'a sans doute pas toujours \'et\'e tr\`es facile, et je te
remercie pour ta disponibilit\'e et ton aide.\\

Un clin d'\oe il \`a Bj\"orn, qui me montra qu'il n'y avait pas
que les collisionneurs dans la vie, mais \'egalement la cosmo.
Parmi nos exploits, il faudra retenir que nous avons prouv\'e que
l'homme est plus malin que Mathematica (ou pas). Fin septembre, je
m'en irai vers de nouvelles contr\'ees, mais la rel\`eve est
tr\`es prometteuse. Bon courage pour ta th\`ese, Jonathan.\\

Bien s\^ur, je ne peux oublier l'\'equipe des physiciens de
Grenoble. Sabine pour la relecture du manuscrit, Guillaume pour
les suggestions concernant le r\'esum\'e en fran\c cais, Beno\^it,
Bertrand, Ingo, Jean-Marc et Ji-Young pour avoir assist\'e aux
r\'ep\'etitions de soutenance, et les membres du groupe de
physique th\'eorique, qui sont autant amateurs de pique-niques que
de physique. Je tiens ensuite \`a remercier l'ensemble des
habitants du royaume du Bidul, fervents adorateurs de la
sacro-sainte pause-caf\'e de midi qui dure des heures. Par ordre
alphab\'etique, Antje, Colas, Florent, Jonathan, Julien, Julien,
Kevin, Lauranne, Marie-Anne, Maud, Pierre-Antoine, St\'ephanie,
Sylvain,
Thibaud, Vincent, Yoann, et ceux d\'ej\`a cit\'es auparavant.\\

Parfois, les rouages de l'administration peuvent para\^itre
simples, surtout avec C\'ecile et France pour vous aider \`a y
voir plus clair. Merci \`a toutes les deux pour votre
disponibilit\'e et votre gentillesse.\\

Je d\'esire \'egalement \'ecrire une petite d\'edicace \`a Daniel
Baye dont le cours d'\'el\'ements de m\'ecanique quantique a
hautement influenc\'e mon choix concernant la voie de la
physique.\\

\'Evidemment, je ne peux clore ces lignes sans une pens\'ee pour
mes proches, ma famille et mes amis. Une attention toute
particuli\`ere pour mes parents, mes grands-parents, et mes
fr\`eres qui me soutiennent et supportent depuis toujours, ainsi
que pour Marraine Vera et Parrain John qui ont fait le
d\'eplacement jusqu'\`a Grenoble pour assister \`a ma soutenance.
Ses encouragements, son soutien et sa bonne humeur n'ont jamais
tenu compte des fronti\`eres; un grand merci \`a toi, Manou. Je
remercie \'egalement mes amis, de Bruxelles, de Grenoble, ou
d'ailleurs, et sp\'ecialement Catherine, Donio et Julien qui n'ont
pas eu peur des kilom\`etres pour venir assister \`a ma
pr\'esentation, ainsi qu'Aline, Beno\^it, Elise, Fab, Floh, Gima,
Jean-Louis, Jonathan, Karo, Lolo, M\'elanie, Myriam, Nico, Sophie,
Wawaa et Yann pour leur soutien sans
faille.\\

Pour ceux que j'aurais oubli\'es, je ne l'ai pas fait expr\`es,
mais merci \`a vous.

\chapter*{Abstract}
\footnotesize \noindent Cross sections for supersymmetric
particles production at hadron colliders have been extensively
studied in the past at leading order and also at next-to-leading
order of perturbative QCD. The radiative corrections include large
logarithms which have to be resummed to all orders in the strong
coupling constant in order to get reliable perturbative results.
In this work, we perform a first and extensive study of the
resummation effects for supersymmetric particle pair production at
hadron colliders. We focus on Drell-Yan like slepton-pair and
slepton-sneutrino associated production in minimal supergravity
and gauge-mediated supersymmetry-breaking scenarios, and present
accurate transverse-momentum and invariant-mass distributions, as
well as total cross sections.\\

\noindent In non-minimal supersymmetric models, novel effects of
flavour violation may occur. In this case, the flavour structure
in the squark sector cannot be directly deduced from the trilinear
Yukawa couplings of the fermion and Higgs supermultiplets. We
perform a precise numerical analysis of the experimentally allowed
parameter space in the case of minimal supergravity scenarios with
non-minimal flavour violation, looking for regions allowed by
low-energy, electroweak precision, and cosmological data. Leading
order cross sections for the production of squarks and gauginos at
hadron colliders are implemented in a flexible computer program,
allowing us to study in detail the dependence of these cross
sections on flavour violation.\\

\vspace{1cm}

\noindent Les sections efficaces de production hadronique de
particules supersym\'etriques ont \'et\'e largement \'etudi\'ees
par le pass\'e, aussi bien \`a l'ordre dominant qu'\`a l'ordre
sous-dominant en QCD perturbative. Les corrections radiatives
incluent de larges termes logarithmiques qu'il faut resommer \`a
tous les ordres afin d'obtenir des pr\'edictions consistantes.
Dans ce travail, nous effectuons une premi\`ere \'etude
d\'etaill\'ee des effets de resommation pour la production
hadronique de particules supersym\'etriques. Nous nous concentrons
sur la production de type Drell-Yan de sleptons et sur la
production associ\'ee d'un slepton et d'un sneutrino dans des
sc\'enarios de supergravit\'e minimale et de brisure de
supersym\'etrie v\'ehicul\'ee par interactions de jauge, et nous
pr\'esentons des distributions d'impulsion transverse et de masse
invariante, ainsi que des sections efficaces totales.\\

\noindent Dans les mod\`eles supersym\'etriques non minimaux, de
nouveaux effets de violation de la saveur peuvent avoir lieu. Dans
ce cas, la structure de saveur dans le secteur des squarks ne peut
pas \^etre d\'eduite directement du couplage trilin\'eaire entre
les supermultiplets de Higgs et de fermions. Nous effectuons une
analyse num\'erique de l'espace des param\`etres permis dans le
cas de sc\'enarios de supergravit\'e minimale avec violation de la
saveur non minimale, cherchant les r\'egions permises par les
mesures de pr\'ecision \'electrofaibles, les observables \`a basse
\'energie et les donn\'ees cosmologiques. La d\'ependance des
sections efficaces \`a l'ordre dominant pour la production
hadronique de squarks et de jauginos par rapport \`a la violation
de la saveur non minimale est \'etudi\'ee en d\'etails.

\normalsize \newpage $~$\\ \newpage \tableofcontents

\chapter*{R\'esum\'e}
Le Mod\`ele Standard (SM) de la physique des particules
\cite{Glashow:1961tr, Weinberg:1967tq, Glashow:1970gm,
Weinberg:1971nd, Gross:1973ju, Gross:1974cs, Politzer:1974fr}
d\'ecrit avec succ\`es un grand nombre de donn\'ees
exp\'erimentales de haute \'energie. Cependant, certaines
questions fondamentales restent sans r\'eponse, comme par exemple
les origines de la brisure de la sym\'etrie \'electrofaible et des
masses des particules, la large hi\'erarchie entre l'\'echelle de
Planck et l'\'echelle \'electrofaible, le m\'ecanisme responsable
des oscillations de neutrinos, les origines de la mati\`ere sombre
et de la constante cosmologique, ou encore le probl\`eme CP li\'e
\`a l'interaction forte. Les tentatives visant \`a relier
diff\'erents param\`etres du SM m\`enent en g\'en\'eral \`a des
th\'eories plus fondamentales qui r\'esolvent naturellement
certains de ces probl\`emes ouverts.\\

La philosophie g\'en\'erale des Th\'eories de Grande Unification
(GUTs) \cite{Georgi:1974sy, Georgi:1974yf, Fritzsch:1974nn} est de
consid\'erer que les groupes de sym\'etrie du SM \'emergent de la
brisure d'un groupe simple de rang plus \'elev\'e. A l'\'echelle
GUT, les trois constantes de couplage de jauge du SM sont
unifi\'ees, et les quarks et les leptons sont d\'ecrits par des
repr\'esentations communes de ce groupe de jauge plus large. Ces
th\'eories pr\'edisent en g\'en\'eral un certain nombre de bosons
de jauge additionnels, menant \'eventuellement \`a des
interactions pouvant violer la conservation des nombres baryonique
et leptonique. Il s'agit de l'un des probl\`emes
ph\'enom\'enologiques les plus importants pour les GUTs, vu que
les baryons sont alors instables, ce qui est contraire aux
donn\'ees exp\'erimentales li\'ees \`a la non observation de la
d\'esint\'egration du proton. Les th\'eories GUTs peuvent
expliquer la quantification de la charge \'electrique et
incorporer des neutrinos massifs, mais  ont quelques difficult\'es
pour reproduire la valeur mesur\'ee de l'angle de m\'elange
\'electrofaible. De plus, un grand nombre de probl\`emes
conceptuels d\'ej\`a pr\'esents dans le SM demeurent sans
r\'eponse.\\

Une approche populaire pour r\'esoudre le probl\`eme de
hi\'erarchie du SM est d'ajouter \`a l'espace-temps des dimensions
suppl\'ementaires \cite{Arkani-Hamed:1998rs, Randall:1999ee}. Dans
ce cadre th\'eorique, les interactions de jauge et
gravitationnelle sont unies \`a une \'echelle proche de
l'\'echelle \'electrofaible, qui est alors la seule \'echelle
fondamentale de la th\'eorie, la valeur importante de l'\'echelle
de Planck \'etant seulement une cons\'equence de la pr\'esence des
nouvelles dimensions. L'espace \`a quatre dimensions habituel est
contenu dans une ``brane'' quadridimensionnelle, elle-m\^eme
incluse dans une structure plus large contenant N dimensions
additionnelles, le ``bulk''. Dans ces th\'eories, chaque champ du
SM poss\`ede une s\'erie d'excitations de Kaluza-Klein avec les
m\^emes nombres quantiques, mais une masse diff\'erente. Au jour
d'aujourd'hui, ces excitations n'ont pas encore \'et\'e
observ\'ees, mais l'ordre de grandeur de leur masse est le TeV, ce
qui les rend tout \`a fait d\'etectables au
futur Grand Collisionneur de Hadrons, le LHC, au CERN.\\

Plus r\'ecemment, d'autres tentatives pour r\'esoudre ce
probl\`eme de hi\'erarchie ont \'et\'e propos\'ees, comme par
exemple les th\'eories ``Little-Higgs'' ou ``Twin-Higgs'', qui
pr\'edisent \'egalement de nouvelles particules avec des masses de
l'ordre du TeV \cite{Arkani-Hamed:2001nc, Schmaltz:2005ky,
Chacko:2005pe}. Ces th\'eories incluent des partenaires
fermioniques pour les quarks et les leptons du SM, et des
partenaires bosoniques pour les bosons de jauge. Cela permet la
stabilisation de la masse du boson de Higgs au-del\`a de l'ordre
dominant gr\^ace \`a la r\'ealisation d'une sym\'etrie non
lin\'eaire reliant les couplages au boson de Higgs d'une fa\c con
telle que les divergences venant des corrections quantiques
s'annulent.\\

Dans cette th\`ese, nous nous concentrons sur une autre extension
attractive du SM, la supersym\'etrie (SUSY) \cite{Wess:1974tw,
Fayet:1976cr, Fayet:1977yc, Farrar:1978xj, Fayet:1979sa}, et plus
pr\'ecis\'ement le Mod\`ele Standard Supersym\'etrique Minimal
(MSSM) \cite{Nilles:1983ge, Haber:1984rc}. La supersym\'etrie \`a
basse \'energie fournit une solution naturelle \`a plusieurs des
probl\`emes conceptuels du SM. Reliant les fermions et les bosons,
elle permet la stabilisation de la hi\'erarchie s\'eparant
l'\'echelle de Planck de l'\'echelle \'electrofaible
\cite{Witten:1981nf, Kaul:1981wp} et l'unification des couplages
de jauge aux hautes \'energies \cite{Ibanez:1981yh,
Dimopoulos:1981yj, Amaldi:1991cn, Carena:1993ag}. De plus, la
\mbox{parti}cule SUSY la plus l\'eg\`ere peut dans certains cas
\^etre vue comme un candidat potentiel pour la mati\`ere sombre
\cite{Goldberg:1983nd, Ellis:1983ew}. Vu que les partenaires
supersym\'etriques des particules du SM n'ont pas encore \'et\'e
observ\'es jusqu'\`a pr\'esent, la supersym\'etrie doit \^etre
bris\'ee \`a basse \'energie, mais de fa\c con douce afin qu'elle
reste une solution viable pour le probl\`eme de la hi\'erarchie.
Les \mbox{parti}cules SUSY sont donc plus massives que leurs
\'equivalents du SM, mais leur masse ne devrait pas exc\'eder
quelques TeV. Une recherche concluante couvrant un large r\'egime
de masses allant jusqu'\`a l'\'echelle du TeV est donc l'un des
points principaux du programme exp\'erimental des collisionneurs
hadroniques pr\'esents et futurs, comme par exemple
le Tevatron \`a Fermilab ou le LHC au CERN.\\

Les sections efficaces de production des particules SUSY aupr\`es
des collisionneurs hadroniques ont \'et\'e \'etudi\'ees en
d\'etail par le pass\'e, aussi bien \`a l'ordre dominant (LO)
\cite{Dawson:1983fw, delAguila:1990yw, Baer:1993ew} qu'\`a l'ordre
sous-dominant (NLO) \cite{Beenakker:1996ch, Beenakker:1997ut,
Berger:1998kh, Berger:1999mc, Berger:2000iu, Baer:1997nh,
Beenakker:1999xh} en QCD perturbative. Il est connu que les
corrections NLO QCD \cite{Baer:1997nh} et NLO SUSY-QCD compl\`etes
\cite{Beenakker:1999xh} pour la production d'une paire de sleptons
augmentent les sections efficaces hadroniques d'environ 35\% au
Tevatron et 25\% au LHC, ce qui \'etend le potentiel de
d\'ecouverte des sleptons de plusieurs dizaines de GeV. Cependant,
les corrections SUSY sont bien plus faibles que leurs analogues
QCD en raison de la pr\'esence de squarks et gluinos tr\`es lourds
dans les boucles.\\

Malgr\'e le succ\`es des premi\`eres collisions proton-proton en
mode polaris\'e au collisionneur RHIC, les sections efficaces
polaris\'ees ont re\c cu bien moins d'attention que leurs
\'equivalents non polaris\'es. Les calculs pionniers pour la
production de squarks et de gluinos non massifs
\cite{Craigie:1983as, Craigie:1984tk} n'ont \'et\'e v\'erifi\'es,
g\'en\'eralis\'es au cas de particules SUSY massives et
appliqu\'es aux collisionneurs actuels que r\'ecemment
\cite{Gehrmann:2004xu}. Concernant la production de sleptons,
seulement des calculs n\'egligeant les m\'elanges entre les
\'etats propres d'h\'elicit\'e \'etaient disponibles, et
appliqu\'es uniquement \`a des exp\'eriences d'anciens
collisionneurs \cite{Chiappetta:1985ku}.\\

En raison de leurs couplages purement \'electrofaibles, les
sleptons sont parmi les particules SUSY les plus l\'eg\`eres dans
de nombreux sc\'enarios de brisure de supersym\'etrie
\cite{Allanach:2002nj, Aguilar-Saavedra:2005pw}. Les sleptons et
sneutrinos se d\'esint\`egrent souvent directement en la particule
SUSY la plus l\'eg\`ere (le neutralino le plus l\'eger dans les
mod\`eles de supergravit\'e minimale (mSUGRA) ou le gravitino pour
la brisure de supersym\'etrie v\'ehicul\'ee par interactions de
jauge (GMSB)) et le partenaire du SM correspondant (un lepton ou
un neutrino).  Ainsi, un signal relatif \`a une paire de sleptons
produite en collisionneur hadronique consistera en une paire de
leptons tr\`es \'energ\'etiques, qui sera facilement
d\'etectable, et de l'\'energie manquante associ\'ee.\\

Dans cette th\`ese, nous avons v\'erifi\'e les calculs pionniers
pour la production polaris\'ee d'une paire de sleptons
\cite{Chiappetta:1985ku}, que nous avons ensuite g\'en\'eralis\'es
afin de prendre en compte le m\'elange des \'etats propres
d'interaction, qui est surtout pertinent pour les sleptons de
troisi\`eme g\'en\'eration. Nous pr\'esentons les r\'esultats
analytiques pour les courants de sleptons neutres et charg\'es, et
pr\'edisons num\'eriquement les asym\'etries simple-spin pour le
collisionneur RHIC et pour d'\'eventuelles am\'eliorations du
Tevatron et du LHC o\`u l'un des faisceaux est polaris\'e. Nous
avons mis en \'evidence la sensibilit\'e de l'asym\'etrie
simple-spin \`a l'angle de m\'elange du slepton tau et la
possibilit\'e de l'utiliser comme moyen pour distinguer le signal
SUSY du bruit de fond du SM correspondant \`a la production
Drell-Yan d'une paire de leptons \cite{Bozzi:2004qq}.\\

Le bruit de fond standard principal li\'e \`a la production
hadronique d'une paire de sleptons vient des d\'esint\'egrations
de paires $WW$ et $t\bar t$ en une paire de leptons et de
l'\'energie manquante \cite{Lytken:22, Andreev:2004qq}. Deux
\'el\'ements cl\'es pour distinguer le signal SUSY du bruit de
fond standard sont la reconstruction de la masse et la
d\'etermination du spin des particules produites. Pour une paire
de sleptons, la masse (s)transverse de Cambridge est une
observable particuli\`erement utile, puisqu'une connaissance
pr\'ecise du spectre en impulsion transverse ($q_T$) suffit alors
pour d\'eterminer la masse \cite{Lester:1999tx} et le spin
\cite{Barr:2005dz} des sleptons.\\

Lorsque que l'on \'etudie la distribution en impulsion transverse
d'un syst\`eme non color\'e produit avec une masse invariante $M$
lors d'une collision hadronique, il est pertinent de s\'eparer les
r\'egions cin\'ematiques relatives aux larges et aux faibles
valeurs de $q_T$. Dans la r\'egion des $q_T$ importants ($q_T \geq
M$), l'utilisation de la th\'eorie perturbative \`a ordre fix\'e
est parfaitement justifi\'ee, vu que le d\'eveloppement en s\'erie
de la distribution en $q_T$ est contr\^ol\'e par un param\`etre
d'expansion de faible valeur, la constante de couplage forte
$\as(M^2)$. Dans la r\'egion des petites valeurs de $q_T$, les
coefficients du d\'eveloppement perturbatif sont amplifi\'es par
des termes logarithmiques importants, $\ln(M^2/q_T^2)$, et les
r\'esultats bas\'es sur des calculs perturbatifs divergent pour
$q_T\to 0$, la convergence de la s\'erie \'etant alors
compl\`etement d\'etruite. Ces logarithmes proviennent de
l'\'emission multiple de gluons mous par l'\'etat initial, et
doivent \^etre syst\'ematiquement resomm\'es \`a tous les ordres
en $\as$ afin d'obtenir des r\'esultats consistants. La m\'ethode
pour effectuer cette resommation est bien connue
\cite{Dokshitzer:1978hw, Parisi:1979se, Curci:1979bg,
Kodaira:1981nh, Collins:1981uk, Collins:1981va, Altarelli:1984pt,
Collins:1984kg, Davies:1984hs, Catani:2000vq, Bozzi:2005wk}. La
resommation des logarithmes dominants a \'et\'e effectu\'ee pour
la premi\`ere fois en \cite{Dokshitzer:1978hw}. Il a \'et\'e
montr\'e en \cite{Parisi:1979se} que la proc\'edure de resommation
est plus naturellement effectu\'ee dans l'espace du param\`etre
d'impact $b$, $b$ \'etant la variable conjugu\'ee \`a $q_T$ via
une transformation de Fourier. En effet, dans ce cas-l\`a, la
cin\'ematique de l'\'emission multiple de gluons factorise
compl\`etement. Dans les cas particuliers de la production
Drell-Yan d'une paire de leptons et de la production d'un boson
\'electrofaible, la resommation dans l'espace $b$ a \'et\'e
effectu\'ee au niveau sous-dominant (NLL) \cite{Kodaira:1981nh},
un formalisme de resommation consistant \`a n'importe quelle
pr\'ecision logarithmique a \'et\'e d\'evelopp\'e
\cite{Collins:1984kg}, et les termes d'ordre sous-sous-dominant
ont \'et\'e calcul\'es \cite{Davies:1984hs}. Pour les valeurs de
$q_T$ interm\'ediaires, le r\'esultat resomm\'e doit \^etre
ajust\'e de fa\c con consistante avec celui bas\'e sur la
th\'eorie perturbative, afin d'obtenir des pr\'edictions d'une
pr\'ecision th\'eorique uniforme sur tout le domaine d'impulsion
transverse consid\'er\'e.\\

Dans ce travail, nous avons impl\'ement\'e le formalisme de
resommation en $q_T$ propos\'e en \cite{Catani:2000vq,
Bozzi:2005wk}, et pr\'edit le spectre en impulsion transverse pour
la production d'une paire de sleptons au LHC. Nous avons combin\'e
le r\'esultat resomm\'e (valide pour les faibles valeurs de
$q_T$), calcul\'e au niveau NLL, avec la section efficace \`a
ordre fix\'e (valide pour les larges valeurs de $q_T$), calcul\'ee
\`a l'$\mathcal{O}(\as)$ en QCD perturbative qui correspond \`a la
production d'une paire de sleptons associ\'ee \`a un jet QCD
\cite{Bozzi:2006fw}. Il s'agit du premier calcul de pr\'ecision
concernant la distribution en impulsion transverse pour un
processus de production d'une paire de particules SUSY aupr\`es
d'un collisionneur hadronique. Dans nos r\'esultats num\'eriques,
nous avons montr\'e l'importance de la resommation aussi bien pour
les faibles valeurs de $q_T$ que pour les valeurs
interm\'ediaires. Par ailleurs, la resommation permet de r\'eduire
la d\'ependance de la distribution en $q_T$ en les \'echelles non
physiques de factorisation et de renormalisation. Nous avons
\'egalement \'etudi\'e l'influence des contributions non
perturbatives sur le r\'esultat resomm\'e, et observ\'e qu'elle
\'etait r\'eduite par rapport \`a l'effet de la resommation. \\

En ce qui concerne les corrections NLO SUSY-QCD, elles ont \'et\'e
calcul\'ees uniquement en n\'egligeant le m\'elange des \'etats
propres d'interaction des squarks apparaissant dans les boucles
\cite{Beenakker:1999xh}. Nous avons g\'en\'eralis\'e ce travail en
incluant ce m\'elange pertinent pour les squarks de troisi\`eme
g\'en\'eration, et avons consid\'er\'e les effets de seuil
provenant de l'\'emission de gluons mous par l'\'etat initial.
Lorsque les partons initiaux ont tout juste assez d'\'energie pour
produire la paire de sleptons dans l'\'etat final, les corrections
virtuelles et l'\'emission de gluons r\'eels supprim\'ee par
l'espace de phase m\`enent \`a l'apparition de termes
logarithmiques importants $\as^n [\ln^{2n-1}(1-z)/(1-z)]_+$ \`a
l'ordre $n$ de la th\'eorie perturbative, o\`u $z = M^2/s$, $s$
est l'\'energie dans le centre de masse partonique et $M$ la masse
invariante de la paire de sleptons. Lorsque $s$ est proche de
$M^2$, ces logarithmes doivent \^etre resomm\'es \`a tous les
ordres en $\as$. Bien que ces divergences apparaissent de fa\c con
explicite dans la section efficace partonique, la section efficace
hadronique n'est en g\'en\'eral pas divergente en raison de la
convolution avec les densit\'es de partons tr\`es faibles pour les
grandes valeurs de la fraction d'impulsion longitudinale du proton
correspondant aux valeurs de $z$ proches de un. La resommation en
seuil est donc plut\^ot une tentative de quantification de l'effet
d'un ensemble de corrections bien d\'efinies qu'une simple somme
de logarithmes d'origine cin\'ematique. Ces effets peuvent
cependant \^etre significatifs m\^eme loin du seuil hadronique, et
l'on s'attend donc \`a des corrections importantes pour la section
efficace de production de type Drell-Yan d'une paire de sleptons
de quelques centaines de GeV au
Tevatron et au LHC.\\

La resommation en seuil \`a tous les ordres en $\as$ \'equivaut
\`a l'exponentiation des radiations de gluons mous, et n'a pas
lieu dans l'espace $z$ directement, mais dans l'espace $N$ de
Mellin, o\`u $N$ est la variable conjugu\'ee \`a $z$ par une
transformation de Mellin, la r\'egion du seuil $z\to 1$
correspondant \`a la limite $N\to \infty$. Ainsi, la section
efficace resomm\'ee dans l'espace $z$ sera obtenue apr\`es une
transformation inverse finale. La resommation en seuil pour le
processus Drell-Yan fut d'abord effectu\'ee en
\cite{Sterman:1986aj, Catani:1989ne} aux niveaux logarithmiques
dominant et sous-dominant (NLL), correspondant \`a la resommation
des termes de type $\as^n \ln^{2n} N$ et $\as^n \ln^{2n-1} N$.
L'extension au niveau NNLL (termes de type $\as^n \ln^{2n-2} N$) a
\'egalement \'et\'e effectu\'ee \`a la fois pour le processus
Drell-Yan \cite{Vogt:2000ci} et pour la production d'un boson de
Higgs \cite{Catani:2003zt}. Il a \'et\'e montr\'e
\cite{Kramer:1996iq, Catani:2001ic} que les contributions dues \`a
l'\'emission de partons colin\'eaires peuvent \'egalement \^etre
incluses de fa\c con consistante dans la formule de resommation.
Cela correspond au formalisme de resommation ``am\'elior\'e
colin\'eairement'', o\`u des termes contenant un facteur
suppressif $1/N$ et une classe de contributions universelles
ind\'ependantes de $N$ sont \'egalement resomm\'es. Tr\`es
r\'ecemment, les contributions \`a l'ordre NNNLL (les termes de
type $\as^n \ln^{2n-3} N$) ont \'et\'e calcul\'ees
\cite{Moch:2005ba, Moch:2005ky, Laenen:2005uz}.\\

Nous pr\'esentons ici une \'etude d\'etaill\'ee des effets de la
resommation en seuil pour la production de type Drell-Yan d'une
paire de sleptons et pour la production associ\'ee d'un slepton et
d'un sneutrino dans le cadre de sc\'enarios mSUGRA et GMSB. Nous
avons ajust\'e les r\'esultats resomm\'es \`a la pr\'ecision NLL,
calcul\'es gr\^ace au formalisme de resommation am\'elior\'e
colin\'eairement, avec les r\'esultats bas\'es sur la th\'eorie
perturbative calcul\'es \`a la pr\'ecision NLO. Num\'eriquement,
nous avons montr\'e une augmentation non n\'egligeable de la
section efficace th\'eorique par rapport aux pr\'edictions NLO, et
une stabilisation de la d\'ependance en les \'echelles non
physiques gr\^ace \`a l'apport des termes d'ordres sup\'erieurs
pris en compte par la resommation \cite{Bozzi:2007qr}.\\

L'origine dynamique des contributions logarithmiques intervenant
dans les formalismes de resommation en impulsion transverse et en
seuil est identique, vu qu'il s'agit de l'\'emission multiple de
gluons mous par l'\'etat initial. Un formalisme de resommation
jointe, prenant en compte simultan\'ement les contributions des
gluons mous dans les deux r\'egions cin\'ematiques concern\'ees
($q_{T} \ll M$ et $M^{2}\sim s$) a \'et\'e d\'evelopp\'e dans la
derni\`ere d\'ecade \cite{Li:1998is, Laenen:2000ij}.
L'exponentiation des termes singuliers dans les espaces de Mellin
et du param\`etre d'impact, pour la resommation en seuil et en
impulsion transverse respectivement, a \'et\'e prouv\'ee, et une
m\'ethode consistante pour effectuer les transformations inverses
a \'et\'e introduite afin d'\'eviter le p\^ole de Landau et les
singularit\'es dues aux densit\'es de partons. Les applications de
ce formalisme \`a la production hadronique d'un photon rapide
\cite{Laenen:2000de}, d'un boson \'electrofaible
\cite{Kulesza:2002rh}, d'un boson de Higgs \cite{Kulesza:2003wn},
et d'une paire de quarks lourds \cite{Banfi:2004xa} montrent les
effets de la resommation sur diff\'erentes distributions.\\

Nous pr\'esentons dans ce travail un traitement joint des
corrections \`a faible impulsion transverse et des contributions
importantes proche du seuil partonique pour la production d'une
paire de sleptons aupr\`es des collisionneurs hadroniques, ce qui
permet une compr\'ehension compl\`ete des effets de gluons mous
pour le spectre en impulsion transverse et pour les distributions
en masse invariante. Avec le travail sur la resommation en
impulsion transverse \cite{Bozzi:2006fw} et la resommation en
seuil \cite{Bozzi:2007qr}, cette \'etude \cite{Bozzi:2007xx}
compl\`ete notre programme ayant pour but de fournir les premiers
calculs de pr\'ecision incluant la resommation de gluons mous pour
la production de sleptons aupr\`es des collisionneurs
hadroniques.\\

Si les particules SUSY existent, elles doivent aussi appara\^itre
dans les boucles de particules virtuelles et affecter les
observables de pr\'ecision \'electrofaibles et les observables \`a
basse \'energie. Plus particuli\`erement, les courants neutres \`a
changement de saveur qui apparaissent seulement au niveau des
boucles dans le SM contraignent s\'ev\`erement les contributions
de nouvelle physique au m\^eme ordre perturbatif. Le MSSM se
lib\`ere de ces contraintes gr\^ace aux hypoth\`eses de Violation
de Saveur Minimale contrainte (cMFV) \cite{Ciuchini:1998xy,
Buras:2000dm} ou de Violation de Saveur Minimale (MFV)
\cite{Hall:1990ac, D'Ambrosio:2002ex, Altmannshofer:2007cs}, o\`u
les particules SUSY peuvent intervenir dans les boucles, mais les
changements de saveur sont soit n\'eglig\'es, soit compl\`etement
dict\'es par la structure des couplages de Yukawa et par la
matrice CKM \cite{Cabibbo:1963yz,
Kobayashi:1973fv}.\\

En SUSY avec MFV, les \'el\'ements des matrices de masse des
squarks violant la saveur d\'ecoulent des couplages trilin\'eaires
de Yukawa entre les supermultiplets de Higgs et de fermions et des
diff\'erentes renormalisations des secteurs des quarks et des
squarks via les \'equations du groupe de renormalisation qui
induisent des violations de saveur suppl\'ementaires \`a
l'\'echelle \'electrofaible \cite{Donoghue:1983mx, Duncan:1983iq,
Bouquet:1984pp, Borzumati:1986qx}. En SUSY avec violation de
saveur non minimale, des sources de violation de saveur
additionnelles sont incluses dans les matrices de masse et leurs
termes non diagonaux qui ne peuvent plus \^etre simplement
d\'eduits \`a partir de la matrice CKM seule doivent \^etre
consid\'er\'es alors comme des param\`etres libres. Dans ce
travail, nous allons consid\'erer le m\'elange des saveurs de
squark de deuxi\`eme et troisi\`eme g\'en\'erations, car d'une
part, les recherches directes de violation de la saveur
d\'ependent des capacit\'es \`a d\'eterminer la saveur, ce qui
n'est exp\'erimentalement bien \'etabli que pour les saveurs
lourdes, et d'autre part, des contraintes exp\'erimentales
s\'ev\`eres pour la premi\`ere g\'en\'eration existent en raison
de mesures tr\`es pr\'ecises des oscillations $K^0-\bar{K}^0$ et
des premi\`eres preuves du m\'elange $D^0-\bar{D}^0$
\cite{Hagelin:1992tc, Gabbiani:1996hi, Ciuchini:2007}.\\

Nous avons analys\'e l'espace des param\`etres NMFV SUSY,
recherchant les r\'egions permises par les contraintes venant des
mesures de pr\'ecision \'electrofaibles, des observables \`a basse
\'energie et des donn\'ees cosmologiques. Nous avons observ\'e que
le m\'elange des chiralit\'es et des saveurs de deuxi\`eme et
troisi\`eme g\'en\'erations est fortement contraint, notamment par
l'erreur exp\'erimentale de plus en plus petite sur le rapport
d'embranchement $b\to s\gamma$ et la densit\'e relique de
mati\`ere sombre. Nous avons d\'efini quatre nouveaux points
typiques avec leur ligne associ\'ee, valides \`a la fois en SUSY
avec cMFV, MFV et NMFV, et pour lesquels nous pr\'esentons la
d\'ependance des masses de squarks et de la d\'ecomposition des
\'etats physiques de squark en la violation de la saveur.\\

Consid\'erant la SUSY avec cMFV (le MSSM habituel), les
corrections SUSY-QCD pour la production de squarks et de gluinos
\cite{Beenakker:1996ch}, de jauginos \cite{Beenakker:1999xh},
ainsi que pour leur production associ\'ee \cite{Berger:1999mc} ont
d\'ej\`a \'et\'e calcul\'ees. En raison de leur couplage fort, les
squarks devraient \^etre produits abondamment aux collisionneurs
hadroniques, et l'espace de phase favorise la production des
\'etats propres de masse les plus l\'egers. Ainsi, les productions
des squarks top \cite{Beenakker:1997ut} et bottom
\cite{Berger:2000mp} avec un grand m\'elange d'h\'elicit\'e ont
re\c cu une attention toute particuli\`ere. Dans cette th\`ese,
nous nous sommes int\'eress\'es \`a l'importance des canaux
\'electrofaibles pour la production de paires de squarks non
diagonales et mixtes de troisi\`eme g\'en\'eration aux
collisionneurs hadroniques \cite{Bozzi:2005sy}. Na\"ivement, l'on
s'attend \`a ce que ces sections efficaces, qui sont d'ordre deux
en la constante de structure fine, $\mathcal{O}(\alpha^2)$, soient
plus faibles que celles concernant la production forte d'une paire
de squarks diagonale d'environ deux ordres de grandeurs. Pour la
production non diagonale, l'importance des canaux QCD est
r\'eduite en raison de la pr\'esence de boucles, et celle des
canaux \'electrofaibles l'est \'egalement en raison du couplage
faible. L'importance relative de ces canaux m\'erite donc une
\'etude approfondie. Si l'on consid\`ere des squarks bottom qui se
m\'elangent, leur contribution au niveau des boucles QCD doit
\'egalement \^etre prise en compte.\\

Ensuite, pour la premi\`ere fois, nous nous sommes concentr\'es
sur les effets possibles de la violation de saveur non minimale
(NMFV) aux collisionneurs hadroniques \cite{Bozzi:2007me}. A cette
fin, nous avons recalcul\'e toutes les amplitudes d'h\'elicit\'e
pour la production et la d\'esint\'egration des squarks et des
jauginos, en prenant en compte les interactions non diagonales des
courants charg\'es des jauginos et les interactions de Yukawa des
Higgsinos, et en g\'en\'eralisant les matrices de m\'elange
d'h\'elicit\'es bidimensionnelles, souvent suppos\'ees r\'eelles,
en matrices de m\'elange d'h\'elicit\'es et de saveurs, complexes
et six-dimensionnelles. Nous avons v\'erifi\'e que nos r\'esultats
reproduisaient ceux de la litt\'erature existant dans
les limites de squarks non m\'elang\'es. \\

Dans notre analyse ph\'enom\'enologique de la production NMFV de
squarks et de jauginos, nous nous sommes concentr\'es sur le LHC
en raison de son \'energie dans le centre de masse \'elev\'ee  et
de sa luminosit\'e importante. Nous avons port\'e une attention
particuli\`ere \`a la comp\'etition entre les effets li\'es aux
densit\'es de partons qui sont domin\'es par les contributions des
quarks l\'egers, les contributions fortes du gluino qui sont
g\'en\'eralement plus importantes que les contributions
\'electrofaibles et qui ne doivent pas n\'ecessairement \^etre
diagonales en saveur, et la pr\'esence de saveurs lourdes dans
l'\'etat final, facilement identifiables exp\'erimentalement et
g\'en\'eralement plus l\'eg\`eres que les saveurs de squark de
premi\`ere et deuxi\`eme g\'en\'erations.

\newpage  \thispagestyle{empty} $~$\\  \newpage
\chapter{Introduction}
\pagenumbering{arabic}

The Standard Model (SM) of particle physics \cite{Glashow:1961tr,
Weinberg:1967tq, Glashow:1970gm, Weinberg:1971nd, Gross:1973ju,
Gross:1974cs, Politzer:1974fr} provides a successful description
of all experimental high energy data. However, despite of its
success many fundamental questions remain unanswered, e.g.\  the
origins of electroweak symmetry breaking and particle masses, the
large hierarchy between the electroweak and the Planck scales, the
mechanism leading to neutrino oscillations, the origins of dark
matter and of the cosmological constant, or the strong CP-problem.
Attempts to relate different SM parameters lead to more
fundamental theories, that may at the same time solve some of the
open problems of the SM. \\

The basic philosophy of Grand Unified Theories (GUTs)
\cite{Georgi:1974sy, Georgi:1974yf, Fritzsch:1974nn} is to
consider the SM symmetry groups as originating from the breaking
of a larger simple group. At the GUT scale, the three SM gauge
coupling constants unify and quarks and leptons are embedded in
common representations of the unifying gauge group. These theories
include then number of additional gauge bosons, leading
potentially to interactions violating the baryon and lepton
numbers. This leads to one of the major phenomenological problems
of GUTs, which predict baryon instability, contrary to the
experimental non-observation of proton decay. GUT theories can
explain the quantization of the electric charge and incorporate
massive neutrinos, but have difficulties in accounting for the
measured value of the electroweak mixing angle. Besides, many
other conceptual SM problems remain unsolved.\\

One popular approach to solve the hierarchy problem of the SM is
to extend space-time to higher dimensions
\cite{Arkani-Hamed:1998rs, Randall:1999ee}. In this framework, the
gravitational and gauge interactions become unified close to the
weak scale, which is then the only fundamental scale of the
theory. The large value of the Planck scale is only a consequence
of the new dimensions. The usual four-dimensional space is
contained in a four-dimensional ``brane'', embedded in a larger
structure with $N$ additional dimensions, the ``bulk''. In these
theories, each field of the SM possesses a tower of Kaluza-Klein
excitations with the same quantum numbers, but different mass,
which have not been observed at the present time, but which should
lie in the TeV-range. They could then be detected at the future
Large Hadron Collider (LHC) at CERN.\\

Recently, other attempts to solve the hierarchy problem have been
proposed, e.g.\ Little-Higgs or Twin-Higgs theories, which predict
new particles with masses in the TeV-range as well
\cite{Arkani-Hamed:2001nc, Schmaltz:2005ky, Chacko:2005pe}. These
theories include fermionic partners for quarks and leptons and
bosonic partners for the SM gauge bosons, which allows for
stabilization of the Higgs mass beyond tree-level thanks to a
non-linearly realized symmetry that relates the couplings to the
Higgs in such a way that the quantum corrections to the Higgs mass
cancel.\\

In this thesis, we focus on another attractive extension of the
SM, supersymmetry (SUSY) \cite{Wess:1974tw, Fayet:1976cr,
Fayet:1977yc, Farrar:1978xj, Fayet:1979sa}, and more precisely the
Minimal Supersymmetric Standard Model (MSSM) \cite{Nilles:1983ge,
Haber:1984rc}. Weak scale supersymmetry provides a natural
solution for a set of conceptual problems of the SM. Linking
fermions with bosons, SUSY allows for a stabilization of the gap
between the Planck scale and the electroweak scale
\cite{Witten:1981nf, Kaul:1981wp} and for a consistent unification
of SM gauge couplings at high energies \cite{Ibanez:1981yh,
Dimopoulos:1981yj, Amaldi:1991cn, Carena:1993ag}. In addition, it
can include a potential dark matter candidate as the stable
lightest SUSY particle \cite{Goldberg:1983nd, Ellis:1983ew}. Since
spin partners of the SM particles have not yet been observed and
in order to remain a viable solution to the hierarchy problem,
SUSY must be broken at low energy via soft mass terms in the
Lagrangian. As a consequence, the SUSY particles must be massive
in comparison to their SM counterparts, but their mass should not
exceed a few TeV. A conclusive search covering a wide range of
masses up to the TeV scale is then one of the main topics in the
experimental program at present and future hadron colliders, such
as the Tevatron at Fermilab and the LHC at CERN.\\

Production cross sections for SUSY particles at hadron colliders
have been extensively studied in the past at leading order (LO)
\cite{Dawson:1983fw, delAguila:1990yw, Baer:1993ew} and also at
next-to-leading order (NLO) of perturbative QCD
\cite{Beenakker:1996ch, Beenakker:1997ut, Berger:1998kh,
Berger:1999mc, Berger:2000iu, Baer:1997nh, Beenakker:1999xh}. The
NLO QCD \cite{Baer:1997nh} and full SUSY-QCD
\cite{Beenakker:1999xh} corrections for slepton pair production
are known to increase the hadronic cross sections by about 35 \%
at the Tevatron and 25\% at the LHC, extending thus the discovery
reaches of sleptons by several tens of GeV. However, the presence
of massive squarks and gluinos in the loops makes the genuine SUSY
corrections considerably smaller than the standard QCD ones.\\

Despite of the first successful runs of the RHIC collider in the
polarized $pp$ mode, polarized cross sections have received much
less attention. Only the pioneering LO calculations for massless
squark and gluino production \cite{Craigie:1983as, Craigie:1984tk}
have recently been confirmed, extended to the massive case, and
applied to current hadron colliders \cite{Gehrmann:2004xu}.
Concerning slepton pair production, only polarized calculations
for non mixing sleptons were available before, and for old
collider experiments only \cite{Chiappetta:1985ku} .\\

Due to their purely electroweak couplings, sleptons are among the
lightest SUSY particles in many SUSY-breaking scenarios
\cite{Allanach:2002nj, Aguilar-Saavedra:2005pw}. Sleptons and
sneutrinos often decay directly into the stable lightest SUSY
particle (the lightest neutralino in minimal supergravity (mSUGRA)
models or the gravitino in gauge-mediated SUSY-breaking models
(GMSB)) plus the corresponding SM partner (lepton or neutrino). As
a result, a slepton signal at hadron colliders will consist in a
highly energetic lepton pair, which will be easily detectable, and
associated missing energy. \\

In this thesis, we verify the pioneering polarized calculations
for slepton pair production \cite{Chiappetta:1985ku} and extend
them by including the mixing of the left- and right-handed
interaction eigenstates relevant for third-generation sleptons. We
present analytical results for neutral and charged current
sleptons and make numerical predictions for longitudinal spin
asymmetries at RHIC and possible upgrades of the Tevatron and the
LHC, where one of the beams is considered to be polarized. We put
particular emphasis on the sensitivity of the asymmetry to the tau
slepton mixing angle as predicted by various SUSY-breaking
mechanisms. Possibilities of using asymmetries to discriminate
between the SUSY signal and the corresponding SM Drell-Yan
background are also discussed \cite{Bozzi:2004qq}.\\

The main SM background to slepton pair production at hadron
colliders is due to $WW$ and $t\bar{t}$ decays to a lepton pair
and missing energy \cite{Lytken:22, Andreev:2004qq}. Two key
features distinguishing the SUSY signal from the SM background are
the reconstruction of the mass and the determination of the spin
of the produced particles. For sleptons, the Cambridge
(s)transverse mass proves to be a particularly useful observable,
requiring only a precise knowledge of the transverse-momentum
($q_T$) spectrum to get their mass \cite{Lester:1999tx} and spin
\cite{Barr:2005dz}.\\

When studying the $q_T$-distribution of a colourless system
produced with an invariant-mass $M$ in a hadronic collision, it is
appropriate to separate the large-$q_T$ and small-$q_{T}$ regions.
In the large-$q_T$ region ($q_T\geq M$) the use of fixed-order
perturbation theory is fully justified, since the perturbative
series is controlled by a small expansion parameter, the strong
coupling constant $\alpha_s(M^2)$. In the small-$q_{T}$ region,
where the coefficients of the perturbative expansion in
$\alpha_s(M^{2})$ are enhanced by powers of large logarithmic
terms, $\ln(M^{2}/q_{T}^{2})$, results based on fixed-order
calculations diverge as $q_T \to 0$, and the convergence of the
perturbative series is spoiled. These logarithms are due to
multiple soft-gluon emission from the initial state and have to be
systematically resummed to all orders in $\alpha_s$ in order to
obtain reliable perturbative predictions. The method to perform
all-order soft-gluon resummation at small $q_{T}$ is well known
\cite{Dokshitzer:1978hw, Parisi:1979se, Curci:1979bg,
Kodaira:1981nh, Collins:1981uk, Collins:1981va, Altarelli:1984pt,
Collins:1984kg, Davies:1984hs, Catani:2000vq, Bozzi:2005wk}. The
resummation of leading logarithms was first performed in
\cite{Dokshitzer:1978hw}. It was shown in \cite{Parisi:1979se}
that the resummation procedure is most naturally performed using
the impact-parameter ($b$) formalism, where $b$ is the variable
conjugate to $q_T$ through a Fourier transformation, to allow the
kinematics of multiple-gluon emission to factorize. In the special
case of Drell-Yan lepton pair or electroweak boson production,
$b$-space resummation was performed at next-to-leading level in
\cite{Kodaira:1981nh}, a resummation formalism consistent at any
logarithmic accuracy was developed in \cite{Collins:1984kg}, and
the next-to-next-to-leading order terms have been calculated in
\cite{Davies:1984hs}. At intermediate $q_T$ the resummed result
has to be consistently matched with fixed-order perturbation
theory in order to obtain predictions with uniform theoretical
accuracy over the entire range of transverse momenta.\\

We implement the universal $q_T$-resummation formalism proposed in
\cite{Catani:2000vq, Bozzi:2005wk} and compute the
$q_T$-distribution of a slepton pair produced at the LHC by
combining resummation at small $q_T$ and the fixed-order cross
section at large $q_T$. The resummed contribution has been
computed at the next-to-leading logarithmic (NLL) accuracy and the
fixed-order cross section at $\mathcal{O}(\as)$ in perturbative
QCD, corresponding to the production of a slepton pair plus a QCD
jet. It is the first precision calculation of the $q_{T}$-spectrum
for SUSY particle pair production at hadron colliders. The
importance of resummed contributions at small and intermediate
values of $q_{T}$, both enhancing the pure fixed-order result and
reducing the scale uncertainty, is shown in our numerical results
\cite{Bozzi:2006fw}.\\

Concerning NLO SUSY-QCD corrections \cite{Beenakker:1999xh}, they
have only been computed for non-mixing squarks appearing in the
loops. We extend this work by including the mixing effects
relevant for the third generation in the squark sector, and we
consider the threshold-enhanced contributions of the QCD
corrections \cite{Baer:1997nh}, also due to soft-gluon emission
from the initial state. They arise when the initial partons have
just enough energy to produce the slepton pair in the final state.
In this case, the mismatch between virtual corrections and
phase-space suppressed real-gluon emission leads to the appearance
of large logarithmic terms $\alpha_s^n[\ln^{2n-1}(1-z)/(1-z)]_+$
at the $n^{{\rm th}}$ order of perturbation theory, where
$z=M^2/s$, $s$ being the partonic centre-of-mass energy and $M$
the slepton pair invariant-mass. When $s$ is close to $M^{2}$,
these large logarithms have to be resummed to all orders in
$\alpha_{s}$. Although they are manifest in the partonic cross
section, they do not generally result in divergences in the
physical cross section since they are smoothed by the convolution
with the steeply falling parton distributions. Threshold
resummation is then not really a summation of kinematic logarithms
in the physical cross section, but rather an attempt to quantify
the effect of a well-defined set of corrections to all orders,
which can be significant even if the hadronic threshold is far
from being reached. Large corrections are thus expected for the
Drell-Yan like production of a slepton pair with invariant-mass of
a few 100 GeV at the Tevatron and LHC.\\

All-order resummation is achieved through the exponentiation of
the soft-gluon radiation, which does not take place in $z$-space
directly, but in Mellin $N$-space, where $N$ is the Mellin
variable conjugate to $z$ and the threshold region $z\rightarrow
1$ corresponds to the limit $N\rightarrow \infty$. Thus, a final
inverse Mellin transform is needed in order to obtain a resummed
cross section in $z$-space. Threshold resummation for the
Drell-Yan process was first performed in \cite{Sterman:1986aj,
Catani:1989ne} at the leading logarithmic and next-to-leading
logarithmic (NLL) levels, corresponding to terms of the form
$\alpha_{s}^{n}\ln^{2n}N$ and $\alpha_{s}^{n}\ln^{2n-1}N$. The
extension to the NNLL level ($\alpha_{s}^{n}\ln^{2n-2}N$ terms)
has been carried out both for the Drell-Yan process
\cite{Vogt:2000ci} and for Higgs-boson production
\cite{Catani:2003zt}. It was shown in \cite{Kramer:1996iq,
Catani:2001ic} that contributions due to collinear parton emission
can be consistently included in the resummation formula, leading
to a ``collinear-improved'' resummation formalism where
$1/N$-suppressed and a class of $N$-independent universal
contributions are resummed as well. Very recently, even the NNNLL
contributions ($\alpha_{s}^{n}\ln^{2n-3}N$ terms) became available
\cite{Moch:2005ba, Moch:2005ky, Laenen:2005uz}.\\

We present here an extensive study on NLL threshold resummation
effects for Drell-Yan like slepton pair and slepton-sneutrino
associated production in mSUGRA and GMSB scenarios, matching the
resummed contributions computed within a collinear-improved
resummation formalism with a fixed-order calculation at NLO
accuracy. Numerically, we show a non-negligible increase of the
theoretical cross sections with respect to the NLO prediction and
a stabilization of the unphysical scale dependences thanks to the
higher order terms taken into account in the resummed component of
the cross section \cite{Bozzi:2007qr}.\\

The dynamical origin of the enhanced contributions is the same
both in transverse-momentum and threshold resummations, since it
comes from the soft-gluon emission by the initial state. A joint
resummation formalism, embodying soft-gluon contributions in both
the delicate kinematical regions ($q_{T} \ll M$, $M^{2}\sim s$)
simultaneously, has been developed in the last decade
\cite{Li:1998is, Laenen:2000ij}. The exponentiation of the
singular terms in the Mellin and impact-parameter spaces, for
threshold and transverse-momentum resummation respectively, has
been proven, and a consistent method to perform the inverse
transforms in order to avoid the Landau pole and the singularities
of the parton distribution functions has been introduced.
Applications to prompt-photon \cite{Laenen:2000de}, electroweak
boson \cite{Kulesza:2002rh}, Higgs boson \cite{Kulesza:2003wn} and
heavy-quark pair \cite{Banfi:2004xa} production at hadron
colliders show the substantial effects of the joint resummation on
the differential cross sections.\\

We present a joint treatment of the recoil corrections at small
$q_T$ and the threshold-enhanced contributions near partonic
threshold for slepton pair production at hadron colliders,
allowing for a complete understanding of the soft-gluon effects in
differential distributions \cite{Bozzi:2007xx}. Together with the
previous papers on transverse-momentum \cite{Bozzi:2006fw} and
threshold \cite{Bozzi:2007qr} resummation, this completes our
program of providing the first precision calculations including
soft-gluon resummation for slepton pair production at hadron
colliders.\\

If SUSY particles exist, they should also appear in virtual
particle loops and affect low-energy and electroweak precision
observables. In particular, flavour-changing neutral currents
(FCNC), which appear only at the one-loop level even in the SM,
put severe constraints on new physics contributions appearing at
the same perturbative order. The MSSM has passed these crucial
tests, largely due to the assumption of constrained Minimal
Flavour Violation (cMFV) \cite{Ciuchini:1998xy, Buras:2000dm} or
Minimal Flavour Violation (MFV) \cite{Hall:1990ac,
D'Ambrosio:2002ex, Altmannshofer:2007cs}, where heavy SUSY
particles may appear in the loops, but flavour changes are either
neglected or completely dictated by the structure of the Yukawa
couplings and thus the CKM-matrix \cite{Cabibbo:1963yz,
Kobayashi:1973fv}.\\

In SUSY with MFV, the flavour violating entries in the squark mass
matrices stem from the trilinear Yukawa couplings of the fermion
and Higgs supermultiplets and the resulting different
renormalizations of the quark and squark mass matrices, which
induce additional flavour violation at the weak scale through
renormalization group running \cite{Donoghue:1983mx,
Duncan:1983iq, Bouquet:1984pp, Borzumati:1986qx}. In non-minimal
flavour violating SUSY, additional sources of flavour violation
are included in the mass matrices at the weak scale, and their
flavour-violating off-diagonal terms cannot be simply deduced from
the CKM matrix alone, and have to be considered as free
parameters. In this work, we consider flavour mixings of second-
and third-generation squarks, since direct searches of flavour
violation depend on the possibility of flavour tagging, which is
established experimentally only for heavy flavours. In addition,
stringent experimental constraints for the first-generation are
imposed by precise measurements of $K^0-\bar{K}^0$ mixing and
first evidence of $D^0-\bar{D}^0$ mixing \cite{Hagelin:1992tc,
Gabbiani:1996hi, Ciuchini:2007}.\\

Considering SUSY with cMFV (the usual MSSM), NLO SUSY-QCD
calculations for the production of squarks and gluinos
\cite{Beenakker:1996ch}, gauginos \cite{Beenakker:1999xh}, as well
as for their associated production \cite{Berger:1999mc} are
available. Due to their strong coupling, squarks should be
abundantly produced at hadron colliders. In addition, phase space
favours the production of the lighter of the squark mass
eigenstates of identical flavour. As a consequence, the production
of top \cite{Beenakker:1997ut} and bottom \cite{Berger:2000mp}
squarks with large helicity mixing has received particular
attention. In this thesis, we investigate the importance of
electroweak channels for non-diagonal and mixed squark pair
production at hadron colliders \cite{Bozzi:2005sy}. Naively, one
expects these cross sections, which are of ${\cal O}(\alpha^2)$ in
the fine structure constant $\alpha$, to be smaller than the
diagonal strong channels by about two orders of magnitude. For
non-diagonal squark production, the interplay between loop
suppression in QCD and coupling suppression in the electroweak
case merits a detailed investigation, and in the presence of the
mixing of bottom squarks, their loop contributions must also be
taken into
account.\\

Then, for the first time, we concentrate on the possible effects
of non-minimal flavour violation (NMFV) at hadron colliders
\cite{Bozzi:2007me}. To this end, we recalculate all squark and
gaugino production and decay helicity amplitudes, keeping at the
same time the CKM-matrix and the quark masses to account for
non-diagonal charged-current gaugino and Higgsino Yukawa
interactions, and generalizing the two-dimensional helicity mixing
matrices, often assumed to be real, to generally complex
six-dimensional helicity and generational mixing matrices.\\

In our phenomenological analysis of NMFV squark and gaugino
production, we concentrate on the LHC due to its larger
centre-of-mass energy and luminosity. We pay particular attention
to the interesting interplay of parton density functions (PDFs),
which are dominated by light quarks, strong gluino contributions,
which are generally larger than electroweak contributions and need
not be flavour-diagonal, and the appearance of third-generation
squarks in the final state, which are easily identified
experimentally and generally lighter than first- and
second-generation squarks.\\

This thesis is organized as follows. In the first part of Chapt.\
\ref{ch:MSSM}, we briefly describe the MSSM within cMFV, showing
several motivating arguments for SUSY and defining the model. In
the second part of this chapter, we set up the notations that we
use in the case of NMFV SUSY, and we perform a precise numerical
analysis of the experimentally allowed NMFV SUSY parameter space
with respect to low-energy constraints, leading to the definition
of four collider-friendly benchmark points for which we
investigate the corresponding helicity and flavour decomposition
of the up- and down-type squarks. Finally, we introduce
generalized couplings in order to compute compact analytical
expression for the various cross sections calculated in this work.
In Chapt.\ \ref{ch:res}, we describe the two $q_T$-resummation
formalisms (CSS and universal), the threshold-resummation
formalism and the joint-resummation formalism that we have used in
the case of slepton pair hadroproduction. Chapts. \ref{ch:slep}
and \ref{ch:NMFV} are devoted to the results, the first one for
slepton pair hadroproduction and the second one to squark and
gaugino production and decays, and we show a large number of
analytical and numerical results. Our conclusion and outlook are
presented in Chapt.\ \ref{ch:concl}.

\chapter{The Minimal Supersymmetric Standard Model}
\label{ch:MSSM}

\section{Motivation}

The Standard Model of particle physics is a gauge field theory
based on the symmetry group $SU(3) \times SU(2) \times U(1)$,
containing the electroweak \cite{Glashow:1961tr, Weinberg:1967tq,
Glashow:1970gm, Weinberg:1971nd} and the strong
\cite{Gross:1973ju, Gross:1974cs, Politzer:1974fr} interactions
and providing a remarkably accurate description of a large class
of phenomena. It is well-established by the discovery of all its
particle content, the Higgs boson excepted, and by precision
measurements at colliders. However, a new framework will certainly
be required, at least at the Planck scale $M_P = 2.8 \times
10^{18}$ GeV, where the quantum gravitational effects become
important, but more probably at a lower scale, since the absence
of new physics between the current experimental limit of several
hundreds of GeV and $M_P$ is highly improbable. Moreover, this
large gap leads to what is called the hierarchy problem
\cite{Witten:1981nf, Kaul:1981wp}.\\

The origin of the masses in the SM is an isodoublet scalar Higgs
field \cite{Goldstone:1961eq, Goldstone:1962es, Higgs:1964ia,
Englert:1964et, Higgs:1964pj, Higgs:1966ev}, yielding electroweak
symmetry breaking and one physical Higgs boson which couples to
each SM fermion of mass $m_f$ with a strength driven by the Yukawa
couplings $\lambda_f$. The quantum corrections $\Delta m_H^2$ to
the Higgs squared mass $m_H^2$, where \bea m_H^2 = (m_H^2)_0 +
\Delta m_H^2,~\eea $(m_H)_0$ being a parameter of the fundamental
theory, are given by \bea \Delta m_H^2
=\frac{\left|\lambda_f\right|^2}{16\, \pi^2}\, \left[-2\,
\Lambda^2 + 6\, m_f^2\, \ln\frac{\Lambda}{m_f} + \ldots
\right],~\eea where $\Lambda$ is an ultraviolet cutoff that
corresponds to the scale at which new physics alters the theory
and that regulates fermionic loop-integrals. If $\Lambda$ is of
order $M_P$, these corrections are some 30 orders of magnitude
larger than the expected value of about (100 GeV)$^2$,
particularly due to the large top Yukawa coupling. Supersymmetry
\cite{Wess:1974tw, Fayet:1976cr, Fayet:1977yc, Farrar:1978xj,
Fayet:1979sa, Nilles:1983ge, Haber:1984rc} provides an elegant
solution to this problem, since a heavy complex scalar particle of
mass $m_S$ can couple to the Higgs boson with a strength
$\lambda_S$. The corresponding quantum corrections to $m_H^2$ are
\bea \Delta m_H^2 =\frac{\lambda_S}{16\, \pi^2}\, \left[\Lambda^2
- 2\, m_S^2\, \ln\frac{\Lambda}{m_S} + \ldots\right].~\eea
Provided that the scalar masses are not too heavy, the systematic
cancellation of fermionic and bosonic corrections to the Higgs
squared mass is then achieved, since each fermion of the SM is now
accompanied by two scalars, and their couplings to the Higgs field
are closely related by $\lambda_S = |\lambda_f|^2$. The remaining
corrections to the squared Higgs mass depend only logarithmically
on the cutoff $\Lambda$ and are thus under control.\\

Furthermore, a set of conceptual problems of the SM can be solved
thanks to supersymmetry, such as the unification of the
fundamental gauge interactions \cite{Ibanez:1981yh,
Dimopoulos:1981yj, Amaldi:1991cn, Carena:1993ag}, since the SUSY
particles modify the renormalization-group evolution of the gauge
couplings with the energy, leading to unification at about
$10^{16}$ GeV. SUSY also provides a potential cold dark matter
candidate, the lightest SUSY particle (LSP) \cite{Goldberg:1983nd,
Ellis:1983ew}, and can even include gravity, in the framework of
local supergravity theories \cite{Freedman:1976xh, Deser:1976eh}.

\section{Definition of the model}

Most present SUSY models are based on the four-dimensional
supersymmetric field theory of Wess and Zumino \cite{Wess:1974tw},
which are free of many of the divergences encountered in similar
SUSY theories of that time \cite{Iliopoulos:1974zv,
Ferrara:1974fv}. The simplest model is the straightforward
supersymmetrization of the SM with the same gauge interactions,
called the Minimal Supersymmetric Standard Model
\cite{Nilles:1983ge, Haber:1984rc}. As for any SUSY model, it
postulates a symmetry between fermionic and bosonic degrees of
freedom in nature, predicting thus the existence of a fermionic
(bosonic) SUSY partner for each bosonic (fermionic) SM particle. A
complete introduction to supersymmetric field theories and the
MSSM can be found in Refs.\ \cite{Martin:1997ns, Ellis:1998eh,
Aitchison:2005cf}.

\subsection{Field content}

Due to the various conserved quantum numbers of the known bosons
and fermions, a minimal supersymmetric model cannot be built up
with the SM particles alone, and new particles have to be
postulated. Quarks and leptons get scalar partners called squarks
and sleptons, while electroweak bosons and gluons get fermionic
partners referred to as gauginos and gluinos. Since in the SM, the
left- and right-handed parts of the fermionic fields transform
differently under the gauge group, it is required that fermions
get two superpartners, named left- and right-handed sfermions.
Finally, to preserve the electroweak symmetry from gauge anomaly
and to give masses to both up- and down-type fermions, the MSSM
requires two Higgs doublets and their fermionic superpartners, the
Higgsinos. The field content of the MSSM is shown in Tab.\
\ref{tab:1}. Let us note that the local version of supersymmetry
includes a spin-2 state and a spin-3/2 state which can be
interpreted as the spin-2 graviton and its spin-3/2 superpartner,
the gravitino.\\

\renewcommand{\arraystretch}{1.4}
\begin{table}\centering
\caption{\label{tab:1}Field content of the MSSM.}\vspace{.2cm}
\begin{tabular}{|c|c|c c|c c|}
\hline \multicolumn{2}{|c|}{Names} & particle & spin &
superpartner & spin \\  \hline\hline (s)quarks & $Q$ &
$(u_L\,d_L)$ & 1/2 & $(\tilde{u}_L \, \tilde{d}_L )$& 0 \\
($\times 3$ families) & $\overline{u}$ & $u^\dagger_R$ & 1/2 &
$\tilde{u}^\ast_R$ & 0 \\ & $\overline{d}$ & $d^\dagger_R$ & 1/2 &
${\tilde d}^\ast_R$ & 0\\ \hline (s)leptons & $L$ &
$(\nu\,e_L)$ &1/2 & $(\tilde{\nu}\, \tilde{e}_L )$ & 0 \\
($\times 3$ families) & $\overline e$ & $e^\dagger_R$ & 1/2 &
$\tilde{e}^\ast_R$ & 0 \\  \hline Higgs(inos) &$H_u$ &$(H_u^+\,
H_u^0 )$& 0 & $(\tilde{H}_u^+ \, \tilde{H}_u^0)$& 1/2 \\
&$H_d$ & $(H_d^0 \, H_d^-)$ & 0 & $(\tilde{H}_d^0 \,
\tilde{H}_d^-)$& 1/2\\  \hline \multicolumn{2}{|c|}{gluon/gluino}
& $g$ & 1 & $\tilde{g}$ & 1/2\\ \hline \multicolumn{2}{|c|}{$W$
bosons/winos} & $W^\pm$, $W^0$ & 1 & $\tilde{W}^\pm$,
$\tilde{W}^0$ & 1/2\\ \hline \multicolumn{2}{|c|}{$B$ boson /
bino} & $B$& 1 & $\tilde{B}$ & 1/2\\ \hline \end{tabular}
\end{table}

All of these particles are organised in chiral and gauge
supermultiplets, containing an equal number of fermionic and
bosonic degrees of freedom. Chiral supermultiplets contain one
on-shell Weyl fermion $\psi$ (i.e.\ the left- or the right-handed
part of a fermionic field) and its associated scalar complex field
$\phi$, corresponding to a total of two fermionic and two bosonic
real degrees of freedom. Off-shell Weyl fermions having two
additional real fermionic degrees of freedom, an auxiliary complex
scalar field $F$ is introduced, preserving supersymmetry
off-shell, but being eliminated when one goes on-shell by imposing
its equations of motion. Gauge supermultiplets contain a massless
gauge boson $A^a_\mu$ and its on-shell associated fermionic
partner $\lambda^a$, which corresponds as well to two fermionic
and two bosonic real degrees of freedom. If one goes off-shell,
the fermionic field gets two additional real degrees of freedom,
while the vector boson only gets one. As for chiral
supermultiplets, an auxiliary field $D$ with one real bosonic
degree of freedom is introduced, preserving SUSY off-shell and
being eliminated on-shell through its equations of motion.

\subsection{Lagrangian density}

The MSSM gauge interactions are the same as those of the SM and
are determined by the gauge group $SU(3) \times SU(2) \times
U(1)$. The full Lagrangian density for a renormalizable SUSY
theory is then given by \bea \label{eq:SUSYlag} \mathcal{L} =
\mathcal{L}_{\rm chiral} + \mathcal{L}_{\rm gauge} - \sqrt{2} g
(\phi^\ast T^a \psi)\lambda^a - \sqrt{2} g \lambda^{\dagger a}
(\psi^\dagger T^a \phi) + g (\phi^\ast T^a \phi) D^a.~ \eea
$\mathcal{L}_{\rm chiral}$ and $\mathcal{L}_{\rm gauge}$ contain
the kinetic terms and the gauge interactions for the chiral and
gauge supermultiplets, \bea \mathcal{L}_{\rm chiral} &=& - \D^\mu
\phi^{\ast i} \D_\mu \phi_i - i \psi^{\dagger i}
\overline{\sigma}^\mu \D_\mu \psi_i -\frac{1}{2} \left( W^{ij}
\psi_i \psi_j  + W^\ast_{ij}
\psi^{\dagger i} \psi^{\dagger j} \right) - W^i W^\ast_i,~~~~~\\
\mathcal{L}_{\rm gauge} &=& -{1\over 4} F_{\mu\nu}^a F^{\mu\nu a}
- i \lambda^{\dagger a} \overline{\sigma}^\mu \D_\mu \lambda^a +
{1\over 2} D^a D^a,~\label{eq:lgauge} \eea where $\sigma$ are the
Pauli matrices \bea \overline{\sigma}_0 \!=\! \sigma_0 \!=\!
\begin{pmatrix}1&0\\ 0&1\\ \end{pmatrix}&,&
\overline{\sigma}_1 \!=\! -\sigma_1 \!=\! \begin{pmatrix} 0&1\\
1&0\\ \end{pmatrix},\nn\\ \overline{\sigma}_2 \!=\! -\sigma_2 \!=\!
\begin{pmatrix} 0&-i\\ i&0\\ \end{pmatrix}&,&
\overline{\sigma}_3 \!=\! -\sigma_3 \!=\! \begin{pmatrix} 1&0\\
0&-1\\\end{pmatrix}.~\eea The last terms of Eq.\
(\ref{eq:SUSYlag}) are interactions whose strength is given by the
usual gauge couplings, but which are not gauge interactions from
the point of view of an ordinary gauge theory. The covariant
derivatives \bea \D_\mu \phi_i = \partial_\mu \phi_i - i g A^a_\mu
(T^a\phi)_i, &~~~& \D_\mu \phi^{\ast i} = \partial_\mu \phi^{\ast
i} + i g A^a_\mu (\phi^\ast T^a)^i,~ \nn \\ \D_\mu \psi_i =
\partial_\mu \psi_i - i g A^a_\mu (T^a\psi)_i,&~~~& \D_\mu \lambda^a
= \partial_\mu \lambda^a + g f^{abc} A^b_\mu \lambda^c \eea make
the Lagrangian gauge-invariant, $T^a$ being hermitian matrices
corresponding to the representation in which the chiral
supermultiplets transform under the gauge group and satisfying
$[T^a,T^b] =i f^{abc} T^c$, where $f^{abc}$ are the totally
antisymmetric structure constants defining the group. Finally, in
Eq.\ (\ref{eq:lgauge}), $F_{\mu\nu}^a$ is the usual Yang-Mills
field strength \bea F^a_{\mu\nu} = \partial_\mu A^a_\nu -
\partial_\nu A^a_\mu+ g f^{abc} A^b_\mu A^c_\nu .~\eea The
interactions between the chiral and the gauge supermultiplets are
embodied in derivatives of the superpotential $W$ \bea W^i =
{\partial W\over \partial \phi_i} {\rm ~~and~~} W^{ij} =
{\partial^2 W \over \partial \phi_i \partial \phi_j},~ \eea $W$
being analytic in the complex fields $\phi$. Let us note that the
auxiliary fields $F_i$ do not explicitly appear in the Lagrangian
since they are expressed in terms of $W_i$, using the equations of
motion $F_i = -W_i^\ast$ and $F^{i\ast} = -W^i$. This Lagrangian
is obviously invariant under a global infinitesimal SUSY
transformation $\epsilon$, the transformation rules being \bea
\delta \phi_i = \epsilon\psi_i, && \delta \psi_{i\alpha} = i
(\sigma^\mu \epsilon^\dagger)_{\alpha} \D_\mu \phi_i +
\epsilon_\alpha F_i,~ \nn  \\ \delta A_\mu^a = {1\over \sqrt{2}}
\left(\epsilon^\dagger \overline{\sigma}_\mu \lambda^a +
\lambda^{\dagger a} \overline{\sigma}_\mu \epsilon \right), &&
\delta \lambda^a_\alpha = {i\over 2\sqrt{2}} (\sigma^\mu
\overline{\sigma}^\nu \epsilon)_\alpha F^a_{\mu\nu} + {1\over
\sqrt{2}} \epsilon_\alpha D^a ,~\nn  \\ \delta F_i = i
\epsilon^\dagger \overline{\sigma}^\mu \D_\mu \psi_i + \sqrt{2} g
(T^a \phi)_i \epsilon^\dagger \lambda^{\dagger a}, &&\delta D^a
= {i\over \sqrt{2}} \left( \epsilon^\dagger \overline{\sigma}^\mu
D_\mu \lambda^a - D_\mu \lambda^{\dagger a} \overline{\sigma}^\mu
\epsilon \right).~~~~~~~~ \eea

The MSSM is specified by the choice of its superpotential \bea W =
- \overline{e}\, {\bf y_e}\, L\, H_d - \overline{u}\, {\bf y_u}\,
Q\, H_u  - \overline{d}\, {\bf y_d}\, Q\, H_d + \mu\, H_u\, H_d,~
\label{eq:superpot}\eea where $\overline{e}$, $\overline{u}$,
$\overline{d}$, $L$, $Q$, $H_u$ and $H_d$ are the chiral and Higgs
superfields described in previous subsection. The $3 \times 3$
Yukawa matrices ${\bf y}$ give rise to the masses of the quarks
and leptons when the Higgs fields acquire their vacuum expectation
values (vevs). As in the SM, different rotations are needed to
diagonalize both the up- and down-type quark Yukawa matrices ${\bf
y_u}$ and ${\bf y_d}$, leading to the usual flavour mixing driven
by the CKM matrix \cite{Cabibbo:1963yz, Kobayashi:1973fv}.
Finally, the $\mu$-term provides the Higgs and Higgsino
squared-mass terms. This superpotential is minimal since it is
sufficient to produce a phenomenologically viable model. However,
other gauge-invariant terms could be included in the
superpotential, \bea W_{\not R} = \frac{1}{2} \lambda L\, L\,
\overline{e} + \lambda^\prime L\, Q\, \overline{d} +
\frac{1}{2}\lambda^{\prime\prime} \overline{u}\, \overline{d}\,
\overline{d} + \mu^\prime\, L\, H_u,~ \label{eq:Rppot}\eea
violating either the total lepton number $L$, or the baryon number
$B$. In principle, we could just postulate $B$ and $L$
conservation, forbidding then the terms of Eq.\ (\ref{eq:Rppot}),
as there are no possible renormalizable terms in the SM Lagrangian
violating $B$ or $L$. But neither $B$ nor $L$ are fundamental
symmetries of nature since they are violated by non-perturbative
electroweak effects \cite{'tHooft:1976up}. Therefore, an
alternative symmetry is rather imposed, forbidding the $B$- and
$L$-violating terms of Eq.\ (\ref{eq:Rppot}), the $R$-parity
\cite{Farrar:1978xj}. It is defined by \bea R =
(-1)^{3B+L+2S},~\eea $S$ being the spin of the particle. The SM
particles then have a positive $R$-parity, while their SUSY
counterparts have a negative one. Due to $R$-parity conservation,
interaction vertices have to contain an even number of SUSY
particles, leading to SUSY particle production by pairs at
colliders and to decays into states containing an odd number of
stable LSPs, which can only interact via annihilation vertices.
Furthermore, if we assume the LSP to be electrically and colour
neutral, it can even be a potential dark matter candidate
\cite{Ellis:1983ew}. In this work, we assume $R$-parity
conservation, but an introduction to SUSY models with $R$-parity
violation can be found in Refs.\ \cite{Dreiner:1997uz,
Barbier:2004ez}.\\

The scalar potential is already included in the Lagrangian, via
the $F$- and $D$-terms, which can be expressed as a function of
the scalar fields, \bea V(\phi, \phi^\ast)=F^{i\ast}\,F_i +
{1\over 2} D^a D^a = W^i\,W_i^\ast + \frac{1}{2} \sum_a g_a^2
(\phi^\ast T^a \phi)^2,~\label{eq:scalpot}\eea where we sum over
the different gauge groups. We should note that $V$ is entirely
determined by the other interactions of the theory, contrary to
the SM potential containing free parameters.

\subsection{Soft SUSY-breaking}

SUSY particles still remain to be discovered, and their masses
must therefore be considerably larger than those of the
corresponding SM particles, so that supersymmetry must be broken.
In order to remain a viable solution to the hierarchy problem,
SUSY can, however, only be broken via soft terms in the
Lagrangian, i.e.\ with positive mass dimension, which prevents us
from introducing new quadratic divergences in the quantum
corrections to the Higgs squared mass. Since we do not know the
SUSY-breaking mechanism and at which scale it occurs, we usually
modify the Lagrangian at low energies by adding all possible terms
breaking explicitly the SUSY. In the MSSM, we get
\cite{Girardello:1981wz} \bea \mathcal{L}_{\rm soft} &=&
-\frac{1}{2} \left( M_3\, \tilde{g}\, \tilde{g} + M_2\,
\tilde{W}\, \tilde{W} + M_1\, \tilde{B}\,
\tilde{B} + {\rm c.c.} \right) \nonumber \\
&-& \left( \tilde{Q}^\dagger \, {\bf m^2_Q}\, \tilde{Q} +
\tilde{L}^\dagger \,{\bf m^2_L}\, \tilde{L} + \tilde{\overline{u}}
\,{\bf m^2_{\overline{u}}}\, \tilde{\overline{u}}^\dagger
-\tilde{\overline{d}} \,{\bf m^2_{\overline{d}}} \,
\tilde{\overline{d}}^\dagger + \tilde{\overline{e}} \,{\bf
m^2_{\overline{e}}}\, \tilde{\overline{e}}^\dagger \right) \nn
\\ &-& \left(m_{H_u}^2\, H_u^\ast\, H_u + m_{H_d}^2\, H_d^\ast\,
H_d + \left( b\, H_u\, H_d\, + {\rm c.c.} \right)\right)\nn \\ &-&
\left( \tilde{\overline{u}} \,{\bf a_u}\, \tilde{Q}\, H_u -
\tilde{\overline{d}} \,{\bf a_d}\, \tilde{Q}\, H_d -
\tilde{\overline{e}} \,{\bf a_e}\, \tilde{L}\, H_d + {\rm c.c.}
\right).~\label{eq:lsoft}\eea The first line contains the gluino,
wino and bino mass terms and the second line the squark and
slepton mass terms, ${\bf m^2_Q}$, ${\bf m^2_L}$, ${\bf
m^2_{\overline{u}}}$, ${\bf m^2_{\overline{d}}}$, ${\bf
m^2_{\overline{e}}}$ being $3\times3$ hermitian matrices in family
space. In the third line of $\mathcal{L}_{\rm soft}$, we have the
mass terms for the Higgs fields contributing to the scalar
potential and in the fourth line the trilinear scalar
interactions, ${\bf a_u}$, ${\bf a_d}$, and ${\bf a_e}$, which are
also $3\times 3$ matrices in generation space. Contrary to the
supersymmetric part of the MSSM Lagrangian, which has only one new
free parameter $\mu$, the SUSY-breaking Lagrangian contain 105
masses, phases and mixing angles, which cannot be rotated away by
redefining the field basis and which have no counterpart in the SM
\cite{Dimopoulos:1995ju}. Most of them are strongly constrained,
since they introduce new sources of flavour mixing and CP
violation, which could enhance processes severely restricted by
experiment, such as $K^0-\bar{K}^0$, $D^0-\bar{D}^0$ and
$B^0-\bar{B}^0$ mixing, flavour-changing neutral-current (FCNC)
$B$, $\mu$ or $\tau$ decays, and so forth
\cite{Masiero:1997bv}.\\

Present models assume that SUSY is broken in a hidden sector,
containing particles that have no or small coupling to the visible
sector, and SUSY-breaking is mediated to the visible sector via an
interaction shared by the two sectors. Let us note that if the
mediating interaction is flavour-blind and the related parameters
are real, we get automatically conditions which evade the flavour
and CP violating terms in $\mathcal{L}_{\rm soft}$ at low
energies. In local SUSY theories, the Goldstone fermion related to
SUSY breaking, the goldstino, is absorbed by the gravitino which
acquires a mass through the super-Higgs mechanism
\cite{Cremmer:1978iv, Cremmer:1978hn}, analogously to the usual
Higgs mechanism where the electroweak gauge bosons acquire a mass
by absorbing the Goldstone bosons associated to electroweak
symmetry breaking. We have studied two SUSY-breaking scenarios,
supergravity \cite{Chamseddine:1982jx, Barbieri:1982eh,
Hall:1983iz} and gauge-mediated SUSY-breaking \cite{Dine:1993yw,
Dine:1994vc, Dine:1995ag, Giudice:1998bp}.\\

In the framework of supergravity, SUSY-breaking is mediated to the
MSSM through gravitational interactions, appearing in the
Lagrangian via non-renormalizable terms suppressed by powers of
the Planck mass. Assuming minimal supergravity, the soft terms in
$\mathcal{L}_{\rm soft}$ are completely determined by five
parameters, the universal scalar and gaugino masses  $m_0$ and
$m_{1/2}$, the universal trilinear coupling $A_0$, the ratio of
the vevs of the two neutral Higgs fields, $\tan\beta$, and the
sign of $\mu$. At the SUSY-breaking scale, we get the relations
\bea &M_3 = M_2 = M_1 = m_{1/2},~&\\& {\bf m^2_Q} = {\bf
m^2_{{\overline{u}}}} = {\bf m^2_{{\overline{d}}}} = {\bf m^2_L} =
{\bf m^2_{{\overline{e}}}} = m_0^2\, {\bf 1};~~~ m_{H_u}^2 =
m^2_{H_d} = m_0^2,~&\\& {\bf a_u} = A_0 {\bf y_u},~~ {\bf a_d} =
A_0 {\bf y_d},~~ {\bf a_e} = A_0 {\bf y_e}.~&\eea Assuming that
all of these parameters are real, the problematic flavour and CP
violation terms of the Lagrangian are automatically suppressed.
Low-energy parameters are deduced from renormalization-group
evolution of the high scale parameters down to the electroweak
scale.\\

SUSY-breaking can also be mediated by usual gauge interactions. To
this aim, new chiral supermultiplets are introduced in the
Lagrangian, the messenger fields, which carry $SU(3) \times SU(2)
\times U(1)$ quantum numbers and couple to the particles of both
the visible and hidden sectors. Virtual loops of messengers
generate gaugino and sfermion masses, i.e.\ the mass terms in
$\mathcal{L}_{\rm soft}$, in a completely renormalizable
framework. At the messenger scale $M_{{\rm mes}}$, the trilinear
couplings are generated via two-loop diagrams, and we can then
neglect them with respect to the masses generated by one-loop
diagrams, \bea& {\bf a_u} = {\bf a_d} = {\bf a_e} = 0.~&\eea Gauge
interactions being flavour-blind and assuming real SUSY-breaking
parameters, undesired large FCNC and CP violation effects are
again avoided. SUSY-breaking is parameterized by one gauge-singlet
chiral superfield, whose scalar and auxiliary components acquire
the two vevs $\langle S \rangle$ and $\langle F_S \rangle$,
respectively, defining the scale $\Lambda = \langle
F_S\rangle/\langle S\rangle$. At the scale $M_{{\rm mes}}$, the
soft SUSY-breaking mass parameters are given by \bea M_i(M_{{\rm
mes}}) &=& \frac{\alpha_i(M_{{\rm mes}})}{4 \pi} \Lambda \,
g\bigg( \frac{\Lambda}{M_{{\rm mes}}} \bigg) (N_5 + 3\, N_{10})\\
m_{\tilde{j}}^2(M_{{\rm mes}}) & =& 2\, (N_5 + 3\, N_{10})\,
\Lambda^2\, f\bigg(\frac{\Lambda}{M_{{\rm mes}}} \bigg) \sum_i
\bigg[ \frac{\alpha_i(M_{{\rm mes}})}{4 \pi} \bigg]^2 C_j^i,~\eea
where the coefficients $C_j^i$ are quadratic Casimir invariants
and $N_i$ is the multiplicity of the messengers in the
$5+\overline{5}$ and $10+\overline{10}$ vector-like
supermultiplets, assuming that messengers come in complete
multiplets of $SU(5)$ to preserve gauge coupling unification. The
threshold functions are given by \cite{Martin:1996zb,
Dimopoulos:1996gy} \bea g(x) &=& \frac{1}{x^2} [(1+x)\log
(1+x)+(1-x) \log (1-x)],~
\\ f(x) &=& \frac{1+x}{x^2} \bigg[\log (1+x) -2 {\rm Li}_2 \bigg(
\frac{x}{1+x} \bigg) + \frac{1}{2} {\rm Li}_2 \bigg(
\frac{2x}{1+x} \bigg) \bigg] + (x \leftrightarrow -x).~~~~~ \eea The
different terms of $\mathcal{L}_{\rm soft}$ are then determined
via the evolution of six parameters, the numbers of (s)quark and
(s)lepton messenger fields $n_{\hat{q}}$ and $n_{\hat{l}}$ needed
for the calculation of the multiplicities, the messenger scale
$M_{{\rm mes}}$, $\Lambda$, $\tan\beta$ and the sign of $\mu$.

\subsection{Electroweak symmetry breaking and particle mixing}

The scalar potential of the MSSM is given by \bea V &=& (|\mu|^2 +
m^2_{H_u}) |H_u|^2 + (|\mu|^2 + m^2_{H_d}) |H_d|^2 + b\, (H_u H_d
+ {\rm c.c.}) \nn\\ &+& {1\over 8} (g^2 + g^{\prime 2}) (|H_u|^2 -
|H_d|^2)^2 + {1\over 2} g^2 |H_u^+ H_d^{0\ast} + H_u^0
H_d^{-\ast}|^2.~ \eea The terms proportional to $\mu$, $g$ and
$g^\prime$ come from the $F$- and $D$-terms of Eq.\
(\ref{eq:scalpot}), while the terms proportional to $m_{H_u}$,
$m_{H_d}$ and $b$ come from the SUSY-breaking Lagrangian of Eq.\
(\ref{eq:lsoft}). $SU(2)$ gauge transformations allow to rotate
away any possible vev of one of the charged Higgs fields, and we
simply take $\langle H_u^+\rangle = 0$. Since the minimum of the
potential satisfies $\partial V/\partial H_u^+ = 0$, we
automatically get $\langle H_d^-\rangle = 0$, which leads to
electric charge conservation in the Higgs sector. As for $U(1)$
gauge transformations, they allow to redefine the phases of $H_u$
and $H_d$, so that all complex phases of the Lagrangian can be
absorbed, making the vevs real and positive. CP is thus not
spontaneously broken by the Higgs scalar potential, and the Higgs
physical states are also CP eigenstates.\\

To get electroweak symmetry breaking, the potential has to have a
minimum, which is the case if \bea (|\mu|^2 + m^2_{H_u} ) +
(|\mu|^2 + m^2_{H_d}) > 2\,b &{\rm~and~}& (|\mu|^2 + m^2_{H_u})
(|\mu|^2 + m^2_{H_d}) < b^2.~~~~~\eea These two conditions can be
satisfied if $m_{H_u}^2 \neq m_{H_d}^2$, implying that electroweak
symmetry breaking is not possible without SUSY-breaking, since in
unbroken SUSY, the two Higgs mass terms do not exist and the
masses are then equal to zero. Imposing the stationary conditions
$\partial V/\partial H_u^0 =
\partial V/\partial H_d^0 = 0$ leads to the relations \bea
|\mu|^2 + m_{H_u}^2  -b\cot\beta - \frac{m_Z^2}{2}\, \cos(2\beta)
\!=\! |\mu|^2 + m_{H_d}^2 -b\tan\beta + \frac{m_Z^2}{2}\, \cos(
2\beta) \!=\! 0.~~\eea

The particle masses are calculated by expanding the potential
about the minimum, once the Higgs fields get their vevs, $v_u$ and
$v_d$, leading to various bilinear terms with fields with the same
quantum numbers, contributing to the off-diagonal terms in the
different mass matrices. Let us first note that the electroweak
gauge bosons mix as in the SM. Among the eight degrees of freedom
of the two Higgs doublets, three of them are the Nambu-Goldstone
bosons $G^{\{0,\pm\}}$, remaining massless after electroweak
symmetry breaking and being absorbed by the electroweak gauge
bosons which become massive. The five remaining ones represent two
CP even (the light $h_0$ and the heavy $H_0$), one CP odd ($A_0$)
and two charged Higgs bosons ($H^\pm$) \cite{Haber:1984rc,
Gunion:1984yn}\bea
\begin{pmatrix}G^0\\ A^0\\
\end{pmatrix}\! &=& \!\sqrt{2} \begin{pmatrix}\sin\beta &
-\cos\beta \\ \cos\beta & \sin\beta \\
\end{pmatrix}\!\! \begin{pmatrix}{\rm Im}(H_u^0)\\
{\rm Im}(H_d^0)\\\end{pmatrix},\\ \begin{pmatrix}G^+\\
H^+\\ \end{pmatrix} \!&=&\! \begin{pmatrix}\sin\beta & -\cos\beta \\
\cos\beta & \sin\beta \\ \end{pmatrix}\!\! \begin{pmatrix}H_u^+\\
H_d^{-\ast}\\\end{pmatrix},~~~~\\ \begin{pmatrix}h^0\\
H^0\\ \end{pmatrix} &=& \sqrt{2}
\begin{pmatrix}\cos\alpha & \sin\alpha \\ -\sin\alpha & \cos\alpha
\\ \end{pmatrix}\, \begin{pmatrix}{\rm Re}(H_u^0)- v_u\\
{\rm Re}(H_d^0)-v_d\\ \end{pmatrix},~ \eea where the mixing angle
$\alpha$ and the tree-level masses are given by \bea \tan(2\alpha)
&=& \tan(2\beta) \left(m_A^2 + m_Z^2 \right)/\left(m_A^2 - m_Z^2
\right),~\\ m_A^2 &=& 2\,b\,/ \sin (2\beta),~\\ m_{H^\pm}^2 &=&
m_A^2 + m_W^2,~\\m_{h^0, H^0}^2 &=& 1/2 \left(m_A^2 + m_Z^2 \mp
\sqrt{(m_A^2 + m_Z^2)^2 - 4\,m_A^2\, m_Z^2\,
\cos(2\beta)}\right),\eea $m_Z$ and $m_W$ being the $Z$ and $W$
boson masses. The negatively charged Goldstone and Higgs bosons
are defined by $G^- \equiv (G^+)^\ast$ and $H^- \equiv
(H^+)^\ast$. Let us note that the upper limit on the lightest
neutral Higgs $h^0$ mass \bea m_{h^0} \leq m_Z |\cos(2\beta)|\eea
has already been exceeded by the current experimental lower bound
of 114.4 GeV \cite{Barate:2003sz}. However, the tree-level Higgs
mass formulas above receive significant one-loop corrections, the
upper limit for $m_{h^0}$ being then shifted to about 140 GeV
\cite{Okada:1990vk, Ellis:1990nz, Haber:1990aw, Barbieri:1990ja,
Okada:1990gg, Ellis:1991zd}. \\

In the most general case, sfermion mass eigenstates are obtained
by diagonalizing $6\times 6$ mass matrices, except for sneutrinos
with their $3\times 3$ mass matrix, since there exist only
left-handed sneutrinos. In cMFV SUSY, the mixing between different
generations, which could enhance strongly constrained FCNC
processes \cite{Gabbiani:1996hi, Ellis:1981ts, Bartl:2005yy}, is
neglected, and the $6\times 6$ matrices are decomposed into
several $2 \times 2$ matrices for squarks and charged sleptons,
describing the mixing of scalars of a specific flavour
\cite{Haber:1984rc, Ellis:1983ed} \bea {\cal M}^2 &=&
\left(\begin{array}{cc} m_{LL}^2 & m_f\,m_{LR} \\
m_f\,m_{RL}& m_{RR}^2 \end{array}\right),~\label{eq:sfermmass}\eea
with \bea m_{LL}^2&=&m_{\tilde{F}}^2 +
(T_f^3-e_f\,\sin^2\theta_W)\,m_Z^2\,\cos2\beta +
m_f^2,\label{eq:mll}\\ m_{RR}^2&=&m_{\tilde{F}^\prime}^2 +
e_f\,\sin^2\theta_W\,m_Z^2\,\cos2\beta + m_f^2,\label{eq:mrr}\\
m_{LR}&=&m_{RL}^\ast = A_f^\ast-\mu\left\{
\begin{array}{l}\cot\beta\hspace*{3.mm}{\rm
for~up-type~sfermions.} \\ \tan\beta\hspace*{2.8mm}{\rm
for~down-type~sfermions.} \end{array}\right.\hspace*{3mm}
\label{eq:mlr}\eea The soft SUSY-breaking mass terms for left- and
right-handed sfermions are $m_{\tilde{F}}$ and
$m_{\tilde{F}^\prime}$ respectively, and $A_f$ is the trilinear
Higgs-sfermion-sfermion interaction, i.e.\ the entries of the
diagonal matrices ${\bf a_f}$ of Eq.\ (\ref{eq:lsoft}). The weak
isospin quantum numbers are $T_f^3=\pm1/2$ for left-handed and
$T_f^3=0$ for right-handed sfermions, their fractional
electromagnetic charges are denoted by $e_f$, and $\theta_W$ is
the weak mixing angle. ${\cal M}^2$ is diagonalized by a unitary
matrix $S^{\tilde{f}}$, $S^{\tilde{f}}\, {\cal
M}^2\,S^{\tilde{f}\dagger}={\rm diag}\,(m_1^2,m_2^2)$ and has the
squared mass eigenvalues \bea m_{1,2}^2={1\over 2}\Big( m_{LL}^2 +
m_{RR}^2 \mp\sqrt{(m_{LL}^2-m_{RR}^2)^2 + 4\,m_f^2\,|m_{LR}|^2}
\Big),~\eea where by convention $m_1 < m_2$. For real values of
$m_{LR}$, the sfermion mixing angle $\theta_{\tilde{f}}$, $0 \leq
\theta_{\tilde{f}} \leq \pi/2$, in \bea
 S^{\tilde{f}} = \left( \begin{array} {cc}~~\,\cos\theta_{\tilde{f}} &
 \sin\theta_{\tilde{f}} \\
 -\sin\theta_{\tilde{f}} & \cos\theta_{\tilde{f}} \end{array} \right)
 \hspace*{1mm} {\rm with} \hspace*{1mm} \left(
 \begin{array}
 {c} \tilde{f}_1 \\
 \tilde{f}_2
 \end{array} \right) =
 S^{\tilde{f}} \left(
 \begin{array}
 {c} \tilde{f}_L \\ \tilde{f}_R
 \end{array} \right) \hspace*{5mm}&& \label{eq:2to2mix}
\eea can be obtained from \beq
 \tan2\theta_{\tilde{f}}={2\,m_f\,m_{LR}\over
 m_{LL}^2-m_{RR}^2}.\label{eq:2to2mix2} \eeq
Let us note that for purely left-handed sneutrino eigenstates a
diagonalizing matrix is not needed.\\

The neutral Higgsinos and gauginos ($\tilde{B}^0$, $\tilde{W}^3$,
$\tilde{H}_d^0$, $\tilde{H}_u^0$) also mix when $SU(2)\times U(1)$
is spontaneously broken. The diagonalization of the generally
complex mass matrix \cite{Haber:1984rc, Gunion:1984yn} \bea Y\!\!
= \!\! \left(\begin{array}{c c c c} \hspace{-1mm} M_1 &
\hspace{-1mm} 0 & \hspace{-1mm} -m_Z\,\sin\theta_W\,\cos\beta &
\hspace{-1mm} m_Z\,\sin\theta_W\,\sin\beta \\ \hspace{-1mm} 0 &
\hspace{-1mm}M_2 & \hspace{-1mm} m_Z\,\cos\theta_W\,\cos\beta
& \hspace{-1mm} -m_Z\,\cos\theta_W\,\sin\beta \\
\hspace{-1mm} -m_Z\,\sin\theta_W\,\cos\beta & \hspace{-1mm}
m_Z\,\cos\theta_W\,\cos\beta & \hspace{-1mm} 0 & \hspace{-1mm}
-\mu \\ \hspace{-1mm} m_Z\,\sin\theta_W\,\sin\beta & \hspace{-1mm}
-m_Z\, \cos\theta_W\, \sin\beta & \hspace{-1mm} -\mu &
\hspace{-1mm} 0 \end{array} \hspace{-2mm}\right)\nn \eea
\vspace{-.8cm}\bea ~ \eea by a unitary matrix $N$ leads to four
neutral mass eigenstates, the neutralinos $\tilde{\chi}^0_i$
($i=1,2,3,4$), $\tilde{\chi}^0_1$ being the lightest one. The
resulting diagonal matrix $N_D = N^\ast \, Y \, N^{-1} = {\rm
diag}(m_{\tilde{\chi}^0_1}, m_{\tilde{\chi}^0_2},
m_{\tilde{\chi}^0_3}, m_{\tilde{\chi}^0_4})$, has four real
entries, and the neutralino fields are given by \bea
\tilde{\chi}_i^0 = N_{ij}\,\psi_j^0,~~~i,j=1,\dots,4,~ {\rm} {\rm
~~with~~ } \psi^0_j = (-i\tilde{B}^0, -i\tilde{W}^3,
\tilde{H}_d^0, \tilde{H}_u^0)^T.~ \eea Analytical expressions for
the diagonalizing matrix $N$ and the mass eigenvalues can be found
in Refs.\ \cite{Guchait:1991ia, ElKheishen:1992yv}. In general,
all the possible complex phases can be absorbed by a redefinition
of the fields, making the masses real and non-negative.\\

The charged analogues of the neutralinos are the two charginos,
$\tilde{\chi}^\pm_1$ and $\tilde{\chi}^\pm_2$, resulting from the
mixing of the charged wino and Higgsino fields ($\tilde{W}^\pm,
\tilde{H}_{u,d}^\pm$). Two unitary matrices $U$ and $V$ are needed
to diagonalize the generally complex mass matrix
\cite{Haber:1984rc, Gunion:1984yn} \bea X &=&
\left(\begin{array}{c c} M_{2}  & m_{W}\, \sqrt{2}\, \sin\beta \\
m_{W}\, \sqrt{2}\, \cos\beta & \mu
\end{array}\right),~\eea since $X \neq X^T$. The eigenvalues of
the diagonal matrix $M_C = U^\ast X V^{-1} = {\rm diag}\,
(m_{\tilde{\chi}_1}^\pm, m_{\tilde{\chi}_2}^\pm)$ can be chosen to
be real and non-negative, absorbing all complex phases by a
suitable redefinition of the fields \bea
m_{\tilde{\chi}_{1,2}^\pm}^2 = \frac{1}{2}\bigg\{ M_2^2 \!+\!
\mu^2 \!+\! 2 m_W^2 \!\mp\! \sqrt{(M_2^2 \!+\! \mu^2 \!+\! 2\,
m_W^2)^2 \!-\! 4\,(\mu\,M_2 \!-\! m_W^2\, \sin(2\beta))^2}
\bigg\},~\eea where by convention $m_{\tilde{\chi}_1^\pm} <
m_{\tilde{\chi}_2^\pm}$, while the $U$ and $V$ matrices \bea U =
{\cal O}_{-} {\rm ~~ and~~ } V = \left\{
\begin{array}{l l} {\cal O}_{+} & {\rm if} ~ {\rm det}\, X\, \geq
0\\ \sigma_{3}{\cal O}_{+} & {\rm if} ~ {\rm det} \,X\, < 0
\end{array}\right., {\rm ~with~} {\cal O}_{\pm}
\!=\!\left(\begin{array}{cc}\!\cos\theta_{\pm}\! &
\!\sin\theta_{\pm}\! \\ \!-\sin\theta_{\pm}\! &
\!\cos\theta_{\pm}\!
\end{array} \hspace{-2mm}\right)~\eea are determined by the mixing angles
$\theta_{\pm}$ with $0 \leq \theta_{\pm} \leq \pi/2$, \bea
\tan(2\theta_+) \!=\! \frac{2 \sqrt{2} m_W\!\left( M_2\sin\beta
\!+\! \mu \cos\beta\right)}{M_2^2 - \mu^2 + 2 m_W^2\cos(2\beta)}
{\rm and }\ \tan(2\theta_-) \!=\! \frac{2 \sqrt{2} m_W\!\left( M_2
\cos\beta \!+\! \mu \sin\beta\right)}{M_2^2 - \mu^2 - 2
m_W^2\cos(2\beta)}.~\nn \eea \vspace{-.8cm}\bea ~ \eea The
chargino mass eigenstates are given by
\bea \left( \begin{array}{c} \tilde{\chi}_1^- \\
\tilde{\chi}_2^- \end{array} \right) = U \left(
\begin{array}{c} -i \tilde{W}^- \\ \tilde{H}_d^-
\end{array} \right) \hspace*{.2mm} {\rm and} \hspace*{.2mm} \left(
\begin{array}{c} \tilde{\chi}_1^+ \\ \tilde{\chi}_2^+
\end{array} \right) =  V  \left( \begin{array}{c} -i \tilde{W}^+
\\ \tilde{H}_u^+ \end{array} \right).~\eea

\section{The MSSM with non-minimal flavour violation}

\subsection{The model}
When SUSY is embedded in larger structures such as grand unified
theories, new sources of flavour violation can appear
\cite{Gabbiani:1988rb}. In addition, SUSY parameter space is today
becoming more and more constrained by electroweak precision
measurements, direct searches for Higgs and SUSY particles at
colliders, and precise data on the cosmological relic density of
(possibly) SUSY dark matter. This encourages the investigation of
non-minimal models, e.g.\ with additional sources of flavour
violation. Non-minimal flavour violation in SUSY is well
parameterized in the super-CKM basis \cite{Hall:1985dx}, where the
up- and down-type squark mass matrices are  \bea M_{\tilde{Q}}^2 =
\left(
\begin{array}{c c c|c c c} m^2_{LL,1} &
\Delta^{12}_{LL} & \Delta^{13}_{LL} & m_1\, m_{LR,1} &
\Delta^{12}_{LR} & \Delta^{13}_{LR}\\
\Delta^{21}_{LL} & m^2_{LL,2} & \Delta^{23}_{LL} &
\Delta^{21}_{RL} & m_2\, m_{LR,2} & \Delta^{23}_{LR}\\
\Delta^{31}_{LL} & \Delta^{32}_{LL} & m^2_{LL,3} &
\Delta^{31}_{RL} & \Delta^{32}_{RL} & m_3\, m_{LR,3}\\
\hline m_1\, m_{RL,1} & \Delta^{12}_{RL} & \Delta^{13}_{RL} &
m^2_{RR,1} & \Delta^{12}_{RR} & \Delta^{13}_{RR}\\
\Delta^{21}_{LR} & m_2\, m_{RL,2} & \Delta^{23}_{RL} &
\Delta^{21}_{RR} &
m^2_{RR,2} & \Delta^{23}_{RR}\\
\Delta^{31}_{LR} & \Delta^{32}_{LR} & m_3\, m_{RL,3} &
\Delta^{31}_{RR} & \Delta^{32}_{RR} & m^2_{RR,3}
\end{array} \right),~ \nn \eea
\vspace{-.8cm}\bea ~ \eea the indices $(1, 2, 3)$ being the
flavour indices $(u, c,  t)$ for the up-type and $(d, s, b)$ for
the down-type mass matrix, and $m_{LL}$, $m_{RR}$, $m_{LR}$ and
$m_{RL}$ being defined by Eqs.\ (\ref{eq:mll}), (\ref{eq:mrr}) and
(\ref{eq:mlr}). The flavour-changing elements of the mass matrices
are usually normalized with respect to their diagonal entries
\cite{Gabbiani:1996hi},  \bea \Delta_{ab}^{f_1\,f_2} =
\lambda^{f_1\,f_2}_{ab} m_{ab,f_1}\, m_{ab,f_2},~\eea and obey to
the relations $\Delta_{LL,RR}^{f_1\, f_2} = \Delta_{LL,RR}^{f_2\,
f_1\ast}$. In this basis, even if the fields are not the mass
eigenstates, all the charged current $H^\pm$ and $W^\pm$
interactions couple with a strength given by the CKM matrix
\cite{Cabibbo:1963yz, Kobayashi:1973fv}, as their SUSY
counterparts do. $M_{\tilde{U}}^2$ and $M_{\tilde{D}}^2$ are then
diagonalized via two additional $6 \times 6$ matrices $R^u$ and
$R^d$, ${\rm diag}(m_{\tilde{u}_1}^2, \ldots, m_{\tilde{u}_6}^2) =
R^u\, M_{\tilde{U}}^2\, R^{u\dag}$ and ${\rm
diag}(m_{\tilde{d}_1}^2, \ldots, m_{\tilde{d}_6}^2) = R^d\,
M_{\tilde{D}}^2\, R^{d\dag}$, where by convention, the masses are
ordered as $m_{\tilde{q}_1} < \ldots < m_{\tilde{q}_6}$. The
physical mass eigenstates are given by \bea (\tilde{u}_1,
\tilde{u}_2, \tilde{u}_3, \tilde{u}_4, \tilde{u}_5, \tilde{u}_6)^T
&=& R^u (\tilde{u}_L, \tilde{c}_L, \tilde{t}_L, \tilde{u}_R,
\tilde{c}_R, \tilde{t}_R)^T,~\nn\\
(\tilde{d}_1,  \tilde{d}_2, \tilde{d}_3, \tilde{d}_4, \tilde{d}_5,
\tilde{d}_6)^T &=& R^d (\tilde{d}_L, \tilde{s}_L, \tilde{b}_L,
\tilde{d}_R, \tilde{s}_R, \tilde{b}_R)^T.~\label{eq:physstate}\eea
In the limit of vanishing off-diagonal parameters, the matrices
$R^q$ become flavour-diagonal, leaving only the well-known
helicity-mixing already present in cMFV.

\subsection{Experimental constraints on flavour violating
SUSY}\label{sec:const}

\begin{table}\centering
 \caption{\label{tab:2}The 95\% probability bounds on
$|\lambda_{ij}^{d_k d_l}|$ obtained in Ref.\
\cite{Ciuchini:2007ha}.}\vspace{.2cm}
\begin{tabular}{c|cccc}
\underline{ij} & LL & LR & RL & RR\\
$kl$           &    &    &    &   \\ \hline 12 &
1.4$\times10^{-2}$ & 9.0$\times10^{-5}$ & 9.0$\times10^{-5}$ &
9.0$\times10^{-3}$ \\ 13 & 9.0$\times10^{-2}$ & 1.7$\times10^{-2}$
& 1.7$\times10^{-2}$ & 7.0$\times10^{-2}$ \\ 23 &
1.6$\times10^{-1}$ & 4.5$\times10^{-3}$ & 6.0$\times10^{-3}$ &
2.2$\times10^{-1}$ \\
\end{tabular}
\end{table}

The scaling of the flavour violating entries $\Delta_{ij}$ with
the SUSY-breaking scale $M_{\rm SUSY}$ implies a hierarchy
$\Delta_{LL} \gg \Delta_{LR,RL} \gg \Delta_{RR}$
\cite{Gabbiani:1988rb, Hikasa:1987db}. Note also that $SU(2)$
gauge invariance relates the $\Delta_{LL}^{qq^\prime}$ of up- and
down-type quarks through the CKM-matrix, implying that a large
difference between them is not allowed. Experimental bounds coming
from the neutral kaon sector (on $\Delta m_K$, $\epsilon$,
$\epsilon^\prime/\epsilon$), on $B$- ($\Delta m_B$) and $D$-meson
oscillations ($\Delta m_D$), various rare decays (BR($b\to
s\gamma$), BR($\mu\to e\gamma$), BR($\tau\to e \gamma$), and
BR($\tau\to\mu\gamma$)), and electric dipole moments ($d_n$ and
$d_e$) can be used to set constraints on non-minimal flavour
mixing in the squark and slepton sectors \cite{Ciuchini:2007ha,
Ciuchini:2007cw, Foster:2006ze, Hahn:2005qi}. As example, we show
the 95\% probability bounds on $|\lambda_{ij}^{d_k d_l}|$ in Tab.\
\ref{tab:2} \cite{Ciuchini:2007ha}. In our own analysis, we take
implicitly into account all of the previously mentioned
constraints by restricting ourselves to the case of only one real
NMFV parameter,\bea \lambda\equiv\lambda_{LL}^{sb} =
\lambda_{LL}^{ct}.~\label{eq:lambda} \eea Let us note that in
addition, direct searches of flavour violation depend on the
possibility of flavour tagging, established experimentally only
for heavy flavours, comforting us in our restriction to consider
only mixing between the second and the third
generations in our analysis.\\

Allowed regions for this parameter are then obtained by imposing
several low-energy electroweak precision and cosmological
constraints. We start by imposing the branching ratio \bea {\rm
BR}(b\to s \gamma) = (3.55\pm 0.26) \times 10^{-4}, \eea obtained
from the combined measurements of BaBar, Belle, and CLEO
\cite{Barberio:2006bi}, which affects directly the allowed squark
mixing between the second and third generation. A second important
consequence of NMFV in the MSSM is the generation of large
splittings between squark-mass eigenvalues. The splitting within
isospin doublets influences the $Z$- and $W$-boson self-energies
at zero-momentum $\Sigma_{Z,W}(0)$ in the electroweak
$\rho$-parameter $\Delta\rho =
\Sigma_Z(0)/M_Z^2-\Sigma_W(0)/M_W^2$ and consequently the
$W$-boson mass $M_W$ and the squared sine of the weak mixing angle
$\sin^2\theta_W$. The latest combined fits of the $Z$-boson mass,
width, pole asymmetry, $W$-boson and top-quark mass constrain new
physics contributions to $T=-0.13\pm0.11$ \cite{Yao:2006px} or
\bea \Delta\rho = -\alpha T = 0.00102\pm0.00086, \eea where we
have used $\alpha(M_Z)=1/127.918$. A third observable sensitive to
SUSY loop-contributions is the anomalous magnetic moment
$a_\mu=(g_\mu-2)/2$ of the muon, for which recent BNL data and the
SM prediction disagree by \cite{Yao:2006px} \bea \Delta
a_\mu=(22\pm10)\times 10^{-10}. \eea For cosmological reasons,
i.e.\ in order to have a suitable candidate for non-baryonic cold
dark matter \cite{Ellis:1983ew}, we require the LSP to be stable,
electrically neutral, and a colour singlet. The related dark
matter relic density is constrained to the region \bea
0.094<\Omega_{CDM} h^2<0.136 \eea at 95\% (2$\sigma$) confidence
level, which has been obtained from the three-year data of the
WMAP satellite, combined with the SDSS and SNLS survey and Baryon
Acoustic Oscillation data and interpreted within an
eleven-parameter inflationary model \cite{Hamann:2006pf}, which is
more general than the usual six-parameter ``vanilla'' concordance
model of cosmology.\\

We impose both the ${\rm BR}(b\to s \gamma)$ bounds on the
two-loop QCD/one-loop SUSY calculation \cite{Hahn:2005qi,
Kagan:1998bh} and the $\Delta\rho$ bounds on the one-loop NMFV and
two-loop cMFV SUSY calculation \cite{Heinemeyer:2004by} at the
2$\sigma$-level. We take into account the SM and SUSY
contributions to $a_\mu$ up to two loops \cite{Heinemeyer:2003dq,
Heinemeyer:2004yq} and require them to agree with the experimental
region within two standard deviations. $\Omega_{CDM}$ is
calculated using a modified version of DarkSUSY 4.1
\cite{Gondolo:2004sc}, that manages NMFV.

\subsection{Scan of the mSUGRA parameter space}\label{sec:scan}

The above experimental limits are now imposed on the minimal
supergravity models with their five free parameters ($m_0$,
$m_{1/2}$, $A_0$, $\tan\beta$, sign$(\mu)$) at the grand
unification scale. Since our scans \cite{Bozzi:2007me} in the
($m_0$, $m_{1/2})$ plane depend very little on the trilinear
coupling $A_0$, we set it to zero. Furthermore, we fix a small
(10), intermediate (30), and large (50) value for $\tan\beta$, and
investigate the impact of the sign of $\mu$ for $\tan\beta=10$
only, before we set it to $\mu>0$ for the two other cases (see
below). We solve the renormalization group equations numerically
to two-loop order using the computer program SPheno 2.2.3
\cite{Porod:2003um}, evolving the five parameters at grand
unification scale in order to compute the soft SUSY-breaking
parameters at the electroweak scale (see Eq.\ (\ref{eq:lsoft}))
with the complete one-loop formulas, supplemented by two-loop
contributions in the case of the neutral Higgs bosons and the
$\mu$-parameter. At this point we generalize the squark mass
matrices, introducing the $\lambda$ parameter, and compute the
low-energy, electroweak precision, and cosmological observables
with the computer programs FeynHiggs 2.5.1
\cite{Heinemeyer:1998yj} and DarkSUSY 4.1 \cite{Gondolo:2004sc}.\\

For the masses and widths of the electroweak gauge bosons and the
mass of the top quark, we use the current values of $m_Z=91.1876$
GeV, $m_W=80.403$ GeV, $m_t = 174.2$ GeV, $\Gamma_Z=2.4952$ GeV,
and $\Gamma_W=2.141$ GeV. The CKM-matrix elements are computed
using the parameterization \bea V = \left(\begin{array}{c c c}
c_{12} c_{13} & s_{12} c_{13} & s_{13} e^{-i \delta}\\ -s_{12}
c_{23} - c_{12} s_{23} s_{13} e^{i \delta}& c_{12} c_{23} - s_{12}
s_{23} s_{13} e^{i \delta}& s_{23} c_{13}\\ s_{12} s_{23} - c_{12}
c_{23} s_{13} e^{i \delta}& -c_{12} s_{23} - s_{12} c_{23} s_{13}
e^{i \delta}& c_{23} c_{13} \end{array}\right), \eea where $s_{ij}
= \sin\theta_{ij}$ and $c_{ij} = \cos\theta_{ij}$ relate to the
mixing of two specific generations $i$ and $j$ and $\delta$ is the
SM CP violating complex phase. The numerical values are given by
\bea s_{12} = 0.2243,~s_{23} = 0.0413,~ s_{13} = 0.0037,~{\rm
and}~ \delta = 1.05. \eea The squared sine of the electroweak
mixing angle $\sin^2\theta_W=1- m_W^2/m_Z^2$ and the
electromagnetic fine structure constant $\alpha= \sqrt{2} G_F
m_W^2\sin^2\theta_W/\pi$ are calculated in the improved Born
approximation using the world average value of $G_F=1.16637\cdot
10^{-5}$ GeV$^{-2}$ for Fermi's coupling constant
\cite{Yao:2006px}.\\

\begin{figure}
 \centering
 \includegraphics[width=0.24\columnwidth]{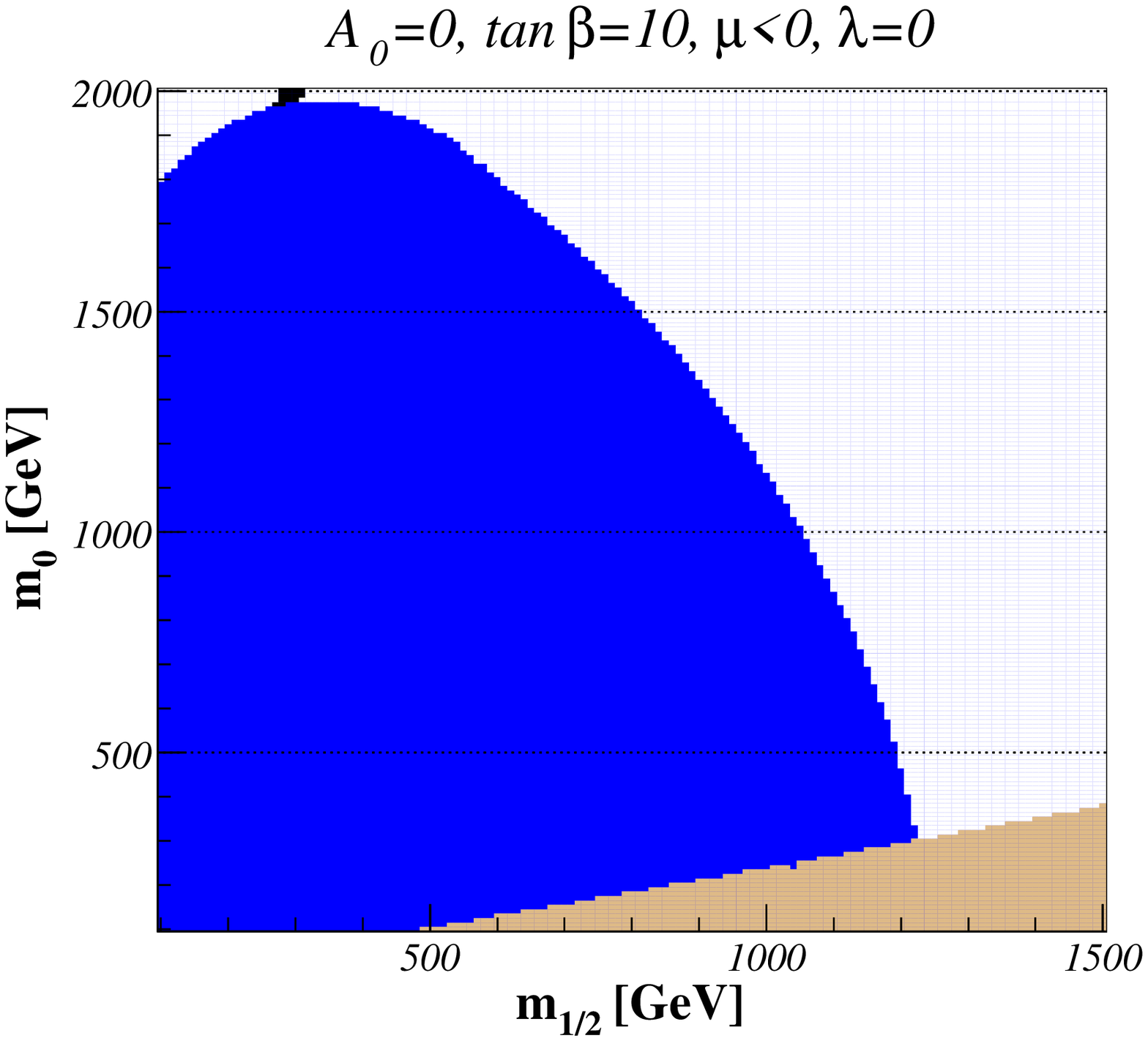}
 \includegraphics[width=0.24\columnwidth]{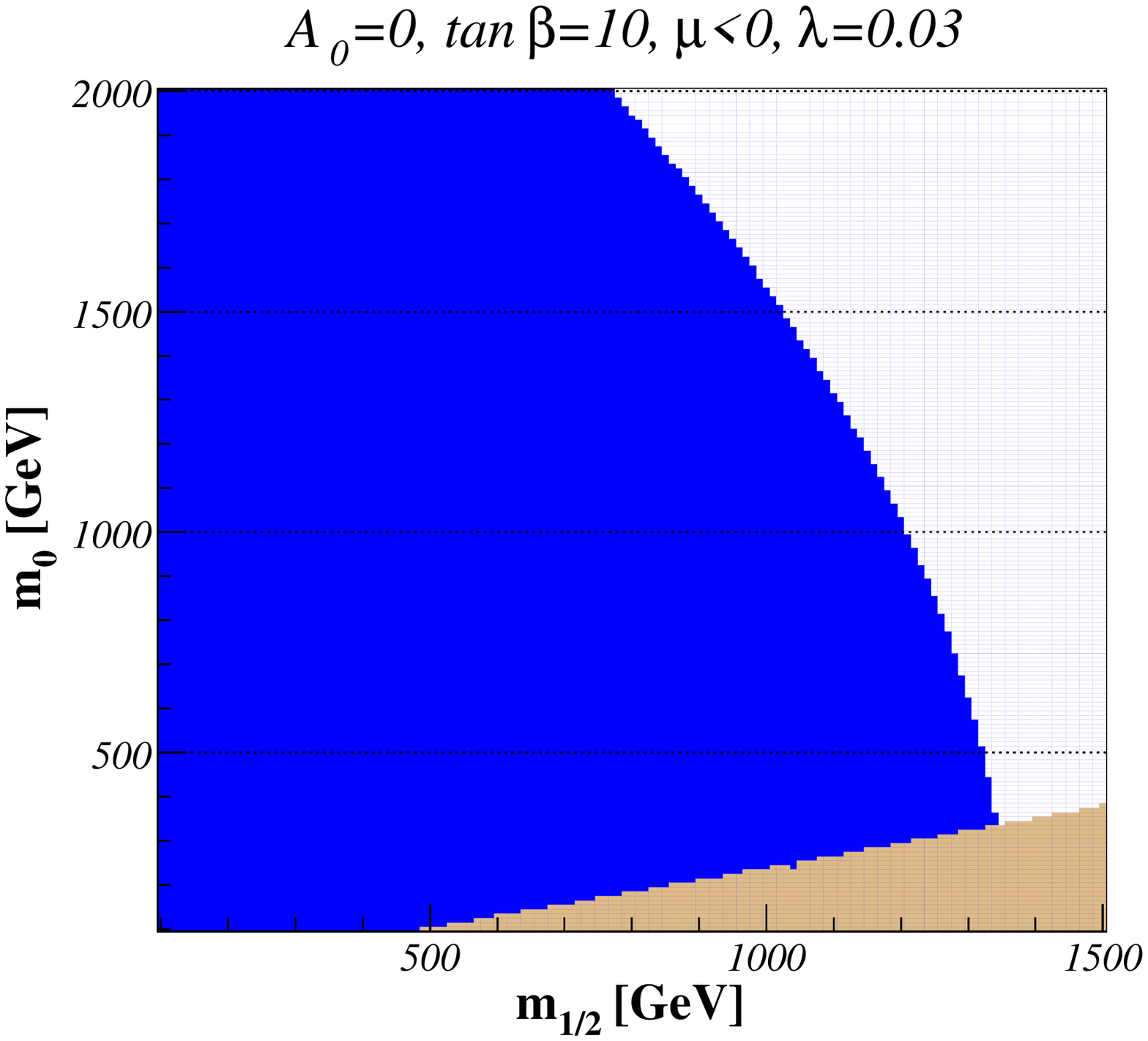}
 \includegraphics[width=0.24\columnwidth]{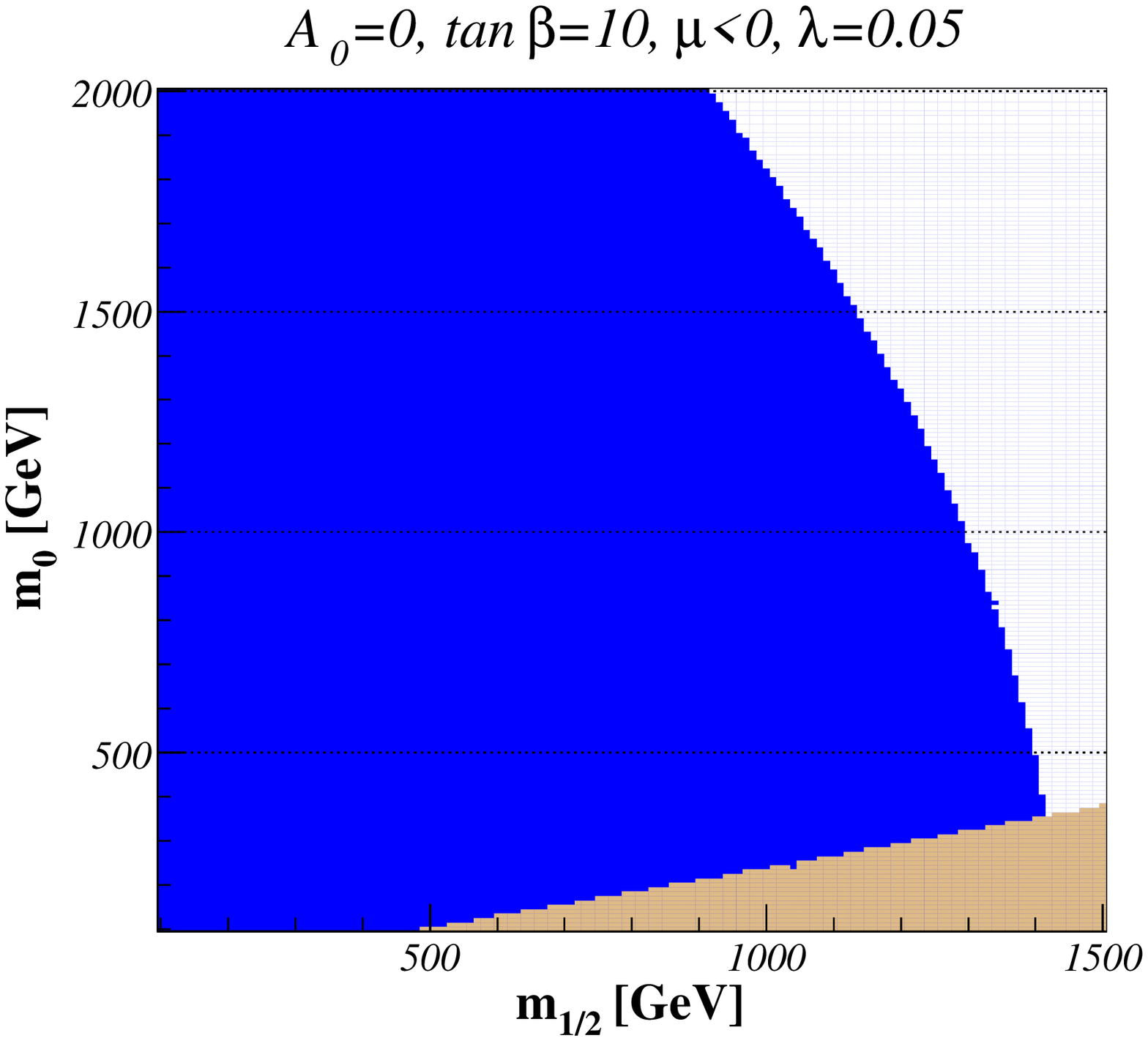}
 \includegraphics[width=0.24\columnwidth]{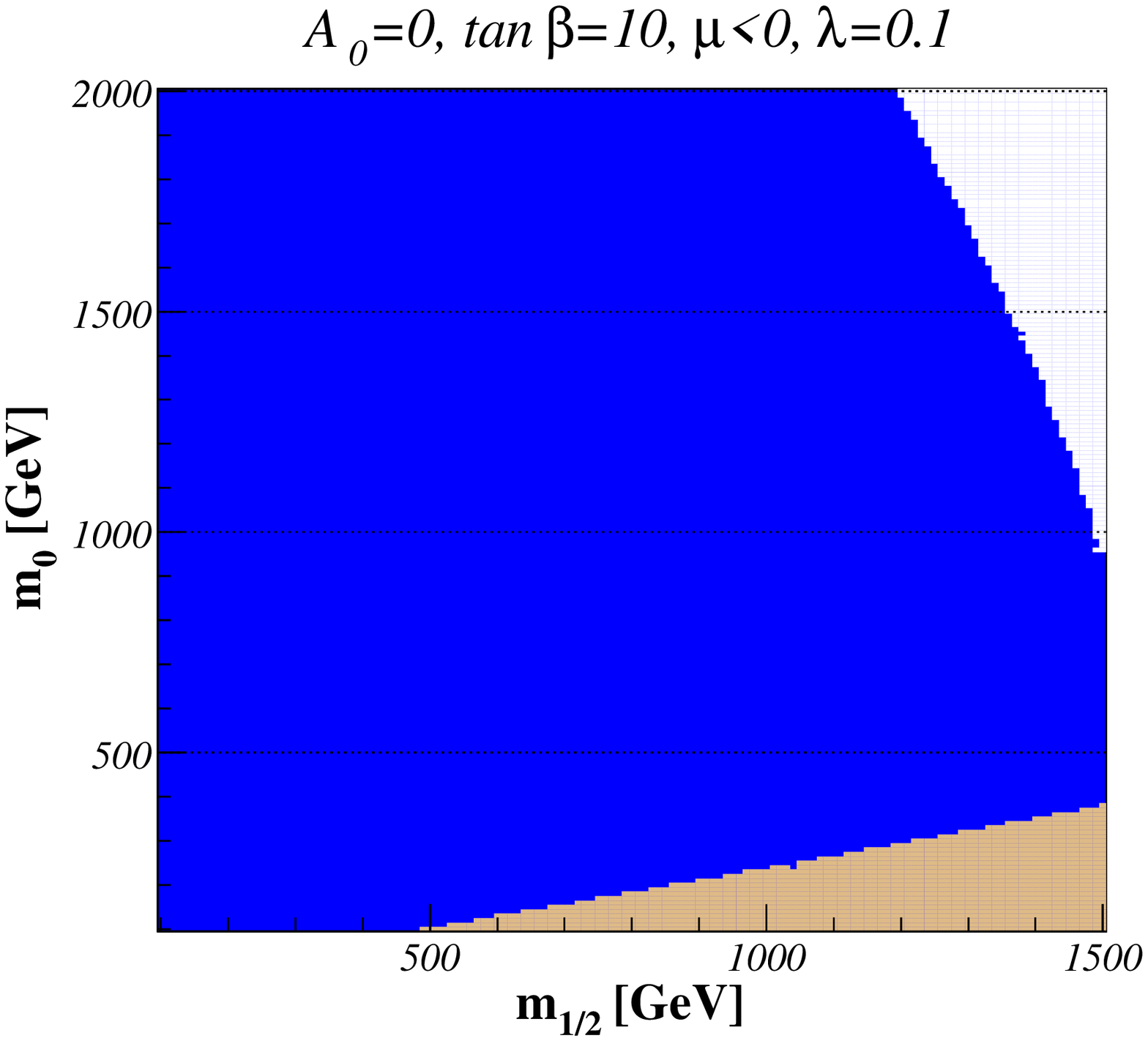}
 \caption{\label{fig:01}The $(m_0,m_{1/2})$-planes for $\tan\beta=10$,
          $A_0=0$ GeV, $\mu<0$, and $\lambda=0$, 0.03, 0.05 and 0.1. We show
          WMAP (black) favoured as well as $b\to s\gamma$ (blue) and charged
          LSP (beige) excluded regions of mSUGRA parameter space in minimal
          ($\lambda=0$) and non-minimal ($\lambda>0$) flavour
          violation.}\vspace{2mm}
 \includegraphics[width=0.24\columnwidth]{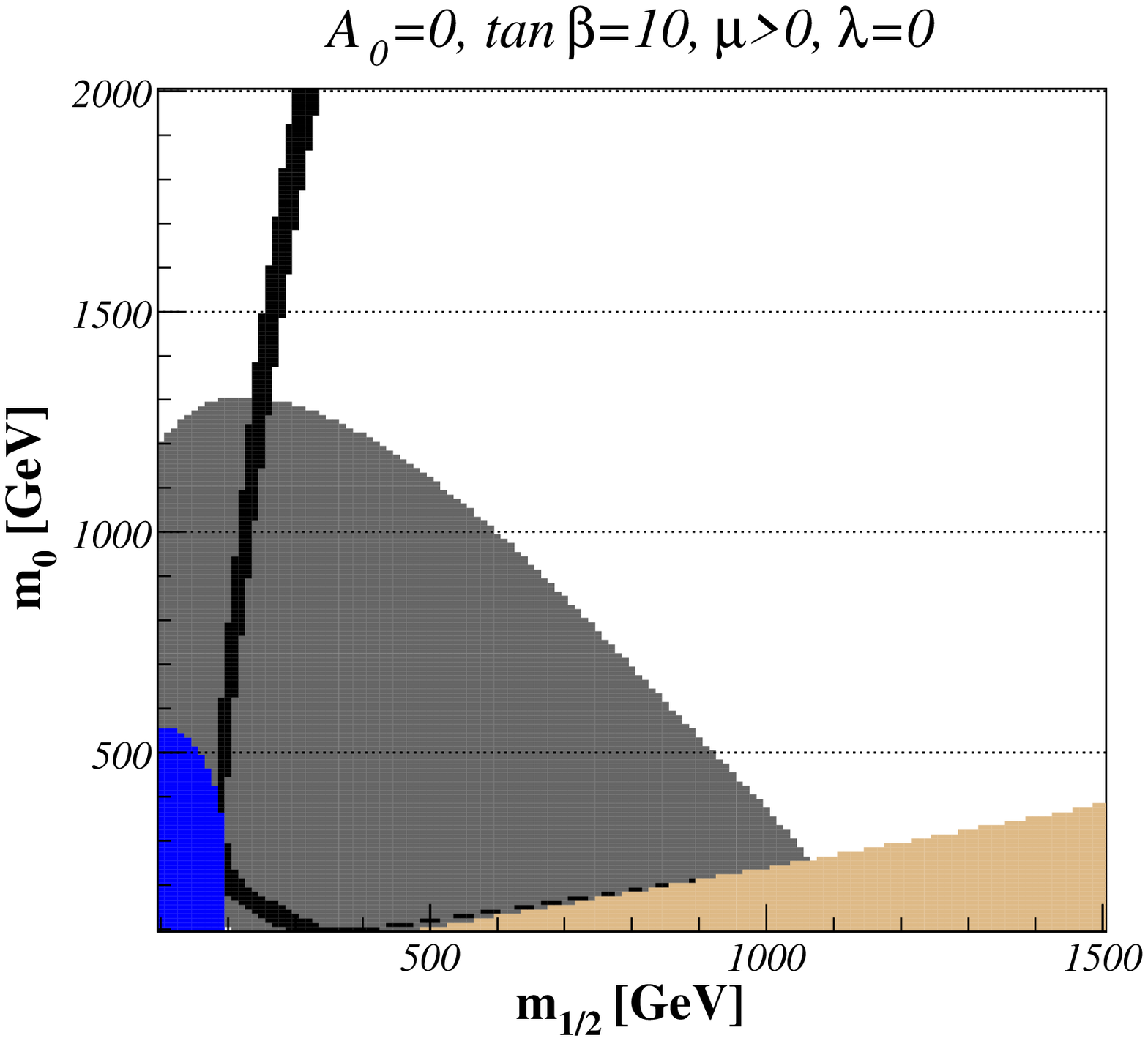}
 \includegraphics[width=0.24\columnwidth]{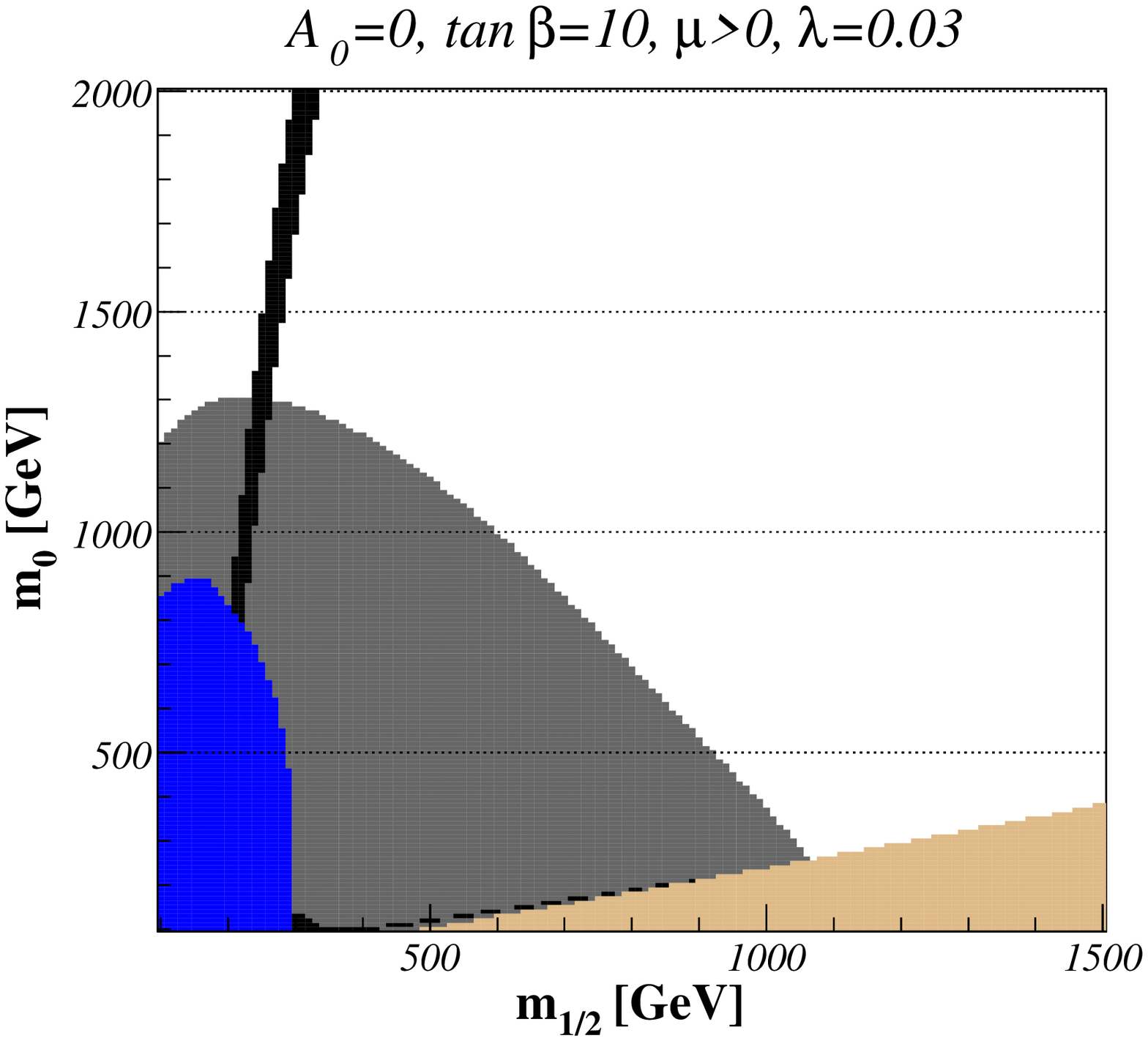}
 \includegraphics[width=0.24\columnwidth]{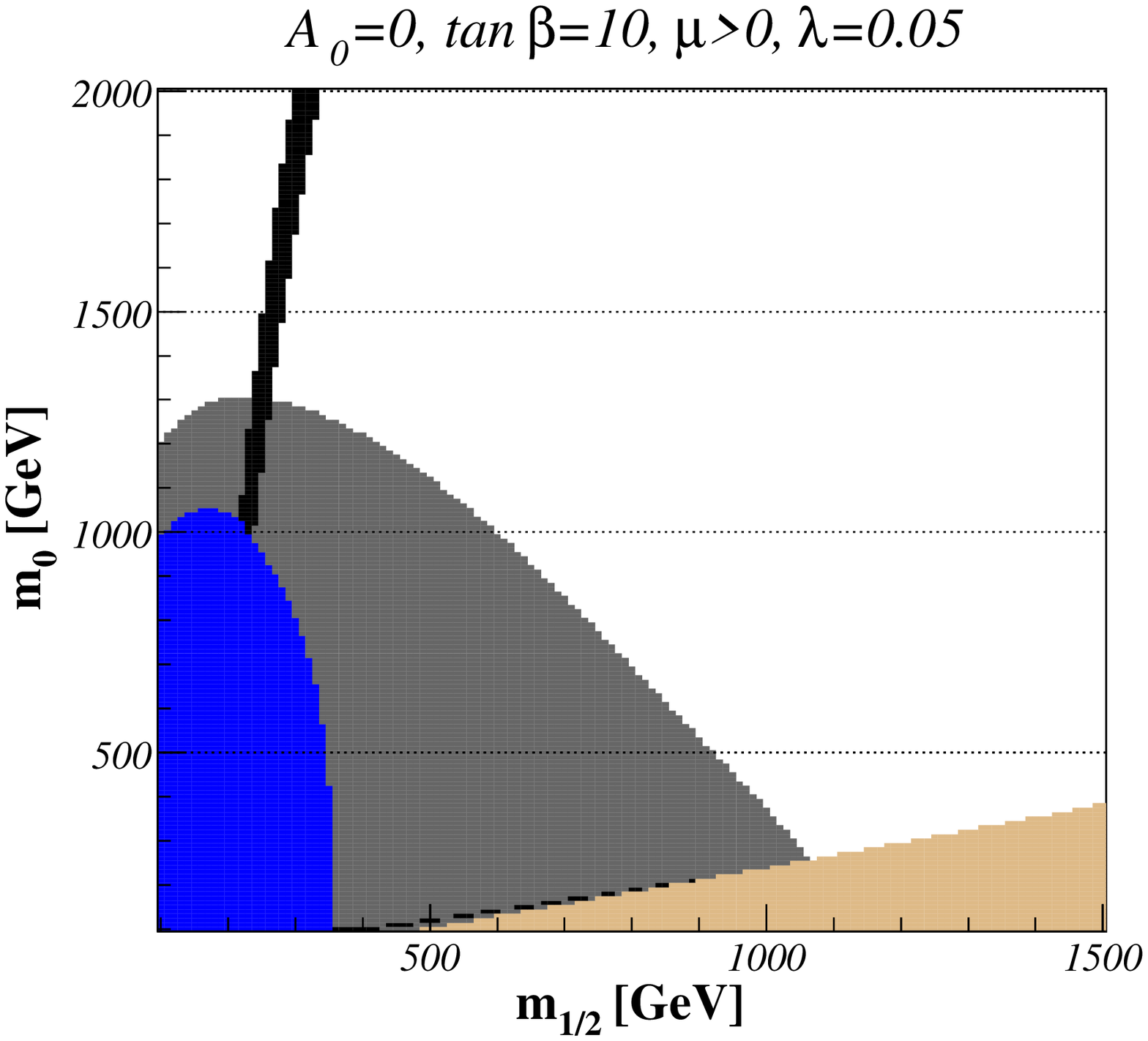}
 \includegraphics[width=0.24\columnwidth]{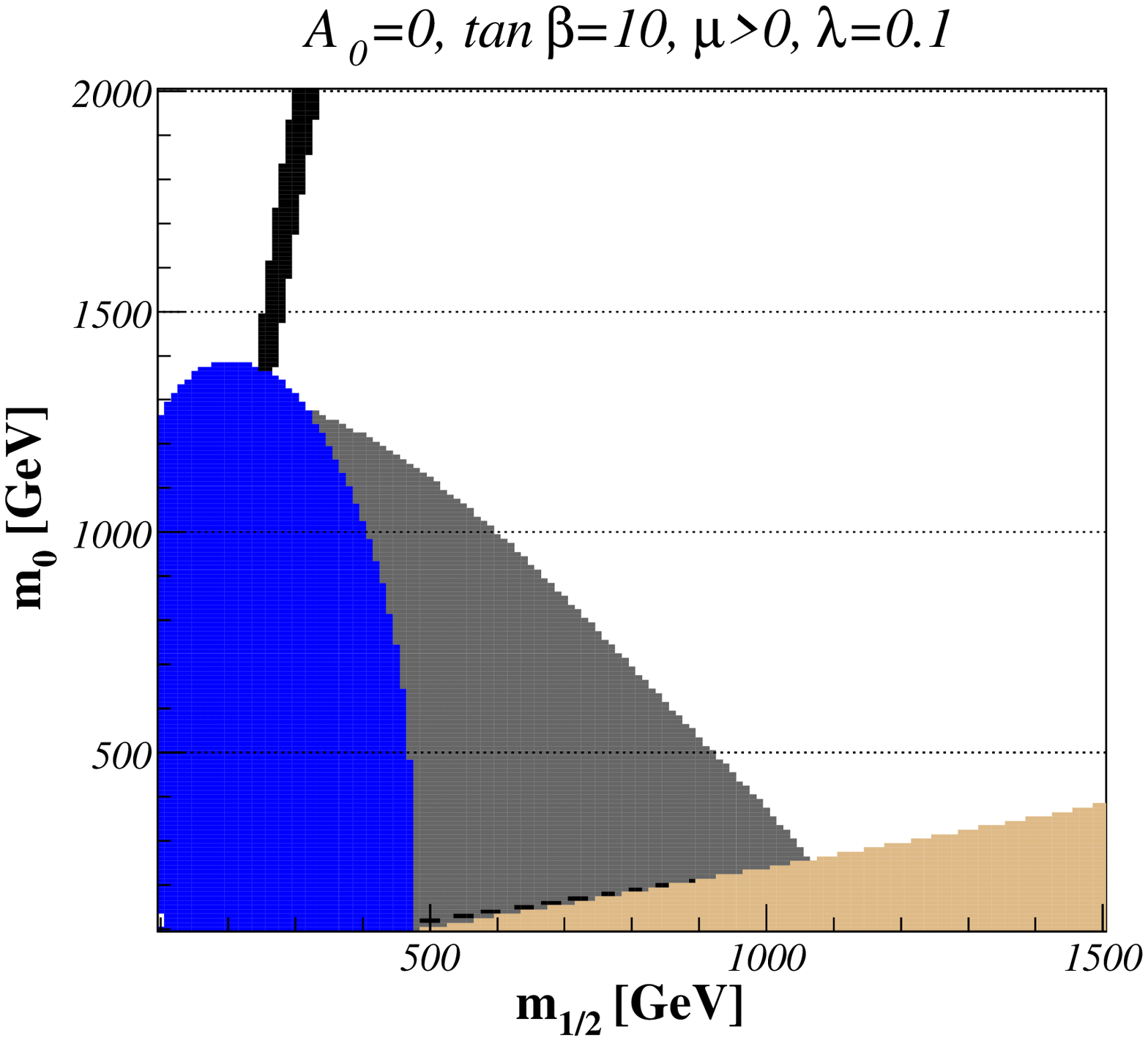}
 \caption{\label{fig:02}Same as Fig.\ \ref{fig:01}, for $\tan\beta=10$,
          $A_0=0$ GeV, $\mu>0$, and $\lambda=0$, 0.03, 0.05 and 0.1.}\vspace{2mm}
 \includegraphics[width=0.24\columnwidth]{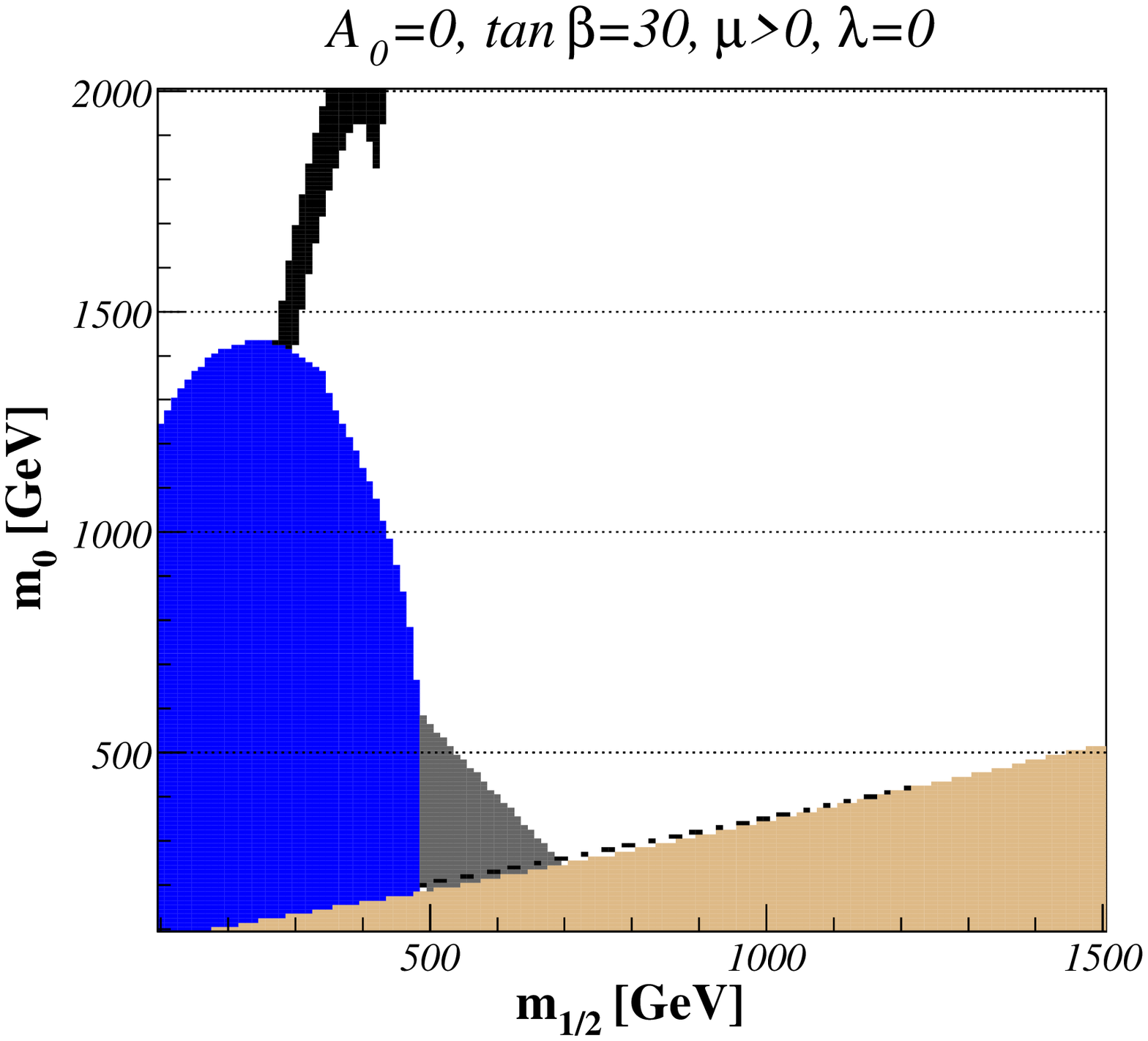}
 \includegraphics[width=0.24\columnwidth]{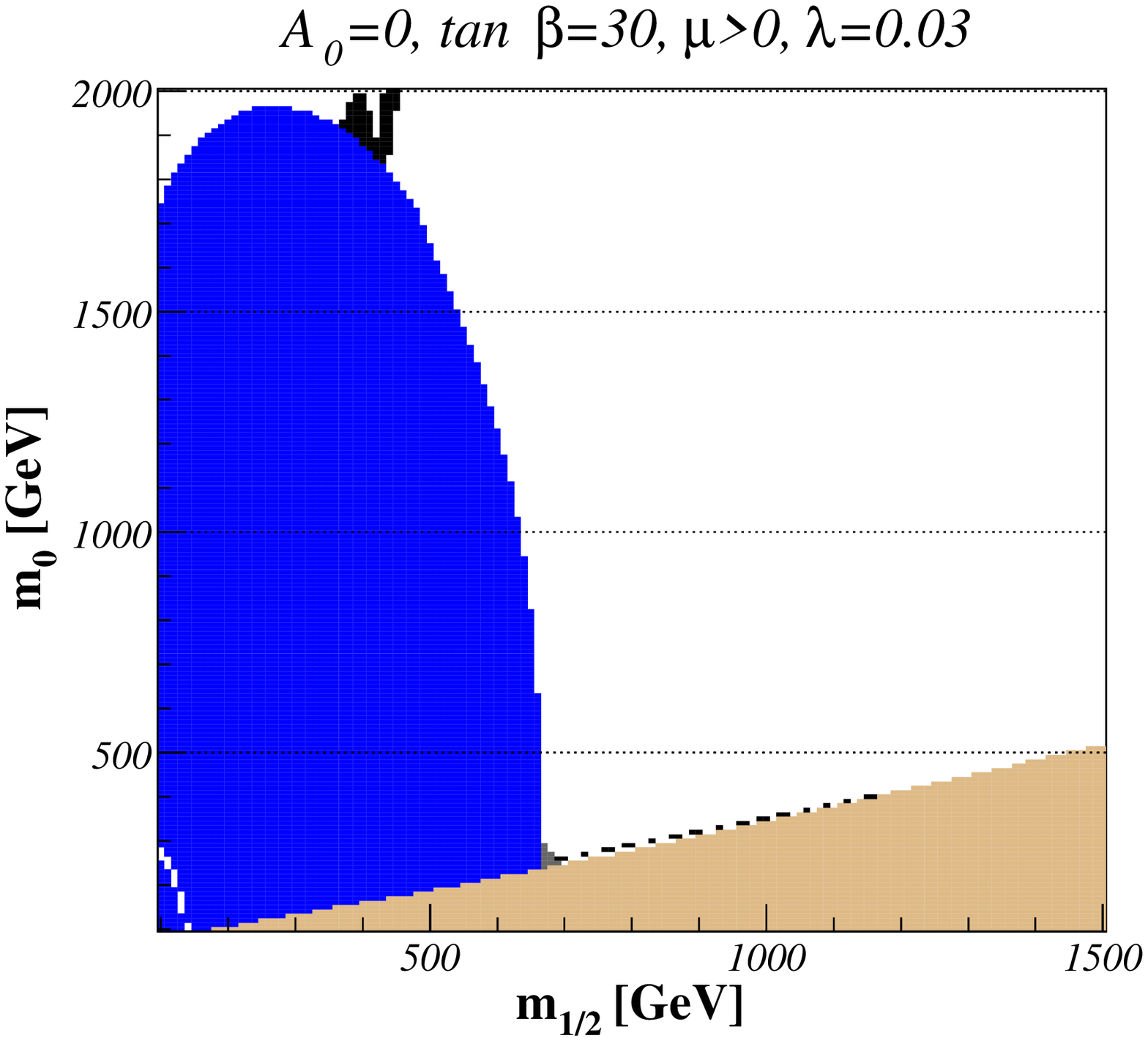}
 \includegraphics[width=0.24\columnwidth]{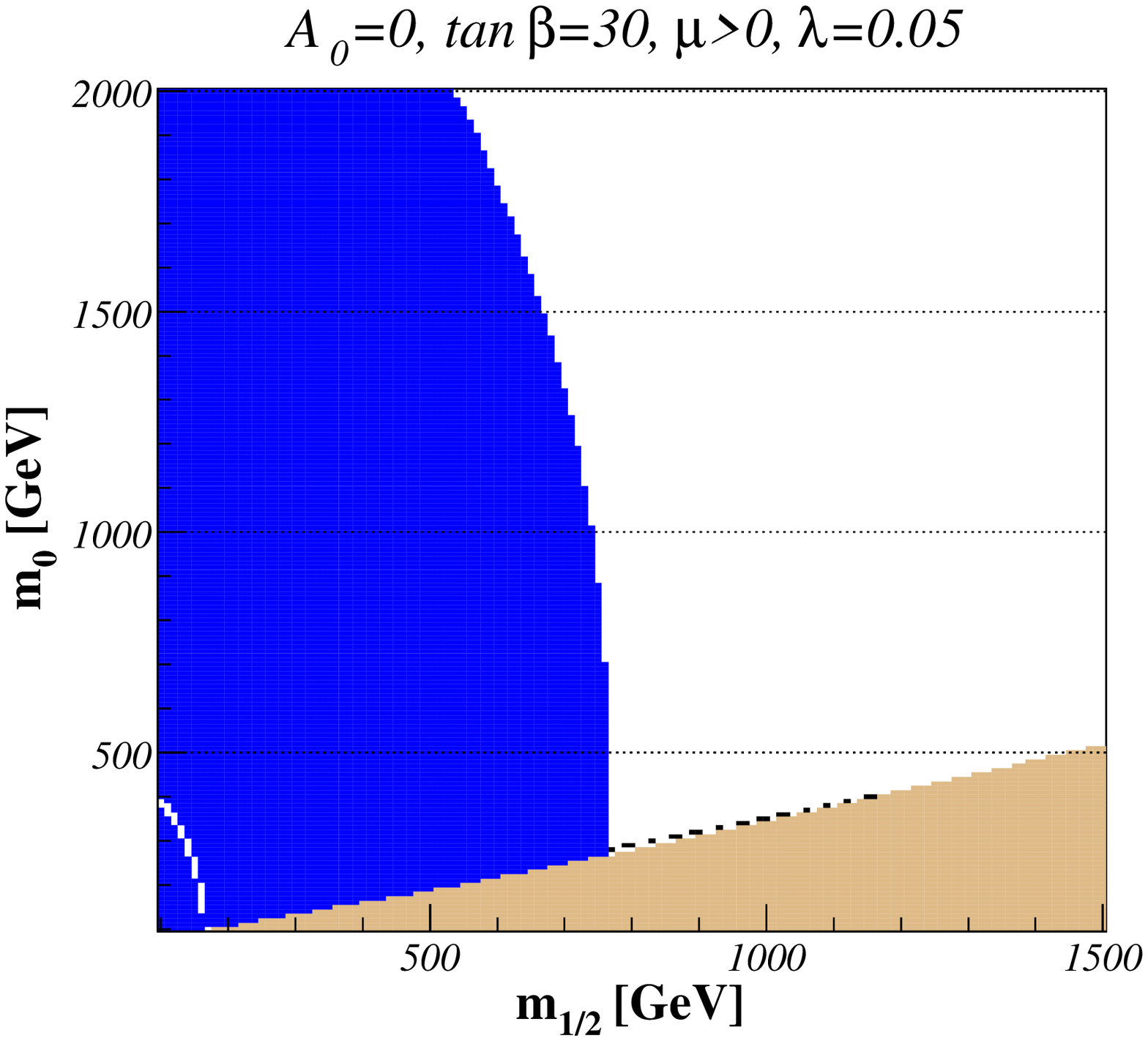}
 \includegraphics[width=0.24\columnwidth]{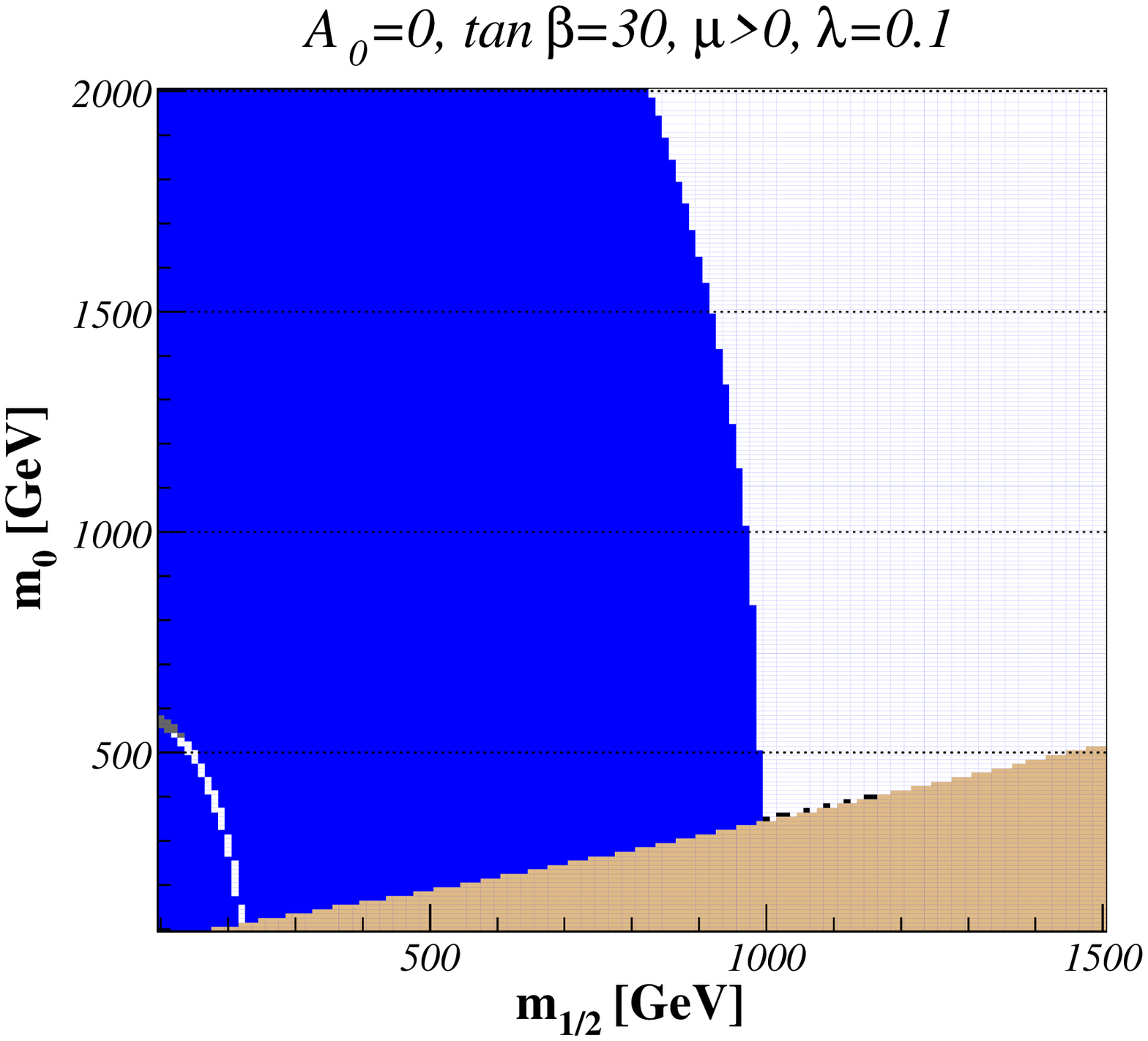}
 \caption{\label{fig:03}Same as Fig.\ \ref{fig:01}, for $\tan\beta=30$,
          $A_0=0$ GeV, $\mu>0$, and $\lambda=0$, 0.03, 0.05 and 0.1.}\vspace{2mm}
 \includegraphics[width=0.24\columnwidth]{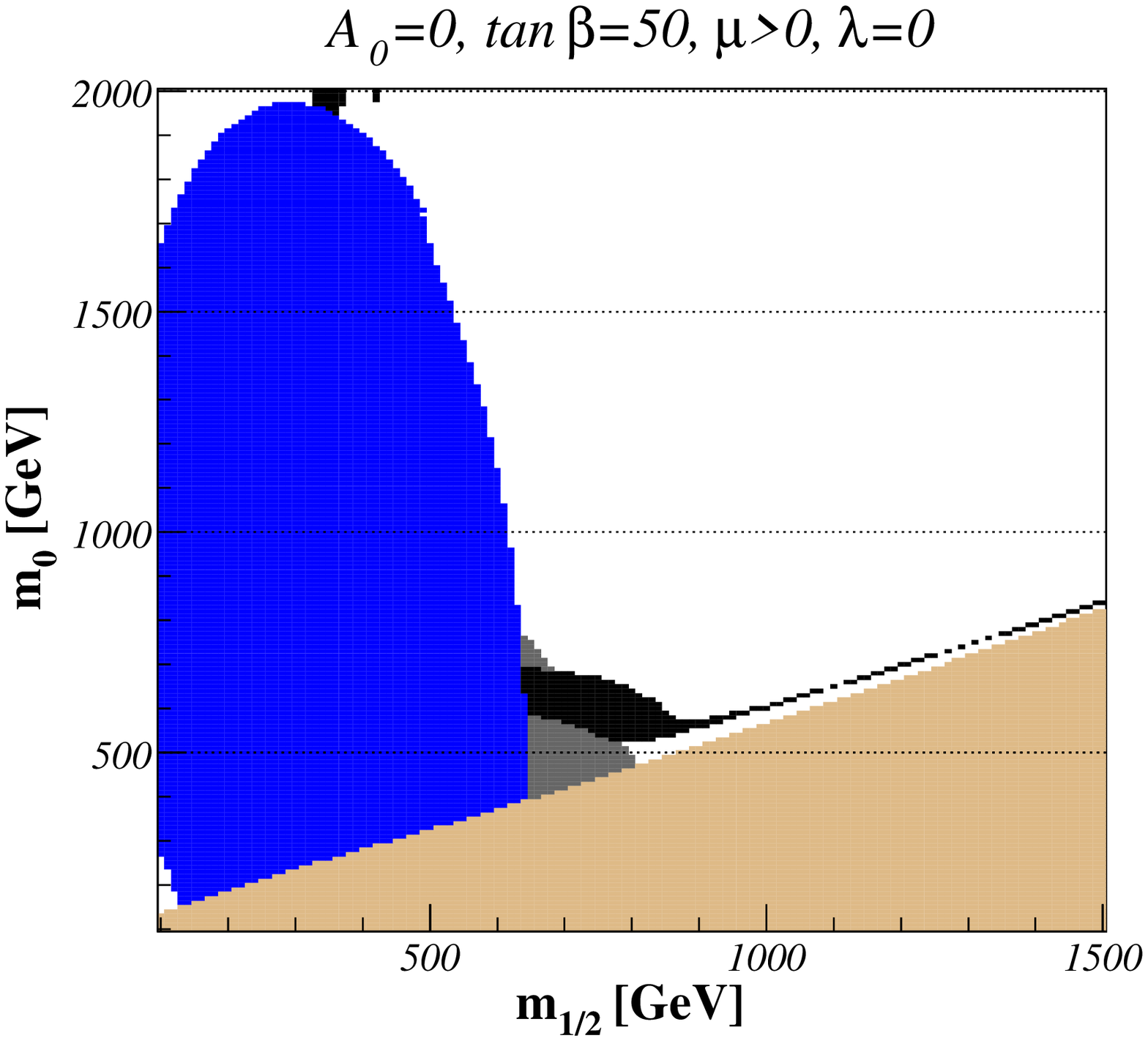}
 \includegraphics[width=0.24\columnwidth]{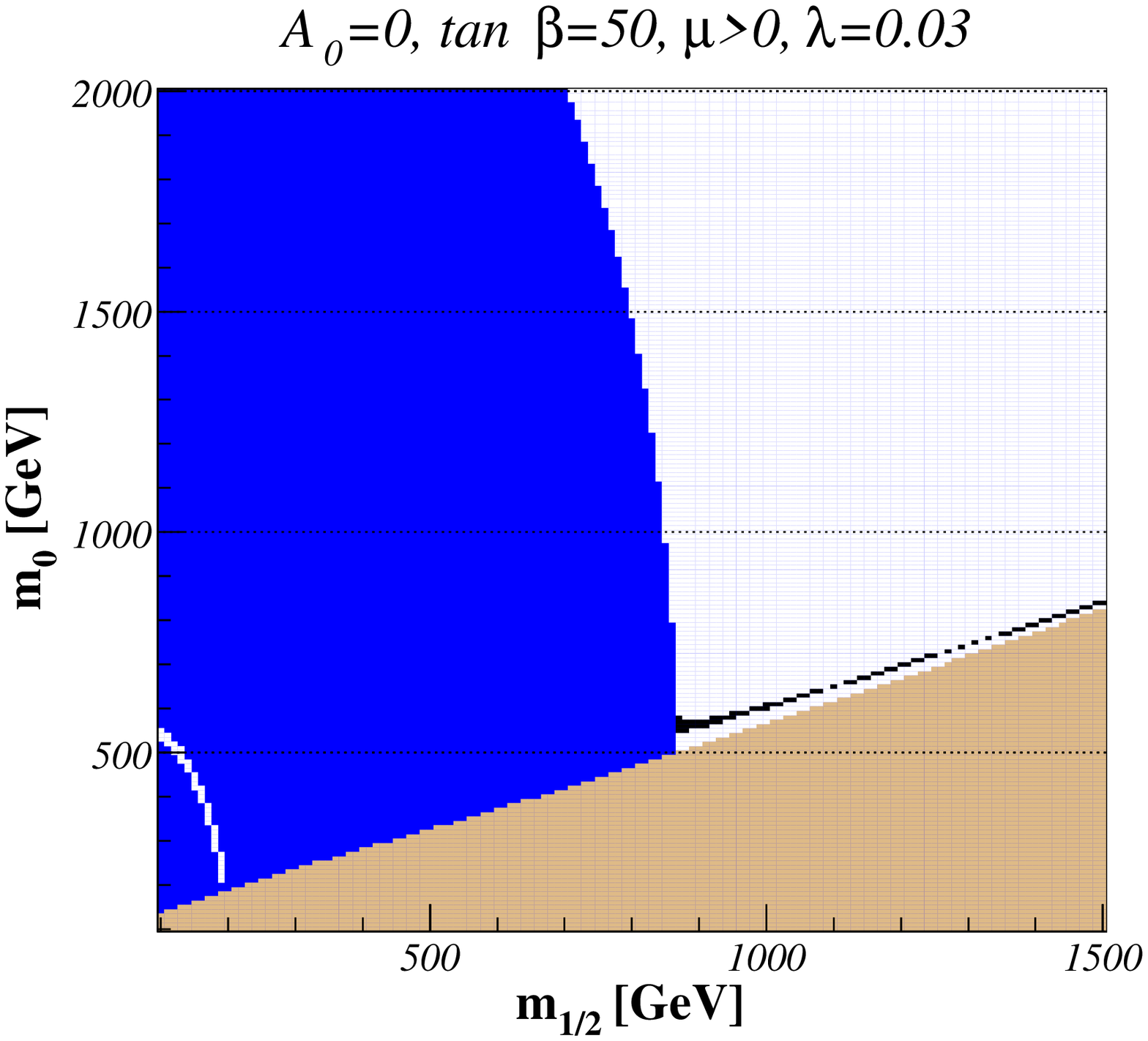}
 \includegraphics[width=0.24\columnwidth]{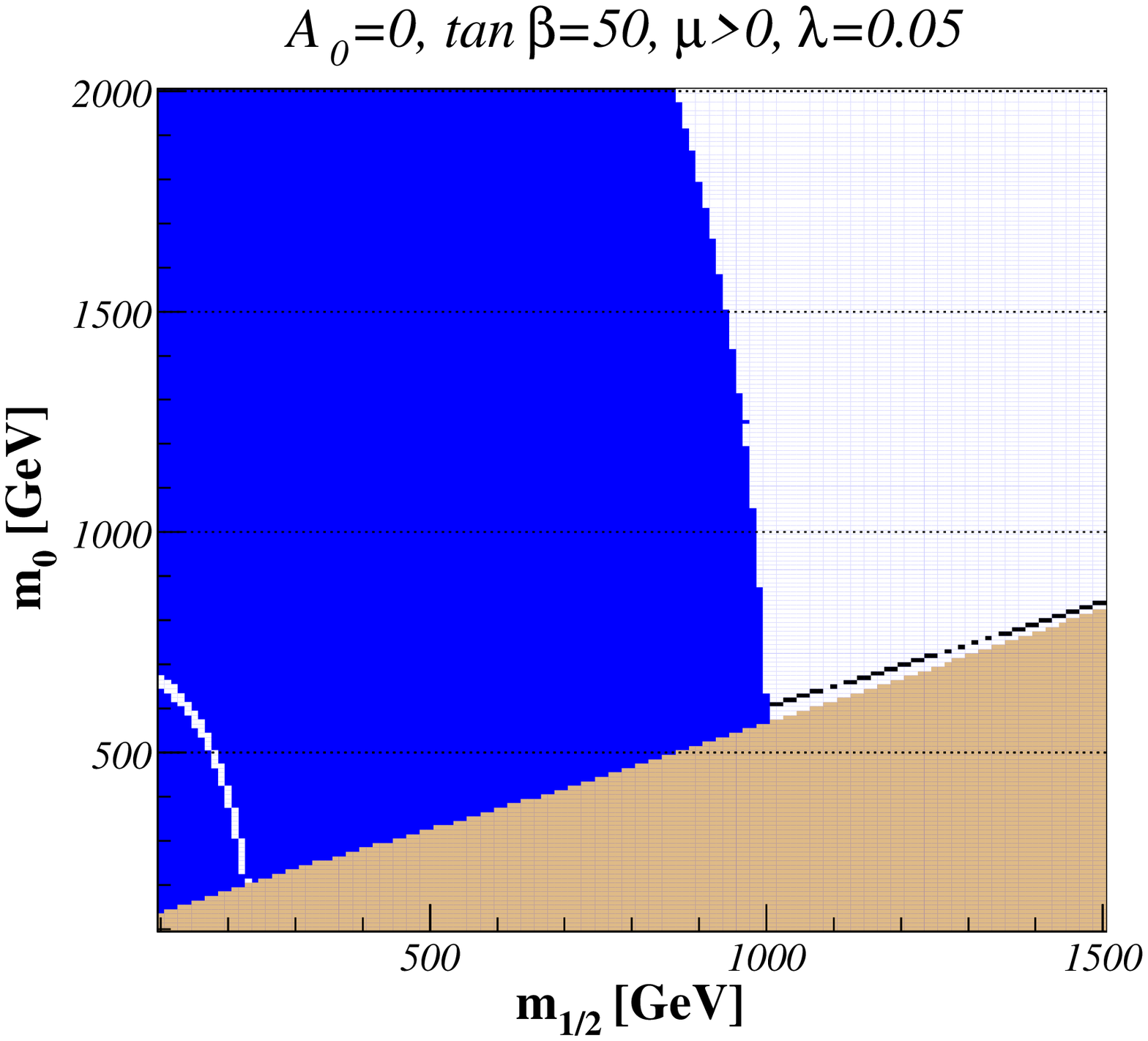}
 \includegraphics[width=0.24\columnwidth]{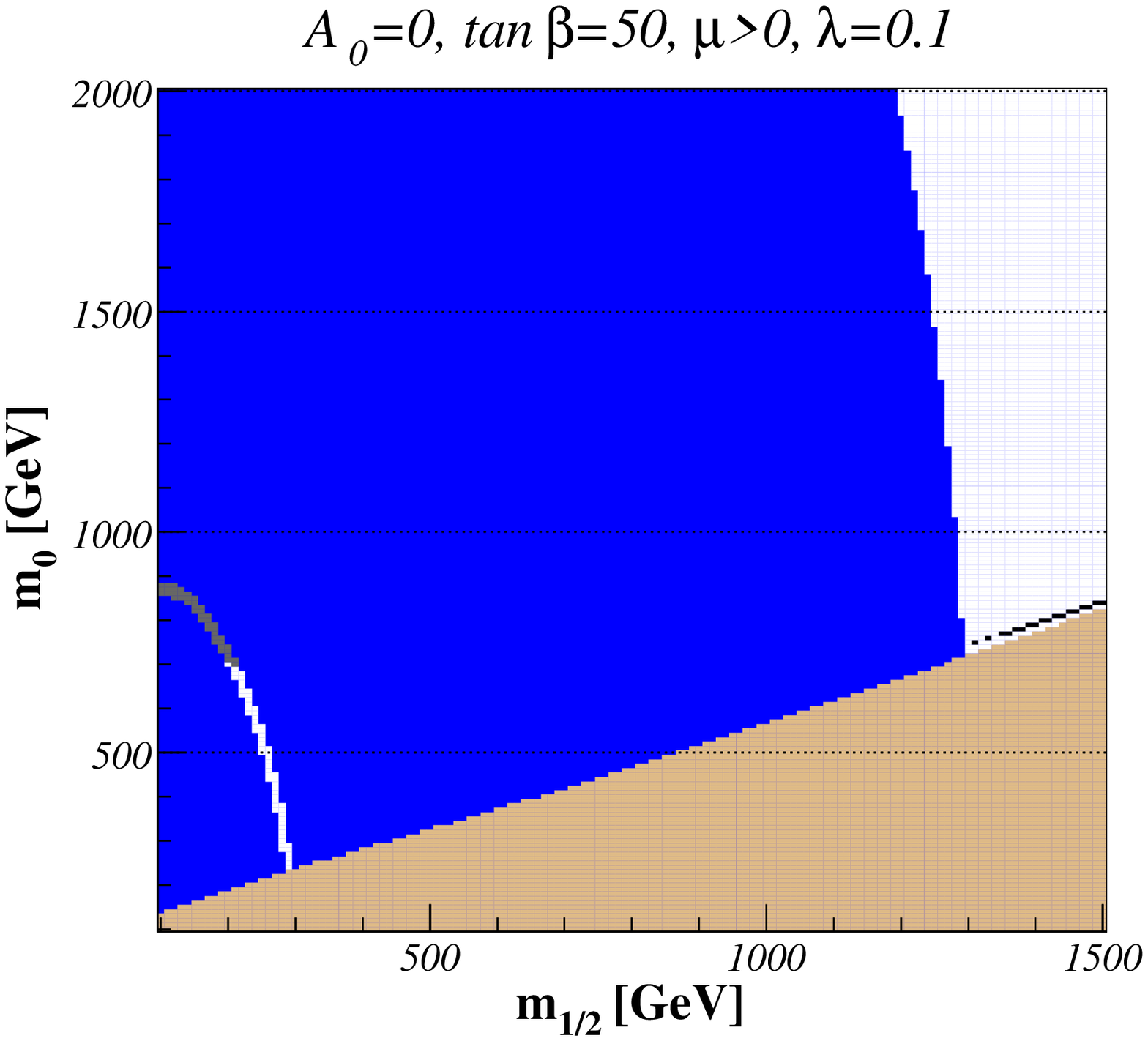}
 \caption{\label{fig:04}Same as Fig.\ \ref{fig:01}, for $\tan\beta=50$,
          $A_0=0$ GeV, $\mu>0$, and $\lambda=0$, 0.03, 0.05 and 0.1.}
\end{figure}

Typical scans of the mSUGRA parameter space in $m_0$ and $m_{1/2}$
with a relatively small value of $\tan\beta = 10$ and $A_0 = 0$
are shown in Figs.\ \ref{fig:01} and \ref{fig:02} for $\mu < 0$
and $\mu > 0$, respectively. All experimental limits described in
the previous section are imposed at the $2\sigma$-level. The $b\to
s \gamma$ excluded region depends strongly on flavour mixing,
while the regions favoured by $g_\mu - 2$ and the dark matter
relic density are quite insensitive to variations of the
$\lambda$-parameter. $\Delta\rho$ constrains the parameter space
only for heavy universal scalar masses $m_0>2000$ GeV and heavy
universal gaugino masses $m_{1/2}>1500$ GeV, so that the
corresponding excluded regions are not shown here. The dominant
SUSY effects in the calculation of the anomalous magnetic moment
of the muon come from induced quantum loops of a neutralino or a
chargino and a slepton, while squarks contribute only at the
two-loop level. This reduces the dependence on flavour violation
in the squark sector considerably. Furthermore, the region $\mu<0$
is disfavoured in all SUSY models, since the one-loop SUSY
contributions are approximatively given by \cite{Moroi:1995yh}
\bea a_\mu^{{\rm SUSY,~1-loop}} \simeq 13 \times 10^{-10}\, \lr
\frac{100~{\rm GeV}}{M_{\rm SUSY}} \rr^2 \tan\beta \ {\rm
sgn}(\mu), \eea where $M_{\rm SUSY}$ is a typical SUSY mass scale.
Negative values of $\mu$ would then increase, not decrease, the
disagreement between the experimental measurements and the
theoretical value of $a_\mu$. Furthermore, the measured $b \to s
\gamma$ branching ratio excludes virtually all of the region
favoured by the dark matter relic density, except for very high
scalar SUSY masses. We therefore do not consider negative values
of $\mu$ in the rest of this work.\\

In Figs.\ \ref{fig:03} and \ref{fig:04}, we show the
($m_0,m_{1/2}$)-planes for larger $\tan\beta$, namely
$\tan\beta=30$ and $\tan\beta=50$, and for $\mu>0$. The regions
which are favoured both by the anomalous magnetic moment of the
muon and by the cold dark matter relic density, and which are not
excluded by the $b\to s \gamma$ measurements, are stringently
constrained and do not allow for large flavour violation.

\subsection{NMFV benchmark points and slopes}\label{sec:bench}

Restricting ourselves to non-negative values of $\mu$, we now
inspect the $(m_0,m_{1/2})$-planes in Figs.\
\ref{fig:02}-\ref{fig:04} for scenarios that are allowed or
favoured by low-energy, electroweak precision, and cosmological
constraints, that permit non-minimal flavour violation among
left-chiral squarks of the second and third generation up to
$\lambda \lesssim 0.1$, and that are at the same time
collider-friendly, with relatively low values of $m_0$ and
$m_{1/2}$. We propose \cite{Bozzi:2007me} the four benchmark
points given in Tab.\ \ref{tab:3}. We also attach a model line
(slope) to each point in Tab.\ \ref{tab:4}. These slopes trace the
allowed/favoured regions from lower to higher masses and can, of
course, also be used in MFV scenarios where the off-diagonal terms
are expressed in function of the CKM matrix and the Yukawa
couplings, and in cMFV with $\lambda=0$.\\

\begin{table}\centering
\caption{\label{tab:3}Benchmark points allowing for flavour
violation among the second and third generations for $A_0=0$,
$\mu>0$, and three different values of $\tan\beta$.}\vspace{.2cm}
\begin{tabular}{c|ccccc} & $m_0$
[GeV] & $m_{1/2}$ [GeV] & $A_0$ [GeV] & $\tan\beta$ & sgn($\mu$)\\
\hline A & 700 & 200 & 0 & 10 & 1\\ B & 100 & 400 & 0 & 10 & 1 \\
C & 230 & 590 & 0 & 30 & 1\\ D & 600 & 700 & 0 & 50 & 1 \\
\end{tabular}
\end{table}

\begin{table}\centering
\caption{\label{tab:4}Model lines allowing for flavour violation
among the second and third generations for $A_0=0$, $\mu>0$, and
three different values of $\tan\beta$.}\vspace{.2cm}
\begin{tabular}{c| c l}
A & $180~{\rm GeV}~\leq~m_{1/2}~\leq~250~{\rm GeV}~,$ &
$m_0~=~~-1936~{\rm GeV}+~12.9\,m_{1/2},~$\\ B & $400~{\rm
GeV}~\leq~m_{1/2}~\leq~900~{\rm GeV}~,$ & $m_0~=~~~~~4.93~{\rm
GeV}+0.229\,m_{1/2},~$\\ C & $500~{\rm
GeV}~\leq~m_{1/2}~\leq~700~{\rm GeV}~,$ & $m_0~=~~~~~~~\,54~{\rm
GeV}+0.297\,m_{1/2},~$\\ D & $575~{\rm
GeV}~\leq~m_{1/2}~\leq~725~{\rm GeV}~,$ & $m_0~=~~~~~~600~{\rm
GeV}.$\\ \end{tabular}
\end{table}

Starting with Fig.\ \ref{fig:02} and $\tan\beta=10$, the bulk
region of equally low scalar and fermion masses is all but
excluded by the $b\to s \gamma$ branching ratio. This leaves us
with two favoured regions. Our benchmark point A lies in the
so-called focus point region of low fermion masses $m_{1/2}$,
where the lightest neutralinos are relatively heavy, have a
significant Higgsino component, and annihilate dominantly into
pairs of electroweak gauge bosons. Our values for the universal
masses are smaller than those of the pre-WMAP point SPS 2
\cite{Allanach:2002nj, Aguilar-Saavedra:2005pw} ($m_0=1450$ GeV,
$m_{1/2}=300$ GeV) and post-WMAP point BDEGOP E'
\cite{Battaglia:2003ab} ($m_0=1530$ GeV, $m_{1/2}=300$ GeV), which
lie outside the region favoured by $a_\mu$ (grey-shaded) and lead
to collider-unfriendly heavy squark and gaugino masses. Our
benchmark point B lies in the co-annihilation branch of low scalar
masses $m_0$, where the lighter $\stau_1$ mass eigenstate is not
much heavier than the lightest neutralino, the two having a
considerable co-annihilation cross section. This point differs
from the points SPS 3 ($m_0=90$ GeV, $m_{1/2}=400$ GeV) and BDEGOP
C' ($m_0=85$ GeV, $m_{1/2}=400$ GeV) only very little in the
scalar mass. \\

At the larger value of $\tan\beta=30$ in Fig.\ \ref{fig:03}, only
the co-annihilation region survives the constraints coming from
$b\to s\gamma$ decays, where we choose our point C, which has
slightly higher masses than both SPS 1b ($m_0=200$ GeV,
$m_{1/2}=400$ GeV) and BDEGOP I' ($m_0=175$ GeV, $m_{1/2}=350$
GeV). Finally for the very large value of $\tan\beta=50$ in Fig.\
\ref{fig:04}, the bulk region reappears at relatively heavy scalar
and fermion masses. Here, the couplings of the heavier scalar and
pseudo-scalar Higgses $H^0$ and $A^0$ to bottom quarks and
tau-leptons and the charged-Higgs coupling to top-bottom pairs are
significantly enhanced, resulting e.g.\ in increased dark matter
annihilation cross sections through $s$-channel Higgs-exchange
into bottom-quark final states. So as $\tan\beta$ increases
further, the so-called Higgs-funnel region eventually makes its
appearance on the diagonal of large scalar and fermion masses. We
choose our point D in the concentrated (bulky) region favoured by
cosmology and $a_\mu$ at masses, that are slightly higher than
those of SPS 4 ($m_0=400$ GeV, $m_{1/2}=300$ GeV) and BDEGOP L'
($m_0=300$ GeV, $m_{1/2}=450$ GeV). In this scenario, squarks and
gluinos are very heavy with masses above 1 TeV.\\

Let us now turn to the dependence of the precision variables
discussed in Sec.\ \ref{sec:const} on the flavour violating
parameter $\lambda$ in our four benchmark scenarios. As already
mentioned, we expect the leptonic observable $a_\mu$ to depend
weakly on the squark sector, which is confirmed by our numerical
analysis of the SUSY contribution $a_\mu^{(SUSY)}$. We find
constant values of 6, 14, 16, and 13$\times10^{-10}$ for the
benchmarks A, B, C, and D, all of which lie well within $2\sigma$
(the latter three even within $1\sigma$) of the experimentally
favoured range $(22\pm10)\times10^{-10}$.\\

\begin{figure}
 \centering
 \includegraphics[width=0.32\columnwidth]{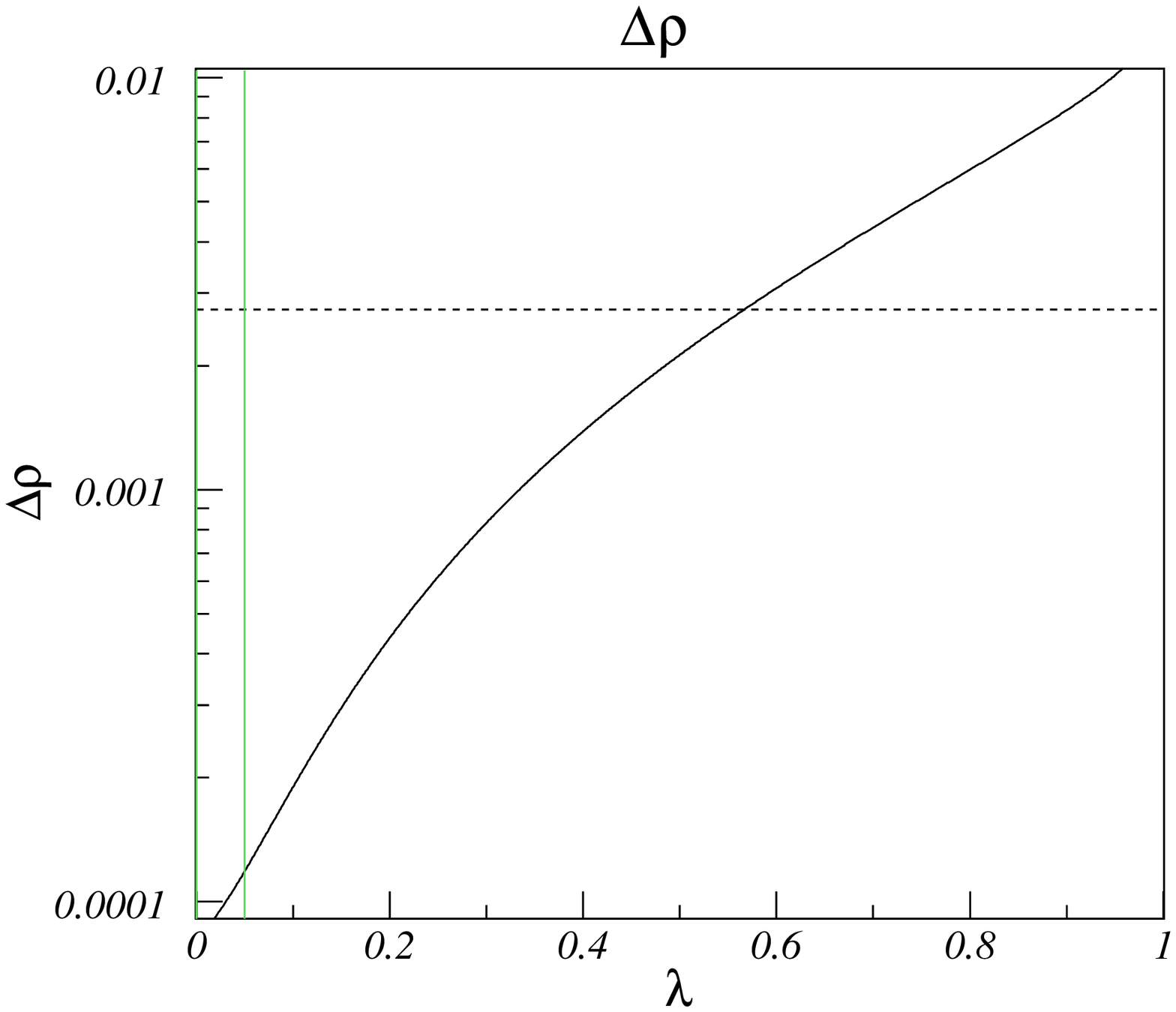}
 \includegraphics[width=0.32\columnwidth]{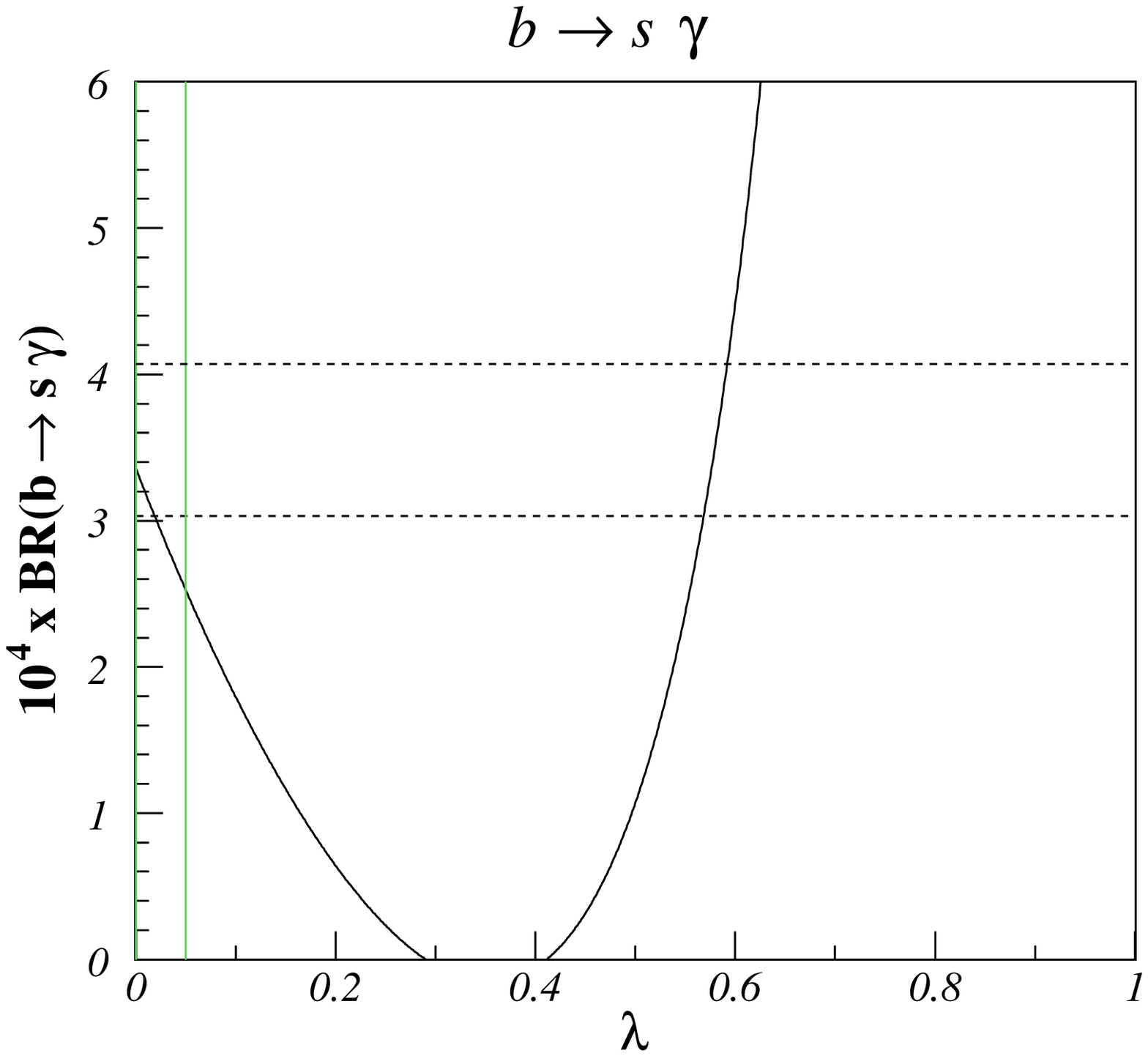}
 \includegraphics[width=0.32\columnwidth]{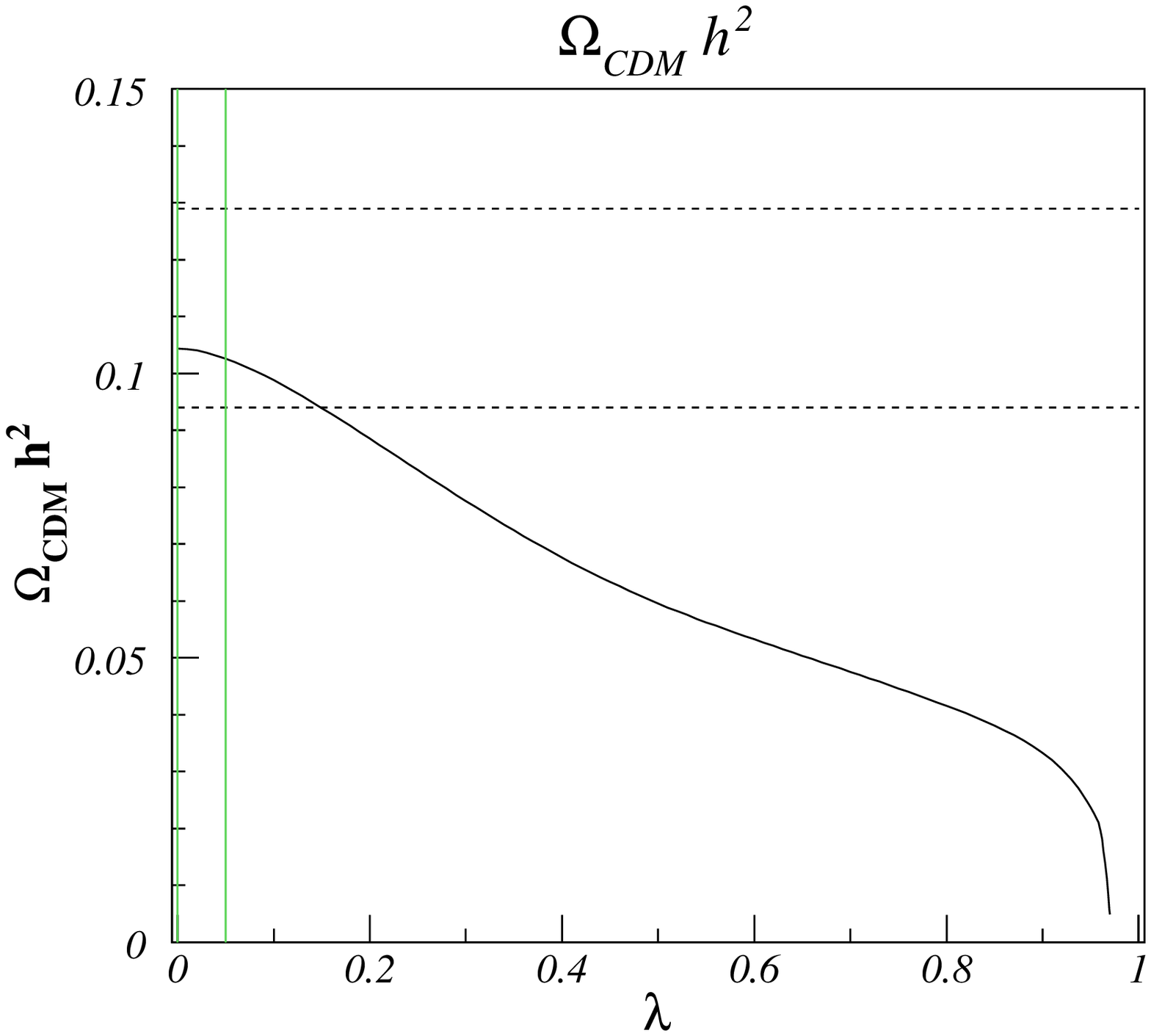}\vspace*{2mm}
 \includegraphics[width=0.32\columnwidth]{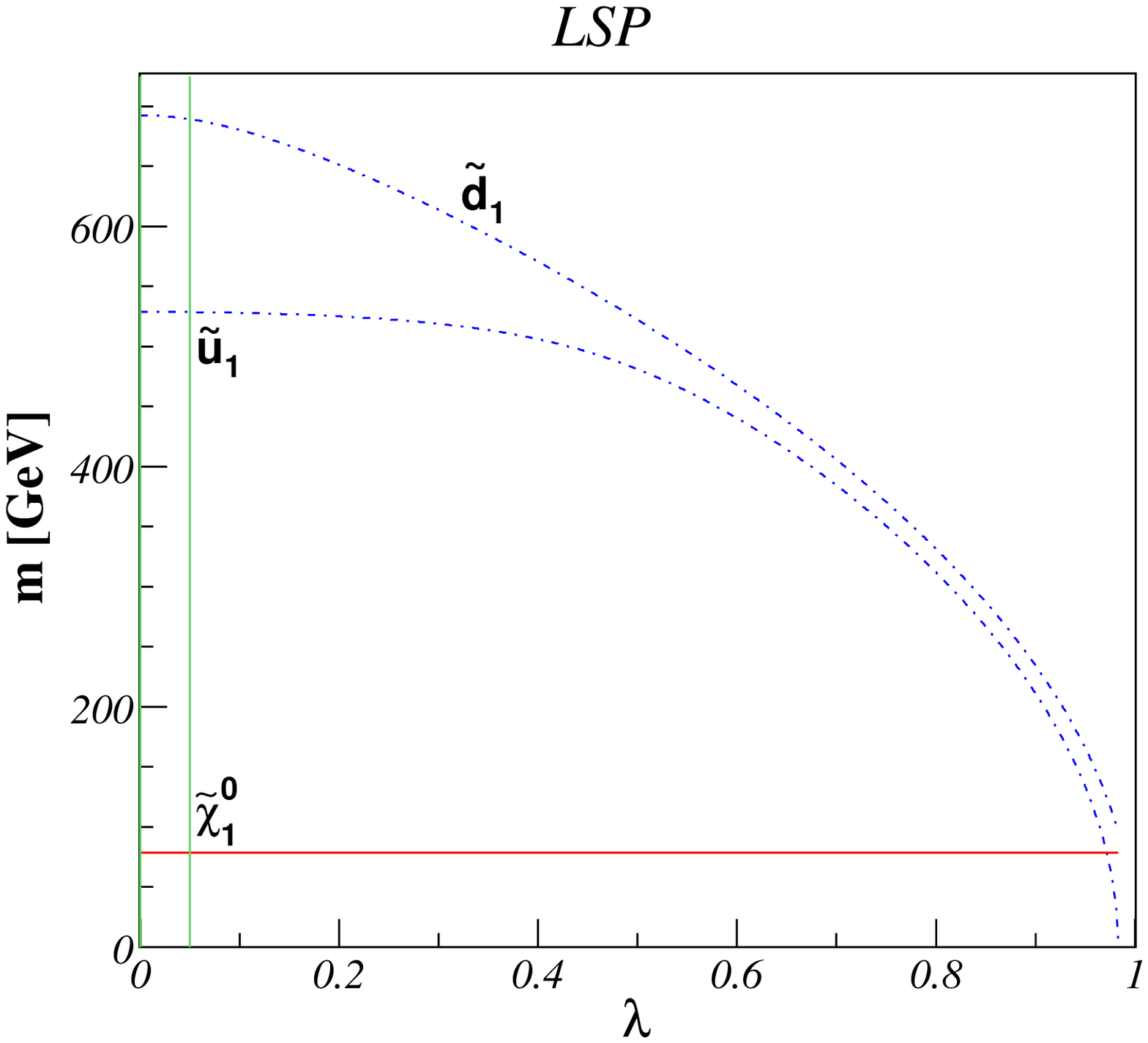}
 \includegraphics[width=0.32\columnwidth]{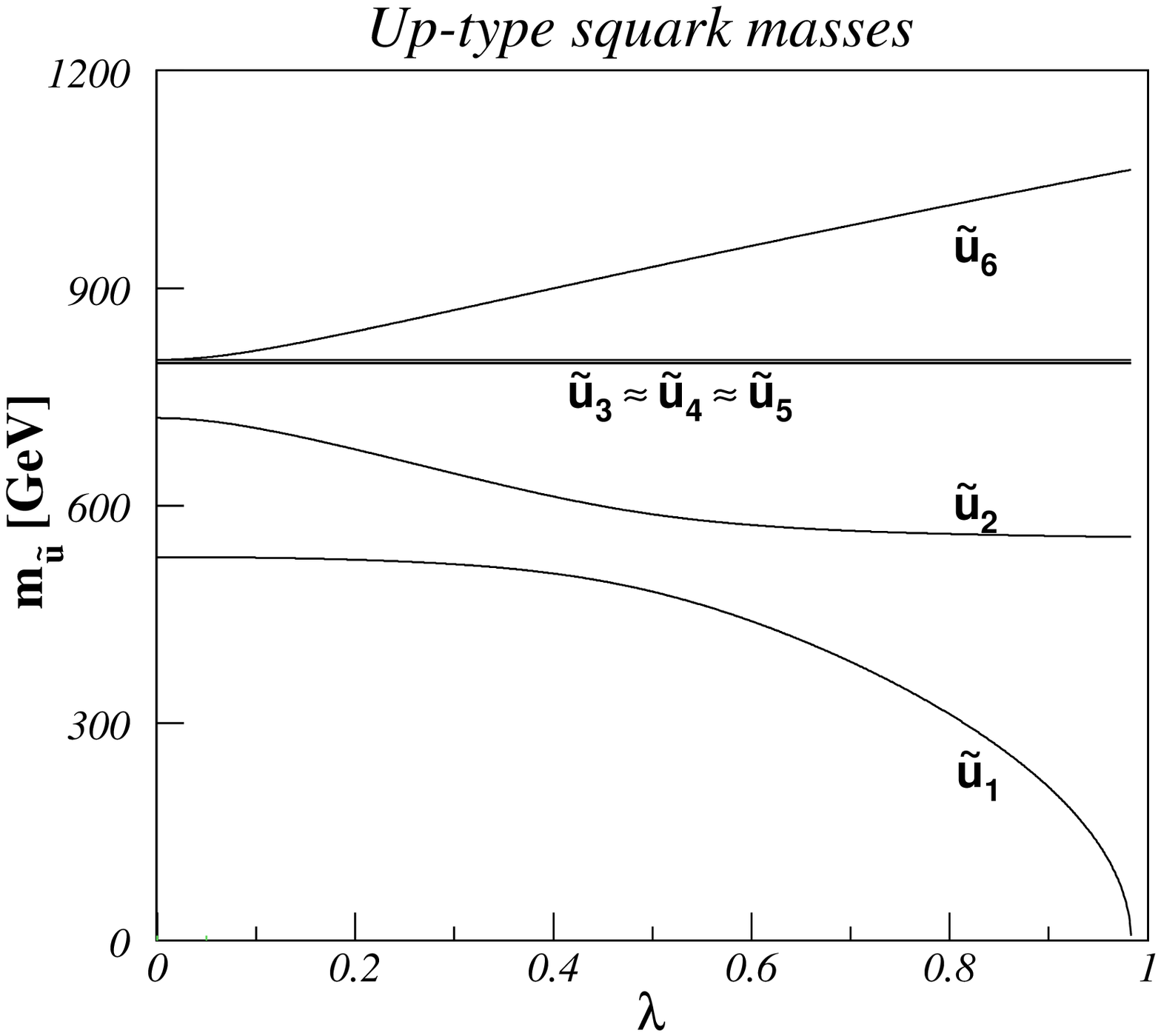}
 \includegraphics[width=0.32\columnwidth]{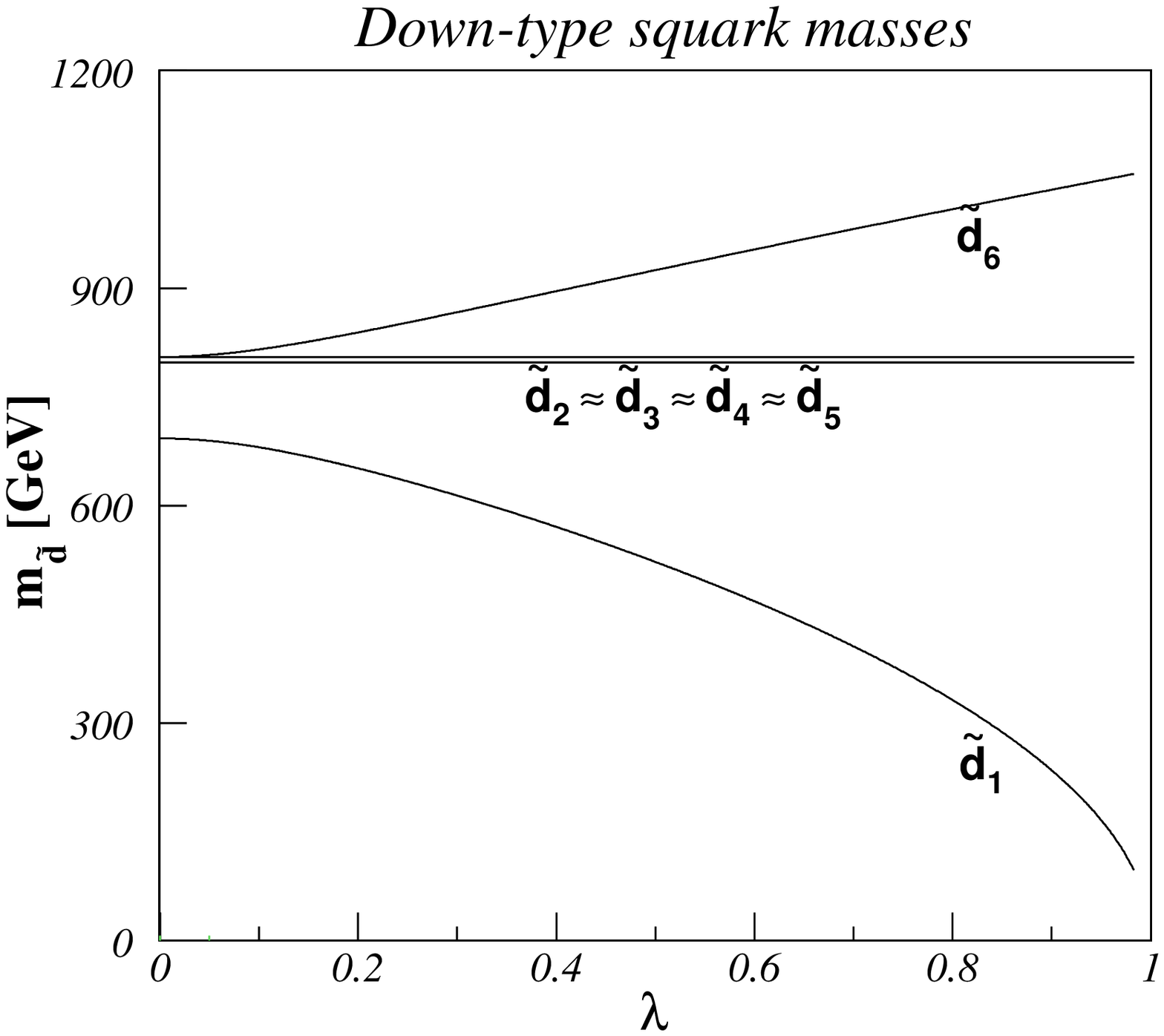}
 \caption{\label{fig:05}Dependence of the precision variables $\Delta\rho$,
          BR$(b\to s\gamma)$, and the cold dark matter relic density
          $\Omega_{CDM}h^2$ (top) as well as of the lightest SUSY particle,
          up- and down-type squark masses (bottom) on the NMFV parameter
          $\lambda$ in our benchmark scenario A. The experimentally allowed
          ranges (within $2\sigma$) are indicated by horizontal dashed
          lines.}\vspace{2mm}
 \includegraphics[width=0.32\columnwidth]{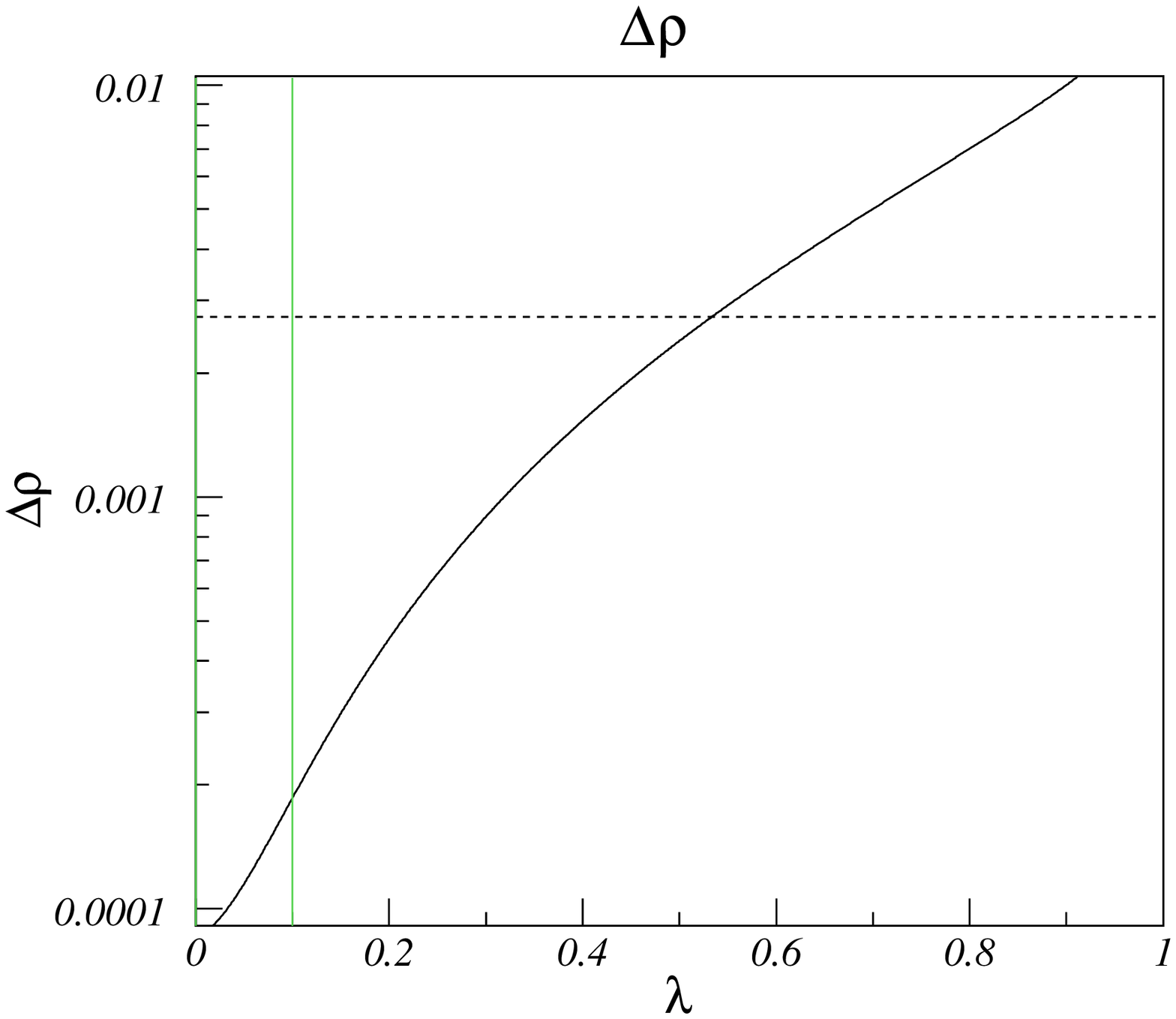}
 \includegraphics[width=0.32\columnwidth]{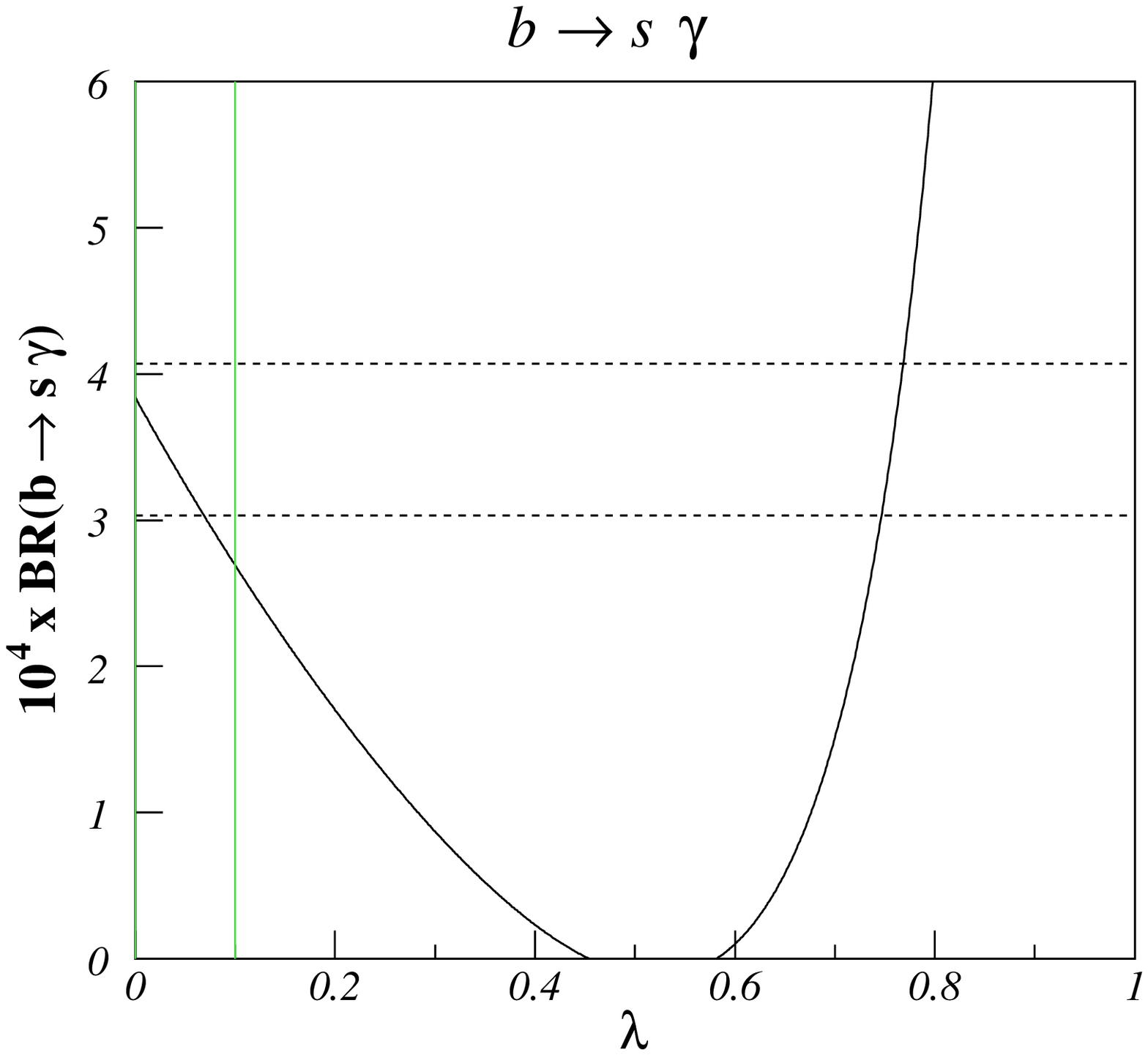}
 \includegraphics[width=0.32\columnwidth]{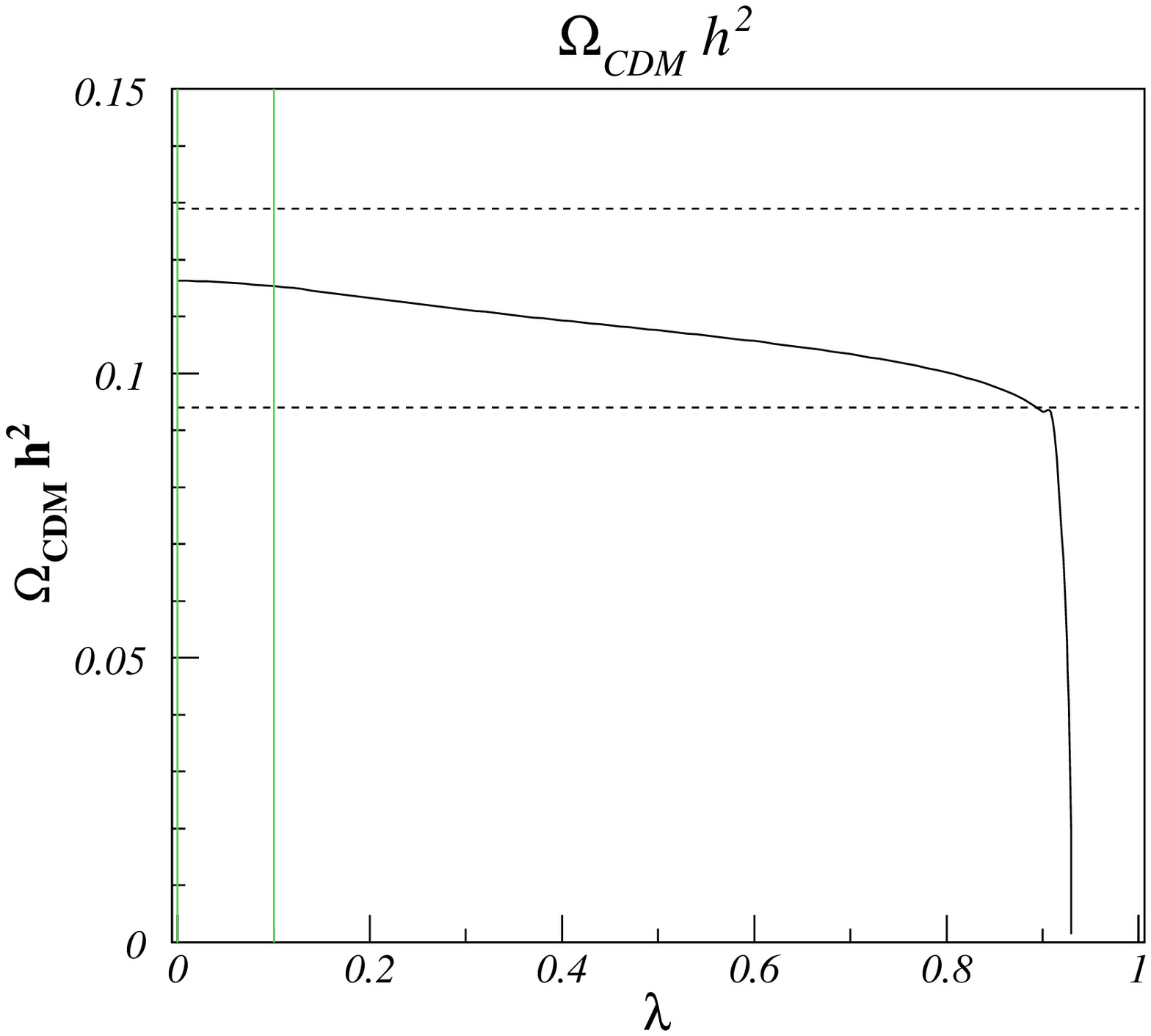}\vspace*{2mm}
 \includegraphics[width=0.32\columnwidth]{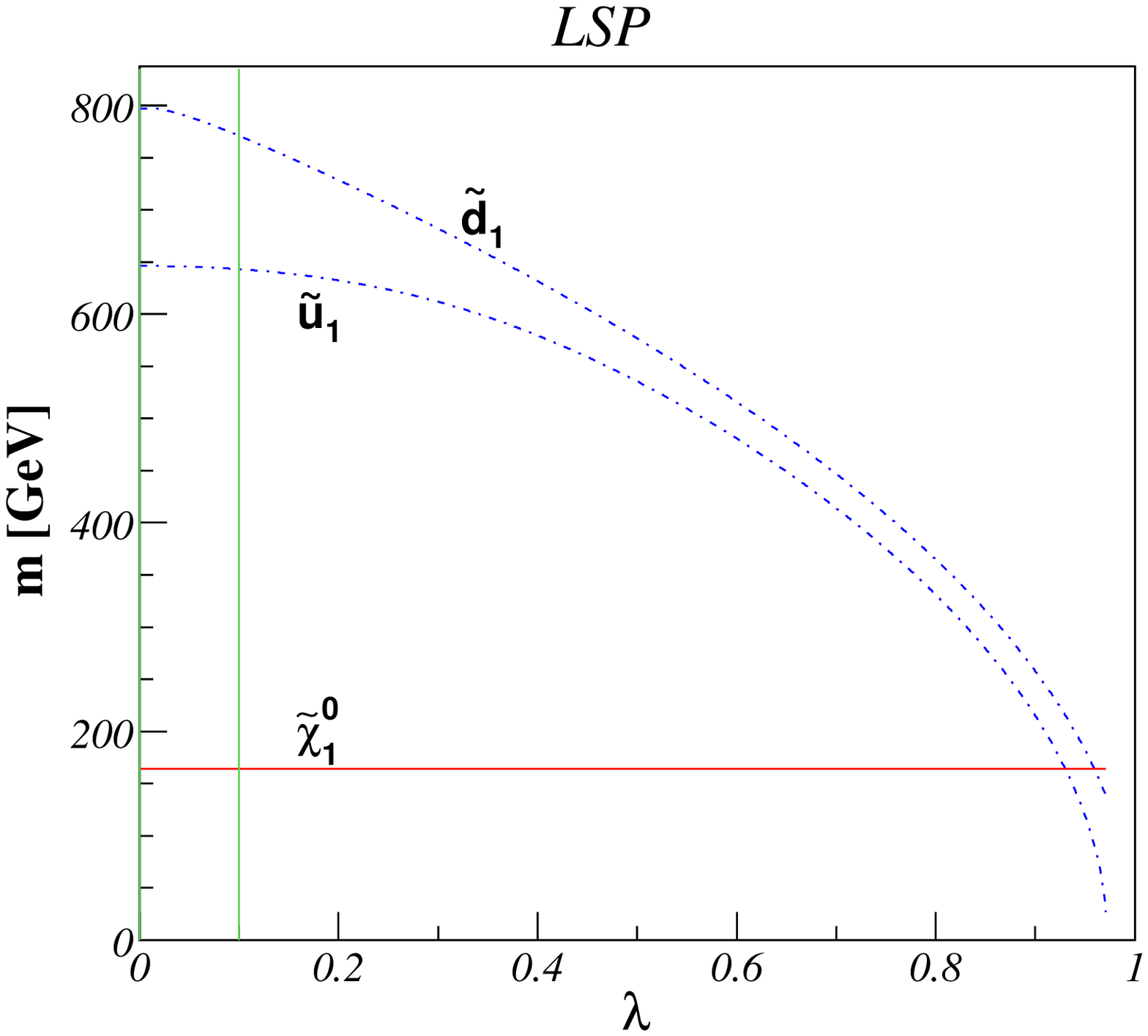}
 \includegraphics[width=0.32\columnwidth]{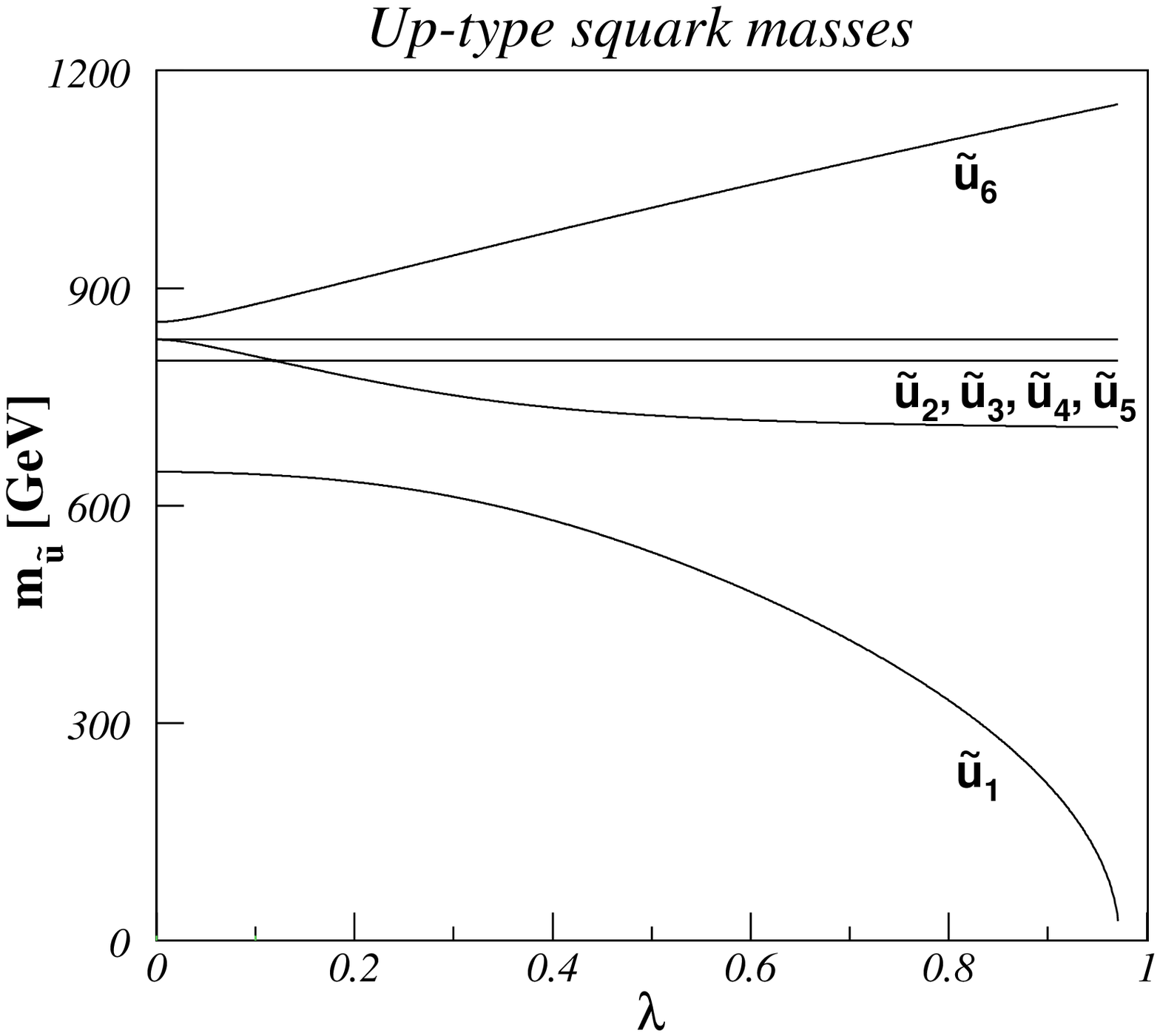}
 \includegraphics[width=0.32\columnwidth]{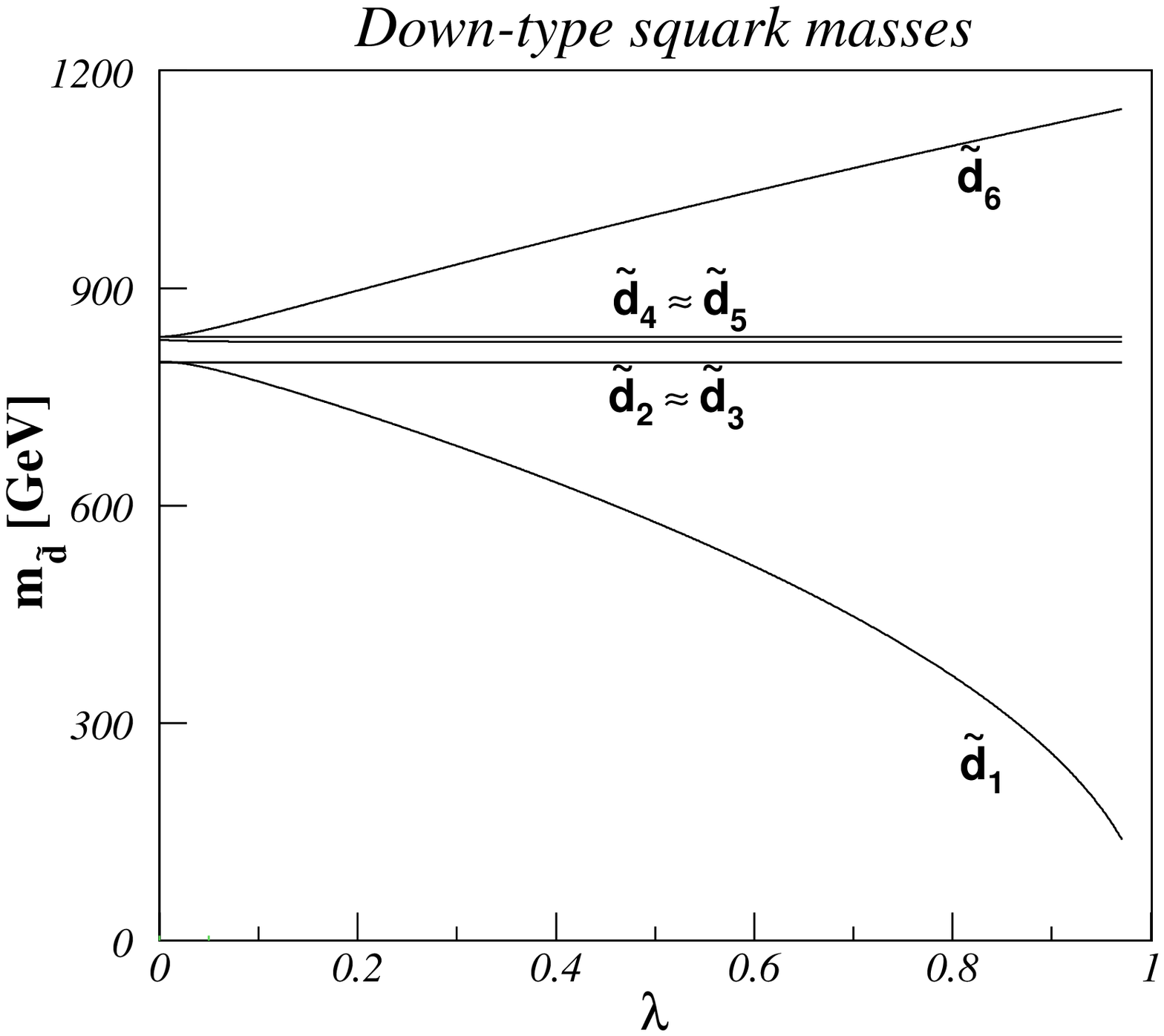}
 \caption{\label{fig:06}Same as Fig.\ \ref{fig:05} for our benchmark
          scenario B.}
\end{figure}

\begin{figure}
 \centering
 \includegraphics[width=0.32\columnwidth]{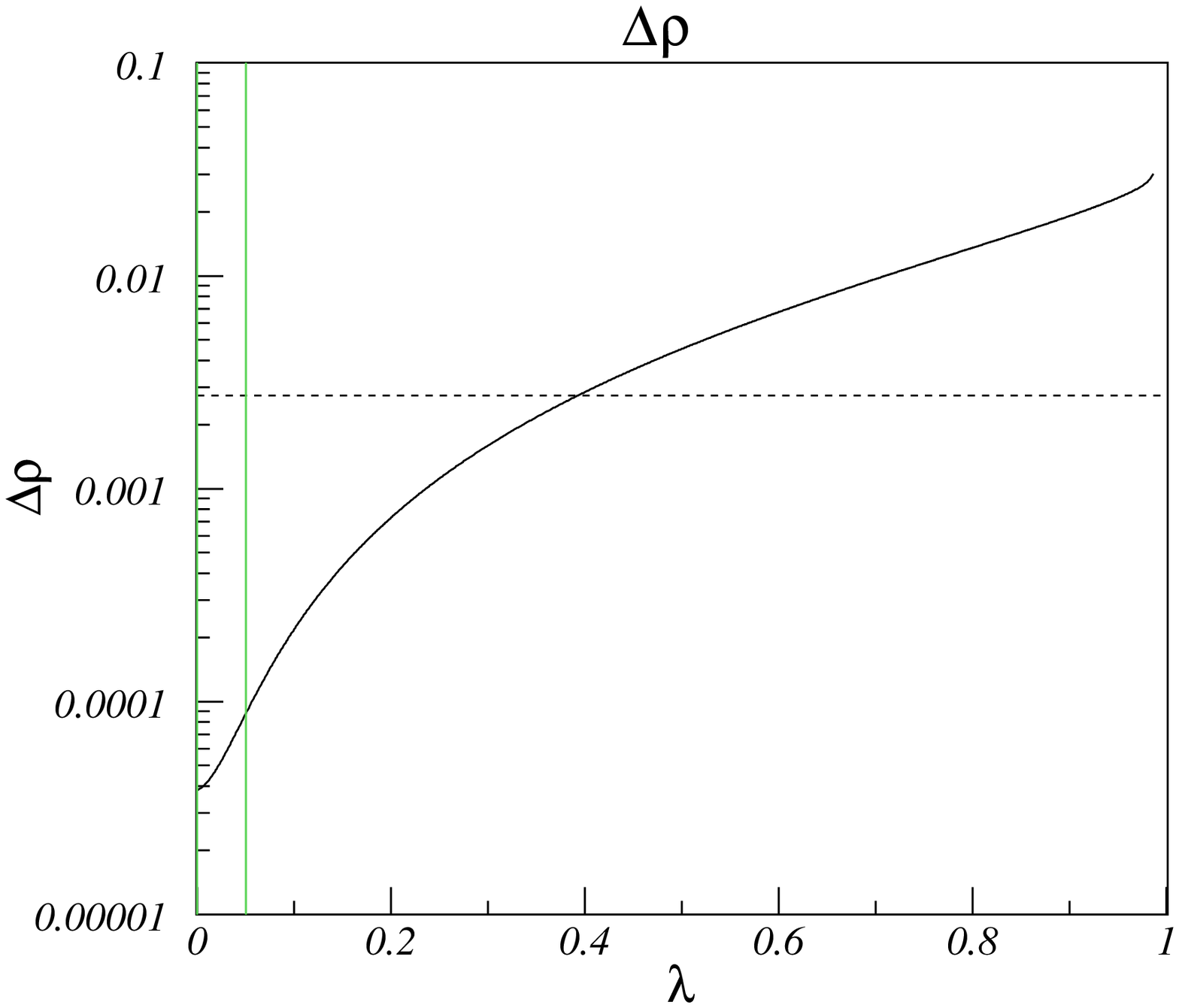}
 \includegraphics[width=0.32\columnwidth]{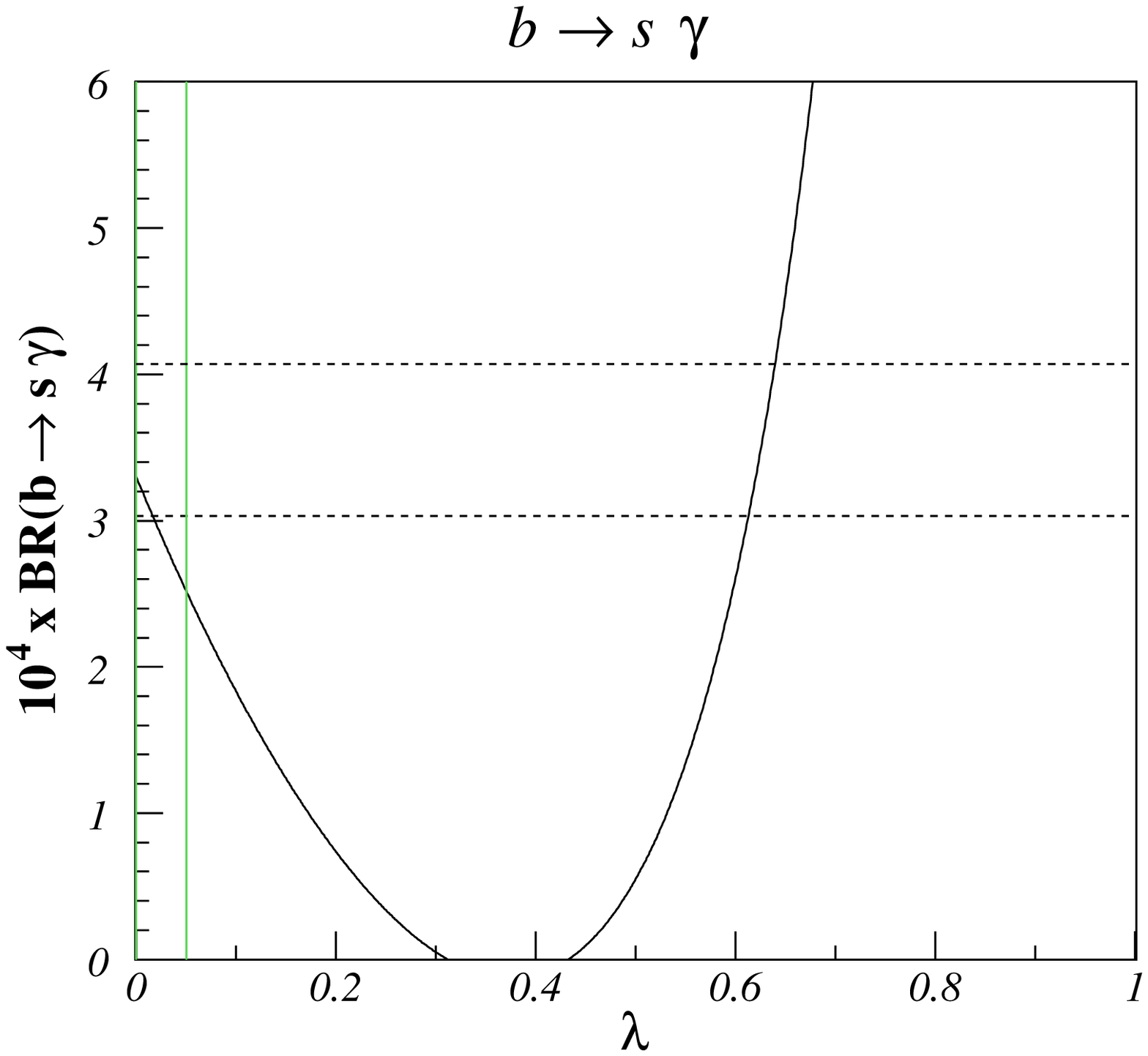}
 \includegraphics[width=0.32\columnwidth]{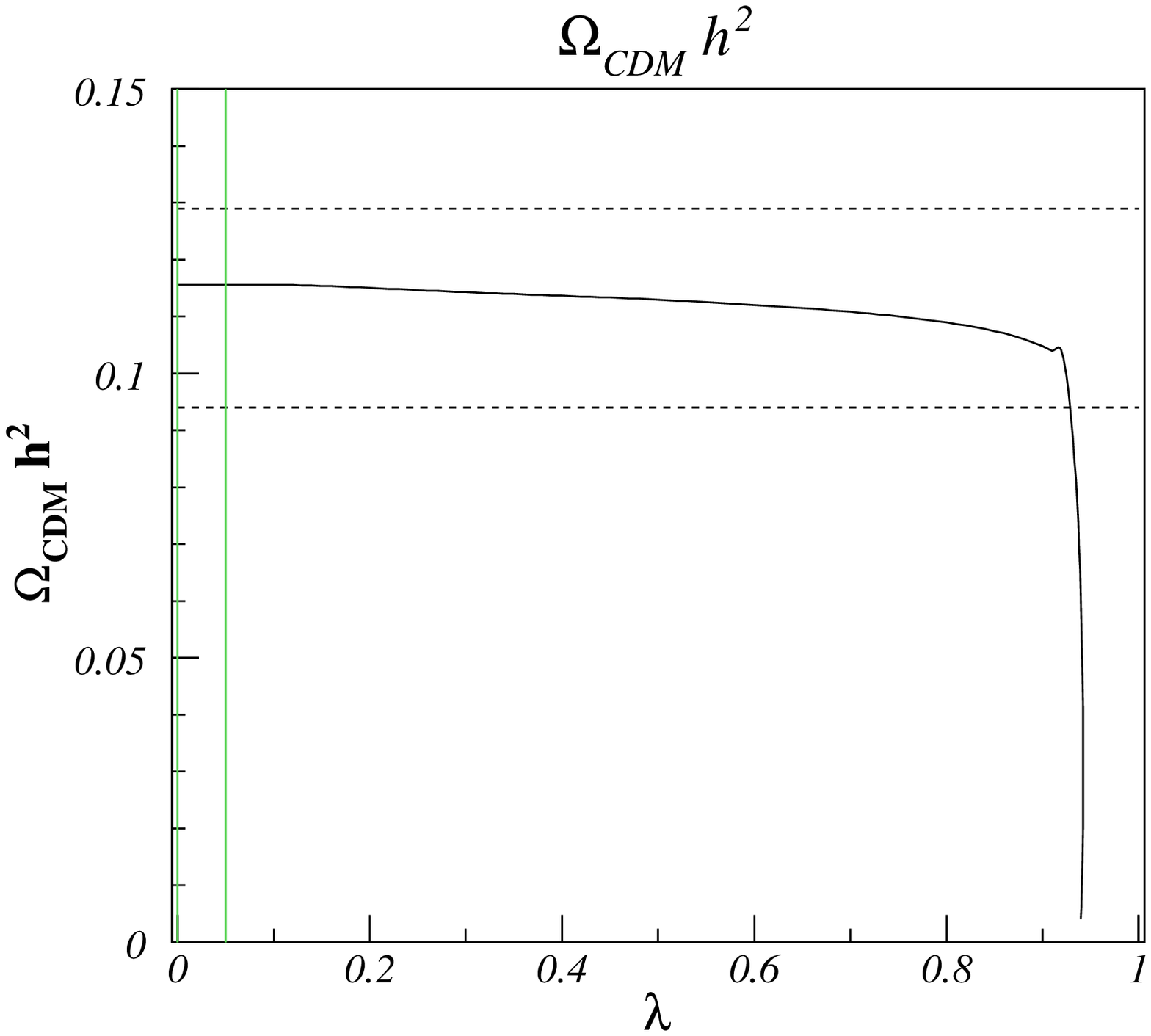}\vspace*{2mm}
 \includegraphics[width=0.32\columnwidth]{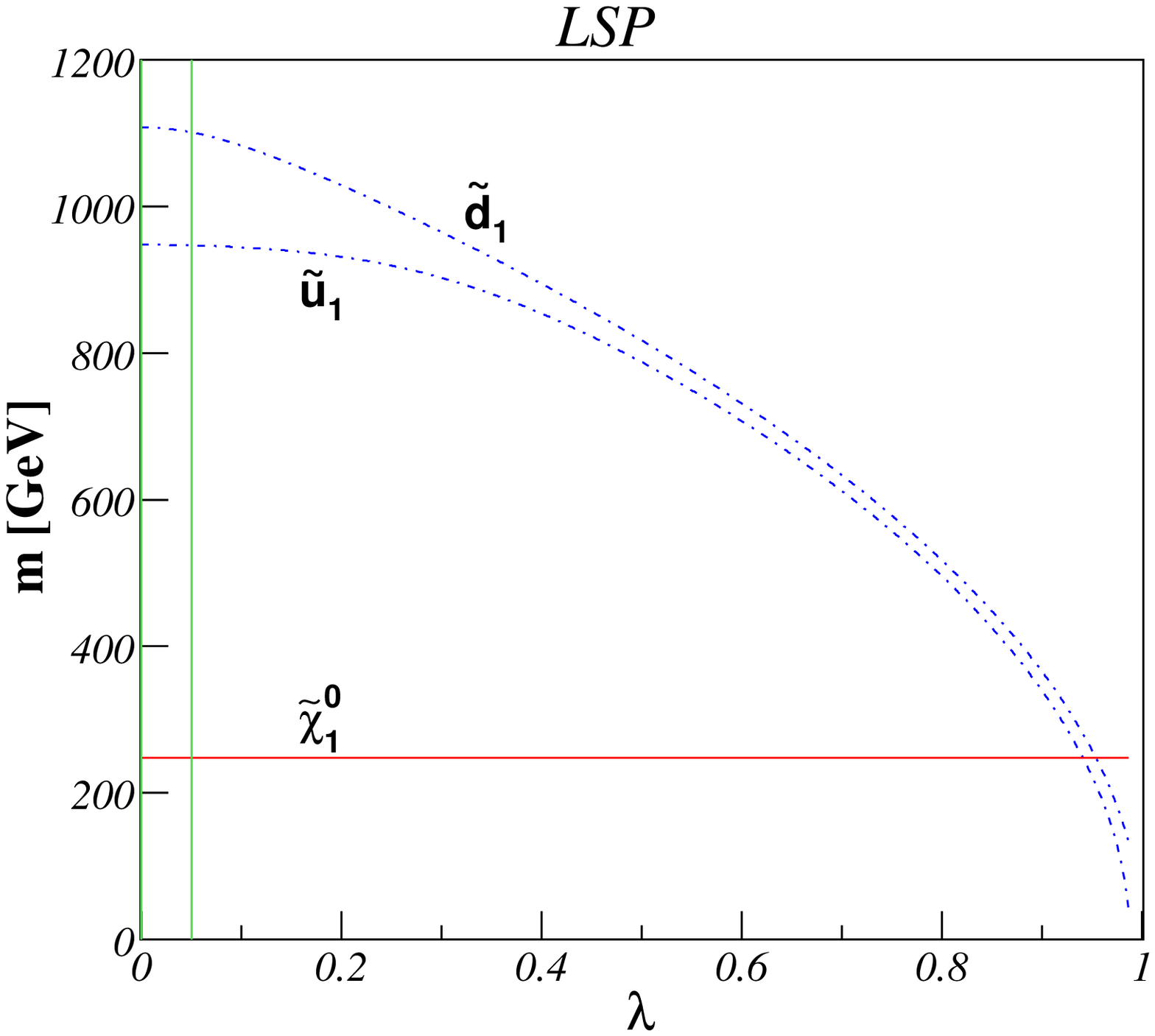}
 \includegraphics[width=0.32\columnwidth]{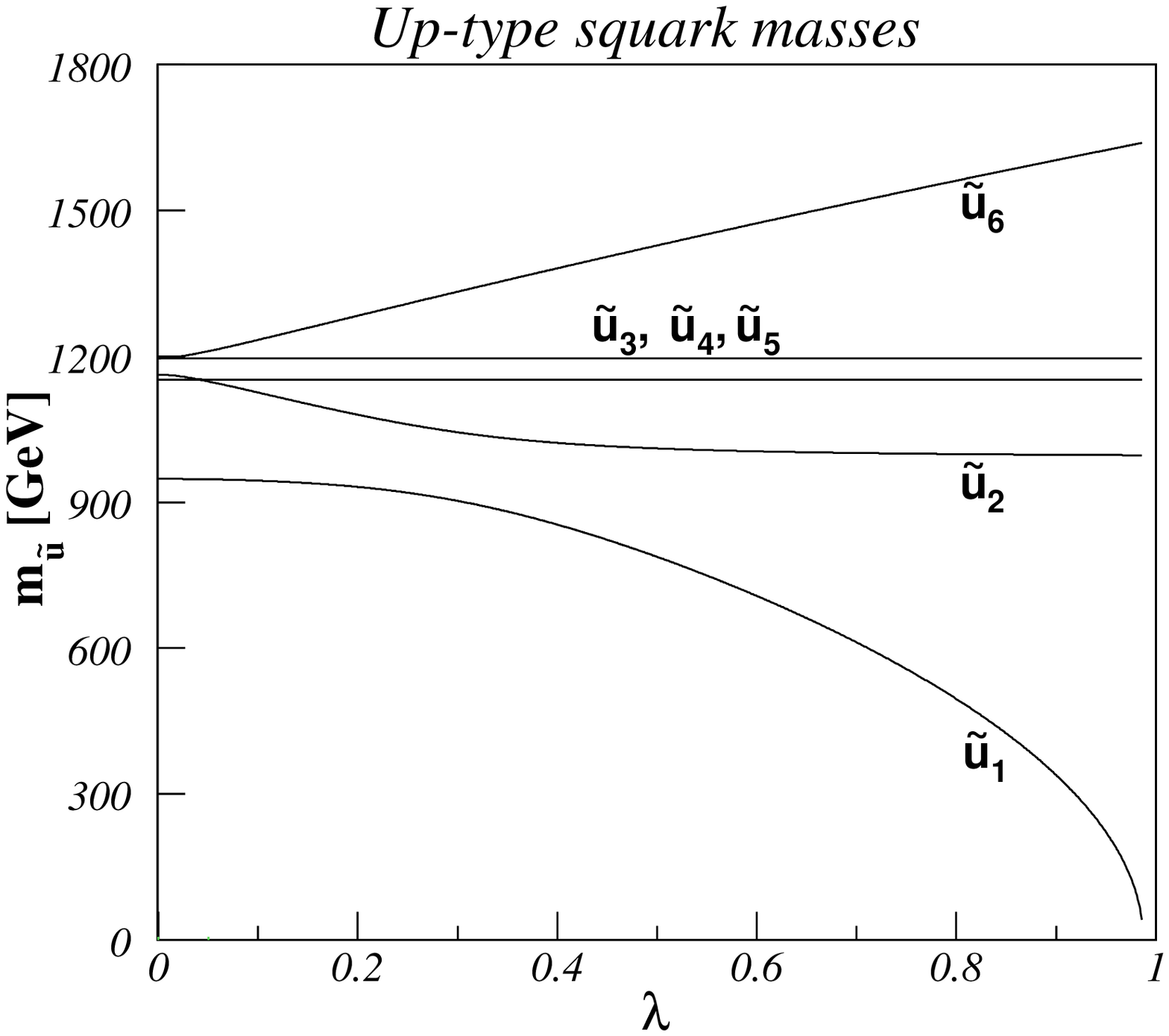}
 \includegraphics[width=0.32\columnwidth]{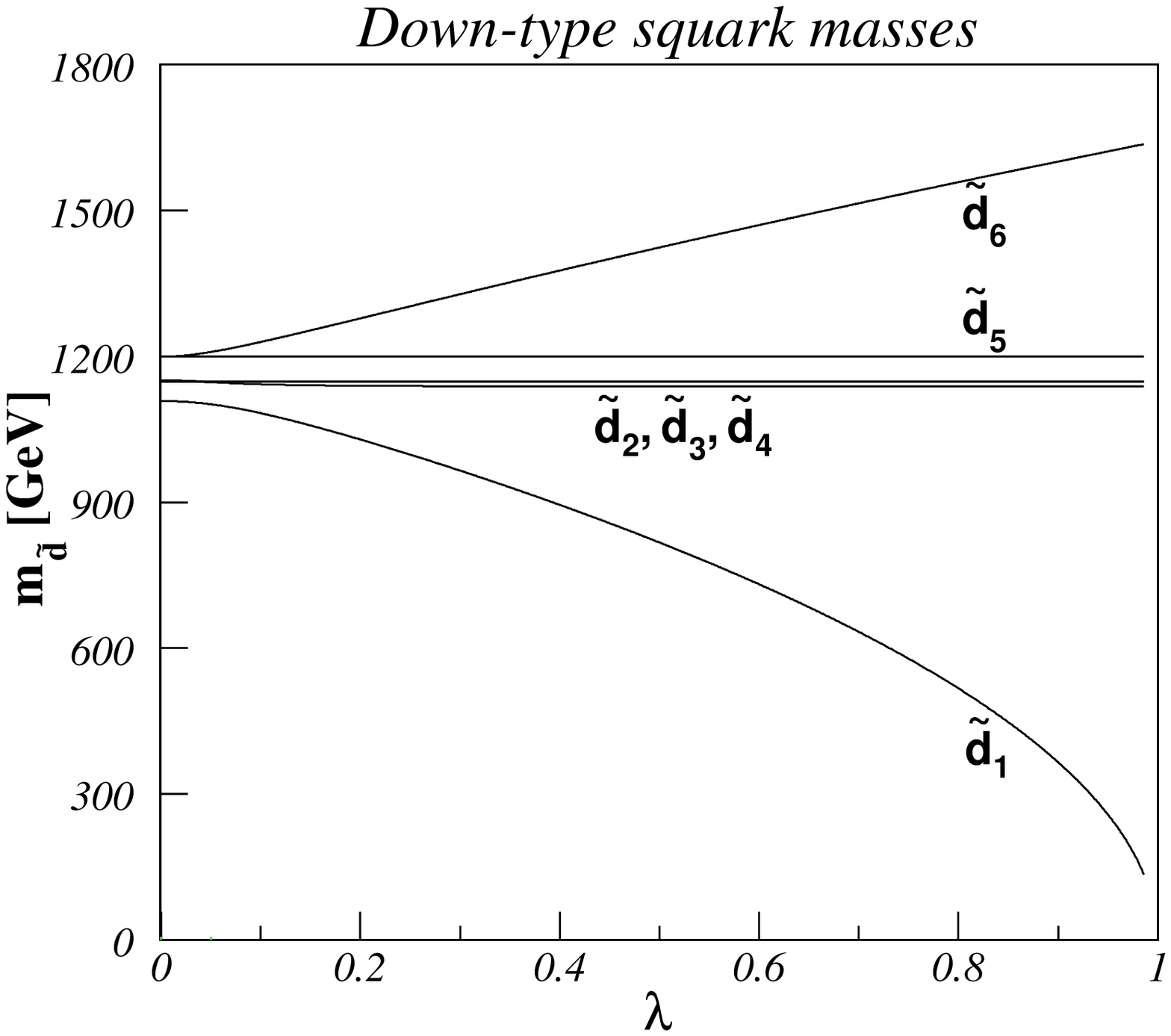}
 \caption{\label{fig:07}Same as Fig.\ \ref{fig:07} for our benchmark
          scenario C.}\vspace{2mm}
 \includegraphics[width=0.32\columnwidth]{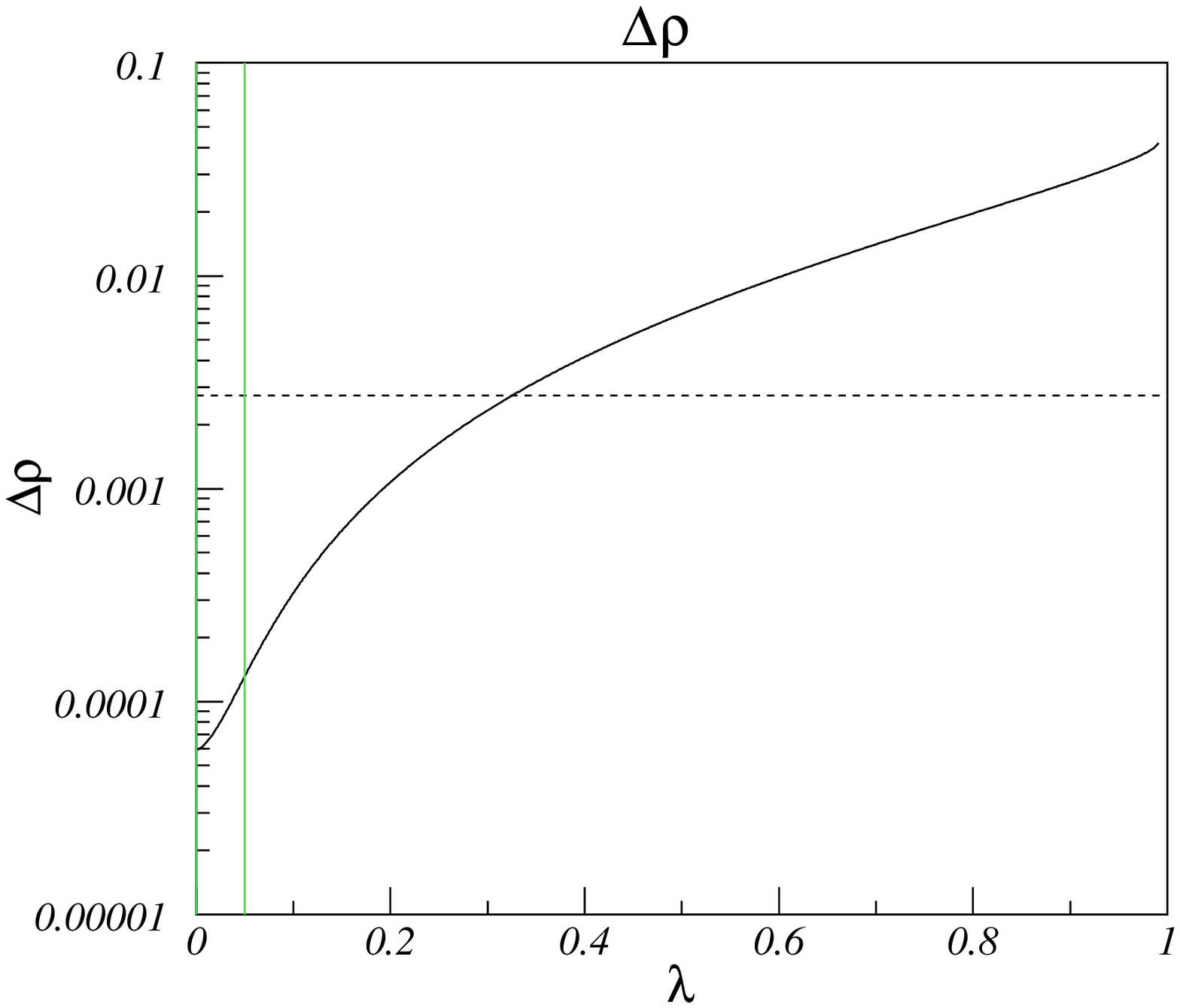}
 \includegraphics[width=0.32\columnwidth]{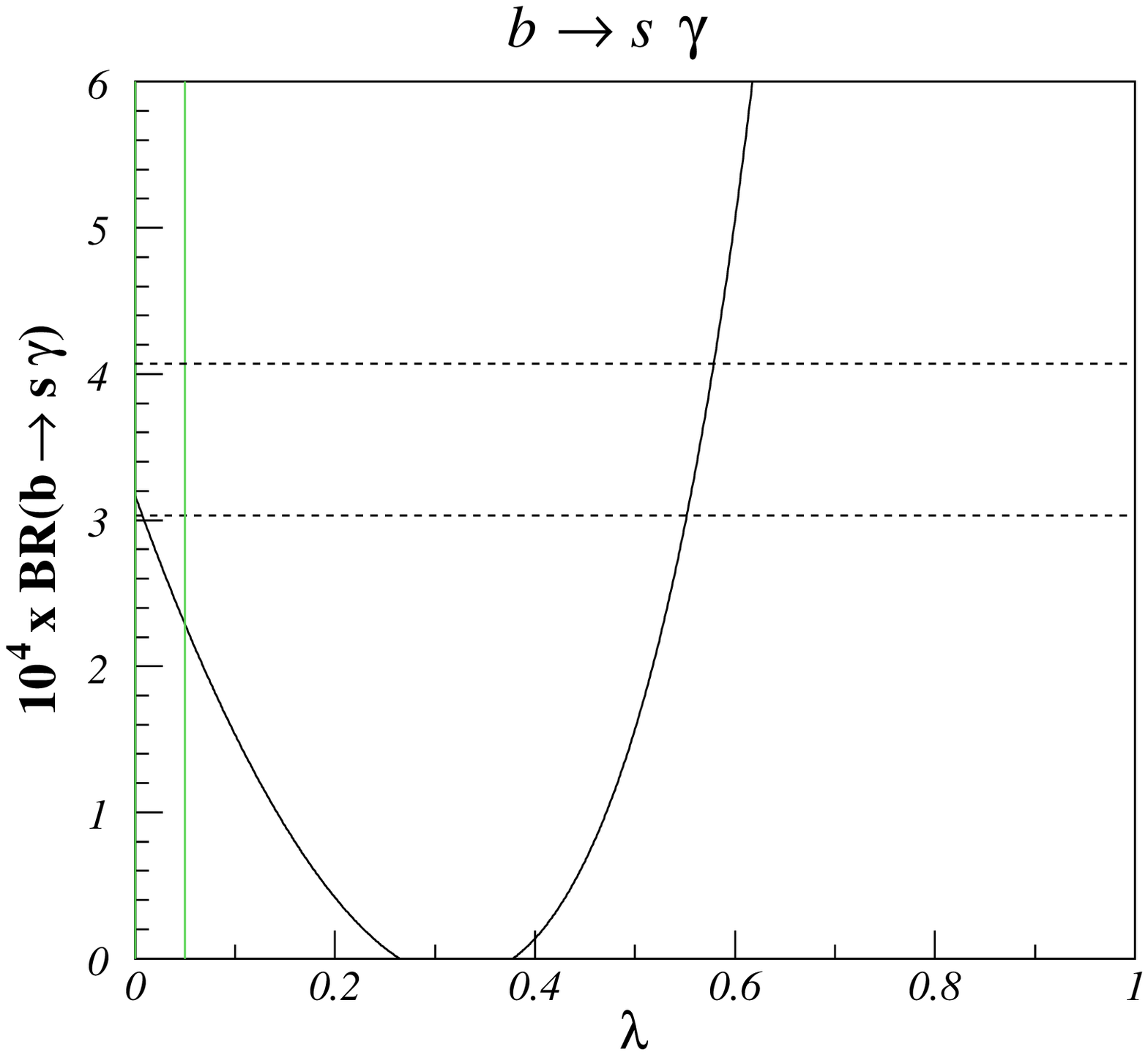}
 \includegraphics[width=0.32\columnwidth]{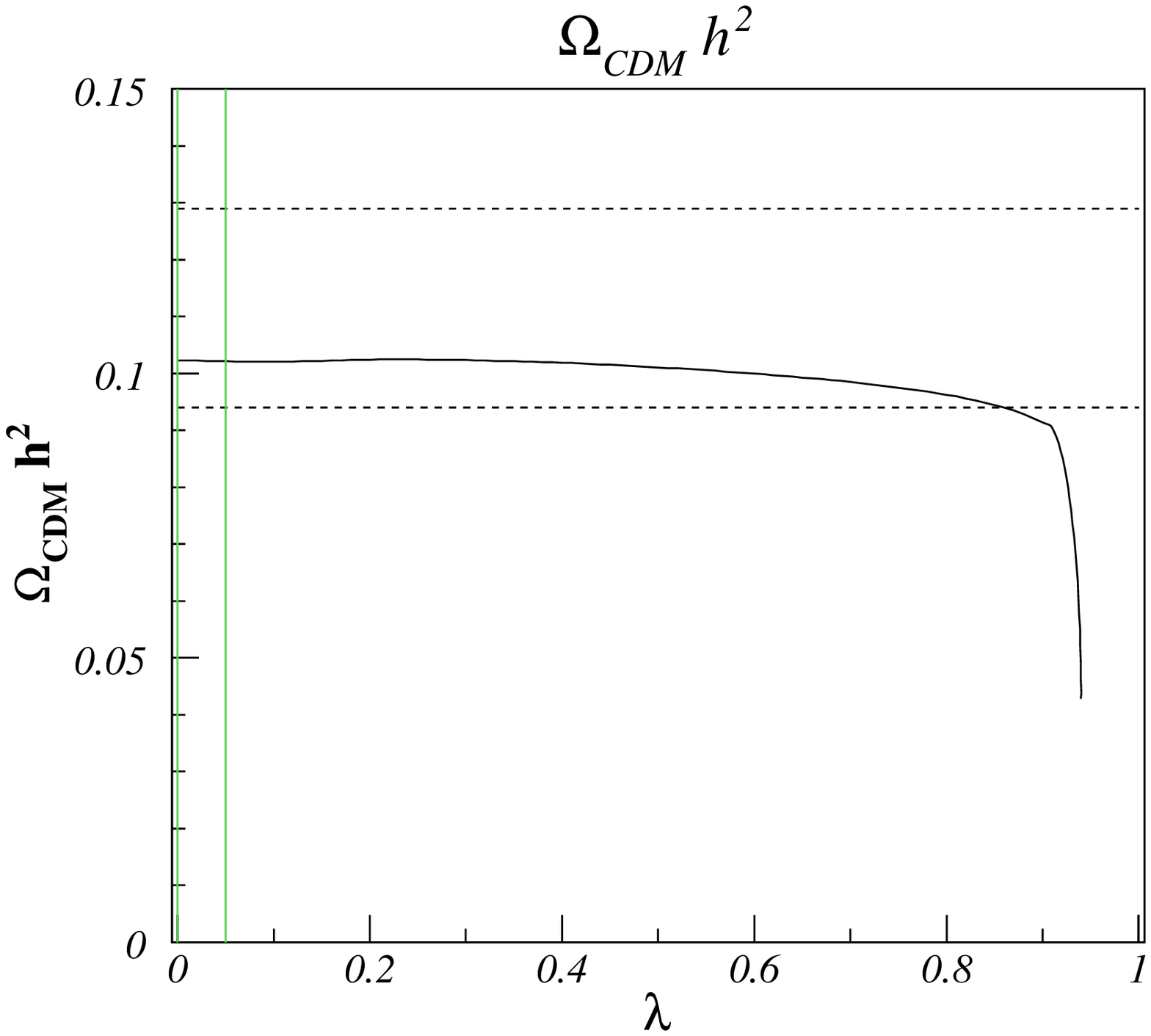}\vspace*{2mm}
 \includegraphics[width=0.32\columnwidth]{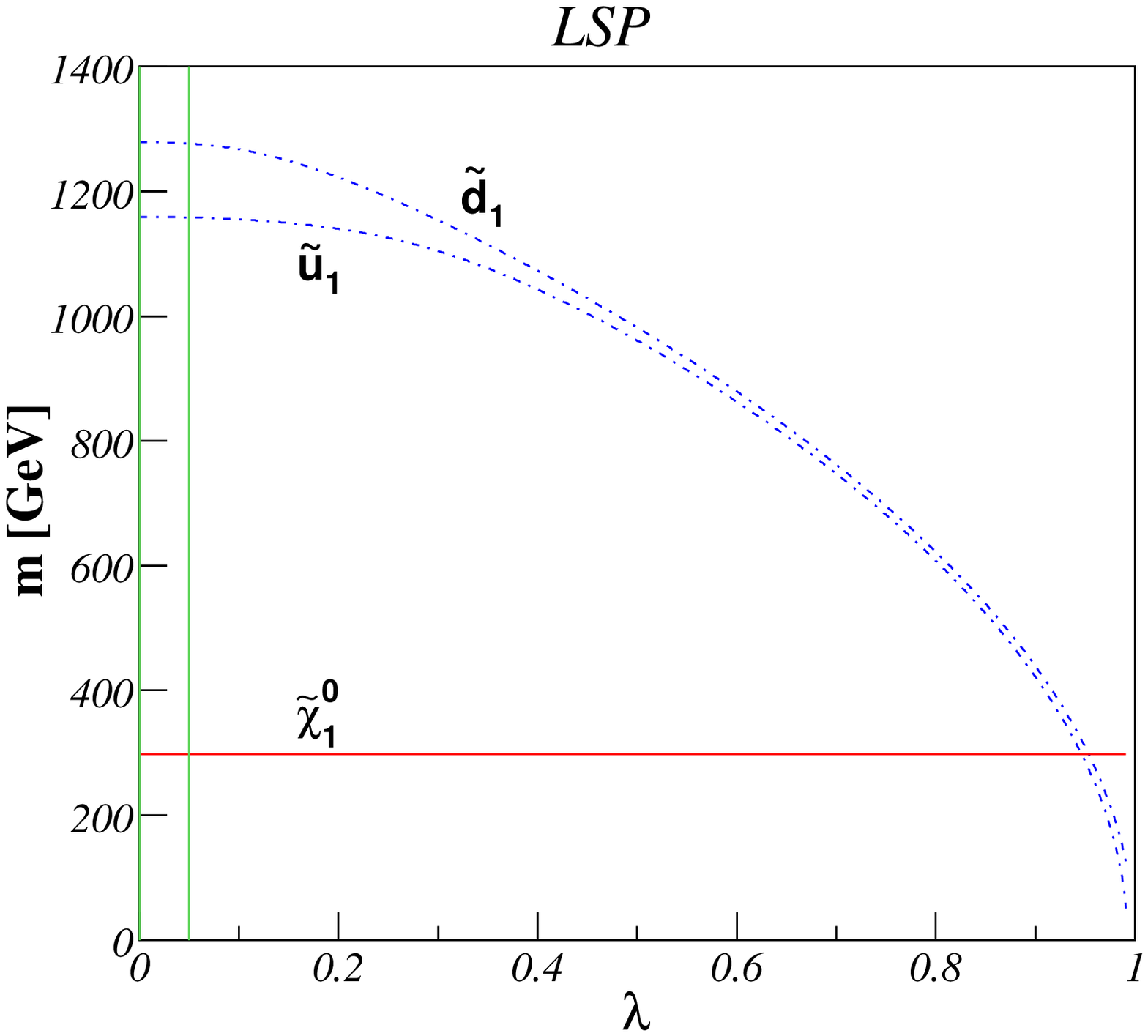}
 \includegraphics[width=0.32\columnwidth]{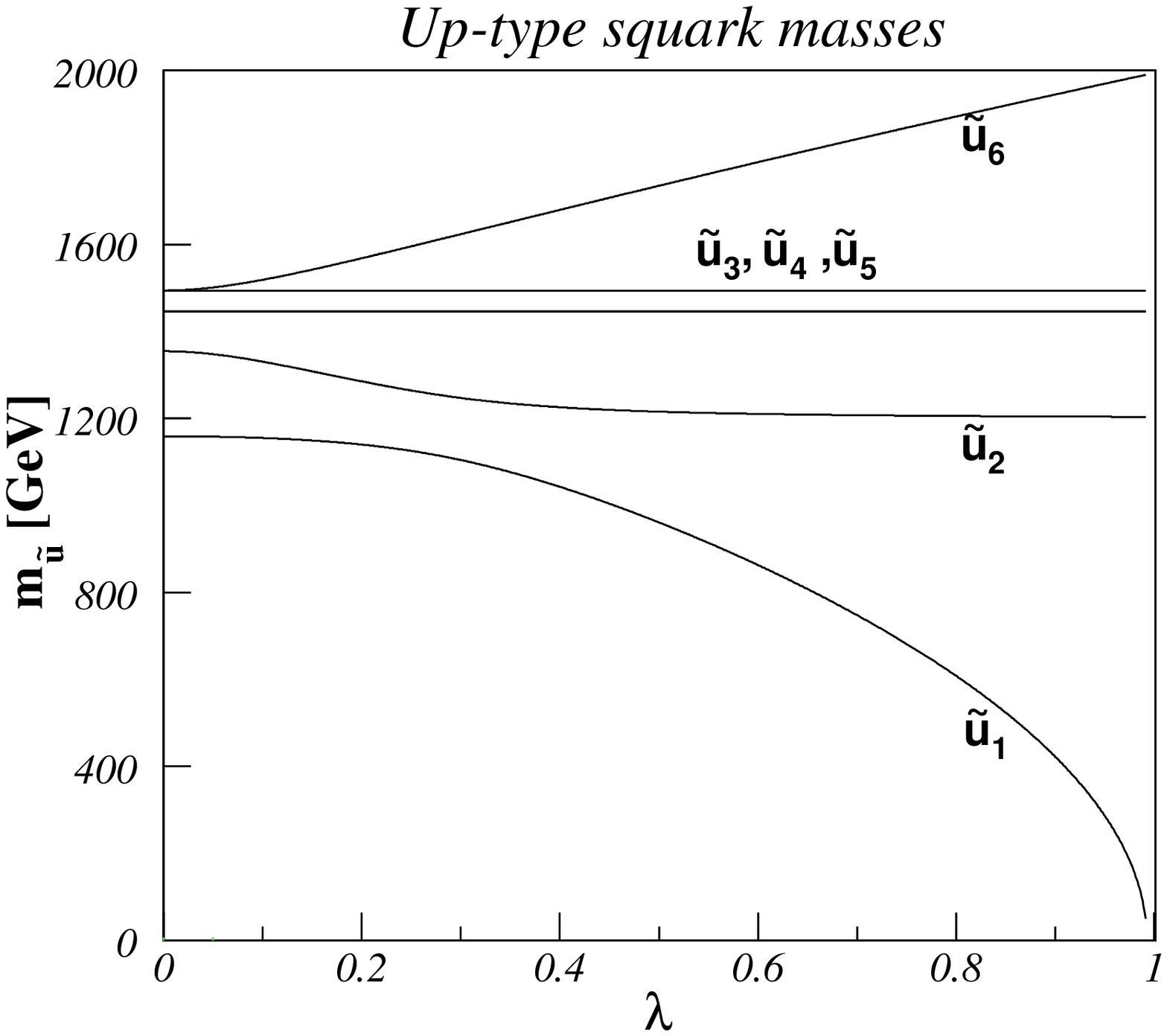}
 \includegraphics[width=0.32\columnwidth]{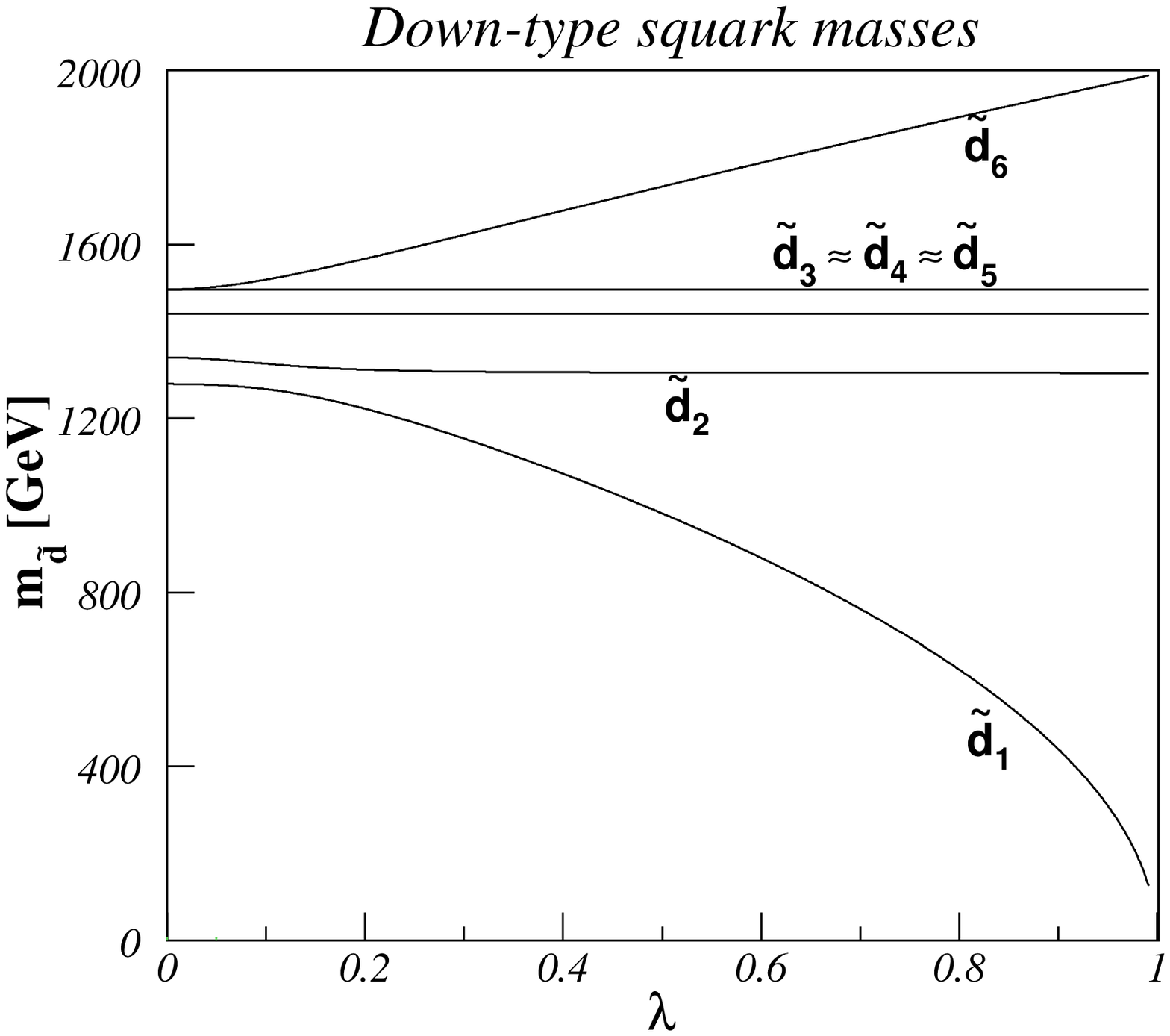}
 \caption{\label{fig:08}Same as Fig.\ \ref{fig:05} for our benchmark
          scenario D.}
\end{figure}

The electroweak precision observable $\Delta\rho$ is shown first
in Figs.\ \ref{fig:05}-\ref{fig:08} for the four benchmark
scenarios, where only the experimental upper bound of the
$2\sigma$-range is visible as a dashed line. While the self-energy
diagrams of the electroweak gauge bosons depend obviously strongly
on the helicities, flavours, and mass eigenvalues of the squarks
in the loop, the SUSY masses in our scenarios are sufficiently
small and the experimental error is still sufficiently large to
allow for relatively large values of $\lambda\leq 0.57$, 0.52,
0.38, and 0.32 for the benchmark points A, B, C, and D,
respectively. As mentioned above, $\Delta\rho$ conversely
constrains SUSY models in cMFV ($\lambda=0$) only for masses above
2000 GeV for $m_0$ and 1500 GeV for $m_{1/2}$.\\

The next diagram in Figs.\ \ref{fig:05}-\ref{fig:08} shows the
dependence of the most stringent low-energy constraint, coming
from the good agreement between the measured $b\to s\gamma$
branching ratio and the two-loop SM prediction, on the NMFV
parameter $\lambda$. The dashed lines of the $2\sigma$-bands
exhibit two allowed regions, one close to $\lambda=0$ (vertical
green line) and a second one around $\lambda\simeq0.57$, 0.75,
0.62, and 0.57, respectively. As is well-known, the latter are,
however, disfavoured by $b\to s\mu^+\mu^-$ data constraining the
sign of the $b\to s \gamma$ amplitude to be the same as in the SM
\cite{Gambino:2004mv}. We will therefore limit ourselves later to
the regions $\lambda\leq0.05$ (points A, C, and D) and
$\lambda\leq0.1$ (point B) in the vicinity of cMFV.\\

The 95\% confidence-level (or $2\sigma$) region for the cold dark
matter density is shown as a dashed band in the upper right part
of Figs.\ \ref{fig:05}-\ref{fig:08}. However, only the lower bound
(0.094) is of relevance, as the relic density falls with
increasing $\lambda$. This is not so pronounced in our model B as
in our model A, where squark masses are light and the lightest
neutralino has a sizable Higgsino-component, so that squark
exchanges contribute significantly to the annihilation cross
sections. For models C and D there is little sensitivity of
$\Omega_{CDM} h^2$ (except at very large $\lambda$), as the squark
masses are generally larger. The rapid fall-off of the relic
density for very large $\lambda \lesssim 1$ can be understood by
looking at the resulting lightest up- and down-type squark mass
eigenvalues in the lower left part of Figs.\
\ref{fig:05}-\ref{fig:08}. For maximal flavour violation, the
off-diagonal squark mass matrix elements are of similar size as
the diagonal ones, leading to one squark mass eigenvalue that
approaches and finally falls below the lightest neutralino (dark
matter) mass. Light squark propagators and co-annihilation
processes thus lead to a rapidly falling dark matter relic density
and finally to cosmologically excluded NMFV SUSY models, since the
LSP is assumed to be electrically neutral and a colour singlet.\\

An interesting phenomenon of level reordering between neighbouring
states can be observed in the lower central diagrams of Figs.\
\ref{fig:05}-\ref{fig:08} for the two lowest mass eigenvalues of
up-type squarks. As $\lambda$ and the off-diagonal entries in the
mass matrix increase, the splitting between the lightest and
highest mass eigenvalues naturally increases. It is first the
second-lowest mass that decreases up to intermediate values of
$\lambda =0.2...0.5$, whereas the lowest mass is constant, and
only at this point the second-lowest mass becomes constant and
takes approximately the value of the until here lowest squark
mass, whereas the lowest squark mass starts to decrease further
with $\lambda$. These ``avoided crossings'' are a common
phenomenon for Hermitian matrices and reminiscent of meta-stable
systems in quantum mechanics. At the point where the two levels
should cross, the corresponding squark eigenstates mix and change
character. This level-reordering phenomenon occurs also for other
scenarios, as well as for down-type squarks, even if it is not so
pronounced for the latter, as shown in the lower right diagrams of
Figs.\ \ref{fig:05}-\ref{fig:08}.\\

\begin{figure}
 \centering
 \includegraphics[width=0.21\columnwidth]{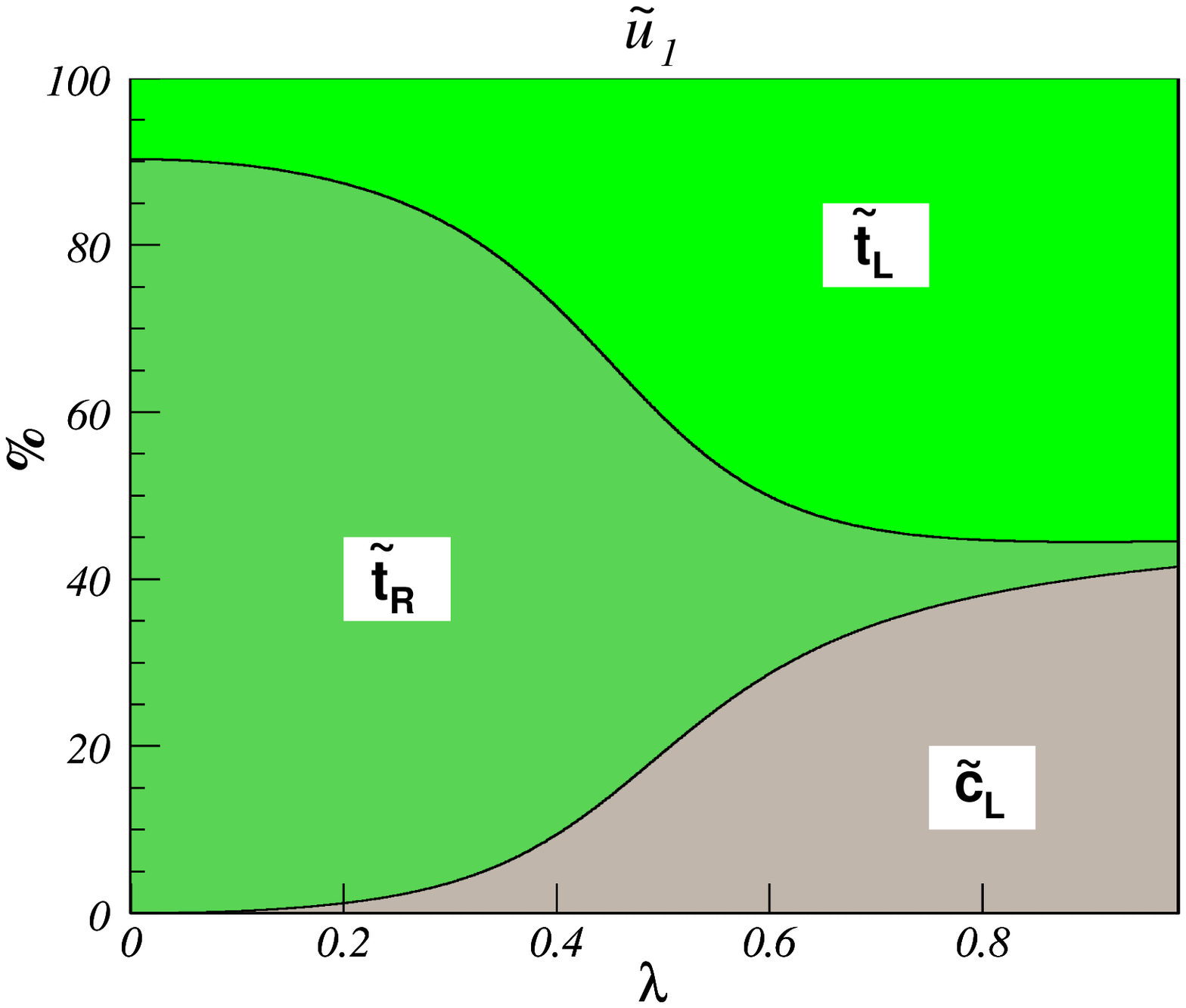}\hspace{1mm}
 \includegraphics[width=0.21\columnwidth]{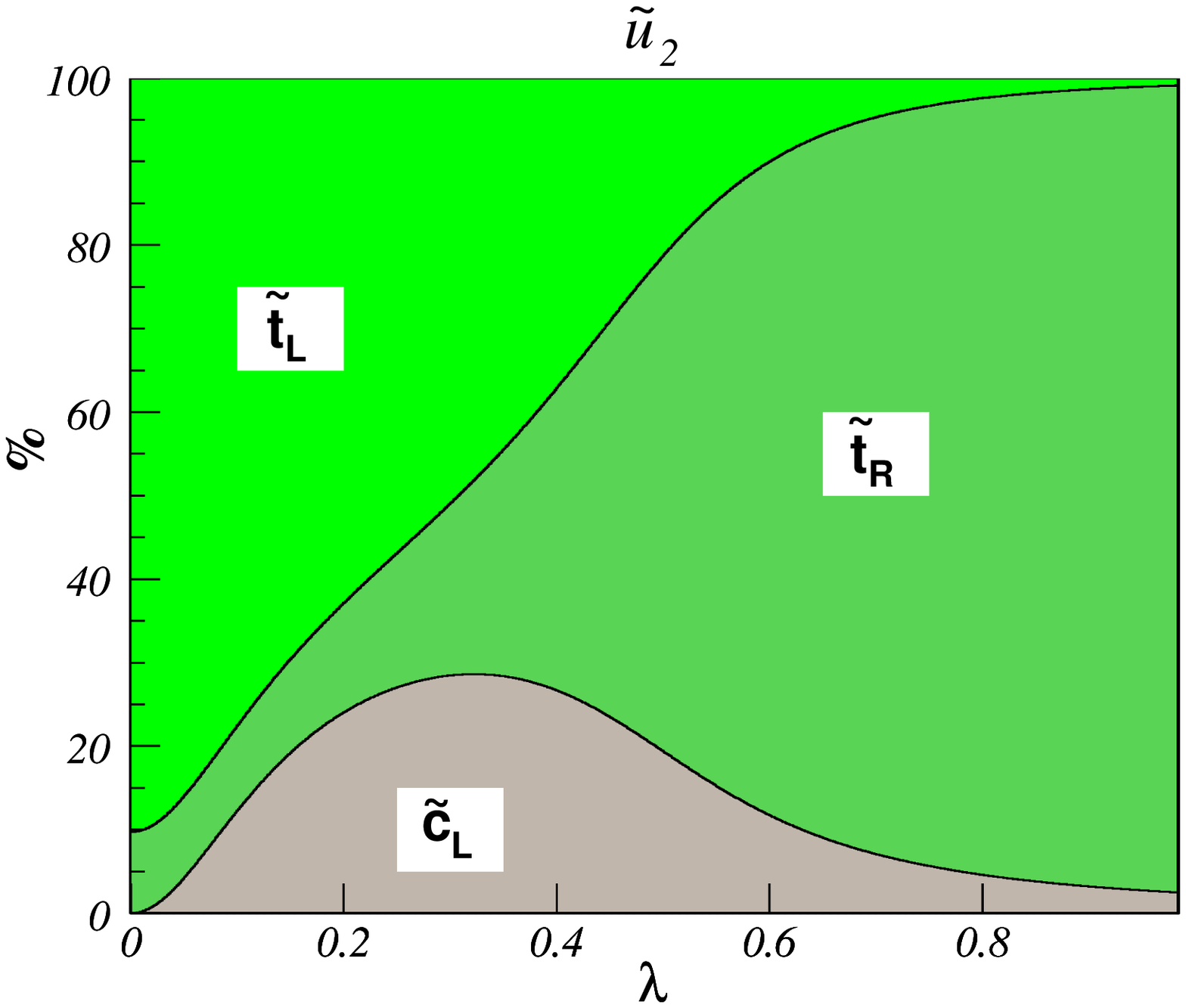}\hspace{1mm}
 \includegraphics[width=0.21\columnwidth]{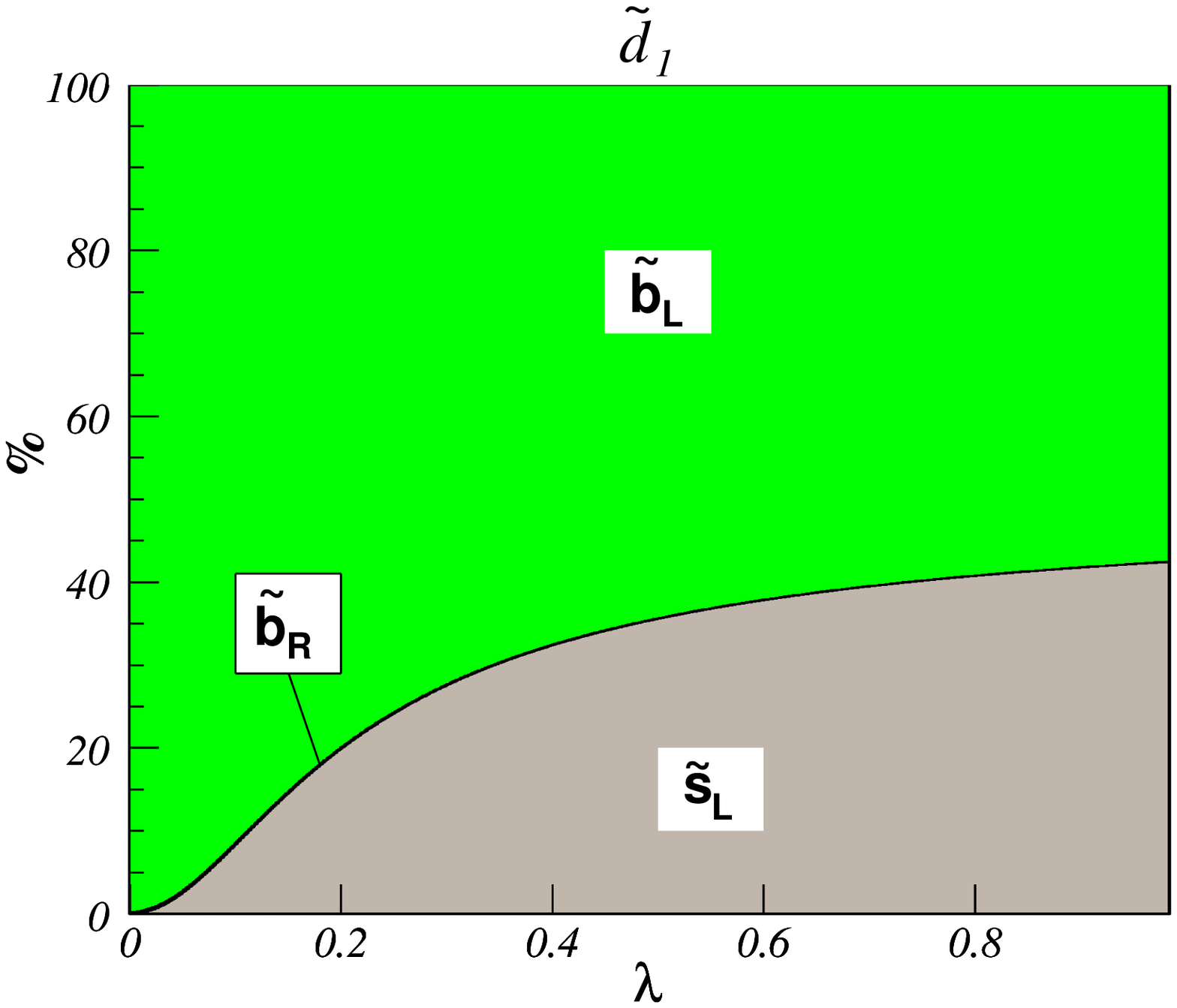}\hspace{1mm}
 \includegraphics[width=0.21\columnwidth]{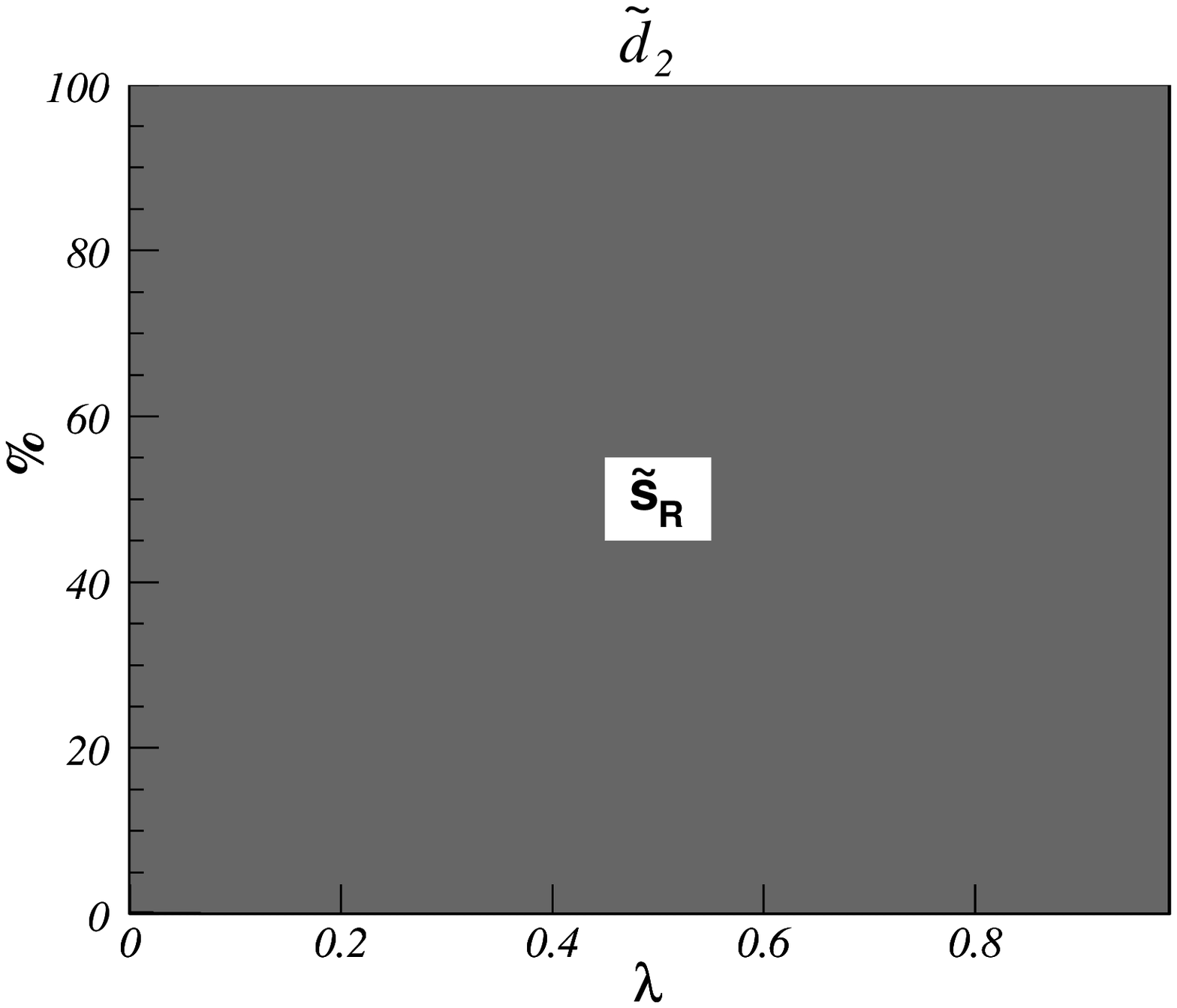}\vspace*{4mm}
 \includegraphics[width=0.21\columnwidth]{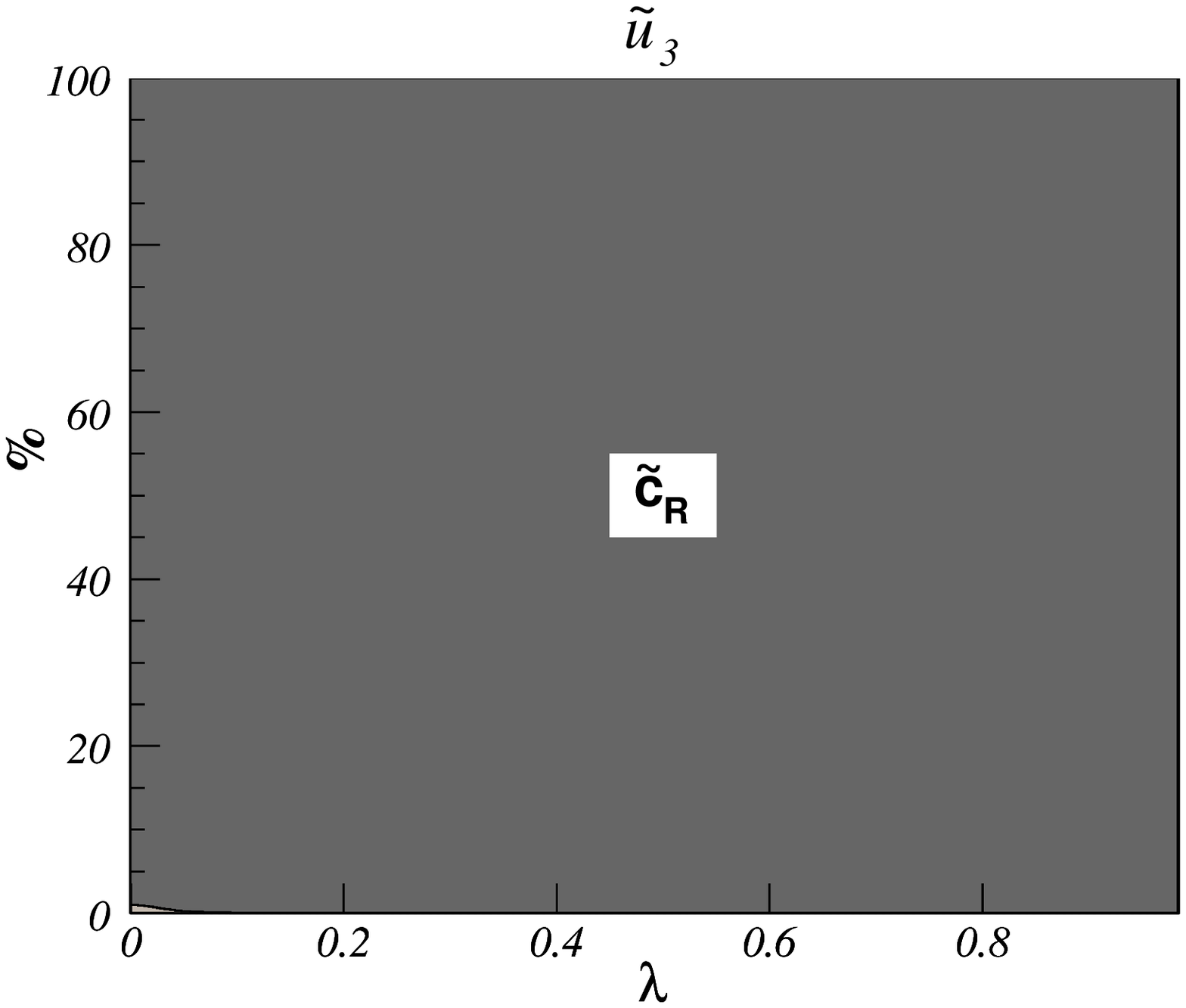}\hspace{1mm}
 \includegraphics[width=0.21\columnwidth]{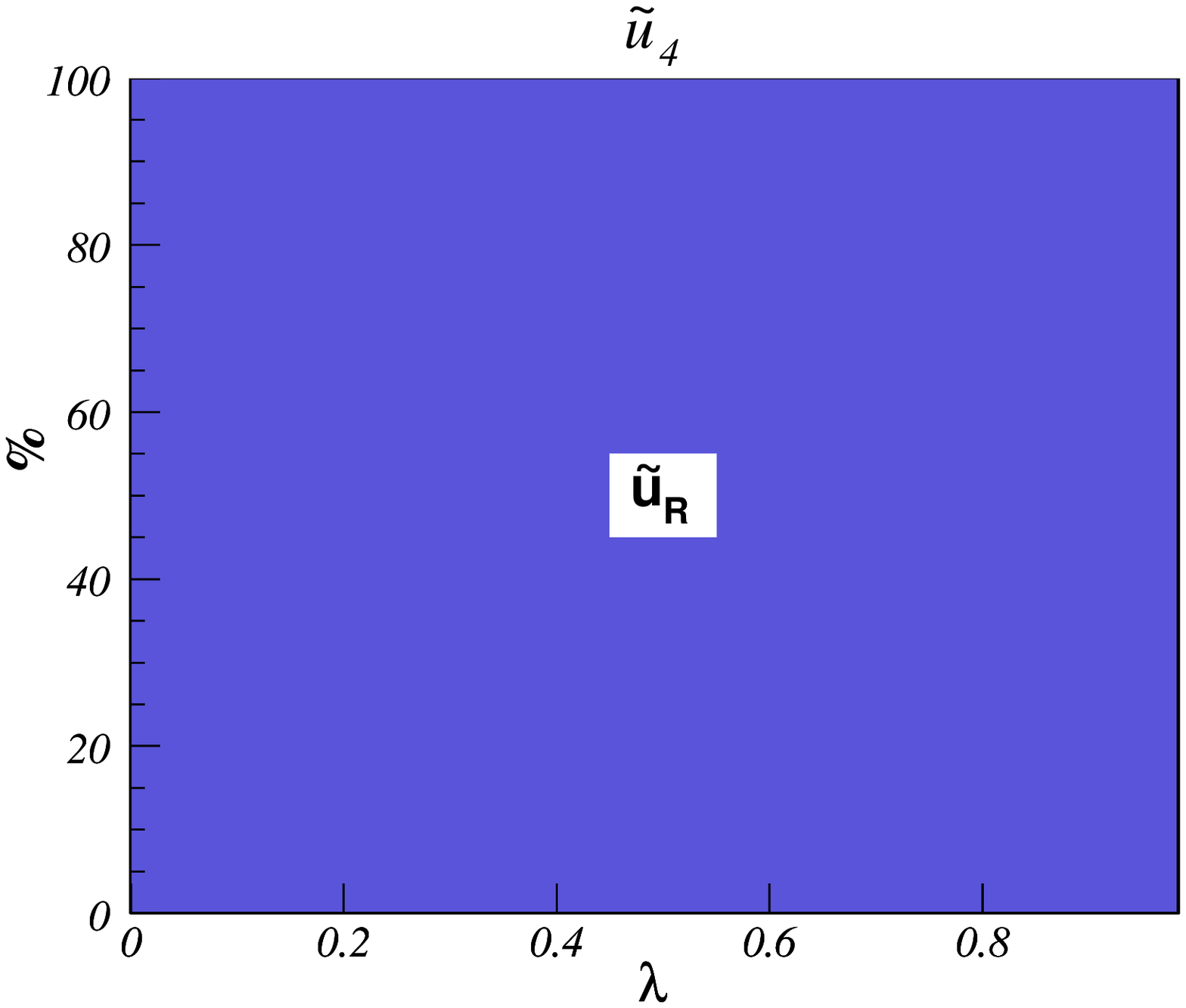}\hspace{1mm}
 \includegraphics[width=0.21\columnwidth]{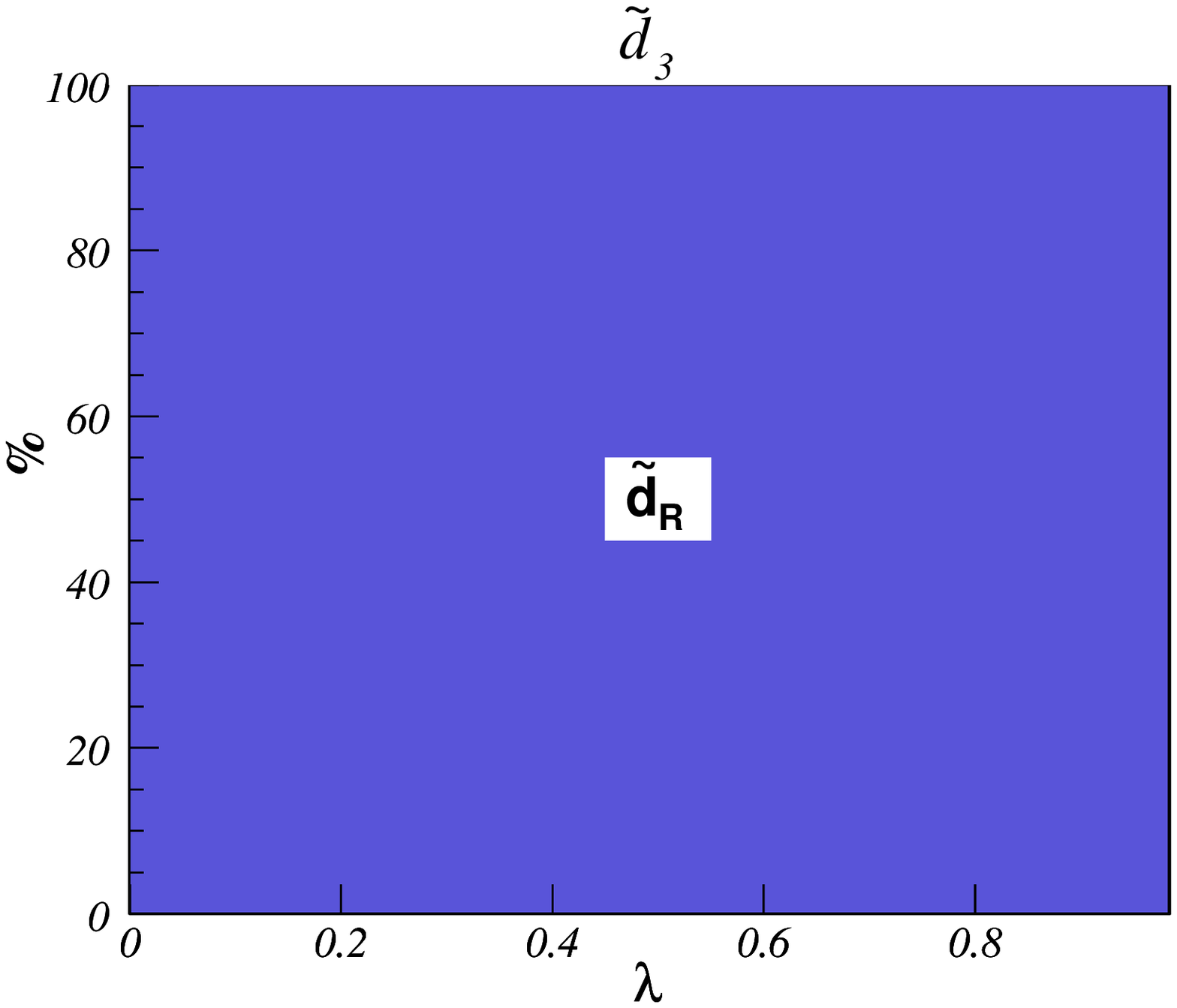}\hspace{1mm}
 \includegraphics[width=0.21\columnwidth]{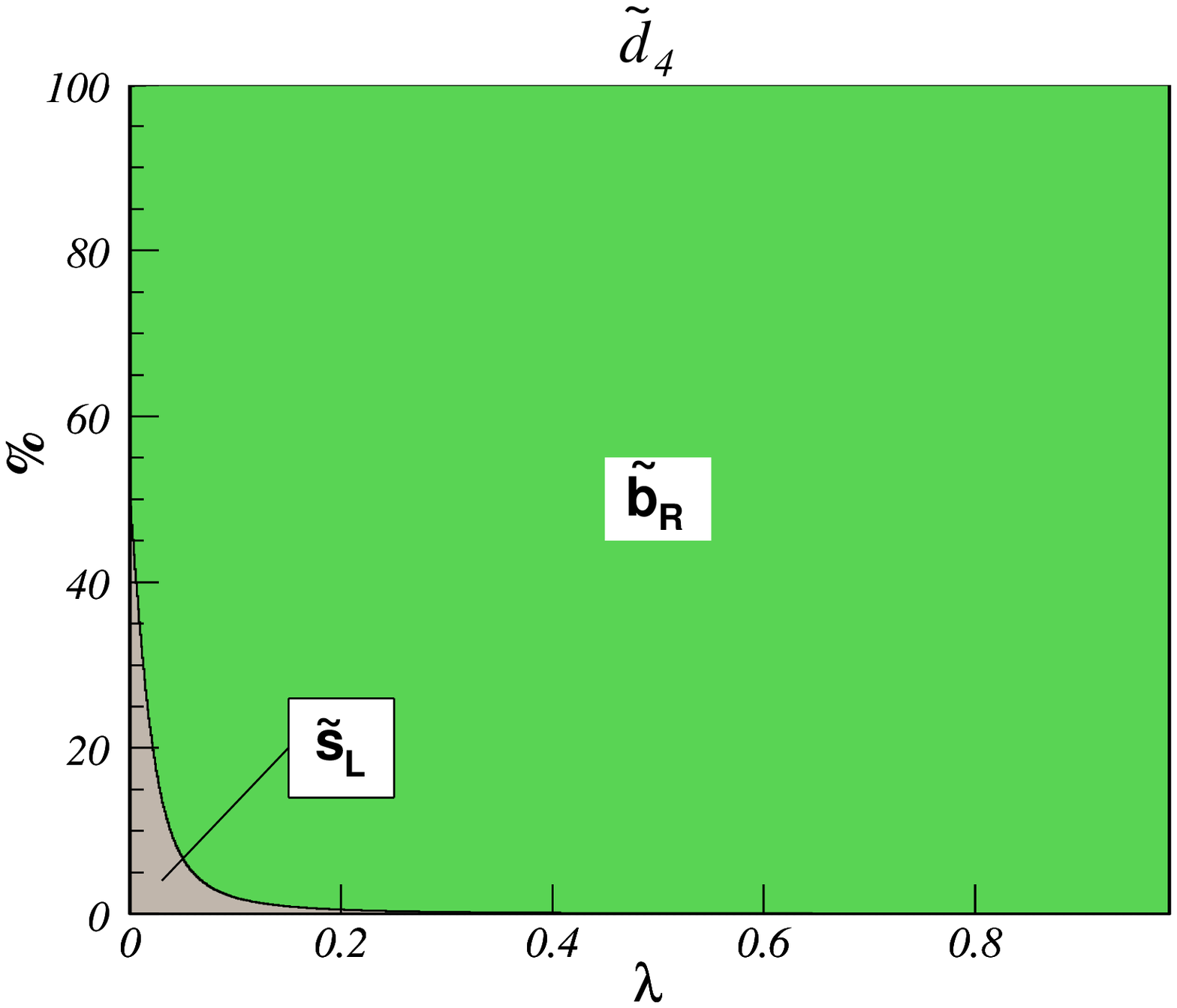}\vspace*{4mm}
 \includegraphics[width=0.21\columnwidth]{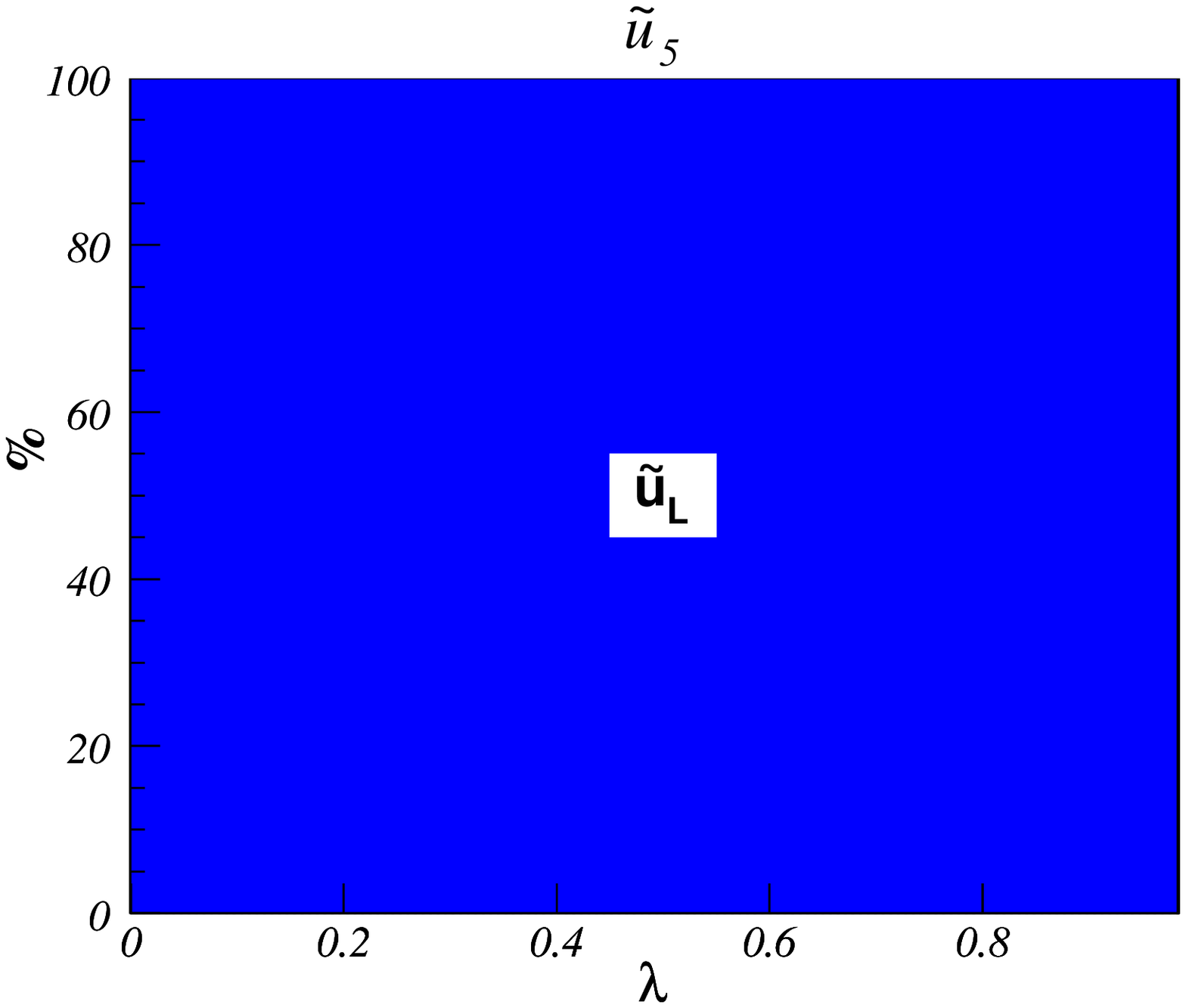}\hspace{1mm}
 \includegraphics[width=0.21\columnwidth]{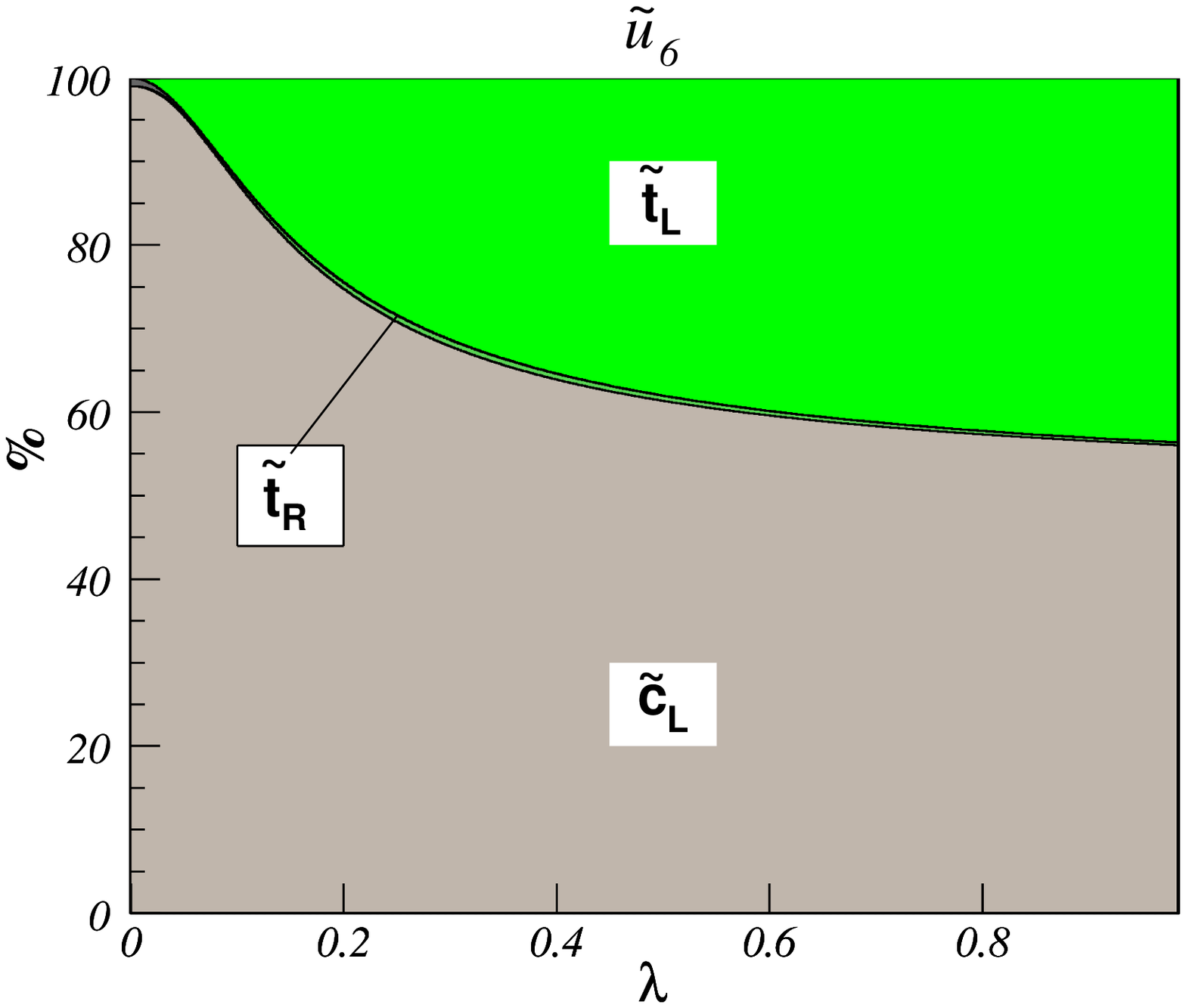}\hspace{1mm}
 \includegraphics[width=0.21\columnwidth]{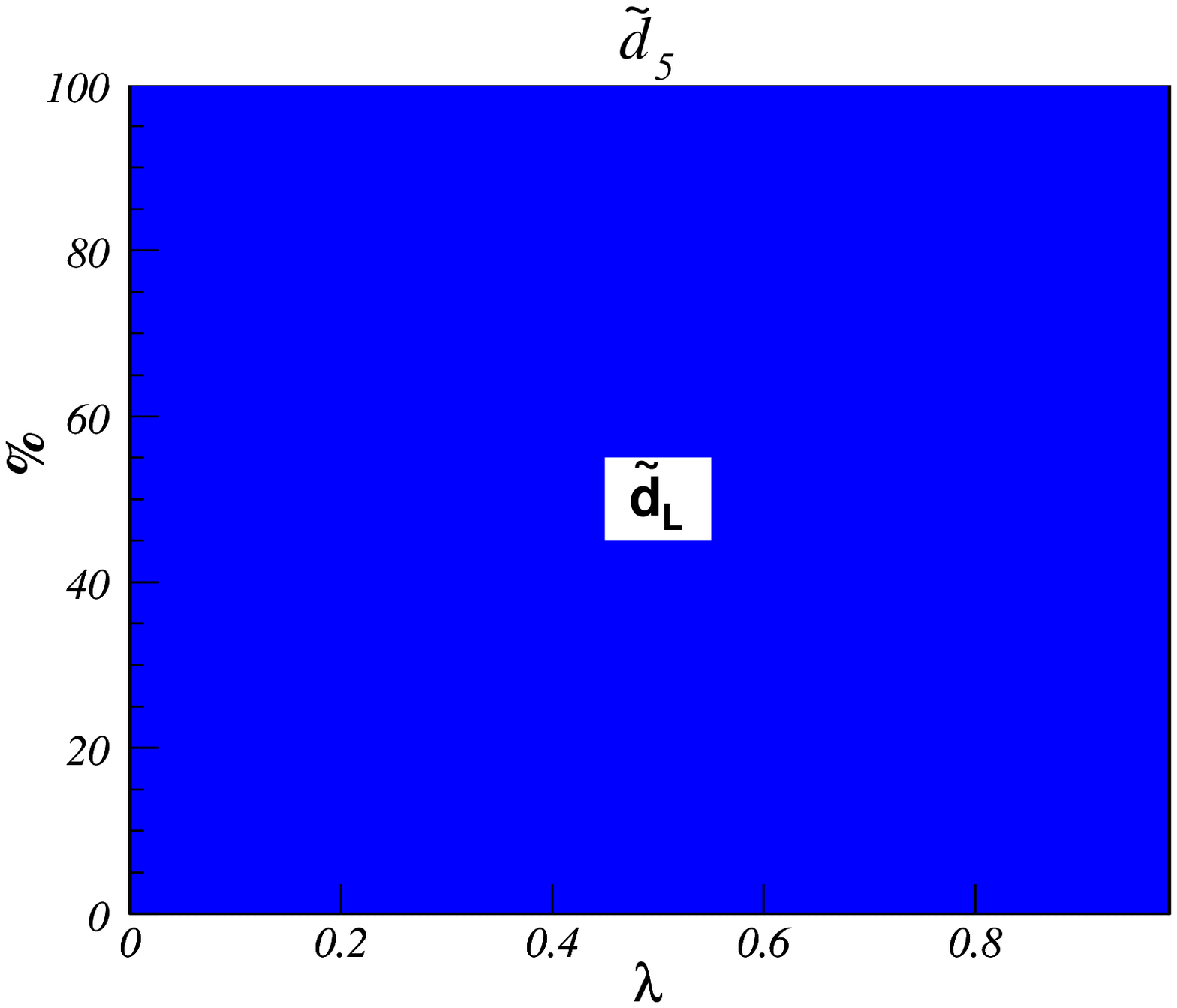}\hspace{1mm}
 \includegraphics[width=0.21\columnwidth]{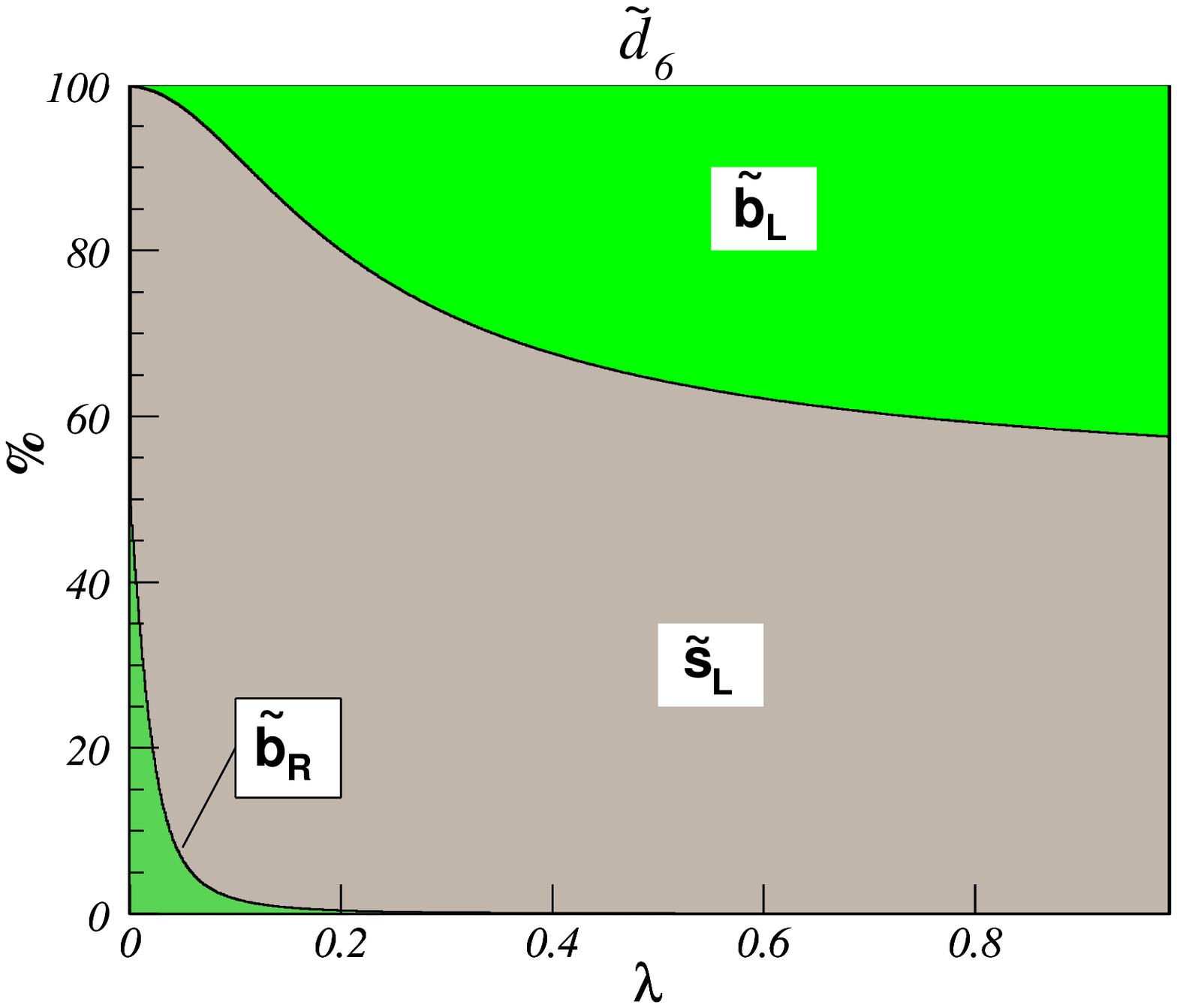}
 \caption{\label{fig:09}Dependence of the chirality (L, R) and flavour
          ($u$, $c$, $t$; $d$, $s$, and $b$) content of up- ($\tilde{u}_i$)
          and down-type ($\tilde{d}_i$) squark mass eigenstates on the NMFV
          parameter $\lambda\in[0;1]$ for benchmark point A.}\vspace{4mm}
 \includegraphics[width=0.21\columnwidth]{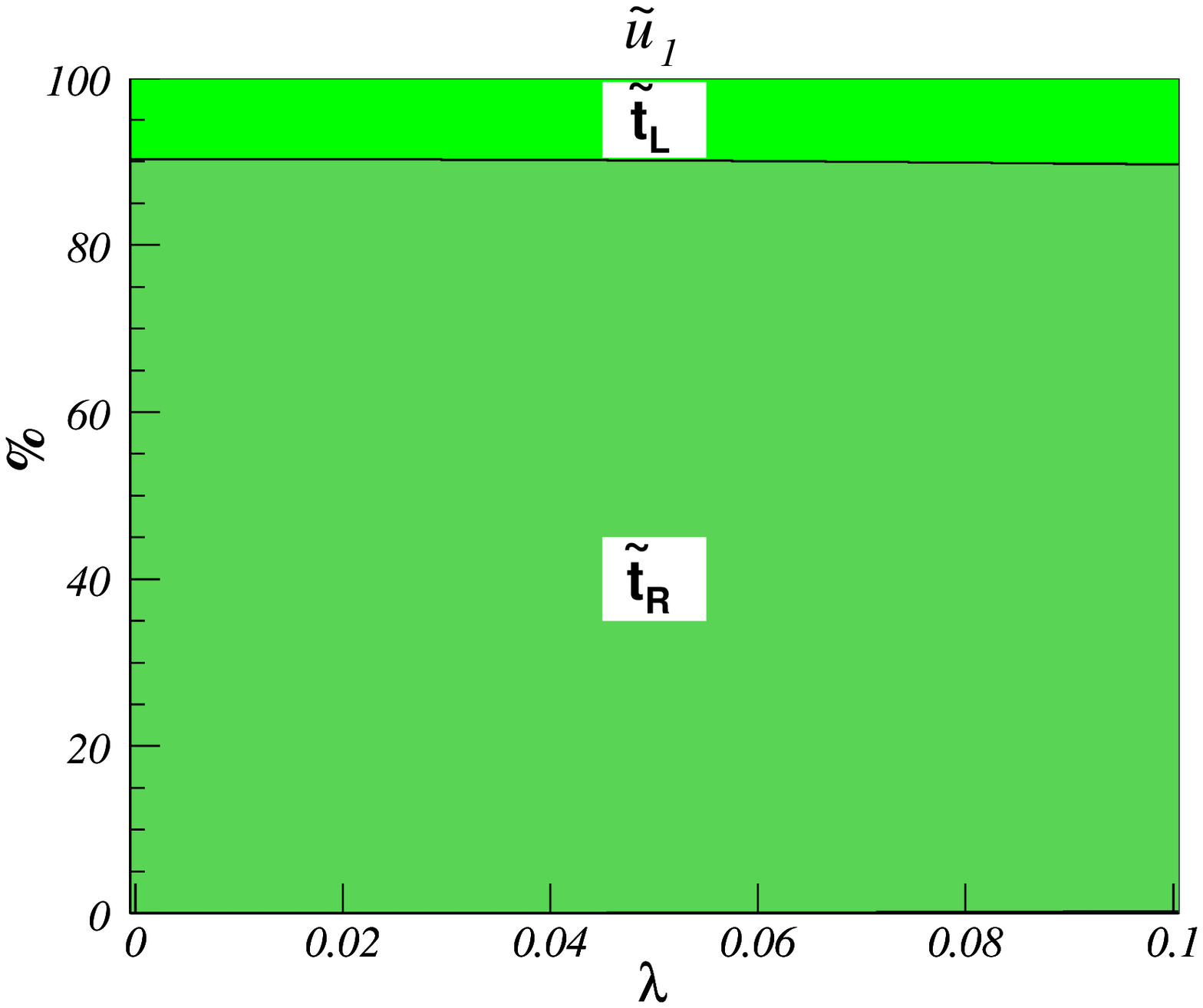}\hspace{1mm}
 \includegraphics[width=0.21\columnwidth]{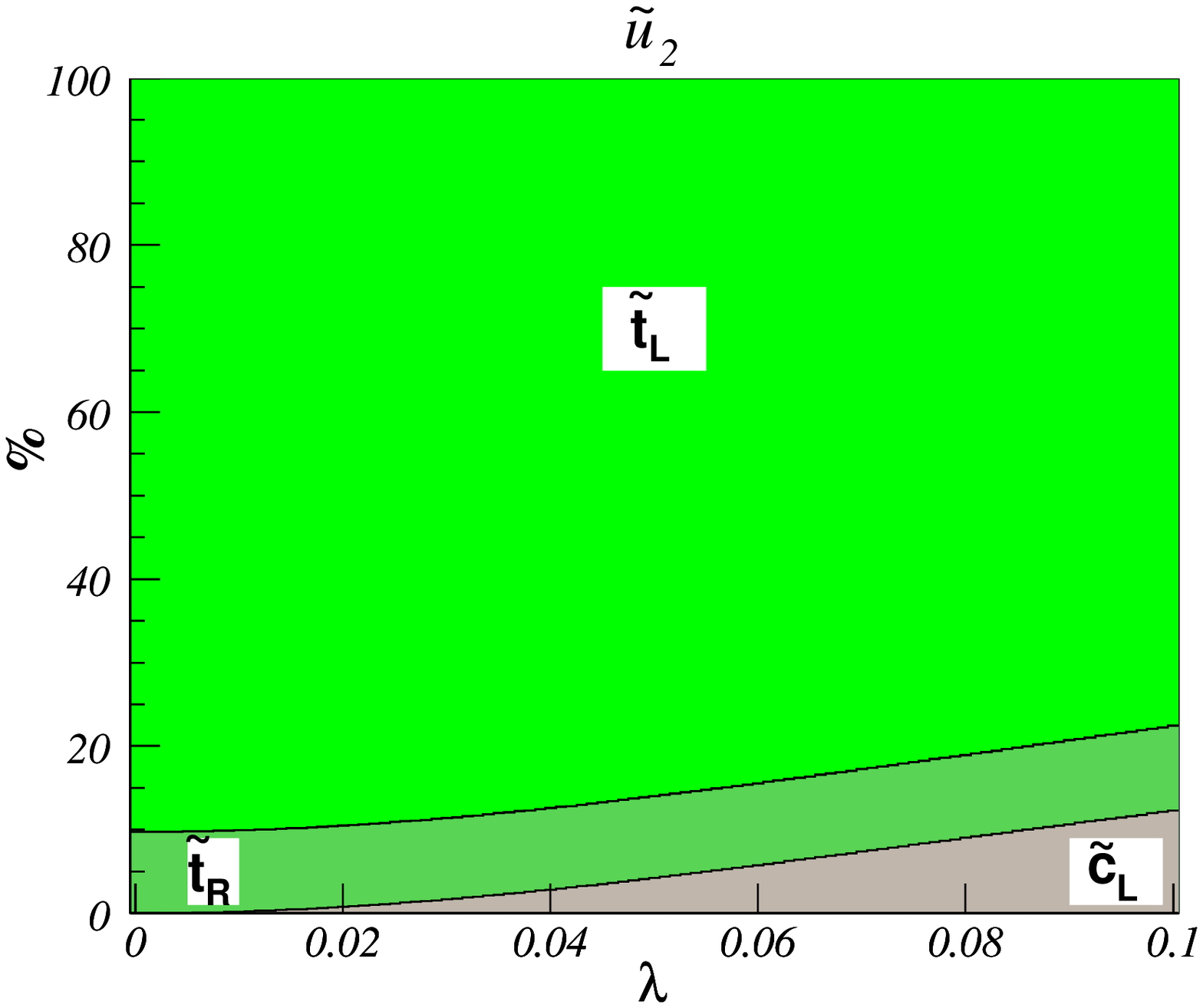}\hspace{1mm}
 \includegraphics[width=0.21\columnwidth]{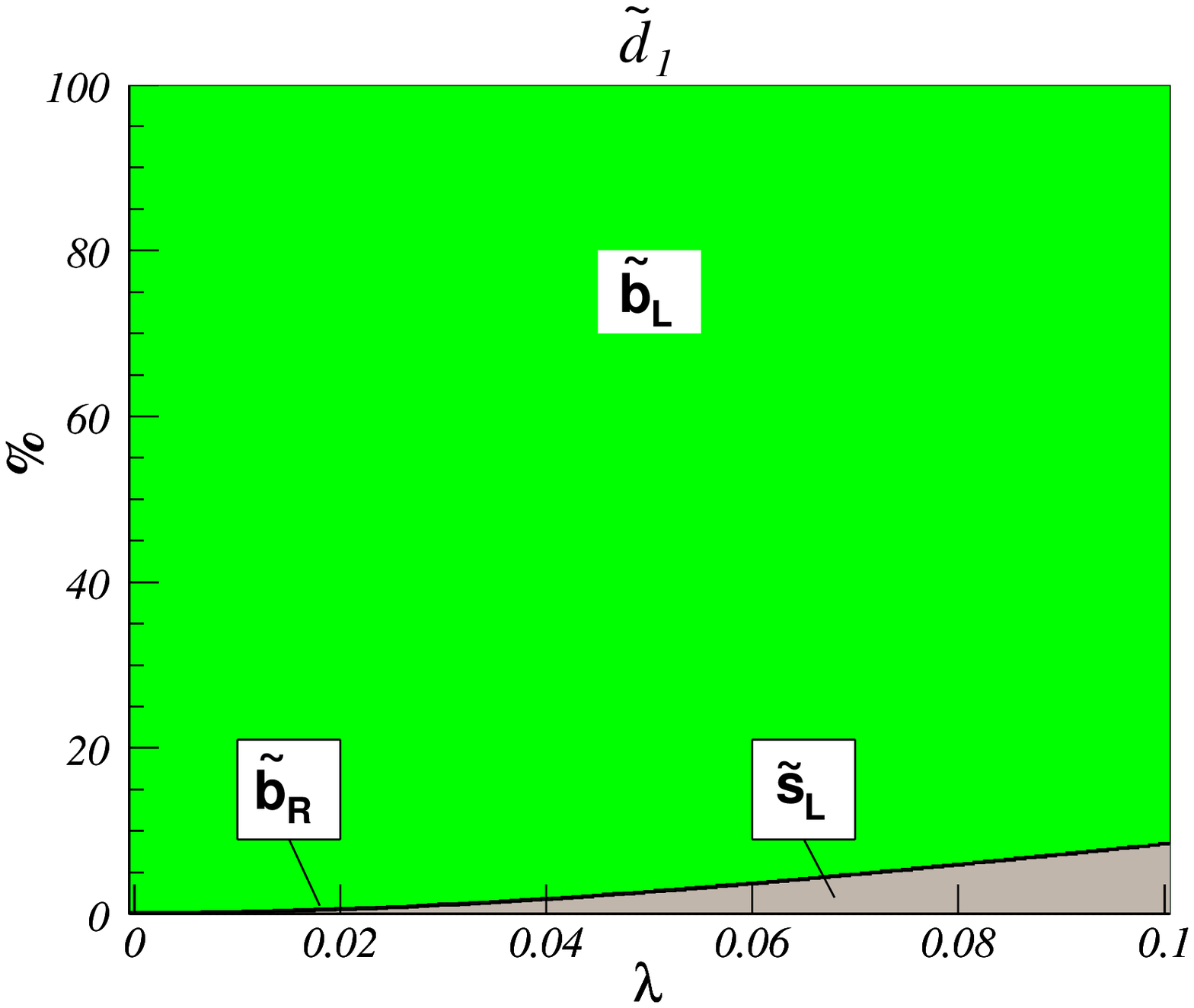}\hspace{1mm}
 \includegraphics[width=0.21\columnwidth]{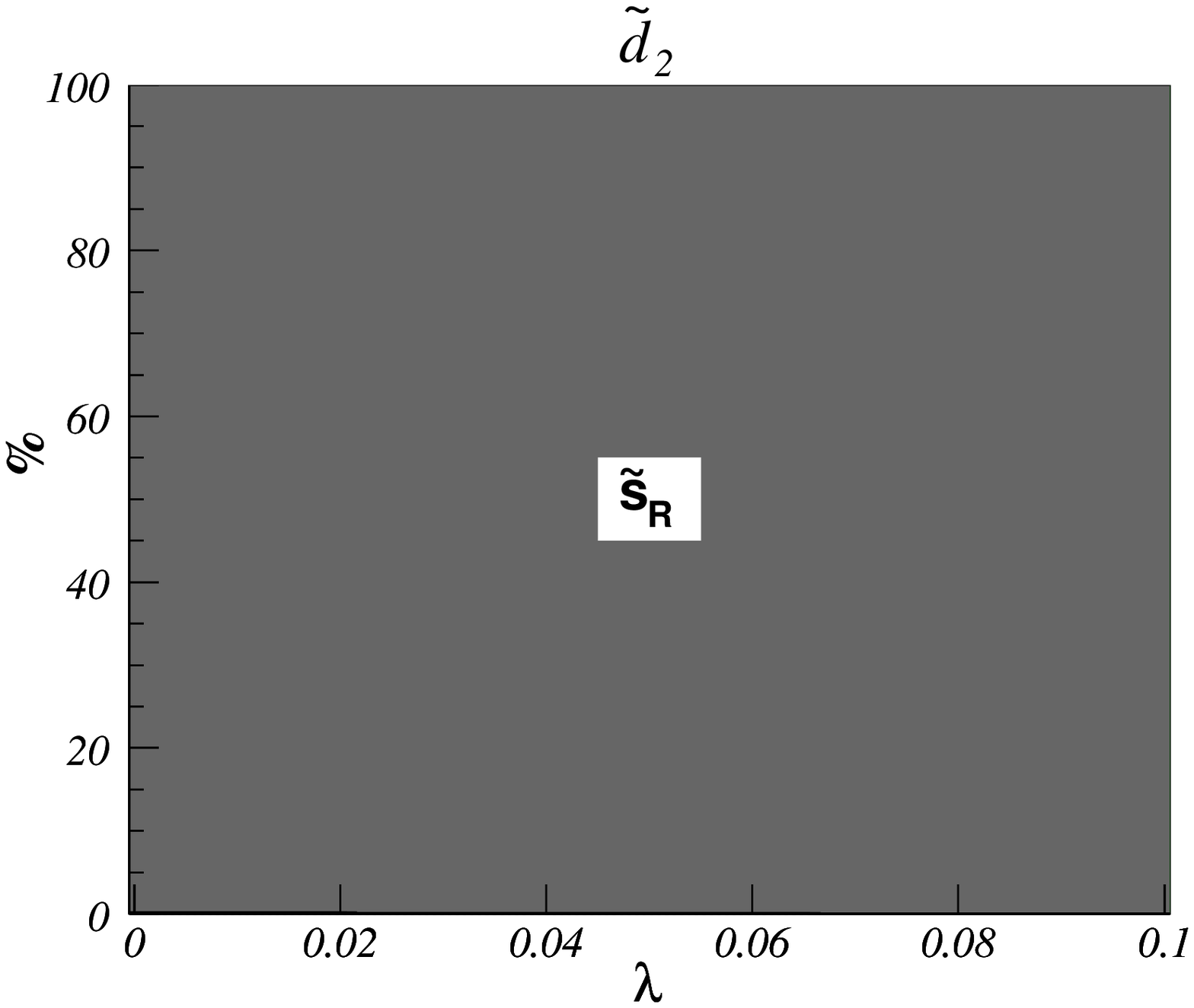}\vspace*{4mm}
 \includegraphics[width=0.21\columnwidth]{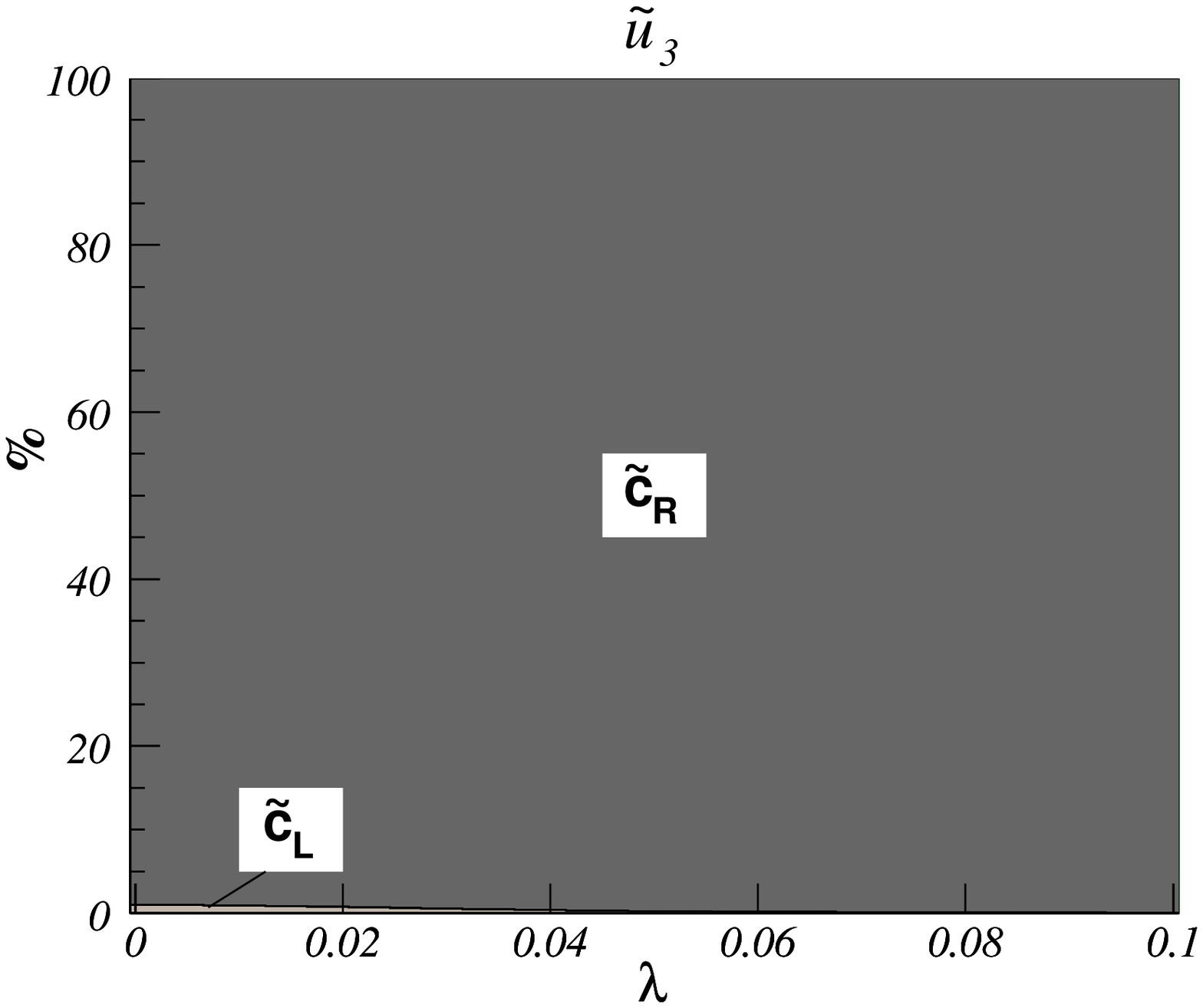}\hspace{1mm}
 \includegraphics[width=0.21\columnwidth]{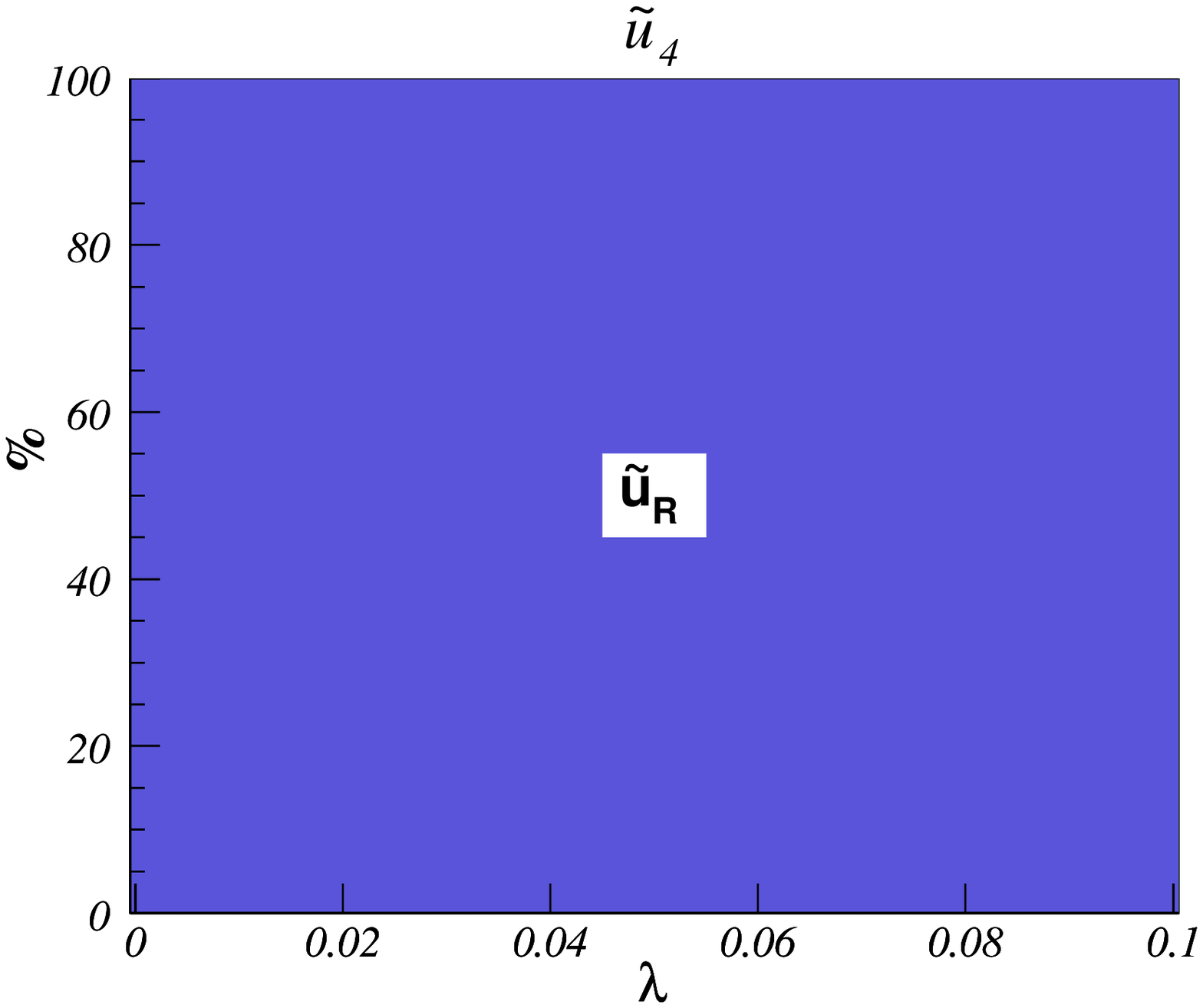}\hspace{1mm}
 \includegraphics[width=0.21\columnwidth]{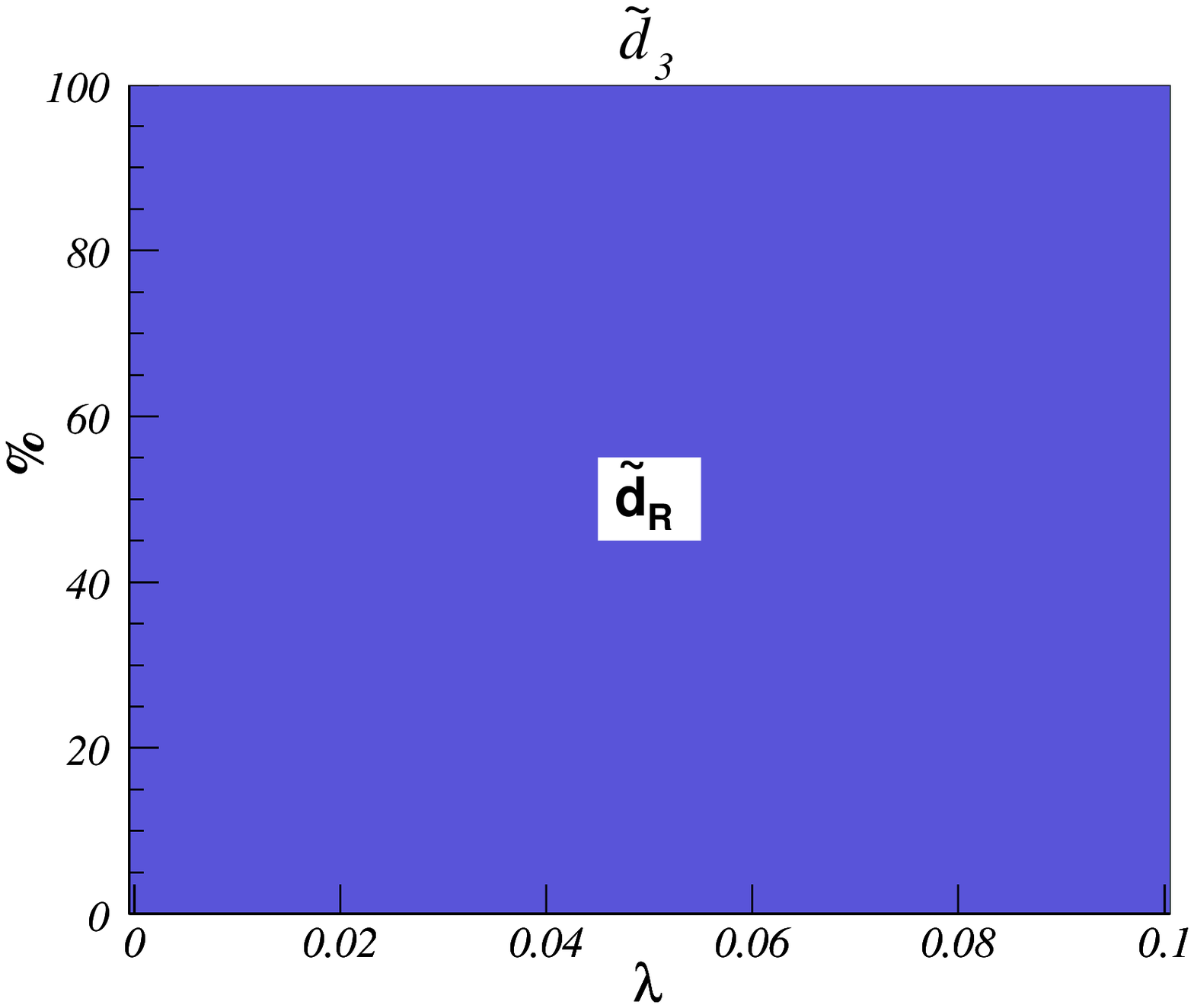}\hspace{1mm}
 \includegraphics[width=0.21\columnwidth]{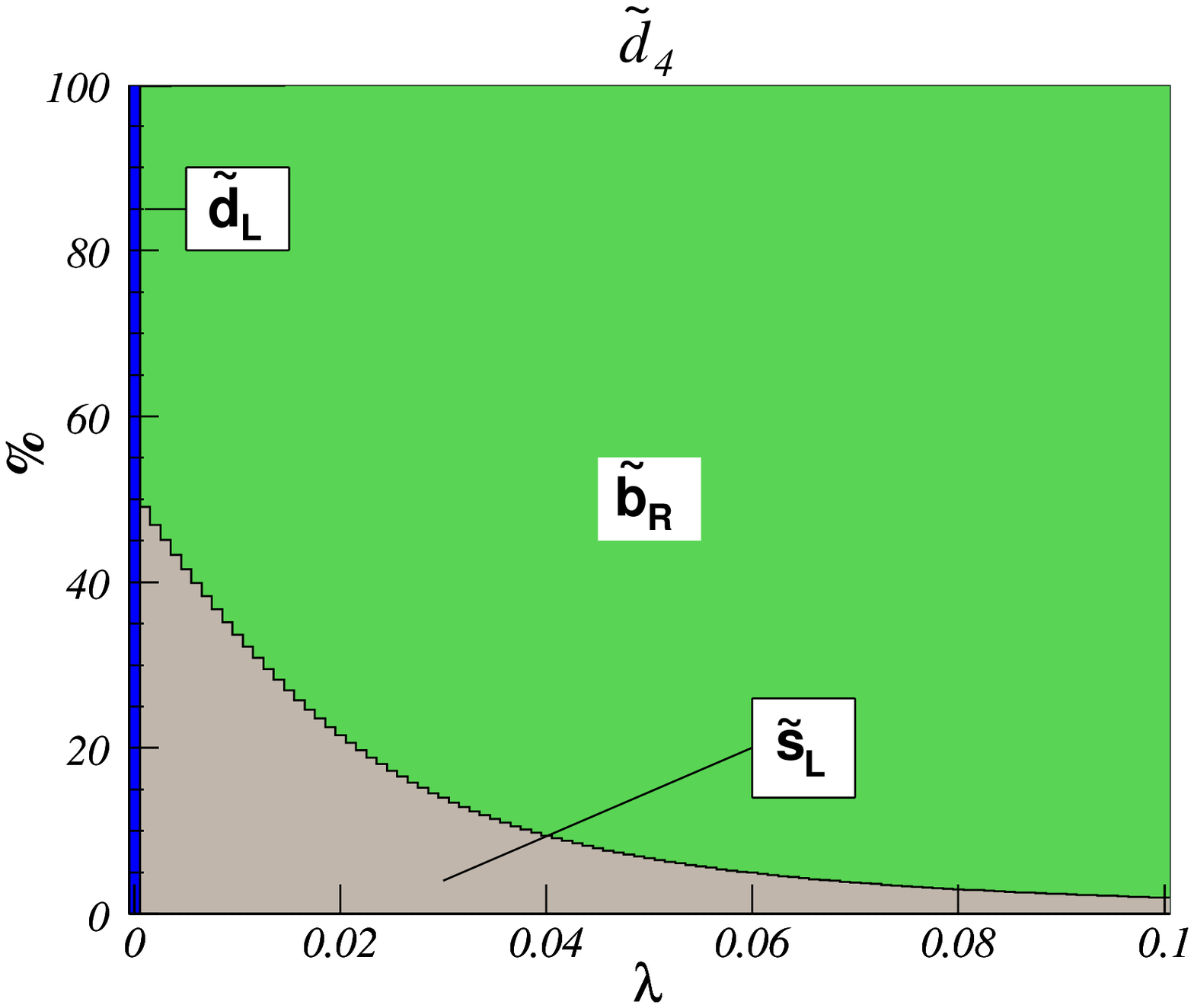}\vspace*{4mm}
 \includegraphics[width=0.21\columnwidth]{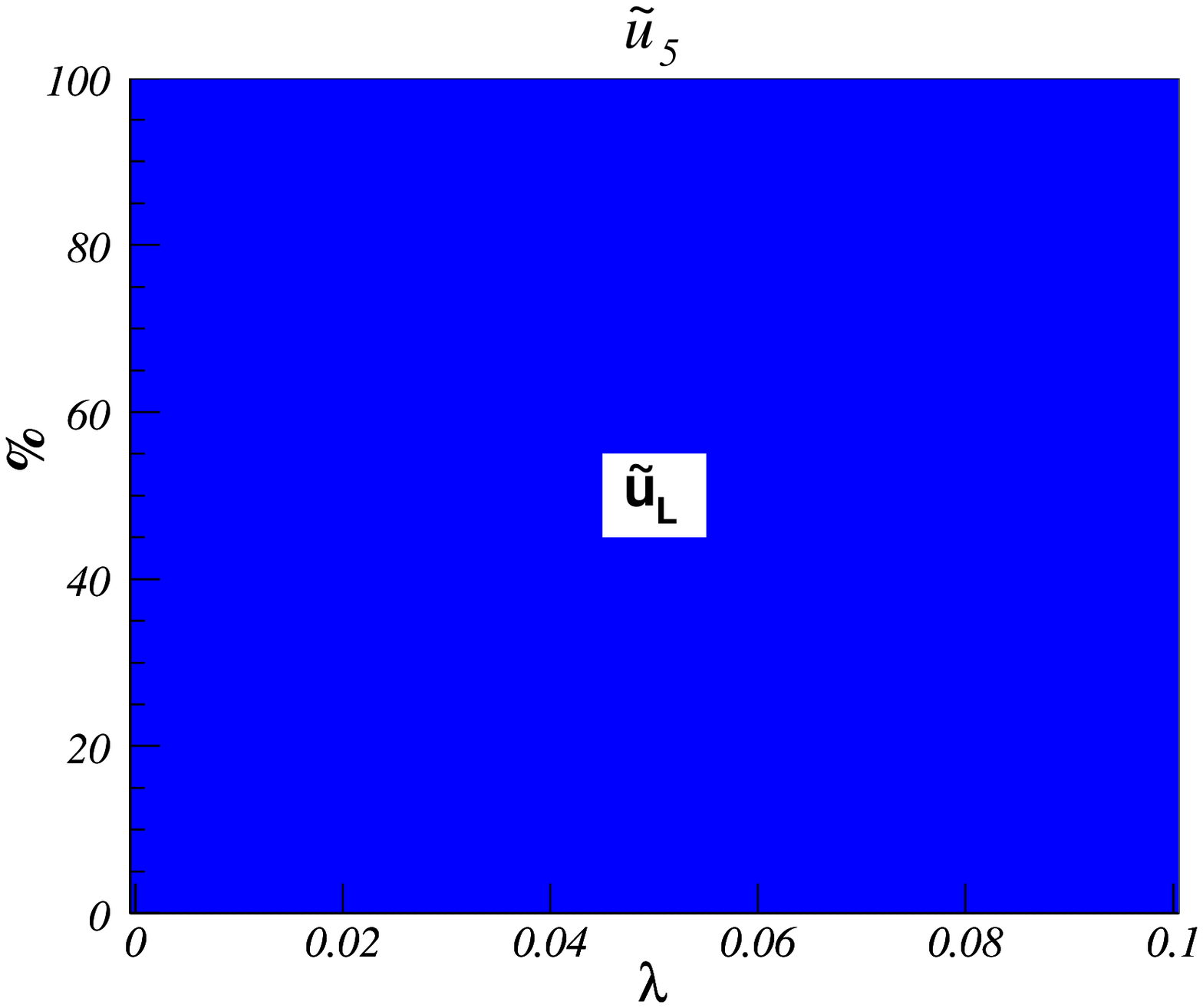}\hspace{1mm}
 \includegraphics[width=0.21\columnwidth]{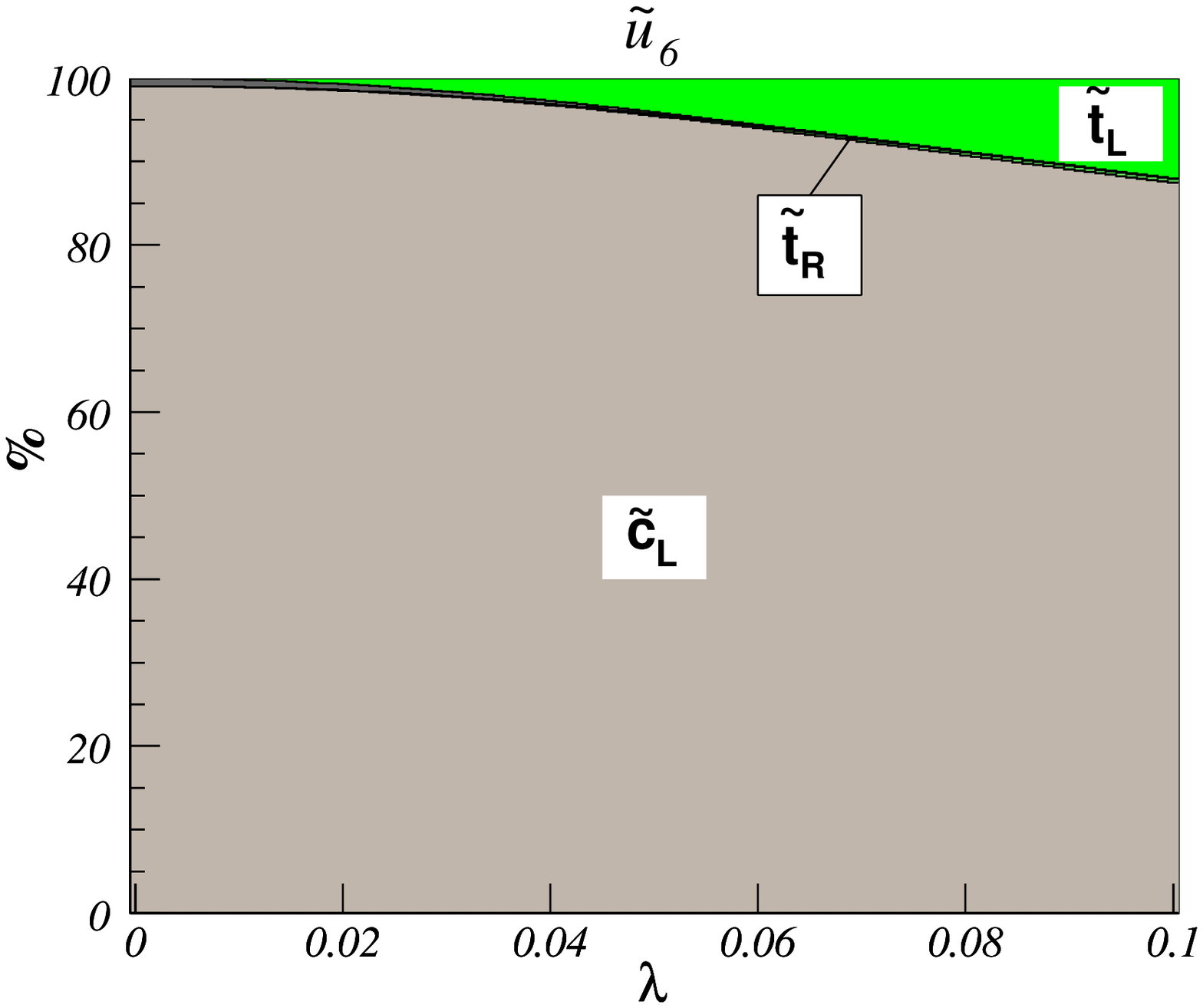}\hspace{1mm}
 \includegraphics[width=0.21\columnwidth]{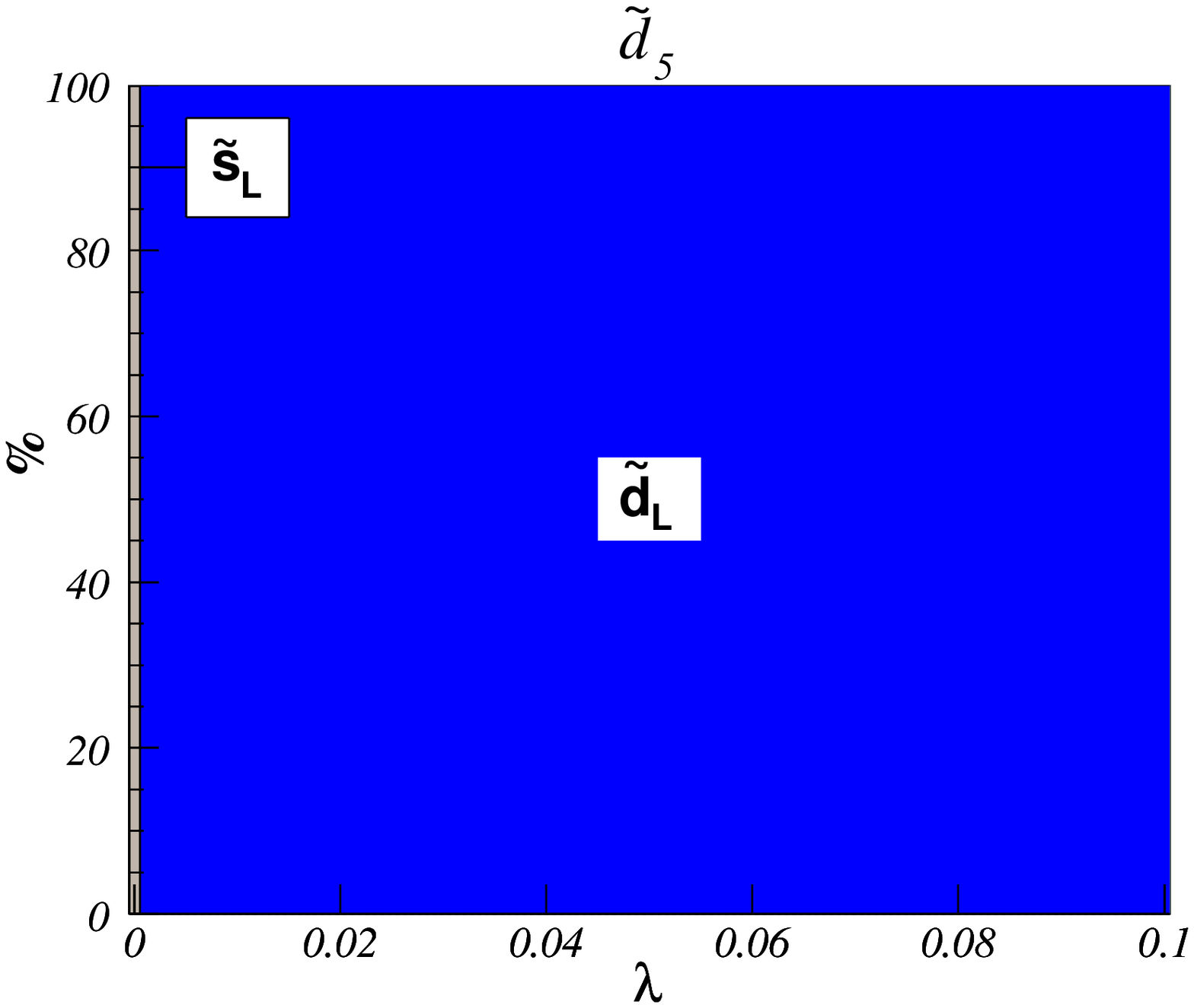}\hspace{1mm}
 \includegraphics[width=0.21\columnwidth]{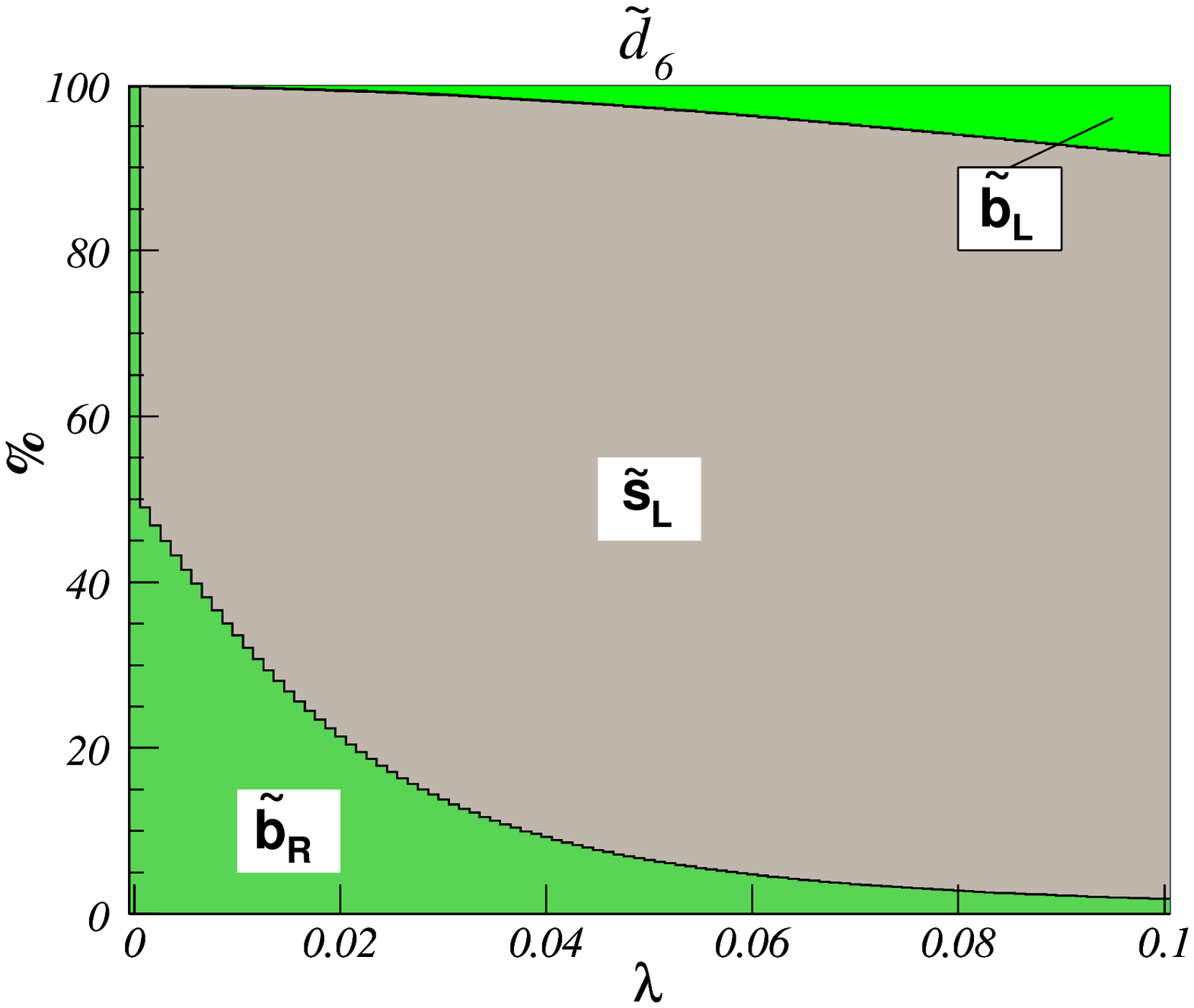}
 \caption{\label{fig:09p}Same as Fig.\ \ref{fig:09} for $\lambda\in
          [0;0.1]$.}
\end{figure}

\begin{figure}
 \centering
 \includegraphics[width=0.21\columnwidth]{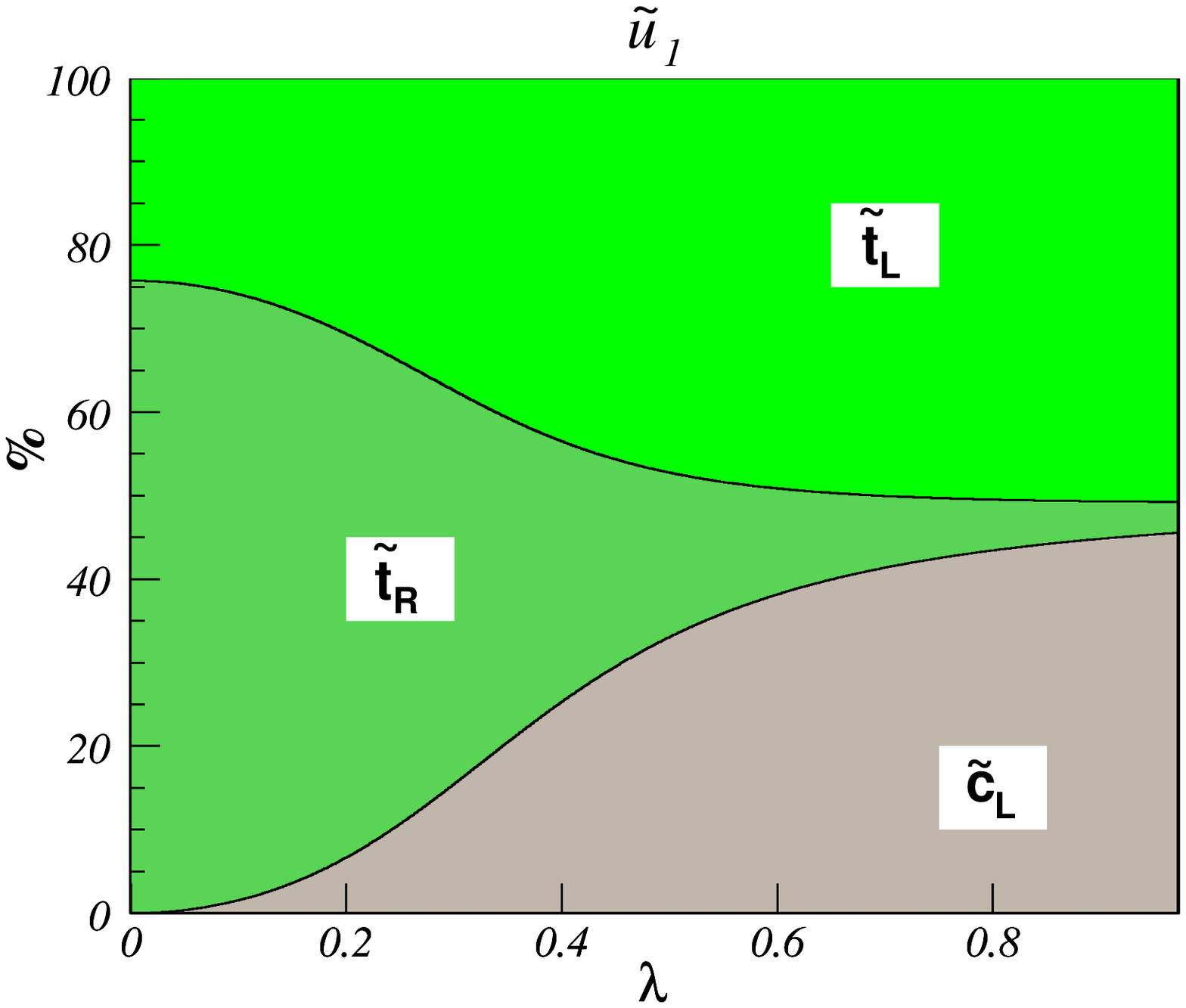}\hspace{1mm}
 \includegraphics[width=0.21\columnwidth]{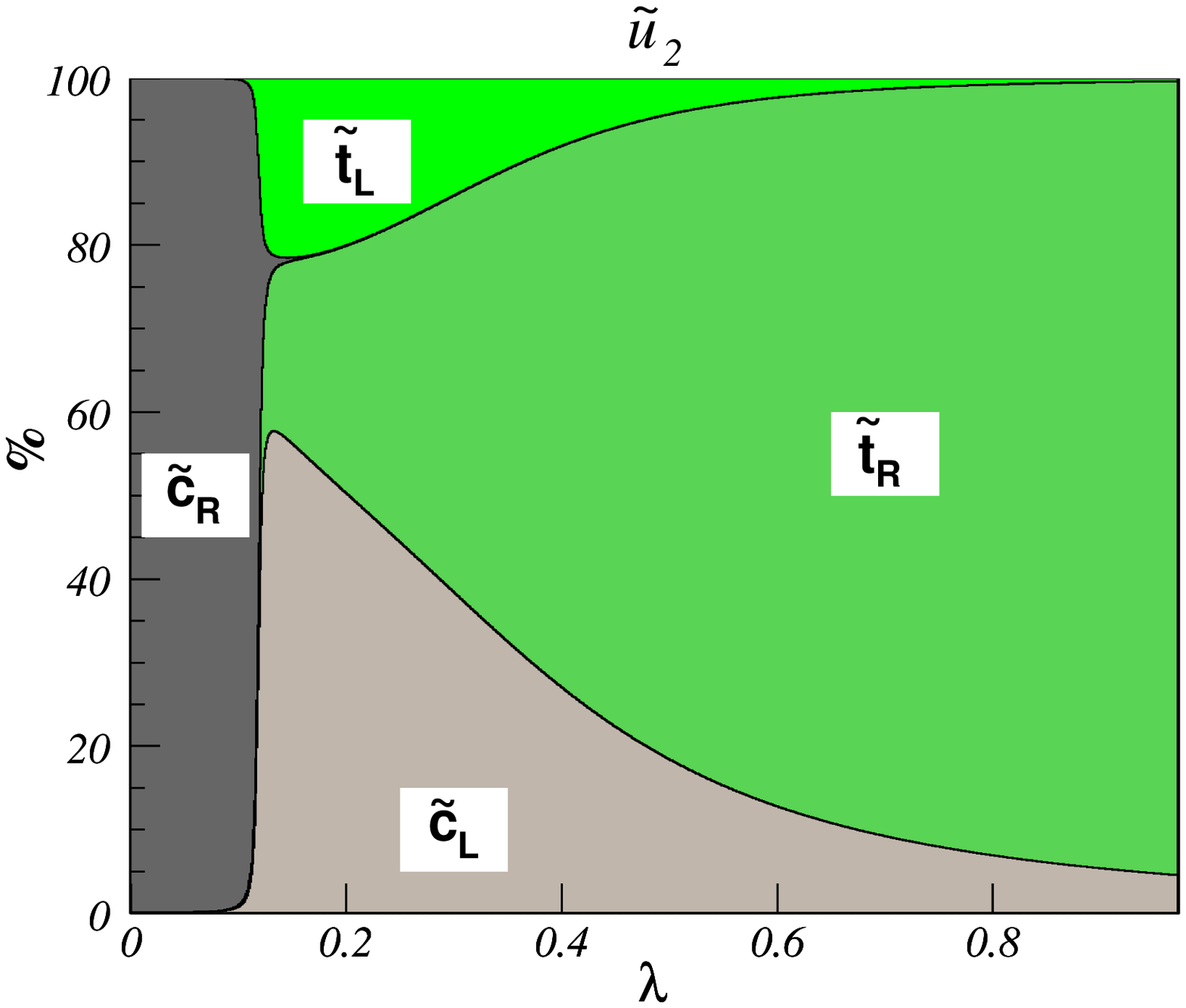}\hspace{1mm}
 \includegraphics[width=0.21\columnwidth]{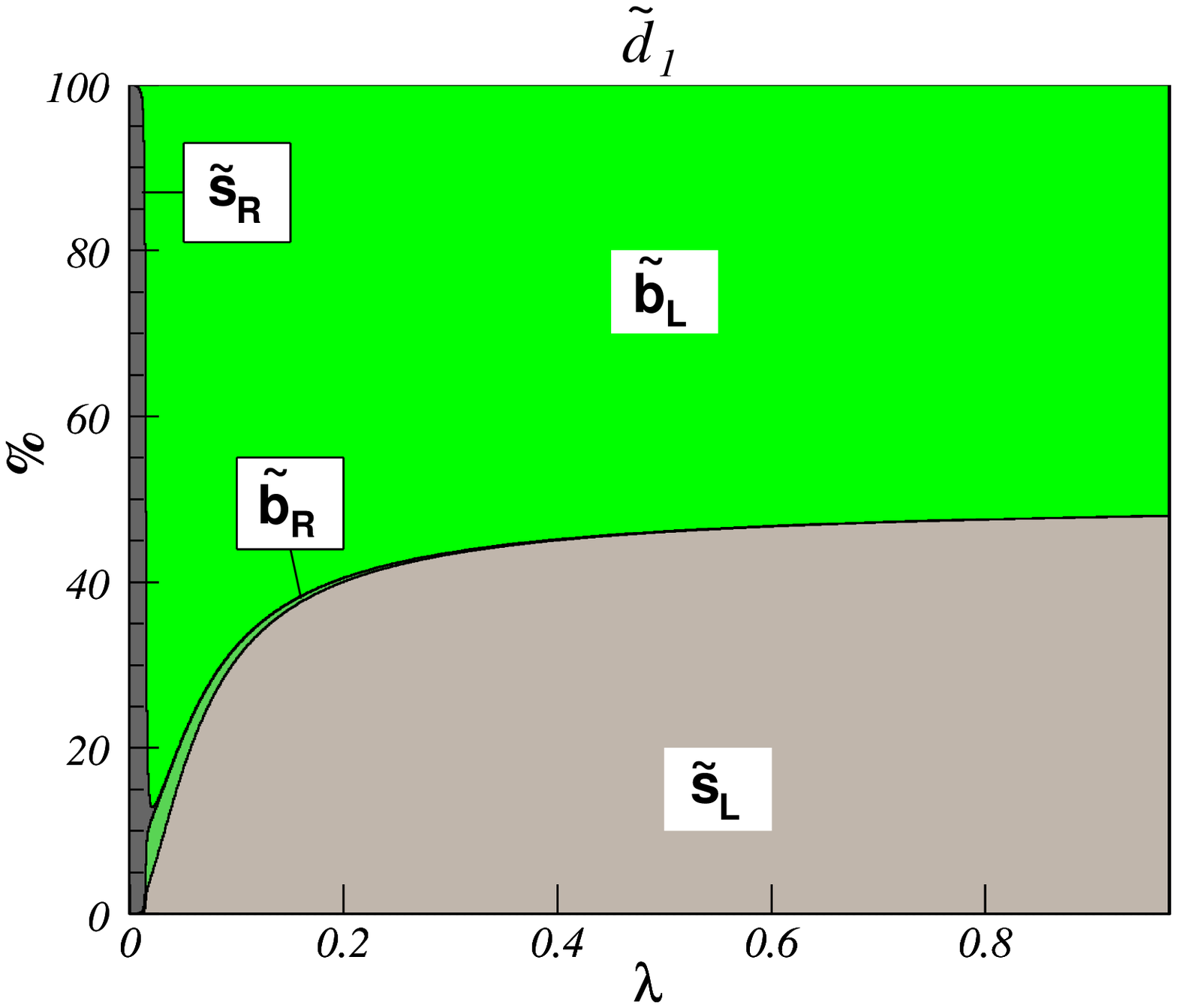}\hspace{1mm}
 \includegraphics[width=0.21\columnwidth]{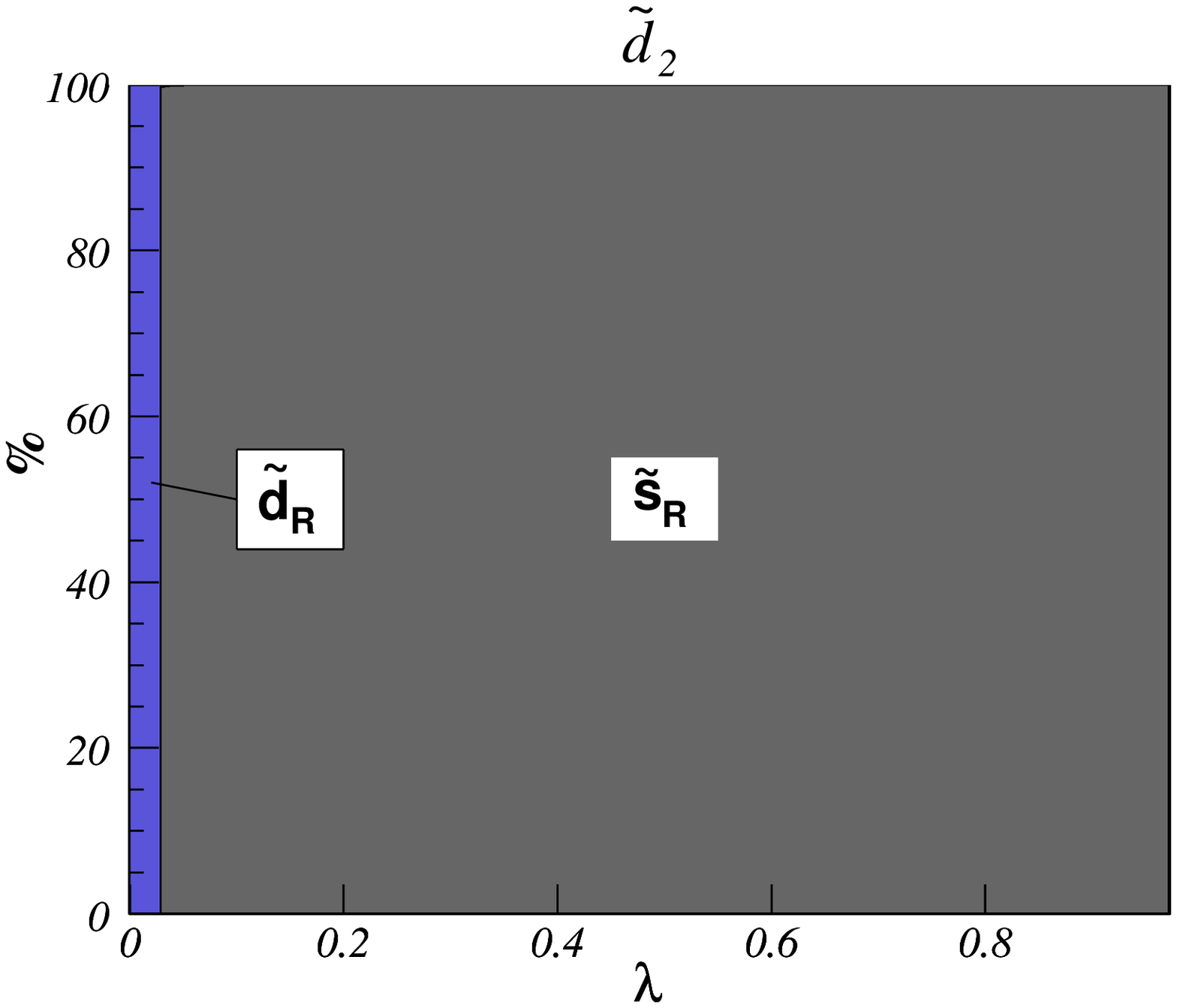}\vspace*{4mm}
 \includegraphics[width=0.21\columnwidth]{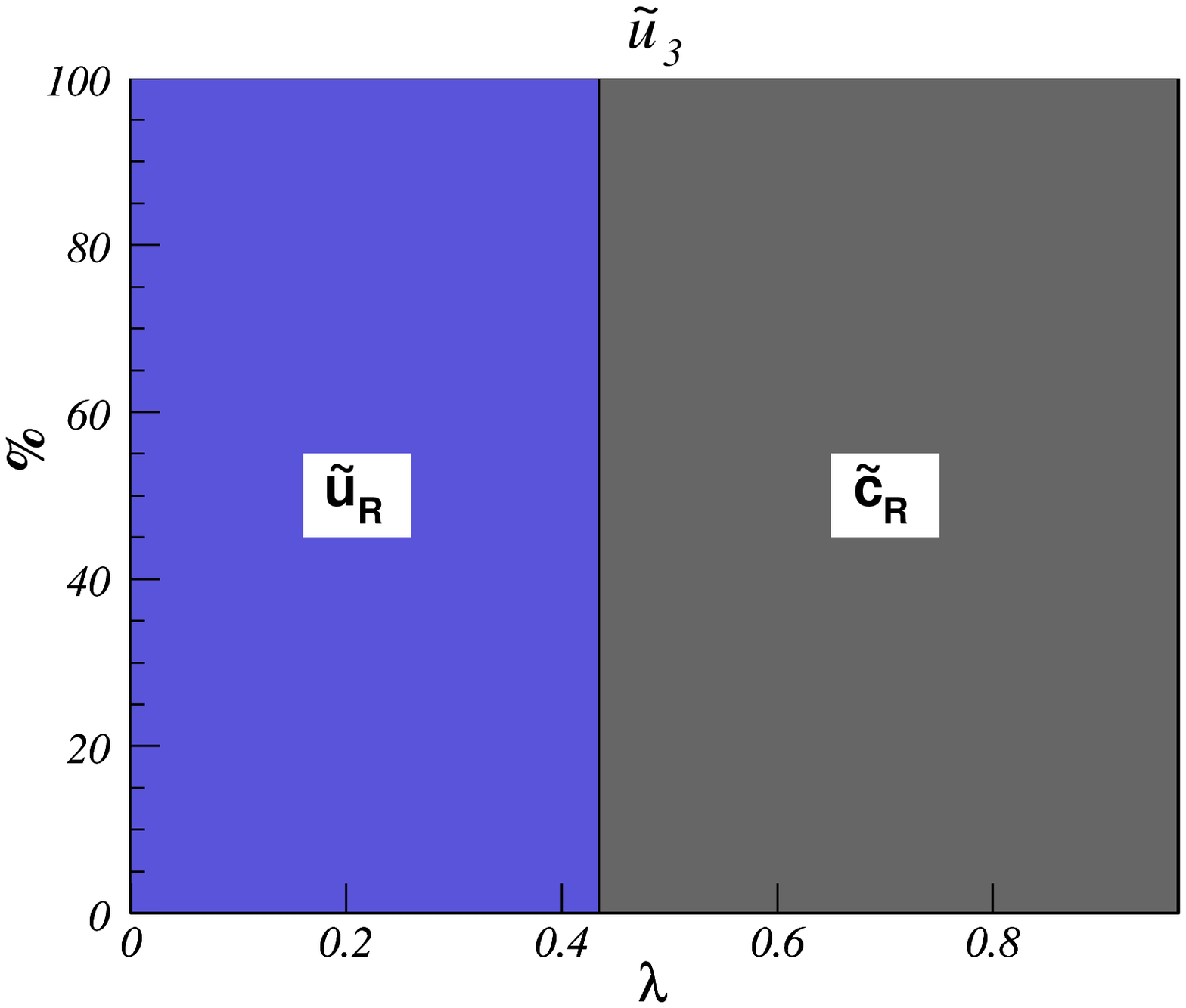}\hspace{1mm}
 \includegraphics[width=0.21\columnwidth]{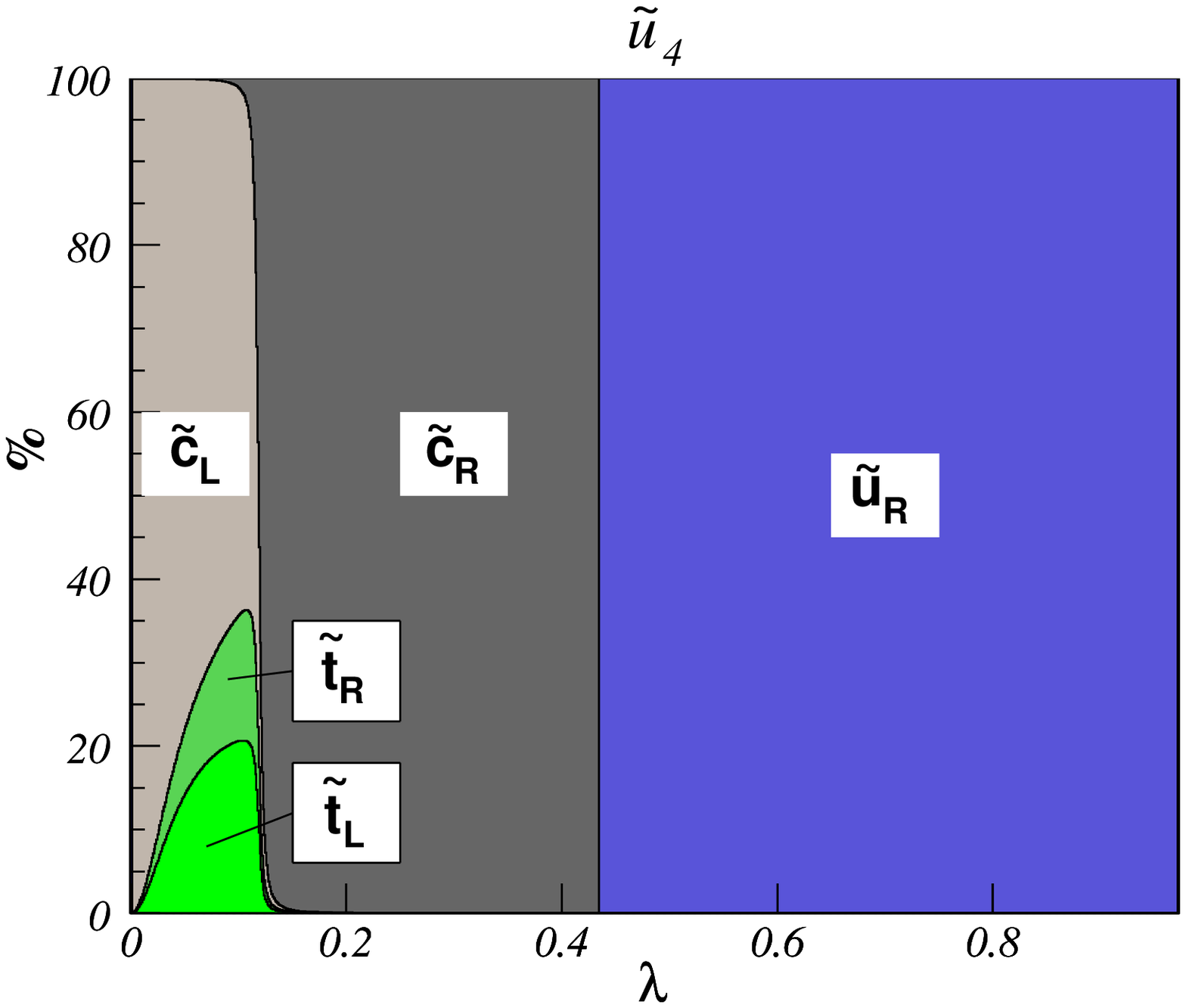}\hspace{1mm}
 \includegraphics[width=0.21\columnwidth]{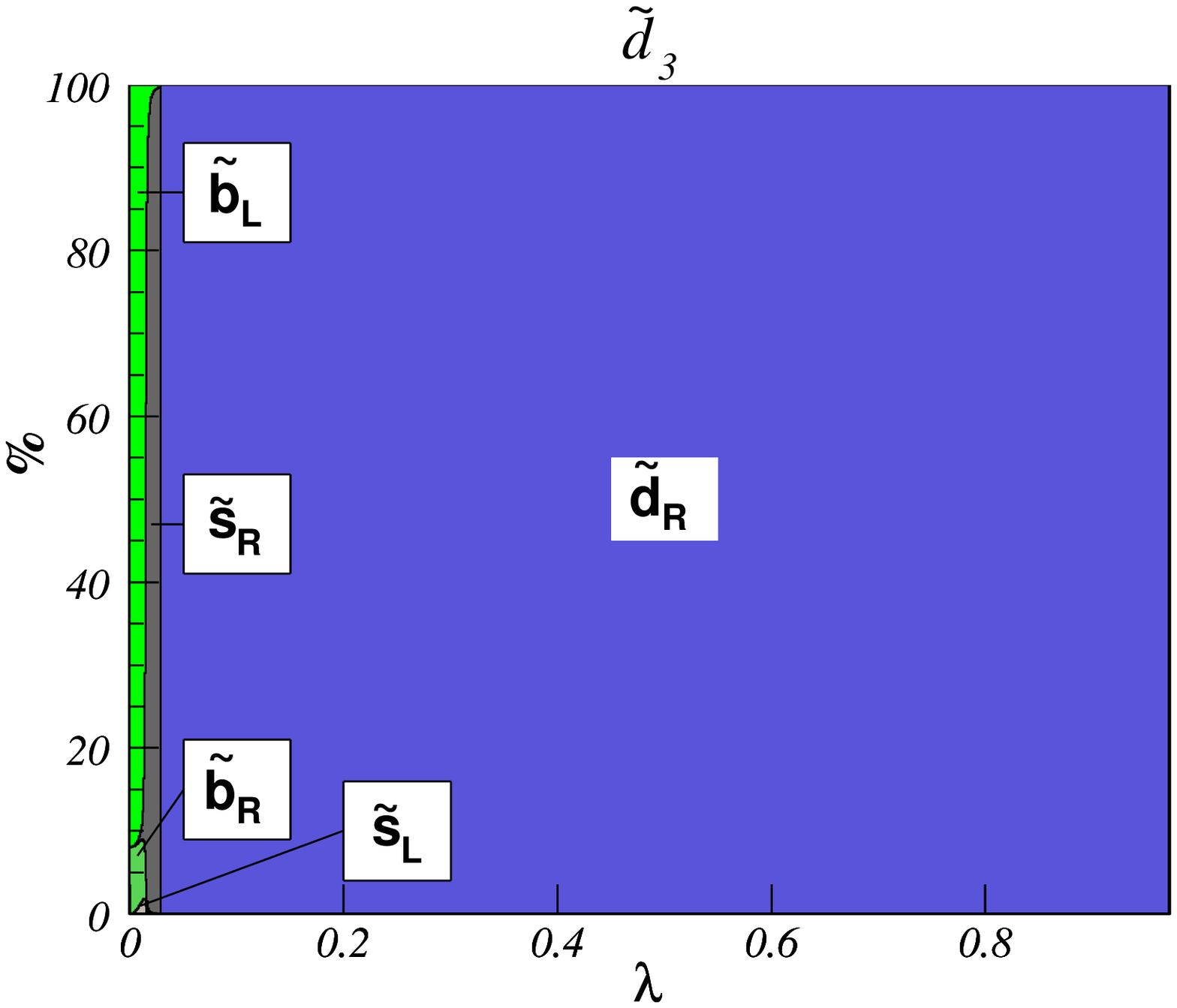}\hspace{1mm}
 \includegraphics[width=0.21\columnwidth]{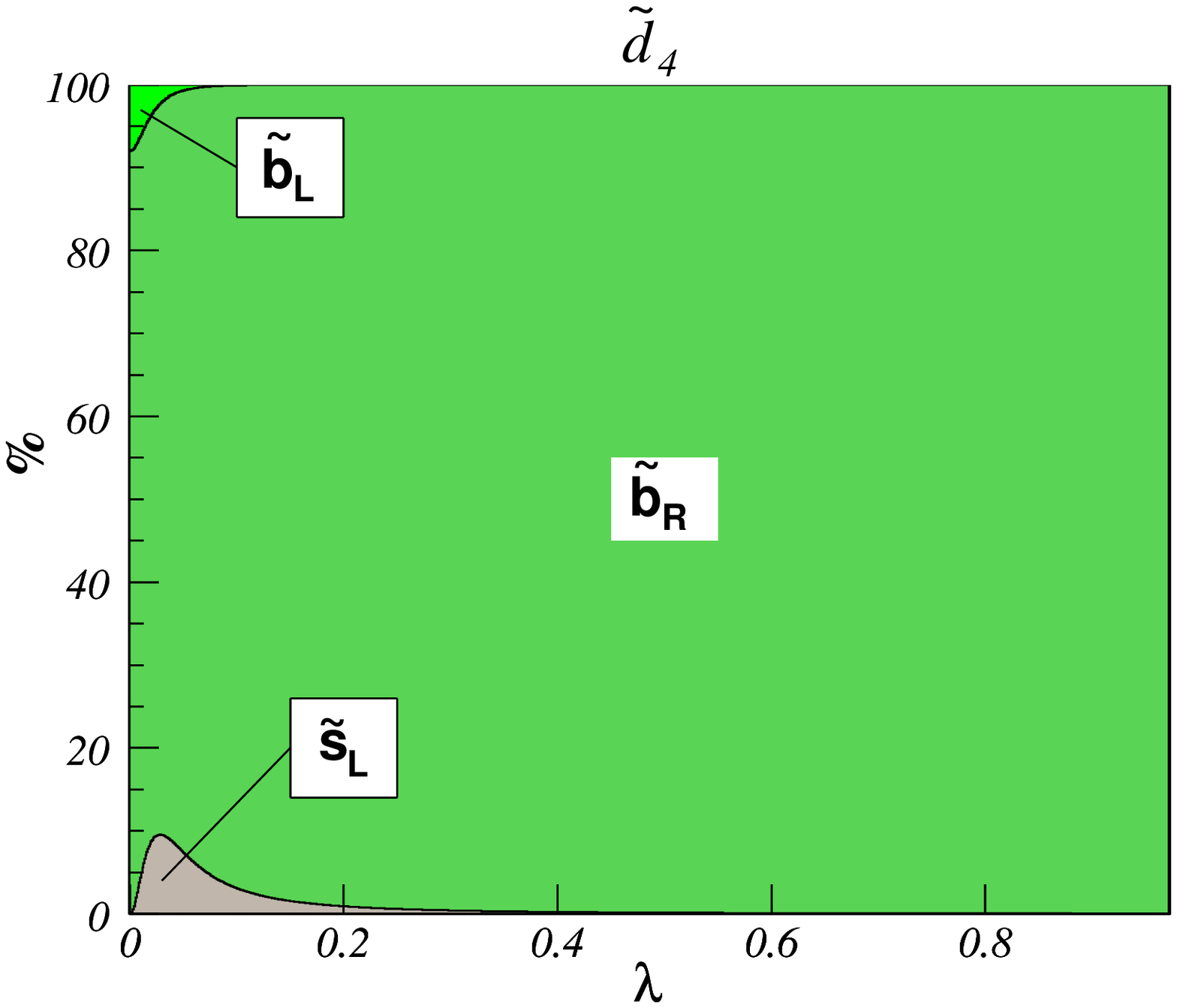}\vspace*{4mm}
 \includegraphics[width=0.21\columnwidth]{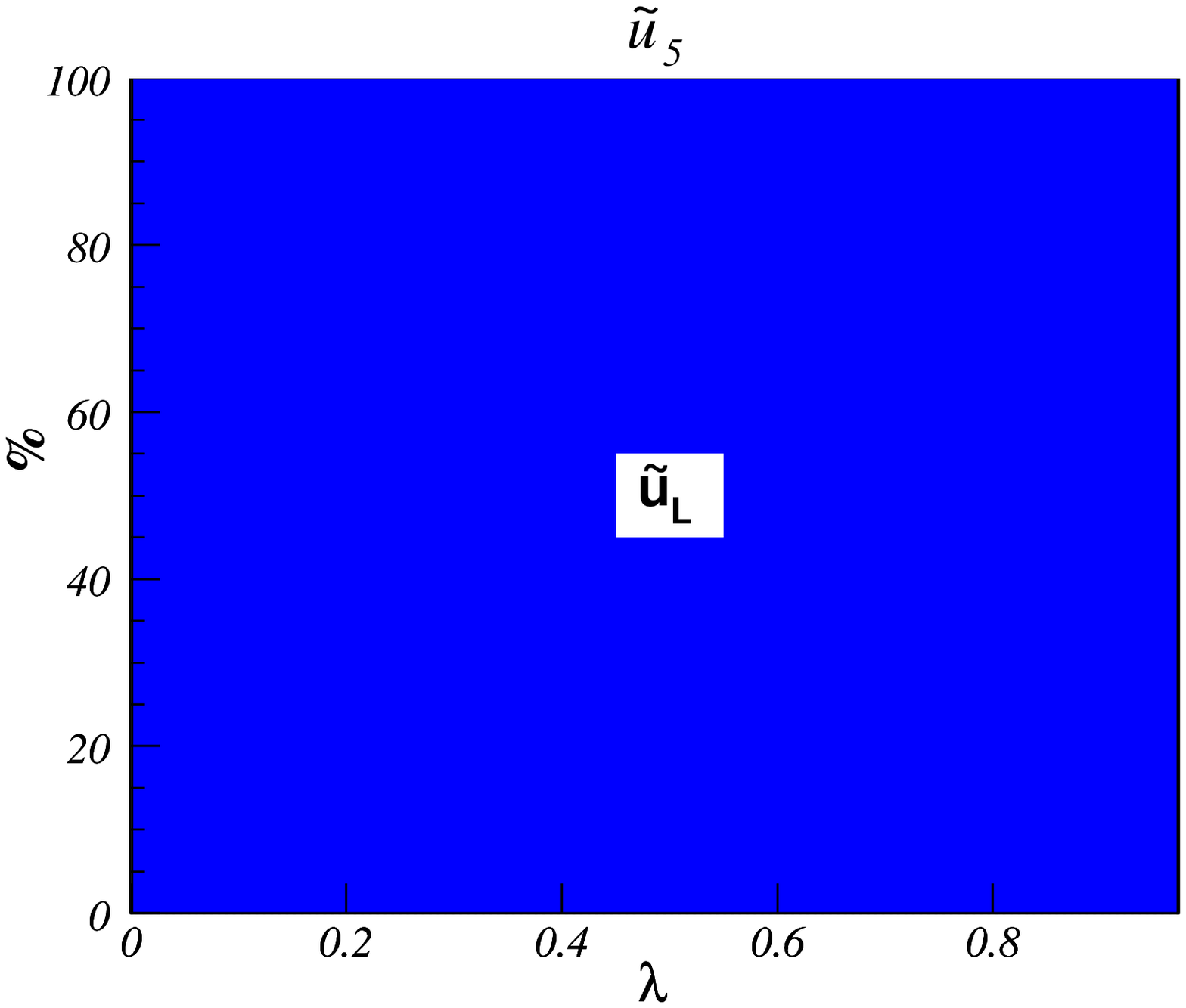}\hspace{1mm}
 \includegraphics[width=0.21\columnwidth]{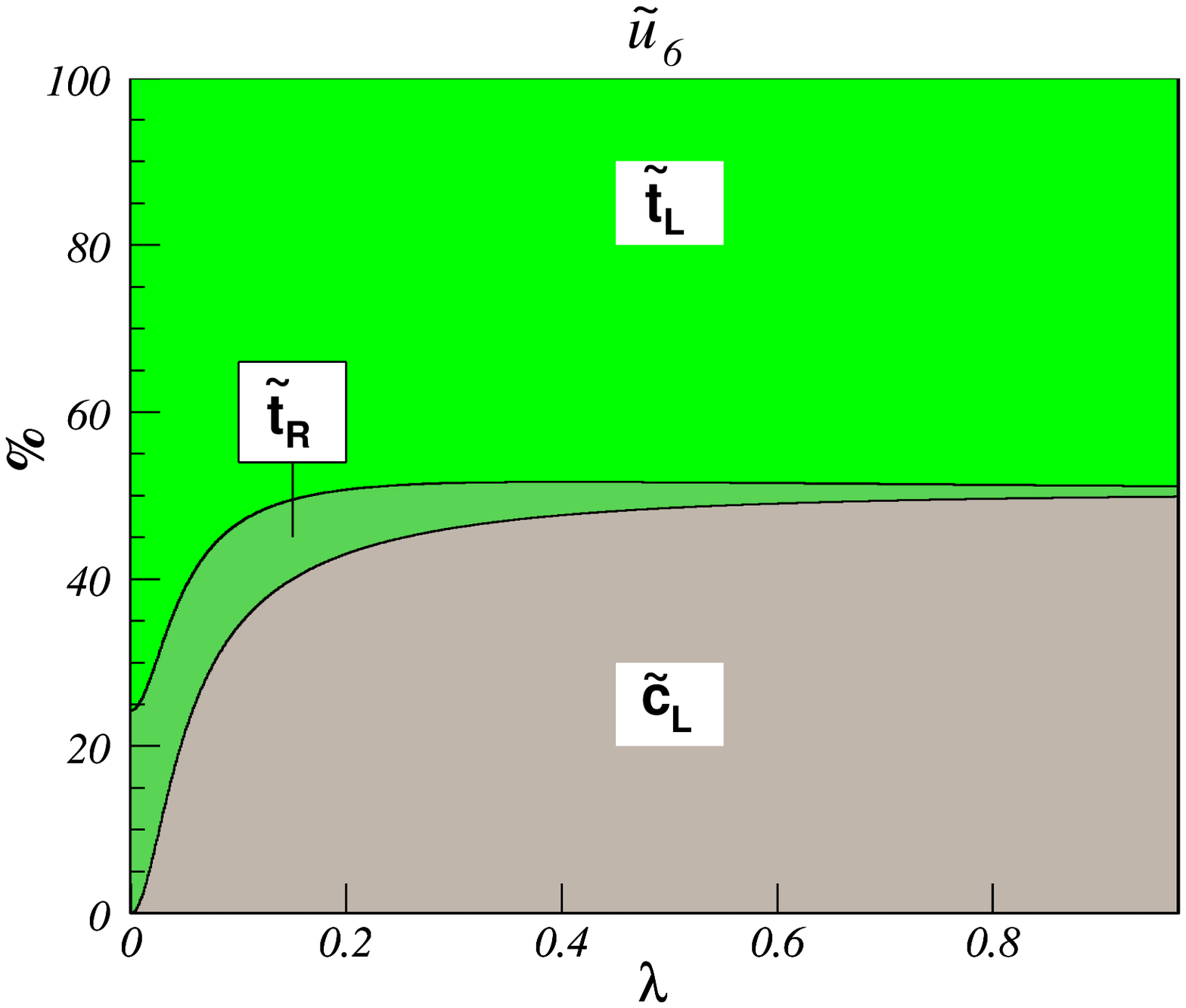}\hspace{1mm}
 \includegraphics[width=0.21\columnwidth]{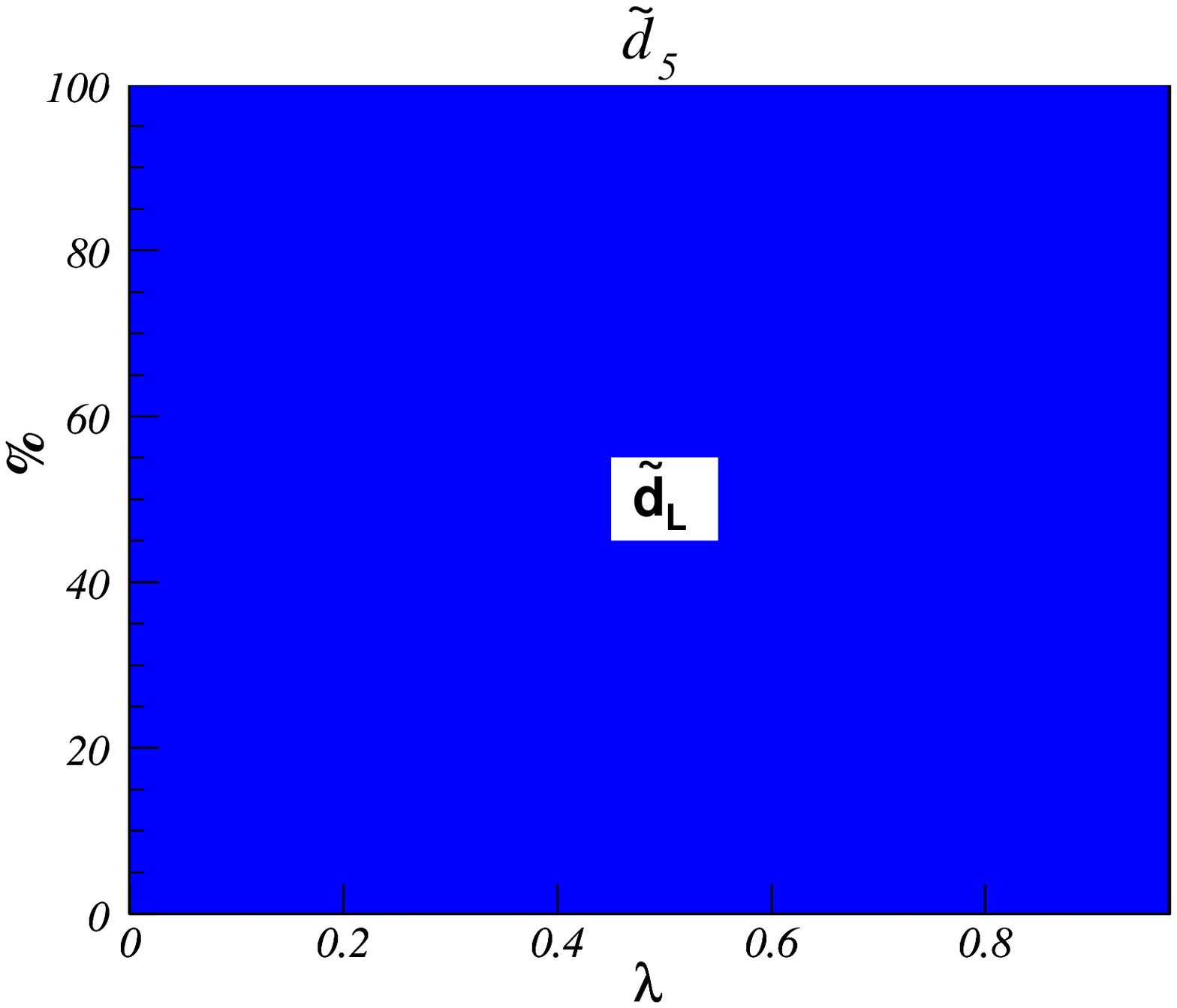}\hspace{1mm}
 \includegraphics[width=0.21\columnwidth]{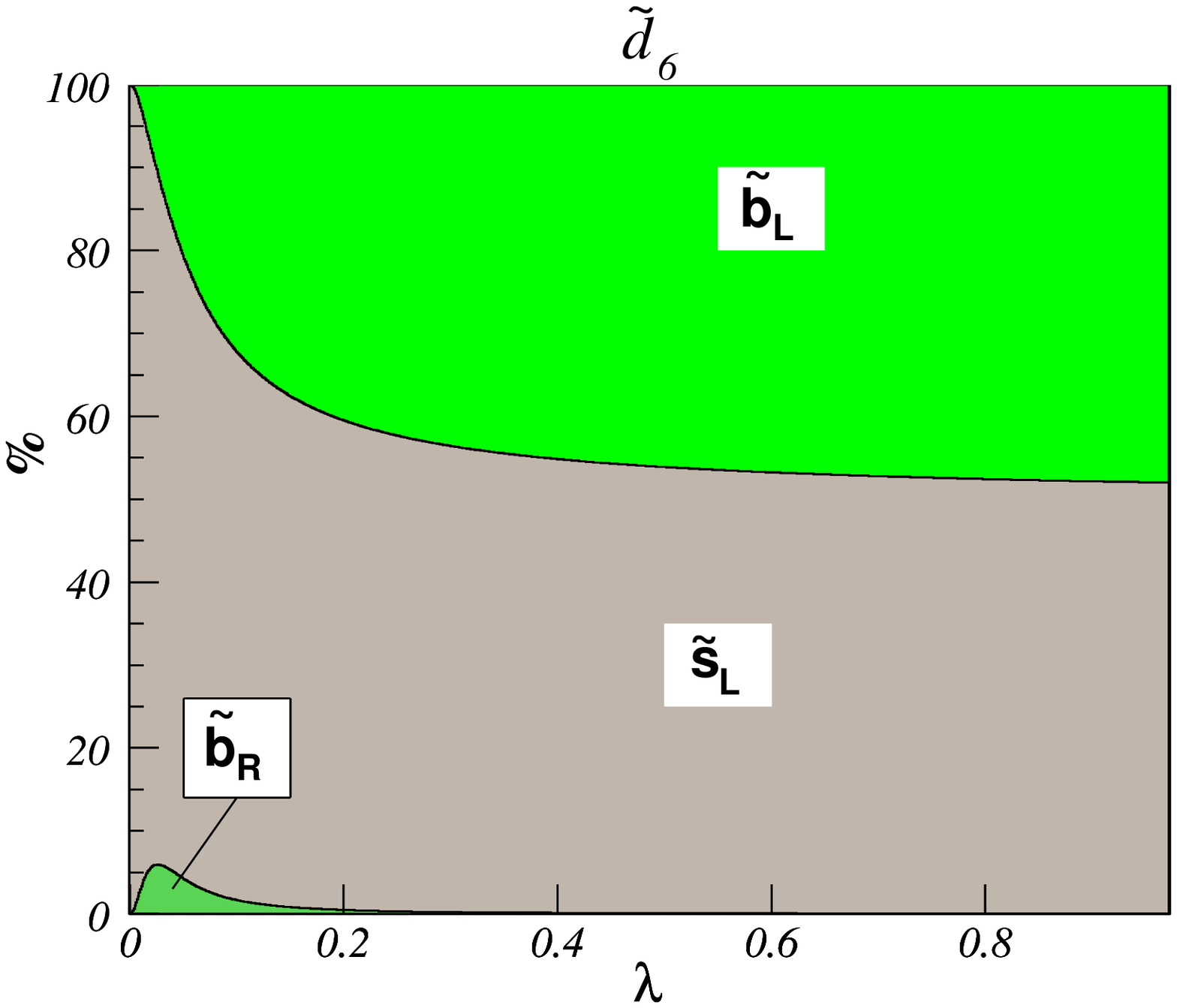}
 \caption{\label{fig:010}Same as Fig.\ \ref{fig:09} for benchmark point B.}\vspace{4mm}
 \includegraphics[width=0.21\columnwidth]{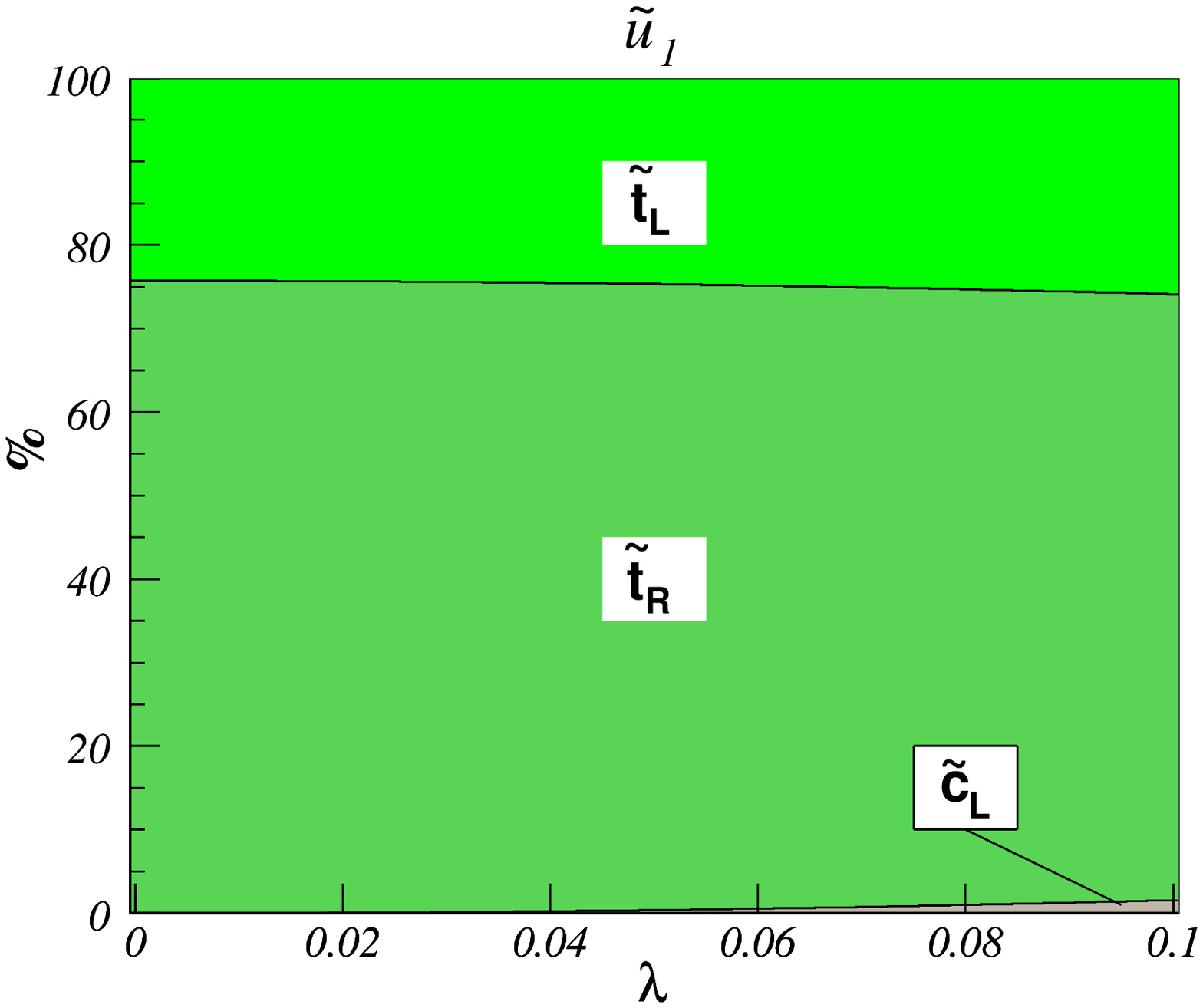}\hspace{1mm}
 \includegraphics[width=0.21\columnwidth]{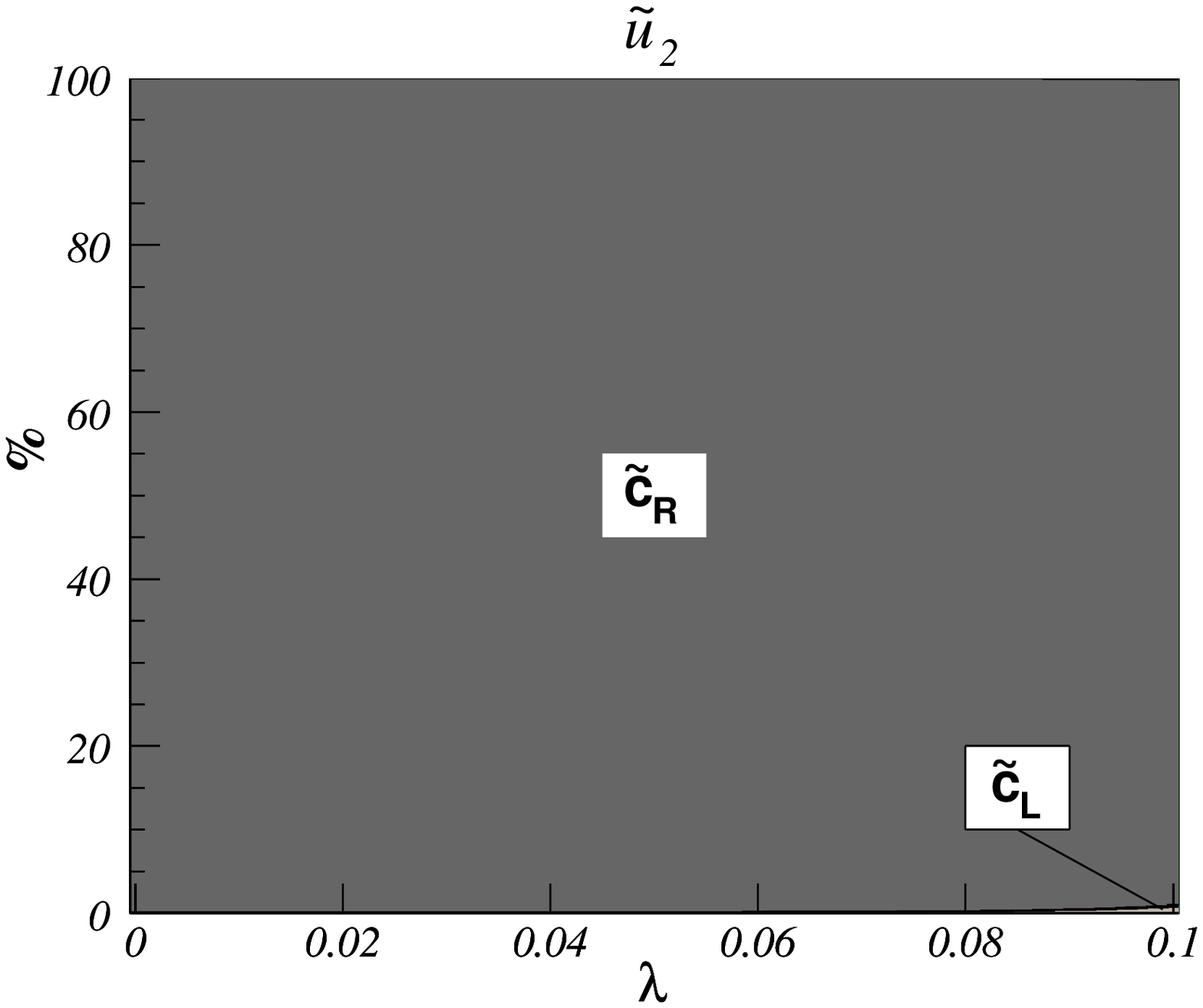}\hspace{1mm}
 \includegraphics[width=0.21\columnwidth]{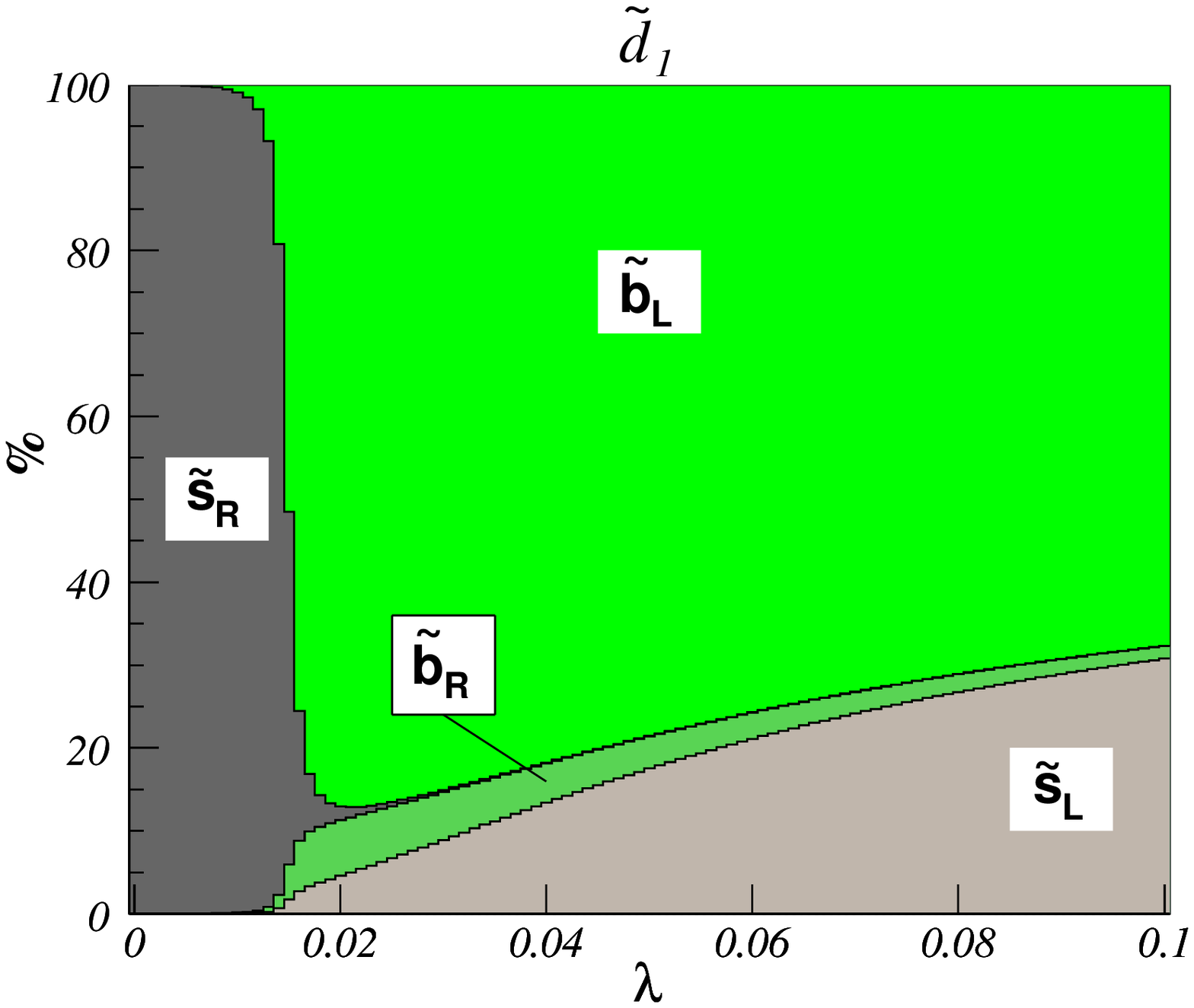}\hspace{1mm}
 \includegraphics[width=0.21\columnwidth]{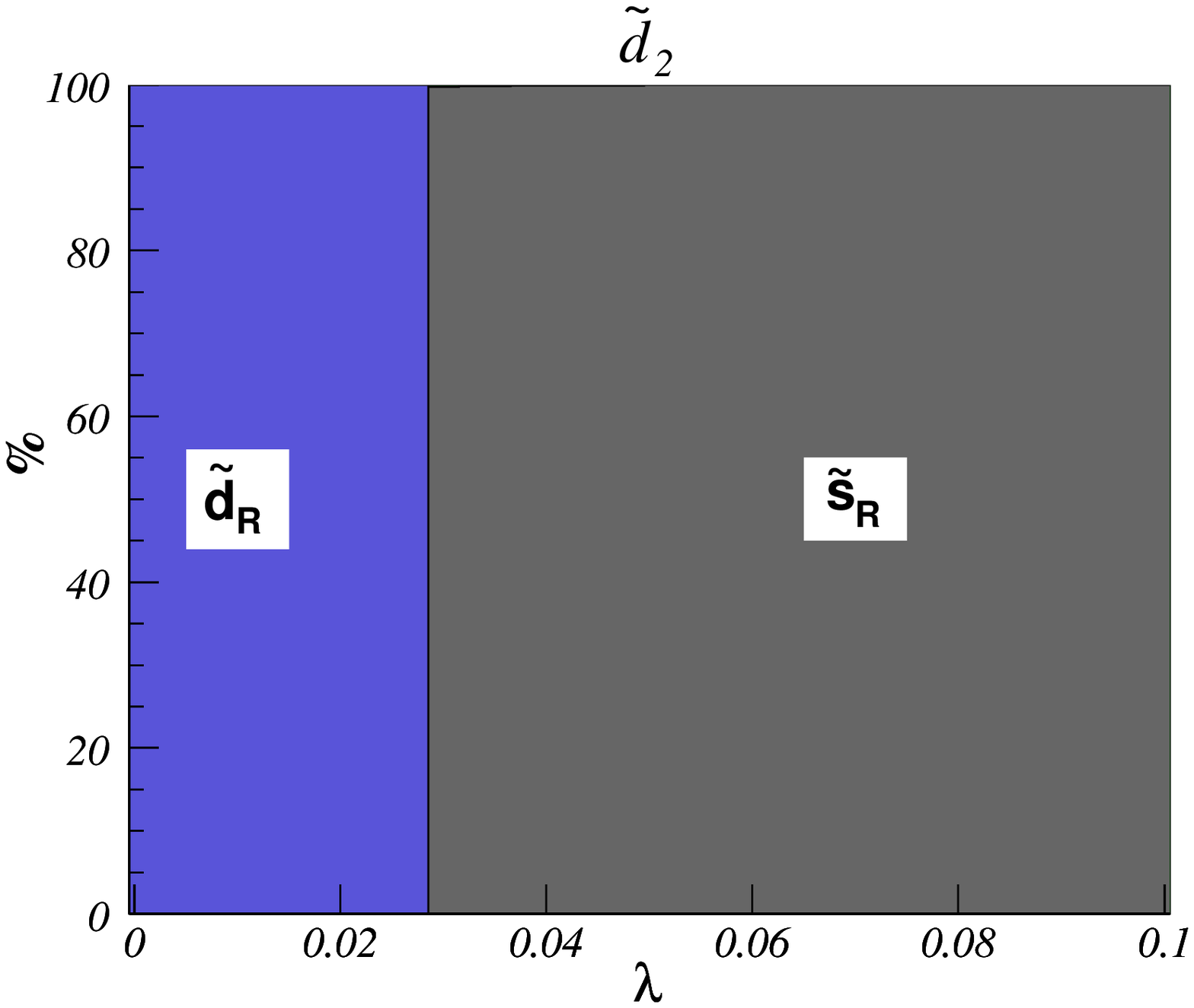}\vspace*{4mm}
 \includegraphics[width=0.21\columnwidth]{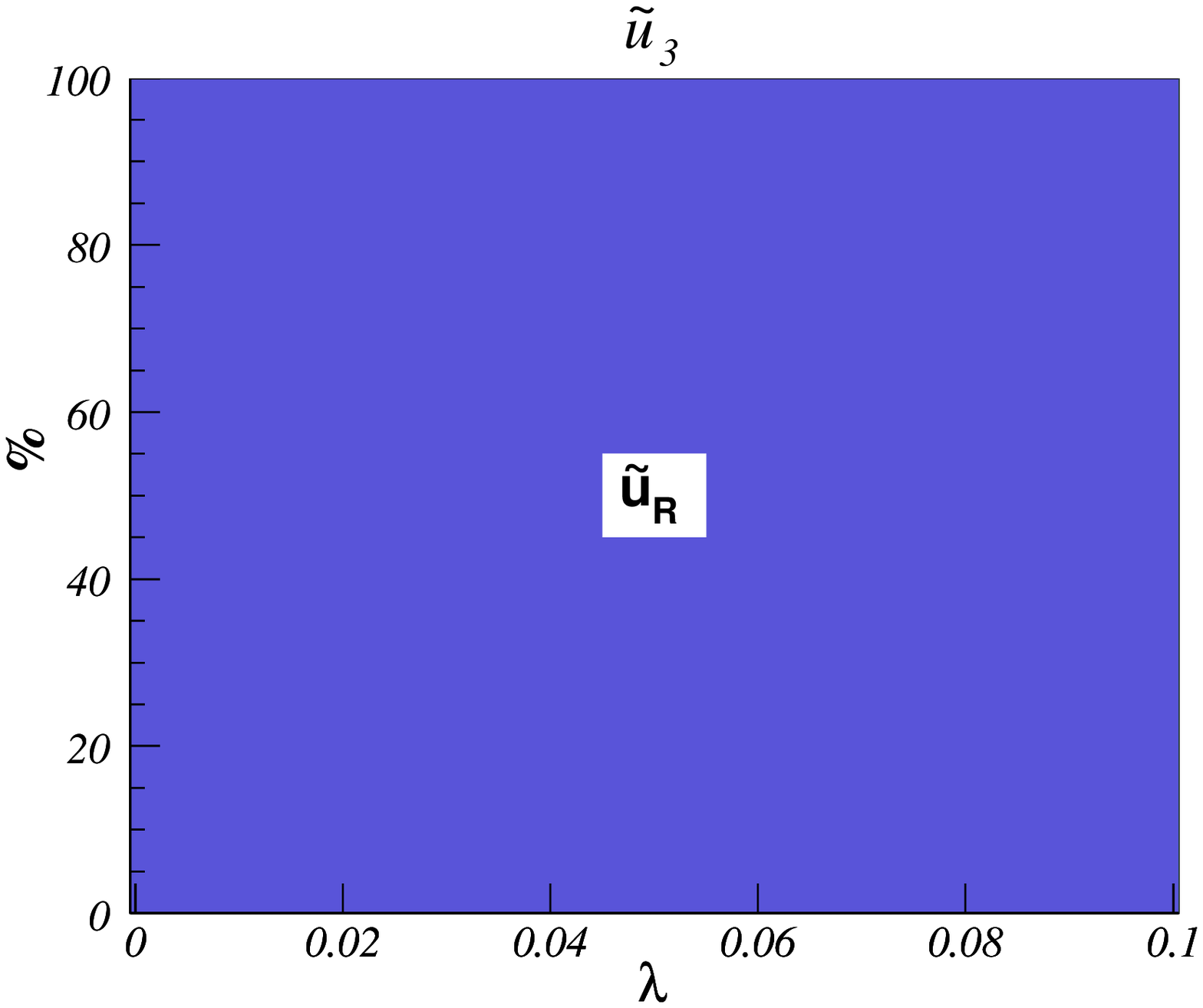}\hspace{1mm}
 \includegraphics[width=0.21\columnwidth]{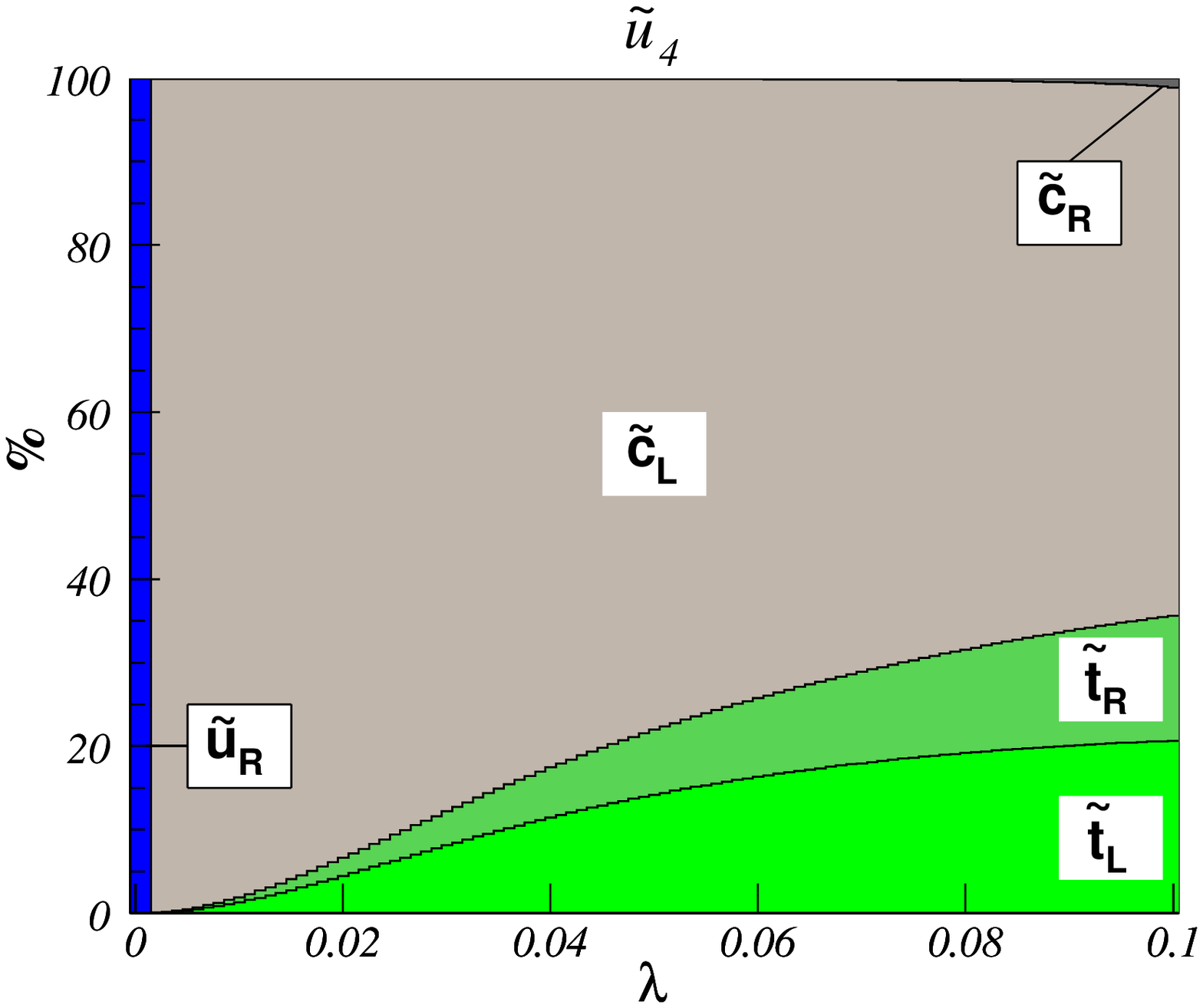}\hspace{1mm}
 \includegraphics[width=0.21\columnwidth]{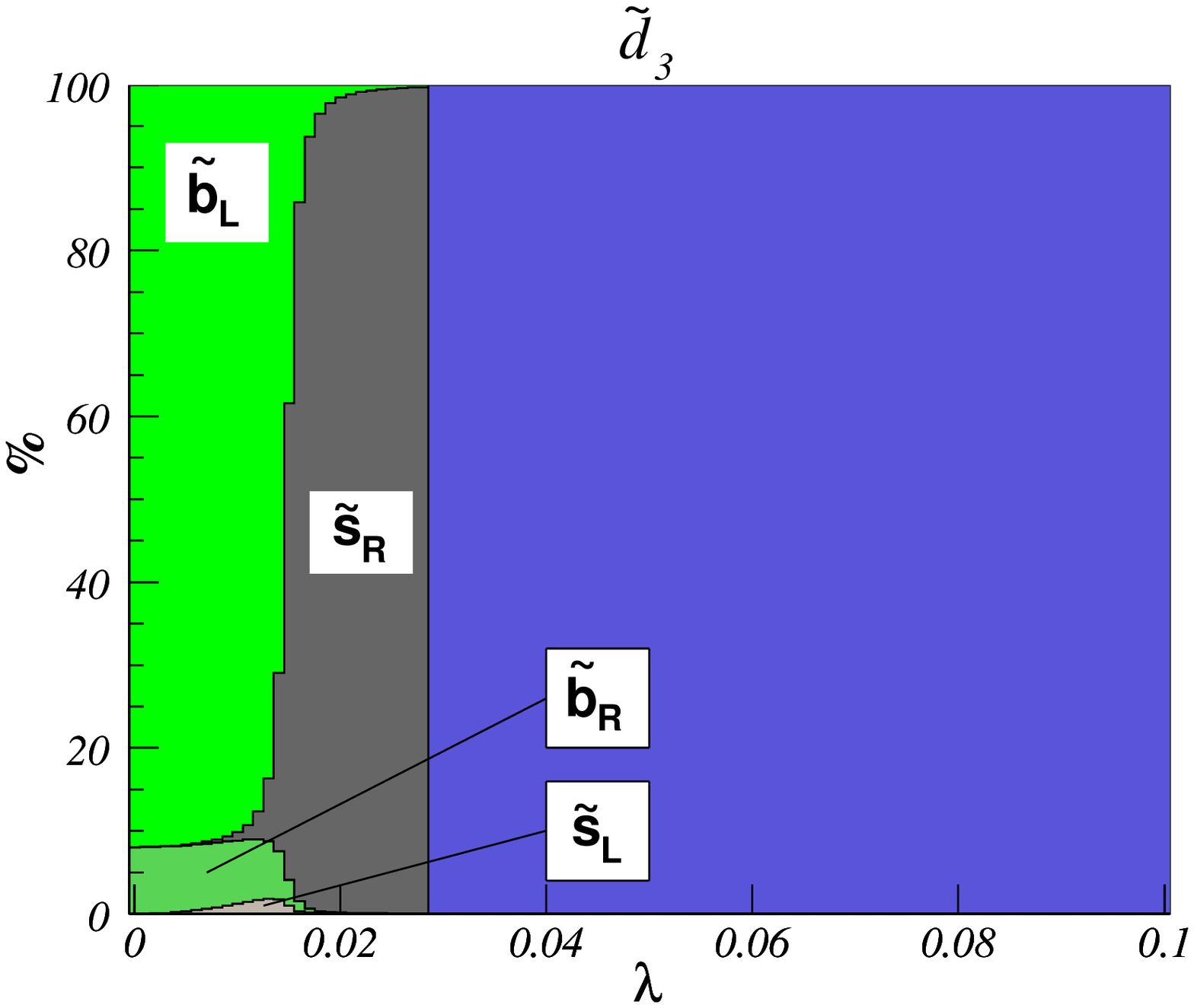}\hspace{1mm}
 \includegraphics[width=0.21\columnwidth]{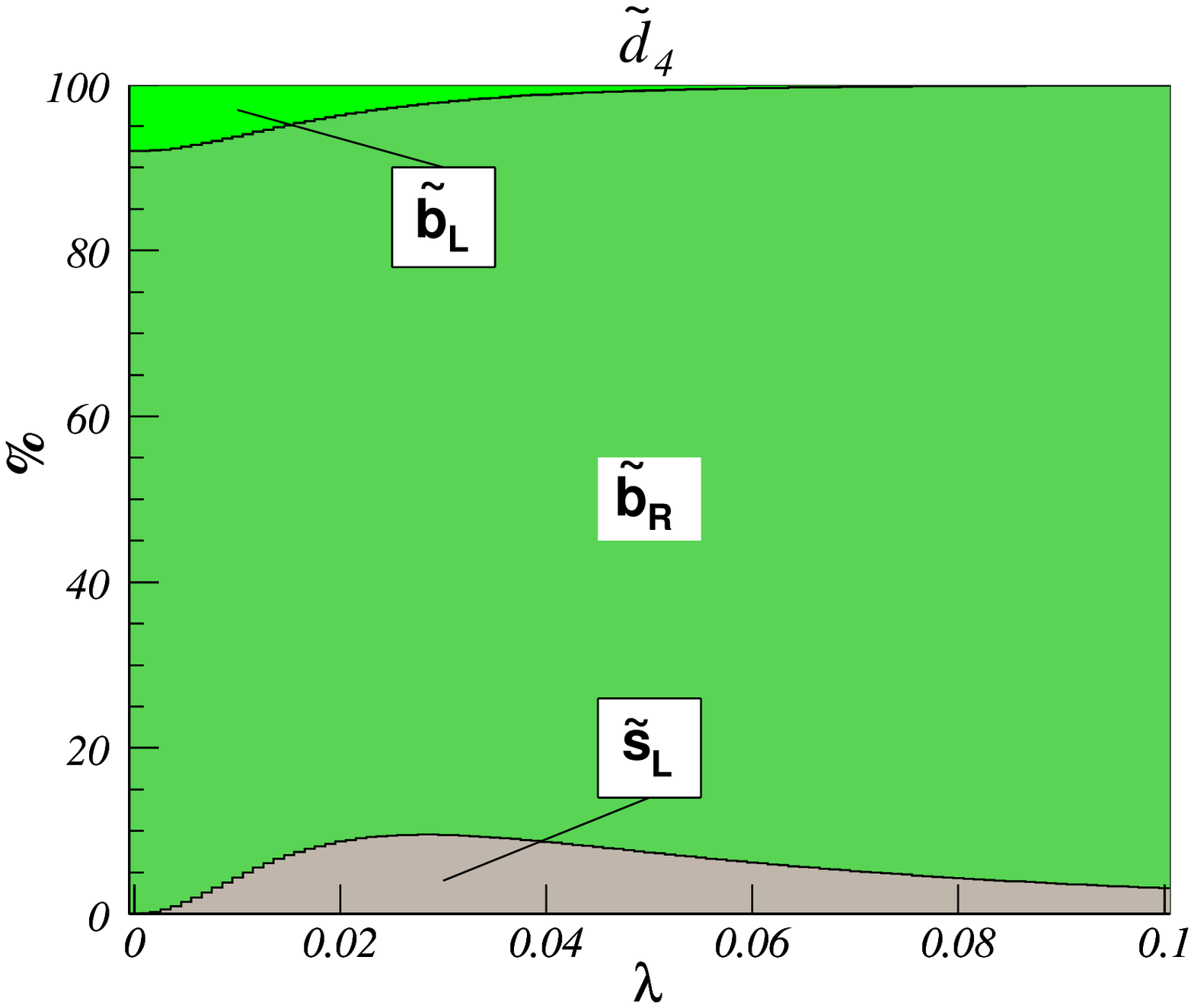}\vspace*{4mm}
 \includegraphics[width=0.21\columnwidth]{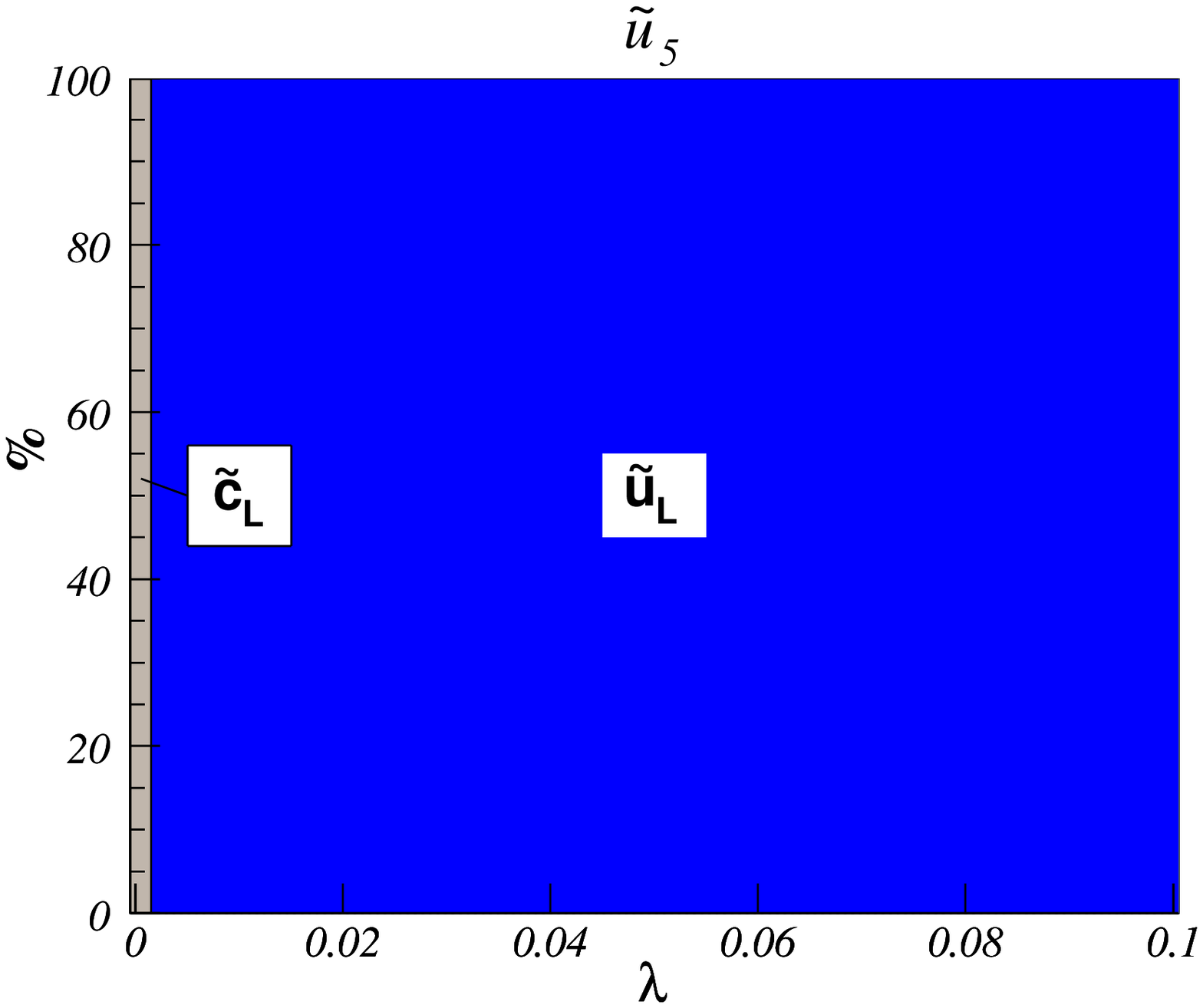}\hspace{1mm}
 \includegraphics[width=0.21\columnwidth]{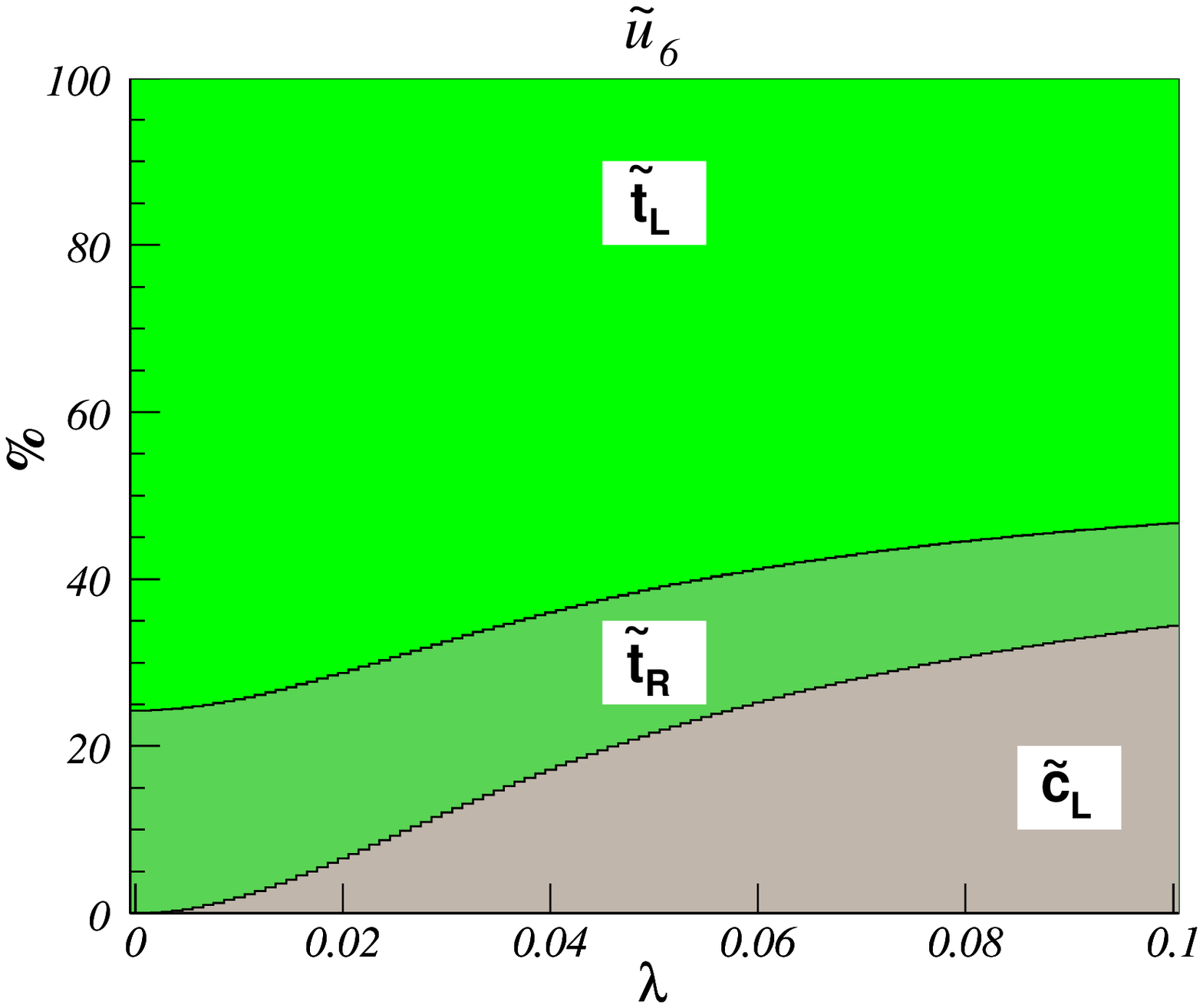}\hspace{1mm}
 \includegraphics[width=0.21\columnwidth]{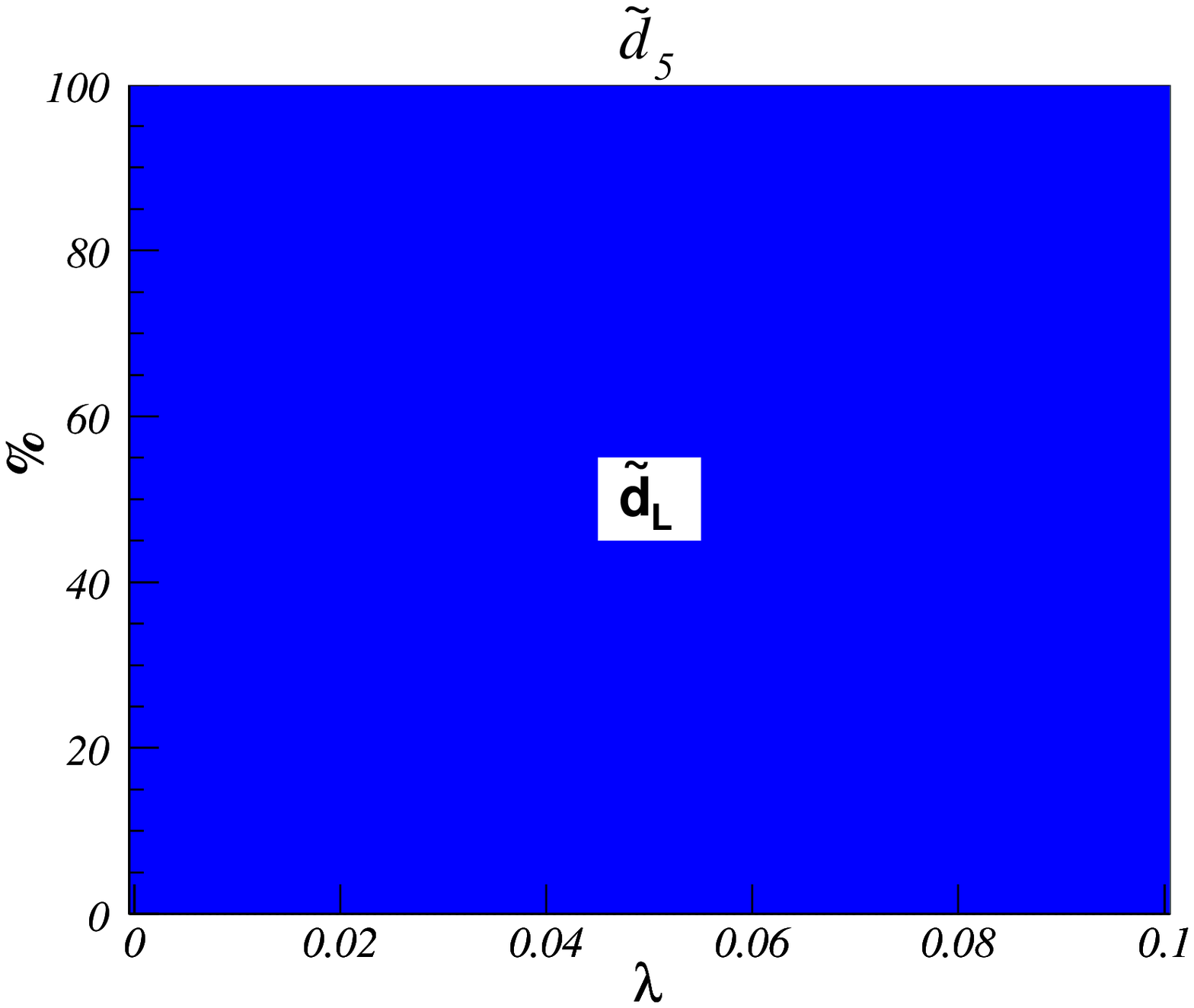}\hspace{1mm}
 \includegraphics[width=0.21\columnwidth]{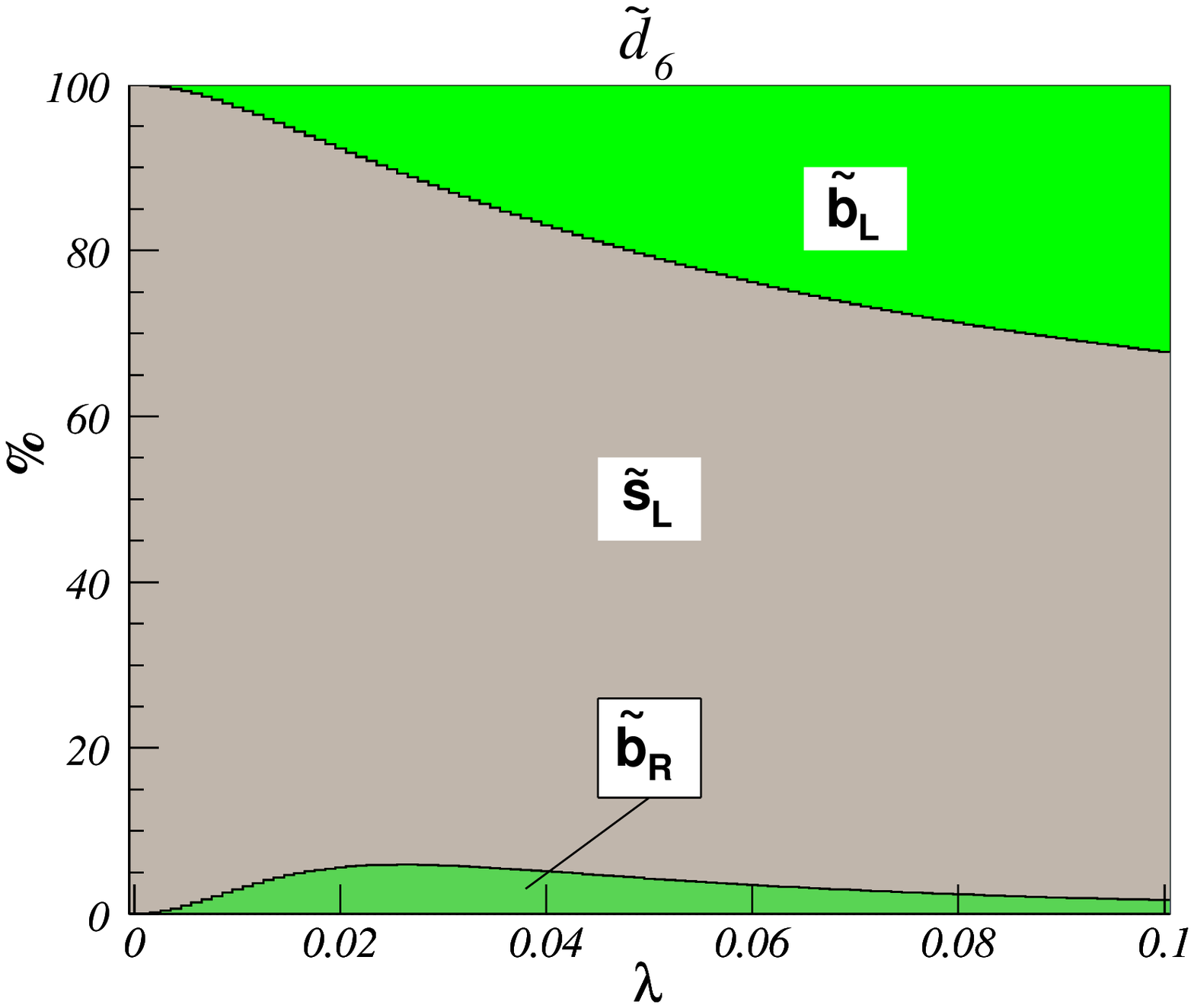}
 \caption{\label{fig:010p}Same as Fig.\ \ref{fig:010} for $\lambda\in
          [0;0.1]$.}
\end{figure}

\begin{figure}
 \centering
 \includegraphics[width=0.21\columnwidth]{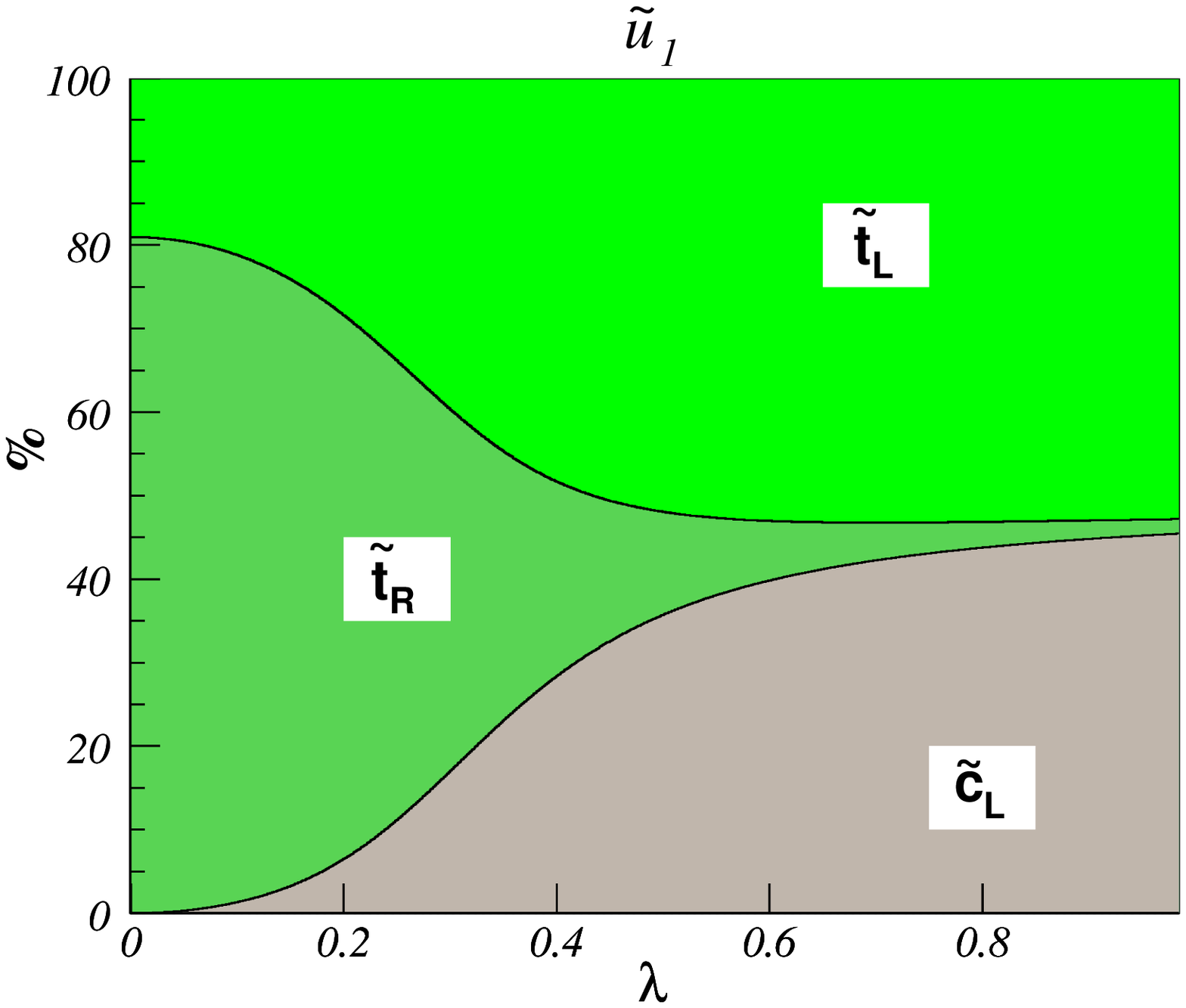}\hspace{1mm}
 \includegraphics[width=0.21\columnwidth]{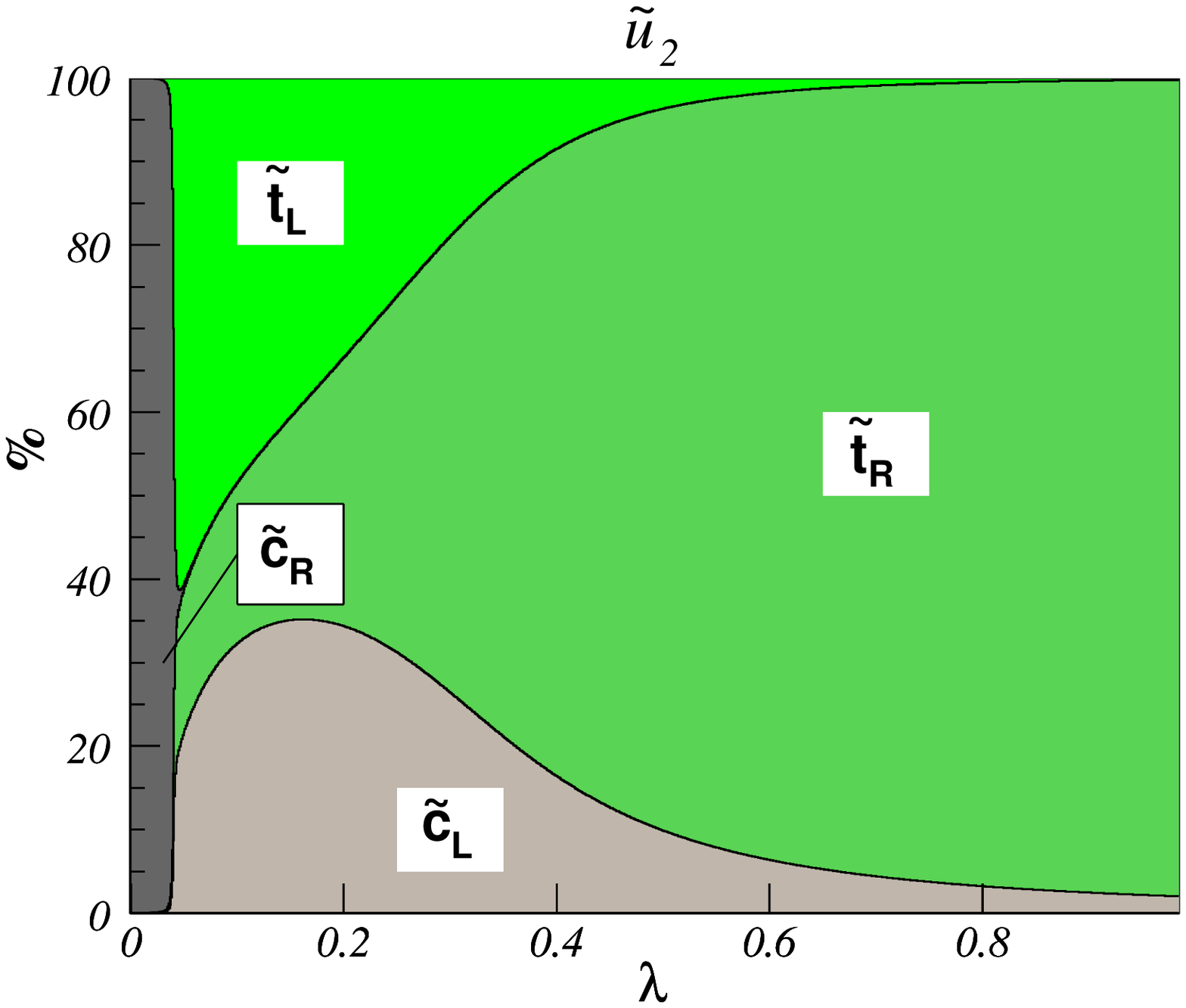}\hspace{1mm}
 \includegraphics[width=0.21\columnwidth]{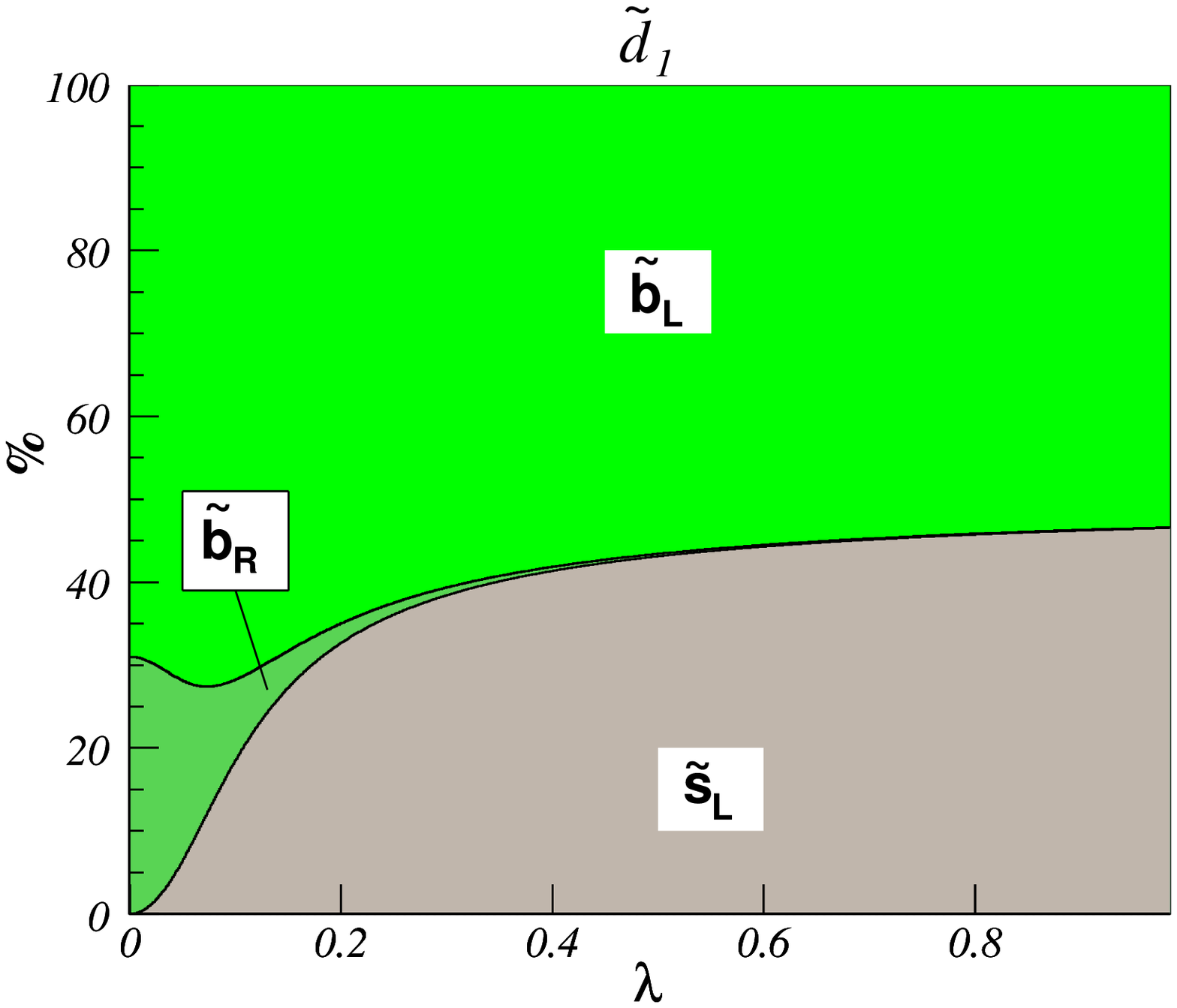}\hspace{1mm}
 \includegraphics[width=0.21\columnwidth]{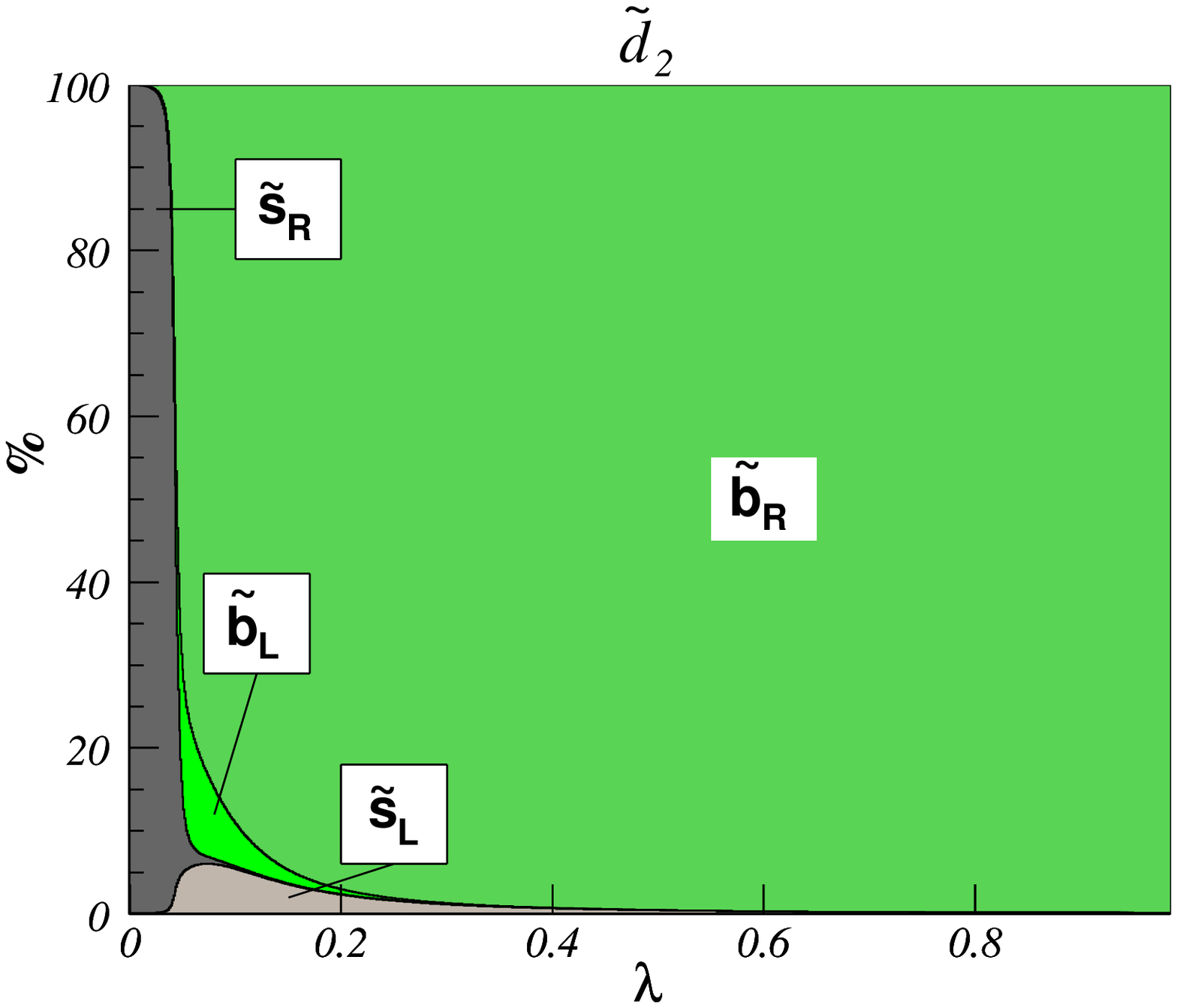}\vspace*{4mm}
 \includegraphics[width=0.21\columnwidth]{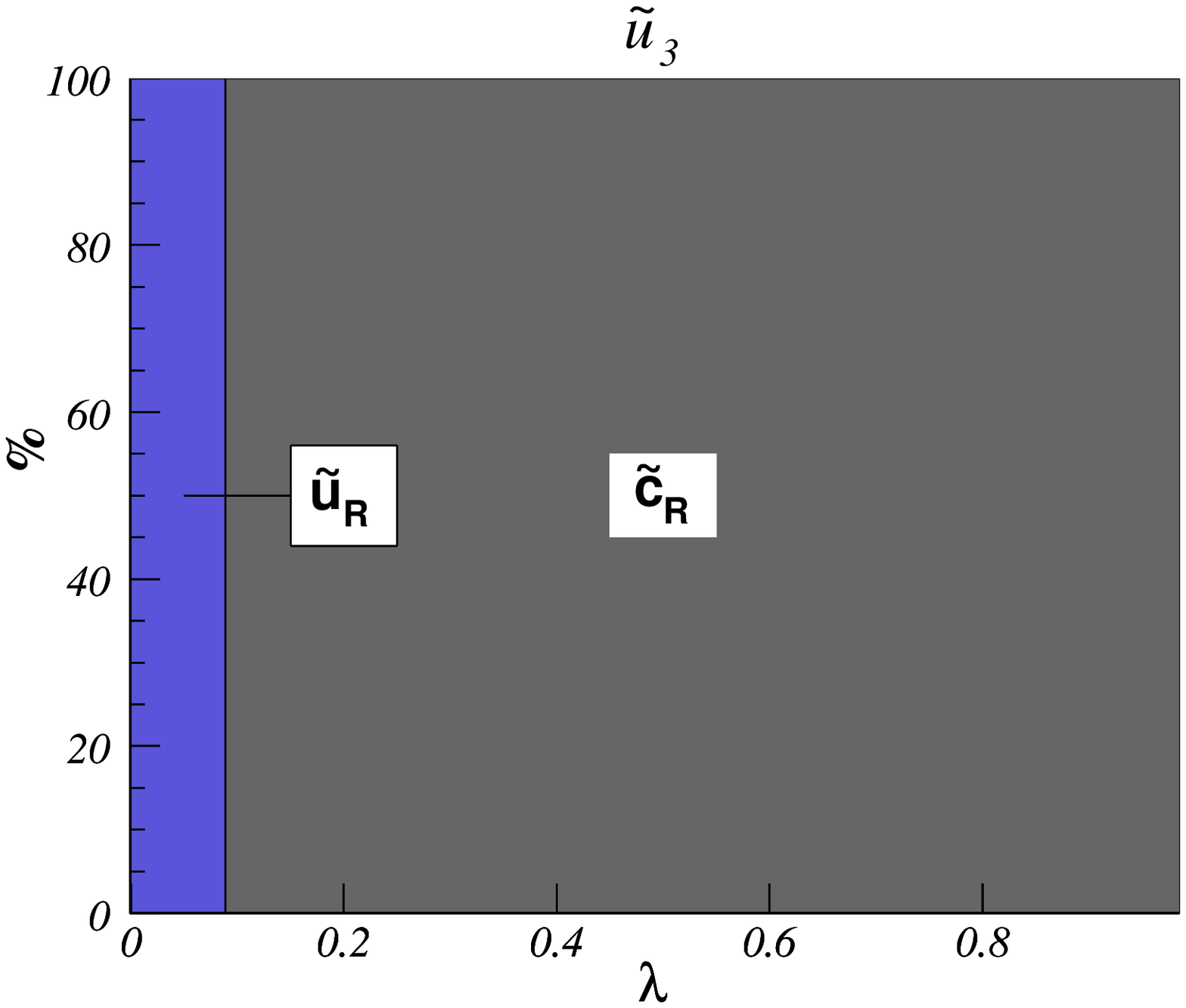}\hspace{1mm}
 \includegraphics[width=0.21\columnwidth]{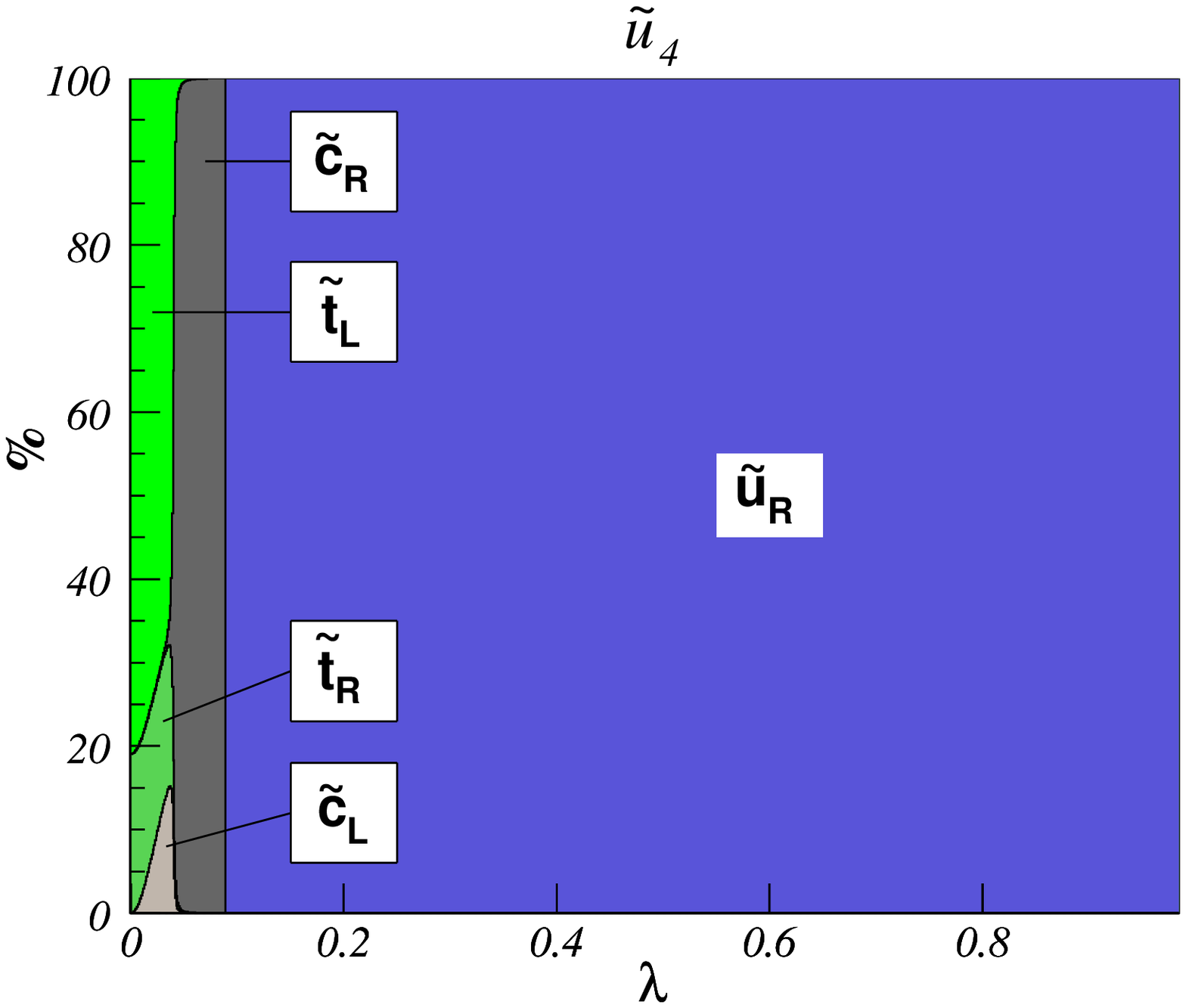}\hspace{1mm}
 \includegraphics[width=0.21\columnwidth]{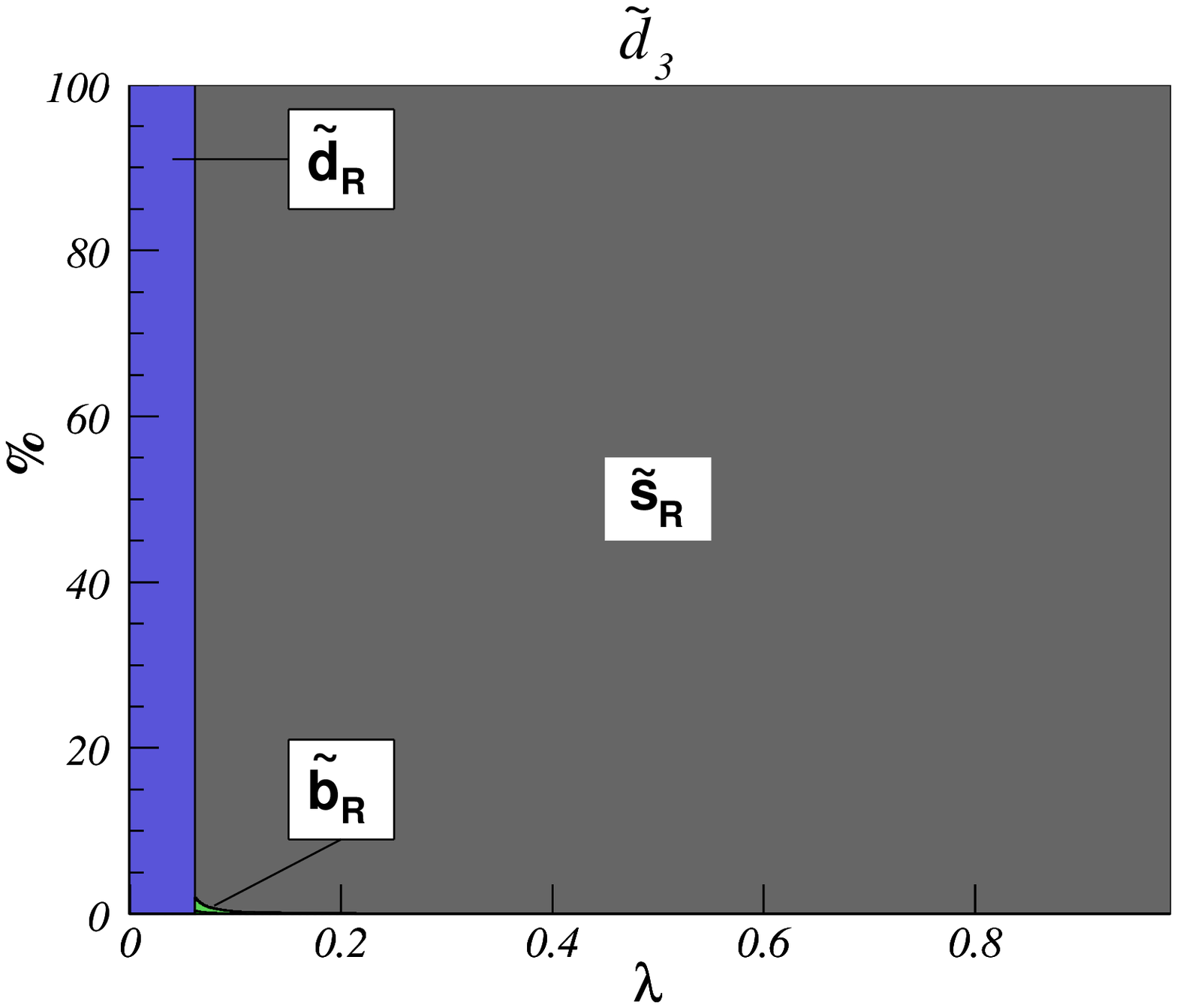}\hspace{1mm}
 \includegraphics[width=0.21\columnwidth]{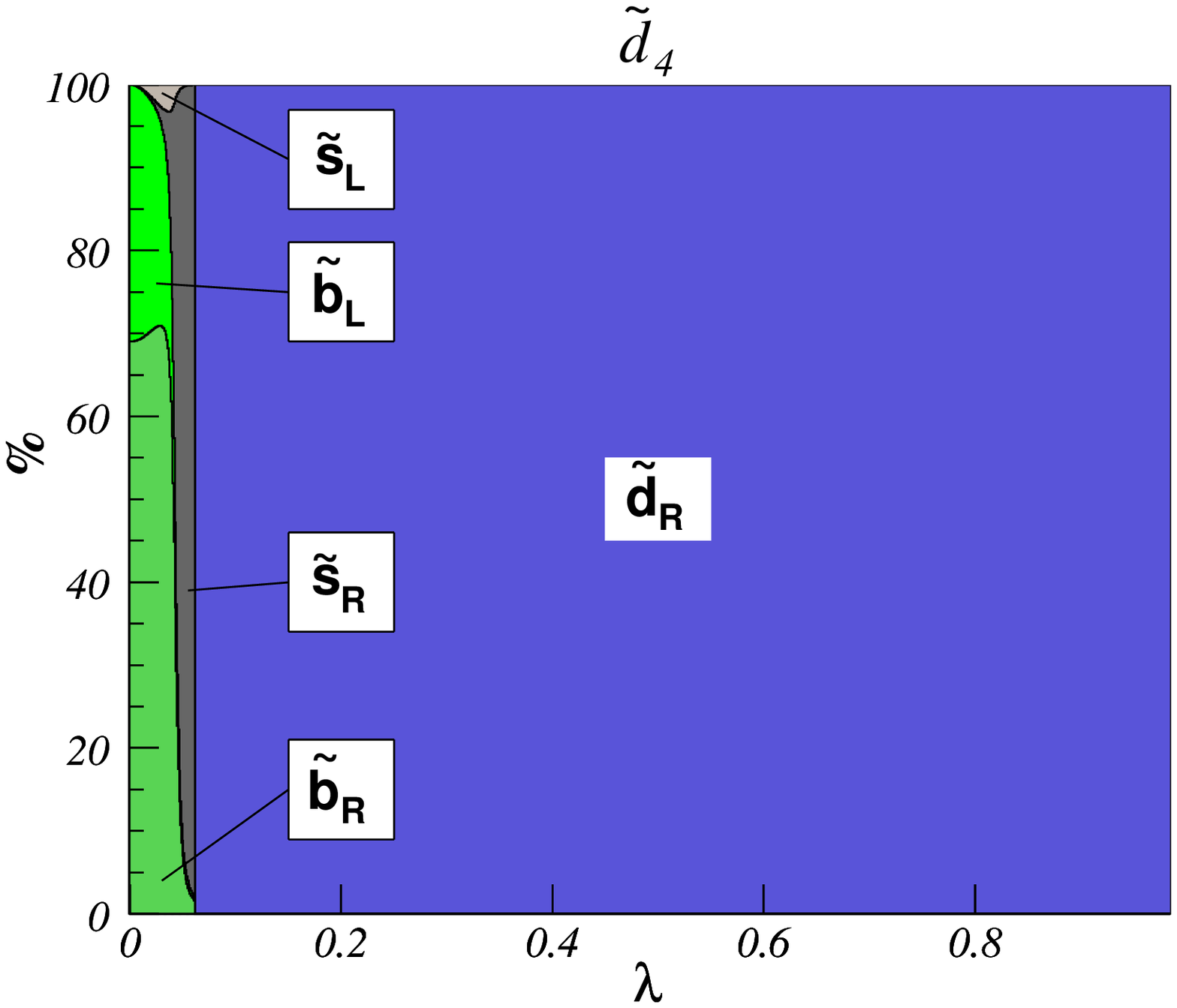}\vspace*{4mm}
 \includegraphics[width=0.21\columnwidth]{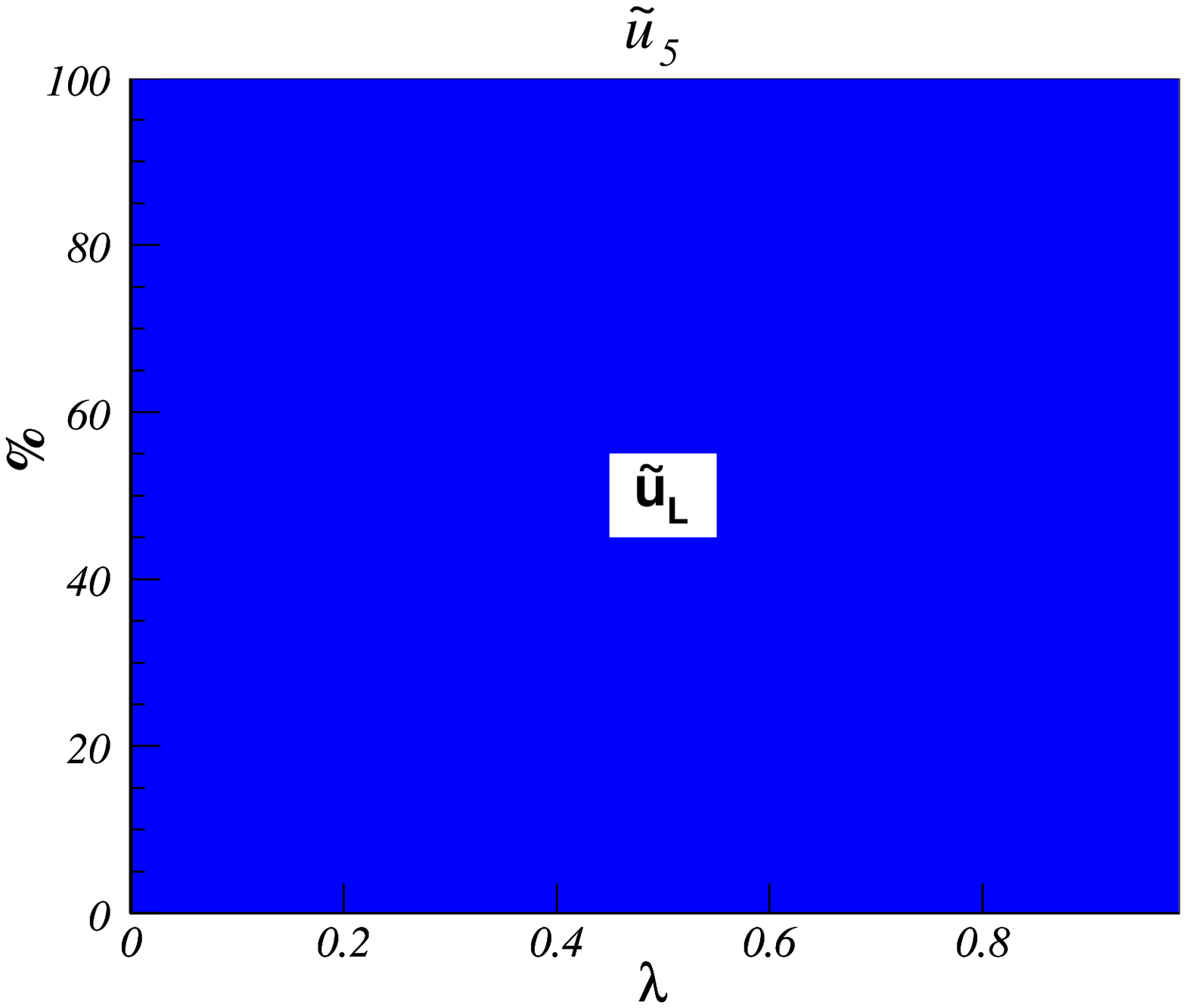}\hspace{1mm}
 \includegraphics[width=0.21\columnwidth]{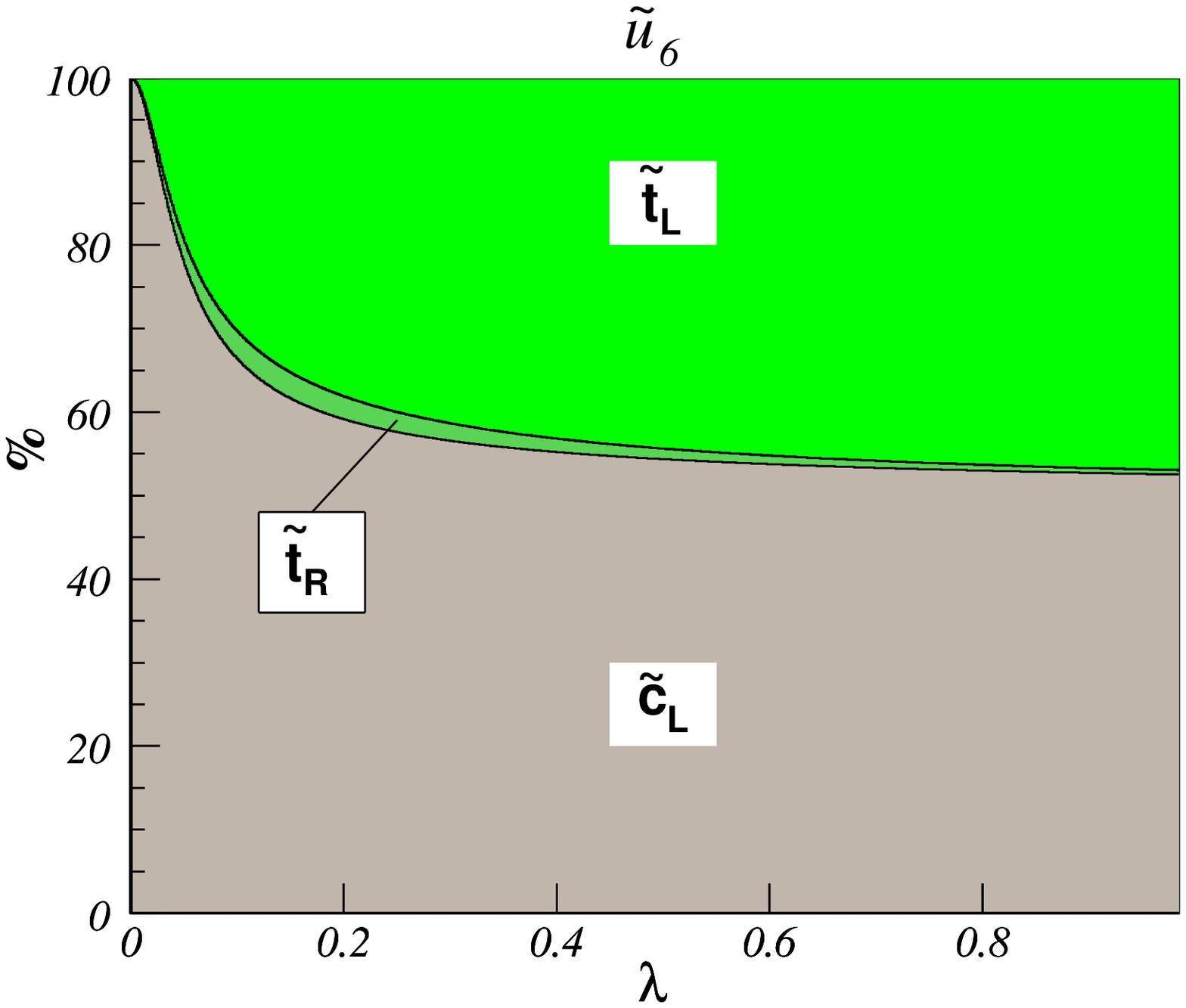}\hspace{1mm}
 \includegraphics[width=0.21\columnwidth]{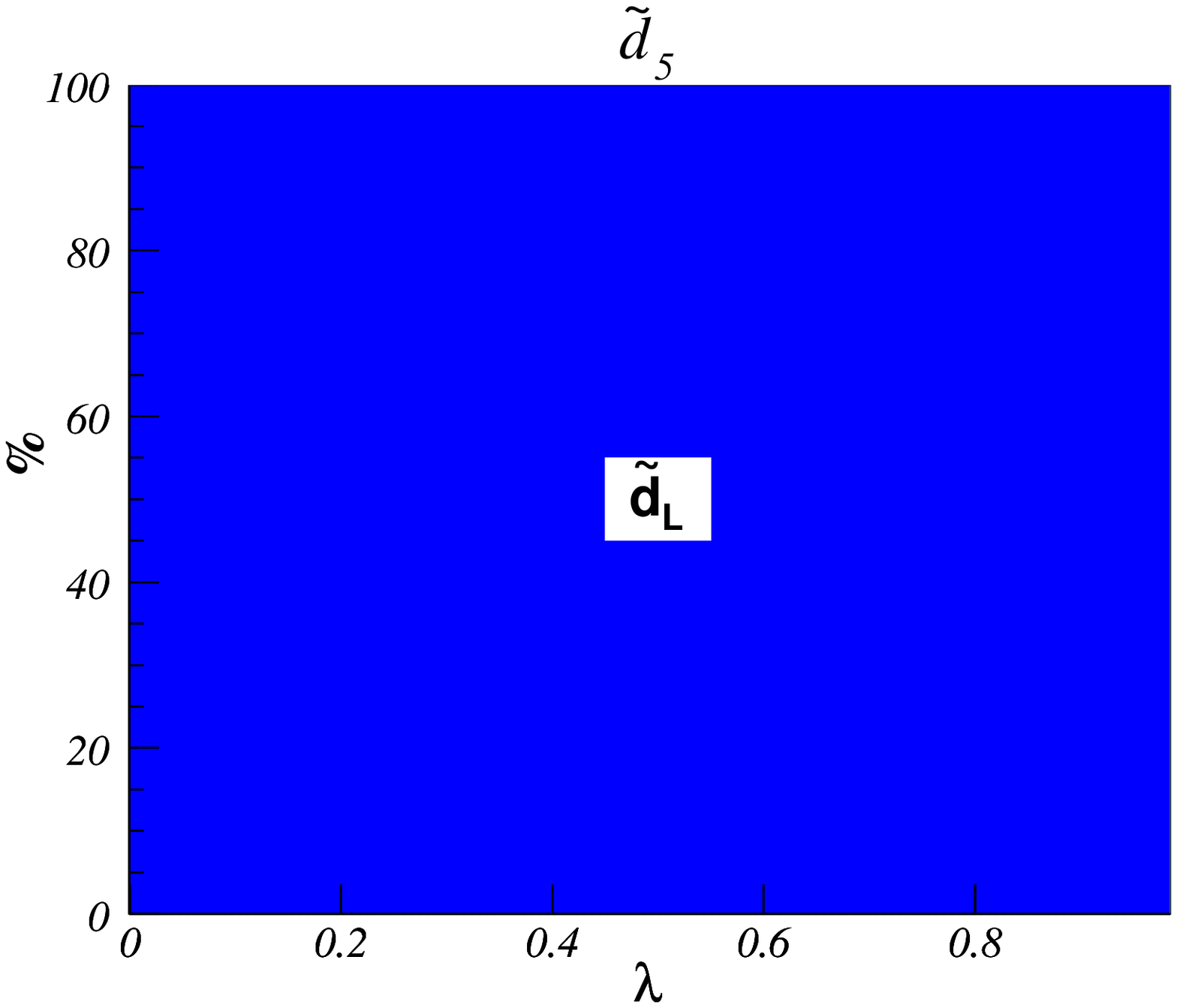}\hspace{1mm}
 \includegraphics[width=0.21\columnwidth]{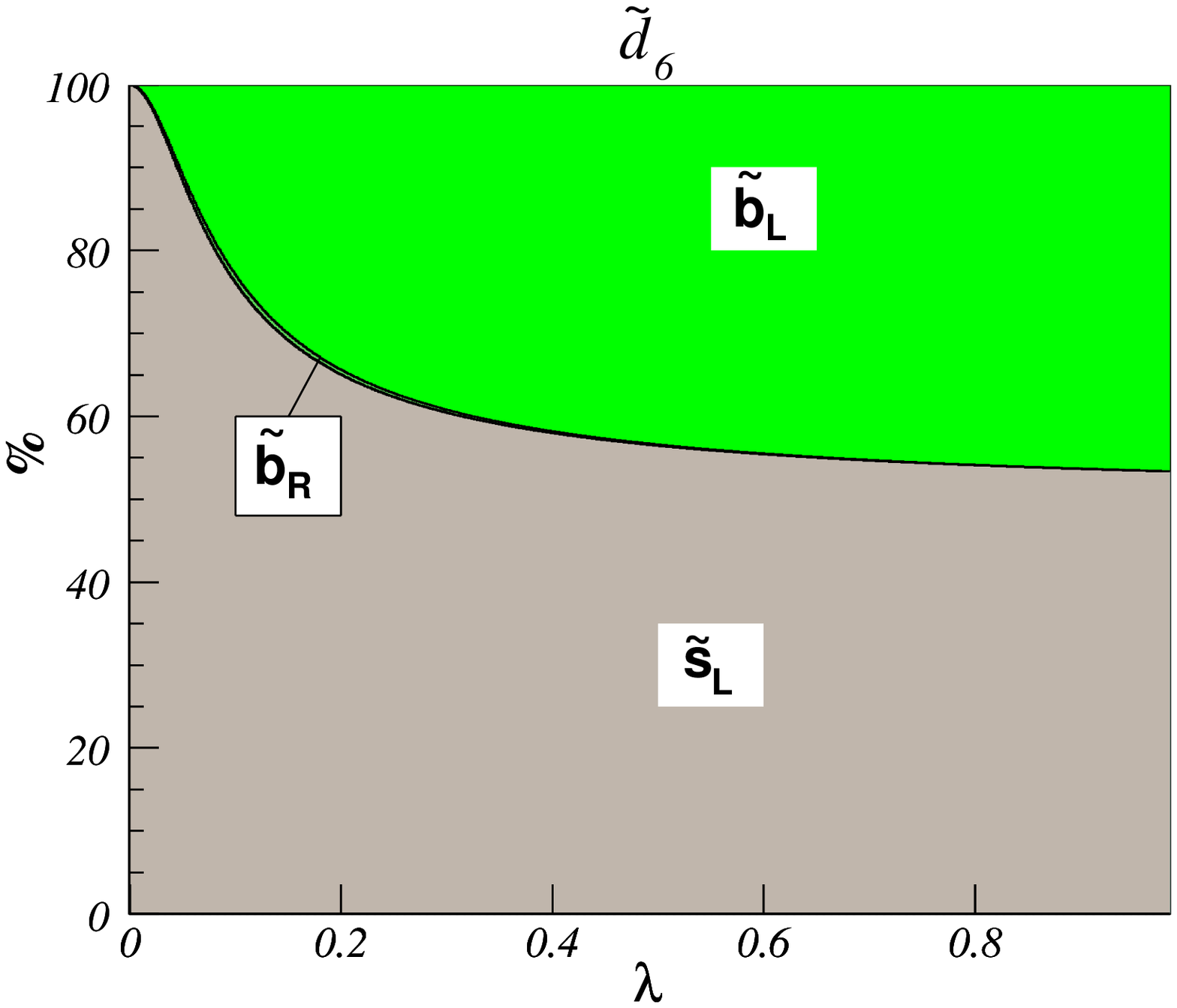}
 \caption{\label{fig:011}Same as Fig.\ \ref{fig:09} for benchmark point C.}\vspace{4mm}
 \includegraphics[width=0.21\columnwidth]{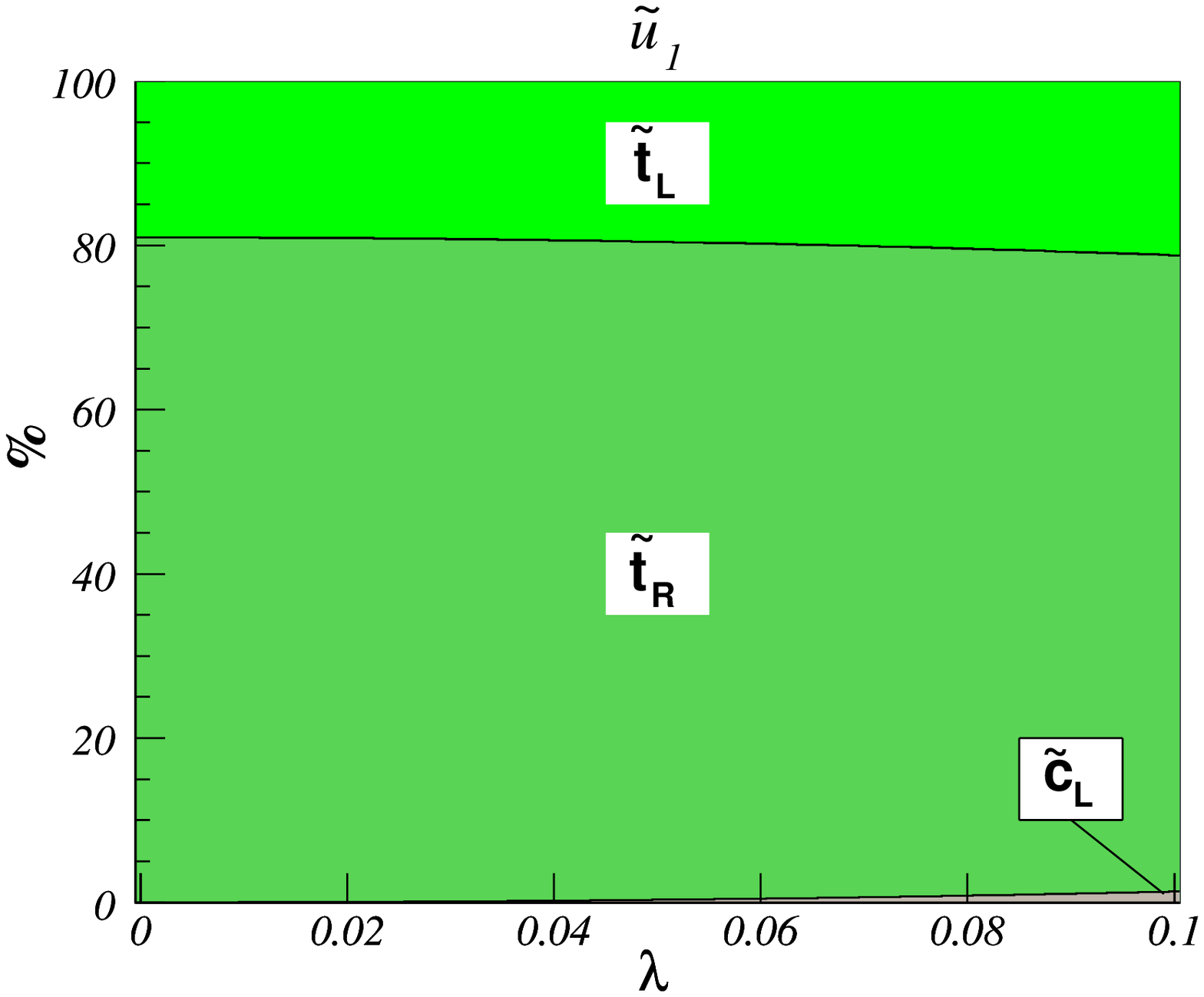}\hspace{1mm}
 \includegraphics[width=0.21\columnwidth]{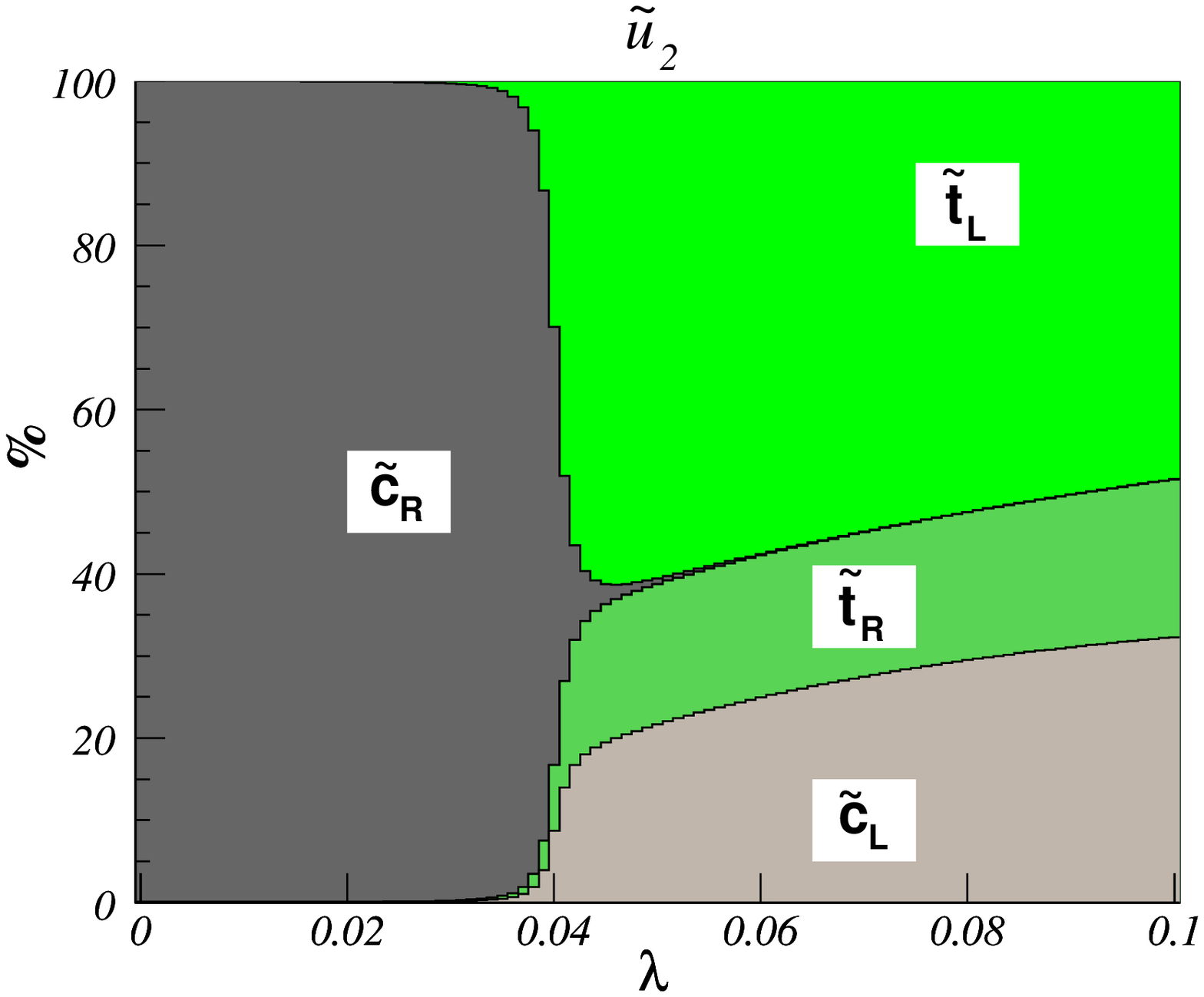}\hspace{1mm}
 \includegraphics[width=0.21\columnwidth]{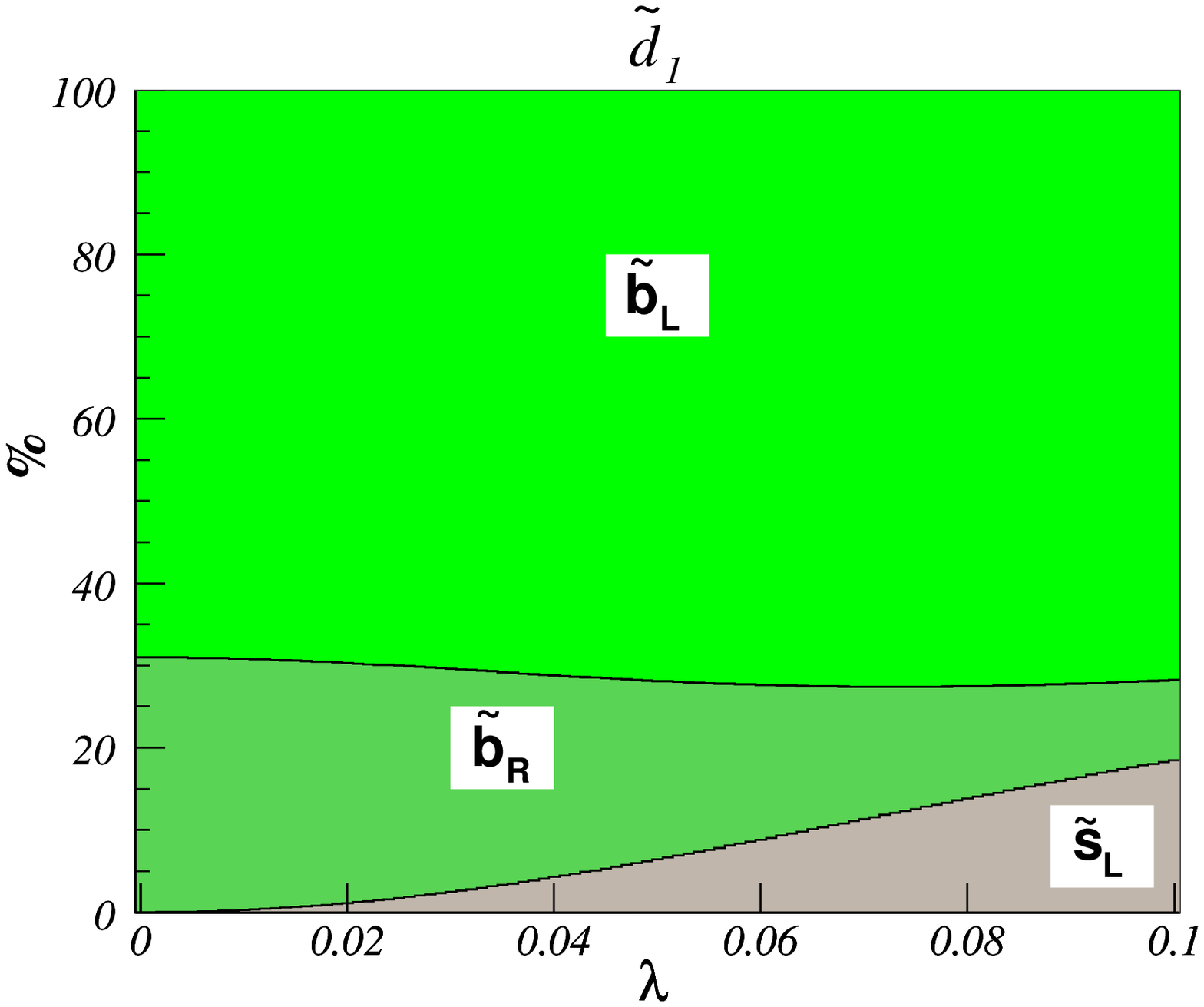}\hspace{1mm}
 \includegraphics[width=0.21\columnwidth]{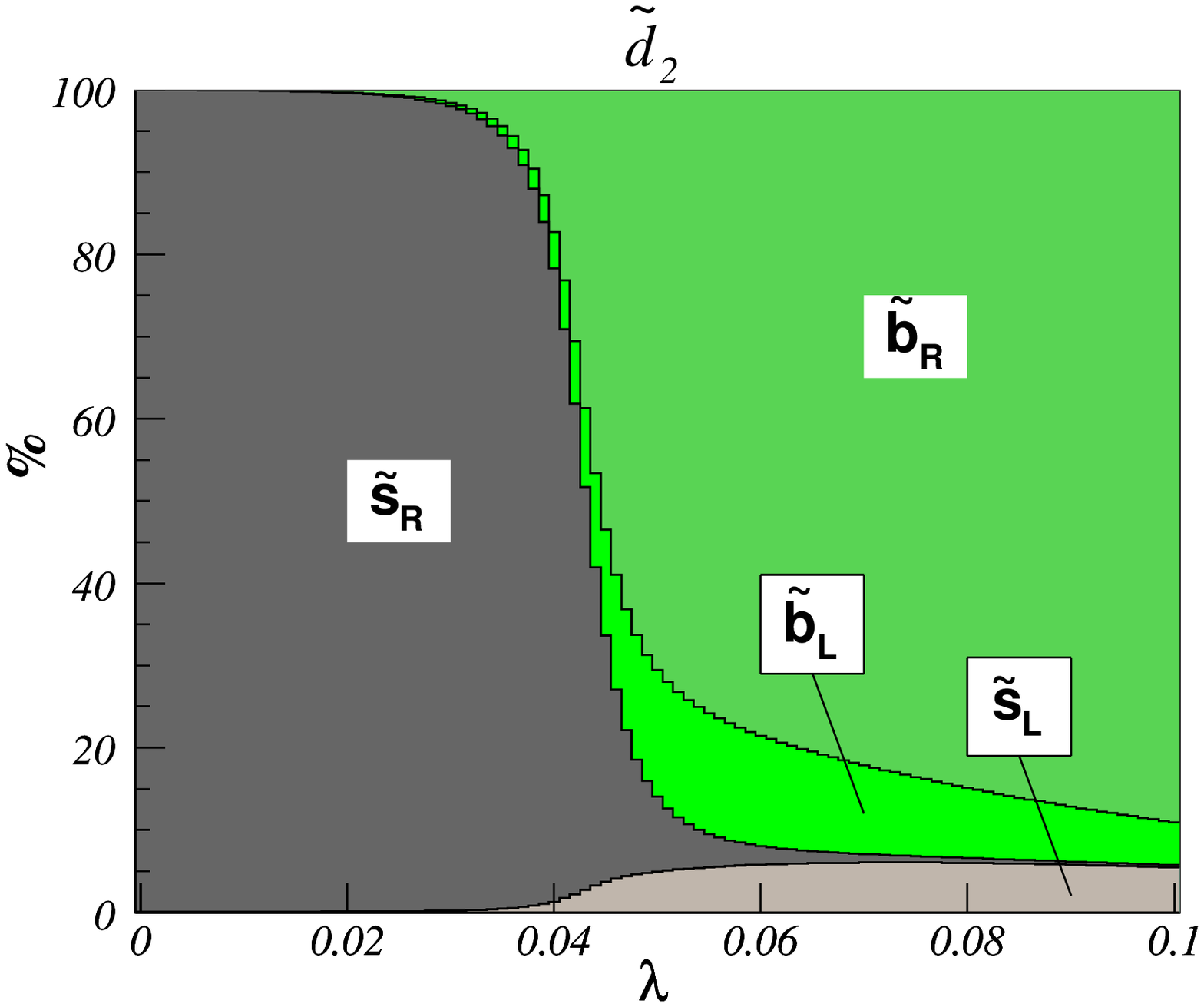}\vspace*{4mm}
 \includegraphics[width=0.21\columnwidth]{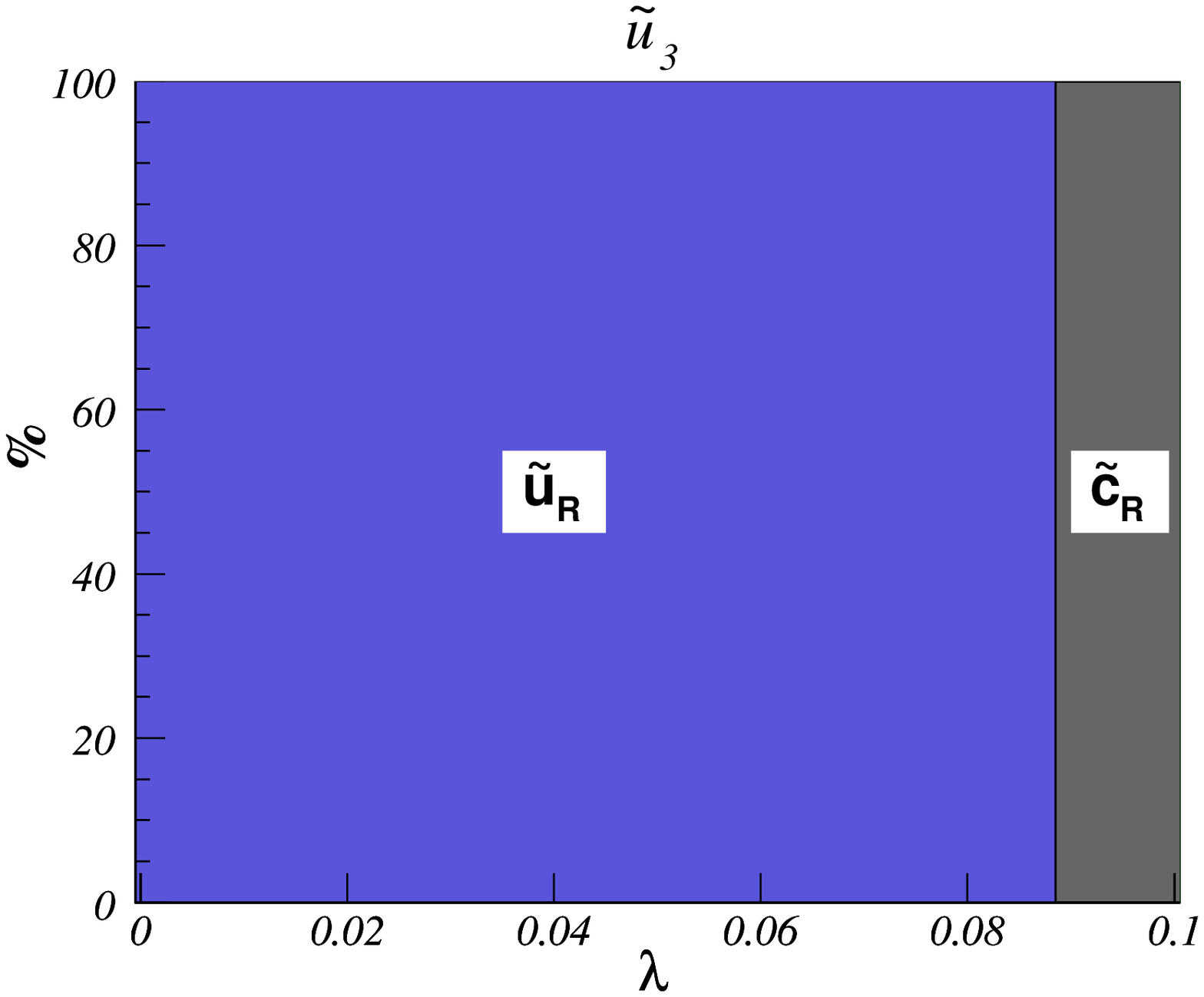}\hspace{1mm}
 \includegraphics[width=0.21\columnwidth]{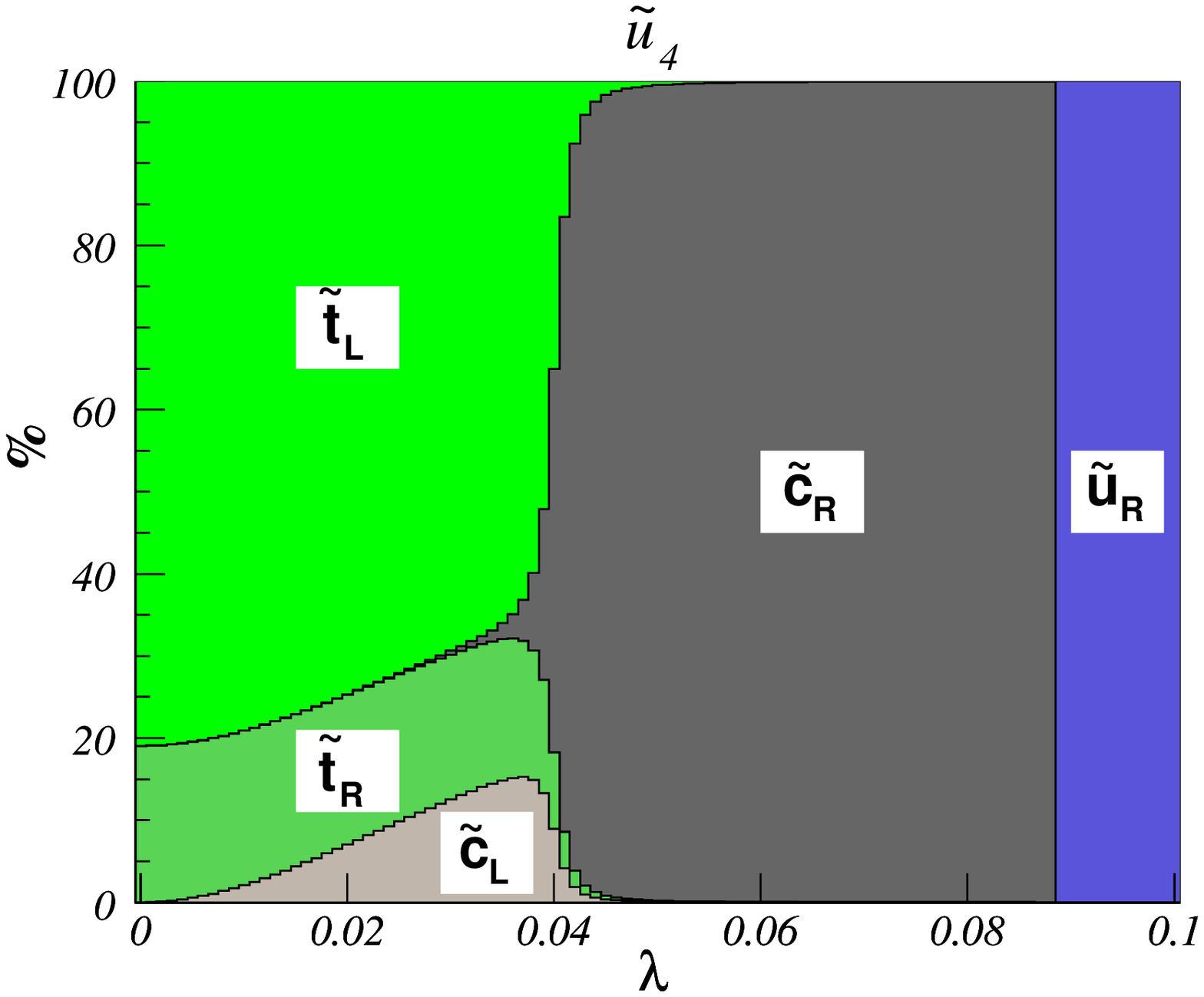}\hspace{1mm}
 \includegraphics[width=0.21\columnwidth]{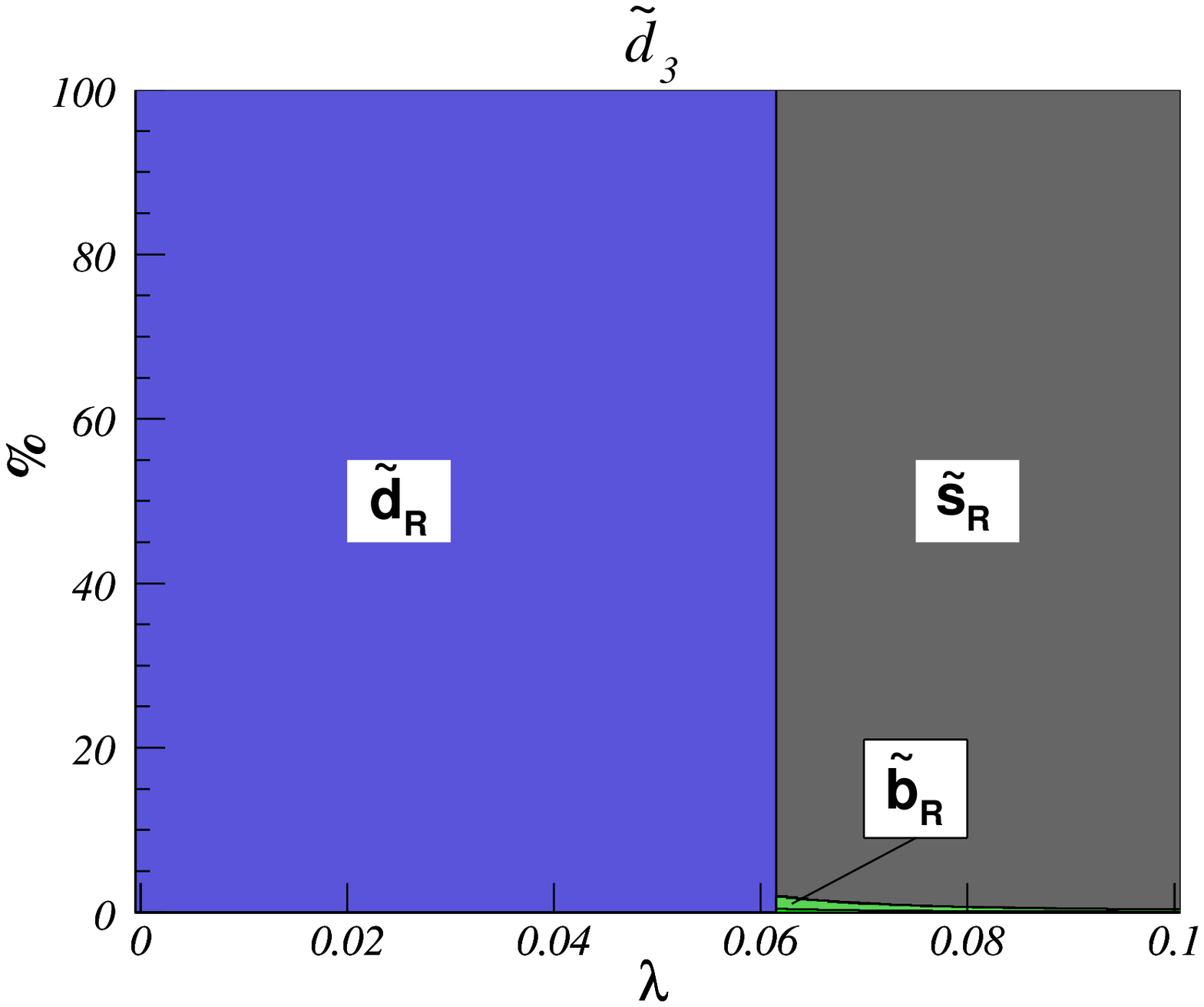}\hspace{1mm}
 \includegraphics[width=0.21\columnwidth]{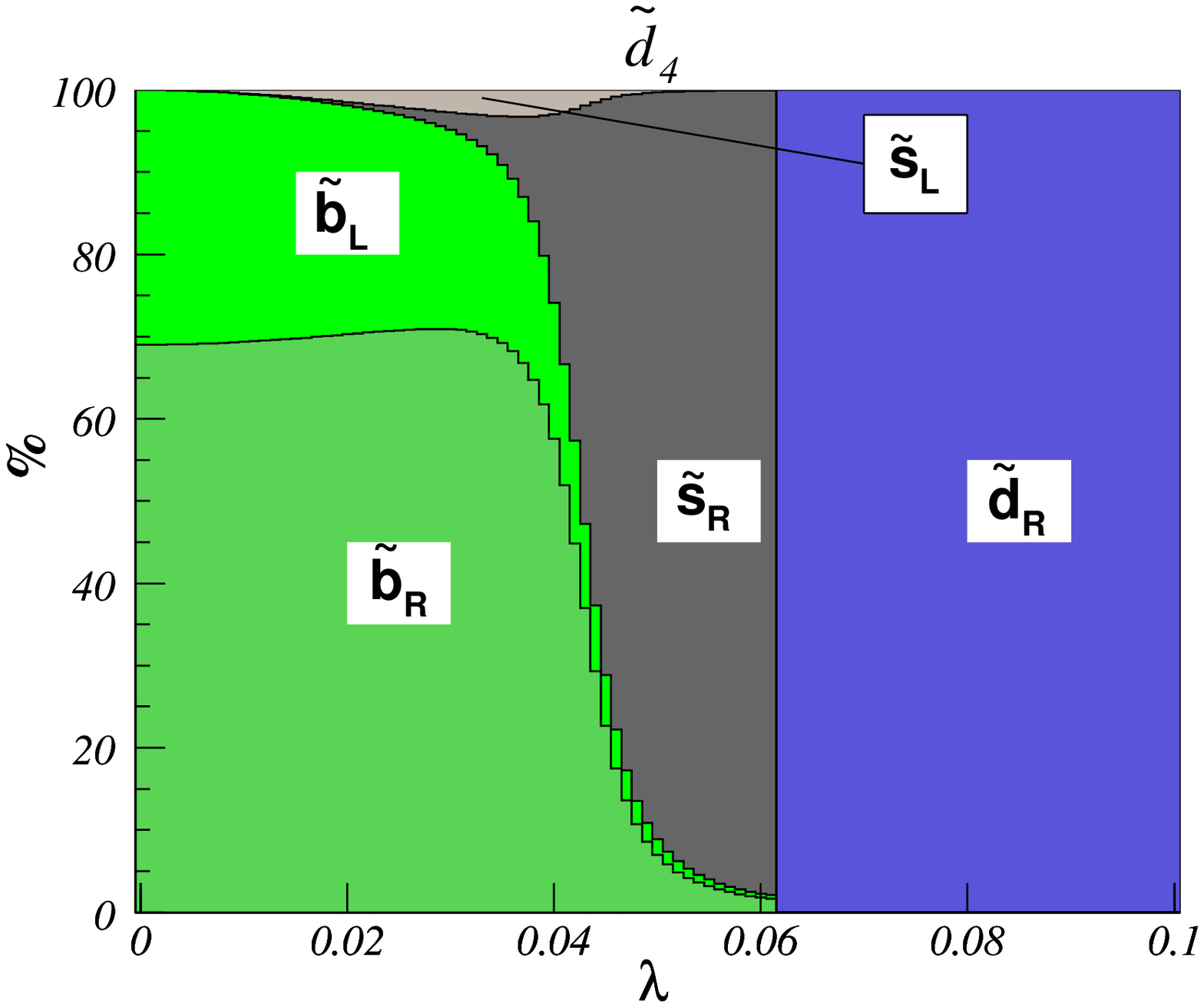}\vspace*{4mm}
 \includegraphics[width=0.21\columnwidth]{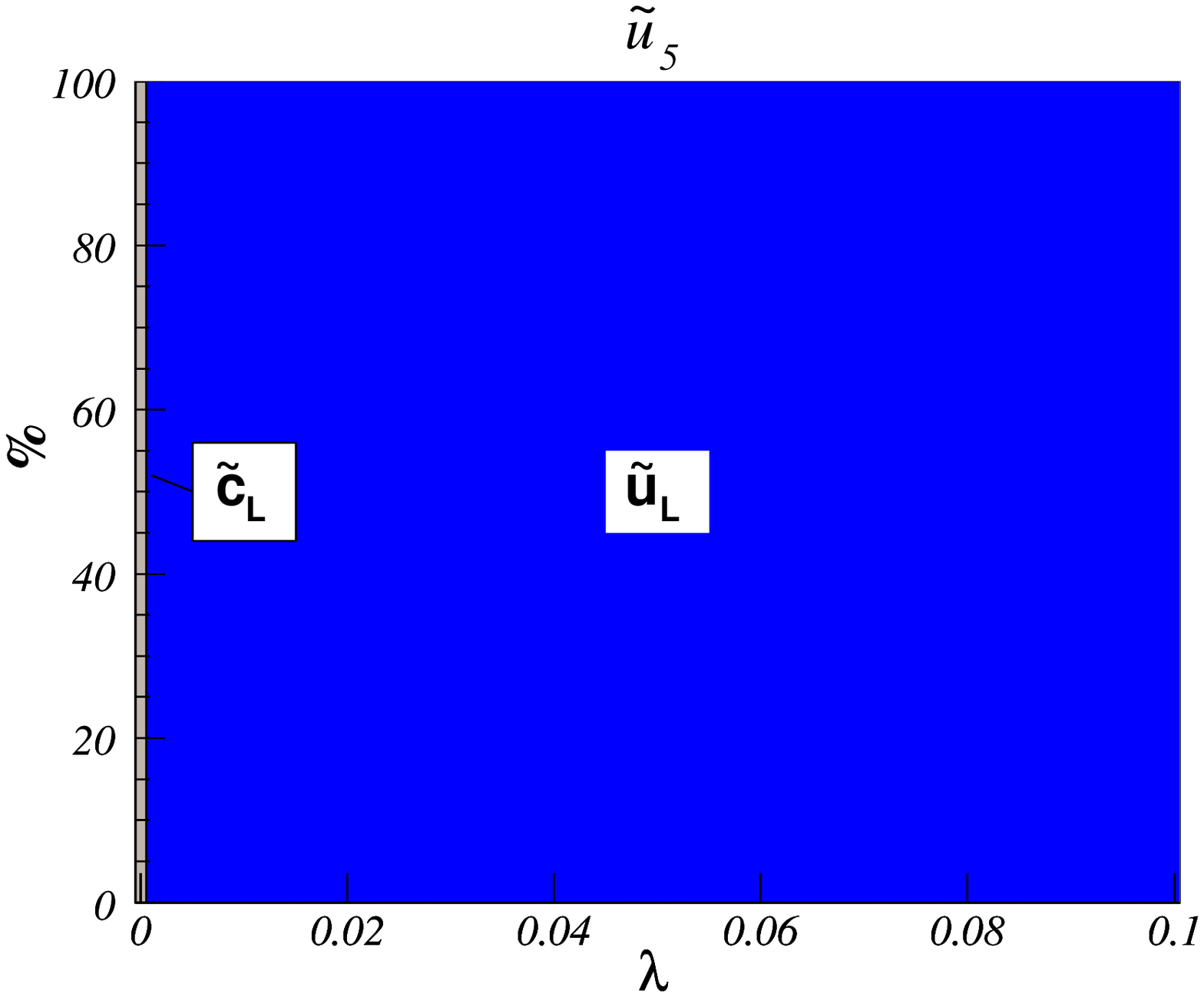}\hspace{1mm}
 \includegraphics[width=0.21\columnwidth]{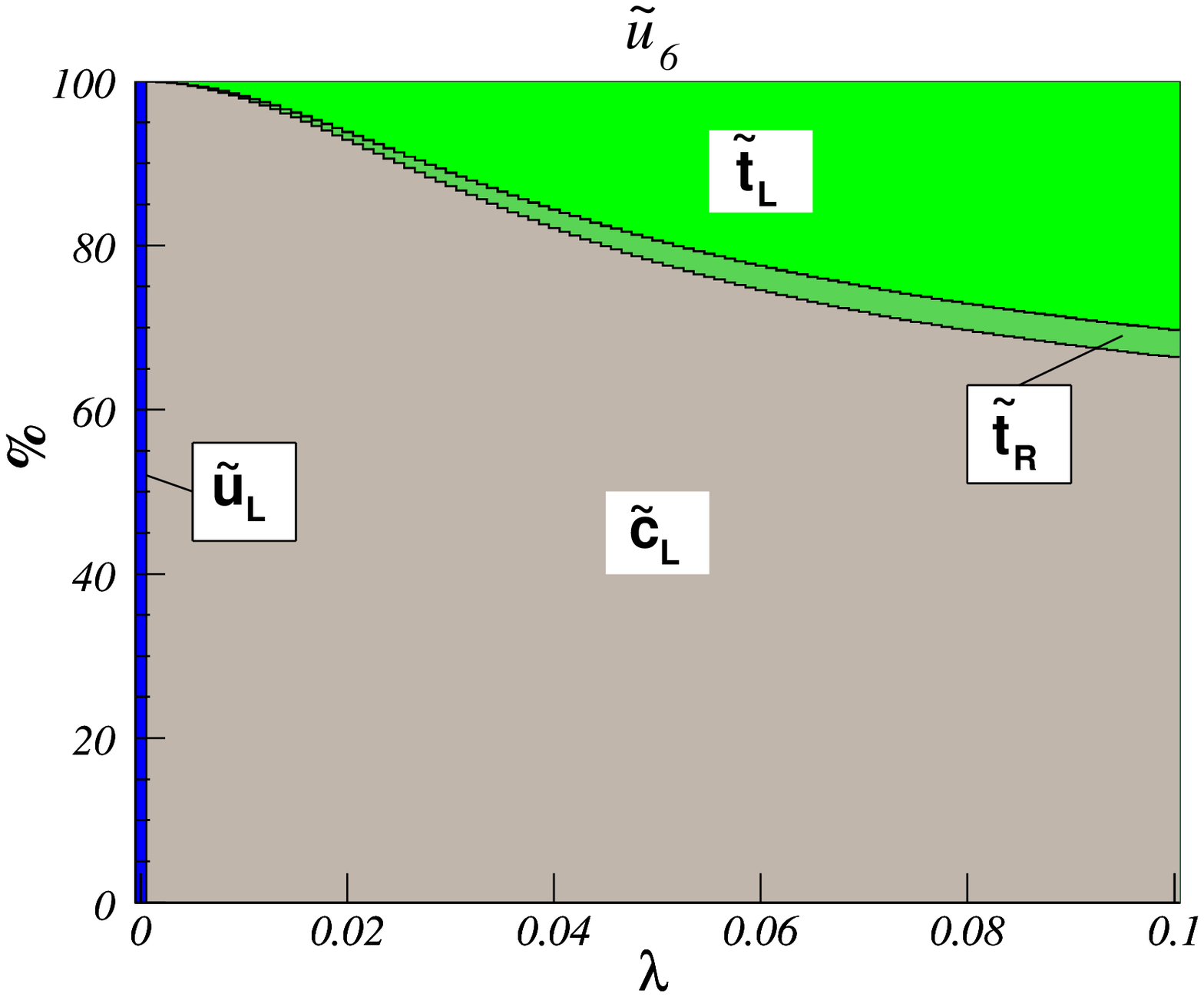}\hspace{1mm}
 \includegraphics[width=0.21\columnwidth]{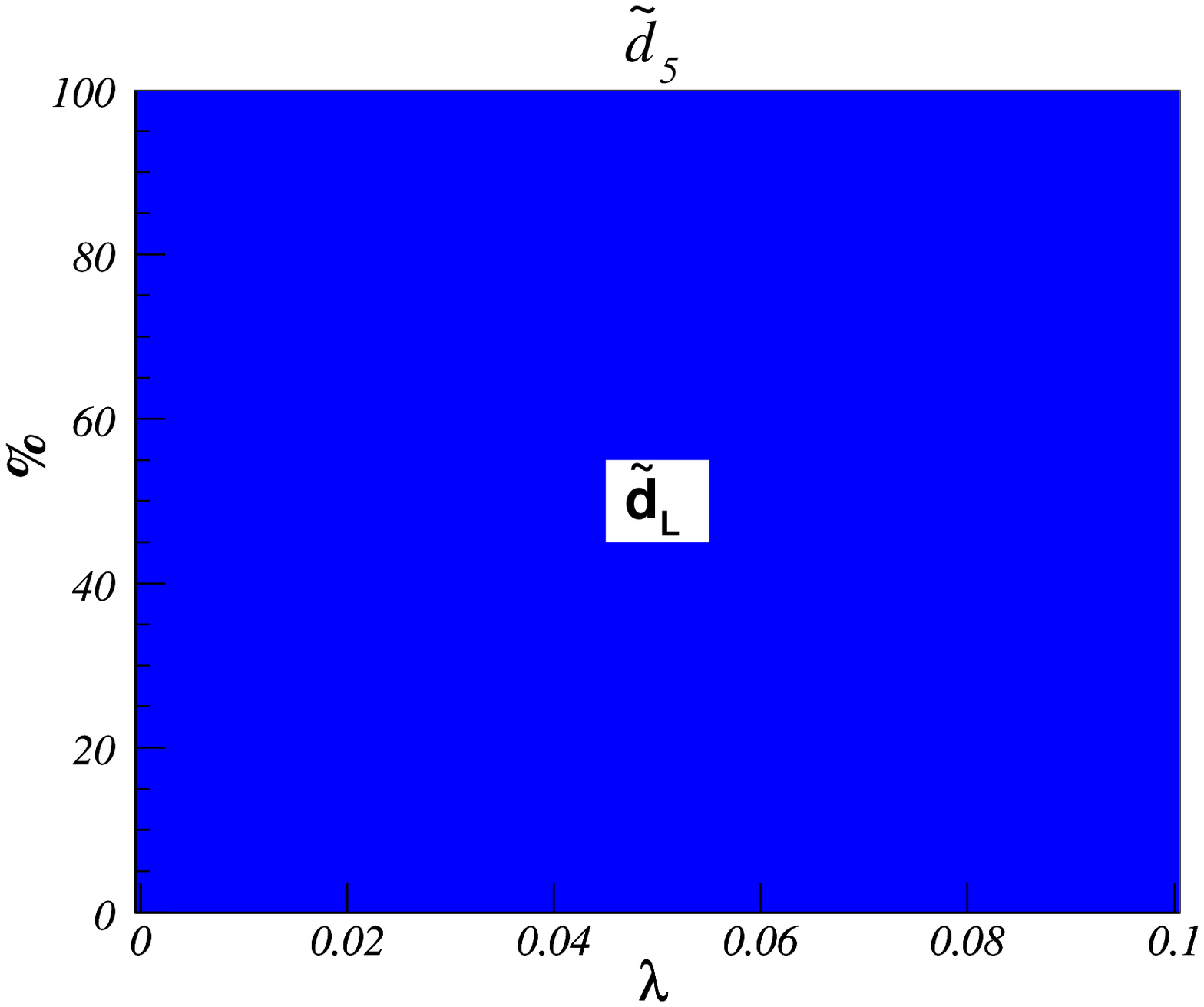}\hspace{1mm}
 \includegraphics[width=0.21\columnwidth]{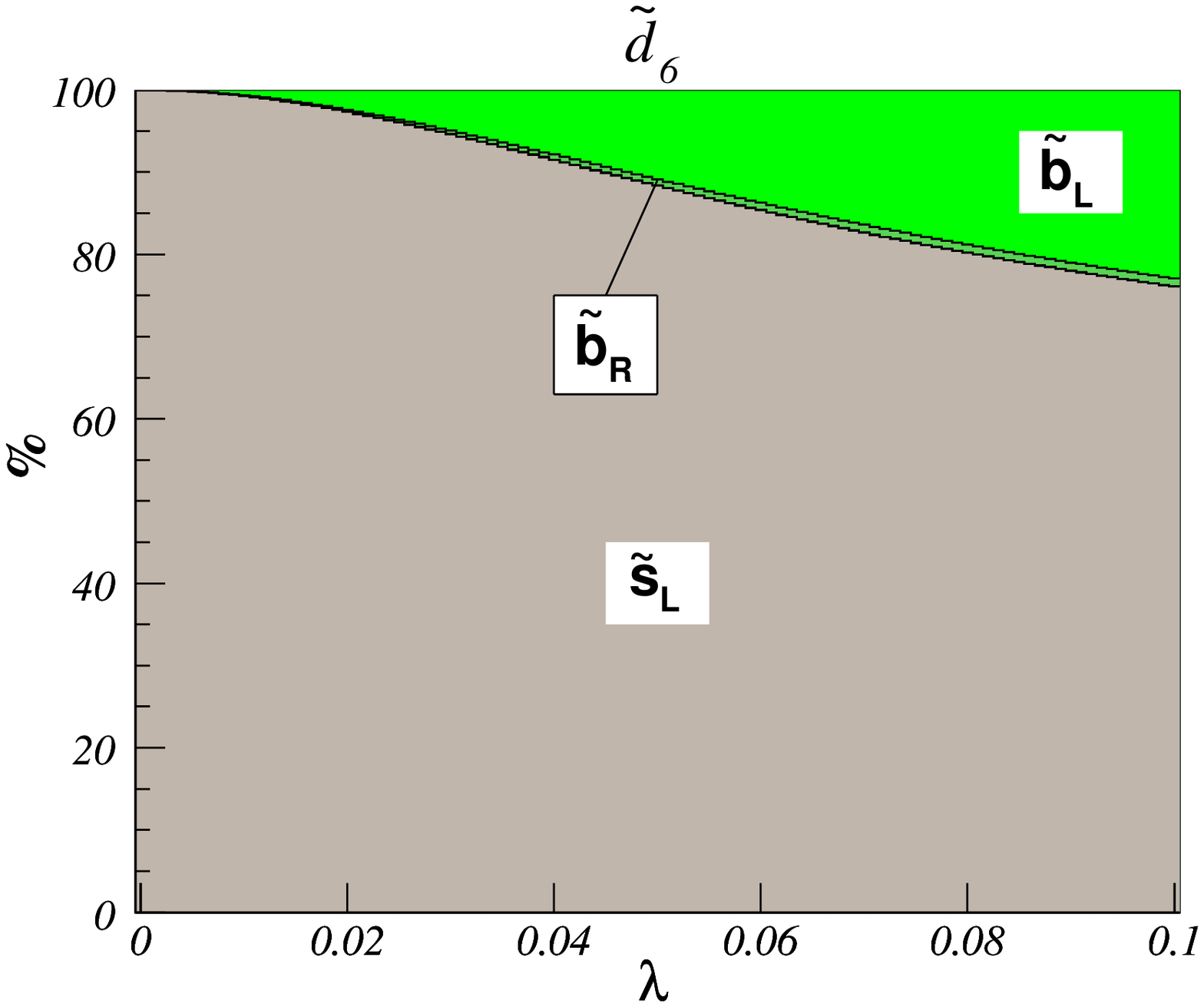}
 \caption{\label{fig:011p}Same as Fig.\ \ref{fig:011} for $\lambda\in
          [0;0.1]$.}
\end{figure}

\begin{figure}
 \centering
 \includegraphics[width=0.21\columnwidth]{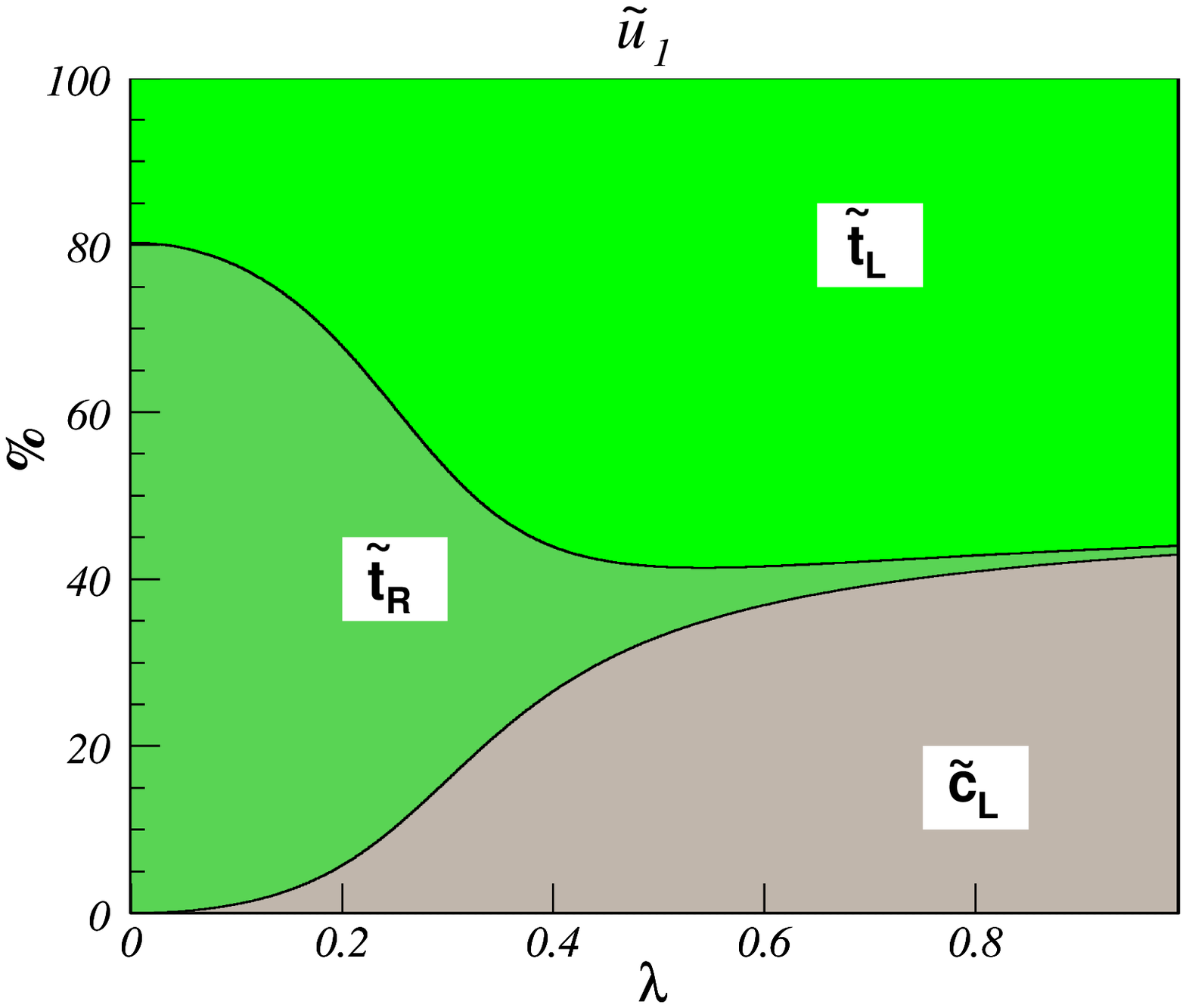}\hspace{1mm}
 \includegraphics[width=0.21\columnwidth]{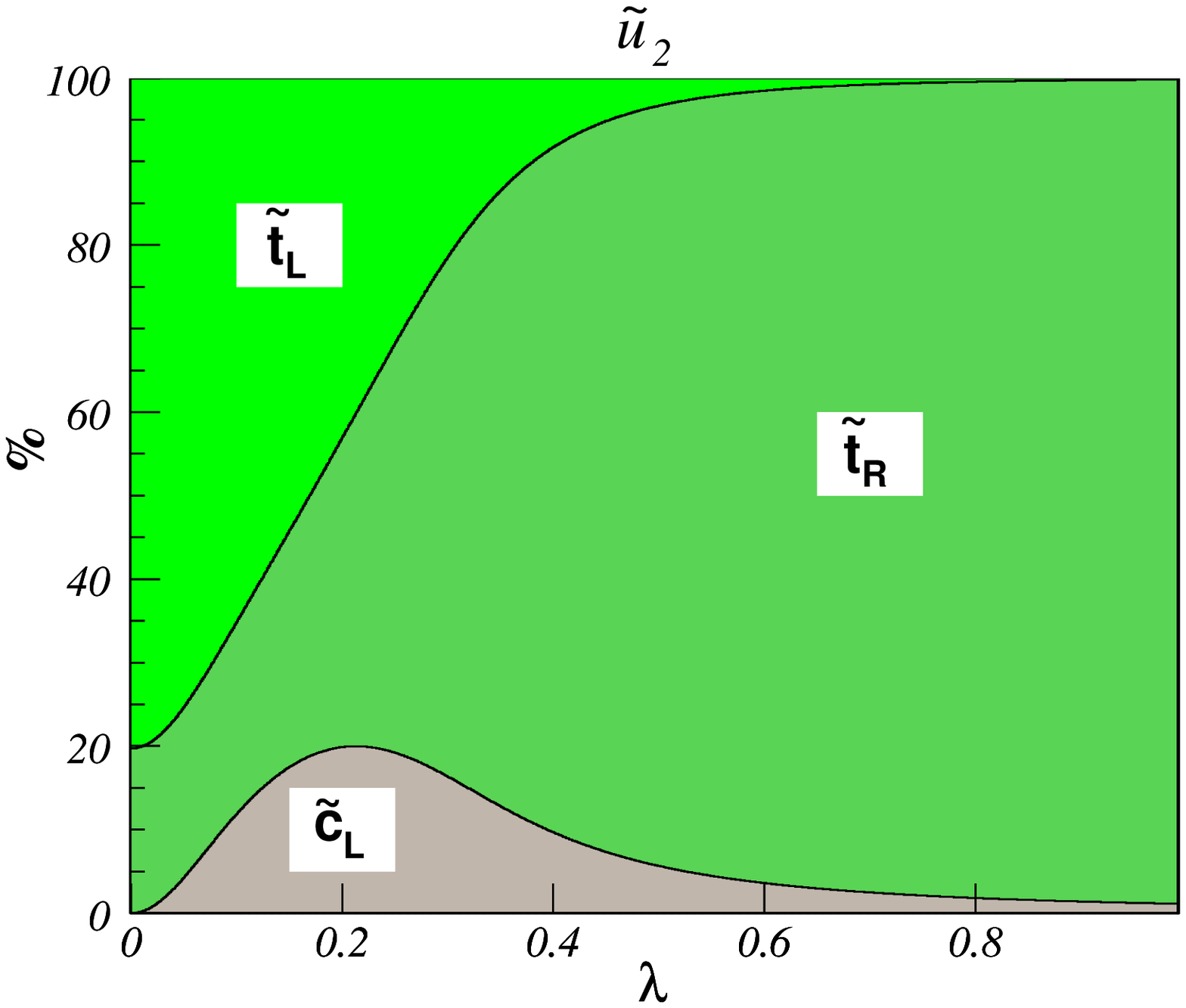}\hspace{1mm}
 \includegraphics[width=0.21\columnwidth]{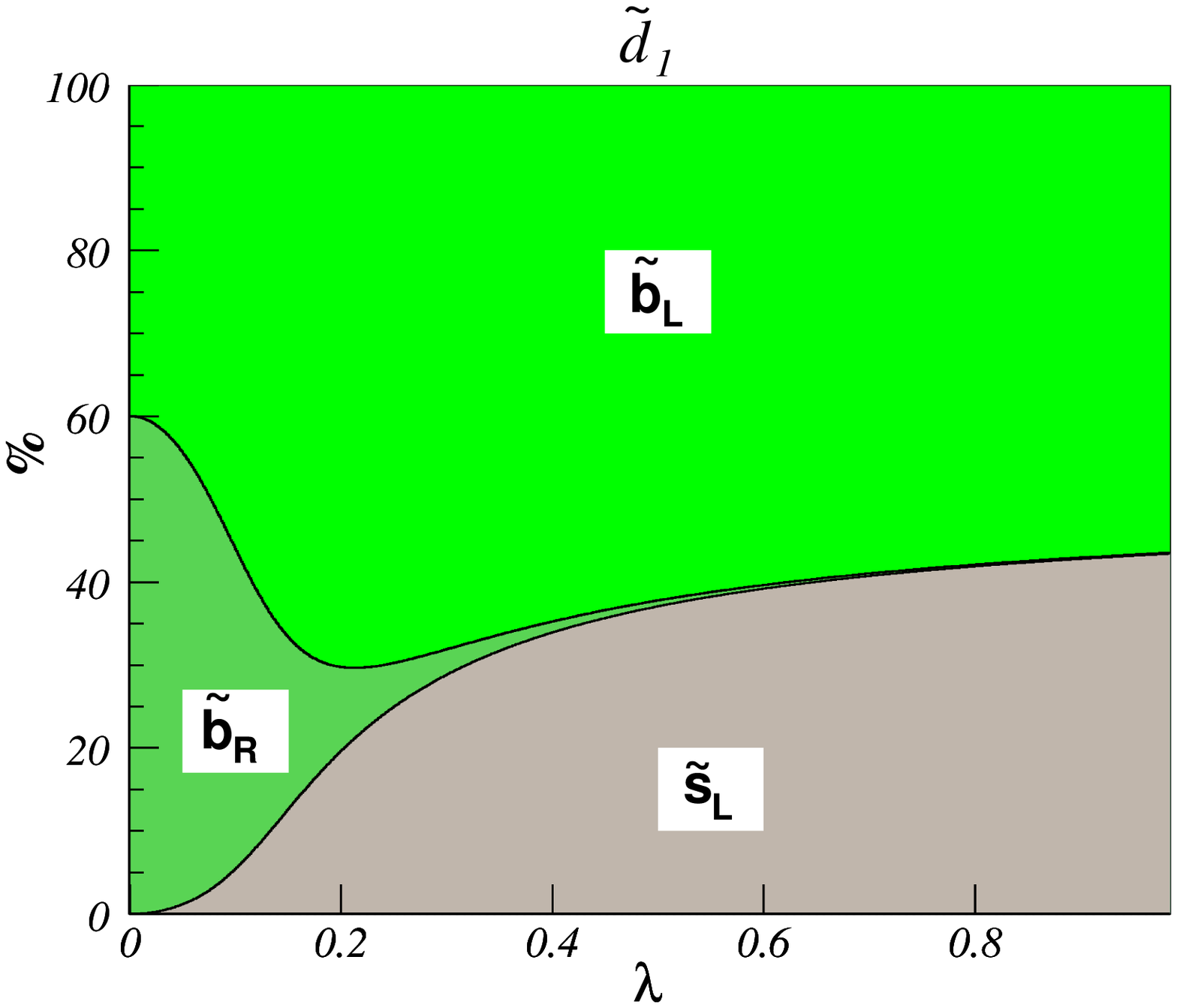}\hspace{1mm}
 \includegraphics[width=0.21\columnwidth]{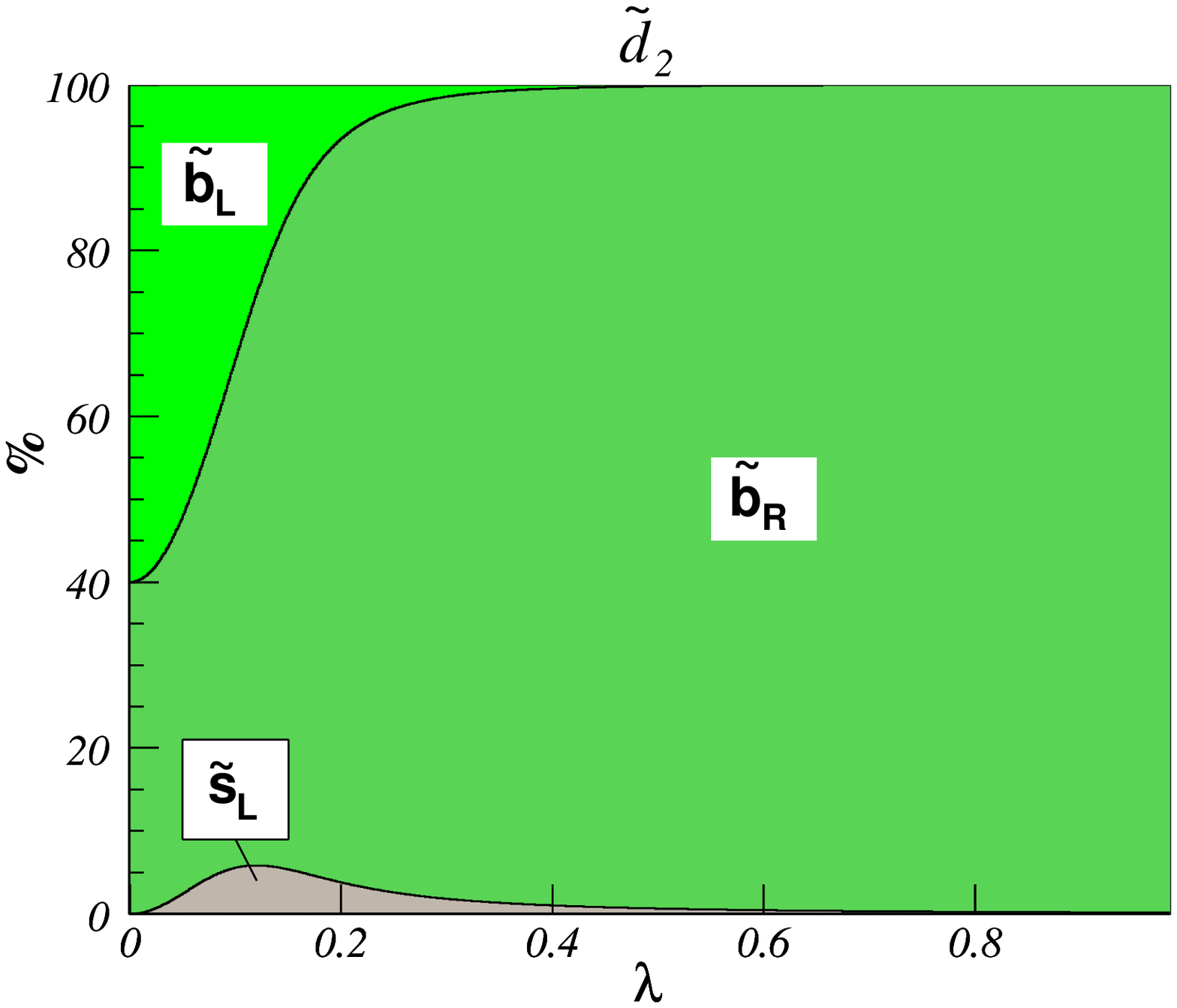}\vspace*{4mm}
 \includegraphics[width=0.21\columnwidth]{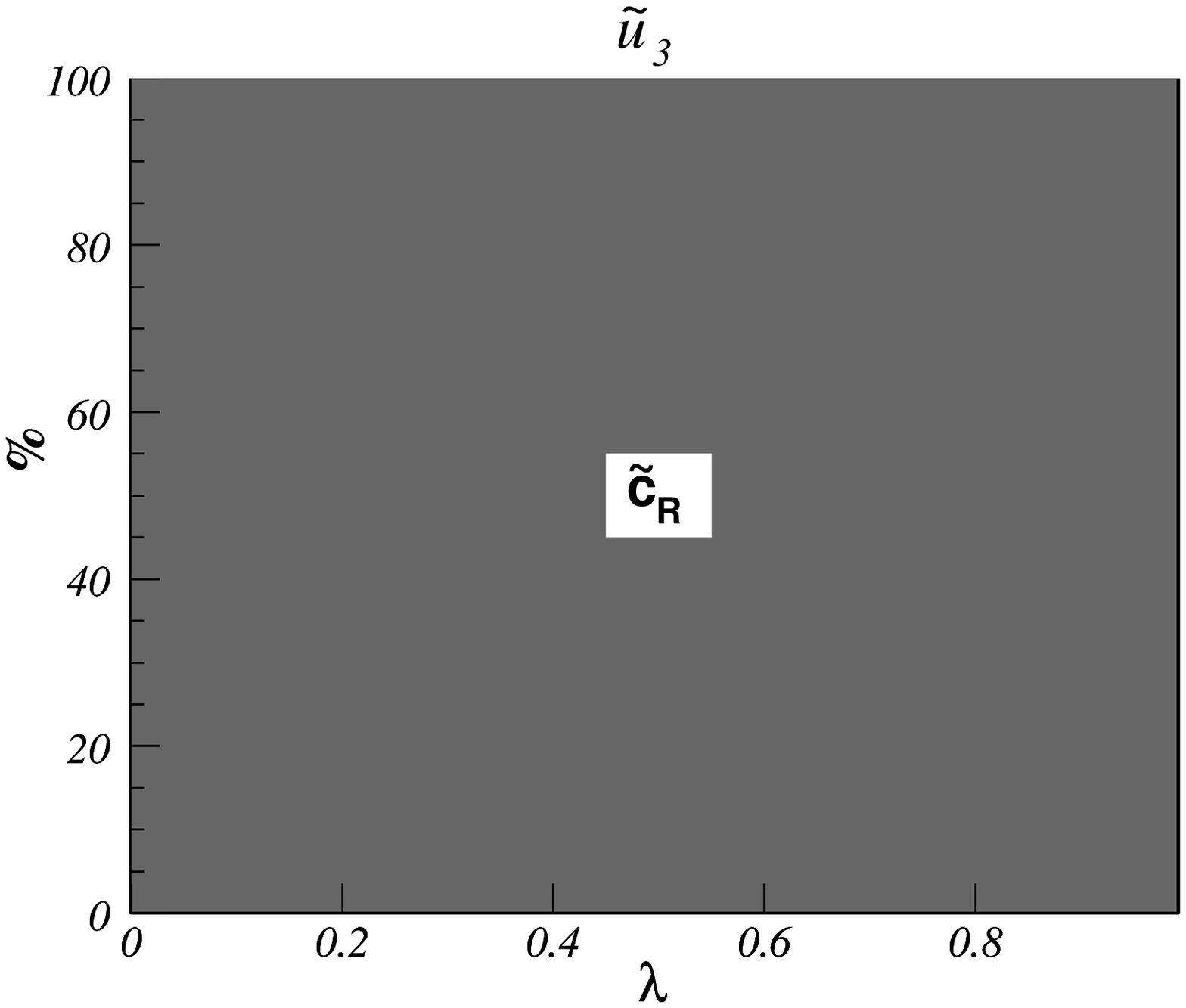}\hspace{1mm}
 \includegraphics[width=0.21\columnwidth]{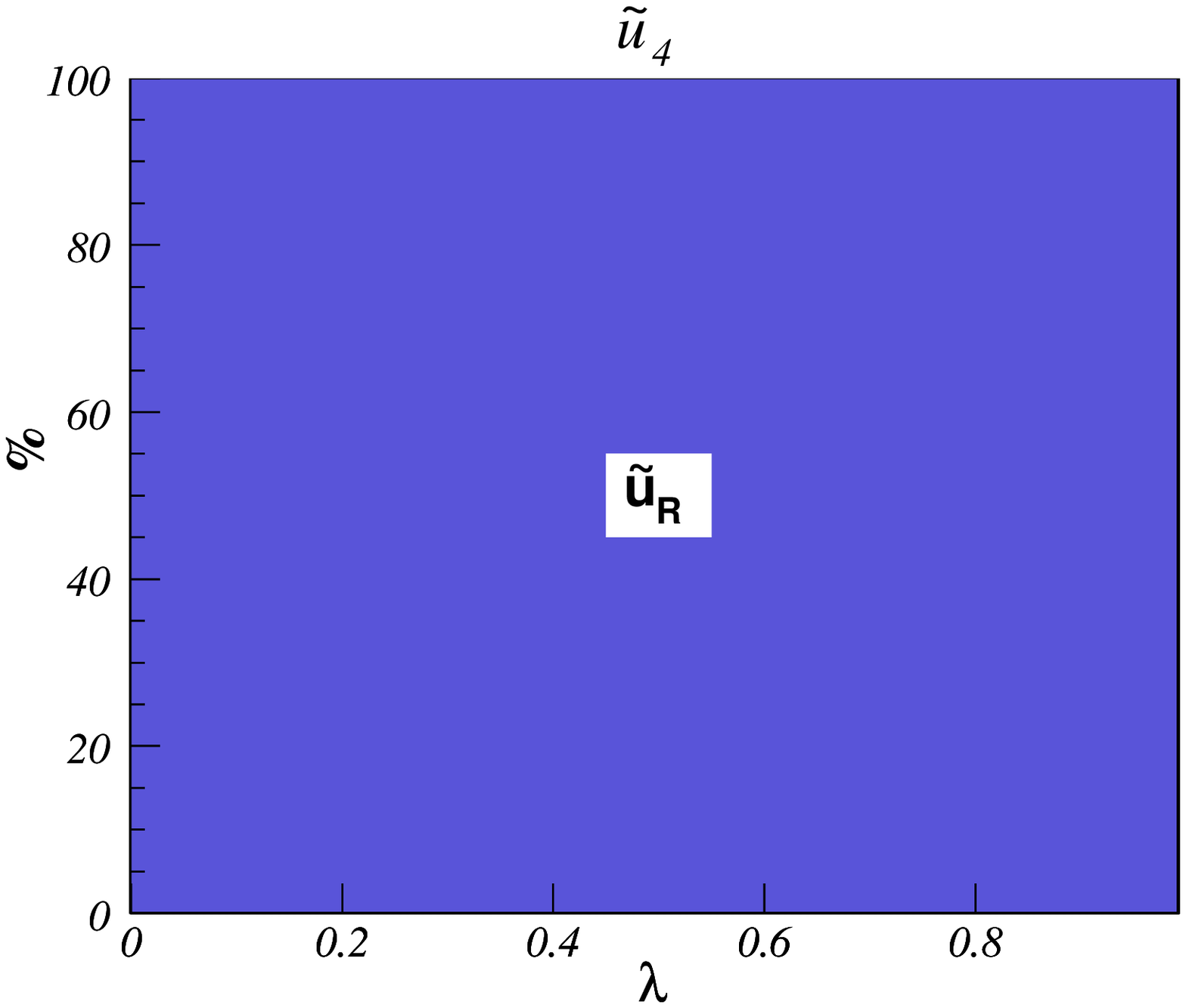}\hspace{1mm}
 \includegraphics[width=0.21\columnwidth]{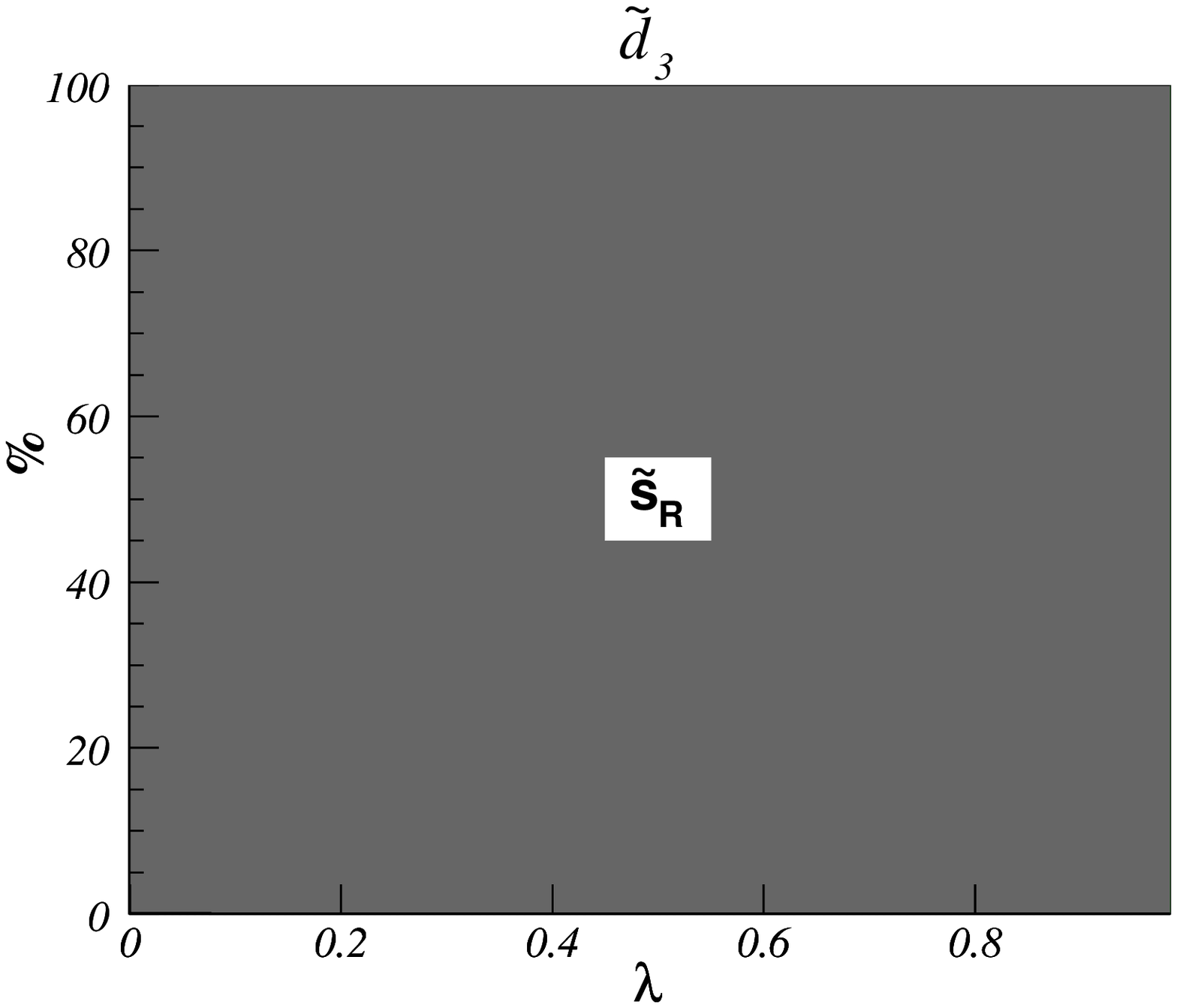}\hspace{1mm}
 \includegraphics[width=0.21\columnwidth]{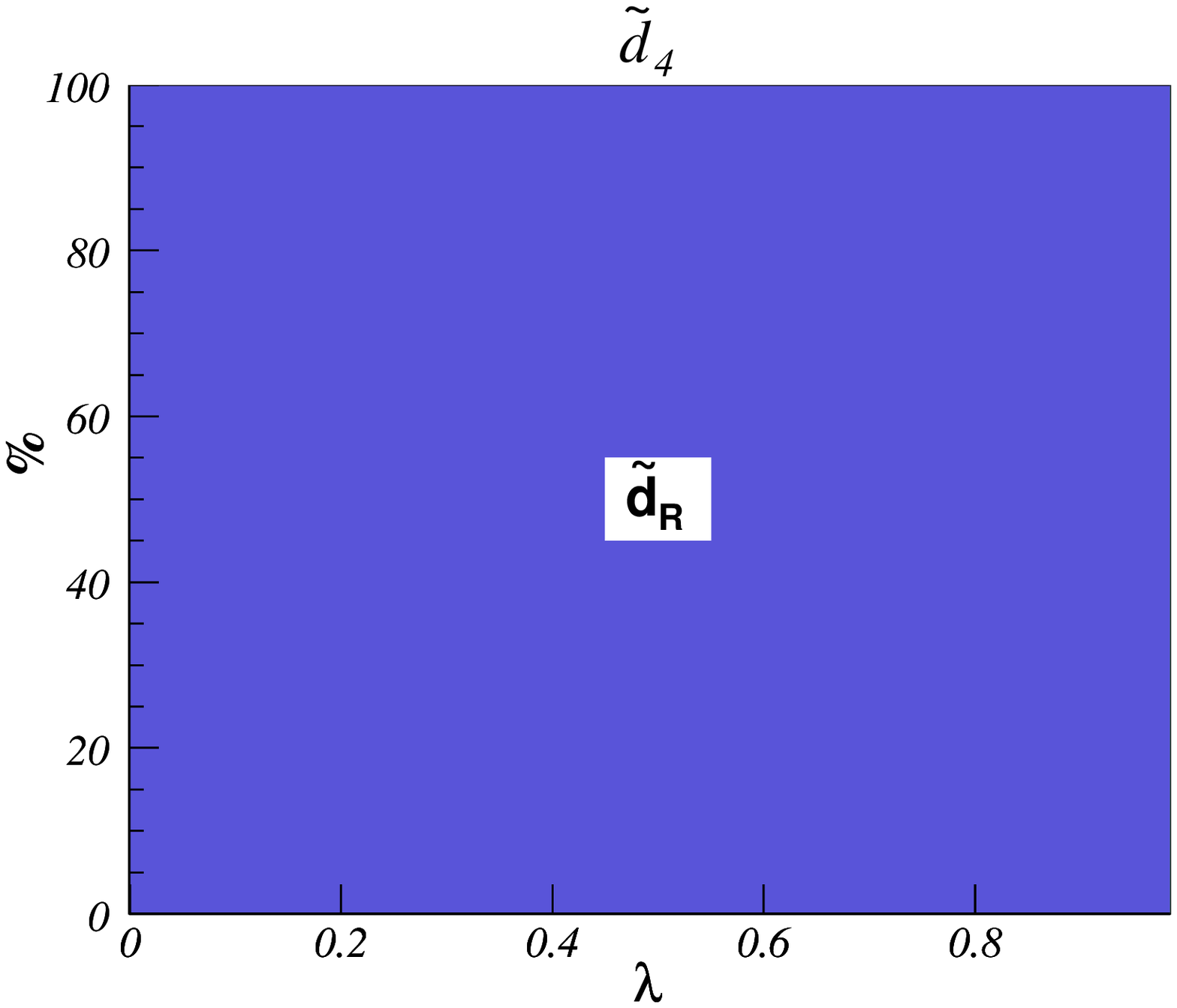}\vspace*{4mm}
 \includegraphics[width=0.21\columnwidth]{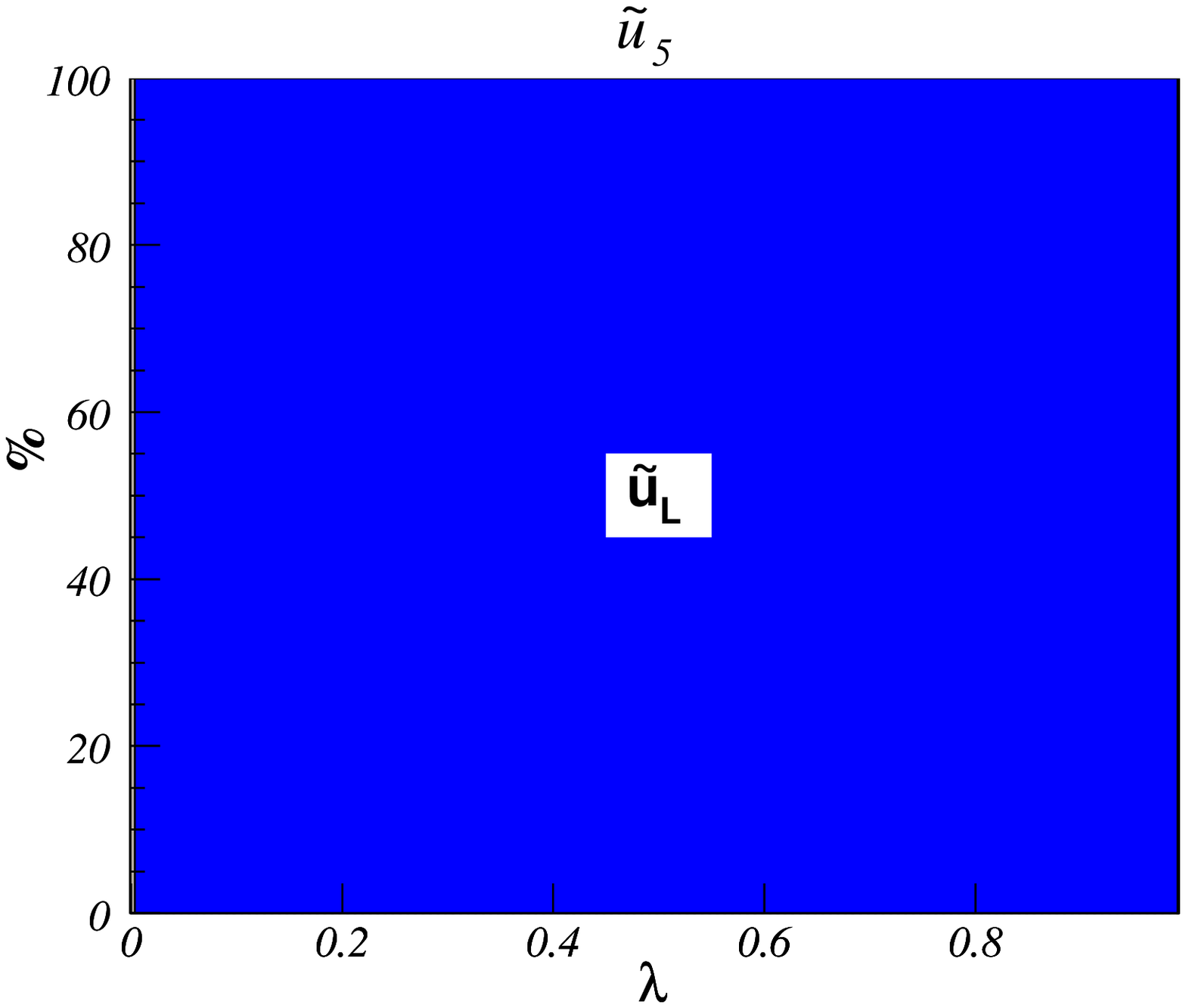}\hspace{1mm}
 \includegraphics[width=0.21\columnwidth]{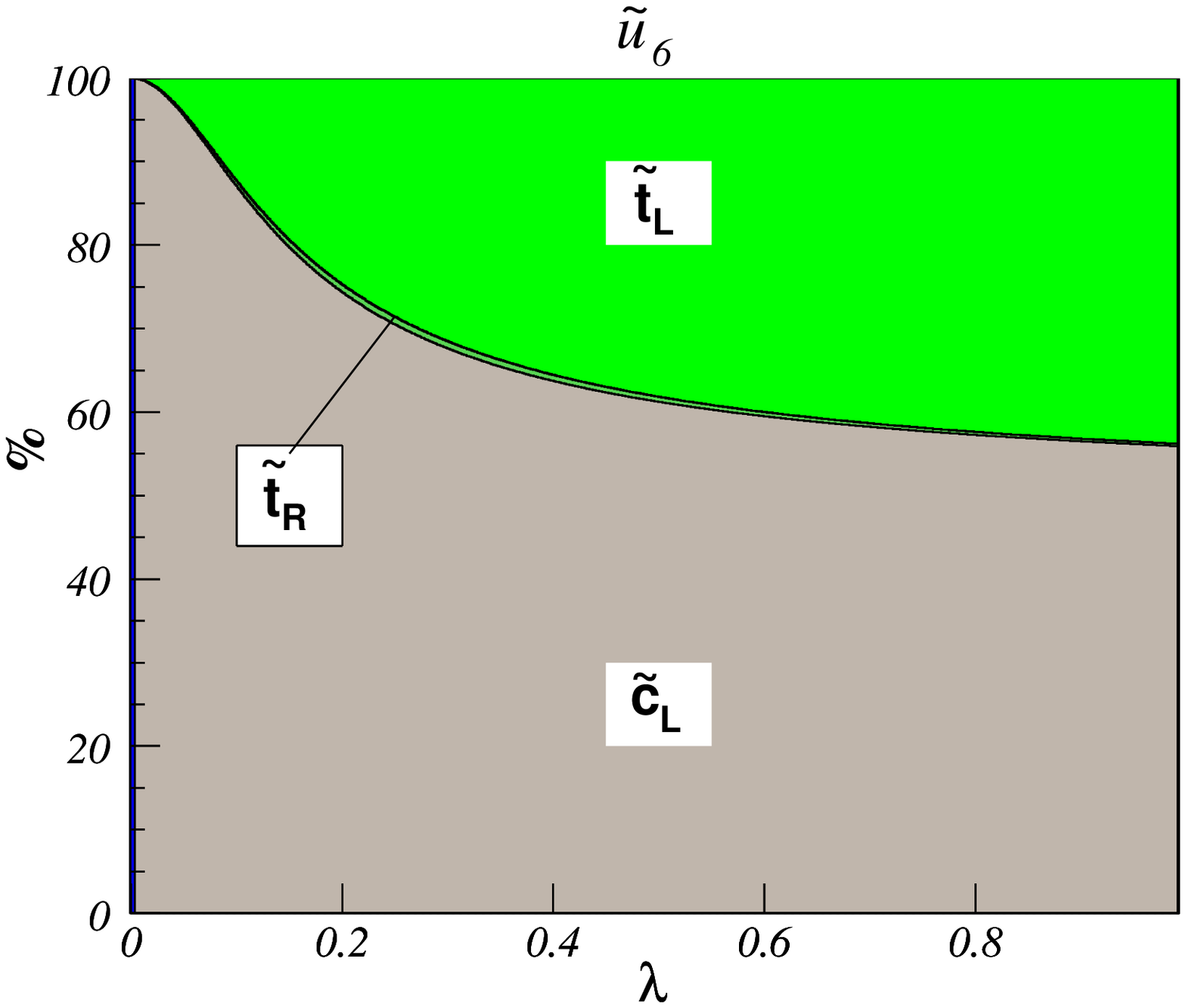}\hspace{1mm}
 \includegraphics[width=0.21\columnwidth]{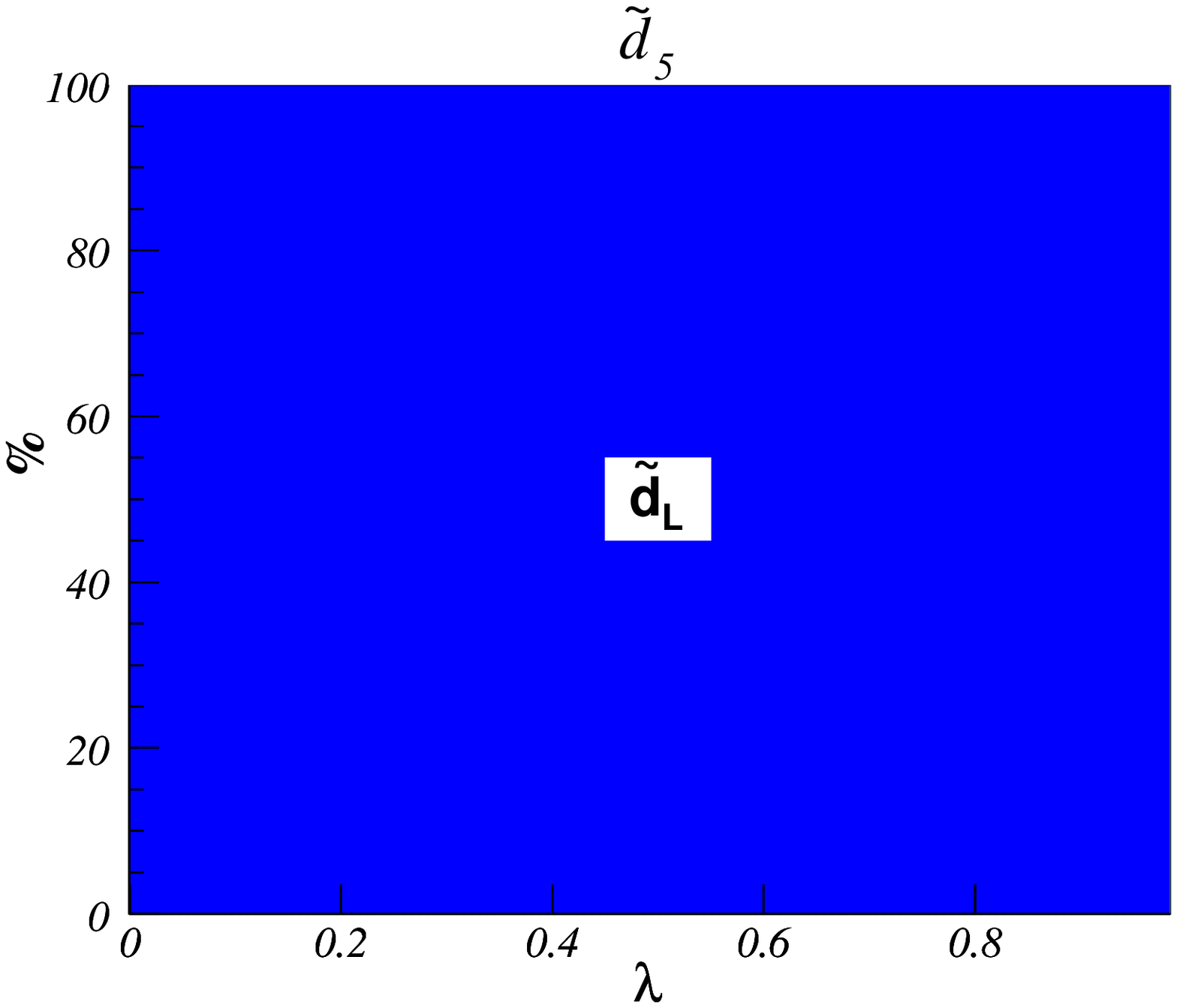}\hspace{1mm}
 \includegraphics[width=0.21\columnwidth]{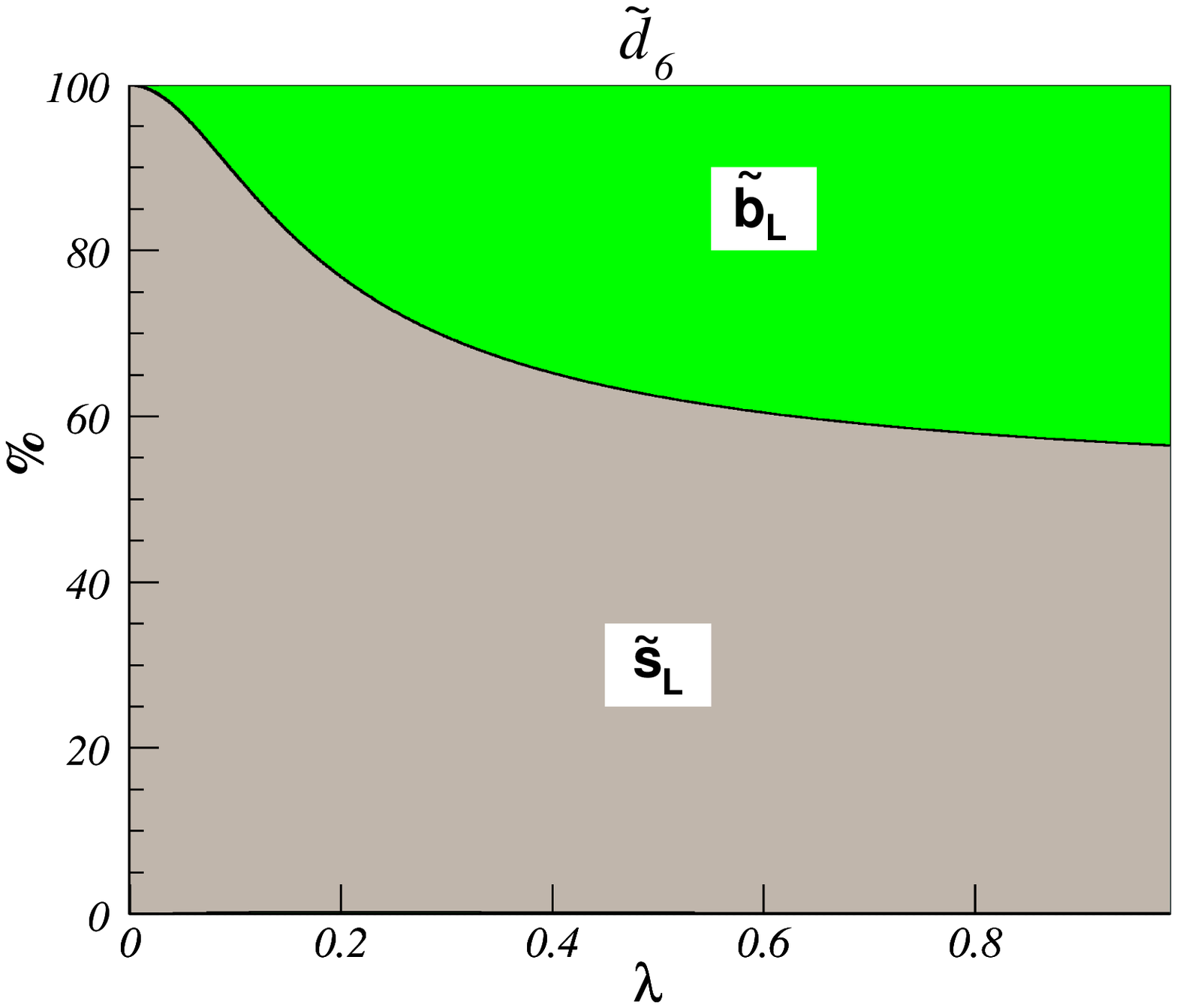}
 \caption{\label{fig:012}Same as Fig.\ \ref{fig:09} for benchmark point D.}\vspace{4mm}
 \includegraphics[width=0.21\columnwidth]{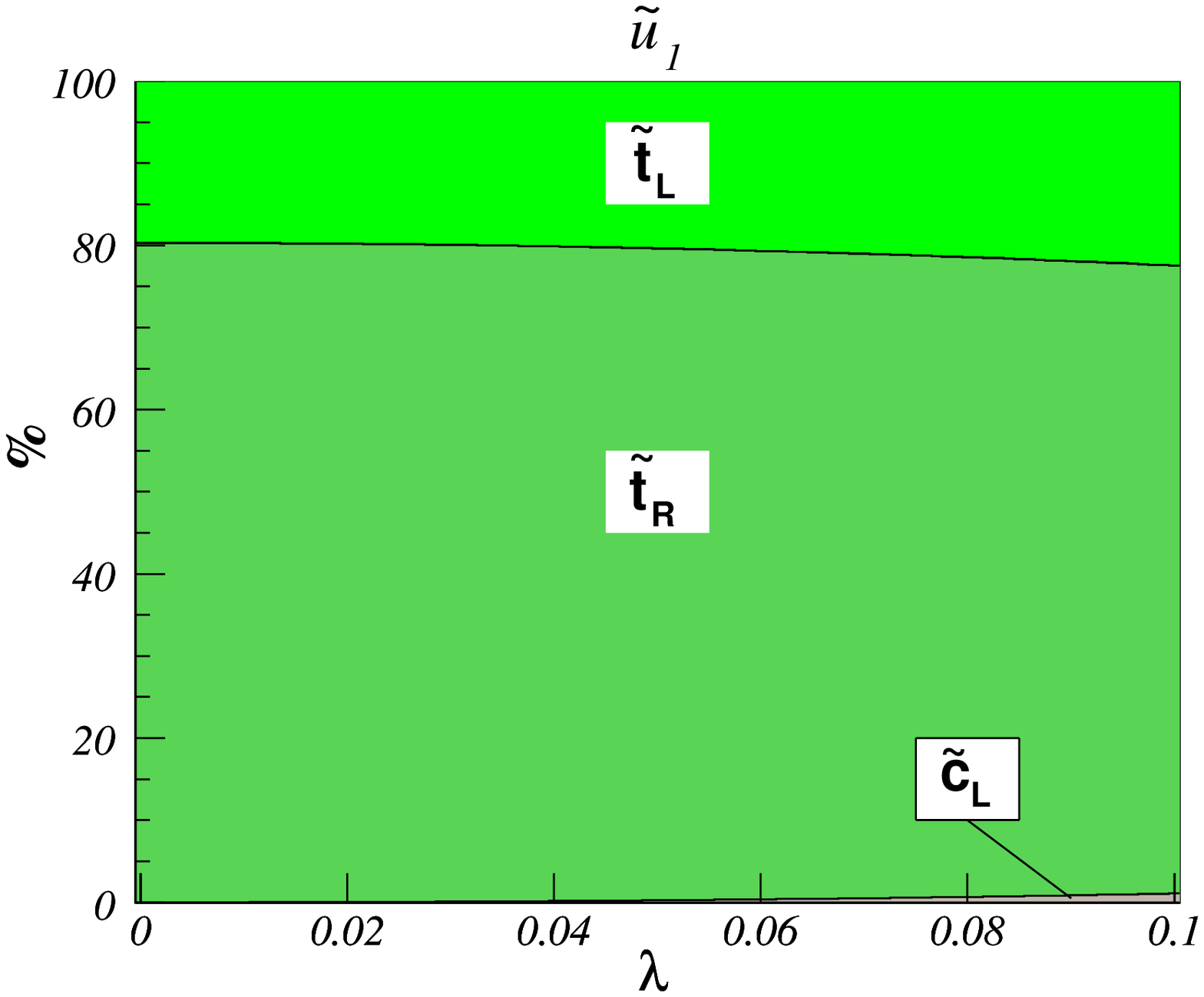}\hspace{1mm}
 \includegraphics[width=0.21\columnwidth]{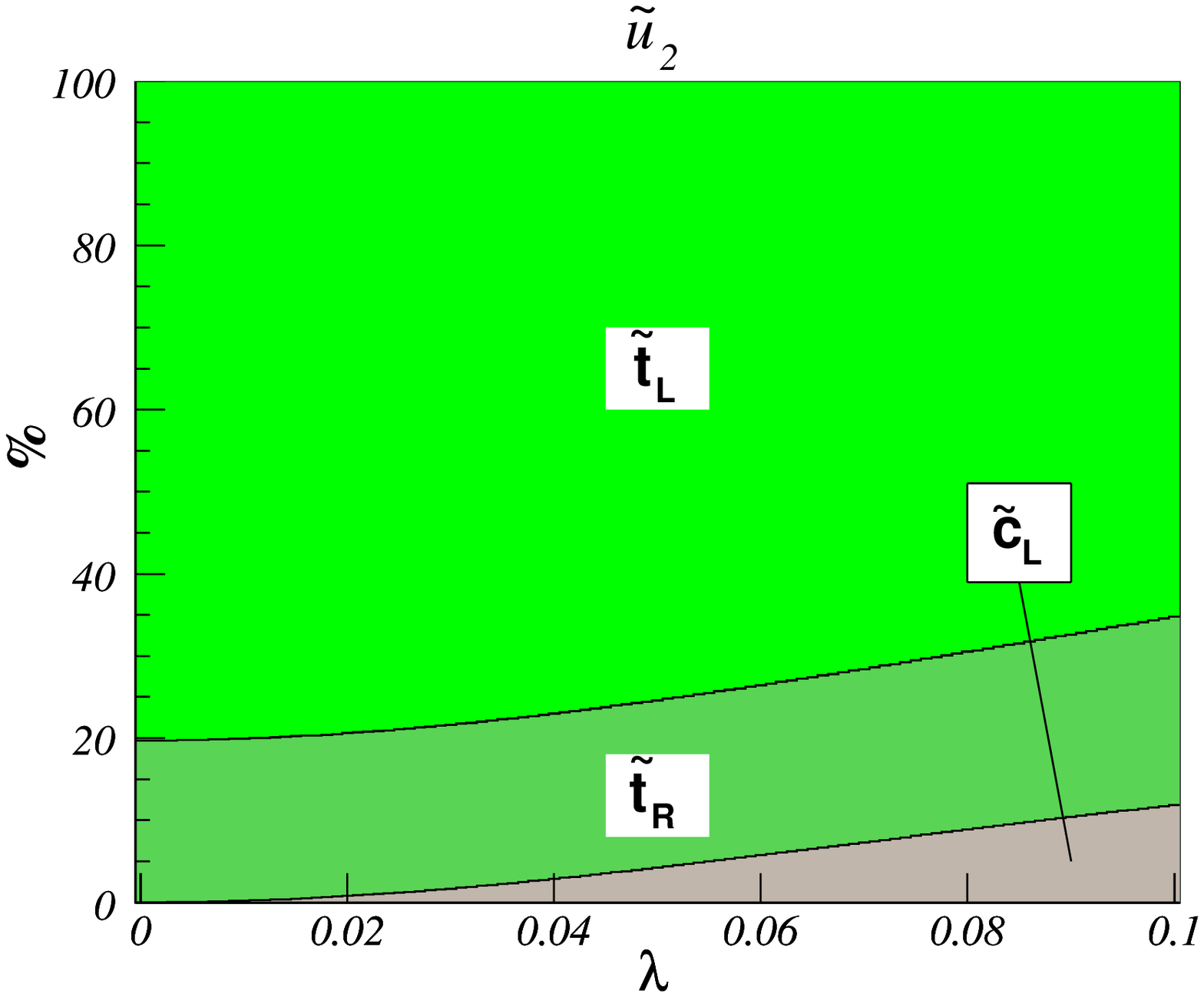}\hspace{1mm}
 \includegraphics[width=0.21\columnwidth]{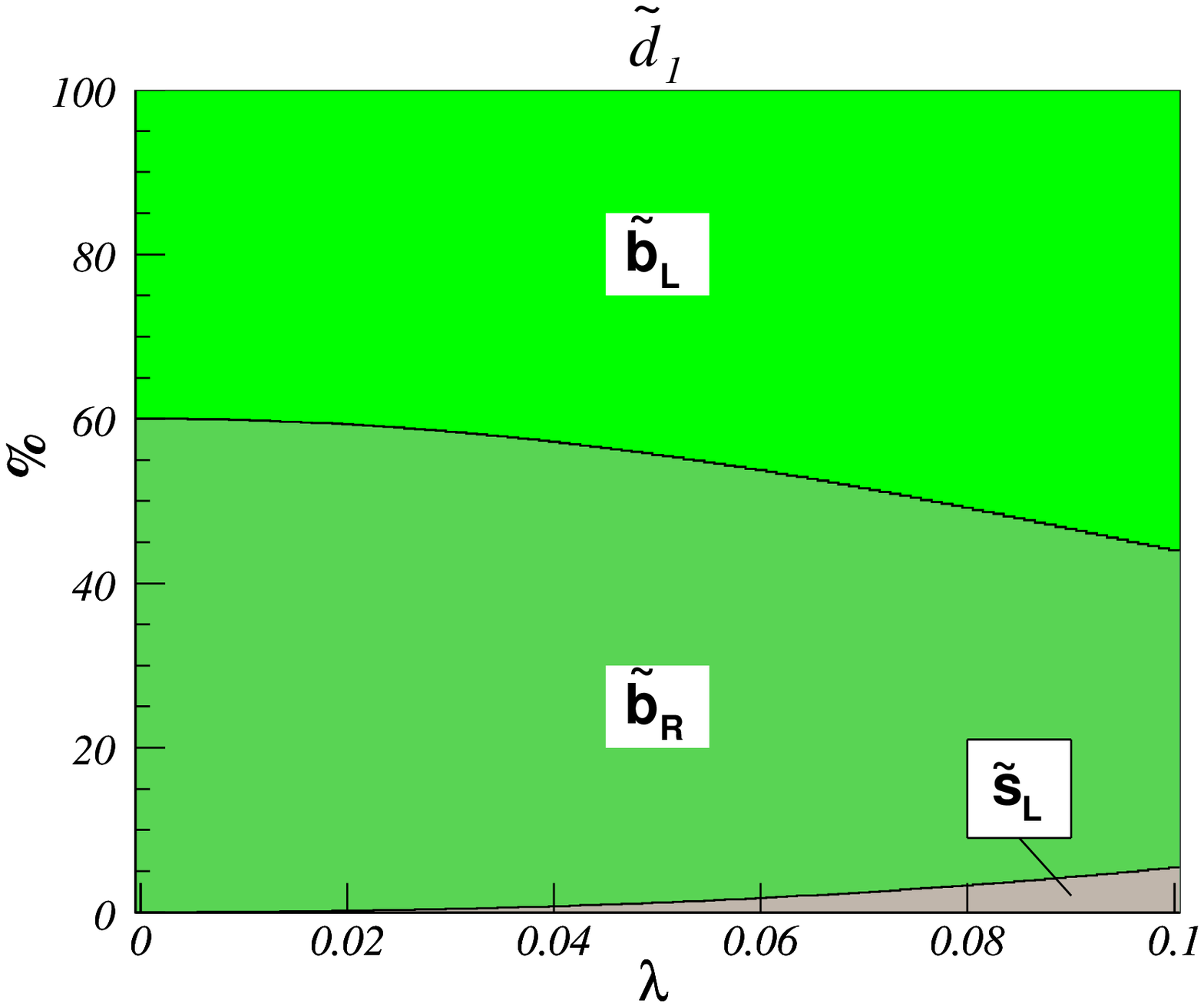}\hspace{1mm}
 \includegraphics[width=0.21\columnwidth]{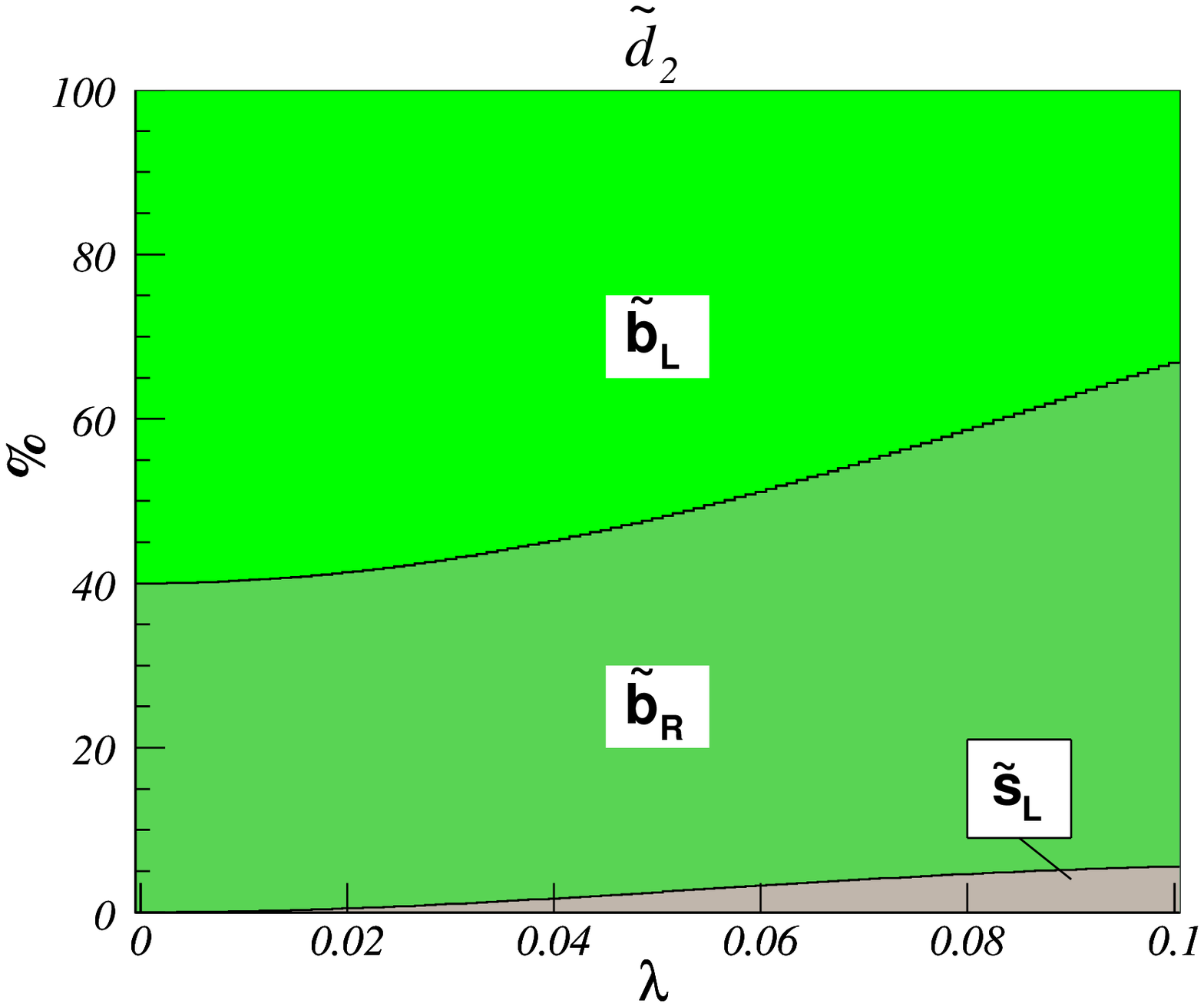}\vspace*{4mm}
 \includegraphics[width=0.21\columnwidth]{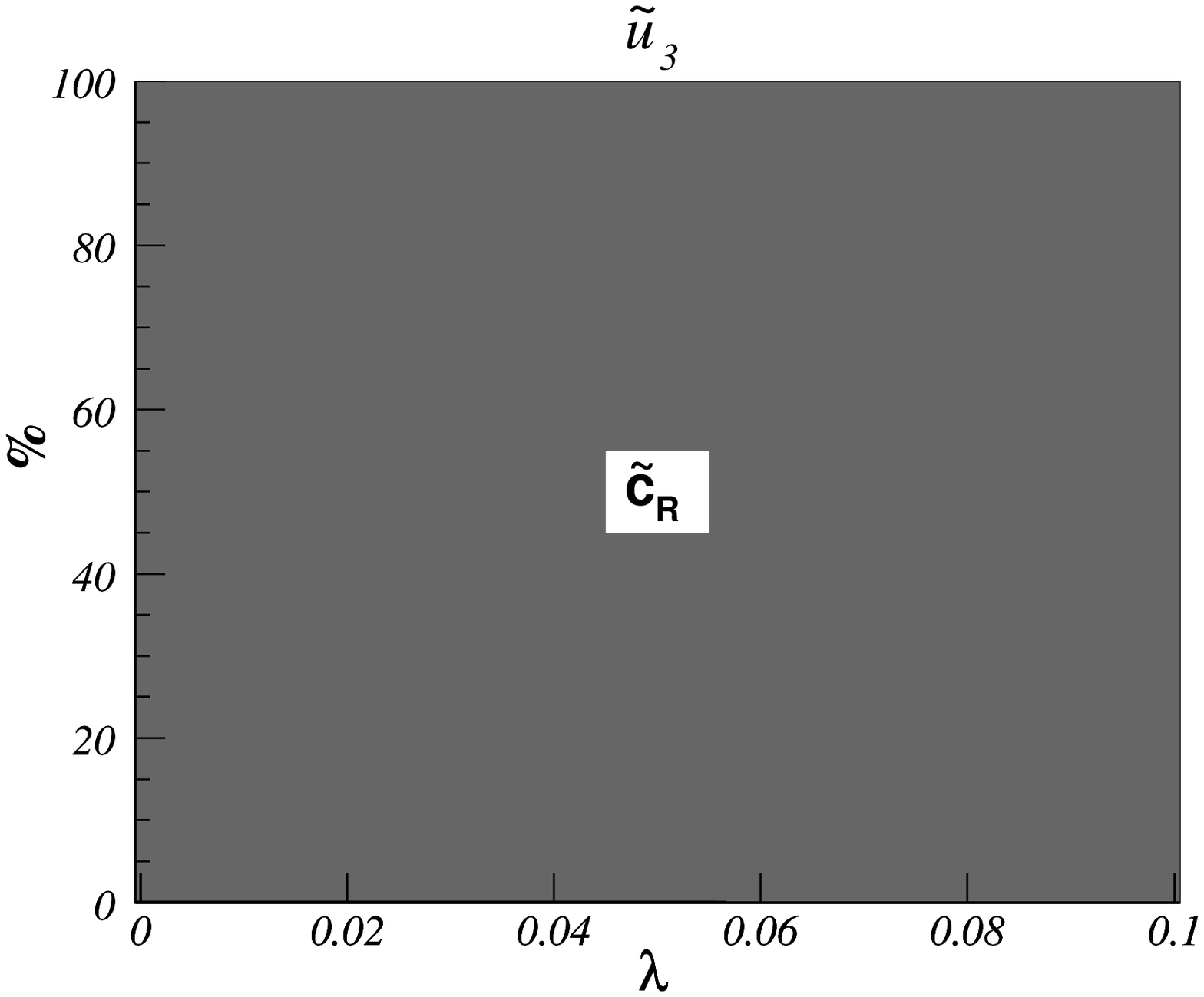}\hspace{1mm}
 \includegraphics[width=0.21\columnwidth]{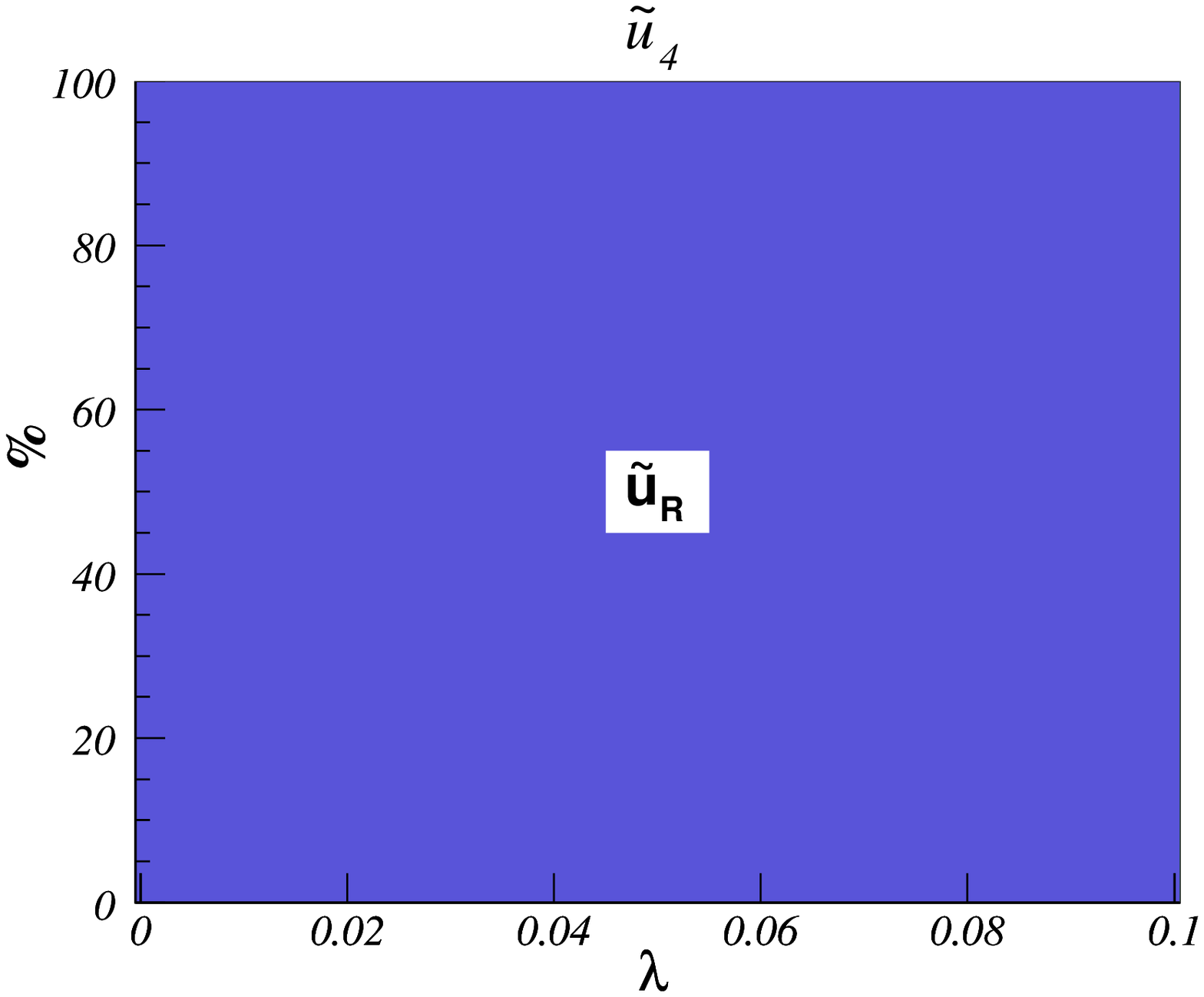}\hspace{1mm}
 \includegraphics[width=0.21\columnwidth]{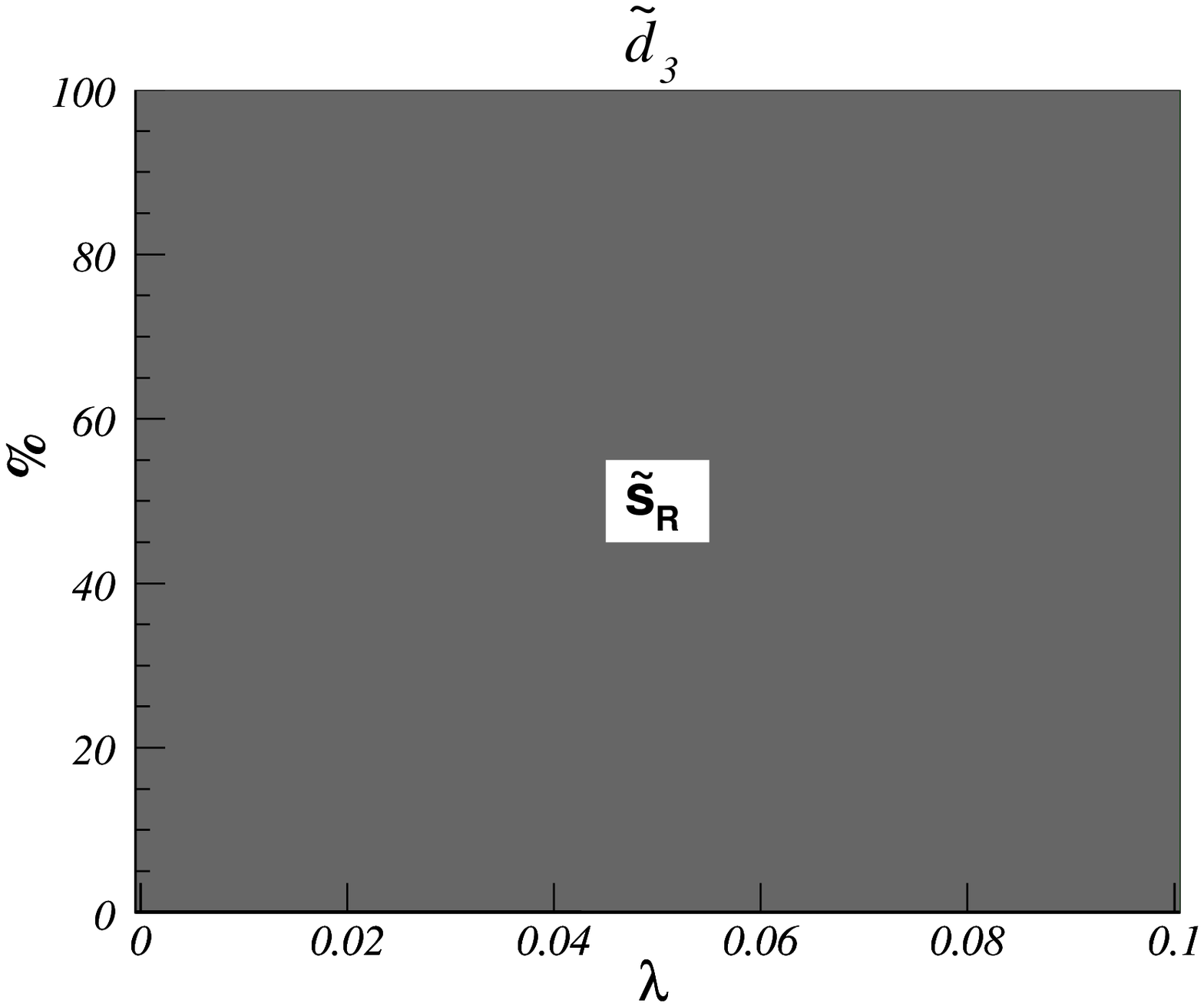}\hspace{1mm}
 \includegraphics[width=0.21\columnwidth]{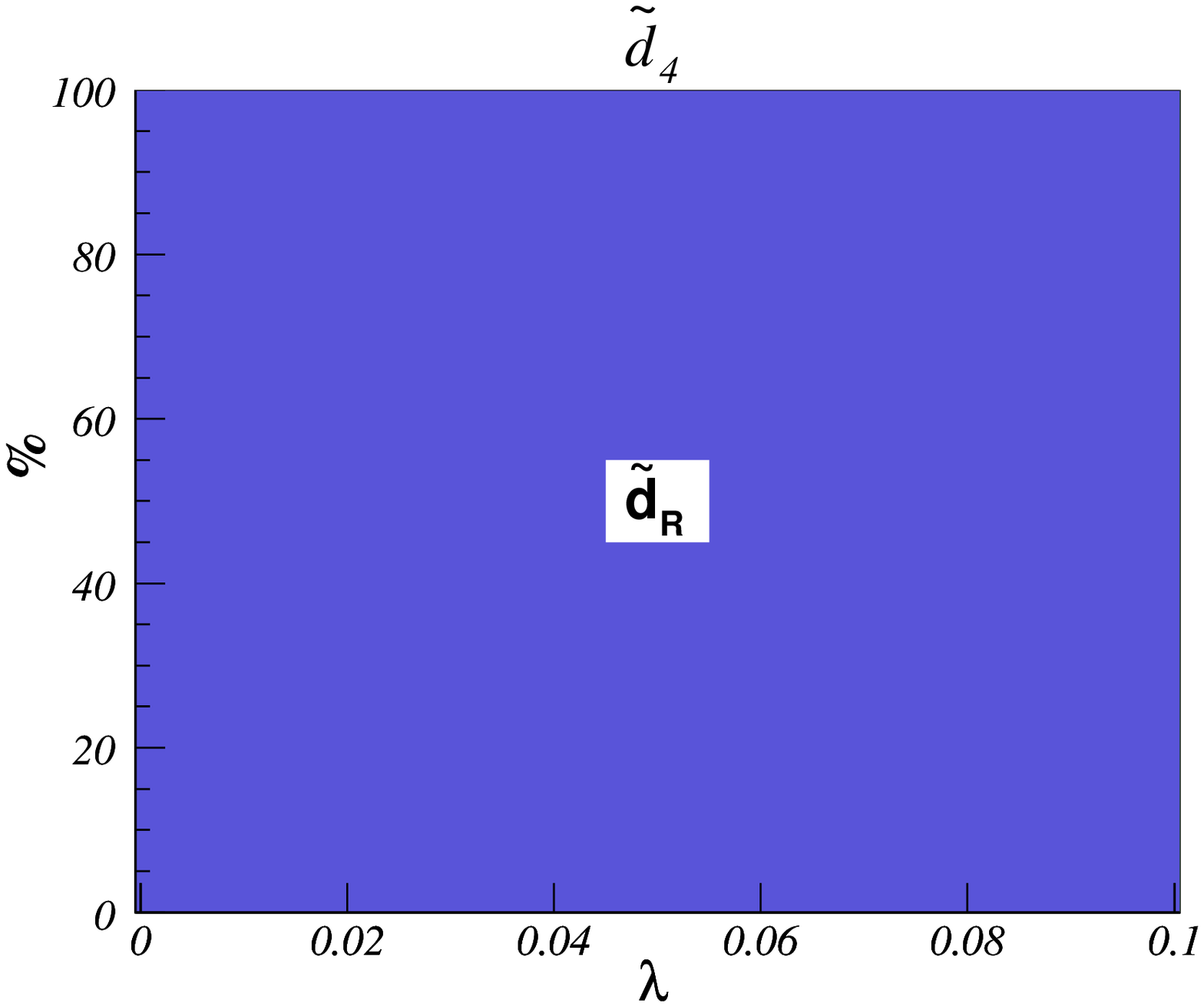}\vspace*{4mm}
 \includegraphics[width=0.21\columnwidth]{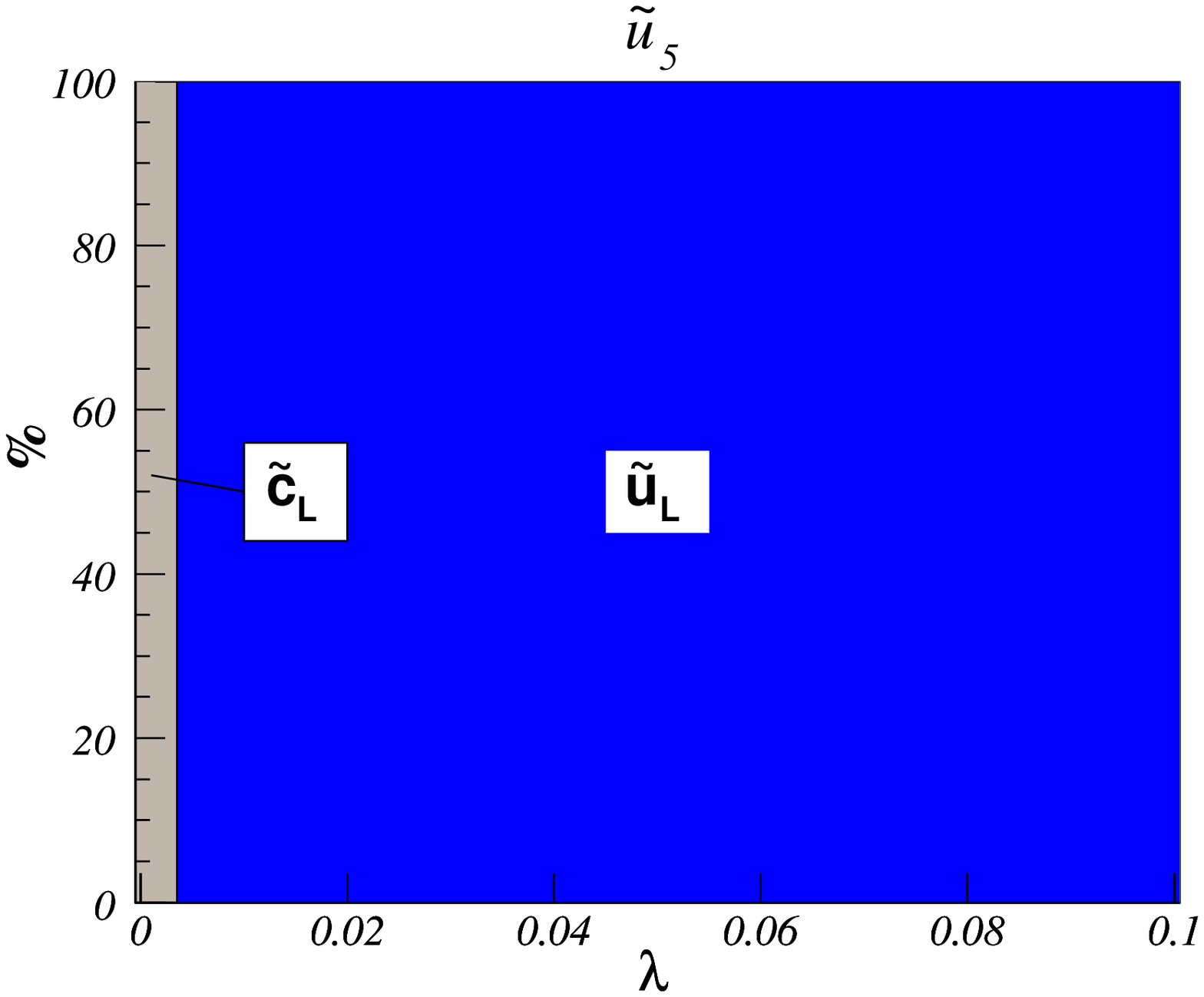}\hspace{1mm}
 \includegraphics[width=0.21\columnwidth]{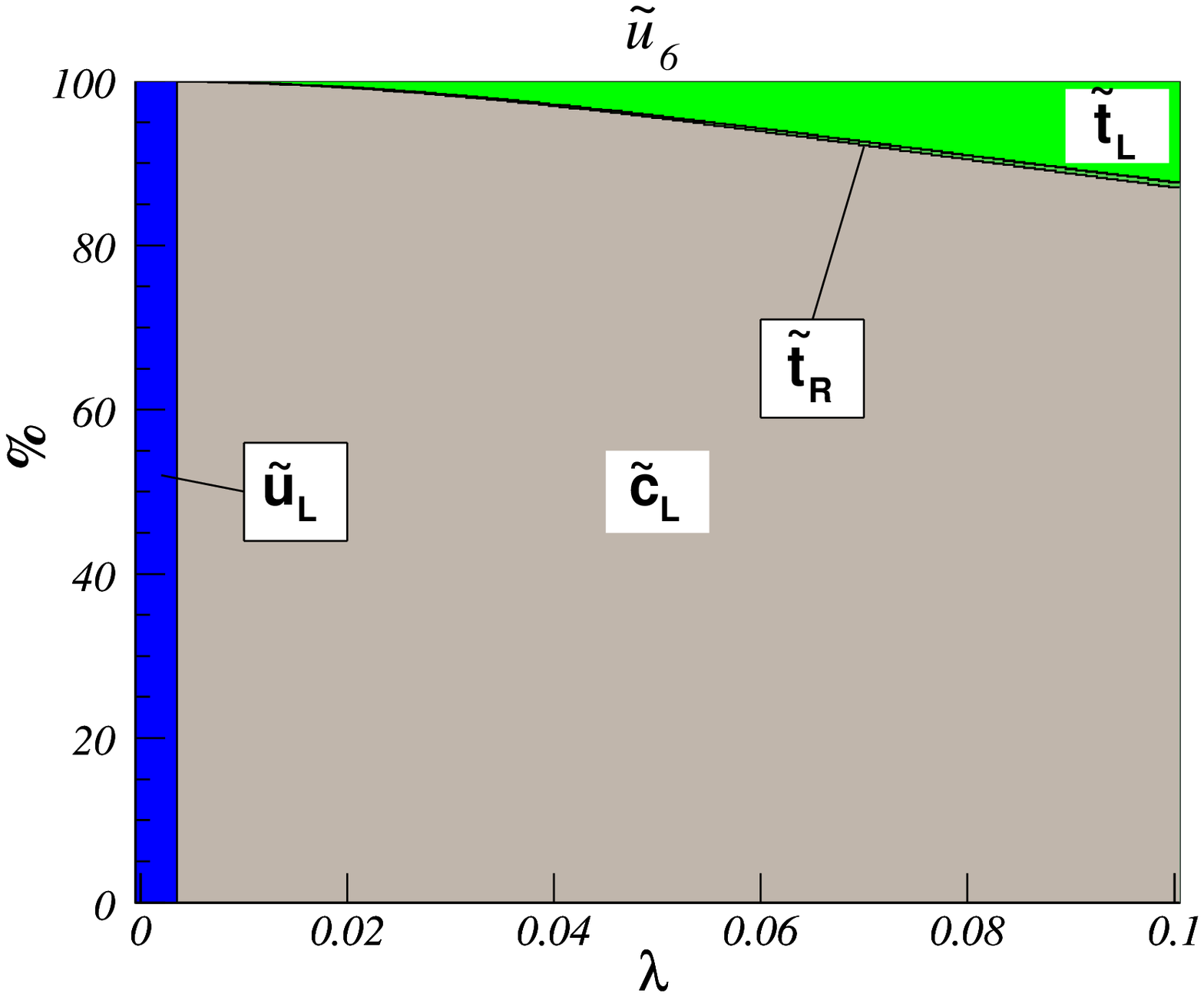}\hspace{1mm}
 \includegraphics[width=0.21\columnwidth]{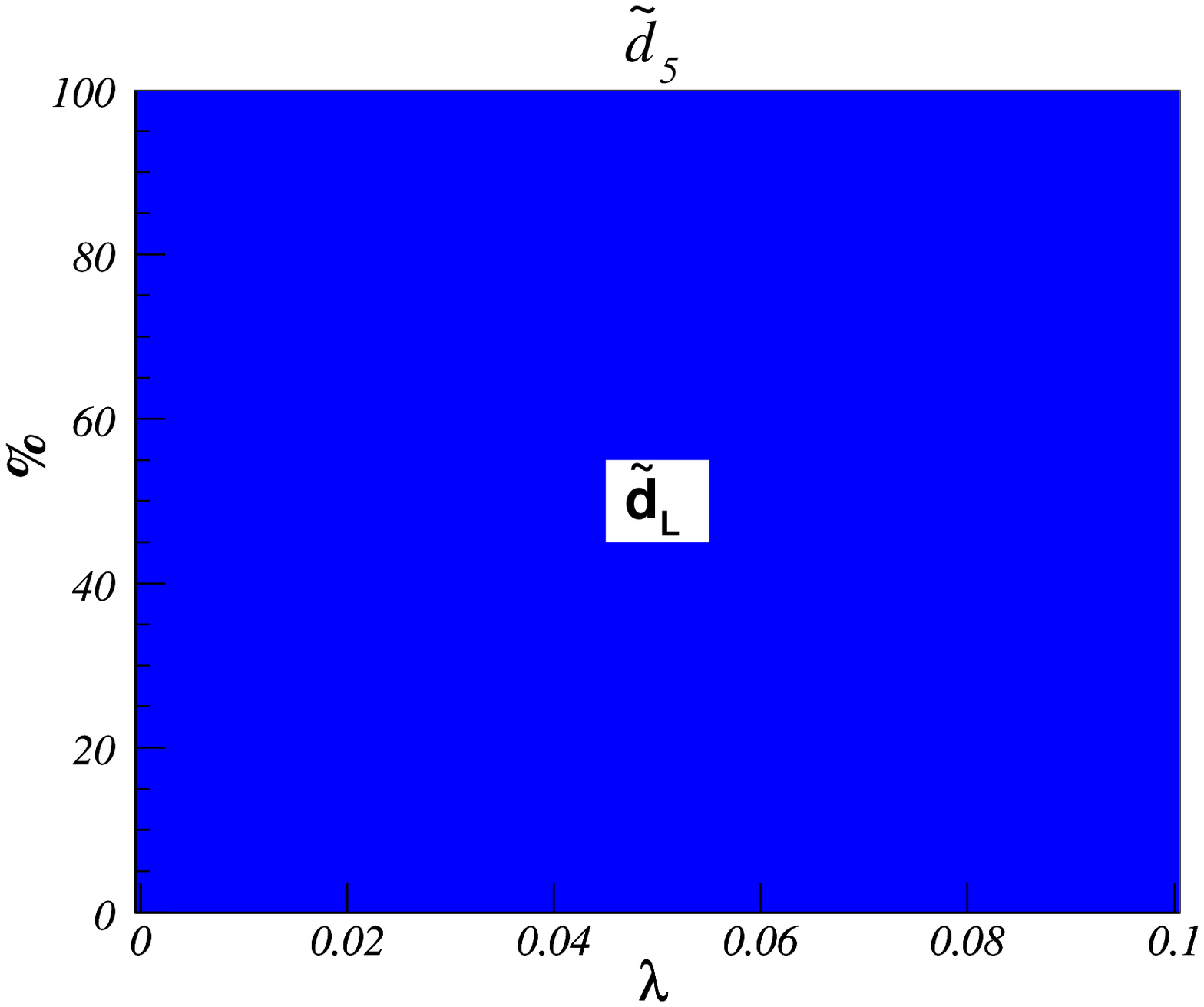}\hspace{1mm}
 \includegraphics[width=0.21\columnwidth]{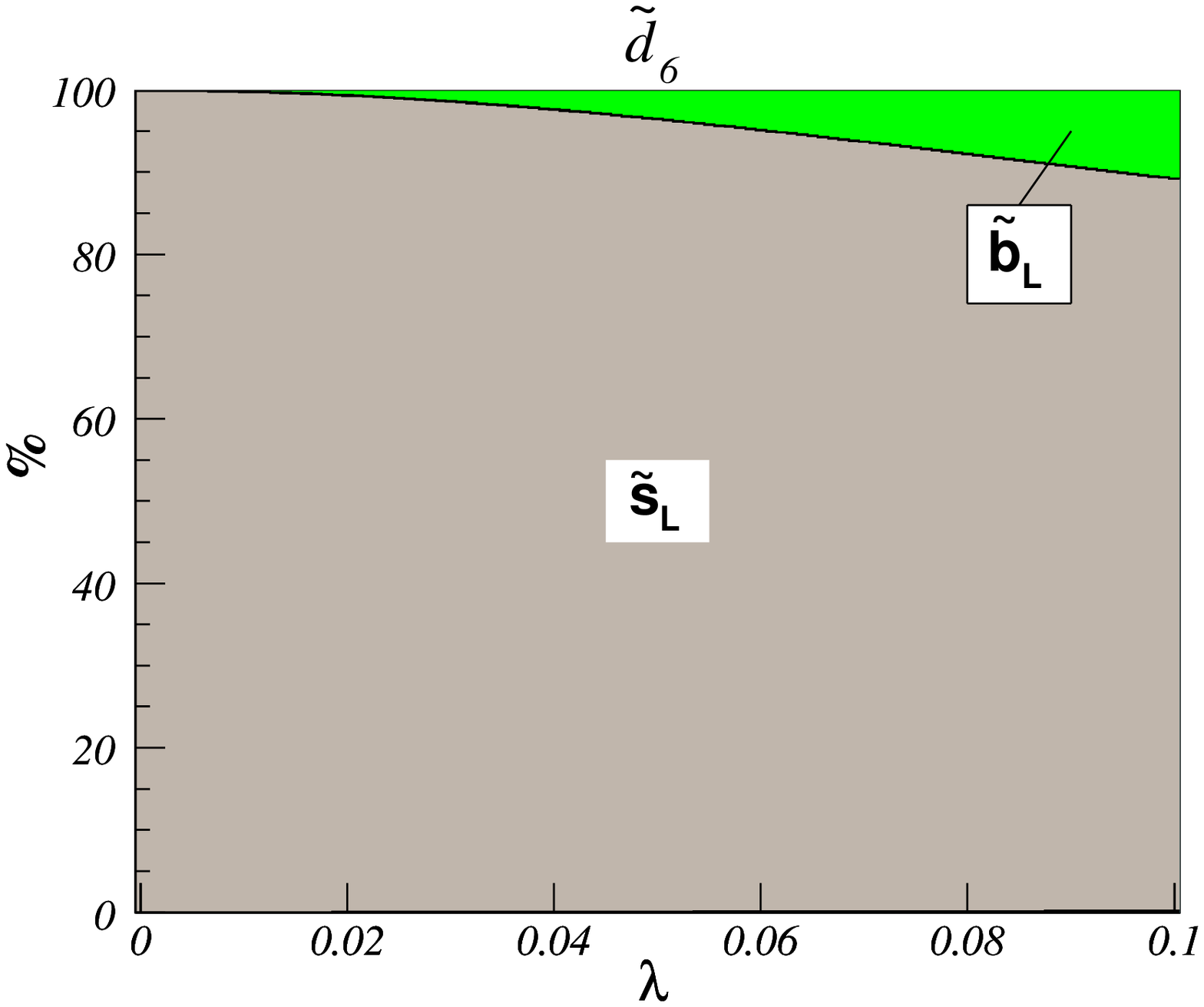}
 \caption{\label{fig:012p}Same as Fig.\ \ref{fig:012} for $\lambda\in
          [0;0.1]$.}
\end{figure}

As shown in Eq.\ (\ref{eq:physstate}) in NMFV, squarks exhibit
mixing of left- and right-handed helicities and the three
generations. For our benchmark scenario A, the helicity and
flavour decomposition of the six up-type (left) and down-type
(right) squark mass eigenstates is shown in Fig.\ \ref{fig:09} for
the full range of the parameter $\lambda\in[0;1]$ and in Fig.\
\ref{fig:09p} for the experimentally favoured range in the
vicinity of MFV, $\lambda\in [0;0.1]$. Left- and right-handed
first-generation and right-handed second-generation squarks
remain, of course, helicity- and flavour-diagonal, with the
left-handed and first-generation squarks being slightly heavier
due their weak isospin coupling (see Eqs.\ (\ref{eq:mll}),
(\ref{eq:mrr}) and (\ref{eq:mlr})) and different
renormalization-group running effects. The lightest up-type squark
$\tilde{u}_1$ remains the traditional mixture of left- and
right-handed stops over a large region of $\lambda\leq0.4$, but it
shows at this point the flavour transition expected from the level
reordering phenomenon (central plot of Fig.\ \ref{fig:05}). The
transition happens, however, above the experimental limit of
$\lambda\leq0.1$. Below this limit, it is the states
$\tilde{u}_2$, $\tilde{u}_6$, $\tilde{d}_1$, and in particular
$\tilde{d}_4$ and $\tilde{d}_6$ that show, in addition to helicity
mixing, the most interesting and smooth variation of second- and
third-generation flavour content (see Fig.\ \ref{fig:09p}). Note
that at very low $\lambda\simeq 0.002$ the states $\tilde{d}_L$
and $\tilde{s}_L$ rapidly switch
levels. \\

For the benchmark point B, whose helicity and flavour
decomposition is shown in Fig.\ \ref{fig:010}, level reordering
occurs at $\lambda\simeq0.4$ for the intermediate-mass up-type
squarks. Close inspection of Fig.\ \ref{fig:010p} shows, however,
that also $\tilde{d}_R$ and $\tilde{s}_R$ switch levels at low
values of $\lambda \simeq0.02$. At $\lambda\simeq0.01$, in
addition $\tilde{s}_R$ and $\tilde{b}_L$ switch levels, and at
$\lambda\simeq0.002$ it is the states $\tilde{u}_L$ and
$\tilde{c}_L$. The lightest up-type squark is again nothing but a
mix of left- and right-handed stops up to $\lambda\leq0.4$.
Phenomenologically smooth transitions below $\lambda\leq0.1$
involving taggable third-generation squarks are observed for
$\tilde{u}_4$, $\tilde{u}_6$, $\tilde{d}_1$, and
$\tilde{d}_6$.\\

For our scenario C, shown in Fig.\ \ref{fig:011}, just below
$\lambda=0.1$, $\tilde{u}_R$ and $\tilde{c}_R$ as well as
$\tilde{d}_R$ and $\tilde{s}_R$ rapidly switch levels, and
$\tilde{u}_L$ and $\tilde{c}_L$ switch levels at very low
$\lambda\simeq0.002$. These changes were already visible upon
close inspection of the lower central and right plots in Fig.\
\ref{fig:07}. On the other hand, the lightest squarks
$\tilde{u}_1$ and $\tilde{d}_1$ only acquire significant flavour
admixtures at relatively large $\lambda\simeq 0.2...0.4$, whereas
they are mostly superpositions of left- and right-handed stops and
sbottoms in the experimentally favourable range of $\lambda\leq
0.1$ shown in Fig.\ \ref{fig:011p}. Here, the heaviest
$\tilde{u}_6$ and $\tilde{d}_6$ show already smooth admixtures of
third-generation squarks as it was the case for the scenarios A
and B discussed above. The most interesting states are
$\tilde{u}_2$, $\tilde{u}_4$, $\tilde{d}_2$, and $\tilde{d}_4$,
respectively, since they represent combinations of up to four
different helicity and flavour states and have a significant,
taggable third-generation
flavour content.\\

The helicity and flavour decomposition for our scenario D, shown
in Fig.\ \ref{fig:012}, is rather similar to the one in scenario
A,  and only the mixed down-type state $\tilde{d}_4$ is now
lighter and becomes $\tilde{d}_2$. The lightest up-type squark
$\tilde{u}_1$ is again mostly a mix of left- and right-handed top
squarks up to $\lambda \simeq0.4$, where the level reordering and
generation mixing occurs (see lower central part of Fig.\
\ref{fig:08}). At the experimentally favoured lower value of
$\lambda\leq 0.1$, the states $\tilde{u}_2$, $\tilde{d}_1$,
$\tilde{d}_2$, and $\tilde{d}_6$ exhibit some smooth variations,
shown in detail in Fig.\ \ref{fig:012p}, albeit to a lesser extent
than in scenario A. At very low $\lambda\simeq0.004$, it is now
the up-type squarks $\tilde{u}_L$ and $\tilde{c}_L$ that rapidly
switch levels.

\section{Generalized couplings} \label{sec:coupling}

Considering the strong interaction first, it is well known that
the interaction of quarks, squarks, and gluinos can in general
lead to flavour violation in the left- and right-handed sectors
through non-diagonal entries in the matrices $R^q$, \bea
\left\{L_{\tilde{q}_j q_k \tilde{g}}, R_{\tilde{q}_j q_k
\tilde{g}} \right\} &=& \left\{R^q_{jk}, -
R^q_{j(k+3)}\right\}.~\eea Of course, the involved quark and
squark both have to be up- or down-type, since the gluino is
electrically neutral.\\

For the electroweak interaction, we define the square of the weak
coupling constant $g_W^2=e^2/\sin^2\theta_W$ in terms of the
electromagnetic fine structure constant $\alpha=e^2/(4\pi)$ and
the squared sine of the electroweak mixing angle
$x_W=\sin^2\theta_W=s_W^2 = 1-\cos^2\theta_W = 1-c_W^2$. The
$W^\pm-\tilde{\chi}^0_i- \tilde{\chi}^\pm_j$,
$Z-\tilde{\chi}^+_i-\tilde{\chi}^-_j$, and $Z-\tilde{\chi}^0_i-
\tilde{\chi}^0_j$ interactions are proportional to
\cite{Haber:1984rc} \bea O^L_{ij} = -\frac{1}{\sqrt{2}} N_{i4}
V^\ast_{j2} + N_{i2} V^\ast_{j1} &{\rm and}& O^R_{ij} =
\frac{1}{\sqrt{2}} N_{i3}^\ast U_{j2} + N_{i2}^\ast U_{j1},~\\
O^{\prime L}_{ij} = -V_{i1} V_{j1}^\ast - \frac{1}{2} V_{i2}
V_{j2}^\ast + \delta_{ij} x_W &{\rm and}& O^{\prime R}_{ij} =
-U_{i1}^\ast U_{j1} -
\frac{1}{2} U_{i2}^\ast U_{j2} + \delta_{ij} x_W,~~~~~~\\
O^{\prime\prime L}_{ij} = -\frac{1}{2} N_{i3} N_{j3}^\ast +
\frac{1}{2} N_{i4}N_{j4}^\ast &{\rm and}& O^{\prime\prime R}_{ij}
= \frac{1}{2} N_{i3}^\ast N_{j3} - \frac{1}{2} N_{i4}^\ast
N_{j4}~. \eea

In NMFV, the coupling strengths of left- and right-handed
(s)quarks to the electroweak gauge bosons are given by \bea \{
L_{q q^\prime Z},R_{qq^\prime Z} \}&=& (2\,T^{3}_q - 2\,e_q\,x_W)
\times \delta_{q q^\prime},~ \\ \{ L_{\tilde{q}_i \tilde{q}_j
Z},R_{\tilde{q}_i \tilde{q}_j Z} \} &=&
(2\,T^{3}_{\tilde{q}}-2\,e_{\tilde{q}}\,x_W) \times \sum_{k=1}^3
\{R^u_{ik}\, R^{u\ast}_{jk}, R^u_{i(3+k)}\, R^{u\ast}_{j(3+k)}
\},~\\ \{L_{qq^{\prime}W},R_{qq^{\prime}W}\} &=&
\{\sqrt{2}\,c_W\,V_{qq^{\prime}}, 0\},~\\ \{L_{\tilde{u}_i
\tilde{d}_j W}, R_{\tilde{u}_i \tilde{d}_j W}\} &=&
\sum_{k,l=1}^3\{\sqrt{2}\,c_W\, V_{u_kd_l}\, R^u_{ik}\,
R^{d\ast}_{jl},\, 0\}, \eea where $V_{kl}$ denotes the elements of
the CKM-matrix. To simplify the notation, we have introduced
flavour indices in the latter, $d_1=d$, $d_2=s$, $d_3=b$, $u_1=u$,
$u_2=c$, and $u_3=t$. The SUSY counterparts of these vertices
correspond to the quark-squark-gaugino couplings, \bea
L_{\tilde{d}_j d_k \tilde{\chi}^0_i} &=& \!\!\bigg[ (e_q \!-\!
T^3_q)\, s_W N_{i1} \!+\! T^3_q\, \cos\theta_W N_{i2} \bigg]
R^{d\ast}_{jk}\! +\! \frac{m_{d_k}\! c_W N_{i3}
R^{d\ast}_{j(k+3)}}{2\, m_W \cos\beta},~~~~~\\
-R_{\tilde{d}_j d_k \tilde{\chi}_i^0}^\ast &=& e_q\, s_W\,
N_{i1}\, R^d_{j(k+3)} - \frac{m_{d_k}
c_W\, N_{i3}\, R^d_{jk}}{2\, m_W\, \cos\beta},~~~~~\\
L_{\tilde{u}_j u_k \tilde{\chi}^0_i} &=& \!\!\bigg[ (e_q \!-\!
T^3_q)\, s_W N_{i1} \!+\! T^3_q\, c_W N_{i2} \bigg] R^{u\ast}_{jk}
\!+\! \frac{m_{u_k}\! c_W N_{i4} R^{u\ast}_{j(k+3)}}{2\, m_W
\sin\beta},~~~~~\\ -R_{\tilde{u}_j u_k \tilde{\chi}_i^0}^\ast &=&
e_q\, s_W\, N_{i1}\, R^u_{j(k+3)} - \frac{m_{u_k}\,
c_W\, N_{i4}\, R^u_{jk}}{2\, m_W\, \sin\beta} ,~\\
L_{\tilde{d}_j u_l \tilde{\chi}_i^\pm}&=& \sum_{k=1}^3\bigg[
U_{i1}\, R^{d\ast}_{jk} - \frac{m_{d_k}\, U_{i2}\,
R^{d\ast}_{j(k+3)}}{\sqrt{2}\, m_W\, \cos\beta} \bigg] V_{u_l d_k}
,~\\ -R^\ast_{\tilde{d}_j u_l \tilde{\chi}_i^\pm} &=& \sum_{k=1}^3
\frac{m_{u_l}\, V_{i2}\,
V_{u_l d_k}^\ast\, R^d_{jk}}{\sqrt{2}\, m_W\, \sin\beta} ,~\\
L_{\tilde{u}_j d_l \tilde{\chi}_i^\pm}&=& \sum_{k=1}^3 \bigg[
V_{i1}^\ast\, R^u_{jk} - \frac{m_{u_k}\, V_{i2}^\ast\,
R^u_{j(k+3)}}{\sqrt{2}\, m_W\, \sin\beta}  \bigg] V_{u_k d_l} ,~\\
-R^\ast_{\tilde{u}_j d_l \tilde{\chi}_i^\pm} &=& \sum_{k=1}^3
\frac{m_{d_l}\, U_{i2}^\ast V_{u_k d_l}^\ast\,
R^{u\ast}_{jk}}{\sqrt{2}\, m_W\, \cos\beta}~. \eea These general
expressions can be simplified by neglecting the Yukawa couplings
except for the one of the top quark, whose mass is not small
compared to $m_W$. The usual MSSM couplings, and the couplings
involving sleptons and sneutrinos are easily obtained by replacing
the mixing matrices, the masses, and the CKM matrix in a proper
way from the above equations.

\chapter{Resummation formalisms}\label{ch:res}
\section{General points}
\subsection{Main features of the resummation procedure}
\label{sec:main}

The finite energy or angular resolution of any particle detector
implies that physical cross sections are always inclusive over
arbitrarily soft produced particles. For example, a single quark
jet cannot be distinguished from a quark jet accompanied by some
collinear gluons or by partons with vanishing momentum. This
implies that after the renormalization procedure, the remaining
divergences in any physical observable are soft (due to parton
radiation with small four-momentum) and/or collinear (due to
emission of partons moving in parallel to the emitting one).\\

In fixed-order perturbation theory, infrared divergences of
virtual gluons are exactly cancelled by the emission of real
undetected ones, but this cancellation can leave large finite
terms. In specific kinematics configurations, real and virtual
contributions can be highly unbalanced, spoiling the cancellation
mechanism, and only resummation to all order restores the balance.
At the exclusive boundary of the phase space, when for example the
tagged final state carries almost all the available energy or when
its transverse momentum tends to zero, the real emission is
strongly suppressed, producing a loss of balance with the virtual
contribution, and spreads out up to the phase space kinematical
limits through finite logarithmic terms $L$ which become large.\\

Let us take a given physical observable $R$, whose perturbative
expansion in the strong coupling constant $\as$ is \bea
\label{eq:usual_pert_exp} R &=&R_0\Bigg[ 1 +
\sum_{n=1}^{\infty}\alpha_s^n\left(c_{2n}^{(n)}L^{2n} +
c_{2n-1}^{(n)}L^{2n-1}+\ldots \right)\Bigg].~\eea The logarithmic
corrections $L$ can be large even if $\as$ is small, enhancing
thus the coefficients of the perturbative expansion. The ratio of
two successive terms in Eq.\ (\ref{eq:usual_pert_exp}) is
$\mathcal{O}(\alpha_s L^2)$, which means that any higher order
contribution is $\mathcal{O}(1)$ with respect to the previous
terms, and the reliability of perturbative predictions is thus
spoiled. However, all-order resummation gives a satisfactory
solution to this problem in the context of perturbation theory.\\

The key features of the resummation procedure are the dynamical
and kinematical factorizations. Dynamical factorization comes from
gauge invariance and unitarity and is completely general. The
multi-gluon radiation amplitude factorizes, in the soft-limit,
into the product of single-gluon emission probabilities. It is
similar to multiple soft-photon emission in QED, apart from the
fact that radiations of photons are uncorrelated, since they are
electrically neutral. In QED, soft divergences related to
degenerate final states cancel out after summation, as stated by
the Bloch-Nordsieck theorem \cite{Bloch:1937pw}. In general, this
theorem is broken in QCD, since the collinear degeneracy leads to
a divergence that does not necessary cancel out in the transition
rates due to colour correlation in gluon interactions. The
Kinoshita-Lee-Nauenberg theorem \cite{Kinoshita:1962ur,
Lee:1964is} asserts that all the infrared divergences, soft and
collinear, exactly cancel if the summation over initial and final
degenerate states is carried out, and the resulting observables
are infrared safe. Although it is not possible to sum over
degenerate initial states, the corresponding uncancelled soft
divergences are under control if the energy scale is high enough,
because they are power-suppressed, while the QCD factorization
theorem guarantees that the collinear divergences can be absorbed
in the scale dependence of the parton densities, and they thus
factorize from the hard-scattering process. This eventually leads
to an exponentiation due to coherent rather than independent
emission as in QED.\\

The second factorization concerns kinematics and occurs if phase
space can be factorized in terms of single-parton phase spaces. It
does not generally occur in the space where the shape variable is
defined, because it depends in a complicated way on the
multi-parton configuration, but rather in a conjugate space
introduced via Mellin or Fourier transformation, in which momentum
conservation is more easily implemented.\\

Both factorizations lead to exponentiation, and the observable $R$
defined in Eq.\ (\ref{eq:usual_pert_exp}) can be rewritten as
\bea\label{eq:expo}R &=&R_0 \exp\Bigg[\sum_{n=1}^{\infty}
\alpha_s^n \sum_{m=1}^{n+1}G_{nm} L^m\Bigg] \nonumber \\ &=&R_0
\exp\Bigg[L\, g_1(\alpha_s L) + g_2(\alpha_s L) + \alpha_s\,
g_3(\alpha_s L) + \ldots\Bigg]~, \eea where each $g_i$ function
corresponds to the resummation of a specific class of logarithms
and contains all QCD effects due to gluon colour charge and
running coupling. The ratio of two successive terms is now
$\mathcal{O}(1/L)$, and the perturbative expansion is converging
in the $L \gg 1$ region, provided that $\alpha_s L \ll 1$. All the
$\alpha_s L^m$ terms with $n+1 < m \leq 2n$ appearing in Eq.\
(\ref{eq:usual_pert_exp}) are taken into account by the
exponentiation of the lowest order terms and reappear in the
expansion of the exponent at a given power of $\alpha_s$.

\subsection{Soft-photon resummation in QED}\label{sec:soft_ph}

%%%%%%%%%%%%%% Begin Figure I %%%%%%%%%%%%%%%%%%%%%%%%%%%%%%%%%%%%
\begin{figure}
 \centering
 \includegraphics[width=.3\columnwidth]{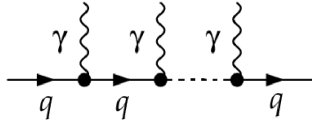}
  \caption{\label{fig:1}Emission of $n$ soft photons of momenta $q_i$.}
\end{figure}
%%%%%%%%%%%%%% End of Figure I %%%%%%%%%%%%%%%%%%%%%%%%%%%%%%%%%%%

Let us first describe multiple soft-photon emission in QED and its
exponentiation \cite{Parisi:1979se, Bassetto:1984ik} with a
generic process $f_1(p_1)\, \bar{f}_2(p_2) \rightarrow F(M) + X$,
describing the production of a final state $F$ of mass $M$, plus
some unobserved states $X$ relative to the number of emitted
photons. The one-photon emission partonic cross section ${\rm d}
\sigma^{(1)}(M;q)$, where $q$ is the photon momentum, is given by
the product of Born cross section ${\rm d} \sigma^{(0)}(M)$ and
the single photon emission probability ${\rm d} w^{(1)}(q)$ \bea
\label{eq:one_phot} {\rm d} \sigma^{(1)}(M;q) &=& \td
\sigma^{(0)}(M)\, {\rm d} w^{(1)}(q) = {\rm d} \sigma^{(0)}(M)
\,\frac{\alpha}{4 \pi^2 \omega} \, \left| j(q) \right|^2 \eea with
restriction to fast fermions (with negligible mass with respect to
their momentum) and to small photon energy $\omega$ compared to
the scale $M$. $j^\mu(q) = \left( p_2^\mu / p_2 \cdot q - p_1^\mu
/ p_1 \cdot q\right)$ is the conserved emission current of a
soft-photon with momentum $q$ by a fermion of momentum $p_1$ and
antifermion of momentum $p_2$. The amplitude for $n$ soft-photon
emissions of momenta $q_i$ (see Fig.\ \ref{fig:1}) is given, in
the soft-limit, by \bea M^{(n)} &=& \frac{e^n}{n!} \prod_{i=1}^n
\frac{p_1 \cdot \varepsilon_i}{q_i\cdot p_1}~, \eea where
$\varepsilon_i$ is the polarization vector of the $i^{{\rm th}}$
photon. By summing over the different polarization states, we see
that the squared matrix element factorizes \bea
\label{eq:n_phot}{\rm d} \sigma^{(n)}(M;q_1,\ldots,q_n) &=& {\rm
d} \sigma^{(0)}(M)\, \frac{1}{n!}\, \prod_{i=1}^n {\rm d}
w^{(1)}(q_i).~\eea This means that soft-photon emission is
uncorrelated and that infrared singularities exponentiate. Indeed,
kinematical factorization occurs naturally in impact-parameter $b$
space, $b$ being conjugate to the shape variable, where the
$\delta$-function implementing four-momentum conservation
exponentiates. Summing Eq.\ (\ref{eq:n_phot}) to all orders, we
get \bea \td\sigma\, (M)&=&{\rm d} \sigma^{(0)}(M) \sum_n \int
\frac{1}{n!}{\rm d} w^{(1)}(q_1) \ldots {\rm d} w^{(n)}(q_n)\,
\delta \left(\sum_i q_i + P - p_1 - p_2 \right)\nonumber \\
&=& {\rm d} \sigma^{(0)}(M) \int \frac{{\rm d}^4 b}{(2 \pi)^4}
e^{i b \cdot (P - p_1 - p_2)} \sum_n \left[ \frac{1}{n!} \int {\rm
d} \tilde{w}^{(1)}(q)\, e^{i q \cdot b} \right]^n\nonumber\\
&=& {\rm d} \sigma^{(0)}(M) \int \frac{{\rm d}^4 b}{(2 \pi)^4}
e^{i b \cdot (P - p_1 - p_2)}\, \exp\left[\int {\rm d}
\tilde{w}^{(1)}(q)\,  e^{i q \cdot b}\right],~ \eea where ${\rm d}
\tilde{w}$ is the Fourier transform of the single-photon emission
probability ${\rm d} w$ and $P$ the four-momentum of the final
state $F$. By using the usual expression of ${\rm d} \tilde{w}$
and performing the integral, we can deduce an exponentiated form
factor, containing all the logarithms.

\subsection{Soft-gluon resummation in QCD}

%%%%%%%%%%%%%% Begin Figure II %%%%%%%%%%%%%%%%%%%%%%%%%%%%%%%%%%%
\begin{figure}
 \centering
 \includegraphics[width=.5\columnwidth]{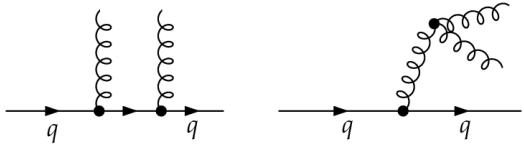}
  \caption{\label{fig:2}Emission of two soft-gluons.}
\end{figure}
%%%%%%%%%%%%%% End of Figure II %%%%%%%%%%%%%%%%%%%%%%%%%%%%%%%%%%

Contrary to photons which are electrically neutral, gluons carry
colour charge. However, the infrared divergences related to gluon
correlations cancel among themselves and an exponentiation is
still possible \cite{Catani:1983bz, Catani:1984dp}. Considering
the same generic process as in the previous section, the cross
section for single-gluon emission can be written as in Eq.\
(\ref{eq:one_phot}), provided that we introduce colour operators,
\bea \label{eq:one_glu} {\rm d} \sigma^{(1)}(M;q) = {\rm d}
\sigma^{(0)}(M)\,{\rm d} w^{(1)}(q) = {\rm d} \sigma^{(0)}(M)\,
\frac{\alpha_s}{4 \pi^2 \omega}\, \left| J^a(q) \right|^2,~ \eea
where $\omega$ ($q$) denotes the gluon energy (momentum) and the
current $J_\mu^a(q)= T^a_{p_2} j_\mu(q)$ includes now the colour
charge operator $T^a_{p_2}$. \\

For the emission of $n$ soft-gluons of momentum $q_i$ and energy
$\omega_i$, diagrams where gluons are emitted by harder ones have
to be taken into account, as in Fig.\ \ref{fig:2} for a two-gluon
emission. The difference with QED comes from the second diagram,
not present in the case of photons. If we assume strong ordering
in gluon energies $\omega_1 \ll \omega_2 \ll \ldots \ll \omega_n
\ll M$, the use of the soft limit is justified, and we can
recursively define the amplitude $M^{(n)}$, since the emission of
the softest gluon factorizes \bea \label{eq:glu_M_N}\langle
a_1,\ldots,a_n \mid M^{(n)} \rangle = \langle a_2,\ldots,a_n \mid
J_{\mu_1}^{a_1}(q_1) \mid b_1,\ldots,b_{n-1} \rangle\, \langle
b_1,\ldots,b_{n-1} \mid M^{(n-1)}\rangle,~\eea where we sum over
repeated indices and where $a_i$ ($\mu_i$) are the colour
(Lorentz) indices of the emitted gluons. $M^{(n)}$ can then be
computed through an iterative insertion scheme, and performing the
summation over $n$ leads to the exponentiation of the gluonic
radiations. As an example, we will show the exponentiation of the
gluon coherent state operator, whose application on the vacuum
leads to the usual resummation formulas, given e.g.\ in
\cite{Collins:1982wa} for transverse-momentum resummation, in
\cite{Sterman:1986aj} for threshold resummation, and in
\cite{Laenen:2000ij} for joint resummation in the case of
Drell-Yan lepton pair production.\\

Let us consider the emission of two gluons. Using the usual
definition of the current $\mathbf{J}_\mu (q_1) = \sum_{i=2}^n
\mathbf{T}_i \frac{{q_i}_\mu}{q_i \cdot q_1} + \mathbf{T}_0\,
j_\mu(q)$ where the index $0$ labels the hard $f_1\bar{f}_2$
contribution, we get \bea \langle a_1,a_2 \!\mid\! M^{(2)} \rangle
&=& \langle a_2 \!\mid\! T_0^{a_1}\, \varepsilon_1\!\cdot\!j(q_1)
\!\mid\! b_1 \rangle \langle b_1 \!\mid\! M^{(1)}\rangle + \langle
a_2 \!\mid\! T^{a_1}_2\frac{\varepsilon_1 \cdot q_2}{q_2 \cdot
q_1} \!\mid\! b_1 \rangle \langle b_1 \!\mid \!M^{(1)}\rangle
\nonumber \\ \label{eq:glu_2_emm}&=& C_{(1)}^{a_1} C_{(1)}^{a_2} +
C_{(2)}^{a_1 a_2},~\eea after the introduction of the polarization
vectors in Eq.\ (\ref{eq:glu_M_N}). The correlation operators $C$
are recursively defined by \bea C_{(1)}^{a_i} &=& \langle a_i \mid
M^{(1)}\rangle = \langle b
\mid T_0^{a_i} \varepsilon_i \cdot j(q) \mid b \rangle\nonumber \\
\label{eq:col_cor_coeff} C_{(n)}^{a_1\ldots a_n} &=& \langle
a_2\ldots a_n \mid \sum_{j=2}^n T^{a_1}_j \frac{\varepsilon_1
\cdot q_j}{q_j \cdot q_1} \mid b_1 \ldots b_{n-1} \rangle\,
C_{(n-1)}^{b_1\ldots b_{n-1}}.~\eea The first term of Eq.\
(\ref{eq:glu_2_emm}) corresponds to the emission of the softest
gluon by one of the external legs (left diagram of Fig.\
\ref{fig:2}), and the second to the emission by the harder gluon
(right diagram of Fig.\ \ref{fig:2}). Generalization to any number
of gluons is straightforward, \bea \label{eq:col_cor_recu} \langle
a_1,\ldots,a_n \!\mid\! M^{(n)} \rangle \!\!=\!\! \sum_{k=1}^n
\sum_{\{n_1, \ldots, n_k\}}\!\!\!\! C_{(n_k)} \Theta_{n_k} \ldots
C_{(n_1)} \Theta_{n_1} \Theta(1,\ldots,k) \!+\! {\rm
perm.},~~~\eea where the dependence on the colour indices is not
explicitly shown in order to simplify the expression. For each
value of $k$, the indices $\{a_1,\ldots,a_n\}$ are partitioned
into $k$ classes of $n_i$ indices, corresponding to $k$ chains of
$n_i$ gluons. Inside a given class, the gluons are ordered in
energy through the $\Theta_{n_i}$ functions so that a specific
gluon can only be emitted by a harder one. Eventually, the
$\Theta(1,\ldots,k)$ function orders the energies of the hardest
gluon of each class.\\

Let us now define the coherent state operator for real emission,
containing the correlation structure between the emitted gluons
\bea U^M(A^+)&=& \sum_{n=0}^\infty \int {\rm d}\phi^n \langle
a_1,\ldots,a_n \mid M^{(n)} \rangle A^{+a_1}_{\alpha_1}(q_1)
\ldots A^{+a_n}_{\alpha_n}(q_n),~\eea where $d\phi^{(n)}= {\rm d}
[q_1]\ldots {\rm d} [q_n]$ is the relativistic $n$-gluon phase
space and $A^{+a_i}_{\alpha_i}(q)$ the creation operator of a
gluon with colour (polarization) index $a_i$ ($\alpha_i$). Eqs.
(\ref{eq:col_cor_coeff}) and (\ref{eq:col_cor_recu}) lead to \bea
U^M(A^+)&=& 1+\sum_{n=1}^\infty \sum_{k=1}^n \sum_{\{n_i\}}\int
{\rm d}\phi^{(n_1)}\ldots {\rm d}\phi^{(n_k)} \left[g_s^{n_k}
C_{(n_k-1)} \Theta_{n_k}
\big(\varepsilon_k \cdot J_{M k}\big)\right] \ldots \nonumber\\
&& \times \left[g_s^{n_1} C_{(n_1-1)} \Theta_{n_1}
\big(\varepsilon_1 \cdot J_{M 1}\big)\right] A^{+ (n_k)} \ldots
A^{+ (n_1)}\, \Theta(1,\ldots,k),~\eea where $A^{+(n_i)}$ is the
product of $n_i$ gluon creation operators and $ \mathbf{J}_{M
i}^{\mu_i} = \sum_{j=2}^{n_i} \mathbf{T}_j \frac{q_j^{\mu_i}}{q_j
\cdot q_i}$. After summing over $\{n_i\}$, each square bracket
factor reproduces the gluon coherent state operator
$\mathcal{U}^{q_i}(A^+)$ where $q_i$ is the momentum of the
fastest gluon in each class. Eventually, we get exponentiation
\bea U_M (A^+) &=& 1+ \sum_{k=1}^\infty g_s^k \int {\rm d}[q_1]
\ldots {\rm d}[q_k] \, \Theta(1,\ldots,k)\,
\big(\varepsilon^{\alpha_k} \cdot J_{M k}^{b_k}\big)\ldots
\big(\varepsilon^{\alpha_1} \cdot J_{M 1}^{b_1}\big) \nonumber
\\ &&\times  \left[\mathcal{U}^{q_k}(A^+)\right]_{a_k b_k} \ldots
\left[\mathcal{U}^{q_1}(A^+)\right]_{a_1 b_1}A^{+
a_1}_{\alpha_1}(q_1) \ldots A^{+ a_k}_{\alpha_k}(q_k) \nonumber \\
\label{eq:insert} &=& \bar{P}_w \exp\Bigg\{ g_s \int {\rm d}[q]\,
\Theta(M-w)\, A^{+ a}_\alpha (q)\,
\left[\mathcal{U}^q(A^+)\right]_{ab} \, \big(\varepsilon^\alpha
\cdot J_{Mq}^b \big) \Bigg\}~,~\nn \eea \vspace{-.8cm}\bea ~ \eea
where $\bar{P}_w$ orders the colour matrices appearing in $J_{Mq}$
in such a way that the
harder operator acts first.\\

Eq.\ (\ref{eq:insert}) is not unitary, since we have only creation
operators. The unitary is restored through the introduction of
destruction operators, replacing $A^+_\alpha (q)$ by
$\mathcal{R}_\alpha (q) = A^+_\alpha (q) - A_\alpha (q)$. It
corresponds to the inclusion of the virtual corrections that we
have neglected in the first place.

\section{Transverse-momentum resummation}
\subsection{The Collins-Soper-Sterman (CSS)
formalism}\label{sec:res_CSS}

Let us consider the process \bea\label{eq:DY_proc} h_1(p_1)\,
h_2(p_2) \rightarrow F(M^2,q_T^2) + X,~\eea where two hadrons
$h_1$ and $h_2$ with momenta $p_1$ and $p_2$ collide with a
centre-of-mass energy $\sqrt{s_h}$ to produce an observed
colourless final state system $F$ with a mass $M$ and a transverse
momentum $q_T$, while $X$ is unobserved. As $F$ does not carry any
colour charge, the lowest order partonic mechanism is initiated
either by $q \bar{q}$ annihilation or by gluon fusion. Isolating
the divergences from the fixed-order production cross section, we
can derive the resummed cross section to get a finite expression
for the $q_T$-spectrum of the system $F$. \\

On the basis of the QCD factorization theorem, the most general
form of the differential cross section, \bea\label{eq:QCDFact}
\frac{{\rm d}^2\sigma}{{\rm d}M^2{\rm d}q_T^2}\! =\! \sum_{a,b}
\int_0^1\!\!{\rm d}x_a \,\int_0^1\!\!{\rm d}x_b\,
f_{a/h_1}(x_a;\mu_F)\, f_{b/h_2}(x_b;\mu_F)\, \hat{\sigma}_{ab}(M,
q_T, s; \mu_R, \mu_F),~~~\eea where $s=x_a\,x_b\,s_h$ is the
partonic centre-of-mass energy, can be computed by convoluting the
hard scattering function $\hat{\sigma}_{ab}$ with universal parton
densities $f_{a/h_1}$ and $f_{b/h_2}$ of partons $a,b$ in the
hadrons $h_1,h_2$. The PDFs depend on the longitudinal momentum
fractions of the two partons $x_{a,b}$ and on the unphysical
factorization scale $\mu_F$. The hard part $\hat{\sigma}_{ab}$ has
a perturbative expansion in $\alpha_s(\mu_R)$, $\mu_R$ being the
unphysical renormalization scale, \bea \hat{\sigma}_{ab}(M, q_T,
s; \mu_R, \mu_F) = \sum_{n=0}^\infty
\left(\frac{\alpha_s(\mu_R)}{\pi}\right)^{n+p}
\hat{\sigma}_{ab}^{(n)}(M, q_T, s; \mu_F),~ \label{eq:fixqt}\eea
where $p$ is the minimum power of $\alpha_s$ needed for the
process to occur at leading order. Isolating the divergent terms
as the transverse momentum $q_T\to 0$, $\hat{\sigma}_{ab}^{(n)}$
can be written as \cite{Dokshitzer:1978hw} \bea \label{eq:Tabn}
\hat{\sigma}_{ab}^{(n)}(M, q_T, s; \mu_F) &=& R_{ab}^{(n)}(M, q_T,
s; \mu_F) + \hat{\sigma}_{ab}^{(n,\delta)}(M, s; \mu_F)\,
\delta(q_T) \nonumber\\ &&+ \sum_{m=0}^{2n-1} \!
\hat{\sigma}_{ab}^{(n,m)}(M, s; \mu_F) \,
\frac{\ln^m(M^2/q_T^2)}{q_T^2},~\label{eq:fixqt2}\eea where
$R_{ab}^{(n)}$ is the regular part of the hard function,
containing all the terms less singular than $\delta(q_T^2)$ and
$q_T^{-2}$ as $q_T \to 0$. All the logarithmically enhanced and
the $\delta(q_T^2)$ contributions are resummed to all orders in
$\alpha_s$, and the remaining terms correspond to the usual
perturbative series after the removal of the logarithmic and
$\delta(q_T^2)$ contributions.\\

We define then the finite component by \bea
\label{eq:resfin}\left[\frac{{\rm d}^2\sigma}{{\rm d}M^2{\rm
d}q_T^2}\right]_{{\rm fin}} &=& \sum_{a,b} \int_0^1{\rm d}x_a\,
\int_0^1{\rm d}x_b \,
f_{a/h_1}(x_a;\mu_F)\, f_{b/h_2}(x_b;\mu_F) \nonumber\\
&&\times \sum_{n=1}^\infty \left(\frac{\alpha_s(\mu_R)}{\pi}
\right)^{n+p} R_{ab}(M, q_T, s; \mu_F),~\eea which is exactly the
difference between the full perturbative result and its asymptote
at low $q_T$, containing the terms which are at least as singular
as $q_T^{-2}$ or proportional to $\delta(q_t^2)$. At low $q_T$, we
are subtracting two terms dominated by their singularities, and
their difference is thus not significant. In the large-$q_T$
region, the asymptote does not play a significant role, and the
fixed-order theory is recovered.\\

For the resummed component, we rather work with its Fourier
transform $\tilde{W}$ with respect to the impact-parameter $b$,
the variable conjugate to $q_T$, because as stated in the previous
section, the kinematic factorization of the multiple gluon
emission is more naturally performed in impact-parameter space
\cite{Parisi:1979se, Curci:1979bg}. Using the singular part of the
partonic cross section $\hat{\sigma}_{ab}$, i.e.\ the two last
terms of Eq.\ (\ref{eq:Tabn}), and Eq.\ (\ref{eq:QCDFact}), we get
\bea \tilde{W}^F(b; M, s; \mu_R, \mu_F) =
\sum_{a,b}\sum_{n=0}^\infty
\left(\frac{\alpha_s(\mu_R)}{\pi}\right)^{n+p}\!\!\!\!
\int_0^1\!\! {\rm d}x_a \!\! \int_0^1\!\! {\rm d}x_b\,
f_{a/h_1}(x_a;\mu_F)\, f_{b/h_2}(x_b;\mu_F)\nonumber
\eea\vspace{-8mm}\bea \!\!\times\!\! \int \!\!{\rm d}^2 q_T \,
e^{-i \, \mathbf{q}_T\cdot\mathbf{b}} \bigg[
\hat{\sigma}_{ab}^{(n,\delta)} (M, s; \mu_F)\, \delta(q_T) \!+\!
\sum_{m=0}^{2n-1} \! \hat{\sigma}_{ab}^{(n,m)}(M, s; \mu_F) \,
\frac{\ln^m(M^2/q_T^2)}{q_T^2}\bigg].\label{eq:wcss}\eea

The dependence in the two partons $a$ and $b$ factorizes, and
$\tilde{W}$ obeys an evolution equation where $b$ and $M$
dependences are separated, so that Eq.\ (\ref{eq:wcss}) can be
simplified to the usual CSS resummation formula
\cite{Collins:1981uk, Collins:1981va, Collins:1984kg}, taking into
account the coherent gluon emission described in section
\ref{sec:main} \cite{Collins:1982wa}. Omitting the dependences on
the unphysical scales, we get \bea \left[\frac{{\rm
d}^2\sigma}{{\rm d}M^2{\rm d}q_T^2}\right]_{{\rm res}}& =&
\sum_{a,b} \int_0^1{\rm d}x_a \!\int_0^1{\rm d}x_b\!\int {\rm d}b
\frac{b}{2} J_0(b\,q_T) \, f_{a/h_1}(x_1;\frac{C_3}{b})\,
f_{b/h_2}(x_2;\frac{C_3}{b})\nonumber \\& &\times
\tilde{W}^F_{ab}(b; M, s),~\label{eq:css1}\\
\tilde{W}^F_{ab}(b; M, s) & = & \sum_c \! \int_0^1 \frac{{\rm
d}z_1}{z_1} \! \int_0^1 \frac{{\rm d}z_2}{z_2} \,
C^F_{ca}(\alpha_s(C_3/b), z_1) \, C^F_{\bar{c} b}(\alpha_s(C_3/b),
z_2) \, S_c^F(M,b)\nonumber \\ &&\times  \sigma_{c \bar{c}}^{({\rm
LO})F}(M)\, \delta(M^2-z_1\, z_2\, s),~ \label{eq:css2}\eea where
$J_0(x)$ is the zeroth order Bessel function of the first kind. By
convention, the constants $C_1$, $C_2$, and $C_3$ are chosen to be
$C_1 = C_3 = b_0 = 2\,e^{-\gamma_E}$ and $C_2 = 1$, where
$\gamma_E = 0.5772...$ is the Euler-Mascheroni constant.
$\tilde{W}^F$ is computed perturbatively and is the
process-dependent partonic cross section. It embodies the
all-order resummation of the large logarithms $\ln (M^2\,b^2)$
(conjugate to $\ln M^2/q_T^2$), and $\sigma_{c\bar{c}}^{({\rm
LO})F}$ is the total cross section for the LO partonic subprocess
$c \bar{c} \to F$. The resummation of the logarithms is completely
achieved by the exponentiation in the quark ($c = q$) or gluon ($c
= g$) Sudakov form factor $S_c^F$, which can be expressed as
\cite{Kodaira:1981nh, Collins:1981uk, Collins:1981va,
Kodaira:1982az}, \bea S^F_c(M,b) = \exp\left\{\! -\!
\int_{C_1^2/b^2}^{C_2^2 M^2} \!\! \frac{{\rm d}q^2}{q^2}\! \left[
A_c(\alpha_s(q); C_1) \ln\frac{C_2^2 M^2}{q^2}\! +\!
B^F_c(\alpha_s(q); C_1, C_2) \right]\! \right\},~~\label{eq:css3}
\eea depending only on the two functions $A_c$, relative to
soft-radiation, and $B_c$, relative to flavour conserving
collinear radiation, and which can be perturbatively computed\bea
A_c(\alpha_s) = \sum_{n=1}^\infty \left( \frac{\alpha_s}{\pi}
\right)^n A_c^{(n)} &{\rm ~~~~and~~~~}& B_c^F(\alpha_s) =
\sum_{n=1}^\infty \left( \frac{\alpha_s}{\pi} \right)^n
B_c^{(n)}.~\eea The coefficients of these series can be obtained
by comparing the expansion of Eq.\ (\ref{eq:css1}) at a given
order in $\alpha_s$ with the expression of the fixed-order cross
section in the small-$q_T$ limit. The lowest order coefficients,
needed for resummation at NLL accuracy, have been computed both
for the quark \cite{Kodaira:1981nh, Kodaira:1982az} and the gluon
\cite{Catani:1988vd} form factors, \bea A_q^{(1)} =C_F&{\rm
~~~~and~~~~}& A_g^{(1)} =C_A \label{eq:a1coeff} \\ A_q^{(2)} =
\frac{1}{2} C_F\, K &{\rm ~~~~and~~~~}& A_g^{(2)} = \frac{1}{2}
C_A\, K \label{eq:a2coeff}\\ B_q^{(1)} = -\frac{3}{2}C_F&{\rm
~~~~and~~~~}& B_g^{(1)} =-\frac{1}{6}\left(11 C_A - 2 n_f\right),~
\label{eq:b1coeff}\eea where $K=\left[\left(\frac{67}{18} -
\frac{\pi^2}{6} \right)C_A - \frac{5}{9} n_f\right]$, and where
$C_F=4/3$ and $C_A=3$ are the usual QCD colour factors. The
process-dependent coefficient $B_c^{(2)F}$ has been computed for
different processes, as for example Drell-Yan pair
\cite{Davies:1984hs} or Higgs production
\cite{deFlorian:2000pr}.\\

Finally, the $C_{ab}^F$ coefficient functions of Eq.\
(\ref{eq:css2}) can as well be computed perturbatively \bea
C_{ab}(\alpha_s,z) &=& \delta_{ab}\, \delta (1-z) +
\sum_{n=1}^\infty \left(\frac{\alpha_s}{\pi}\right)^n
C_{ab}^{(n)}(z),~\eea where $z=M^2/s$. They contain a collinear
contribution which has its origin in the particularities of the
$\overline{{\rm MS}}$ scheme, where the full splitting functions
are not factorized into the parton densities, and a hard
process-dependent contribution coming from the finite part of
virtual-loop corrections. The first coefficient $C^{(1)}$ is known
for a large number of processes \cite{Altarelli:1984pt,
Davies:1984hs, deFlorian:2000pr, Kauffman:1991cx,Davies:1984sp,
Balazs:1997xd, Balazs:1997hv, Balazs:1998bm}.

\subsection{Non-perturbative effects in the CSS formalism}
\label{sec:CSS_pert}

The $q_T$-distribution is affected by non-perturbative effects
associated with the large-$b$ region, $b \gtrsim 1/ \Lambda_{{\rm
QCD}}$, where the strong coupling $\alpha_s$ and the PDFs at the
scale $b_0/b$ enter the non-perturbative regime. To take these
effects into account, the function $\tilde{W}_{ab}^F$ is evaluated
at a new variable $b_\ast= b/\sqrt{1 + (b/b_{{\rm max}})^2}$, far
from the region where non-perturbative effects are relevant, the
latter being included in a function $F^{{\rm NP}}$. Practically,
the $\tilde{W}^F$ function of Eq.\ (\ref{eq:css1}) is replaced by
\bea \tilde{W}_{ab}^{{\rm NP}F}(b;M,x_1,x_2) =
\tilde{W}_{ab}^F(b_\ast;M,s)\, F_{ab}^{{\rm
NP}}(b;M,x_1,x_2).~\eea The value of $b_{{\rm max}}$ and the shape
of $F^{{\rm NP}}$ have to be chosen so that $\tilde{W}_{ab}^{{\rm
NP}}(b;M,x_1,x_2) \simeq \tilde{W}_{ab}(b;M,s)$ when $b \lesssim
b_{{\rm max}}$. In the original formalism \cite{Collins:1981uk,
Collins:1984kg} \bea F_{ab}^{{\rm NP}}(b,M,x_1,x_2) =
\exp\left[-\ln(M^2\,b_{{\rm max}}^2)\, g_1(b) - g_{a/h_1}(x_1,b) -
g_{b/h_2}(x_2,b)\right] ,~\eea where the three $g$-functions are
assumed to vanish as $b \to 0$. The predictive power of the CSS
formalism relies on the universality and the scale dependence of
these functions. As the PDFs, they can be evolved from a given
energy until the required scale. \\

The main source of non-perturbative effects comes from partons
with a non-zero intrinsic transverse-momentum already inside the
hadron and from unresolved gluons with  $q_T < 1/b_{{\rm max}}$.
Global fits of experimental Drell-Yan data allow for different
forms for this non-perturbative function \cite{Davies:1984hs,
Ladinsky:1993zn, Qiu:2000ga, Landry:2002ix, Konychev:2005iy} \bea
F_{ab}^{{\rm NP}(dws)}(b,M,x_1,x_2) \!\!\!&=&\!\!\!
\exp\!\!\left[-b^2\!\!\left(g_1 + g_2 \ln\frac{b_{{\rm
max}}M}{2}\right)\right],~~~\\ \label{eq:NP1}F_{ab}^{{\rm
NP}(ly)}(b,M,x_1,x_2)\!\!\! &=&\!\!\!
\exp\!\!\left[-b^2\!\!\left(\bar{g}_1 \!+\! \bar{g}_2
\ln\frac{b_{{\rm max}}M}{2} \right) \!-\! b\, \bar{g}_1
\,\bar{g}_3\, \ln(100\,x_1 x_2) \!\right]\!\!,~~~\\\label{eq:NP2}
F_{ab}^{{\rm NP}(blny)}(b,M,x_1,x_2) \!\!\!&=& \!\!\! \exp \!\!
\left[-b^2\!\!\left(\tilde{g}_1 + \tilde{g}_2 \ln\frac{b_{{\rm
max}}M}{2} + \tilde{g}_1\,
\tilde{g}_3 \ln(100\,x_1 x_2) \right)\! \right]\!\!,~~~\\
\label{eq:NP3} F_{ab}^{{\rm NP}(kn)}(b,M,x_1,x_2) \!\!\!&=& \!\!\!
\exp \!\!\left[-b^2\!\! \left(a_1 + a_2 \ln\frac{M}{3.2 {\rm GeV}}
+ a_3 \ln(100\, x_1\, x_2)\right)\right].~~~\eea The most recent
values of these parameters can be found in \cite{Landry:2002ix,
Konychev:2005iy}.

\subsection{Disadvantages of the CSS formalism}

Although the Collins-Soper-Sterman resummation formalism has been
used in several processes at various levels of perturbative
accuracy, it presents some disadvantages, in particular regarding
the process dependence of the various coefficients of Eqs.\
(\ref{eq:css1}, \ref{eq:css2}, \ref{eq:css3}). For instance, the
first coefficient of the $C-$functions, $C_{ab}^{(1)F}$ can be
written in terms of the one-loop matrix element of the considered
process \cite{Balazs:1997xd}. Even the Sudakov form factor,
supposed to be universal and to depend only on the quark or gluon
nature of the emitting particles, is process-dependent through the
$B_c^F$ coefficients involving also loop diagrams
\cite{deFlorian:2000pr}. Moreover, the dependence on
renormalization and factorization scales, parameterized through
the arbitrary coefficients $C_1$, $C_2$ and $C_3$ does not
correspond to the usual procedure followed in fixed-order
perturbative calculations, and the parton distribution functions
called at the scale $b_0/b$ rather than $\mu_F$ involve
extrapolation in the non-perturbative region. Eventually,
expanding the resummed expression at a given power in $\as$ in the
large-$q_T$ region leads to unwanted factorially growing
coefficients, with oscillating signs \cite{Frixione:1998dw}.\\

Some of the difficulties can be overcome by performing resummation
in $q_T$-space instead of $b$-space \cite{Ellis:1997ii,
Kulesza:1999gm}, but at the cost of non fulfilment of the
transverse-momentum conservation \cite{Parisi:1979se}. However,
all these problems can be avoided by using the recently proposed
universal resummation formalism in $b$-space \cite{Catani:2000vq,
Bozzi:2005wk, Bozzi:2003jy}.

\subsection{Universal resummation formalism} \label{sec:qtresuni}

General expressions of the first $C$-function coefficient and of
the second $B$-function coefficient show explicitly the
process-dependence of these terms \cite{deFlorian:2000pr} \bea
C^{(1)F}_{ab} (z) &=& - {\hat P}^\varepsilon_{ab}(z) +
\delta_{ab}\, \delta(1-z) \left( C_a\, \frac{\pi^2}{6}+\frac{1}{2}
{\cal A}_a(\phi) \right),~\\ B_a^{(2)F}&=&-2\, \delta P_{aa}^{(2)}
+ \beta_0 \left( \frac{2}{3} C_a \pi^2 + {\cal A}_a(\phi)
\right),~\eea where ${\hat P}^\varepsilon_{ab}(z)$ is the ${\cal
O}(\varepsilon)$-term in the one-loop Altarelli-Parisi splitting
kernel ($\varepsilon = (4-D)/2$), $\delta P_{aa}^{(2)}$ the
coefficient of the $\delta(1-z)$ term in the two-loop splitting
functions, and ${\cal A}_a(\phi)$ is the finite part of the
one-loop virtual contributions. Contrary to the universal
coefficients $A_a^{(1)}$ and $B_a^{(1)}$ determined only by the
Altarelli-Parisi splitting functions, $B_a^{(2)F}$ and
$C_{ab}^{(1)F}$ contain both collinear and hard contributions, for
which $\as$ should be evaluated at two different scales, the same
scale as the parton densities $b_0/b$ for the process-independent
collinear contribution and the hard scale $M$ for the
process-dependent hard contribution.\\

Making the replacement\bea \sigma_{c \bar{c}}^{({\rm LO})F}(M) \to
\sigma_{c \bar{c}}^{({\rm LO})F}(M, \alpha_s(M)) = \sigma_{c
\bar{c}}^{({\rm LO})F}(M)\, H_c^F(\alpha_s(M)),~\eea we include
now all the hard process-dependent contributions in $ \sigma_{c
\bar{c}}^{({\rm LO})F}$ through the function $H_c^F$, which has a
perturbative expansion \bea H_c^F(\alpha_s) = 1 +
\sum_{n=1}^\infty \left( \frac{\alpha_s}{\pi} \right)^n H_c^{(n)
F}.~\eea

Eqs.\ (\ref{eq:css2}) and (\ref{eq:css3}) can then be rewritten in
their universal form, \bea \tilde{W}^F_{ab}(b; M, s) &=& \sum_c \,
\int_0^1 \frac{{\rm d}z_1}{z_1} \, \int_0^1 \frac{{\rm d}z_2}{z_2}
\, C_{ca}(\alpha_s(b_0/b), z_1) \,
C_{\bar{c} b}(\alpha_s(b_0/b), z_2) \, S_c(M,b)\nonumber \\
\label{eq:uni2} && \times \sigma_{c \bar{c}}^{({\rm LO})F}(M)\,
\delta(M^2-z_1\, z_2\, s),~\\ \label{eq:uni3} S_c(M,b) &=&
\exp\left\{-\int_{b_0^2/b^2}^{M^2} \! \frac{{\rm d}q^2}{q^2}
\left[ A_c(\alpha_s(q)) \ln\frac{M^2}{q^2} + B_c(\alpha_s(q))
\right] \right\},~\eea where the subscript $F$ denotes the
process-dependent factors. The universal Sudakov form factor $S_c$
contains only real and virtual contributions due to soft and
flavour conserving collinear radiation at scales $M \gtrsim q_T
\gtrsim 1/b$, the $C_{ab}$ coefficients include real and virtual
contributions due to collinear radiations at very low $q_T \gtrsim
1/b$, the hard contributions produced by virtual corrections at
scales $q_T \sim M$ are embodied in $H_c^F$, and eventually, soft
radiations at very low $q_T \lesssim 1/b$ are not considered,
since real and virtual soft contributions cancel in this kinematic
range due to infrared safety. All of these coefficients depend now
only on the flavour and the colour charges of the radiating
partons.\\

The two versions of the resummation formula can be related by the
use of the renormalization-group identity \bea H_c^F(\alpha_s(M))
&=& \exp\left[\int^{M^2}_{b_0^2/b^2} \frac{{\rm d}q^2}{q^2}
h_c^F(\alpha_s(q))\right] H_c^F(\alpha_s(b_0/b)),~ \\
h_c^F(\alpha_s(M)) &=& \beta(\alpha_s) \frac{{\rm d}\ln
H_c^F(\alpha_s) }{{\rm d}\ln\alpha_s},~\eea where
$\beta(\alpha_s)$ is the QCD $\beta$-function. The relations
between the process-dependent and process-independent coefficients
are then given by \bea C_{ab}^F(\alpha_s,z) &=&
\sqrt{H_c^F(\alpha_s)}\, C_{ab}(\alpha_s,z)\\ B_c^F(\alpha_s) &=&
B_c(\alpha_s) - \beta(\alpha_s) \frac{{\rm d}\ln H_c^F(\alpha_s)
}{{\rm d}\ln\alpha_s}.~\eea This identity can be use to show that
the universal resummation formula is invariant under the
transformation \bea H_c^{F}(\alpha_s(M)) & \to &
H_c^{F}(\alpha_s(M)) \, \left[ g(\alpha_s(M)) \right]^{-1},~ \eea
where $g$ is an arbitrary perturbative function. The universal
resummation coefficients are then not unambiguously determined,
which is a consequence of the fact that the transverse-momentum
distribution is not a collinear-safe observable. This ambiguity is
similar to the one encountered in the parton distribution
functions, renormalized by fixing an arbitrary factorization
scheme. The resummation coefficients then have to be defined after
choosing a specific resummation scheme. In this work, we set
$H^{F}_q(\alpha_s) = H^{DY}_q(\alpha_s) \equiv 1$, corresponding
to the Drell-Yan resummation scheme.\\

In Eq.\ (\ref{eq:css1}), the PDFs are called at a scale embodying
a $b$-dependence, which eventually leads to an extrapolation into
the non-perturbative regime. Besides, the dependence on the
factorization scale $\mu_F$ cannot be directly computed as for the
fixed-order theory. These problems can be solved
\cite{Catani:1991kz, Catani:1992ua, Bonciani:1998vc,
Catani:1999hs} by using the scale dependence relation for parton
densities \bea f_{a/h}(x,b_0/b) = \sum_d\int_x^1\frac{{\rm
d}z}{z}\, U_{ad}(z;b_0/b,\mu_F)\, f_{d/h}(x/z,\mu_F),~\eea where
$U_{ab}$ is the evolution operator matrix obtained by solving the
DGLAP evolution equations to the required perturbative accuracy.
To avoid dealing with convolution integrals, we rather express the
resummation formula in Mellin $N$-space, taking moments with
respect to $\tau = M^2/s_h$. Eq.\ (\ref{eq:css1}) can then be
rewritten, restoring the dependence on the factorization scale, as
\bea \label{eq:uni1}\left[\frac{{\rm d}^2\sigma_N}{\td M^2\,\td
q_T^2}\right]_{{\rm res}} \!\!\!=\!\!\sum_{ab}
f_{a/h_1}(N\!\!+\!\!1; \mu_F)\, f_{b/h_2}(N\!\!+\!\!1; \mu_F)\!\!
\int_0^\infty\!\!\! \td b\, \frac{b}{2}\, J_0(b\,q_T)
\tilde{W}^F_{ab}(N, b; M, \mu_F).~~~\label{eq:resterm1}\nn \eea
\vspace{-.8cm}\bea ~ \eea  The moments of the $\tilde{W}$-function
are given by \cite{Catani:2000vq} \bea \tilde{W}^F_{ab}(N, b; M,
\mu_F) &=& \sum_c \sigma_{c \bar{c}}^{({\rm LO})F}(M)\,
H_c^F(\alpha_s(M))\,
S_c(M, b) \, \sum_{de} \Big(C_{cd}(N; \alpha_s(b_0/b))\nonumber\\
& \times& U_{da}(N; b_0/b, \mu_F)\, C_{\bar{c}e}(N;
\alpha_s(b_0/b))\, U_{eb}(N; b_0/b, \mu_F)\Big),~\eea where we
consider only a single parton species to simplify. The
generalization to the multiflavour case is treated in App.\ A of
\cite{Bozzi:2005wk}. This expression can be written in a resummed
form, where constant and logarithmic terms factorize, after
restoring all scale dependences, \bea \tilde{W}^F_{ab}(N, b; M,
\mu_F, \mu_R)\! = \! \mathcal{H}^F_{ab}\Big(N, \alpha_s(\mu_R);
\frac{M}{\mu_R},\! \frac{M}{\mu_F},\! \frac{M}{Q}\Big)\! \exp\!
\Big\{\mathcal{G}\big(N, L, \alpha_s(\mu_R); \frac{M}{\mu_R},\!
\frac{M}{Q}\big)\Big\},~ \label{eq:resterm2}\nn \eea
\vspace{-.8cm}\bea ~ \eea  with $L=\ln Q^2b^2/b_0^2$. The
resummation scale $Q$ is introduced due to the degree of
arbitrariness involved by the factorization
\cite{Dasgupta:2001eq}, since the argument of the large logarithms
can be rescaled as $\ln(M^2 b^2) = \ln(Q^2 b^2) + \ln(M^2/Q^2)$,
provided that $Q$ is independent of $b$ and $\ln(M^2/Q^2) =
\mathcal{O}(1)$ when $b\,M\gg1$. Similarly to the case of
$\mu_{R}$ and $\mu_{F}$, one should set $Q=M$ and estimate
uncalculated subleading logarithmic corrections by varying $Q$
around the central value $M$.\\

The function $\mathcal{H}^F_{ab}$ does not depend on the
impact-parameter $b$ and therefore contains all the perturbative
terms that behave as constants in the limit $b \to \infty$. Its
evaluation can be done perturbatively \bea {\cal H}_{ab}^F(N,
\alpha_s; \frac{M}{\mu_R},\! \frac{M}{\mu_F},\! \frac{M}{Q})\! =
\! \sum_c \sigma_{c \bar{c}}^{({\rm LO})F}(M) \le
\delta_{ac}\delta_{b\bar{c}} \!+\! \sum_{n=1}^{\infty}\!\!
\lr\frac{\alpha_s}{\pi}\rr^n \,{\cal H}_{ab\to c\bar{c}}^{F
(n)}(N; \frac{M}{\mu_R},\! \frac{M}{\mu_F},\!
\frac{M}{Q})\re\label{eq:H}~. \nn \eea \vspace{-.8cm}\bea ~ \eea
Let us remark that the notation $ab\to c\bar{c}$ is a compact way
of writing the full subprocess $a b \to c\bar{c} + X \to F + X$.
The first coefficients for Drell-Yan pair production, i.e.\ those
corresponding to the emission of one gluon, ${\cal H}_{q\bar{q}\to
q\bar{q}}^{DY \,(1)}$, and of one quark, ${\cal H}_{qg\to
q\bar{q}}^{DY \,(1)}$, are given by \cite{Bozzi:2005wk} \bea {\cal
H}_{q\bar{q}\to q\bar{q}}^{DY \,(1)}(N; \frac{M}{\mu_R},\!
\frac{M}{\mu_F},\! \frac{M}{Q}) &=& H_q^{DY\, (1)} - \lr B_q^{(1)}
+ \frac{1}{2}\,A_q^{(1)}\, \ln \frac{M^2}{Q^2}\rr \ln
\frac{M^2}{Q^2} \nonumber\\ && + 2\, C_{qq}^{(1)}(N) + 2\,
\gamma_{qq}^{(1)}(N)\, \ln
\frac{Q^2}{\mu_F^2},~\label{eq:H1DYqq}\\ {\cal H}_{qg\to
q\bar{q}}^{DY \,(1)}(N; \frac{M}{\mu_R},\! \frac{M}{\mu_F},\!
\frac{M}{Q}) &=& C_{qg}^{(1)}(N) + \gamma_{qg}^{(1)}(N)\, \ln
\frac{Q^2}{\mu_F^2},~\label{eq:H1DYqg} \eea with \bea
C_{qq}^{(1)}(N) =\frac{2}{3\, N\, (N+1)} + \frac{\pi^2-8}{3} {\rm
~~~~and~~~~} C_{qg}^{(1)}(N) = \frac{1}{2\, (N+1)\,
(N+2)}.~\label{eq:Ccoeff} \eea The exponent $\cal G_{N}$ includes
all the logarithmically divergent terms when $b\to\infty$ and can
be systematically expanded as \vspace{.8cm}\bea \label{eq:g} {\cal
G}(N, L; \alpha_s, \frac{M}{\mu_R},\! \frac{M}{Q}) = L\,
g^{(1)}(\frac{1}{\pi} \,\beta_0 \,\alpha_s(\mu_R^2) \, L) +
g^{(2)}(N, \frac{1}{\pi} \,\beta_0 \,\alpha_s(\mu_R^2) \, L;
\frac{M}{\mu_R},\! \frac{M}{Q})\, + \ldots ,~~~\nn \eea
\vspace{-.8cm}\bea ~ \eea  where $\beta_0 = (11\,C_A - 2\,n_f)/12$
is the first coefficient of the QCD $\beta$ function, $n_f$ being
the number of flavours. The term $L g^{(1)}$ collects the leading
logarithmic (LL) contributions, the function $g^{(2)}$ the NLL
ones, and so forth. The explicit expressions for the $g_{i}$
functions needed to perform NLL resummation are given by \bea
\label{eq:gdef1} g^{(1)}(\lambda) \!\! &=& \!\!
\frac{A^{(1)}}{\beta_0}
\frac{\lambda + \ln(1-\lambda)}{\lambda},~ \\
g^{(2)}\lr N, \lambda;\frac{M}{\mu_R},\! \frac{M}{Q}\rr\!\! &=&
\!\! \frac{B^{(1)}+2\,\gamma_{qq}^{(1)}}{\beta_0} \ln(1-\lambda) -
\frac{A^{(2)}}{\beta_0^2} \lr \frac{\lambda}{1-\lambda} + \ln(1 -
\lambda)\rr \nn \\ &+& \frac{A^{(1)}\, \beta_1}{\beta_0^3} \lr
\frac{1}{2} \ln^2(1-\lambda)+ \frac{\ln(1 - \lambda)}{1 - \lambda}
+ \frac{\lambda}{1-\lambda}\rr \nonumber \\ \label{eq:gdef2} &+&
\frac{A^{(1)}}{\beta_0} \lr \frac{\lambda}{1-\lambda} +
\ln(1-\lambda)\rr \ln\frac{Q^2}{\mu_R^2},~ \eea  where $\beta_1 =
(17\,C_A^2 - 5\,C_A\,n_f-3\,C_f\,n_f)/24$ is the second
coefficient of the QCD $\beta$ function. In the small-$b$
(large-$q_T$) region, resummation should not be applied since the
perturbation theory is reliable. A slight modification of the
expansion parameter $L$ is then introduced \cite{Bozzi:2003jy,
Catani:1992ua}, \bea L \to {\tilde L} \equiv \ln
\lr\frac{Q^2\,b^2}{b_0^2} + 1\rr,~\eea so that the logarithmic
terms are now suppressed for small $b$-values, reducing the impact
of the unjustified resummed logarithms in the small-$b$ region,
but leading to an equivalent behaviour in the large-$b$ region.

\subsection{Inverse transforms and matching procedure}
Once resummation has been achieved in $N$- and $b$-space, inverse
transforms have to be performed in order to get back to the
physical $x$- and $q_T$-space. Special attention has to be paid to
the singularities in the resummed exponent, related to the
divergent behaviour near $\lambda = 0$ and $\lambda = 1$. For
$N$-space, an inverse Mellin transform is performed following a
contour inspired by the minimal prescription \cite{Catani:1996yz}
and the principal value resummation \cite{Contopanagos:1993yq},
\bea \left[\frac{{\rm d}^2\sigma}{\td M^2\,\td q_T^2}\right]_{{\rm
res}}(\tau) = \oint_{C_N} \frac{\td N}{2\pi i}\, \tau^{-N}
\left[\frac{{\rm d}^2\sigma_N}{\td M^2\,\td q_T^2}\right]_{{\rm
res}} ,~\eea where the contour $C_N$ is chosen in such a way that
all the singularities related to the $N$-moments of the PDFs are
to the left of the integration contour in the complex $N$-plane
\bea N = C + z\,e^{\pm i\phi},~\eea with $0 \leq z \leq \infty$,
$\pi > \phi > \pi/2$ and $C > 0$. This leads to an exponentially
convergent integral. \\

For the inverse transform of $b$-space, we should first note that
the functions $g_N^{(i)}$ are singular when $b^2 = (b_0^2/Q^2\,
\exp\{\pi/(\beta_0 \as)\}$, which is related to the divergent
behaviour of the running perturbative coupling near the Landau
pole, corresponding to very large values of $b$. In Sec.\
\ref{sec:CSS_pert}, these singularities were regularized through
the variable $b_\ast$, preventing the variable $b$ to become too
large. Alternatively, we can deform the integration contour of the
$b$ integral in Eq.\ (\ref{eq:uni1}), continuing it in the complex
plane \cite{Laenen:2000de, Kulesza:2002rh}. We define then two
integration branches \bea b = (\cos\varphi \pm i \sin \varphi)
t\eea with $0 \leq t \leq \infty$, and the Bessel function is
replaced by the auxiliary functions $h_{1,2}(z,v)$ \bea h_1(z,v)
&\equiv& - {1\over\pi}\ \int_{-iv\pi}^{-\pi+iv\pi}\, \td\theta\,
{\rm e}^{-iz\, \sin\theta},~ \nonumber\\ h_2(z,v) &\equiv& -
{1\over\pi}\ \int^{-iv\pi}_{\pi+iv\pi}\, \td\theta\, {\rm e}^{-iz\,
\sin\theta}.~\label{eq:branch}\eea Their sum is always
$h_1(z,v)+h_2(z,v)=2\, J_0(z)$, but they distinguish positive and
negative phases of the $b$-contour, being then associated with
only one of the branches.\\

The parameter $C$ is chosen in such a way that all the
singularities related to the $N$-moments of the parton densities
are to the left of the integration contour. It has to lie within
the range $0 < C < \exp[\pi/(2 b_0\as)-\gamma_E]$ in order to have
convergent inverse transform integrals for any choice of $\phi$
and $\varphi$.\\

The resummed component of the cross section dominates in the
small-$q_T$ region, and the finite component defined in Eq.\
(\ref{eq:resfin}) dominates at large values of $q_T$. In the
intermediate-$q_T$ region, both components have to be consistently
matched in order to obtain uniformly accurate theoretical
predictions. To this aim, we express the finite component as the
difference between the usual fixed-order result at a specific
order {\it f.o.}\ in $\as$ and the expansion of the resummed
component at the same order \bea \left[\frac{{\rm d}^2\sigma}{{\rm
d}M^2{\rm d}q_T^2}\right]_{{\rm fin}} = \left[\frac{{\rm
d}^2\sigma}{{\rm d}M^2{\rm d}q_T^2}\right]_{{\rm f.o.}} -
\left[\frac{{\rm d}^2\sigma}{{\rm d}M^2{\rm d}q_T^2}\right]_{{\rm
res}} \Bigg|_{{\rm f.o.}}.~\label{eq:qtfin}\eea Besides we impose
the condition \bea \left[\frac{{\rm d}^2\sigma}{{\rm d}M^2{\rm
d}q_T^2}\right]_{{\rm res, l.a.}} \Bigg|_{{\rm f.o.}} =
\left[\frac{{\rm d}^2\sigma}{{\rm d}M^2{\rm d}q_T^2}\right]_{{\rm
res}} \Bigg|_{{\rm f.o.}},~\eea meaning that the expansion at a
given order {\it f.o.}\ of the resummed component evaluated at a
specific logarithmic accuracy {\it l.a.}\ equals the expansion of
the full resummed component at the same order {\it f.o.}\ in
$\as$. The complete cross section at a given logarithmic accuracy
is then given by \bea \frac{{\rm d}^2\sigma}{{\rm d}M^2{\rm
d}q_T^2} = \left[\frac{{\rm d}^2\sigma}{{\rm d}M^2{\rm
d}q_T^2}\right]_{{\rm res, l.a.}} + \left[\frac{{\rm
d}^2\sigma}{{\rm d}M^2{\rm d}q_T^2}\right]_{{\rm fin}}.~
\label{eq:qtxsec}\eea

Our matching procedure guarantees that we retain the full
information of the perturbative calculation up to the specified
order, plus the resummation of logarithmically enhanced
contributions from higher orders, without double-counting any
term. Moreover, after integration over $q_T$, it allows us to
exactly reproduce the fixed-order calculation of the total cross
section, which is not the case for the CSS-formalism due to
non-vanishing contributions of the resummed component in the
high-$q_T$ region.\\

The fixed-order truncation of the resummed component
$\left[\td^2\sigma\right]_{{\rm res}}|_{{\rm f.o.}}$ is obtained by
perturbatively expanding the $\tilde{W}$-coefficient in Eq.\
(\ref{eq:uni2}) and performing the $b$-integral of Eq.\
(\ref{eq:uni1}), \bea \left[\frac{{\rm d}^2\sigma_{ab}}{{\rm
d}M^2{\rm d}q_T^2}\right]_{{\rm exp}}\!\! &=& \!\! \sum_c
\sigma_{c{\bar c}}^{(0)}(M) \Bigg\{ \delta_{ca} \delta_{{\bar c}b}
\!+\! \sum_{n=1}^{\infty}\Bigg[ \left(\frac{\as(\mu_R)}{\pi}
\right)^n \mathcal{H}_{ab\to c\bar{c}}^{(n)}\Big(N;
\frac{M}{\mu_R}, \frac{M}{\mu_F}, \frac{M}{Q}\Big) \nn\\
\!\!&+&\!\! {\tilde \Sigma}_{ab\to c{\bar c}}^{(n)}\left(N,
\tilde{L}; \frac{M}{\mu_R}, \frac{M}{\mu_F}, \frac{M}{Q}\right)
\Bigg]\Bigg\},~\eea where $ab=q\bar{q}, qg$. Considering Drell-Yan
lepton pair production, we get for the NLL resummation matched
with the $\mathcal{O}(\as)$ fixed-order calculations
\cite{Bozzi:2005wk} \bea {\tilde \Sigma}_{ab\to c{\bar
c}}^{(1)}\left(N, \tilde{L}; \frac{M}{Q}\right) = \sum_{i=1}^2
\left[ \Sigma_{c{\bar c} \to ab}^{F \,(1;i)}\left(N;
\frac{M}{Q}\right) \,{\tilde I}_i(q_T/Q) \right],~\eea with \bea
\Sigma_{q\bar{q} \to q\bar{q}}^{F \,(1;2)}\left(N\right) = -
\,\frac{1}{2} A_q^{(1)} &{\rm ~~and~~}& \Sigma_{q\bar{q} \to
qg}^{F \,(1;2)}\left(N\right) = 0,~\eea\bea \Sigma_{q\bar{q} \to
q\bar{q}}^{F \,(1;1)}\left(N; \frac{M}{Q}\right)\! =\! - \! \left[
\left( B_q^{(1)} \!+\! A_q^{(1)} \ln\frac{M^2}{Q^2} \right) \!+\!
2\, \gamma_{qq}^{(1)} \right] {\rm and~} \Sigma_{q\bar{q} \to
qg}^{F \,(1;1)}\left(N\right)\! =\! - \gamma_{qg}^{(1)},~\eea\bea
{\tilde I}_n(q_T/Q) = Q^2 \int_0^\infty \td b \, \frac{b}{2} \,
J_0(b\, q_T) \, \ln^n\left( \frac{Q^2\,b^2}{b_0^2}+1 \right)
.~\eea

\section{Threshold resummation} \label{sec:thresh}
\subsection{Formalism}

Integrating Eq.\ (\ref{eq:QCDFact}) over $q_T$, the differential
cross section can be written as \bea\frac{{\rm d}\sigma}{{\rm
d}M^2}\! =\! \sum_{a,b} \! \int_0^1\!\! {\rm d}x_a \! \int_0^1
\!\! {\rm d}x_b\, f_{a/h_1}(x_a;\mu_F)\, f_{b/h_2}(x_b;\mu_F)\,
\hat{\sigma}_{ab}(z, M; \as(\mu_R), \mu_R,
\mu_F),~\label{eq:QCDfact2}\eea where the partonic cross section
$\hat{\sigma}_{ab}$ is expanded in powers of $\as$ \bea
\hat{\sigma}_{ab} \lr z, M; \alpha_s(\mu_R), \frac{M}{\mu_F},
\frac{M}{\mu_R}\rr &=&
\sum_{n=0}^\infty\left(\frac{\alpha_s(\mu_R)}{\pi}\right)^n
\sigma_{ab}^{(n)}\lr z,M; \frac{M}{\mu_F},
\frac{M}{\mu_R}\rr.~~~~~\eea At the $n^{{\rm th}}$ order, the
mismatch between virtual corrections and phase-space suppressed
real-gluon emission leads to the appearance of large logarithmic
terms $\alpha_s^n[\ln^{2n-1}(1-z)/(1-z)]_+$, which have to be
resummed when $s$ is close to $M^{2}$. Although these large
logarithms are manifest in the partonic cross section, they do not
generally result in divergences in the physical cross section,
contrary to the $q_T$-spectrum, because they are smoothed by the
convolution with the steeply falling parton distributions in Eq.\
(\ref{eq:QCDfact2}). Threshold resummation is then not a summation
of kinematic logarithms in the physical cross section, but rather
an attempt to quantify the effect of a well-defined set of
corrections to all orders, which can be significant even if the
hadronic threshold is far from being reached. \\

The hadronic cross section of Eq. (\ref{eq:QCDfact2}) is more
conveniently written in Mellin $N$-space \cite{Catani:1990rp},
\bea \sigma(N,M) = \sum_{ab} f_{a/h_a}(N+1, \mu_F)\,
f_{b/h_b}(N+1, \mu_F)\, \hat\sigma_{ab}(N, M;\alpha_s,
\frac{M}{\mu_R}, \frac{M}{\mu_F}).~\eea After performing the
resummation of the radiative corrections, the moments of the
partonic cross section are given by \bea \hat\sigma^{({\rm
res})}_{a b}(N, \alpha_s) = \sigma^{(LO)}\, C_{a b}(\alpha_s)\,
\exp\Big[ S(N,\alpha_s)\Big],~\label{eq:thres}\eea where the scale
dependences are suppressed for brevity and where $\sigma^{(LO)}$
represents the Born cross section. In Mellin space, the $C_{a
b}$-functions, collecting the hard contributions, i.e.\ the
$N$-independent terms in Mellin space or the terms proportional to
$\delta(1-z)$ in the physical space, can be written as a
perturbative series in the strong coupling, \bea C_{ab}(\alpha_s)
&=& \delta_{ab} + \sum_{n=1}^{\infty}\lr
\frac{\alpha_s}{\pi}\rr^{n}\, C^{(n)}_{ab}.~\eea For Drell-Yan
pair production, the first coefficients are given by \bea
C^{(1)}_{q\bar{q}} = C_F\lr\frac{2\, \pi^2}{3} - 4 +
\frac{3}{2}\ln\frac{M^2}{\mu^2_F}\rr {\rm ~~~~and~~~~} C^{(1)}_{q
g} = 0.~ \label{eq:ccoeff}\eea The universal Sudakov form factor
$S$ is given by integrals over functions of the running coupling,
\bea \label{eq:sud} S(N,\alpha_{s})=2\, \int_{0}^{1}{\rm d}z
\frac{z^{N-1} - 1}{1 - z} \int_{\mu^2_F}^{(1-z)^2\,M^2} \frac{{\rm
d}q^2}{q^2}\, A(\alpha_s(q^2)).~\eea This form is valid for
processes where the final state particles do not carry any colour
charge. The function $A$ embodies the contributions related to the
collinear emission of soft gluons by initial-state partons. It is
a series expansion in the strong coupling constant, \bea
\label{eq:acoeff} A(\alpha_s) = \sum_{n=1}^{\infty}\lr
\frac{\as}{\pi}\rr^{n}\, A_n, \eea whose coefficients are
perturbatively computable through a fixed-order calculation. In
particular, it has been proven \cite{Korchemsky:1988si} that in
the $\overline{\mathrm {MS}}$ factorization scheme the
coefficients $A_n$ are exactly equal to the large-$N$ coefficients
of the diagonal splitting function \bea \gamma_{qq}(\alpha_s)~=~
\int_{0}^{1} {\rm d}z\, z^{N-1}\, P_{qq}(z)~=~-A(\alpha_s)\,
\ln\bar{N} + \mathcal{O}(1),\label{eq:split} \eea where $\bar{N} =
N\,\exp[\gamma_E]$. Performing the integration in Eq.\
(\ref{eq:sud}) and using Eq.\ (\ref{eq:acoeff}), we obtain the
form factor up to NLL accuracy, \bea S(N, \alpha_s) =
g_1(\lambda)\, \ln \bar{N} +
g_2(\lambda).~\label{eq:sudthresh}\eea The functions $g_1$ and
$g_2$ resum the LL ($\alpha_s^n\ln^{n+1} N$) and NLL ($\alpha_s^n
\ln^n N$) contributions, respectively, and are given by
\cite{Vogt:2000ci, Moch:2005ba} \bea g_1(\lambda) &=&
\frac{A^{(1)}}{\beta_0\lambda}\, \le 2\, \lambda + (1 - 2\,
\lambda)\ln(1 - 2\, \lambda) \re,~\label{eq:tg1}\\
g_2(\lambda) &=& \frac{A^{(1)}\beta_1}{\beta_0^3}\, \le 2\,
\lambda + \ln(1 - 2\, \lambda) + \frac{1}{2}\ln^2(1 - 2\,\lambda)
\re - \frac{A^{(2)}}{\beta_0^2}\le 2\, \lambda + \ln(1 - 2\,
\lambda)\re\nonumber\\ &+& \frac{A^{(1)}}{\beta_0}\le 2\, \lambda
+ \ln(1 - 2\, \lambda)\re \ln\frac{M^2}{\mu^2_R} - \frac{2\,
A^{(1)}\, \lambda}{\beta_0} \ln\frac{M^2}
{\mu^2_F},~\label{eq:tg2} \eea where $\lambda = [\beta_0\,
\alpha_s\, \ln \bar{N}]/\pi$. Thus, the knowledge of the first two
coefficients of the function $A(\alpha_s)$ \cite{Kodaira:1981nh,
Catani:1988vd}, \bea A^{(1)} = C_F ~~{\rm and}~~ A^{(2)} =
\frac{1}{2}\, C_F\, \le C_A\lr\frac{67}{18} - \frac{\pi^2}{6}\rr -
\frac{5}{9}\, N_f\re, \eea together with the first coefficients of
the $C$-functions allows us to perform resummation up to NLL.

\subsection{Improvements of the resummation
formalism}\label{sec:thresh2}

In the limit of large $N$, the cross section is clearly dominated
by terms of $\mathcal{O}(\ln^{2} N)$, $\mathcal{O}(\ln N)$ and
$\mathcal{O}(1)$. It seems thus reasonable to neglect terms
suppressed by powers of $1/N$ in the resummation formalism.
Actually these last terms are multiplied by powers of $\ln N$ and
could as well provide a non-negligible effect in the threshold
limit. In \cite{Kramer:1996iq, Catani:2001ic} it has been shown
that these contributions are due to collinear parton emission and
can be consistently included in the resummation formula, leading
to a ``collinear-improved'' resummation formalism. The
modification simply amounts to the introduction of an
$N$-dependent term in the $C^{(1)}_{q \bar{q}}$ and $C^{(1)}_{q
g}$ coefficients. For the Drell-Yan case, Eq.\ (\ref{eq:ccoeff})
is replaced by \bea  C^{(1)}_{q \bar{q}}&\rightarrow&
\tilde{C}^{(1)}_{q \bar{q}} = C^{(1)}_{q \bar{q}} + 2\, A^{(1)}\,
\frac{\ln{\bar{N}} - \frac{1}{2} \ln{\frac{M^2}{\mu_F^2}}}{N}~,
\label{eq:c1p}
\\  \label{eq:c2p}  C^{(1)}_{q g} &\rightarrow&
\tilde{C}^{(1)}_{q g} = C^{(1)}_{qg} - T_R\,\frac{\ln{\bar{N}} -
\frac{1}{2}\ln{\frac{M^2}{\mu_F^2}}}{N}.~ \eea

Furthermore, for Drell-Yan processes and deep inelastic
scattering, the exponentiation of the contributions embodied in
the $C$-function has been proven in \cite{Eynck:2003fn}, leading
to the following modification in Eq.\ (\ref{eq:thres}): \bea
\hat\sigma^{({\rm res})}_{a b}(N,\alpha_{s}) = \sigma^{(LO)}\,
\exp\Big[C^{(1)}_{q\bar{q}}(\alpha_s)\Big] \,\exp
\Big[S(N,\alpha_{s})\Big]~.\label{eq:thr_ELM} \eea As the authors
of Ref.\ \cite{Eynck:2003fn} recognize, this exponentiation of the
$N$-independent terms is not comparable to the standard threshold
resummation in terms of predictive power. While in the latter case
a low-order calculation can be used to predict the behaviour of
full towers of logarithms, in the former case it is not possible
to directly get information on the behaviour of constant terms at,
say, $n$ loops, but a complete calculation at the $n^{{\rm th}}$
perturbative order is still necessary. Nonetheless, the comparison
of the numerical results obtained with and without the
exponentiation of the constant terms can at least provide an
estimate of the errors due to missing higher-order corrections.

\subsection{Inverse Mellin transform and matching procedure}

As for transverse-momentum resummation, once resummation has been
achieved in Mellin space, an inverse transform has to be performed
in order to get back to the physical $x$-space. The customary way
to perform this inversion, avoiding the singularities of the
$N$-moments, is the ``Minimal Prescription'' of
\cite{Catani:1996yz}, \bea \sigma=\frac{1}{2\, \pi\,
i}\int_{C_{MP} - i\, \infty}^{C_{MP} + i\, \infty} \td N\,
\tau^{-N}\, \sigma(N, M). \eea The constant $C_{MP}$ has to be
chosen so that all the poles in the integrand are to the left of
the integration contour in the complex $N$-plane except for the
Landau pole at $N = \exp[\pi/(2\, \beta_0\, \alpha_s)]$, which
should lie far to the right on the real axis.\\

Finally, a matching procedure of the NLL resummed cross section to
the NLO result has to be performed in order to keep the full
information contained in the fixed-order calculation and to avoid
possible double-counting of the logarithmic enhanced
contributions. A correct matching is achieved through \bea \sigma
= \sigma^{({\rm F.O.})} + \frac{1}{2\,  \pi\, i} \int_{C_{MP} -
i\, \infty}^{C_{MP} + i\, \infty} \td N \, \tau^{-N} \le
\sigma^{({\rm res})}(N, M) - \sigma^{({\rm exp})}(N,
M)\re,~\label{eq:mtcth} \eea where $\sigma^{({\rm F.O.})}$ is the
fixed-order perturbative result, $\sigma^ {({\rm res})}$ is the
resummed cross section, and $\sigma^{({\rm exp})}$ is the
truncation of the resummed cross section to the same perturbative
order as $\sigma^{({\rm F.O.})}$. In the Drell-Yan case, and
taking into account the improvement of Eqs.\ (\ref{eq:c1p}) and
(\ref{eq:c2p}), the expansion of the resummed partonic cross
section up to order $\alpha_s$ reads \bea \label{eq:expqq}
\hat\sigma^{({\rm exp})}_{q \bar{q}}(N,M) &=& \sigma^{(LO)} \le 1
+ \frac{\as}{\pi}\, \lr C_{F}\,\Big( 2\, \ln^2\bar{N} - 2\, \ln
\bar{N} \ln\frac{M^{2}}{\mu^{2}_{F}} \Big)
+ \tilde{C}^{(1)}_{q \bar{q}} \rr\re,~~~~~ \\
\label{eq:expqg} \hat\sigma^{({\rm exp})}_{q g}(N, M)
&=&\sigma^{(LO)}\le \frac{\as}{\pi}\, \tilde{C}^{(1)}_{q g}
\re.~\eea Let us note that in Mellin space, the fixed-order NLO
cross sections for Drell-Yan read \cite{Martin:1997rz} \bea
\hat\sigma^{({\rm F.O.})}_{q \bar{q}}(N, M)&=& \sigma^{(LO)}
\Bigg[ 1 + \frac{\alpha_{s}}{\pi}\, C_{F}\, \left( 4\, S_1^2(N) -
\frac{4}{N\,(N+1)}\, S_1(N) + \frac{2}{N^2} - 8 \right. \nonumber \\
&+& \!\frac{4\,\pi^2}{3} \!+\! \frac{2}{(N+1)^2} \left. \!+\!
\left[ \frac{2}{N\, (N+1)} \!+ \!3 \!- \!4\,  S_1(N)\right] \,\ln
\frac{M^2}{\mu_F^2} \right)\Bigg],~\\ \hat\sigma^{({\rm F.O.})}_{q
g}(N, M) &=& \sigma^{(LO)}\! \Bigg[ \!\frac{\as}{\pi}\, T_R\,
\left(\! \frac{N^4 + 11\, N^3 + 22\, N^2 + 14\,N + 4}{N^2\,
(N+1)^2 (N+2)^2}  \nonumber \right.\\ &-& 2\, \frac{N^2\! +\! N
\!+\! 2}{N\, (N\!+\!1) (N\!+\!2)}\, S_1(N) \!+\! \left.
\frac{N^2\!+\!N\!+\!2}{N\, (N\!+\!1) (N\!+\!2)}\,
\ln\frac{M^2}{\mu_F^2}\! \right)\! \Bigg]\eea with
$S_1(N)=\sum_{j=1}^N 1/j$. In the large-$N$ limit, $S_1(N) \simeq
\ln\bar{N} + 1/(2\,N)$, and we get \bea \label{eq:exaqq}
\hat\sigma^{({\rm F.O.})}_{q \bar{q}}(N,M) &=& \sigma^{(LO)} \le 1
\!+\! \frac{\alpha_{s}}{\pi}\, \lr C_{F}\,\Big( 2\, \ln^2\bar{N}
\!-\! 2\, \ln \bar{N} \ln\frac{M^{2}}{\mu^{2}_{F}}
\Big) \!+\! \tilde{C}^{(1)}_{q \bar{q}} \rr\re ,~~~\\
\label{eq:exaqg} \hat\sigma^{({\rm F.O.})}_{q g}(N, M) &=&
\sigma^{(LO)} \le \frac{\alpha_{s}}{\pi}\, \tilde{C}^{(1)}_{q g}
\re .~\eea Comparing Eqs.\ (\ref{eq:expqq}), (\ref{eq:expqg}),
(\ref{eq:exaqq}), and (\ref{eq:exaqg}), we see that the expansion
of the resummed cross section at order $\alpha_s$ correctly
reproduces the fixed-order result in the large-$N$ limit,
including even terms that are suppressed by $1/N$.

\section{Joint resummation} \label{sec:joint}

\subsection{Formalism}

Let us go back to the QCD factorization theorem of Eq.\
(\ref{eq:QCDFact}), writing the unpolarized hadronic cross section
as \bea \frac{\td\sigma}{\td M^2\, \td q^2_T} &=&
\sum_{a,b}\int_{\tau}^1\! \td x_a \int_{\tau/x_a}^1\! \td x_b\,
f_{a/h_1}(x_a,\mu_F)\, f_{b/h_2}(x_b,\mu_F)\nn\\ && \times
\frac{\td\hat{\sigma}_{ab}}{\td M^2\, \td q^2_T} (z,
M;\alpha_s(\mu_R),\frac{M}{\mu_F}, \frac{M}{\mu_R}) \eea as the
convolution of the partonic cross section $\hat{\sigma}_{ab}$ with
the universal distribution functions $f_{a,b/ h_{a,b}}$. We have
explicitly shown the lower integration limits depending on the
quantity $\tau=M^2/s_h$, which approaches the limiting value
$\tau=1$ when the process is close to the hadronic threshold. In
Mellin $N$-space, the hadronic cross section naturally factorizes
\bea \frac{\td\sigma}{\td M^2\, \td q^2_T} = \sum_{ab} \oint_{\cal
C}\! \frac{\td N}{2 i \pi} \tau^{-N}\! f_{a/h_1}(N+1, \mu_F)
f_{b/h_2}(N+1, \mu_F) \hat\sigma_{ab}(N; \alpha_s(\mu_R),
\frac{M}{\mu_F}, \frac{M}{\mu_R}),~~~\nn \eea \vspace{-.8cm}\bea ~
\eea  where the contour ${\cal C}$ in the complex $N-$space will
be defined later and the $N$-moments of the various quantities are
defined according to the Mellin transform \bea F(N) = \int_0^1 \td
x\, x^{N-1}\, F(x),~\eea for $x=x_{a,b},z,\tau$ and
$F=f_{a/h_a,b/h_b}, \hat{\sigma}, \sigma$. The jointly resummed
cross section in $N$-space is usually written, at the NLL
accuracy, as \cite{Laenen:2000ij, Kulesza:2002rh, Kulesza:2003wn}
\bea \hat\sigma^{{\rm (res)}}(N) &=& \sum_c \sigma_{c
\bar{c}}^{(0)}(M)\, H_c(\as(\mu_R)) \int \frac{\td^2 {\bf b}}{4\,
\pi} e^{i {\bf b} \cdot {\bf q}_T}\,
{\mathcal C}_{c/h_1}(M, b, N; \alpha_s(\mu_R), \mu_F) \nn\\
&\times& \exp\left[E_c^{\rm (PT)}(N,b;\alpha_s(\mu_R),
\frac{M}{\mu_R})\right]\, {\mathcal C}_{\bar{c}/h_2}(M, b, N;
\alpha_s(\mu_R), \mu_F),~\label{eq:ksv}\eea where the parton
densities and the sums over the different possible initial
partonic states are included in the definition of the
$\mathcal{C}$-coefficients. The exponent is given by \bea E_c^{\rm
(PT)}(N,b; \alpha_s(\mu_R), \frac{M}{\mu_R}) = -
\int_{M^2/\chi^2}^{M^2} {\td \mu^2 \over \mu^2} \, \le
A_c(\as(\mu)) \ln \frac{M^2}{\mu^2} + B_c(\as(\mu))
\re.~\label{eq:expojo}\eea The function $\chi$ organizes the
logarithms of $N$ and $b$ in joint resummation and can be defined
as \bea \chi(\nbar, \bbar)=\bbar + \frac{\nbar}{1+\eta\,\bbar/
\nbar},~ \label{eq:chi} \eea where $\bbar \equiv b\, M\,
e^{\gamma_E}/2$, $\nbar \equiv N e^{\gamma_E}$ and $\eta = 1/4$.
The only requirement to be satisfied when introducing a particular
form of $\chi$ is that the leading- and next-to-leading logarithms
in $\nbar$ and $\bbar$ are correctly reproduced in the limits
$\bar{N}\to \infty$ or $\bar{b}\to\infty$, respectively. The
choice of Eq.\ (\ref{eq:chi}) avoids the introduction of sizeable
subleading terms into perturbative expansions of the resummed
cross section at a given order in $\as$ which are not present in
fixed-order calculations \cite{Kulesza:2002rh}. As for
$q_T$-resummation, the $A$-term resums the soft radiation, while
the $B$-term accounts for the difference between the eikonal
approximation and the full partonic cross section in the threshold
region, i.e. the flavour-conserving collinear contributions. In
the large-$N$ limit, these coefficients are directly connected to
the leading terms in the one-loop diagonal anomalous dimension,
calculated in the $\overline{{\rm MS}}$ factorization scheme (see
Eq.\ \ref{eq:split}) \cite{Korchemsky:1988si} \bea
\gamma_{cc}(N,\as) = - A_c(\as) \ln \nbar - \frac{B_c(\as)}{2} +
\mathcal{O}(1/N),~ \eea and can be expressed as perturbative
series in $\as$ \bea A_c(\alpha_s) = \sum_{n=1}^{\infty}\lr
\frac{\alpha_{s}}{\pi}\rr^{n}\, A^{(n)}_c,~\\
B_c(\alpha_s) = \sum_{n=1}^{\infty}\lr
\frac{\alpha_{s}}{\pi}\rr^{n}\, B_c^{(n)}.~\eea

Performing the integration in Eq.~(\ref{eq:expojo}), we obtain the
form factor up to NLL, \bea E_c^{\rm (PT)}(N,b; \alpha_s(\mu_R),
\frac{M}{\mu_R}) = g_c^{(1)}(\lambda)\,\ln\chi +
g_c^{(2)}(\lambda; \frac{M}{\mu_R}) \label{eq:ept}\eea with
$\lambda = \beta_0/\pi \as(\mu_R) \ln\chi$, and the $g$-functions
are given by \bea g_c^{(1)}(\lambda)&=& \frac{A_c^{(1)}}{\beta_0}
\frac{2\, \lambda +
\ln\big(1 - 2\, \lambda\big)}{\lambda},~\nn\\
g_c^{(2)}(\lambda;\frac{M}{\mu_R}) &=& \frac{A_c^{(1)}\,
\beta_1}{\beta_0^3} \left[ \frac{1}{2} \ln^2 \big(1 - 2\, \lambda
\big) + \frac{2\, \lambda + \ln\big(1 - 2\, \lambda \big)}{1 - 2\,
\lambda} \right] + \frac{B_c^{(1)}}{\beta_0}\, \ln \big(1 - 2\,
\lambda \big)\nn\\ &+& \left[ \frac{A_c^{(1)}}{\beta_0}
\ln\frac{M^2}{\mu_R^2} - \frac{A_c^{(2)}}{\beta_0^2} \right]
\left[ \frac{2\, \lambda}{1 - 2\, \lambda}+ \ln\big(1 - 2\,
\lambda \big) \right].~\label{eq:jointg1}\eea

The $\mathcal{C}$-coefficients are chosen to correspond to
transverse-momentum resummation for $b\to\infty$, $N$ being fixed,
and are defined by \bea {\mathcal C}_{d/H}(M, b, N;
\alpha_s(\mu_R), \mu_F) \!=\! \sum_{a,a_1} C_{d/a_1}(N;
\alpha_s(\frac{M}{\chi})) U_{a_1a}(N; \frac{M}{\chi}, \mu_F)
f_{a/H}(N\!\!+\!\!1, \mu_F).~~~~\label{eq:ccal}\nn \eea
\vspace{-.8cm}\bea ~ \eea The $C$-functions contain contributions
to the singular behaviour at vanishing $q_{T}$ (but not at
threshold) of the fixed order cross section, while $U_{jk}$
represents the evolution of the parton densities from scale
$\mu_F$ to scale $M/\chi$ and the $H$-function of Eq.\
(\ref{eq:ksv}) absorbs hard virtual contributions. The $C$- and
$H$-functions can be expressed perturbatively in powers of
$\alpha_{s}$, \bea H_c(\alpha_s) = 1 + \sum_{n=1}^{\infty}\lr
\frac{\alpha_{s}}{\pi}\rr^{n}\, H_c^{(n)},~\\ C_{a/b}(\alpha_s) =
\delta_{ab} + \sum_{n=1}^{\infty}\lr
\frac{\alpha_{s}}{\pi}\rr^{n}\, C_{a/b}^{(n)}.~\eea

\subsection{Reorganization of the resummed cross section}
In order to explicitly factorize the dependence on the parameter
$\chi$, it is possible to organize the resummation of the
logarithms in analogy to the case of transverse-momentum
resummation \cite{Catani:2000vq, Bozzi:2005wk}. The hadronic
resummed cross section can be written as \bea
\frac{\td\sigma^{({\rm res})}}{\td M^2\, \td q^2_T} &=& \sum_{a,b}
\oint_{\cal C}\, \frac{\td N}{2\, \pi\, i}\,
\tau^{-N}\,f_{a/h_a}(N+1, \mu_F)\, f_{b/h_b}(N+1, \mu_F)
\int_{0}^{\infty} \frac{b\, \td b}{2} \, J_0(b\, q_T)\nn\\ &\times&
\sum_c\! \mathcal{H}_{ab\to c\bar{c}}\Big(N; \alpha_s(\mu_R),
\frac{M}{\mu_R}, \frac{M}{\mu_F}\Big) \exp\Big\{
\mathcal{G}_c(\ln\chi; \alpha_s(\mu_R), \frac{M}{\mu_R})\Big\}.~~~
\label{eq:joint} \eea The function $\mathcal{H}_{ab\to c\bar{c}}$
does not depend on the parameter $\chi$, and contains all the
terms that behave as constant in the limits $b\to\infty$ or
$N\to\infty$, \bea \mathcal{H}_{ab\to c\bar{c}}\Big(N;
\alpha_s(\mu_R), \frac{M}{\mu_R}, \frac{M}{\mu_F}\Big) \!=\!
\sigma_{c \bar{c}}^{(0)}(M)\! \left[\delta_{ca} \delta_{{\bar c}b}
\!+\! \sum_{n=1}^{\infty}\lr \frac{\alpha_{s}}{\pi}\rr^{n}\!
\mathcal{H}_{ab\to c\bar{c}}^{(n)}\Big(N; \frac{M}{\mu_R},
\frac{M}{\mu_F}\Big)\!\right]\!.\nn \eea \vspace{-.8cm}\bea ~ \eea
Using Eqs.\ (\ref{eq:ksv}), (\ref{eq:ccal}) and the QCD evolution
operator $U_{ab}(N; \mu,\mu_0)$, fulfilling the evolution equation
\bea \frac{\td U_{ab}(N; \mu, \mu_0)}{\td\ln\mu^2} = \sum_c
\gamma_{ac}(\as(\mu)) \, U_{cb}(N; \mu, \mu_0),~\eea which has the
solution \bea U_{ab}(N;\mu,\mu_0) = \exp \left[ \int_{\mu_0}^\mu
\frac{\td q^2}{q^2} \gamma(\as(q)) \right],~\eea we get the first
order coefficient \bea \mathcal{H}_{ab\to c\bar{c}}^{(1)}\Big(N;
\frac{M}{\mu_F}\Big)\! = \! \delta_{ca} \delta_{{\bar c}b}\,
H_c^{(1)} \!+\! \delta_{ca}\, C_{{\bar c}b}^{(1)}(N) \!+\!
\delta_{{\bar c}b}\, C_{ca}^{(1)}(N) \!+\! \left( \delta_{ca}
\gamma_{{\bar c}b}^{(1)} \!+\! \delta_{{\bar c}b}
\gamma_{ca}^{(1)} \right) \ln\frac{M^2}{\mu_F^2}.~~~\nn \eea
\vspace{-.8cm}\bea ~ \eea The $\chi$-dependence appearing in the
$C$-coefficients and in the evolution operator of Eq.\
(\ref{eq:ccal}) is factorized and included in the exponent
$\mathcal{G}_c$, which is equal to the exponent $E_c^{\rm (PT)}$
defined in Eq.\ (\ref{eq:expo}), provided we make the replacement
\bea B_c(\as) \to {\tilde B}_c(\as) = B(\as) + 2 \beta(\as)\,
\frac{\td \ln C(N; \as)}{\td \ln\as} + 2\, \gamma(\as),\eea At NLL
accuracy, Eq.\ (\ref{eq:ept}) remains almost unchanged, since only
the coefficient $g_c^{(2)}$ of Eq.\ (\ref{eq:jointg1}) has to be
slightly modified, through the replacement \bea B_c^{(1)} \to
{\tilde B}_c^{(1)}(N) = B_c^{(1)} + 2 \gamma_{cc}^{(1)}.~\eea

Although the first-order coefficients $C_{ab}^{(1)}$ and
$H_{c}^{(1)}$ are in principle resummation-scheme dependent
\cite{Catani:2000vq}, this dependence cancels in the perturbative
expression of $\mathcal{H}_{ab\to c\bar{c}}$ \cite{Bozzi:2005wk}.
In the numerical code we developed for slepton pair production, we
chose to implement the Drell-Yan resummation scheme and take
$H_q(\as)\equiv 1$, the $C$-coefficients being the same as for
$q_T$-resummation (see Eq.\ (\ref{eq:Ccoeff})) \bea
C_{qq}^{(1)}(N) = \frac{2}{3\, N\, (N+1)} +
\frac{\pi^2-8}{3}~~{\rm and}~~ C_{qg}^{(1)}(N) = \frac{1}{2\,
(N+1)\, (N+2)}.~\eea

\subsection{Inverse transform and matching}
Once resummation has been achieved in $N$- and $b$-space, inverse
transforms have to be performed in order to get back to the
physical spaces. Again, special attention has to be paid to the
singularities in the resummed exponent, related to the divergent
behaviour near $\chi = \exp[\pi/(2 \beta_0\as)]$, i.e. the Landau
pole of the running strong coupling, and near $\bbar = -2\nbar$
and $\bbar = -4\nbar$, where $\chi=0$ and infinity respectively.
The integration contours of the inverse transforms in Mellin and
impact-parameter space must therefore avoid all of these poles.\\

As for $q_T$-resummation, the $b-$integration is performed by
deforming the integration contour with a diversion into the
complex $b$-space \cite{Laenen:2000de}, defining two integration
branches \bea b = (\cos\varphi \pm i \sin \varphi) t, ~~~~ {\rm
with} ~~ 0 \leq t \leq \infty,~ \eea under the condition that the
integrand decreases sufficiently rapidly for large $|b|$-values.
The Bessel function $J_0$ is replaced by two auxiliary functions
$h_{1,2}(z,v)$ related to the Hankel functions (see Eq.\
(\ref{eq:branch})) \bea h_1(z,v) &\equiv& - {1\over\pi}\
\int_{-iv\pi}^{-\pi+iv\pi}\, \td\theta\, {\rm e}^{-iz\,
\sin\theta},~ \nonumber\\ h_2(z,v) &\equiv& - {1\over\pi}\
\int^{-iv\pi}_{\pi+iv\pi}\, \td\theta\, {\rm e}^{-iz\,
\sin\theta}.~\eea Their sum is always $h_1(z,v)+h_2(z,v)=2\,
J_0(z)$, but they distinguish positive and negative phases of the
$b$-contour, being then associated with only one of the two
branches.\\

The inverse Mellin transform is performed following a contour
inspired by the Minimal Prescription \cite{Catani:1996yz} and the
Principal Value Resummation \cite{Contopanagos:1993yq}, where one
again defines two branches \bea N = C + z\,e^{\pm i\phi}~~ {\rm
with} ~~ 0 \leq z \leq \infty, \pi > \phi > \frac{\pi}{2}.~\eea
The parameter $C$ is chosen in such a way that all the
singularities related to the $N$-moments of the parton densities
are to the left of the integration contour. It has to lie within
the range $0 < C < \exp[\pi/(2 \beta_0\as)-\gamma_E]$ in order to
have convergent inverse transform integrals for any choice of
$\phi$ and $\varphi$.\\

Finally, a matching procedure of the NLL resummed cross section to
the NLO result has to be performed in order to keep the full
information contained in the fixed-order calculation and to avoid
possible double-counting of the logarithmically enhanced
contributions. A correct matching is achieved through the formula
\bea \frac{\td^2\sigma}{\td M^2\,\td q_T^2}(\tau) &=&
\frac{\td^2\sigma^{({\rm F.O.})}}{\td M^2\,\td q_T^2}(\tau; \as) +
\oint_{C_N} \frac{\td N}{2\pi i}\, \tau^{-N} \int \frac{b \td b}{2}
J_0(q_T\,b)\nn\\ &&\times \left[\frac{\td^2\sigma^{{\rm (res)}}}{\td
M^2\,\td q_T^2}(N, b; \as) - \frac{{\rm d}^2\sigma^{{\rm
(exp)}}}{\td M^2\,\td q_T^2}(N, b; \as) \right],~ \label{eq:mtchjt}
\eea where $\td^2\sigma^{({\rm F.O.})}$ is the fixed-order
perturbative result at a given order in $\as$, $\td^2\sigma^ {({\rm
res})}$ is the resummed cross section at a given logarithmic
accuracy, $\td^2\sigma^{({\rm exp})}$ is the truncation of the
resummed cross section to the same perturbative order as
$\td^2\sigma^{({\rm F.O.})}$, and where
we have removed the scale dependences for brevity. \\

The expansion of the resummed results reads \bea
\frac{\td\sigma^{({\rm exp})}}{\td M^2\, \td q^2_T}(N,b; \as(\mu_R),
\frac{M}{\mu_R}, \frac{M}{\mu_F}) &=& \sum_{a,b} f_{a/h_a}(N+1,
\mu_F)\, f_{b/h_b}(N+1, \mu_F)\, \nn\\ &&\times \sigma^{{\rm
(exp)}}_{ab}(N,b; \alpha_s(\mu_R), \frac{M}{\mu_R},
\frac{M}{\mu_F}),~\eea where $\sigma_{ab}^{{\rm (exp)}}$ is
obtained by perturbatively expanding the resummed component \bea
\sigma^{{\rm (exp)}}_{ab}(N,b; \alpha_s(\mu_R), \frac{M}{\mu_R},
\frac{M}{\mu_F}) = \sum_c \sigma_{c{\bar c}}^{(0)}(M) \Bigg\{
\delta_{ca} \delta_{{\bar c}b} +
\sum_{n=1}^{\infty}\left(\frac{\as(\mu_R)}{\pi} \right)^n\nn \\
\hspace{2cm} \times \Bigg[{\tilde \Sigma}_{ab\to c{\bar
c}}^{(n)}\left(N, \ln\chi; \frac{M}{\mu_R},
\frac{M}{\mu_F}\right)\!+\! \mathcal{H}_{ab\to
c\bar{c}}^{(n)}\Big(N; \frac{M}{\mu_R},
\frac{M}{\mu_F}\Big)\Bigg]\Bigg\}.~\label{eq:exp_joint}\eea The
perturbative coefficient ${\tilde \Sigma}^{(n)}$ is a polynomial
of degree $2n$ in $\ln\chi$ and $\mathcal{H}^{(n)}$ embodies the
constant part of the resummed cross section in the limit of
$b\to\infty$ or $N\to\infty$. The first order coefficient ${\tilde
\Sigma}^{(1)}$ is given by \bea {\tilde \Sigma}_{ab\to c{\bar
c}}^{(1)}\left(N,\ln\chi\right) = {\tilde \Sigma}_{ab\to c{\bar
c}}^{(1;2)} \ln^2\chi + {\tilde \Sigma}_{ab\to c{\bar c}}^{(1;1)}
\ln\chi,~\eea with \bea {\tilde \Sigma}_{ab\to c{\bar c}}^{(1;2)}
= - 2\, A^{(1)}_c \delta_{ca} \delta_{{\bar c}b} ~~~~{\rm and}~~~~
{\tilde \Sigma}_{ab\to c{\bar c}}^{(1;1)} = -2\, \big(B^{(1)}_c
\delta_{ca} \delta_{{\bar c}b} + \delta_{ca} \gamma_{{\bar
c}b}^{(1)} + \delta_{{\bar c}b} \gamma_{ca}^{(1)}\Big).~\eea The
integrals \bea I_n(q_T, N) = \int_0^\infty \frac{b\td b}{2}
J_0(b\,q_T)\ln^n\left(\bbar + \frac{\nbar}{1+\eta\,\bbar/
\nbar}\right),~\eea appearing when performing the inverse
$b-$transform, must be computed numerically.

\section{Comparison between $q_T$, threshold and joint
resummations}

In this section, we summarize the main differences between the
three resummation formalisms described in the previous sections.
For all resummations, the Sudakov form factor at the NLL accuracy
can be written as \bea \mathcal{G}(N,L) = g^{(1)}(\lambda)\,L +
g^{(2)}(\lambda),~ \eea where we have removed all scale
dependences for brevity. The logarithm $L$, $\lambda$ and the two
$g$-functions are given in Tab.\ \ref{tab:5}.\\

\begin{table} \centering
\renewcommand{\arraystretch}{2.}
\small \begin{tabular}{|c||l|l|l|}\hline & $q_T$ & Joint & Threshold \\
\hline $L = \ln(...)$ & $1+\frac{M^2\,b^2}{b_0^2}$ &
$\frac{M\,b}{b_0} + \frac{\nbar}{1+\frac{M\,b}{4\,b_0\,\nbar}}$ &
$\nbar$\\\hline $\lambda$ & $\frac{\beta_0}{\pi} \as(\mu_R) L$ &
$2 \frac{\beta_0}{\pi} \as(\mu_R) L$ & $\frac{\beta_0}{\pi}
\as(\mu_R) L$\\ \hline $g^{(1)}(\lambda)$ &
$\frac{A^{(1)}}{\beta_0} \frac{\lambda + \ln(1-\lambda)}{\lambda}$
&  $\frac{2\, A^{(1)}}{\beta_0} \frac{\lambda + \ln\big(1 -
\lambda\big)}{\lambda}$ & $\frac{A^{(1)}}{\beta_0\lambda}\, \le
2\, \lambda + (1 - 2\, \lambda)\ln(1 - 2\, \lambda) \re$ \\ \hline
\multirow{2}{*}{ \centering $g^{(2)}(\lambda)$} &
\multicolumn{2}{l|}{ $\frac{B^{(1)}+2\,\gamma_{qq}^{(1)}}{\beta_0}
\ln(1-\lambda)$} & $\frac{A^{(1)}\beta_1}{\beta_0^3}\! \le 2\,
\lambda \!+\! \ln(1 \!-\! 2\, \lambda) \!+\! \frac{1}{2}\ln^2(1
\!-\! 2\,\lambda) \re$ \\ & \multicolumn{2}{l|}{$+\! \lr
\frac{A^{(1)}}{\beta_0}\ln\frac{M^2}{\mu_R^2} \!-\!
\frac{A^{(2)}}{\beta_0^2}\rr\! \!\lr \frac{\lambda}{1\!-\!\lambda}
\!+\! \ln(1 \!-\! \lambda)\rr$} & $ - \frac{A^{(2)}}{\beta_0^2}\le
2\, \lambda \!+\! \ln(1 \!-\! 2\, \lambda)\re \!-\!
\frac{2\,A^{(1)} \lambda}{\beta_0} \ln\frac{M^2}{\mu^2_F} $\\ &
\multicolumn{2}{l|}{$+ \frac{A^{(1)}\, \beta_1}{\beta_0^3} \lr
\frac{1}{2} \ln^2(1-\lambda)+ \frac{\lambda + \ln(1 - \lambda)}{1
- \lambda} \rr$} & $+\frac{A^{(1)}}{\beta_0}\le 2\, \lambda +
\ln(1 - 2\, \lambda)\re \ln\frac{M^2}{\mu^2_R}$\\ \hline
\end{tabular}\caption{\label{tab:5}Comparison of the exponent in the
(universal) transverse-momentum, joint, and threshold resummation
formalisms.}
\end{table}

Let us first note that $M$ corresponds to the final state
invariant-mass in the joint resummation formalism, but to the
resummation scale in the $q_T$-resummation formalism (see Eq.\
(\ref{eq:resterm2})). Such a scale is not needed in joint
resummation, since the logarithm cannot be rescaled with a factor
that is $N$- and $b$-independent. Finally, $b_0$ is defined as for
$q_T$-resummation, $b_0 = 2e^{-\gamma_E}$.\\

For all three resummations, the exponent is taken to be universal
and process-independent, since the enhanced logarithmic
contributions have the same dynamical origin, i.e.\ the soft-gluon
emission from the initial state, which cannot depend on the
hard-scattering process. All the process-dependence is factorized
outside of the exponent, i.e.\ in the $\mathcal{H}$-coefficient
for joint and transverse-momentum resummation and in the
$C$-coefficient for the threshold resummation. In the conjugate
Mellin- and $b$-spaces, the resummed partonic cross section can
then be written as \bea \sigma^{{\rm (qt/j)}}_{ab}(N, L) &=&
\mathcal{H}^F_{ab}(N)
\exp \Big\{\mathcal{G}(N, L) + F_{ab}^{{\rm NP}}\Big\},~\\
\sigma^{{\rm (th)}}_{ab}(N, L) &=& \sigma^{(LO)}\, \tilde{C}_{a
b}(\alpha_s)\, \exp\Big\{\mathcal{G}(N, L)\Big\}.~\eea
$F_{ab}^{{\rm NP}}$ is the non-perturbative form factor
parameterizing the non-perturbative effects relevant for
soft-gluon emission with very small $q_T$. The expressions for all
the coefficients can be found in the previous sections. \\

As a side remark, let us note that the `$+1$'-term introduced in
the logarithm of the transverse-momentum resummation in order to
minimize the impact of the resummation in the small-$b$ region
(where the perturbative theory is reliable) is not necessary for
joint resummation, since for small $b$ the corresponding
logarithm is not tending towards zero.\\

\newpage $~$\\ \newpage

\chapter{Slepton-pair production at hadron colliders}
\label{ch:slep}

\begin{figure}
 \centering
 \includegraphics[width=.33\columnwidth]{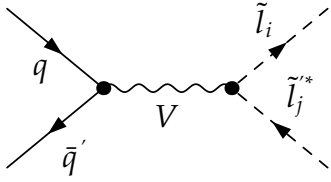}
 \caption{\label{fig:3}Feynman diagram for slepton pair
 ($V=\gamma, Z^0$) and slepton-sneutrino associated ($V=W^\mp$)
 production at leading order in perturbative QCD.}
\end{figure}

The LO cross section for the production of non-mixing slepton
pairs (see Fig.\ \ref{fig:3}) has been calculated in
\cite{Dawson:1983fw, delAguila:1990yw, Baer:1993ew,
Chiappetta:1985ku}. We generalize these results, including the
mixing effects relevant for third generation sleptons, and we
calculate single- and double-spin asymmetries, for both neutral
($\gamma$, $Z^0$) and charged ($W^\pm$) currents
\cite{Bozzi:2004qq}, making numerical predictions for the RHIC
collider  and for possible upgrades of the Tevatron
\cite{Baiod:1995eu} and the LHC \cite{deroeck:priv} as well.\\

Sleptons being among the lightest supersymmetric particles in many
SUSY-breaking scenarios \cite{Aguilar-Saavedra:2005pw}, they often
decay into the corresponding SM partner and the lightest stable
SUSY particle, a possible signal for slepton pair production at
hadron colliders consisting thus in a highly energetic lepton pair
and associated missing energy. The corresponding SM background is
mainly due to $WW$ and $t \bar{t}$ production \cite{Lytken:22,
Andreev:2004qq}. An accurate calculation of the
transverse-momentum spectrum allows us to use the Cambridge
(s)transverse mass to measure the slepton mass
\cite{Lester:1999tx} and spin \cite{Barr:2005dz} and to
distinguish then the signal from the background. When studying the
transverse-momentum distribution of a slepton pair produced with
an invariant-mass $M$ in a hadronic collision, it is appropriate
to separate the large-$q_T$ and small-$q_{T}$ regions. In the
large-$q_T$ region ($q_T\geq M$), the use of fixed-order
perturbation theory is fully justified, since the perturbative
series is controlled by a small expansion parameter,
$\alpha_s(M^2)$, but in the small-$q_{T}$ region, where the
coefficients of the perturbative expansion in $\alpha_s(M^{2})$
are enhanced by powers of large logarithmic terms,
$\ln(M^{2}/q_{T}^{2})$, results based on fixed-order calculations
are completely spoiled. However, a precise $q_T$-spectrum can be
obtained by systematically resumming these logarithms to all
orders in $\alpha_s$, using the universal formalism described in
Sec.\ \ref{sec:qtresuni}.\\

The NLO QCD corrections have been calculated in
\cite{Baer:1997nh}, and the full SUSY-QCD corrections have been
added in \cite{Beenakker:1999xh}, considering however only massive
non-mixing squarks and gluinos in the loops. We extend this last
work by including the mixing effects relative to the squarks
appearing in the loops \cite{Bozzi:2007qr}. We also consider the
threshold-enhanced contributions, due to soft-gluon emission from
the initial state, which arise when the initial partons have just
enough energy to produce the slepton pair in the final state. In
this case, the mismatch between virtual corrections and
phase-space suppressed real-gluon emission leads to the appearance
of large logarithmic terms $\alpha_s^n[\ln^{2n-1}(1-z)/(1-z)]_+$
at the $n^{{\rm th}}$ order of perturbation theory. When $s$ is
close to $M^2$, the large logarithms have to be resummed, which is
achieved through the exponentiation of the soft-gluon radiation,
within the collinear-improved formalism described in Sec.\
\ref{sec:thresh}.\\

Since a complete understanding of the soft-gluon effects in
differential distributions requires a study of the relation
between the recoil corrections at small $q_T$ and the
threshold-enhanced contributions, we finally present a joint
treatment of these, within the method described in Sec.\
\ref{sec:joint}.\\

%%%%%%%%%%%%%% Begin Figure III %%%%%%%%%%%%%%%%%%%%%%%%%%%%%%%%%%%
%%%%%%%%%%%%%% End of Figure III %%%%%%%%%%%%%%%%%%%%%%%%%%%%%%%%%%

\section{LO unpolarized cross section and spin asymmetries}
\label{sec:slLO}

\subsection{Analytical results}

In the following, we define the square of the weak coupling
constant $g^2=e^2/\sin^2\theta_W$ in terms of the electromagnetic
fine structure constant $\alpha=e^2/(4\pi)$ and the squared sine
of the electroweak mixing angle $x_W=\sin^2\theta_W$. The coupling
strengths of left- and right-handed (s)leptons to the neutral
electroweak current given in Sec.\ \ref{sec:coupling} are \bea \{
L_{f f^\prime Z}, R_{f f^\prime Z} \} &=& (2\,T^{3}_f -
2\,e_f\,x_W) \times \delta_{f f^\prime},~\\ \{ L_{\tilde{l}_i
\tilde{l}_j^\prime Z}, R_{\tilde{l}_i \tilde{l}_j^\prime Z} \} &=&
\{ L_{l l^\prime Z}\, S^{\tilde{l}}_{j1}\,
S^{\tilde{l}^\prime\ast}_{i1}, R_{l l^\prime Z}\,
S^{\tilde{l}}_{j2}\, S_{i2}^{\tilde{l}^\prime\ast} \},~\\
\{L_{q q^\prime W}, R_{q q^\prime W}\} &=& \{\sqrt{2}\,c_W\, V_{q
q^\prime}, 0\},~\\ \{L_{\tilde{l}_i \tilde{\nu}_l W},
R_{\tilde{l}_i \tilde{\nu}_l W}\}
&=&\{\sqrt{2}\,c_W\,S^{\tilde{l}\ast}_{i1} ,\, 0\}.~\eea We
express (un)polarized cross sections for the electroweak $2\to2$
scattering processes \bea\begin{array} {l}q\bar{q} \to
\gamma, Z^0 \to \tilde{l}_i\,\tilde{l}^\ast_j,\\
q\bar{q}^\prime \to W^\mp \to \tilde{l}_i\, \tilde{\nu}_l^\ast,\,
\tilde{l}^\ast_i\, \tilde{\nu}_l, \end{array}\eea in terms of the
conventional Mandelstam variables, \bea s=(p_a+p_b)^2 &~~,~~&
t=(p_a-p_1)^2 \mbox{~~,~~and~~} u=(p_a-p_2)^2 \eea and the masses
of the neutral
and charged electroweak gauge bosons $m_Z$ and $m_W$.\\

The neutral current differential cross section for the production
of mixing slepton pairs in collisions of quarks with definite
helicities $h_{a,b}$ is given by \cite{Bozzi:2004qq} \bea
{\td\hat{\sigma}_{h_a,h_b}\over\td t} &=& {2 \pi \alpha^2 \over 3
s^2} \le {u t - m_{\tilde{l}_i}^2 m_{\tilde{l}_j}^2 \over
s^2} \re \Bigg[e_q^2\, e_l^2\, \delta_{ij} (1 - h_a h_b) \nn \\
&+& {e_q\, e_l\, \delta_{ij}\, {\rm Re}\left[L_{\tilde{l}_i
\tilde{l}_j Z} \!+\! R_{\tilde{l}_i \tilde{l}_j Z}\right]
\bigg[(1\!-\!h_a) (1\!+\!h_b)\, L_{q q Z} \!+\! (1\!+\!h_a)
(1\!-\!h_b)\, R_{q q Z}\bigg] \over 4\, x_W (1 \!-\! x_W)
(1\!-\!m_Z^2/s)} \nn
\\  &+& {\left|L_{\tilde{l}_i \tilde{l}_j Z} \!+\! R_{\tilde{l}_i
\tilde{l}_j Z}\right|^2 \bigg[(1\!-\!h_a) (1\!+\!h_b)\, L_{q q
Z}^2 \!+\! (1\!+\!h_a) (1\!-\!h_b)\, R_{q q Z}^2\bigg] \over 32\,
x_W^2\, (1 \!-\! x_W)^2 (1\!-\!m_Z^2/s)^2}
\Bigg],~\label{eq:slep_polZ}\eea the three terms representing the
squared photon-contribution, the photon-$Z^0$ interference and the
squared $Z^0$-contribution, respectively, while the purely
left-handed, charged current cross section is \bea
{\td\hat{\sigma}^\prime_{h_a,h_b}\over\td t} &=& {2 \pi \alpha^2
\over 3 s^2} \le {u t - m_{\tilde{l}_i}^2 m_{\tilde{\nu}_j}^2
\over s^2} \re \le {(1 - h_a) (1 + h_b) \left|L_{q q^\prime W}\,
L_{\tilde{l}_i \tilde{\nu}_l W} \right|^2 \over 32\, x_W^2\, (1 -
m_W^2/s)^2} \re.~\label{eq:slep_polW}\eea Averaging over initial
helicities, \bea \td\hat{\sigma}^{(\prime)} &=&
{\td\hat{\sigma}^{(\prime)}_{ 1, 1} + \td\hat{\sigma}^{(\prime)}_{
1,-1} + \td\hat{\sigma}^{(\prime)}_{-1, 1} +
\td\hat{\sigma}^{(\prime)}_{-1,-1} \over 4}, \eea we obtain the
unpolarized partonic cross sections \bea {\td\hat{\sigma}\over\td t}
&=& {2 \pi \alpha^2 \over 3 s^2} \le {u t \!-\! m_{\tilde{l}_i}^2
m_{\tilde{l}_j}^2 \over s^2} \re \Bigg[e_q^2\, e_l^2\, \delta_{ij}
\!+\! {e_q\, e_l\, \delta_{ij}\, {\rm Re}\left[L_{\tilde{l}_i
\tilde{l}_j Z} \!+\! R_{\tilde{l}_i \tilde{l}_j Z}\right]
\Big[L_{q q Z} \!+\! R_{q q Z}\Big] \over 4\, x_W (1 \!-\! x_W)
(1\!-\!m_Z^2/s)} \nn\\ &+& {\left|L_{\tilde{l}_i \tilde{l}_j Z} +
R_{\tilde{l}_i \tilde{l}_j Z}\right|^2 \Big[L_{q q Z}^2 + R_{q q
Z}^2\big] \over 32\, x_W^2\,
(1 - x_W)^2 (1-m_Z^2/s)^2} \Bigg],~\label{eq:slslLO}\\
{\td\hat{\sigma}^\prime\over\td t} &=& {2 \pi \alpha^2 \over 3 s^2}
\le {u t - m_{\tilde{l}_i}^2 m_{\tilde{\nu}_j}^2 \over s^2} \re
\Bigg[ {\left|L_{q q^\prime W}\, L_{\tilde{l}_i \tilde{\nu}_l W}
\right|^2 \over 32\, x_W^2\, (1 - m_W^2/s)^2}
\Bigg],~~\label{eq:slsnLO}\eea which agrees for mass-degenerate
non-mixing sleptons with the neutral current result of Ref.\
\cite{Dawson:1983fw} and with the charged current result of Ref.\
\cite{Baer:1993ew}. Note that for invariant final state masses
close to the $Z^0$- or $W$-pole, the $Z^0$- and $W$-propagators
must be modified to include the decay width of the corresponding
electroweak boson.\\

From Eqs.\ (\ref{eq:slep_polZ}) and (\ref{eq:slep_polW}), one can
easily calculate the polarized cross sections \bea
\td\Delta\hat{\sigma}^{(\prime)}_{LL} =
{\td\hat{\sigma}^{(\prime)}_{1,1} -
\td\hat{\sigma}^{(\prime)}_{1,-1} -
\td\hat{\sigma}^{(\prime)}_{-1,1} +
\td\hat{\sigma}^{(\prime)}_{-1,-1} \over 4} ,~\\
\td\Delta\hat{\sigma}^{(\prime)}_L =
{\td\hat{\sigma}^{(\prime)}_{1,1} +
\td\hat{\sigma}^{(\prime)}_{1,-1} -
\td\hat{\sigma}^{(\prime)}_{-1,1} -
\td\hat{\sigma}^{(\prime)}_{-1,-1} \over4},~\eea which then read
\cite{Bozzi:2004qq} \bea \td\Delta\hat{\sigma}^{(\prime)}_{LL} &=&
-\hat{\sigma}^{(\prime)},~\\ \td\Delta\hat{\sigma}_L &=& {2 \pi
\alpha^2 \over 3 s^2} \le {u t - m_{\tilde{l}_i}^2
m_{\tilde{l}_j}^2 \over s^2} \re \Bigg[- {e_q\, e_l\,
\delta_{ij}\, {\rm Re}\left[L_{\tilde{l}_i \tilde{l}_j Z} +
R_{\tilde{l}_i \tilde{l}_j Z}\right] \Big[L_{q q Z} - R_{q q
Z}\Big] \over 4\, x_W (1 - x_W) (1-m_Z^2/s)}\nn\\ &-&
{\left|L_{\tilde{l}_i \tilde{l}_j Z} + R_{\tilde{l}_i \tilde{l}_j
Z}\right|^2 \Big[L_{q q Z}^2 - R_{q q Z}^2\Big]
\over 32\, x_W^2\, (1 - x_W)^2 (1-m_Z^2/s)^2} \Bigg],\\
\td\Delta\hat{\sigma}^\prime_L &=& - \hat{\sigma}^\prime.~\eea Let
us note that the single-polarized cross section corresponds to a
polarized initial quark, the corresponding quantity for a
polarized initial antiquark being obtained from the following
definition, \bea \td\Delta\hat{\sigma}^{(\prime)}_L =
{\td\hat{\sigma}^{(\prime)}_{1,1} -
\td\hat{\sigma}^{(\prime)}_{1,-1} +
\td\hat{\sigma}^{(\prime)}_{-1,1} -
\td\hat{\sigma}^{(\prime)}_{-1,-1} \over4}.~\eea These expressions
show that it will be interesting to study the dependence of the
neutral current single-spin asymmetry \bea A_L =
\frac{\td\Delta\hat{\sigma}_L}{\td\hat{\sigma}} \eea on the
SUSY-breaking parameters, since it is the single quantity
remaining sensitive to these. Furthermore the squared photon
contribution, insensitive to them, is eliminated. Finally, this
scenario may also be easier to implement experimentally, e.g.\ at
the Tevatron, since protons are much more easily polarized than
antiprotons \cite{Baiod:1995eu}. Let us note that our polarized
results agree with those of Ref.\ \cite{Chiappetta:1985ku} for
non-mixing sleptons after integration over $t$, provided that we
put parentheses around the $\hat{s}|D_Z|^2$ terms of Eqs.\ (5) and
(7) of Ref.\ \cite{Chiappetta:1985ku}, and if we replace the index
$q$ by $e$ in the first occurrence of $(a_q+\epsilon b_q)$ in Eq.\
(7) of Ref.\ \cite{Chiappetta:1985ku}.

\subsection{Unpolarized cross section} \label{sec:unpo}

For the masses and widths of the electroweak gauge bosons, we use
here older values of $m_Z=91.1876$ GeV, $m_W=80.425$ GeV,
$\Gamma_Z=2.4952$ GeV, and $\Gamma_W=2.124$ GeV
\cite{Eidelman:2004wy}, which are slightly different from the
current ones (see Sec.\ \ref{sec:scan}). Since the sfermion mixing
is proportional to the mass of the corresponding SM partner (see
Eq.\ (\ref{eq:2to2mix2})), it is numerically only important for
third generation sleptons. Consequently, the lightest slepton is
the lighter stau mass eigenstate $\tilde{\tau}_1$ and we focus
our numerical studies on its production.\\

%%%%%%%%%%%%%% Begin Figure 4 %%%%%%%%%%%%%%%%%%%%%%%%%%%%%%%%%%%%%%%%%%
\begin{figure}
 \centering
 \includegraphics[width=.7\columnwidth]{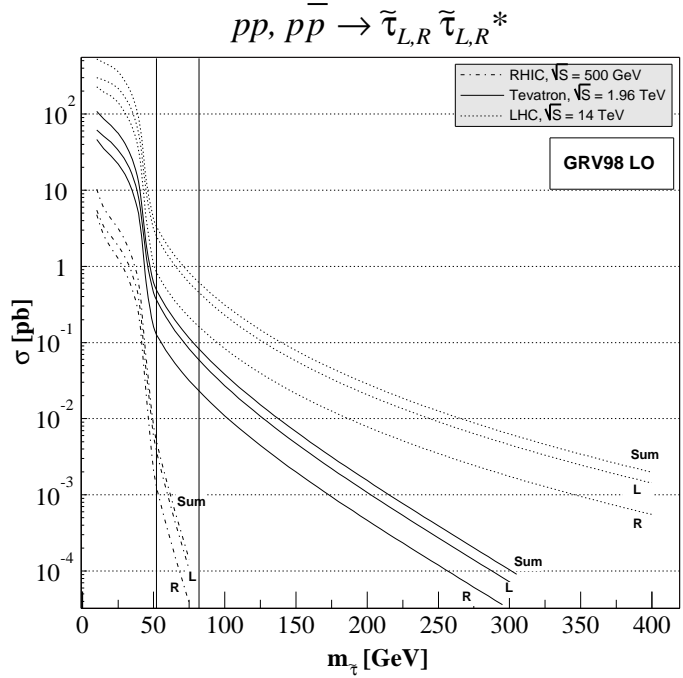}
 \caption{\label{fig:4}Unpolarized hadronic cross sections for
 non-mixing $\tilde{\tau}$ pair production of at the RHIC,
 Tevatron, and LHC colliders as a function of $m_{\stau}$.
 The vertical lines indicate the two different $\stau$ mass limits of 52
 \cite{Barate:1998zp} and 81.9 GeV \cite{Abdallah:2003xe}.}
\end{figure}
%%%%%%%%%%%%%% End of Figure 4 %%%%%%%%%%%%%%%%%%%%%%%%%%%%%%%%%%%%%%%%%

The $\stau$ mass limits imposed by LEP depend strongly on the
assumed SUSY-breaking mechanism, the mass difference between the
$\stau$ and the LSP, and the mixing angle $\theta_{\stau}$. The
weakest limit of 52 GeV is found for GMSB models and stau decays
to gravitinos, if no constraints on their mass difference are
imposed \cite{Barate:1998zp}. This is the scenario that we will
study for the RHIC collider, which has the most restricted
hadronic centre-of-mass energy, equal to 500 GeV. For the Tevatron
and the LHC, with their considerably larger centre-of-mass
energies of 1.96 and 14 TeV, respectively, we will, however,
impose the stricter mass limit of 81.9 GeV \cite{Abdallah:2003xe},
which is valid for staus decaying into neutralinos, with a mass
difference of at least 15 GeV. \\

Thanks to the QCD factorization theorem, the unpolarized hadronic
cross section \bea \sigma = \sum_{a,b}\int_{\tau}^1 {\rm d} x_a
\int_{\tau/x_a}^1 {\rm d}x_b\, f_{a/h_a}(x_a,\mu_F)\, \
f_{b/h_b}(x_b,\mu_F) \,\hat{\sigma}_{ab} \label{eq:QCDFact_th}\eea
can be written as the convolution of the relevant partonic cross
section $\hat{\sigma}_{ab}$ (integrated here over $t$) with the
universal distribution functions $f_{a,b/ h_{a,b}}$ of partons
$a,b$ inside the hadrons $h_{a,b}$, which depend on the
longitudinal momentum fractions of the two partons $x_{a,b}$ and
on the unphysical factorization scale $\mu_F$. In order to employ
a consistent set of unpolarized and polarized parton densities
(see next subsection), we choose the LO set of GRV98
\cite{Gluck:1998xa} for our unpolarized predictions at the
factorization scale $\mu_F = m_{\tilde{\tau}_1}$.\\

In Fig.\ \ref{fig:4}, we show the unpolarized hadronic cross
sections for non-mixing $\stau$ pair production at the RHIC,
Tevatron, and LHC colliders as a function of the $\stau$ physical
mass. The observation of tau sleptons will be difficult at RHIC,
which is the only existing polarized hadron collider, but they
will be detectable at the Tevatron, extending considerably the
discovery reach beyond the current limits. In contrast, at the
LHC, $\stau$ pair production will be visible up to masses of 1
TeV. Before application of any experimental cuts, the SUSY signal
cross section is at least three orders of magnitude smaller than
the corresponding SM background coming from tau lepton pair
production. Evaluated using the physical tau mass of
$m_\tau=1.77699$ GeV as factorization scale for the GRV98 LO
parton densities, this cross section is equal to 1.7, 3.4, and 8.3
nb for the RHIC, Tevatron, and LHC colliders, respectively.
Imposing an invariant-mass cut on the observed lepton pair and a
minimal missing transverse energy will, however, greatly improve
the signal-to-background ratio. In addition, as we will see in the
next section, asymmetries may provide an important tool to further
distinguish the SUSY signal from the corresponding SM process.

\subsection{Single-spin asymmetries}

%%%%%%%%%%%%%% Begin Figure 5 %%%%%%%%%%%%%%%%%%%%%%%%%%%%%%%%%%%%%%%%%%
\begin{figure}
 \centering
 \includegraphics[width=.7\columnwidth]{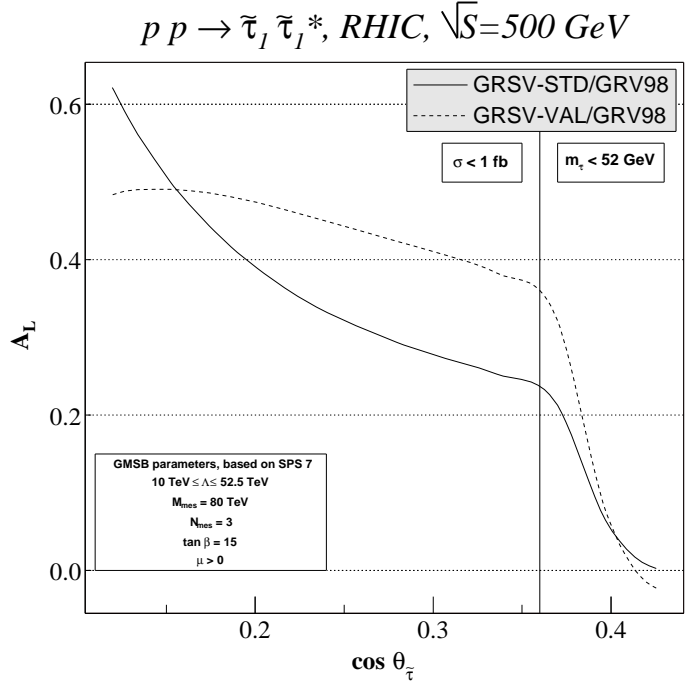}
   \caption{\label{fig:5}Longitudinal single-spin asymmetry $A_L$
   as a function of the cosine of the stau mixing angle for
   $\tilde{\tau}_1$ pair production at RHIC.}
\end{figure}
%%%%%%%%%%%%%% End of Figure 5 %%%%%%%%%%%%%%%%%%%%%%%%%%%%%%%%%%%%%%%%%

%%%%%%%%%%%%%% Begin Figure 6 %%%%%%%%%%%%%%%%%%%%%%%%%%%%%%%%%%%%%%%%%%
\begin{figure}
 \centering
 \includegraphics[width=.7\columnwidth]{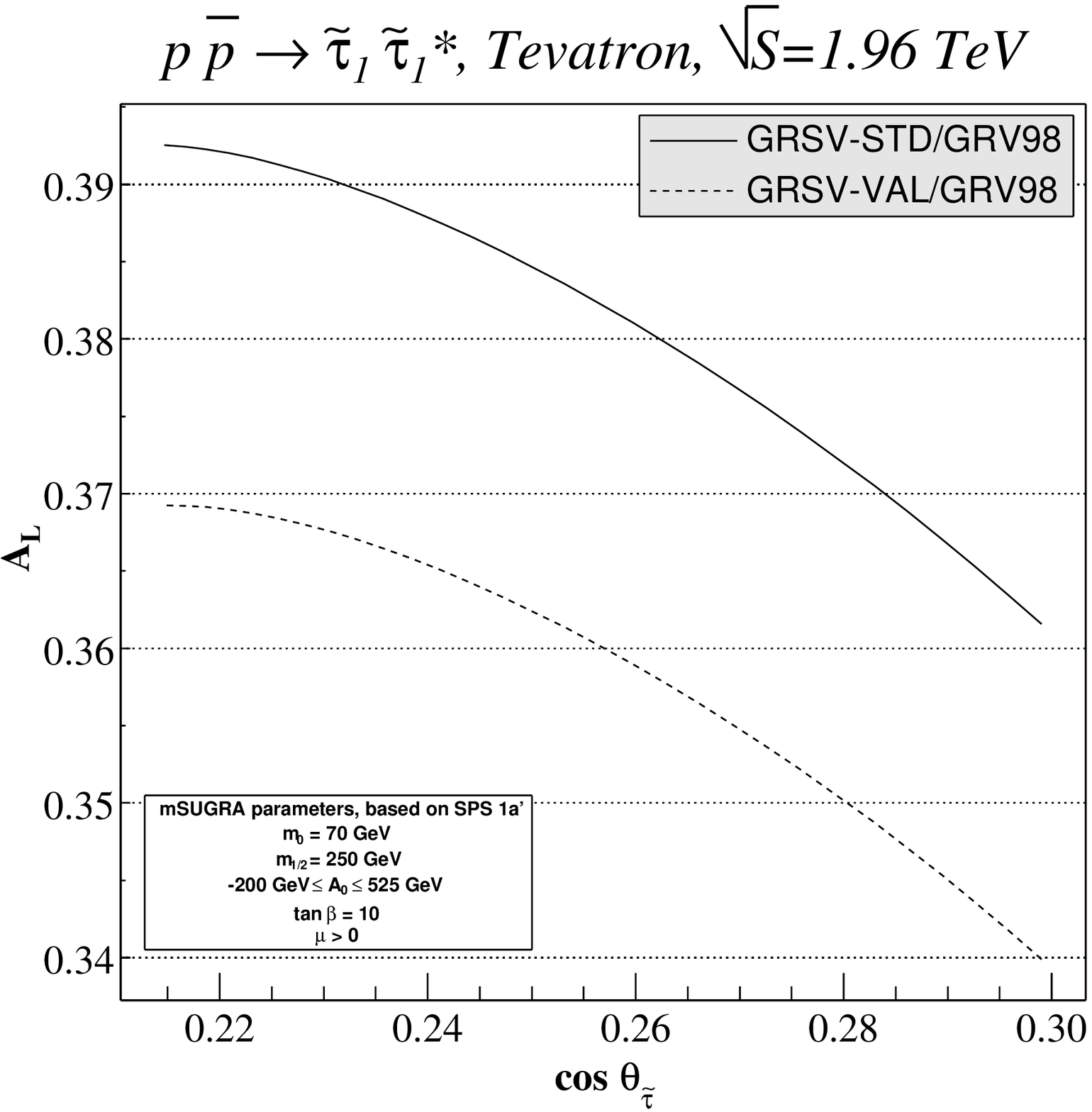}
 \includegraphics[width=.7\columnwidth]{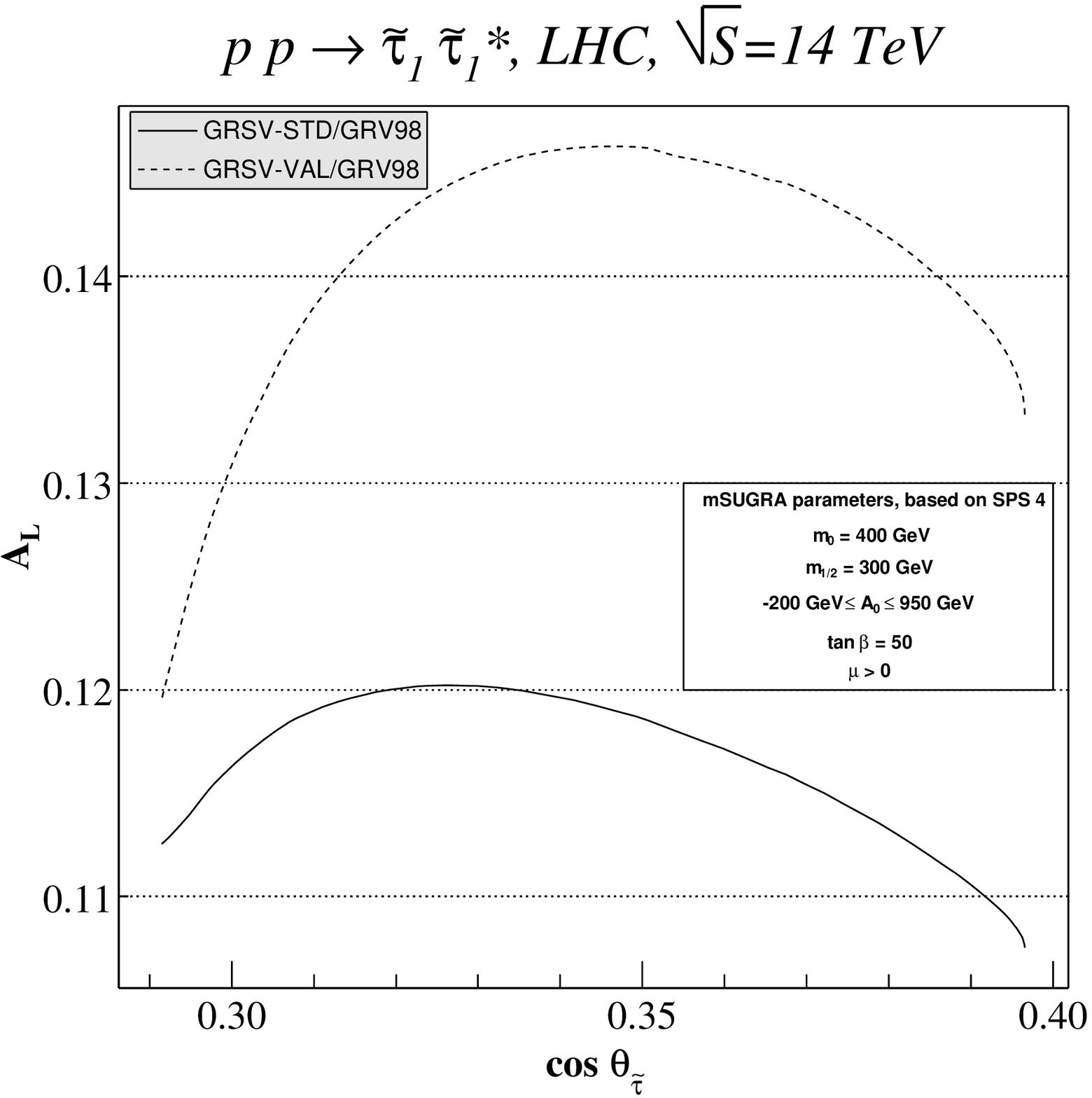}
  \caption{\label{fig:6}Dependence of the longitudinal single-spin
  asymmetry $A_L$ on the cosine of the stau mixing angle for
  $\tilde{\tau}_1$ pair production at the Tevatron (top) and at the
  LHC (bottom).}
\end{figure}
%%%%%%%%%%%%%% End of Figure 6 %%%%%%%%%%%%%%%%%%%%%%%%%%%%%%%%%%%%%%%%%

Using again the QCD factorization theorem, we calculate the
hadronic cross section for longitudinally polarized hadrons $h_a$
with unpolarized hadrons $h_b$ \bea \Delta\sigma_L =
\sum_{a,b}\int_{\tau}^1 {\rm d} x_a \int_{\tau/x_a}^1 {\rm d}x_b\,
\Delta f_{a/h_a}(x_a,\mu_F)\, \ f_{b/h_b}(x_b,\mu_F)
\,\td\Delta\hat{\sigma}_{L,ab} \eea through a convolution of
polarized ($\Delta f_{a/A}$) and unpolarized ($f_{b/B}$) parton
densities with the singly polarized partonic cross section
$\Delta\hat{\sigma}_L$ (integrated over $t$). As mentioned above,
we employ a consistent set of unpolarized \cite{Gluck:1998xa} and
polarized \cite{Gluck:2000dy} LO parton densities. The theoretical
uncertainty due to the less well known polarized parton densities
is estimated by showing our numerical predictions for both the
GRSV2000 LO standard (STD) and valence (VAL) parameterizations.\\

The physical masses of the SUSY particles and the mixing angles
are computed with the program SuSpect 2.3 \cite{Djouadi:2002ze},
including a consistent calculation of the Higgs mass, with all
one-loop and the dominant two-loop radiative corrections in the
renormalization group equations that link the restricted set of
SUSY-breaking parameters at the gauge coupling unification scale
to the complete set of observable SUSY masses and mixing angles at
the electroweak scale. We choose three SPS benchmark points
\cite{Allanach:2002nj, Aguilar-Saavedra:2005pw}, the GMSB point
SPS 7 with a light tau slepton decaying to a gravitino for RHIC
and its very limited mass range, the typical mSUGRA point SPS 1a'
with an intermediate value of $\tan\beta=10$ and a slightly
reduced common scalar mass of $m_0=70$ GeV for the Tevatron, and
the mSUGRA point SPS 4 with a large scalar mass of $m_0=400$ GeV
and large $\tan\beta=50$, which enhances mixing, for the LHC with
its larger mass range. Since $\theta_{\stau}$ depends directly on
the universal soft SUSY-breaking mass scale $\Lambda$ in the GMSB
model and on the trilinear coupling $A_0$ in the mSUGRA models, we
test the sensitivity of the single-spin asymmetry on these
parameters  (see Figs.\ \ref{fig:5}-\ref{fig:6}).\\

In Fig.\ \ref{fig:5}, we show the single-spin asymmetry for
$\tilde{\tau}_1$ pair production at RHIC as a function of the
cosine of the stau mixing angle. The asymmetry is quite large and
depends strongly on the stau mixing angle. However, very large
values of $\cos\theta_{\tilde{\tau}}$, corresponding to $\stau$
masses below 52 GeV may already be excluded by LEP
\cite{Barate:1998zp}, while small values of
$\cos\theta_{\tilde{\tau}}$ may be unaccessible to RHIC limited
luminosity, which is not expected to exceed 1 fb$^{-1}$.
Polarization of the proton beam will also not be perfect, and the
calculated asymmetries should be multiplied by the degree of beam
polarization $P_L \simeq 0.7$. The uncertainty introduced by the
polarized parton densities increases considerably to the left of
the plot, where the stau mass 41 GeV $\leq m_{\tilde{\tau}}\leq$
156 GeV and the associated values of the parton momentum fractions
$x_{a,b} \simeq 2m_{\tilde{\tau}}/\sqrt{s_h}$ become large. While
the SM background cross section is then still two orders of
magnitude larger than the SUSY signal after imposing an
invariant-mass cut on the observed tau lepton pair, the SM
asymmetry of -0.04 (-0.10) for STD (VAL) polarized parton
densities can clearly be
distinguished due to its different sign.\\

An asymmetry measurement at an upgraded Tevatron would be
extremely valuable, since the predicted asymmetry is very sizeable
in the entire viable SUSY parameter range, and depends strongly on
the parameter $A_0$, and then the stau mixing angle (top figure of
Fig.\ \ref{fig:6}). Unfortunately, the parton density
uncertainties are still large, but will be reduced considerably in
the future through more precise measurements at the COMPASS,
HERMES, PHENIX, and STAR experiments. As a recent experimental
study demonstrates, events with tau leptons with associated
missing energy larger than 20 GeV can be identified with the
CDF-II detector, considering hadronic tau decays
\cite{Anastassov:2003vc}. Again, the negative SM asymmetry of
-0.09 (for both polarized parton densities) would
be clearly distinguishable due to its opposite sign.\\

For the LHC, where studies of tau slepton identification with the
ATLAS detector \cite{Hinchliffe:2002se} and tau tagging with the
CMS detector \cite{Gennai:2002qq} have recently been performed,
SUSY masses should be observable up to the TeV-range. We show the
predicted asymmetry for a possible polarization upgrade of this
collider in the bottom panel of Fig.\ \ref{fig:6}. It is still
comfortably large and has again the opposite sign with respect to
the SM asymmetry of -0.02 (for both polarized parton densities).
The dependence of the asymmetry on the stau mixing angle is,
however, also reduced, while the uncertainties from the polarized
parton densities, which are not yet well known at the small $x$
values relevant for the large LHC centre-of-mass energy, are quite
enhanced.

\section{Transverse-momentum spectrum}
\label{sec:QTresult}

\subsection{Analytical results}

As said in the introduction, the main SM background to slepton
pair production comes from $WW$ and $t \bar{t}$ production.
However, an accurate calculation of the transverse-momentum
spectrum would allow us to distinguish the signal from the
background. A transverse-momentum distribution for slepton pair
production at leading order is induced by a final state containing
one slepton pair and one QCD jet. The relative partonic cross
section can be divided into a resummed and a finite part, as in
Eq.\ (\ref{eq:qtxsec}), \bea \frac{\td{\hat \sigma}_{ab}}{\td M^2 \td
q_T^2}= \frac{\td{\hat \sigma}_{ab}^{(\rm res.)}}{\td M^2 \td q_T^2}
+\frac{\td{\hat \sigma}_{ab}^{(\rm fin.)}}{\td M^2 \td q_T^2},~
\label{eq:qtxsec2}\eea where $a,b$ label the partons which take
part in the hard process. In Mellin $N$-space, the (partonic)
resummed component is deduced from Eqs.\ (\ref{eq:resterm1}) and
(\ref{eq:resterm2}), \bea \frac{\td{\hat \sigma}_{ab}^{(\rm
res.)}}{\td M^2 \td q_T^2}(q_T,M,s; \alpha_s(\mu_R),\mu_R,\mu_F) =
\int_0^\infty \td b \; \frac{b}{2} \, J_0(b q_T) \,
\tilde{W}^F_{ab}(N, b; M, \mu_F, \mu_R) ,\nn \eea
\vspace{-.8cm}\bea ~ \eea \vspace{-.8cm}\bea \tilde{W}^F_{ab}(N,
b; M, \mu_F, \mu_R)\! = \! \mathcal{H}^F_{ab}\Big(N,
\alpha_s(\mu_R); \frac{M}{\mu_R},\! \frac{M}{\mu_F},\!
\frac{M}{Q}\Big)\! \exp\! \Big\{\mathcal{G}\big(N, \tilde{L},
\alpha_s(\mu_R); \frac{M}{\mu_R},\!
\frac{M}{Q}\big)\Big\}\label{eq:W},~\nn \eea \vspace{-.8cm}\bea ~
\eea  where at the NLL accuracy, the functions ${\cal H}$ and
${\cal G}$ of Eqs.\ (\ref{eq:H}) and (\ref{eq:g}) read \bea {\cal
H}_{ab}^F(N, \alpha_s; \frac{M}{\mu_R},\! \frac{M}{\mu_F},\!
\frac{M}{Q})\! = \! \sum_c \sigma_{c \bar{c}}^{({\rm LO})F}(M) \le
\delta_{ac}\delta_{b\bar{c}} \!+\! \lr\frac{\alpha_s}{\pi}\rr
\,{\cal H}_{ab\to c\bar{c}}^{F (1)}(N; \frac{M}{\mu_R},\!
\frac{M}{\mu_F},\! \frac{M}{Q})\re~, \label{eq:Hslsl} \nn \eea
\vspace{-.8cm}\bea ~ \eea \vspace{-.8cm}\bea{\cal G}(N,\tilde{L};
\alpha_s, \frac{M}{\mu_R},\! \frac{M}{Q}) = \tilde{L}\,
g^{(1)}(\frac{1}{\pi} \,\beta_0 \,\alpha_s(\mu_R) \, L) +
g^{(2)}(N, \frac{1}{\pi} \,\beta_0 \,\alpha_s(\mu_R) \, \tilde{L};
\frac{M}{\mu_R},\! \frac{M}{Q}),~\eea with \bea {\tilde L} = \ln
\left(\frac{Q^2\,b^2}{b_0^2} + 1\right),~ \eea this definition of
$\tilde{L}$ allowing us to avoid the unwanted resummed
contributions in the small-$b$ region. The explicit expression of
$g^{(1)}$ and $g^{(2)}$ in terms of the universal perturbative
coefficients $A^{(1)}_q$, $A^{(2)}_q$, and $B^{(1)}_q$ (see Eqs.\
(\ref{eq:a1coeff}), (\ref{eq:a2coeff}) and \ref{eq:b1coeff})) are
given by Eqs.\ (\ref{eq:gdef1}) and (\ref{eq:gdef2}). Slepton pair
production being similar to SM lepton pair production, except for
the final state, we can use the Drell-Yan coefficients ${\cal
H}^{(1)}$ from Eqs.\ (\ref{eq:H1DYqq}) and (\ref{eq:H1DYqg}), but
the Born cross section $\sigma^{({\rm LO})F}(M)$ in Eq.\
(\ref{eq:Hslsl}) corresponds obviously to the slepton pair
production one, which can be obtained for neutral and charged
currents by integrating Eqs.\ (\ref{eq:slslLO}) and
(\ref{eq:slsnLO}) over $t$, respectively, \bea
\sigma_{q\bar{q}}^{(LO)}(M) &=& \frac{\alpha^2\, \pi\,
\beta^3}{9\,M^2} \Bigg[e_q^2\, e_l^2\, \delta_{ij} +
\frac{e_q\,e_l\,\delta_{ij} (L_{q q Z} + R_{q q Z})\, {\rm
Re}\left[L_{\tilde{l}_i \tilde{l}_j Z} + R_{\tilde{l}_i
\tilde{l}_j Z}\right]}{4\, x_W\,(1-x_W)\, (1-m_Z^2/M^2)}
\nonumber\\ &+& \,\frac{(L_{q q Z}^2 + R_{q q Z}^2)
\left|L_{\tilde{l}_i \tilde{l}_j Z} + R_{\tilde{l}_i \tilde{l}_j
Z}\right|^2}{32\,x_W^2\, (1-x_W)^2
(1-m_Z^2/M^2)^2}\Bigg],~\label{eq:sig0Z}\\
\sigma_{q\bar{q}^\prime}^{(LO)}(M) &=& \frac{\alpha^2\, \pi\,
\beta^3}{9\, M^2} \Bigg[\frac{\left| L_{q q^\prime W}
L_{\tilde{l}_i \tilde{\nu}_l W} \right|^2}{32\, x_W^2\, (1-x_W)^2
(1-m_W^2/M^2)^2}\Bigg]\label{eq:sig0W},~\eea where the
slepton-mass dependence is factorized in the velocity \bea \beta
&=& \sqrt{1 + m_i^4/M^4 + m_j^4/M^4 - 2(m_i^2/M^2 + m_j^2/M^2 +
m_i^2\,m_j^2/M^4)},~\eea $m_i$ and $m_j$ being the two
masses of the final state particles.\\

The second term ($\td{\hat \sigma}_{ab}^{(\rm fin.)}/\td M^2 \td
q_T^2$) in Eq.\ (\ref{eq:qtxsec2}) is free of divergent
contributions and can be computed by fixed-order truncation of the
perturbative series, as in Eq.\ (\ref{eq:qtfin}). In order to be
consistently matched with the resummed contribution at
intermediate $q_T$ ($q_{T}\simeq M$), this term is evaluated
starting from the usual perturbative calculation of the partonic
cross section from which we subtract the expansion of the resummed
component of the cross section at the same perturbative order. In
order to consistently perform a NLL+LO matching, we need the
fixed-order cross section relative to the production of a slepton
pair with non-vanishing transverse-momentum at order $\as$, i.e.\
one slepton pair plus one parton \cite{Baer:1997nh,
Beenakker:1999xh}. The different channels read \bea {\td{\hat
\sigma}_{qg} \over\td M^2 \td q_T^2} &=& \frac{\as(\mu_R)\, T_R}{2\,
\pi\, s} A_{qg}(s,t,u; M^2)\,
\sigma_{q\bar{q}^{(\prime)}}^{(LO)}(M),~\\ {\td{\hat
\sigma}_{g\bar{q}}\over \td M^2 \td q_T^2} &=& \frac{\as(\mu_R)\,
T_R}{2\, \pi\, s} A_{qg}(s,u,t;M^2)\,
\sigma_{q\bar{q}^{(\prime)}}^{(LO)}(M),~\\ {\td{\hat
\sigma}_{q\bar{q}^{(\prime)}}\over \td M^2 \td q_T^2} &=&
\frac{\as(\mu_R)\, C_F}{2\, \pi\, s} A_{qq}(s,t,u;M^2)\,
\sigma_{q\bar{q}^{(\prime)}}^{(LO)}(M),~ \eea where $s$, $t$, and
$u$ denote the partonic Mandelstam variables relative to the $2\to
2$ subprocess $a b \to \gamma,Z^0, W^{\mp} + {\rm one~parton}$,
\bea s &=& x_{a}x_{b}s_h,\\ t &=&
M^2-\sqrt{s_h(M^2+q^2_T)}x_{b}e^{y},\\ u &=&
M^2-\sqrt{s_h(M^2+q^2_T)}x_{a}e^{-y}, \eea $y$ being the rapidity
of the slepton pair. We have defined the $A_{ab}$-functions as
\cite{Gonsalves:1989ar}, \bea A_{qg}(s,t,u;M^2) &=& -\lr
\frac{s}{t} + \frac{t}{s} + \frac{2u\,M^2}{s t} \rr,~\\
A_{qq}(s,t,u;M^2) &=& -A_{qg}(u,t,s;M^2).~\eea\\

It is known \cite{Collins:1981va} that the transverse-momentum
distribution is affected by non-perturbative effects which become
important in the large-$b$ region. In the case of electroweak
boson production, these contributions are usually parameterized by
multiplying the function $\tilde{W}^F$ in Eq.\ (\ref{eq:W}) by a
NP form factor $F^{NP}(b)$ \cite{Davies:1984sp, Ladinsky:1993zn,
Qiu:2000ga, Landry:2002ix, Konychev:2005iy}, whose coefficients
are obtained through global fits to DY data. We include in our
analysis three different parameterizations of non-perturbative
effects corresponding to three different choices of the form
factor: the Ladinsky-Yuan (LY-G) \cite{Ladinsky:1993zn},
Brock-Landry-Nadolsky-Yuan (BLNY) \cite{Landry:2002ix}, and the
recent Konychev-Nadolsky (KN) \cite{Konychev:2005iy} form factor.
The explicit expressions can be found in Eqs.\ (\ref{eq:NP1}),
(\ref{eq:NP2}) and (\ref{eq:NP3}).

\subsection{Numerical results}\label{sec:qtnum}

\begin{figure}
 \centering
 \includegraphics[width=.7\columnwidth]{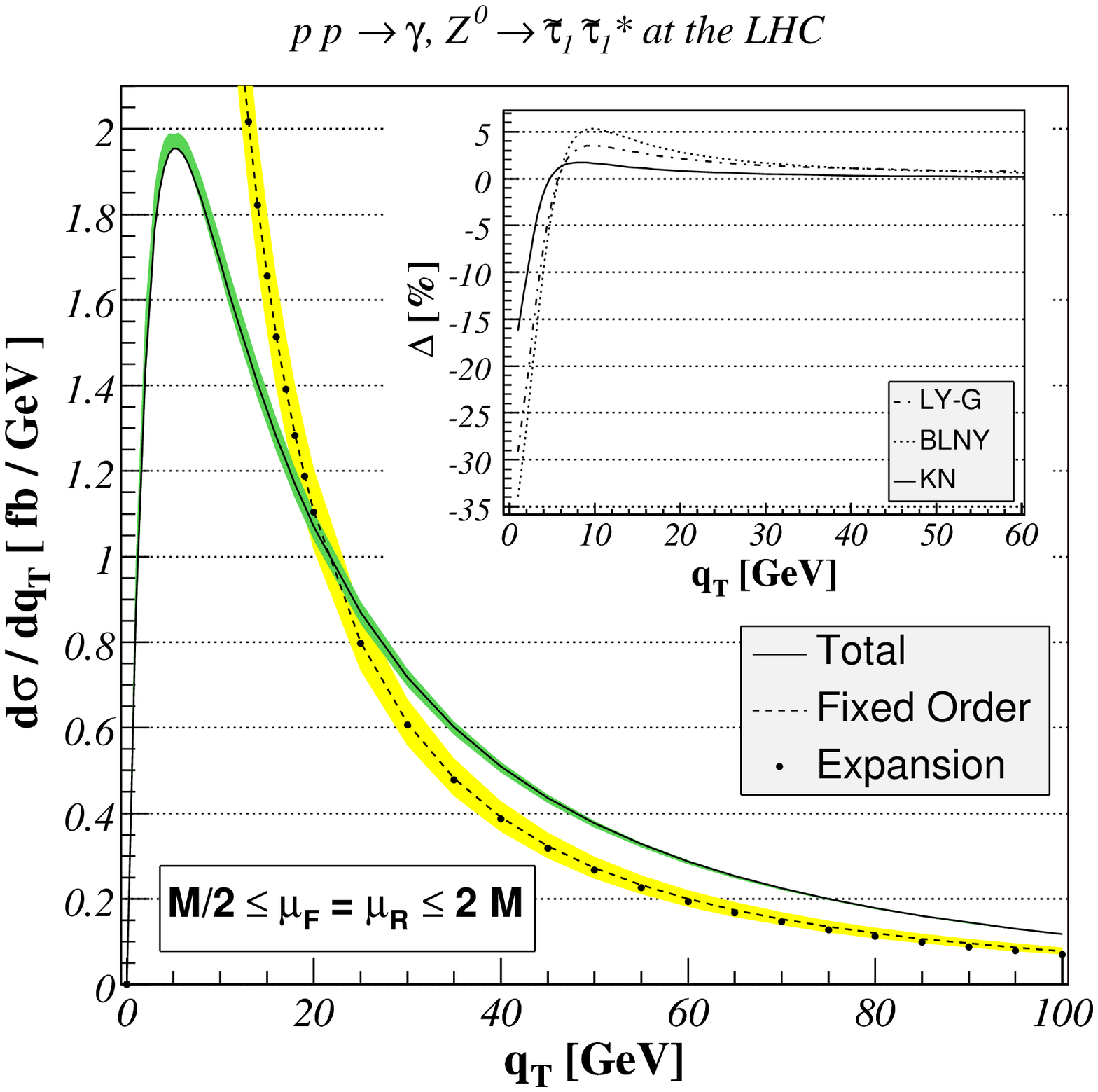}
 \includegraphics[width=.7\columnwidth]{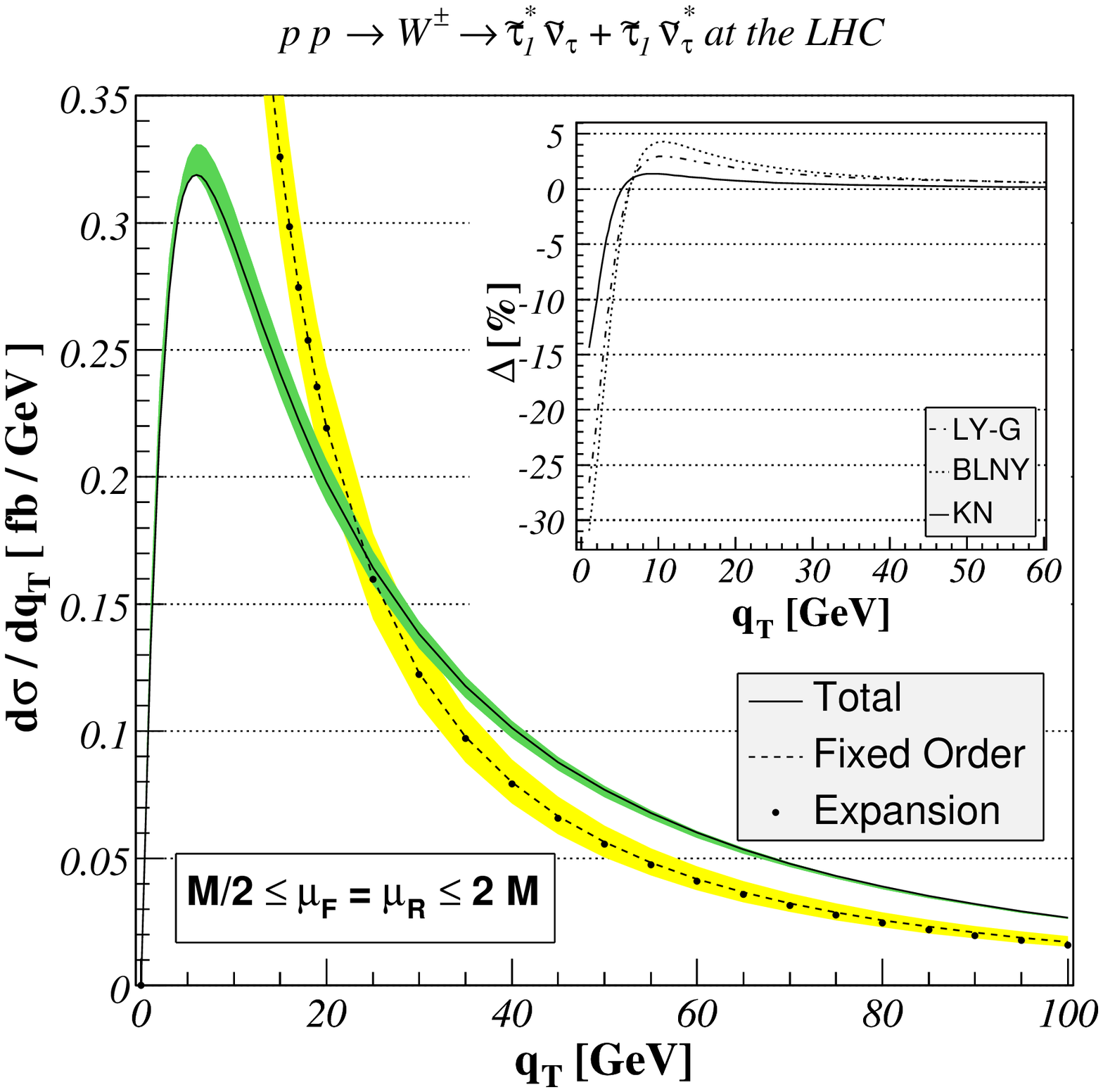}
   \caption{\label{fig:8} Differential cross section for the
   processes $p p \to \tilde{\tau}_1
   \tilde{\tau}_1^\ast$ (top)
   and $p p \to \tilde{\tau}_1\tilde{\nu}_{\tau}^\ast +
   \tilde{\tau}_1^\ast \tilde{\nu}_{\tau}$ (bottom).
   NLL+LO matched result, LO result, asymptotic expansion of
   the resummation formula and $\Delta$-parameter are shown.}
\end{figure}

We present quantitative results for the $q_T$-spectrum of slepton
pair (slepton-sneutrino associated) production at NLL+LO accuracy
at the LHC collider. For the masses and widths of the electroweak
gauge bosons, the electroweak mixing angle and the electromagnetic
fine structure constant, we use the same values as those given in
Sec.\ \ref{sec:unpo}, while the CKM-matrix elements are computed
as in Sec.\ \ref{sec:scan}. As in the previous section, we focus
our study on the lightest slepton mass eigenstate ${\tilde
\tau}_1$ and thus we consider the processes \bea
\begin{array}{c} q \bar{q} \to \gamma,Z^0 \to
\tilde{\tau}_1 \tilde{\tau}^\ast_1, \\
q \bar{q}^\prime \to W^{\mp} \to
\tilde{\tau}_1\tilde{\nu}_{\tau}^\ast,
\tilde{\tau}^*_1\tilde{\nu}_{\tau}. \end{array} \eea We use the
MRST (2004) NLO set of parton distribution functions
\cite{Martin:2004ir} and $\alpha_s$ is evaluated at two-loop
accuracy. We fix the resummation scale $Q$ (see Eq.\
(\ref{eq:resterm2})) equal to the invariant-mass $M$ of the
slepton (slepton-sneutrino) pair and we allow $\mu=\mu_F=\mu_R$ to
vary between $M/2$ and $2M$ to estimate the perturbative
uncertainty. We also integrate Eq.\ (\ref{eq:qtxsec2}) with
respect to $M^2$, taking as lower limit the energy threshold for
$\tilde{\tau}_{1} \tilde{\tau}_1^\ast$ $(\tilde{\tau}_1
\tilde{\nu}_{\tau})$ production and as upper limit the hadronic
energy ($\sqrt s_h$ = 14 TeV at the
LHC).\\

In the case of $\tilde{\tau}_1 \tilde{\tau}_1^*$ production
(neutral current process, see the top panel of Fig.\ \ref{fig:8}),
we choose the SPS 7 GMSB benchmark point \cite{Allanach:2002nj}
which gives, after the renormalization group evolution of the
SUSY-breaking parameters performed by the SuSpect computer program
\cite{Djouadi:2002ze}, a light ${\tilde \tau}_{1}$ of mass
$m_{{\tilde \tau}_{1}}=114$ GeV. In the case of $\tilde{\tau}_1
\tilde{\nu}_{\tau}^\ast + \tilde{\tau}_1^\ast \tilde{\nu}_{\tau}$
production (charged current process, see the bottom panel of Fig.\
\ref{fig:8}), we use instead the SPS 1a mSUGRA benchmark point
which gives a light ${\tilde \tau}_1$ of mass $m_{{\tilde \tau}_1}
= 136$ GeV as well as a light $\tilde{\nu}_{\tau}$ of mass
$m_{{\tilde \nu}_\tau} =196$
GeV.\\

In both cases we plot the LO result (dashed line) and the total
NLL+LO matched result (solid line) with their uncertainty band
relative to scale variation (yellow band for LO result, and green
band for NLL+LO result). The expansion of the resummation formula
at LO (dotted line) is also shown. We can see that the LO result
diverges, as expected from the generic form of the fixed-order
cross section in Eq.\ (\ref{eq:fixqt2}), for both processes as
$q_{T}\to 0$. The asymptotic expansion of the resummation formula
at LO is in very good agreement with LO both at small and
intermediate values of $q_T$, from which we can conclude that the
cross section is clearly dominated by the logarithms that we are
resumming in this kinematical region. The effect of resummation is
clearly visible at small and intermediate values of $q_T$, the
resummation-improved result being nearly 39\% (36\%) higher at
$q_T=50$ GeV than the pure fixed-order result in the neutral
(charged) current case. When integrated over $q_T$, the former
leads to a total cross section of 66.8 fb (12.9 fb) in good
agreement (within 3.5\%) with the QCD-corrected total cross
section at ${\cal O}(\alpha_s)$ \cite{Beenakker:1999xh}.\\

The scale dependence is clearly improved in both cases with
respect to the pure fixed-order calculations. In the small and
intermediate $q_T$-region (up to 100 GeV) the effect of scale
variation is 10\% for the LO result, while it is always less than
5\% for the NLL+LO curve.\\

Finally, we study the dependence of the total NLL+LO matched
result on non-perturbative effects. We show the quantity $\Delta$
\bea \Delta = \frac{d\sigma^{\rm (res.+NP)}(\mu=M)-d\sigma^{(\rm
res.)}(\mu=M)}{d\sigma^{(\rm res.)} (\mu=M)},\label{eq:delta}\eea
as a function of the transverse momentum of the slepton pair. The
parameter $\Delta$ gives thus an estimate of the contributions
from the different NP parameterizations (LY-G, BLNY, KN) that we
included in the resummed formula, which are under good control,
since they are always less than 5\% for $q_{T}>$ 5 GeV and thus
considerably smaller than the resummation effects.

\section{Invariant-mass distributions}

\subsection{Next-to-leading order calculations}

Restoring the various scale dependences in the QCD factorization
theorem of Eq.\ (\ref{eq:QCDFact_th}), the invariant-mass
distribution of a given process can be written as \bea \sigma =
\sum_{a,b}\int_{\tau}^1 {\rm d}x_a \int_{\tau/x_a}^1 {\rm d}x_b\,
f_{a/h_a}(x_a,\mu_F)\, \ f_{b/h_b}(x_b,\mu_F) \,\hat{\sigma}_{ab}
\lr z, M;\alpha_s(\mu_R),\frac{M}{\mu_F}, \frac{M}{\mu_R}\rr,~ \nn
\eea \vspace{-.8cm}\bea ~ \eea where $M$ is the invariant-mass.
The partonic cross section is usually expanded in powers of
$\alpha_s$ \bea \hat{\sigma}_{ab} \lr z, M; \alpha_s(\mu_R),
\frac{M}{\mu_F}, \frac{M}{\mu_R}\rr &=&
\sum_{n=0}^\infty\left(\frac{\alpha_s(\mu_R)}{\pi}\right)^n
\sigma_{ab}^{(n)}\lr z,M; \frac{M}{\mu_F}, \frac{M} {\mu_R}\rr.~
\label{eq:pert}\eea Concerning slepton pair hadroproduction, we
have computed the LO ($n=0$) coefficients in the previous
sections, which read, showing explicitly the dependence on the
scaling variable $z=M^2/s$, \bea
\sigma_{q\bar{q}^{(\prime)}}^{(0)}\lr
z,M;\frac{M}{\mu_F},\frac{M}{\mu_R}\rr &=&
\sigma_0^{(\prime)}(M)\,\delta (1-z),~\eea $\sigma_0$ and
$\sigma_0^\prime$ being defined in Eqs.\ (\ref{eq:sig0Z}) and
(\ref{eq:sig0W}) for the neutral and charged currents,
respectively, \bea \sigma_0(M) &=& \frac{\alpha^2\, \pi\,
\beta^3}{9\,M^2} \Bigg[e_q^2\, e_l^2\, \delta_{ij} +
\frac{e_q\,e_l\,\delta_{ij} (L_{q q Z} + R_{q q Z})\, {\rm
Re}\left[L_{\tilde{l}_i \tilde{l}_j Z} + R_{\tilde{l}_i
\tilde{l}_j Z}\right]}{4\, x_W\,(1-x_W)\, (1-m_Z^2/M^2)}
\nonumber\\ &+& \,\frac{(L_{q q Z}^2 + R_{q q Z}^2)
\left|L_{\tilde{l}_i \tilde{l}_j Z} + R_{\tilde{l}_i \tilde{l}_j
Z}\right|^2}{32\,x_W^2\, (1-x_W)^2
(1-m_Z^2/M^2)^2}\Bigg],~\label{eq:sig0Z2}\\
\sigma_0^\prime(M) &=& \frac{\alpha^2\, \pi\, \beta^3}{9\, M^2}
\Bigg[\frac{\left| L_{q q^\prime W} L_{\tilde{l}_i \tilde{\nu}_l
W} \right|^2}{32\, x_W^2\, (1-x_W)^2
(1-m_W^2/M^2)^2}\Bigg]\label{eq:sig0W2}.~\eea

\begin{figure}
 \centering
 \includegraphics[width=.9\columnwidth]{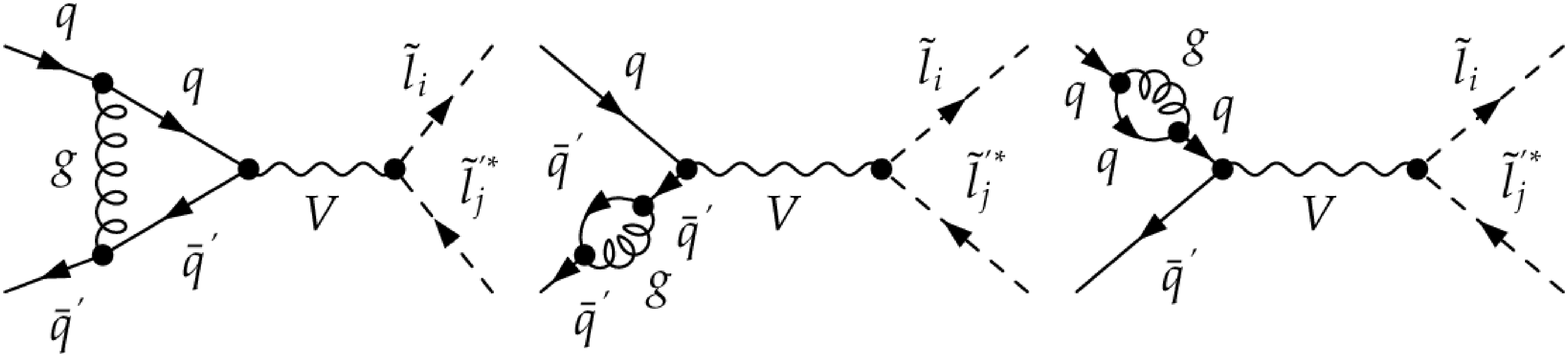}
 \includegraphics[width=.9\columnwidth]{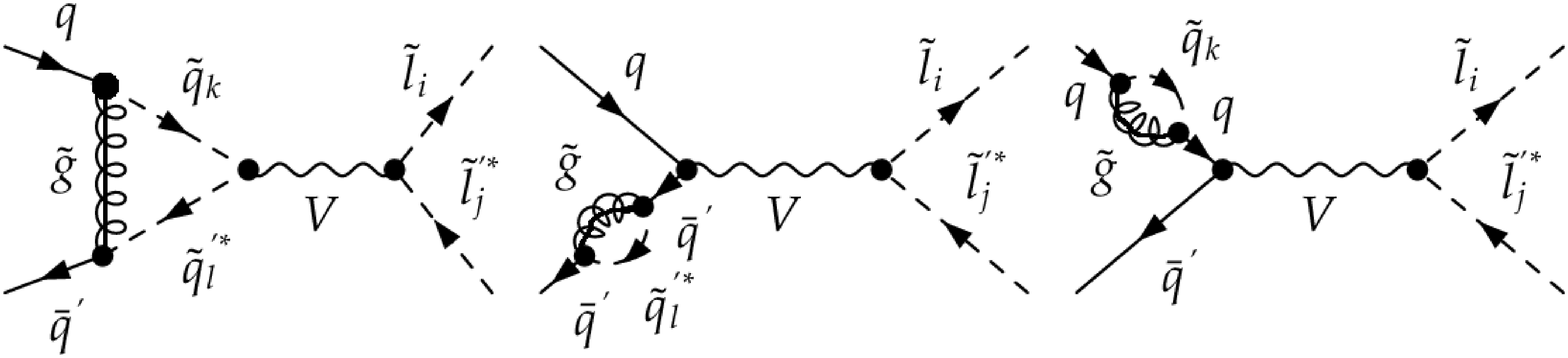}
   \caption{\label{fig:10} Contributions of virtual diagrams for
   slepton pair ($V=\gamma, Z^0$) and slepton-sneutrino associated
   ($V=W^\mp$) production at next-to-leading order in perturbative
   QCD. The first and second lines show the QCD and SUSY-QCD
   corrections, respectively. In the SUSY-QCD case, one has to sum
   over squark mass-eigenstates $k,l=1,2$.}
 \includegraphics[width=.6\columnwidth]{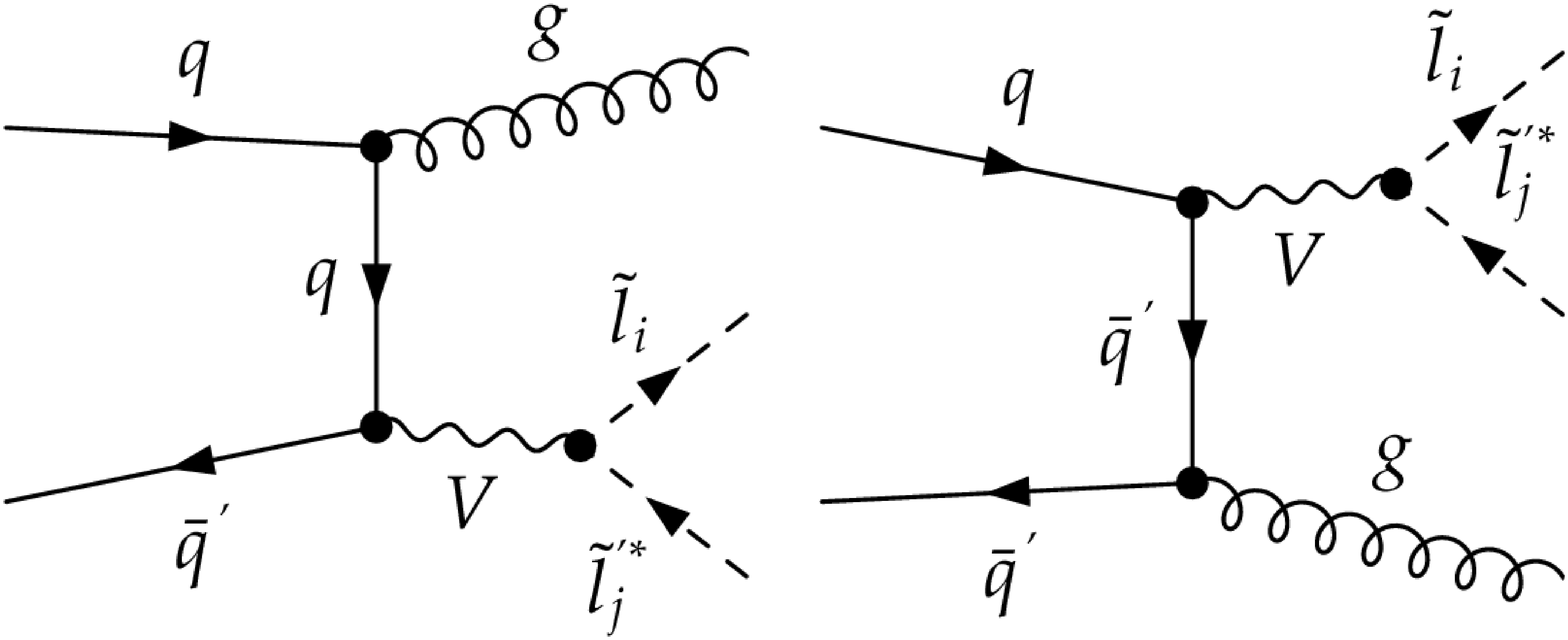}
  \caption{\label{fig:12} Contributions from real gluon emission
  diagrams for slepton pair ($V=\gamma, Z^0$) and slepton-sneutrino
  associated ($V=W^\mp$) production at next-to-leading order in
  perturbative QCD.}
 \includegraphics[width=.6\columnwidth]{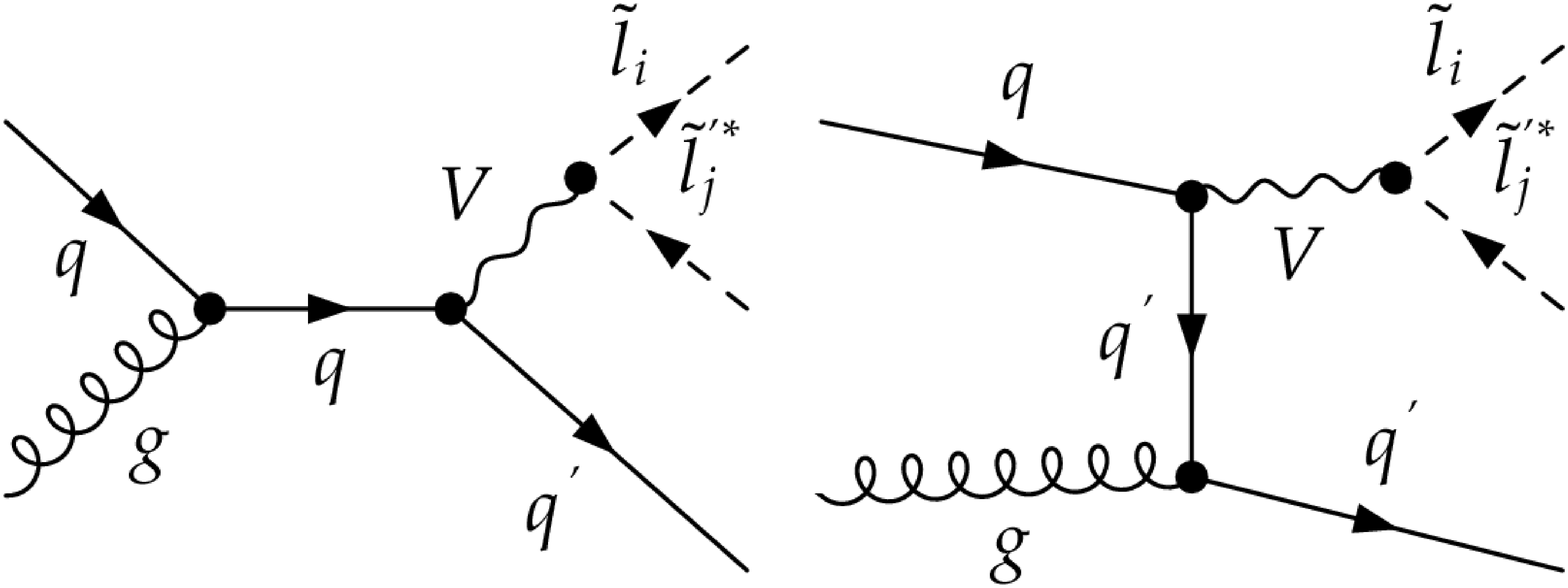}
  \caption{\label{fig:13} Contributions from $qg$ diagrams for
  slepton pair ($V=\gamma, Z^0$) and slepton-sneutrino associated
  ($V=W^\mp$) production at next-to-leading order in perturbative
  QCD.}
\end{figure}

The NLO QCD and SUSY-QCD corrections, i.e.\ the $n=1$-coefficients
in the perturbative expansion of Eq.\ (\ref{eq:pert}), have been
studied for non-mixing sleptons in \cite{Baer:1997nh,
Beenakker:1999xh}. At NLO, the quark-antiquark annihilation
process receives contributions from virtual gluon exchange (see
upper part of Fig.\ \ref{fig:10}), real gluon emission (see Fig.\
\ref{fig:12}) diagrams, and we also have to take into account the
quark-gluon initiated subprocesses (see
Fig.\ \ref{fig:13}).\\

The infrared and collinear singularities of the three-parton cross
sections are extracted using the dipole subtraction formalism
\cite{Catani:1996vz}, and the virtual corrections have been
evaluated in the $\overline{{\rm MS}}$ renormalization scheme. For
the SM QCD diagrams one has the well-known results
\cite{Baer:1997nh, Furmanski:1981cw} \bea
\sigma_{q\bar{q}^{(\prime)}}^{(1;{\rm QCD})}\lr
z,M;\frac{M}{\mu_F}, \frac{M}{\mu_R}\rr &=&
\sigma_0^{(\prime)}(M)\, C_F\, \Bigg[\left(\frac{\pi^2}{3} -
4\right)\, \delta (1-z) + 4 \left( \frac{\ln (1-z)}{1-z}\right)_+
\nonumber \\ & \!-\! & \frac{1+z^2}{1-z} \, \ln z \!-\! 2\, (1+z)
\ln (1-z) \!+ \! \frac{2 P_{qq}^{(0)}(z)}{C_F}\,
\ln{\frac{M^2}{\mu_F^2}} \Bigg],~\label{eq:NLOqq}\nn \\&~&\\
\sigma_{qg}^{(1;{\rm QCD})}\lr z,M;\frac{M}{\mu_F},\frac{M}
{\mu_R}\rr &=& \sigma_0^{(\prime)}(M)\, T_R\,
\Bigg[\left(\frac{1}{2}-z+z^2\right)\ln \frac{(1-z)^2}{z} +
\frac{1}{4} + \frac{3z}{2}  \nonumber \\ & -& \frac{7 z^2}{4} +
\frac{P_{qg}^{(0)}(z)}{T_R}\ln\frac{M^2}{\mu_F^2} \Bigg],~
\label{eq:NLOqg} \eea which expose the LO cross-sections
$\sigma_0^{(\prime)}(M)$ in factorized form. $C_F=4/3$ and
$T_R=1/2$ are the usual QCD colour factors, and $P_{qq,qg}^{(0)}$
are the Altarelli-Parisi splitting functions
\cite{Altarelli:1977zs} \bea P_{qq}^{(0)}(z) &=&
\frac{C_F}{2}\left[\frac{3}{2}\,\delta (1-z) + \frac{2}{(1-z)_+} -
(1+z)\right]~~{\rm and} \\ P_{qg}^{(0)}(z) &=& \frac{T_R}{2}
\left[z^2 + (1-z)^2\right]. \eea We remind the reader that our
normalization corresponds to a perturbative expansion in powers of
$\alpha_s/\pi$.\\

The three lower diagrams of Fig.\ \ref{fig:10} contain SUSY-QCD
corrections. We generalize the results from
\cite{Beenakker:1999xh} to the case of mixed squark mass
eigenstates $k,l=1,2$ in the virtual loop diagrams, obtaining for
neutral and charged currents \cite{Bozzi:2007qr} \bea
\sigma_{q\bar{q}}^{(1;{\rm SUSY})}\lr z, M;\frac{M}{\mu_F},
\frac{M}{\mu_R}\rr &=& \frac{\alpha^2\,\pi\, \beta^3\,C_F}{36\,
M^2} \Bigg[f_\gamma\, e_q^2\, e_l^2\, \delta_{ij} \nn\\ &+&
f_{\gamma Z}\, \frac{e_q\,e_l\, \delta_{ij} {\rm Re} \left[L_{Z
\tilde{l}_i \tilde{l}_j} + R_{Z \tilde{l}_i \tilde{l}_j}
\right]}{4\, x_W\, (1-x_W)\, (1-m_Z^2/M^2)} \nonumber\\ &+& f_Z\,
\frac{\left|L_{Z \tilde{l}_i \tilde{l}_j} \!+\! R_{Z \tilde{l}_i
\tilde{l}_j}\right|^2}{32\, x_W^2\, (1\!-\!x_W)^2
(1\!-\!m_Z^2/M^2)^2}\Bigg]\, \delta(1\!-\!z),~\\
\sigma_{q\bar{q}^\prime}^{(1;{\rm SUSY})}\lr z, M;\frac{M}{\mu_F},
\frac{M}{\mu_R}\rr &=& \frac{\alpha^2\,\pi\, \beta^3\,C_F}{36\,
M^2} \Bigg[\frac{f_W\, \left|L_{W \tilde{l}_i\tilde{\nu}_l}
\right|^2\, \delta(1\!-\!z)}{32\, x_W^2(1\!-\!x_W)^2
(1\!-\!m_W^2/M^2)^2}\Bigg],~~~~~~~\eea where now only the diagonal
squared photon contribution to the Born cross section factorizes.
The virtual loop coefficients $f_\gamma$, $f_{\gamma Z}$, $f_Z$
and $f_W$ are \bea f_\gamma &=& 2 + \sum_{i=1,2} \Bigg[\frac{2\,
m_{\tilde{g}}^2 - 2\, m_{\tilde{q}_i}^2 + M^2}{M^2}\,
\Big(B_{0f}\left(M^2, m_{\tilde{q}_i}^2, m_{\tilde{q}_i}^2\right)
- B_{0f} \left(0, m_{\tilde{g}}^2, m_{\tilde{q}_i}^2\right)\Big)
\nn\\ &+& \left(m_{\tilde{q}_i}^2 - m_{\tilde{g}}^2\right)\,
B^\prime_{0f}\left(0,
m_{\tilde{g}}^2, m_{\tilde{q}_i}^2\right) \nonumber \\
&+&2 \, \frac{m_{\tilde{g}}^4 + (M^2-2\, m_{\tilde{q}_i}^2)\,
m_{\tilde{g}}^2 + m_{\tilde{q}_i}^4}{M^2} C_{0f}\left(0, M^2, 0,
m_{\tilde{q}_i}^2, m_{\tilde{g}}^2, m_{\tilde{q}_i}^2\right)
\Bigg],~\eea \bea f_{\gamma Z} &=& 2\, (L_{Z q q} + R_{Z q q})
\nonumber\\ &+& \sum_{i=1,2}\Bigg[2\,\frac{\left(2\,
m_{\tilde{g}}^2 - 2\, m_{\tilde{q}_i}^2 + M^2\right)\,{\rm
Re}\Big[L_{Z \tilde{q}_i \tilde{q}_i} + R_{Z \tilde{q}_i
\tilde{q}_i}\Big]}{M^2}\,B_{0f}\left(M^2, m_{\tilde{q}_i}^2,
m_{\tilde{q}_i}^2\right) \Bigg] \nonumber\\
&-&\sum_{i=1,2}\!\Bigg[2\,\frac{\left(2\, m_{\tilde{g}}^2 \!-\!
2\, m_{\tilde{q}_i}^2 \!+\! M^2\right)\,\Big(L_{Z q q}\,
\Big|S^{\tilde{q}}_{i1}\Big|^2 \!+\! R_{Z q q}\,
\Big|S^{\tilde{q}}_{i1}\Big|^2 \Big)}{M^2}\, B_{0f}\left(0,
m_{\tilde{g}}^2, m_{\tilde{q}_i}^2\right)\Bigg] \nonumber\\
&+& \sum_{i=1,2}\Bigg[\left(m_{\tilde{q}_i}^2 - m_{\tilde{g}}^2
\right) (L_{Z q q}+R_{Z q q}) \, B_{0f}^\prime\left(0,
m_{\tilde{g}}^2, m_{\tilde{q}_i}^2\right)\Bigg]\nonumber\\
&+& \sum_{i=1,2}\Bigg[ 4 \, \frac{\left(m_{\tilde{g}}^4 + (M^2 -
2\, m_{\tilde{q}_i}^2)\, m_{\tilde{g}}^2 +
m_{\tilde{q}_i}^4\right)\,{\rm Re}\Big[L_{Z \tilde{q}_i
\tilde{q}_i} + R_{Z \tilde{q}_i \tilde{q}_i}\Big]}{M^2}\nn
\\&& \hspace{12mm} \times  C_{0f}\left(0, M^2, 0, m_{\tilde{q}_i}^2,
m_{\tilde{g}}^2, m_{\tilde{q}_i}^2\right)\Bigg],~\eea \bea f_Z &=&
2\, \left(L_{Z q q}^2+R_{Z q q}^2\right) \nonumber \\ &+&
\sum_{i,j=1,2} \left[2\, \frac{\left(2\,m_{\tilde{g}}^2 \!-\!
m_{\tilde{q}_i}^2 \!-\! m_{\tilde{q}_j}^2 \!+\!
M^2\right)}{M^2}\,\Big|L_{Z \tilde{q}_i \tilde{q}_j} \!+\! R_{Z
\tilde{q}_i \tilde{q}_j}\Big|^2\, B_{0f}\left(M^2,
m_{\tilde{q}_i}^2, m_{\tilde{q}_j}^2\right) \right]\nonumber
\eea\bea &-&\!\sum_{i=1,2}\Bigg[2\, \frac{\left(2\,
m_{\tilde{g}}^2 \!-\! 2\, m_{\tilde{q}_i}^2 \!+\!
M^2\right)\,\Big(L^2_{Z q q}\,\Big|S^{\tilde{q}}_{i1}\Big|^2 \!+\!
R^2_{Z q q}\, \Big|S^{\tilde{q}}_{i2}\Big|^2
\Big)}{M^2}\,B_{0f}\left(0, m_{\tilde{g}}^2,
m_{\tilde{q}_i}^2\right) \Bigg] \nonumber \\ &+& \!
\sum_{i=1,2}\Bigg[\left(m_{\tilde{q}_i}^2 - m_{\tilde{g}}^2\right)
\left(L_{Z q q}^2 + R_{Z q q}^2\right)\, B_{0f}^\prime\left(0,
m_{\tilde{g}}^2, m_{\tilde{q}_i}^2\right)\Bigg]\nonumber\\
&+& \! \sum_{i,j=1,2}\Bigg[4\,\frac{\left(m_{\tilde{g}}^4 +
\left(M^2 - m_{\tilde{q}_i}^2 - m_{\tilde{q}_j}^2\right)\,
m_{\tilde{g}}^2 + m_{\tilde{q}_i}^2\, m_{\tilde{q}_j}^2
\right)}{M^2}\, \Big|L_{Z \tilde{q}_i \tilde{q}_j} + R_{Z
\tilde{q}_i \tilde{q}_j}\Big|^2\nn
\\ && \hspace{12mm} \times C_{0f}\left(0, M^2, 0, m_{\tilde{q}_i}^2,
m_{\tilde{g}}^2, m_{\tilde{q}_j}^2\right) \Bigg],~ \\
f_W &=& 2\,\Big|L_{W q q^\prime}\Big|^2 \nonumber \\
&+& \! \sum_{i,j=1,2} \Bigg[ \frac{2 \left(2\, m_{\tilde{g}}^2
\!-\! m_{\tilde{q}_i}^2 \!-\! m_{\tilde{q}^\prime_j}^2 \!+\!
M^2\right) \Big|L_{W \tilde{q}_i
\tilde{q}_j^\prime}\Big|^2}{M^2}\, B_{0f}\left(M^2,
m_{\tilde{q}_i}^2, m_{\tilde{q}^\prime_j}^2\right)\!\Bigg]
\nonumber \\ &-& \! \sum_{\tilde{Q}=\tilde{q},\tilde{q}^\prime} \!
\sum_{i=1,2} \Bigg[\frac{\left(2\, m_{\tilde{g}}^2 \!-\! 2\,
m_{\tilde{Q}_i}^2 \!+\! M^2\right) B_{0f}\left(0, m_{\tilde{g}}^2,
m_{\tilde{Q}_i}^2 \right)\Big|L_{W q q^\prime} \,
S^{\tilde{Q}}_{i1}\, \Big|^2 }{M^2} \Bigg]\nonumber \\ &+& \!
\sum_{\tilde{Q}=\tilde{q},\tilde{q}^\prime} \, \sum_{i=1,2}
\Bigg[\frac{1}{2} \left(m_{\tilde{Q}_i}^2 -
m_{\tilde{g}}^2\right)\, \Big|L_{W q q^\prime}\Big|^2\,
B_{0f}^\prime\left(0, m_{\tilde{g}}^2, m_{\tilde{Q}_i}^2\right)
\Bigg]\nonumber \\ &+& \!\sum_{i,j=1,2}\Bigg[\frac{4
\left(m_{\tilde{g}}^4 - \left(m_{\tilde{q}_i}^2 +
m_{\tilde{q}^\prime_j}^2 - M^2\right)\, m_{\tilde{g}}^2 +
m_{\tilde{q}_i}^2\, m_{\tilde{q}^\prime_j}^2\right) \Big|L_{W
\tilde{q}_i \tilde{q}_j^\prime}\Big|^2}{M^2}\nn
\\ && \hspace{12mm} \times C_{0f}\left(0, M^2, 0,
m_{\tilde{q}^\prime_i}^2, m_{\tilde{g}}^2,
m_{\tilde{q}_j}^2\right)\Bigg].~\eea The functions $B_{0f}(p^2,
m_1^2, m_2^2)$, $B_{0f}^\prime(p^2,m_1^2,m_2^2)$ and
$C_{0f}(p_1^2, (p_1 + p_2)^2, p_2^2, m_1^2,m_2^2, m_3^2)$ are the
finite parts of the scalar two- and three-point functions \bea
B_0(p^2, m_1^2,m_2^2) &=& \mu_R^{2\varepsilon}\, \int \frac{{\rm
d}^Dq} {i \pi^2}\, \frac{1}{q^2 \!-\! m_1^2}\frac{1}{(q\!+\!p)^2
\!-\!m_2^2},~\\ B_0^\prime(p^2, m_1^2, m_2^2) &=& \frac{{\rm
d}B_0(k^2, m_1^2, m_2^2)}{{\rm d}k^2}\Big|_{k^2=p^2},~\\
C_0(p_1^2, (p_1 + p_2)^2, p_2^2, m_1^2, m_2^2, m_3^2) &=&
\mu_R^{2\varepsilon}\, \int \frac{{\rm d}^Dq}{i \pi^2}\,
\frac{1}{q^2 - m_1^2} \frac{1}{(q+p_1)^2 - m_2^2}\nn\\ &&\times
\frac{1}{(q+p_1+p_2)^2 - m_3^2}. \eea Our results agree with those
of Ref.\ \cite{Djouadi:1999ht} in the case of mass-degenerate
non-mixing squarks. Note that the quark mass, which appears in the
off-diagonal mass matrix elements of the squarks running in the
loops (see Eq.\ (\ref{eq:sfermmass})), corresponds to a linear
Yukawa coupling in the superpotential and can not be neglected,
even if it is much smaller than the total centre-of-mass energy of
the colliding partons allowing for a massless factorization inside
the outer hadrons. The full NLO contributions to the cross section
are then given by \bea&& \hspace{-12mm}
\sigma_{q\bar{q}^{(\prime)}}^{(1)} \!\lr z,M;\frac{M}{\mu_F},
\frac{M}{\mu_R}\rr \!\!=\! \sigma_{q\bar{q}^{(\prime)}}^{(1;{\rm
QCD})}\lr z, M;\frac{M}{\mu_F}, \frac{M}{\mu_R}\rr \!+\!
\sigma_{q\bar{q}^{(\prime)}}^{(1;{\rm SUSY})}\lr
z,M;\frac{M}{\mu_F}, \frac{M}{\mu_R}\rr,~~~~\label{eq:SUSYZ}\\
&&\hspace{-12mm} \sigma_{q g}^{(1)}\lr z, M;\frac{M}{\mu_F},
\frac{M}{\mu_R}\rr = \sigma_{q g}^{(1;{\rm QCD})}\lr z,M;
\frac{M}{\mu_F},\frac{M}{\mu_R}\rr.~ \label{eq:SUSYW}\eea

\subsection{Threshold-enhanced contributions}

Eqs.\ (\ref{eq:NLOqq}) and (\ref{eq:NLOqg}) explicitly show
logarithmic terms of the form $\alpha_s [\ln(1-z)/(1-z)]_+$. When
the initial partons have just enough energy to produce the slepton
pair in the final state, i.e.\ $z$ is close to one, these terms
can become large and have to be resummed to all order in $\as$ in
order to correctly quantify the effect of this set of corrections,
corresponding to soft-collinear gluon emission. Collinear
parton-emission can also be taken into account, using the
collinear-improved resummation formalism which is described in
Sec.\ \ref{sec:thresh} for Drell-Yan pair production, which can
directly be applied to slepton pair production by replacing the
Born cross sections by those of Eqs.\ (\ref{eq:sig0Z2}) and
(\ref{eq:sig0W2}). In Mellin space, we have then \bea
\hat\sigma^{({\rm res})}_{a b}(N, \alpha_s) =
\sigma_0^{(\prime)}\, \tilde{C}_{a b}(\alpha_s)\, \exp\Big[
S(N,\alpha_s)\Big],~\label{eq:th_sl}\eea with \bea
\tilde{C}^{(1)}_{q \bar{q}} &=& C_F\lr\frac{2\, \pi^2}{3} - 4 +
\frac{3}{2}\ln\frac{M^2}{\mu^2_F}\rr + 2\, A^{(1)}\,
\frac{\ln{\bar{N}} - \frac{1}{2} \ln{\frac{M^2}{\mu_F^2}}}{N}~, \\
\tilde{C}^{(1)}_{q g} &=& - T_R\,\frac{\ln{\bar{N}} - \frac{1}{2}
\ln{\frac{M^2}{\mu_F^2}}}{N},~ \eea the Sudakov form factor $S$
being defined by Eqs.\ (\ref{eq:sudthresh}), (\ref{eq:tg1}) and
(\ref{eq:tg2}). The errors due to missing high-order corrections
can be estimated by comparing our results with those obtained by
taking the resummed contribution given by Eq.\ (\ref{eq:thr_ELM}),
where the hard contributions exponentiate as well, \bea
\hat\sigma^{({\rm res})}_{a b}(N,\alpha_{s}) = \sigma^{(LO)}\,
\exp\Big[C^{(1)}_{q\bar{q}}(\alpha_s)\Big] \,\exp
\Big[S(N,\alpha_{s})\Big]~.\label{eq:th_eyn}\eea The matching with
the fixed-order results is achieved through Eq.\ (\ref{eq:mtcth})
\bea \sigma = \sigma^{({\rm F.O.})} + \frac{1}{2\, \pi\, i}
\int_{C_{MP} - i\, \infty}^{C_{MP} + i\, \infty} \td N \, \tau^{-N}
\le \sigma^{({\rm res})}(N, M) - \sigma^{({\rm exp})}(N, M)\re.~
\label{eq:mtcth2}\eea The truncation of the resummed cross section
to the same perturbative order as $\sigma^{({\rm F.O.})}$, reads
for a NLL+NLO matching (i.e.\ at order $\as$) as in Eqs.\
(\ref{eq:expqq}) and (\ref{eq:expqg}) \bea \label{eq:expqq2}
\hat\sigma^{({\rm exp})}_{q \bar{q}}(N,M) &=& \sigma_0^{(\prime)}
\le 1 + \frac{\as}{\pi}\, \lr C_{F}\,\Big( 2\, \ln^2\bar{N} - 2\,
\ln \bar{N} \ln\frac{M^{2}}{\mu^{2}_{F}} \Big)
+ \tilde{C}^{(1)}_{q \bar{q}} \rr\re,~ \\
\label{eq:expqg2} \hat\sigma^{({\rm exp})}_{q g}(N, M) &=&
\sigma_0^{(\prime)}\le \frac{\as}{\pi}\, \tilde{C}^{(1)}_{q g}
\re.~\eea

Let us note that in Mellin-space, the large-$N$ limit of the fixed
order NLO cross sections of Eqs.\ (\ref{eq:NLOqq}) and
(\ref{eq:NLOqg}) is correctly reproduced by the expansion of the
resummed cross section at order $\alpha_s$, including even terms
that are suppressed by $1/N$, as expected from the
collinear-improved resummation formalism (see Eqs.\
(\ref{eq:exaqq}) and (\ref{eq:exaqg})).

\subsection{Numerical results}

\begin{figure}
\centering
\includegraphics[width=.7\columnwidth]{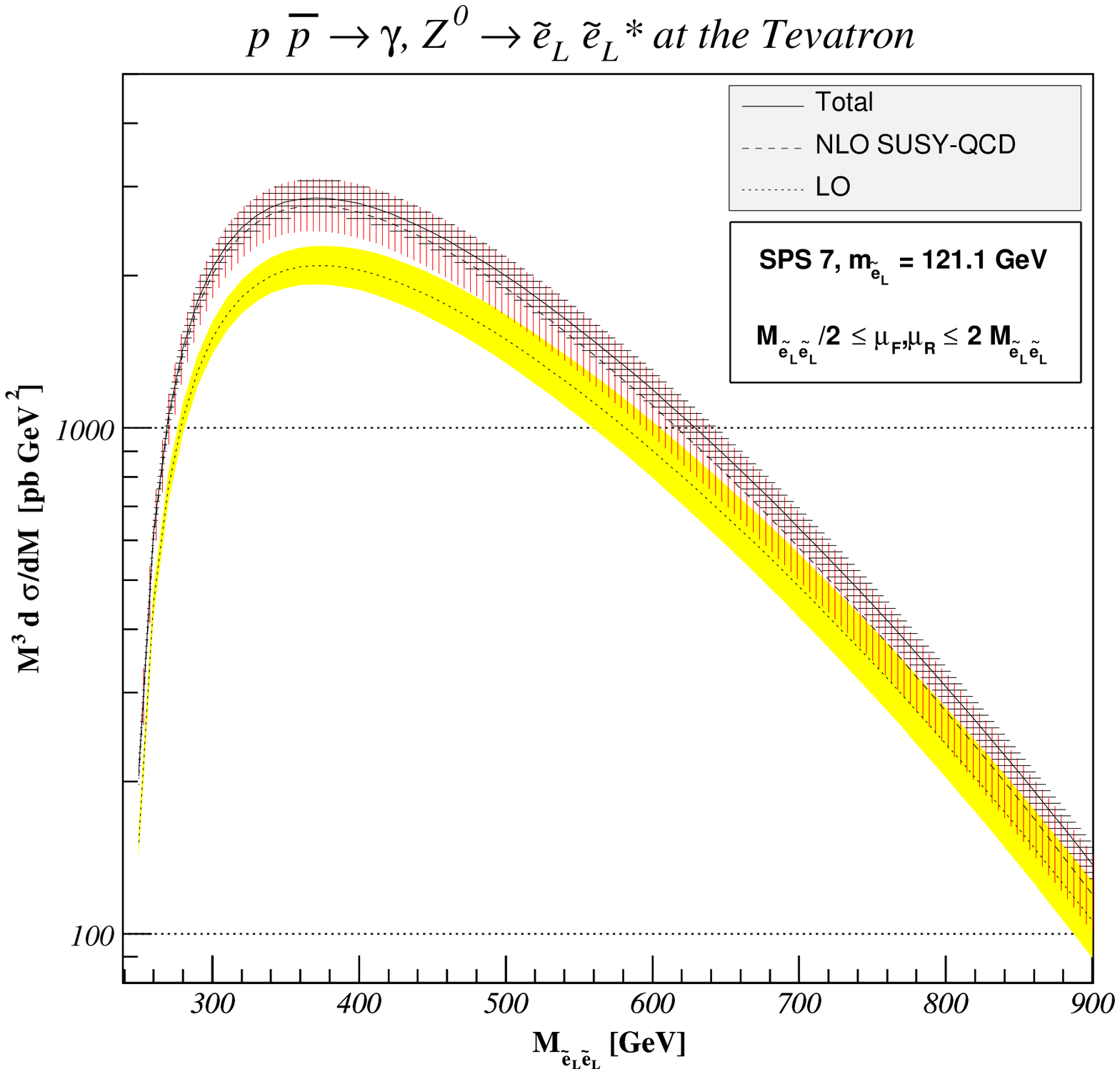}
\includegraphics[width=.7\columnwidth]{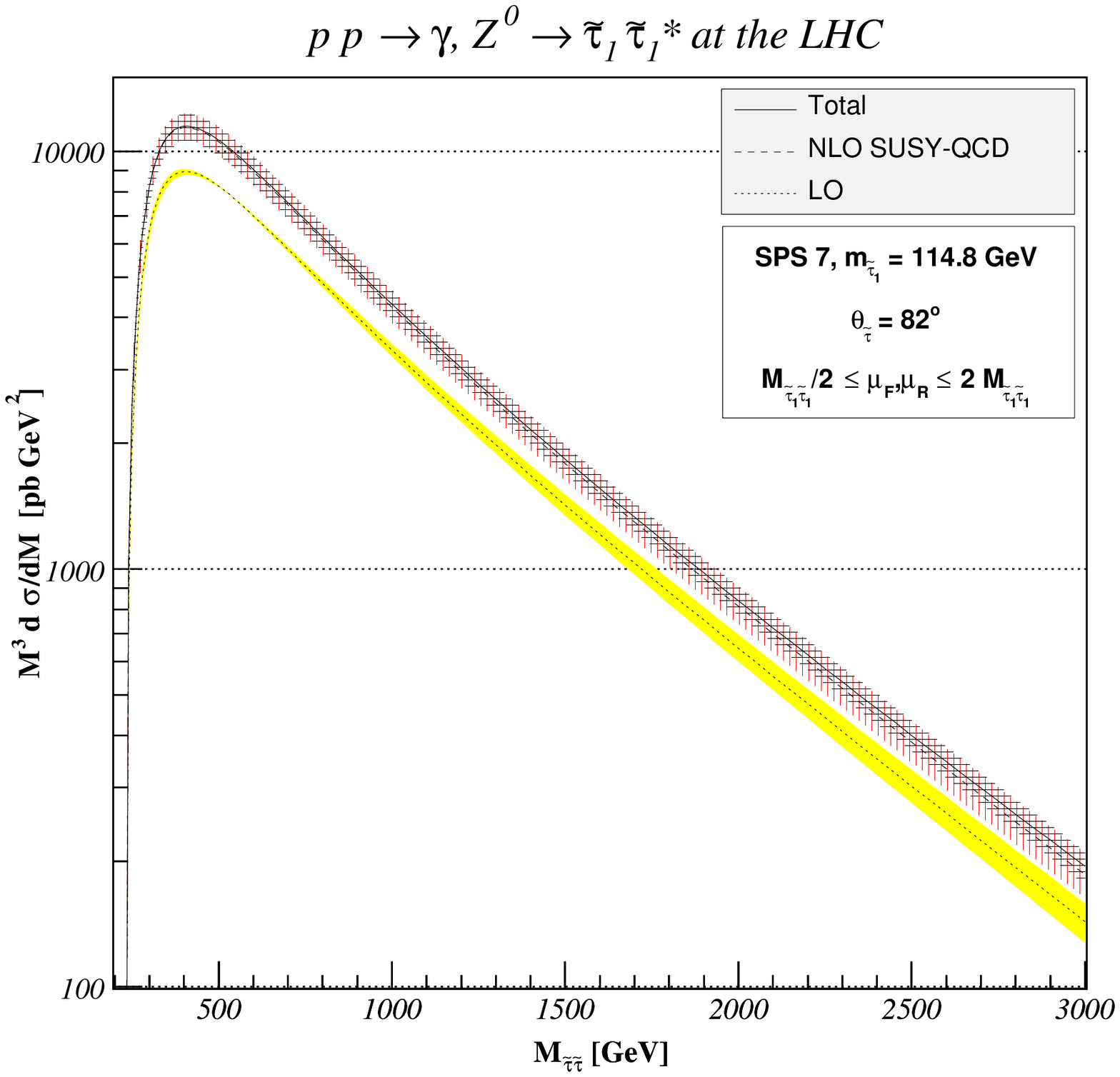}
  \caption{\label{fig:14}Invariant-mass distribution
  $M^3\,\td\sigma/\td M$ of $\tilde{e}_L$ pairs at the Tevatron (top)
  and $\tilde{\tau}_1$ pairs at the LHC (bottom) for the benchmark
  point SPS 7. We show the total NLL+NLO matched and the fixed-order
  NLO SUSY-QCD and LO QCD results, including the respective scale
  uncertainties as horizontally hatched, vertically hatched and
  shaded bands.}
\end{figure}

We take current masses and widths of the electroweak gauge bosons,
electroweak mixing angle, electromagnetic fine structure constant
and CKM matrix elements, as in Sec.\ \ref{sec:scan}
\cite{Yao:2006px}, and the physical masses of the SUSY particles
and the mixing angles are obtained with the computer program
SuSpect \cite{Djouadi:2002ze}. We choose one mSUGRA point, SPS 1a,
and one GMSB point, SPS 7, with their associated model lines, as
benchmarks for our numerical study \cite{Allanach:2002nj}. SPS 1a
is a typical mSUGRA point with an intermediate value of
$\tan\beta= 10$ and $\mu>0$, and the model line attached to it is
specified by $m_0=-A_0=0.4~m_{1/2}$. For $m_{1/2}=250$ GeV, this
SUSY-breaking scenario leads to light sleptons $\tilde{\tau}_1$,
$\tilde{e}_1$, $\tilde{\tau}_2$, $\tilde{e}_2$, $\tilde{\nu}_\tau$
and $\tilde{\nu}_e$ with masses of 136.2, 146.4, 216.3, 212.3,
196.1 and 197.1 GeV and to heavy squarks with masses around
500-600 GeV. However, even if the top-squark mass eigenstate
$\tilde{t}_1$ is slightly lighter, it does nonetheless not
contribute to the virtual squark loops due to the negligible
top-quark density in the proton (we consider flavour-conserving
SUSY-loops in Fig.\ \ref{fig:10}). SPS 7 is a GMSB scenario with a
$\tilde{\tau}_1$ as the next-to-lightest SUSY particle and an
effective SUSY-breaking scale $\Lambda=40$ TeV, $N_{\rm mes}=3$
messenger fields of mass $M_{\rm mes}=80$ TeV, $\tan\beta=15$, and
$\mu>0$, which leads again to light sleptons with masses of 114.8,
121.1, 263.9, 262.1, 249.5 and 249.9 GeV, respectively, and even
heavier squarks with masses around 800-900 GeV. Its model line is
defined by $M_{\rm mes} = 2\, \Lambda$. The slepton masses and
mixing angles are actually quite similar for the SPS 1a mSUGRA and
SPS 7 GMSB points, so that the corresponding production cross
sections will not differ significantly. Slepton detection will,
however, be slightly different in both scenarios, as the sleptons
decay to a relatively massive neutralino LSP at SPS 1a, but to a
very light gravitino LSP at SPS 7. The lightest tau slepton thus
decays into
a tau lepton and missing transverse energy. \\

Our cross sections are calculated for the Tevatron
$p\bar{p}$-collider, as well as for the LHC $pp$-collider. For the
LO (NLO and NLL) predictions, we use the LO 2001
\cite{Martin:2002dr} (NLO 2004 \cite{Martin:2004ir}) MRST-sets of
parton distribution functions. For the NLO and NLL predictions,
$\alpha_s$ is evaluated with the corresponding value of
$\Lambda_{\overline{\rm MS}}^{n_f=5}=255$ MeV at two-loop
accuracy. We fix the unphysical scales $\mu_{F}$ and $\mu_{R}$
equal to the invariant-mass $M$ of the slepton (slepton-sneutrino)
pair.\\

The invariant-mass distribution $M^3\td\sigma/\td M$ for first- (and
equal-mass second-) generation sleptons at the Tevatron is shown
in the top panel of Fig.\ \ref{fig:14}, and the one for (slightly
lighter) third-generation sleptons at the LHC in the bottom panel
of the same figure. In both cases, we have chosen the SPS 7 GMSB
benchmark point. The differential cross section $\td\sigma/\td M$
has been multiplied by a factor $M^3$ in order to remove the
leading mass dependence of propagator and phase space factors. As
is to be expected for $P$-wave production of scalar particles, the
distributions rise above the threshold at $\sqrt{s}=2
m_{\tilde{l}}$ with the third power of the slepton velocity
$\beta$, see Eq.\ (\ref{eq:sig0Z}), and peak at about 100 GeV
above threshold (at 370 GeV for $M^3\td\sigma/\td M$ and 310 GeV for
$\td\sigma/\td M$ for the Tevatron; 410 GeV and 300 GeV for the
LHC), before falling off steeply due to the $s$-channel propagator
and the decreasing parton luminosity. Furthermore, it can also be
seen that the QCD corrections do not alter the $P$-wave velocity
dependence close to threshold. At the Tevatron, the total and NLO
SUSY-QCD predictions exceed the maximal LO cross section by 36 and
31\%, respectively, whereas at the LHC, the maximal cross section
increases by 28 and 27\%. Threshold resummation effects are thus
clearly more important at the Tevatron, where the hadronic
centre-of-mass energy is limited and the scaling variable
$\tau=M^2/s_h$ is closer to one, and they increase with $M$ to the
right of both plots.\\

The maximal theoretical error is estimated in Fig.\ \ref{fig:14}
by an independent variation of the factorization and
renormalization scales between $M/2$ and $2M$. It is indicated as
a shaded, vertically, and horizontally hatched band for the LO,
NLO SUSY-QCD, and the total prediction. At LO, the only dependence
comes from the factorization scale. It increases with the
momentum-fraction $x$ of the partons in the proton or anti-proton
and is therefore already substantial for small $M$ at the
Tevatron, but only for larger $M$ at the LHC. At NLO, this
dependence is reduced due to the factorization of initial-state
singularities, but a strong additional dependence is introduced by
the renormalization scale in the coupling $\alpha_s(\mu_R)$. After
resummation, this dependence is reduced as well, so that the total
scale uncertainty at the Tevatron diminishes from 20\%--35\% for
NLO to only 16\%--17\% for the matched resummed result. The
reduction is, of course, more important in the large-$M$ region.
At the LHC, where $\alpha_s$ is evaluated at a larger
renormalization scale and is thus less sensitive to it, the
corresponding numbers are 18\%--25\% and 15\%--17\%.\\

\begin{figure}
\centering
\includegraphics[width=.7\columnwidth]{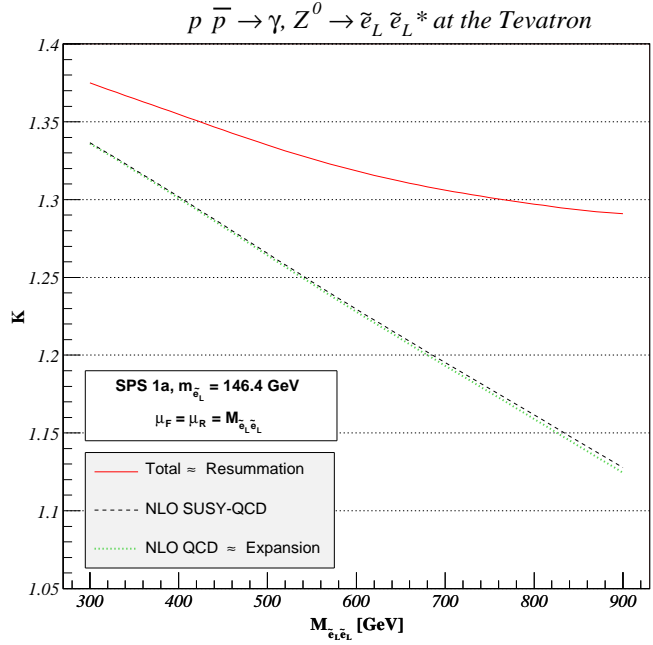}
\includegraphics[width=.7\columnwidth]{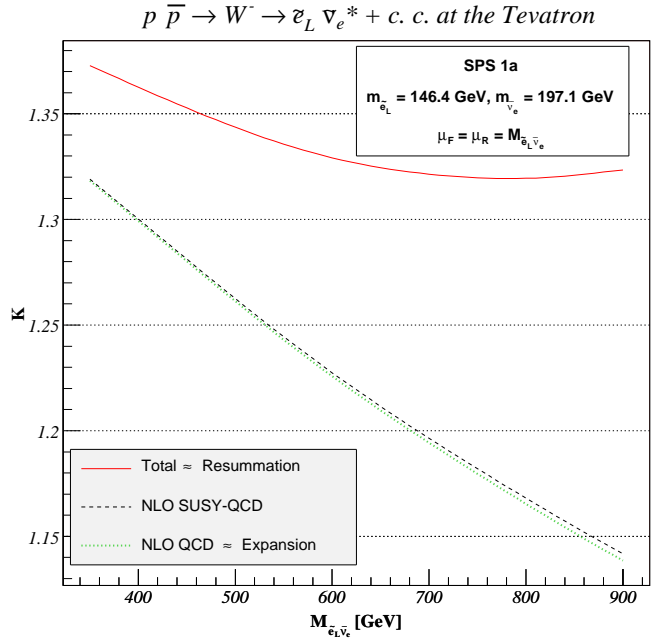}
  \caption{\label{fig:16}$K$-factors as defined in Eq.\ (\ref{eq:K})
  for $\tilde{e}_L$ pair (top) and associated
  $\tilde{e}_L\tilde{\nu}_e^\ast$ production (bottom) at the
  Tevatron for the benchmark point SPS 1a. We show the total NLL+NLO
  matched result, which is almost identical to the purely resummed
  result at NLL, as well as the fixed-order NLO SUSY-QCD and QCD
  results. The latter practically coincides with the resummed result
  expanded up to NLO.}
\end{figure}
\begin{figure}
\centering
\includegraphics[width=.7\columnwidth]{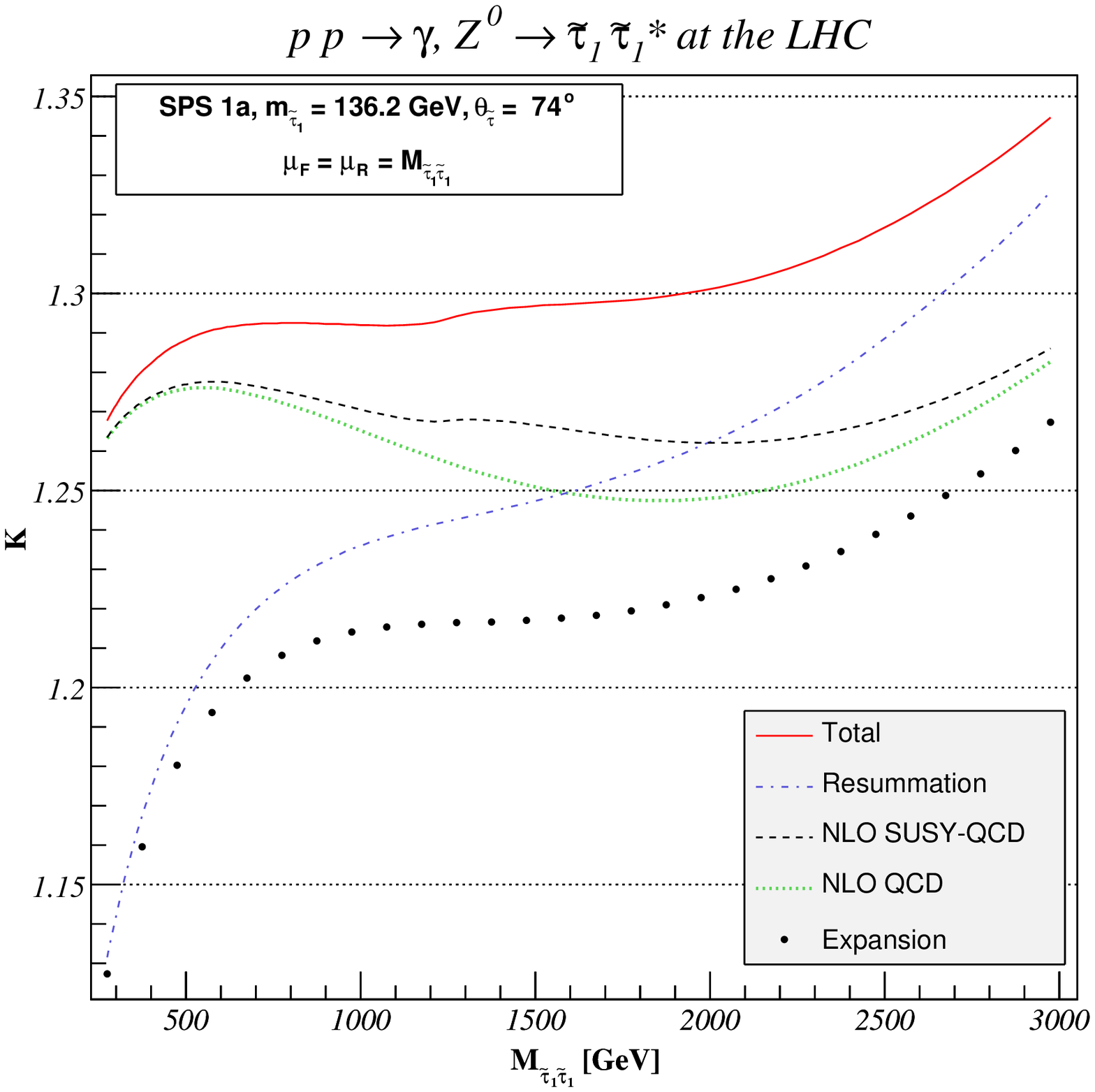}
\includegraphics[width=.7\columnwidth]{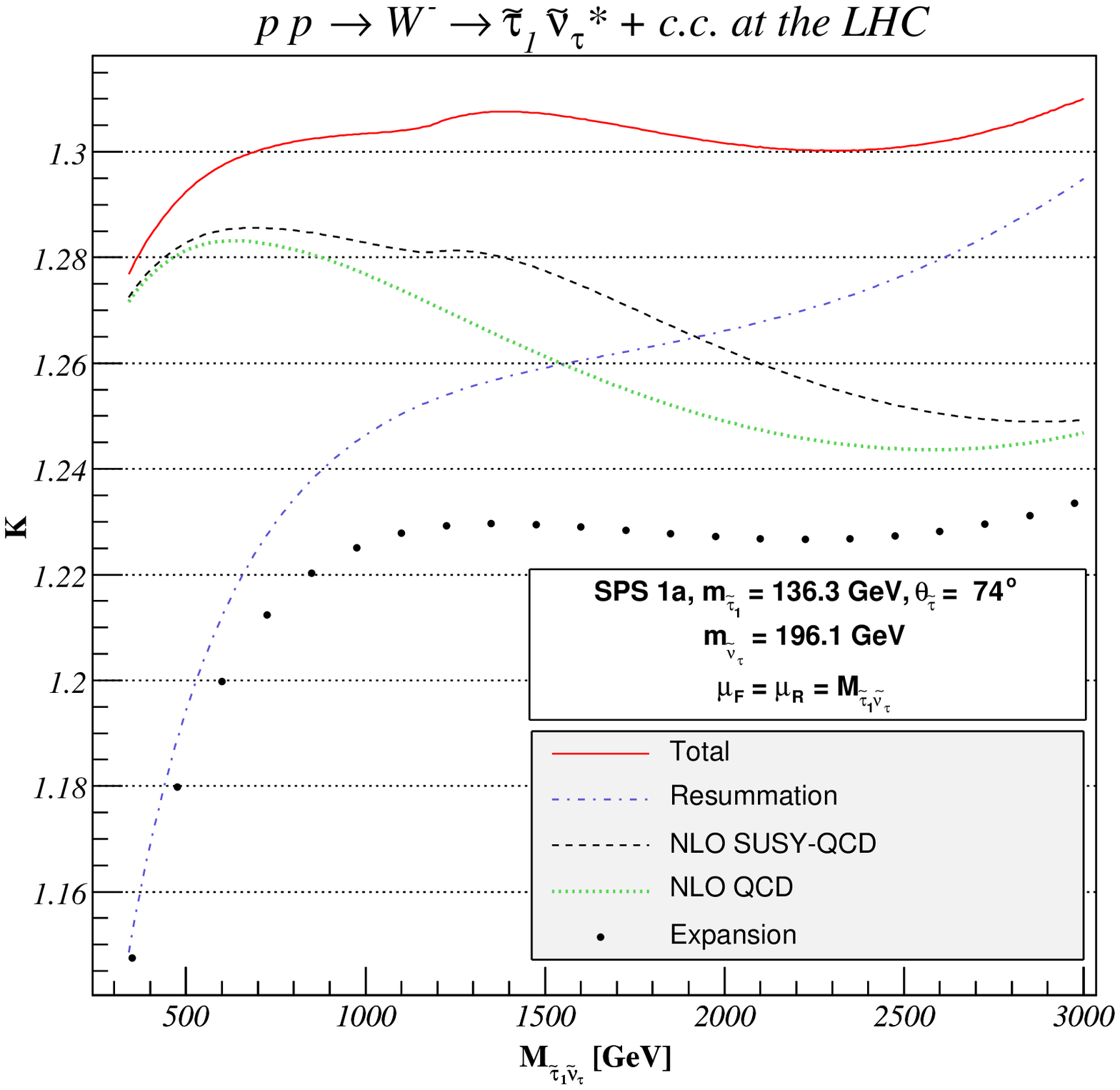}
  \caption{\label{fig:18}$K$-factors as defined in Eq.\ (\ref{eq:K})
  for $\tilde{\tau}_1$ pair (top) and associated
  $\tilde{\tau}_1\tilde{\nu}_\tau ^\ast$ production (bottom) at the
  LHC for the benchmark point SPS 1a. We show the total NLL+NLO
  matched result, the resummed result at NLL, the fixed-order NLO
  SUSY-QCD and QCD results, and the resummed result expanded up to
  NLO.}
\end{figure}

For the mSUGRA scenario SPS 1a, we show in Figs.\ \ref{fig:16} and
\ref{fig:18} the cross section correction factors \bea
\label{eq:K} K^i = \frac{{\rm d}\sigma^i / {\rm d}M}{{\rm
d}\sigma^{\rm LO} / {\rm d}M}, \eea where $i$ labels the
corrections induced by NLO QCD (Eqs.\ (\ref{eq:NLOqq}) and
(\ref{eq:NLOqg})), additional NLO SUSY-QCD (Eqs.\ (\ref{eq:SUSYZ})
and (\ref{eq:SUSYW})), resummation (Eq.\ (\ref{eq:th_sl})), and
the matched total contributions (Eq.\ (\ref{eq:mtcth2})) as well
as the fixed-order expansion (Eqs.\ (\ref{eq:expqq2}) and
(\ref{eq:expqg2})) of the resummation contribution as a function
of the invariant-mass $M$. As one can see immediately, the
mass-dependence of these corrections for charged-current
associated production of sleptons and sneutrinos (lower parts of
Figs.\ \ref{fig:16} and Fig.\ \ref{fig:18}) does not differ
substantially from the mass-dependence of the neutral-current
production of slepton pairs (upper parts). \\

At the Tevatron (Fig.\ \ref{fig:16}), where we are close to the
threshold, resummation effects are already important at low $M$
(4\%) and increase to sizeable 16\% at large $M$. The NLO QCD
result is thus dominated by large logarithms and coincides with
the expanded result at the permille level. In addition, the
relative importance of the (finite) SUSY-QCD contributions is
reduced, and the total prediction coincides with the resummed
prediction, since fixed-order and expanded contributions cancel
each other in Eq.\ (\ref{eq:mtcth2}). We have also verified that
exponentiating the finite ($N$-independent) terms collected in the
coefficient function $C^{(1)}_ {q\bar{q}^{(\prime)}}$, as proposed
in \cite{Eynck:2003fn} (see Eq.\ \ref{eq:th_eyn}), leads only to a
0.6\%--0.8\% increase of the matched resummed result. The Tevatron
being a $p\bar{p}$-collider, the total cross section is dominated
by $q\bar{q}$-annihilation, and $qg$-scattering contributes at
most 1\% at small $M$ (or small $x$), where the gluon density is
still appreciable. Integration over $M$ leads to total cross
sections for the neutral (charged) current processes in Fig.\
\ref{fig:16} of 4.12 (3.92) fb in LO, 5.3 (4.96) fb in NLO
(SUSY-)QCD, and 5.55 (5.28) fb for the matched resummed
calculation. The corresponding (global) $K$-factors \bea
\label{eq:kglob} K^i_{{\rm glob}} &=& \frac{\sigma^{i~~}}
{\sigma^{\rm LO}} ~=~ \frac{\int {\rm d}M\, {\rm d}\sigma^{i~~} /
{\rm d}M}{\int {\rm d}M\,{\rm d}\sigma^{\rm LO} / {\rm d}M} \eea
are then 1.29 (1.27) at fixed-order and 1.35 (1.35) with
resummation.\\

At the LHC, sleptons can be produced with relatively small
invariant-mass $M$ compared to the total available centre-of-mass
energy, so that $z=\tau/(x_ax_b)=M^2/s\ll 1$ and the resummation
of ($1-z$)-logarithms is less important. This is particularly true
for the production of the light mass-eigenstates of mixing
third-generation sleptons, as shown in Fig.\ \ref{fig:18}. In the
low-$M$ (left) parts of these plots, the total result is less than
0.5\% larger than the NLO (SUSY-)QCD result. Only at large $M$ the
logarithms become important and lead to a 7\% increase of the
$K$-factor with resummation over the fixed-order result. In this
region, the resummed result approaches the total prediction, since
the NLO QCD calculation is dominated by large logarithms and
approaches the expanded resummed result. However, we are still far
from the hadronic threshold region, and a consistent matching of
both resummed and fixed-order contributions is needed. At low $M$,
where finite terms dominate, the resummed contribution is close to
its fixed-order expansion and disappears with $M$. In the
intermediate-$M$ region, one can observe the effect of SUSY-QCD
contributions, in particular the one coming from the
$\tilde{q}\tilde{q}\tilde{g}$-vertex correction (lower left
diagram in Fig.\ \ref{fig:10}). As $M \geq 2 m_{\tilde{q}}$, one
crosses the threshold for squark pair production and observes a
resonance in Fig.\ \ref{fig:18}. As for the Tevatron,
exponentiating the finite ($N$-independent) terms collected in the
coefficient function $C^{(1)}_ {q\bar{q}^{(\prime)}}$ leads only
to a 1\% increase of the matched resummed result. The LHC being a
high-energy $pp$-collider, it has a significant gluon-luminosity,
in particular at small $M$ (or $x$), and indeed the
$qg$-subprocess changes (lowers) the total cross section by 7\% at
small $M$ and 3\% at large $M$. After integration over $M$, we
obtain total cross sections of 27 (9.59) fb in LO, 34.3 (12.3) fb
in NLO SUSY-QCD, and 34.6 (12.5) fb for the resummed-improved
result, corresponding to global $K$-factors of 1.28 for
fixed-order and 1.29 for the matched resummed cross section for
both processes. Resummation of large logarithms is thus not as
important as for the Tevatron at the benchmark point SPS 1a.

\section{Total cross section at NLO}

\begin{figure}
\centering
\includegraphics[width=.7\columnwidth]{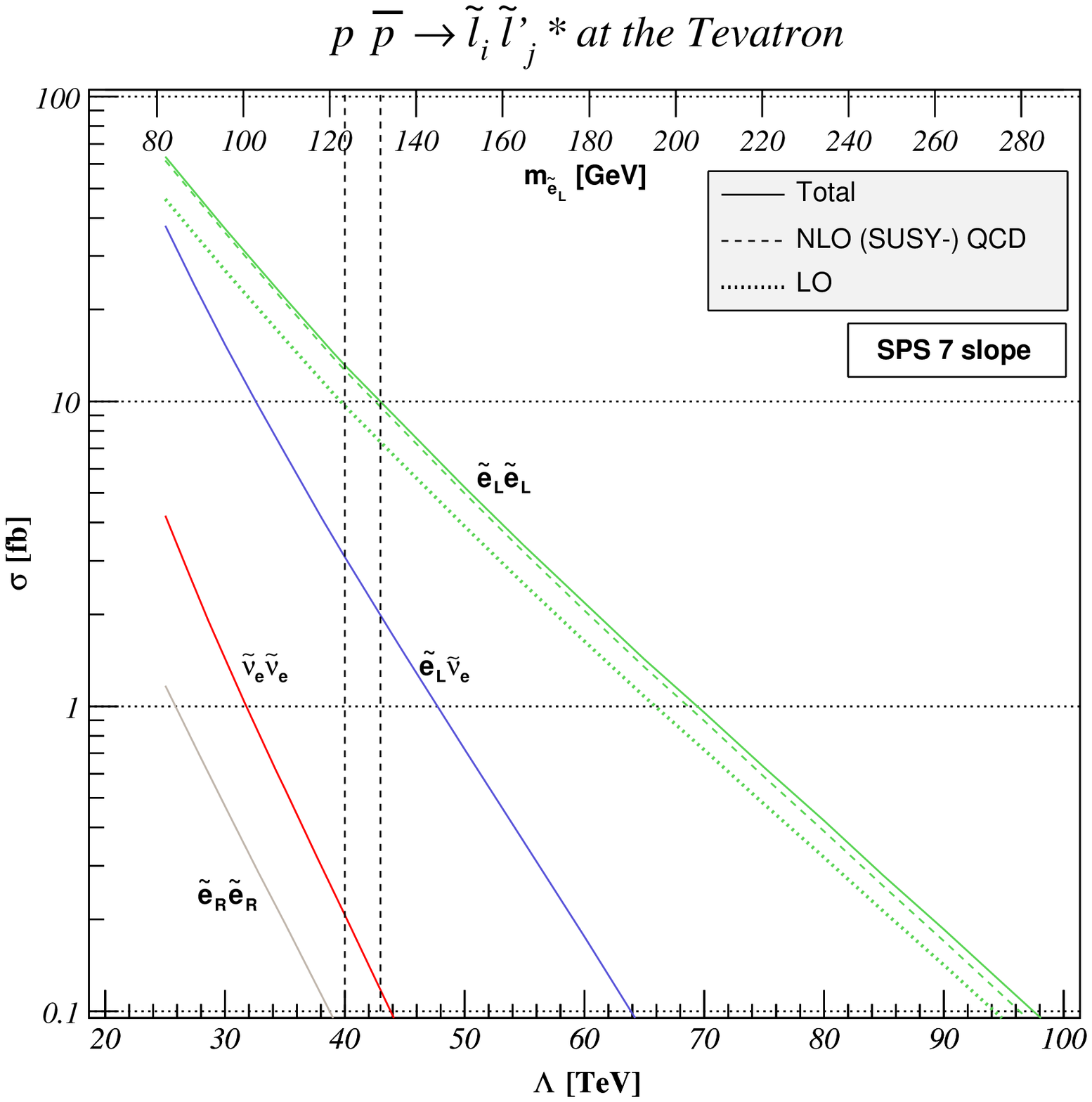}
\includegraphics[width=.7\columnwidth]{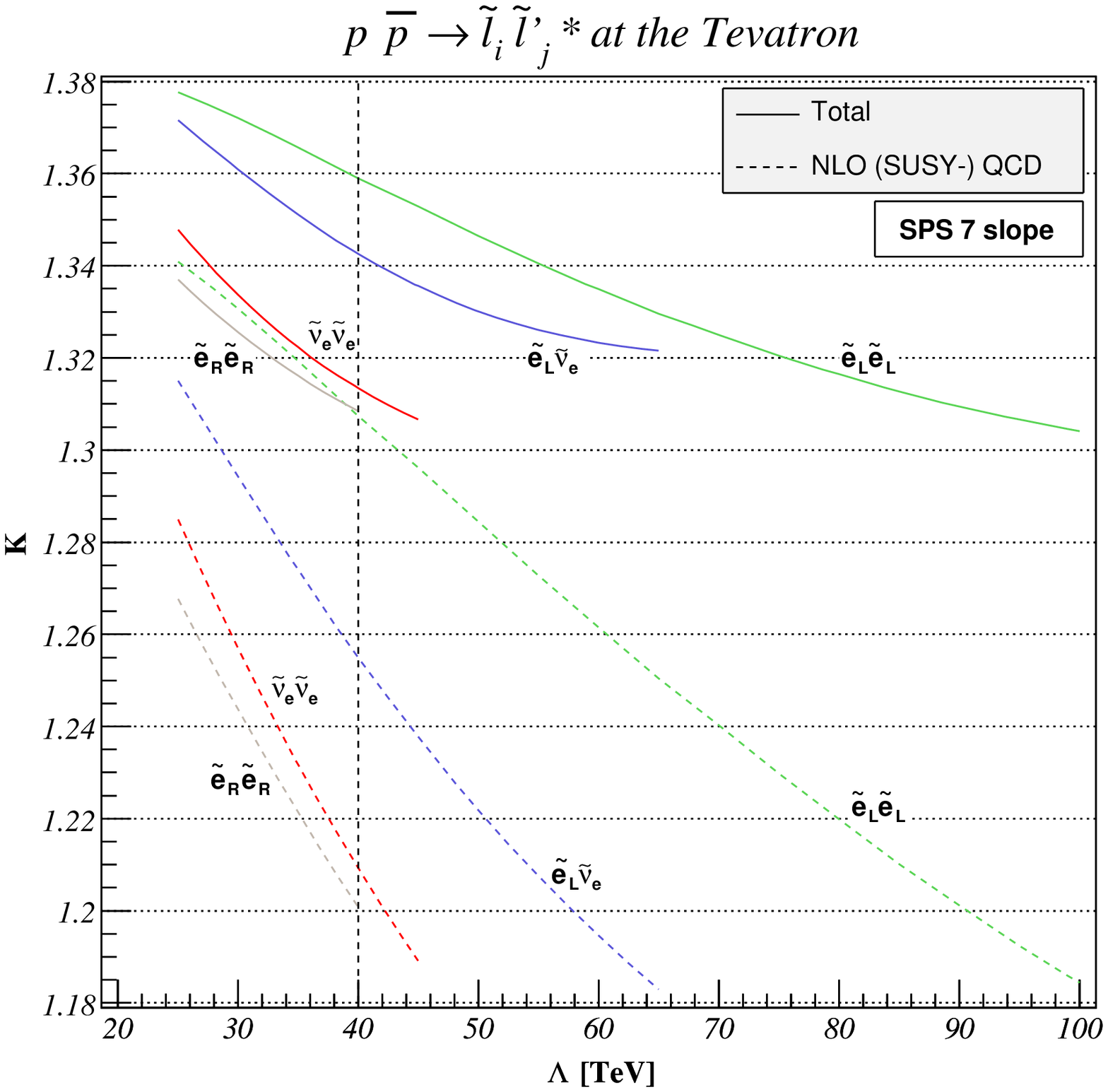}
  \caption{\label{fig:20}Total cross sections (top) and $K$-factors
  as defined in Eq.\ (\ref{eq:kglob}) (bottom) for first- (and
  second-) generation slepton pair and slepton-sneutrino associated
  production at the Tevatron along the model line attached to the
  SPS 7 benchmark point (vertical dashed line). We show the total
  NLL+NLO matched and the fixed-order NLO (SUSY-)QCD and LO QCD
  results.}
\end{figure}

\begin{figure}
\centering
\includegraphics[width=.7\columnwidth]{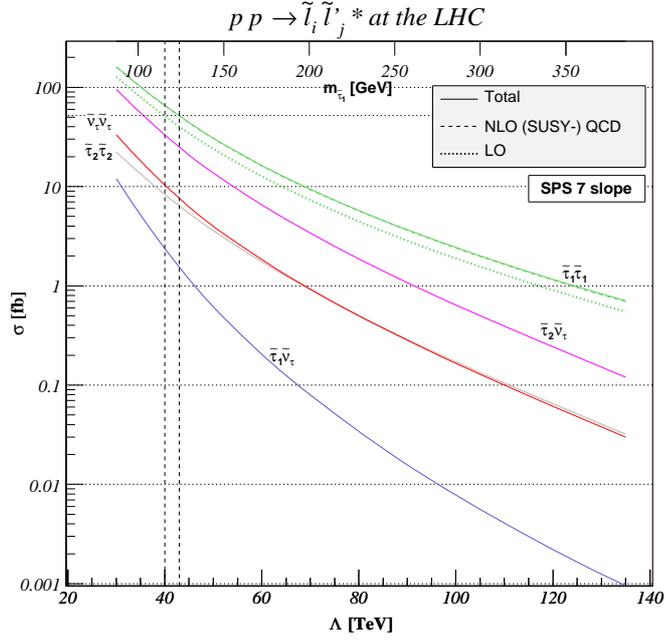}
\includegraphics[width=.7\columnwidth]{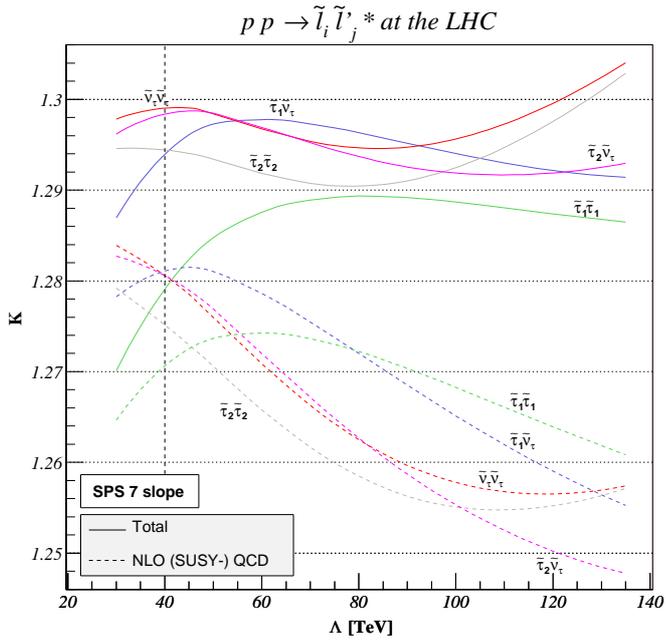}
  \caption{\label{fig:22}Total cross sections (top) and $K$-factors
  as defined in Eq.\ (\ref{eq:kglob}) (bottom) for third-generation
  slepton pair and slepton-sneutrino associated production at the
  LHC along the model line attached to the SPS 7 benchmark point
  (vertical dashed line). We show the total NLL+NLO matched and the
  fixed-order NLO (SUSY-)QCD and LO QCD results.}
\end{figure}

\begin{figure}
\centering
\includegraphics[width=.7\columnwidth]{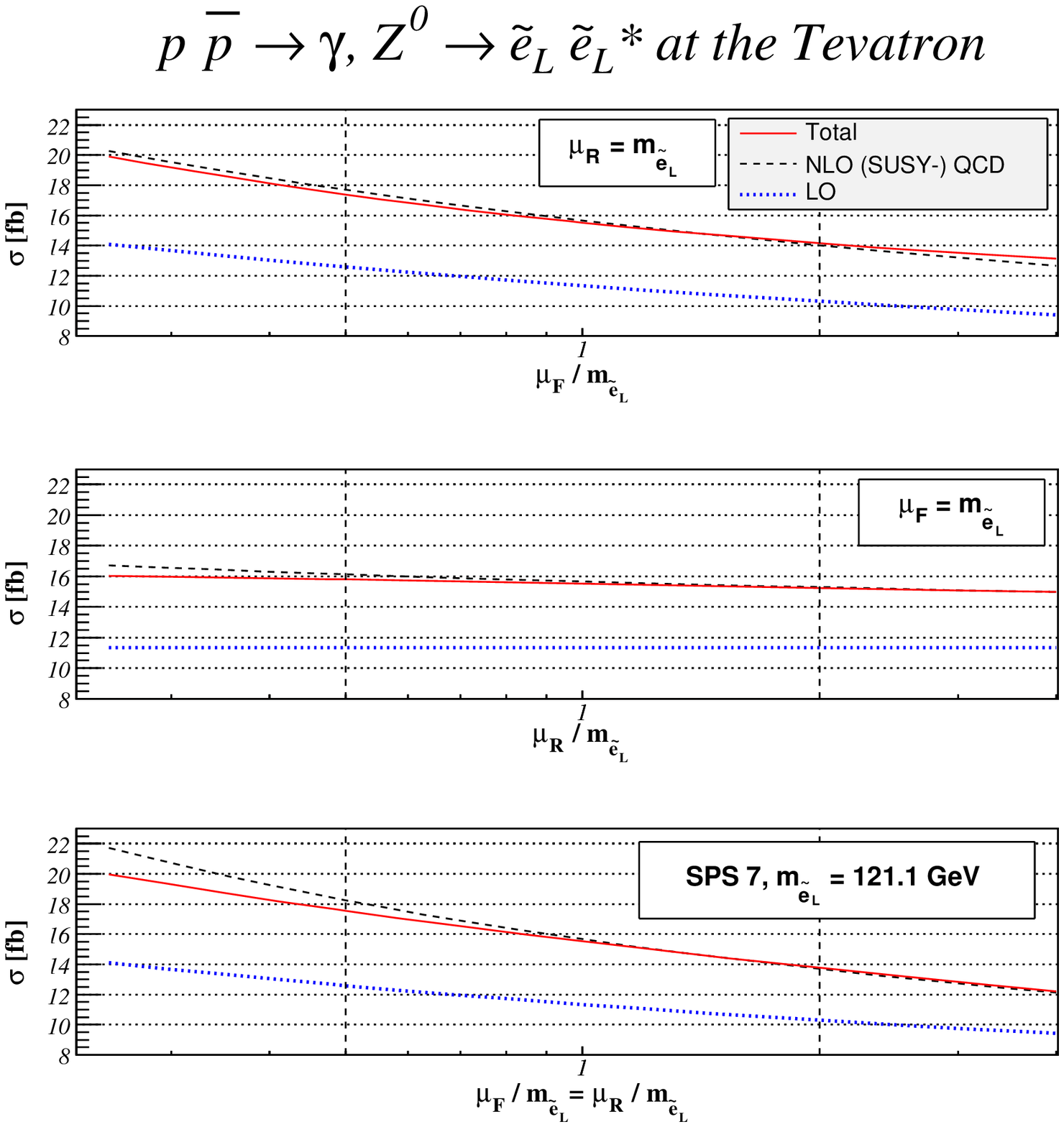}
\caption{\label{fig:24}Dependence of the total cross section for
  first- (and second-) generation slepton pairs at the Tevatron on
  the factorization scale (top), renormalization scale (middle), and
  both scales (bottom) for the SPS 7 benchmark point.}
\includegraphics[width=.7\columnwidth]{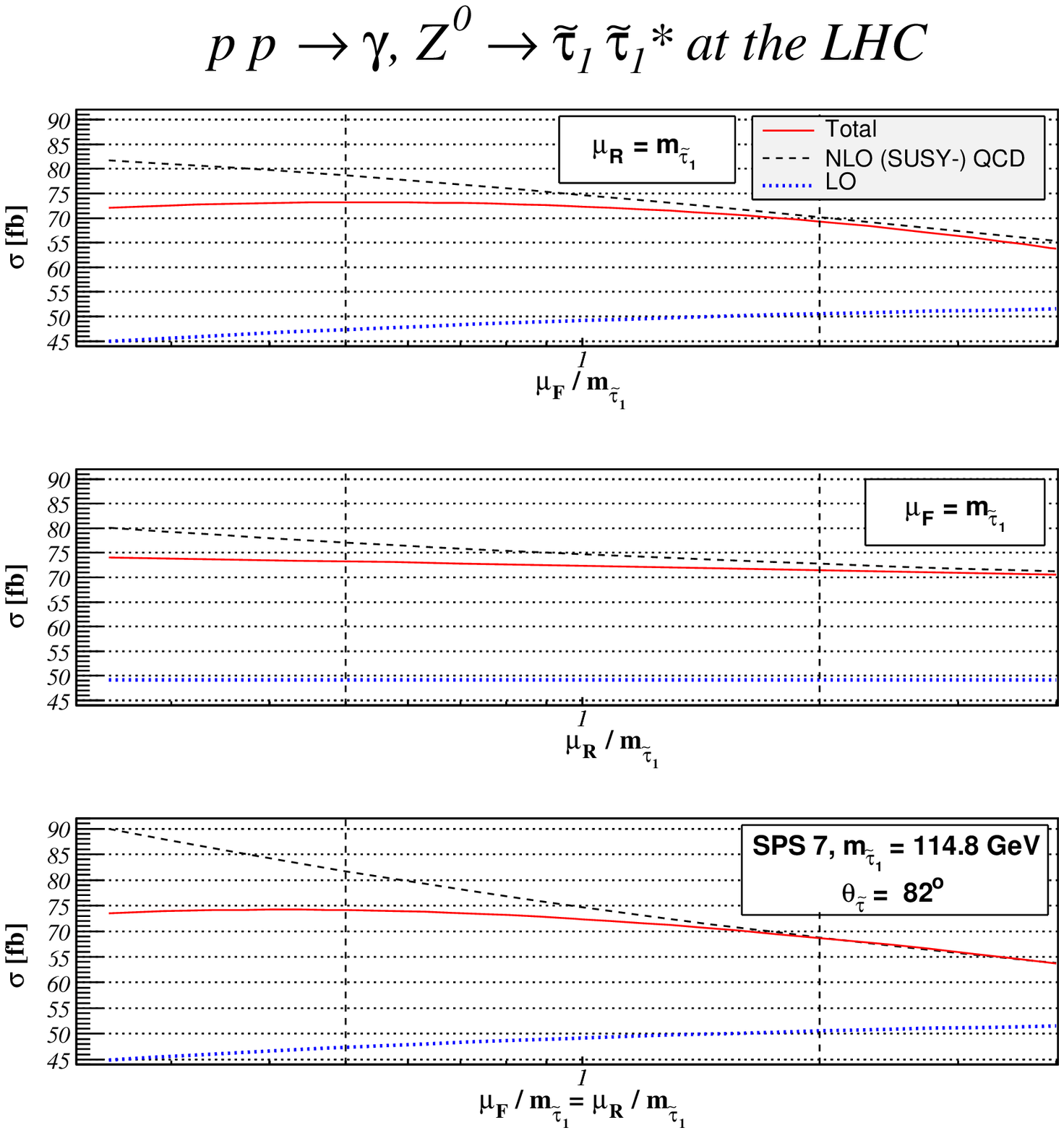}
\caption{\label{fig:25}Dependence of the total cross section for
 third-generation slepton pairs at the LHC on the factorization
 scale (top), renormalization scale (middle), and both scales (bottom) for
 the SPS 7 benchmark point.}
\end{figure}

In this section, we study the dependence of total, i.e.\
invariant-mass integrated, slepton pair and slepton-sneutrino
cross sections on three different scales: first the dependence on
the effective SUSY-breaking scale $\Lambda$ as defined in the GMSB
model line SPS 7, and then the dependence on the renormalization
and factorization scales $\mu_{R,F}$ at the point SPS 7. The
benchmark point at $\Lambda=40$ TeV will be indicated in the
figures where $\Lambda$ varies as a vertical dashed line, and we
show in addition to the scale $\Lambda$ the mass scale of the
produced charged slepton, ranging from 80 (87.5) to 280 (385) GeV
for $\tilde{e}_L$ ($\tilde{\tau}_1$).\\

In Fig.\ \ref{fig:20}, total cross sections (top) and $K$-factors
(bottom) are shown at the Tevatron, which is expected to produce a
total integrated luminosity of 4--8 fb$^{-1}$. Even for
$\tilde{e}_L$ pair production the mass-range is limited to masses
below 280 GeV, where the cross section reaches 0.1 fb and at most
one event would be produced. NLO and resummation corrections are
clearly important, as they increase the LO prediction by 18 to
28\% (lower part of Fig.\ \ref{fig:20}). At the SPS 7 benchmark
point, the corrections would thus induce a shift in the selectron
mass as deduced from a total cross section measurement by about 8
GeV (cf.\ the two dashed lines in the upper part of Fig.\
\ref{fig:20}). By comparing the NLO and total predictions, one
observes an increased importance of threshold
resummation for heavier sleptons, as expected.\\

We show in Fig.\ \ref{fig:22} the total cross sections in the
restricted range $\Lambda\leq135$ TeV. The NLO and resummed
corrections are again large (25--30\%), but the resummation
corrections only become appreciable for large SUSY-breaking scales
(or slepton masses). The largest cross section is obtained for
pair production of the light stau mass eigenstate, even though it
has a large right-handed component. Conversely, the heavier stau
mass eigenstate has a large left-handed component, so that its
cross section is less suppressed. At the SPS 7 benchmark point,
the corrections would again induce a shift in the slepton mass as
deduced from a total cross section measurement by about 8 GeV, as
can be seen for $\tilde{\tau}_1$ on the upper part of Fig.\
\ref{fig:22}, cf.\ the two dashed lines.\\

Finally, we consider the theoretical uncertainty of invariant-mass
integrated total cross sections at the Tevatron (Fig.\
\ref{fig:24}) and the LHC (Fig.\ \ref{fig:25}) as induced by
variations of the factorization scale (top), renormalization scale
(middle), or both (bottom). The $\mu_R$-dependence (middle), which
is absent in LO, is first introduced in NLO, but then tamed by the
resummation procedure. On the other hand, the logarithmic
$\mu_F$-dependence (top), already present through the PDFs at LO,
is overcompensated (reduced) at NLO for the LHC (Tevatron) and
then (further) stabilized by resummation. This works considerably
better at the LHC, where at least one quark PDF is sea-like and
the PDFs are evaluated at lower $x$, than at the Tevatron, where
both PDFs can be valence-like and are evaluated at relatively
large $x$. In total, the theoretical uncertainty at the Tevatron
(LHC), defined by the ratio of the cross section difference at
$\mu_F=\mu_R=m_{\tilde{l}}/ 2$ and $\mu_F=\mu_R=2\, m_{\tilde{l}}$
over their sum, increases from 20 (7) \% in LO to 29 (17) \% in
NLO, but is then reduced again to 23 (8) \% for the
resummed-improved prediction.

\section{Jointly resummed results}

We present here a joint treatment of the recoil corrections at
small $q_T$ and the threshold-enhanced contributions near partonic
threshold, allowing a complete understanding of the soft-gluon
effects in differential distributions for slepton pair production
at hadron colliders \cite{Bozzi:2007xx}. We use the joint
formalism described in Sec.\ \ref{sec:joint} and compare it to the
$q_T$- and threshold resummation formalisms of Secs.\
\ref{sec:qtresuni} and \ref{sec:thresh2}. The fixed-order
perturbative and the $q_T$- and threshold-resummed results for the
transverse-momentum and invariant-mass distributions are those
presented in the previous subsections, while the jointly-resummed
matched cross section (see Eq.\ (\ref{eq:mtchjt})) is \bea
\frac{\td^2\sigma}{\td M^2\,\td q_T^2}(\tau) &=&
\frac{\td^2\sigma^{({\rm F.O.})}}{\td M^2\,\td q_T^2}(\tau; \as) +
\frac{1}{2\, \pi\, i} \oint_{C_N} \frac{\td
N}{2\pi i}\, \tau^{-N} \int \frac{b \td b}{2} J_0(q_T\,b)\nn\\
&&\times \left[\frac{\td^2\sigma^{{\rm (res)}}}{\td M^2\,\td
q_T^2}(N, b; \as) - \frac{{\rm d}^2\sigma^{{\rm (exp)}}}{\td
M^2\,\td q_T^2}(N, b; \as) \right].~ \eea The resummed contribution
is given by Eq.\ (\ref{eq:joint}) \bea \frac{\td\sigma^{({\rm
res})}}{\td M^2\, \td q^2_T} &=&
\sum_{a,b} f_{a/h_a}(N+1, \mu_F)\, f_{b/h_b}(N+1, \mu_F) \,\nn\\
&\times&\! \sum_c\! \mathcal{H}_{ab\to c\bar{c}}\Big(N;
\alpha_s(\mu_R), \frac{M}{\mu_R}, \frac{M}{\mu_F}\Big) \exp\Big\{
\mathcal{G}_c(\ln\chi; \alpha_s(\mu_R),
\frac{M}{\mu_R})\Big\},~~~~~ \eea while the expansion of the
resummed component is (see Eq.\ (\ref{eq:exp_joint})) \bea
\frac{\td\sigma^{({\rm exp})}}{\td M^2\, \td q^2_T} &=& \sum_{a,b}
f_{a/h_a}(N+1, \mu_F)\, f_{b/h_b}(N+1, \mu_F) \,\nn\\ &\times &
\sum_c \sigma_{c{\bar c}}^{(0)}(M) \Bigg\{ \delta_{ca}
\delta_{{\bar c}b}
+ \sum_{n=1}^{\infty}\left(\frac{\as(\mu_R)}{\pi} \right)^n\nn \\
& \times &\Bigg[{\tilde \Sigma}_{ab\to c{\bar c}}^{(n)}\left(N,
\ln\chi; \frac{M}{\mu_R}, \frac{M}{\mu_F}\right)\!+\!
\mathcal{H}_{ab\to c\bar{c}}^{(n)}\Big(N; \frac{M}{\mu_R},
\frac{M}{\mu_F}\Big)\Bigg]\Bigg\}.~\eea The $\mathcal{O}(\as)$
coefficients of the perturbative functions $\mathcal{H}$,
$\mathcal{G}$ and ${\tilde \Sigma}$, can be found in Sec.\
\ref{sec:joint}. We remind the reader only of the definition of
the argument in the logarithm (see Eq.\ (\ref{eq:chi})), \bea
\chi(\nbar, \bbar)=\bbar + \frac{\nbar}{1+\eta\,\bbar/ \nbar}
.~\eea

For the masses and widths of the electroweak gauge bosons, the
electroweak mixing angle and the electromagnetic fine structure
constant, we use the same values as in the previous section, i.e.\
those given by the latest version of the PDG review
\cite{Yao:2006px}. We focus our study on the production of a
right-handed selectron pair at the LHC, \bea
\begin{array}{c} q \bar{q} \to \gamma,Z^0 \to
\tilde{e}_R\, \tilde{e}^\ast_R. \end{array} \eea We use the MRST
(2004) NLO set of parton distribution functions
\cite{Martin:2004ir}, $\alpha_s$ is evaluated at two-loop
accuracy, and we allow $\mu_F$ and $\mu_R$ to vary between $M/2$
and $2M$ to estimate the perturbative uncertainty. We choose the
mSUGRA benchmark point BFHK B (see Tab.\ \ref{tab:3}) which gives,
after the renormalization group evolution of the SUSY-breaking
parameters performed by the SPheno computer program
\cite{Porod:2003um}, a light ${\tilde e}_{R}$ of mass $m_{{\tilde
e}_{R}}=186$ GeV and rather heavy squarks with masses around
800-850 GeV, except for top-squark mass eigenstate $\tilde{t}_1$
which is slightly lighter, but which nevertheless does not
contribute to the virtual squark loops due
to the negligible top-quark density in the proton.\\

\begin{figure}
\centering
\includegraphics[width=.7\columnwidth]{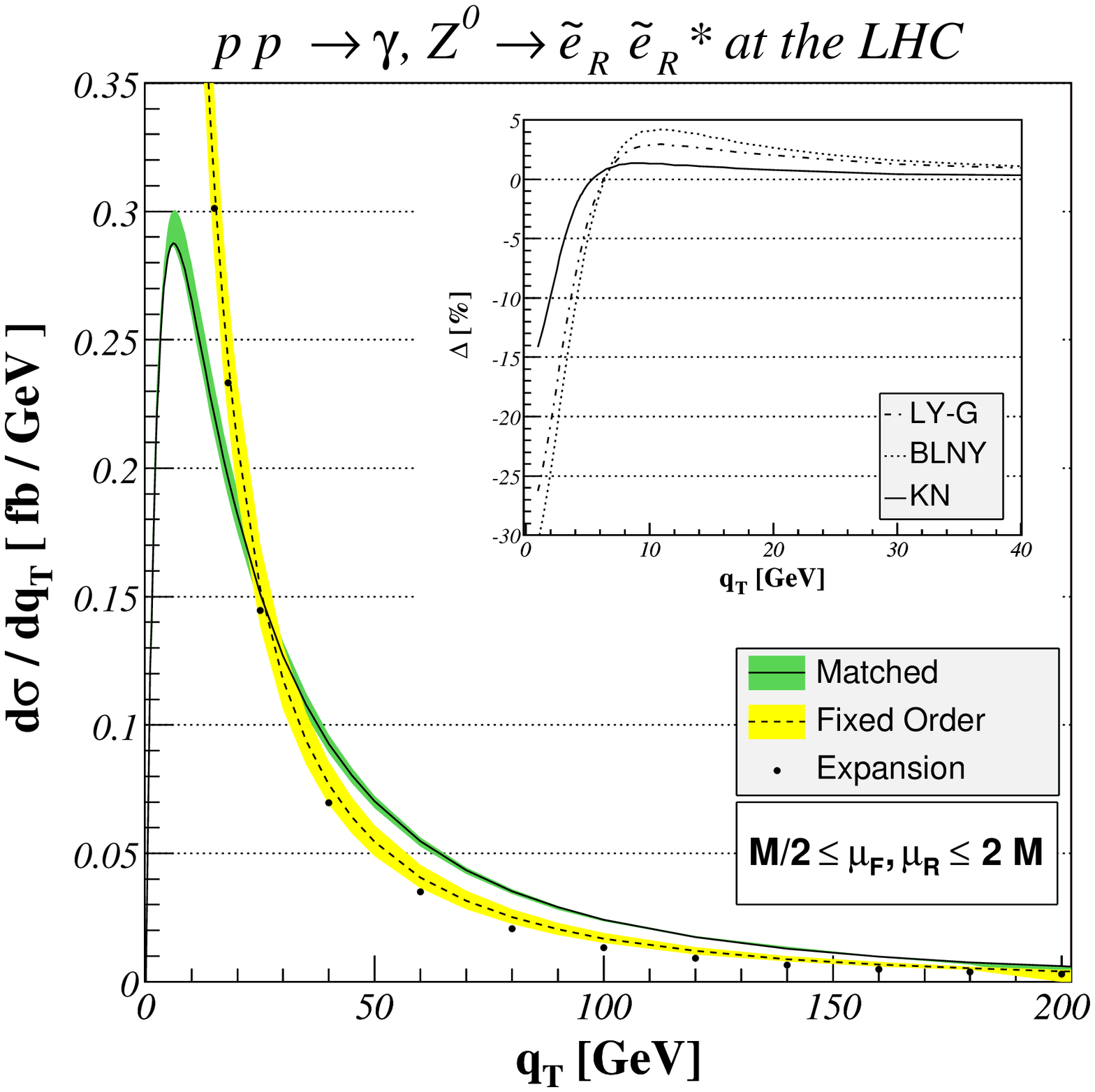}
\caption{\label{fig:43}Transverse-momentum distribution for the
process $p p \to \tilde{e}_R\, \tilde{e}_R^\ast$ at the LHC.
NLL+LO matched result, LO result, asymptotic expansion of the
resummation formula and $\Delta$-parameter are shown.}
\includegraphics[width=.7\columnwidth]{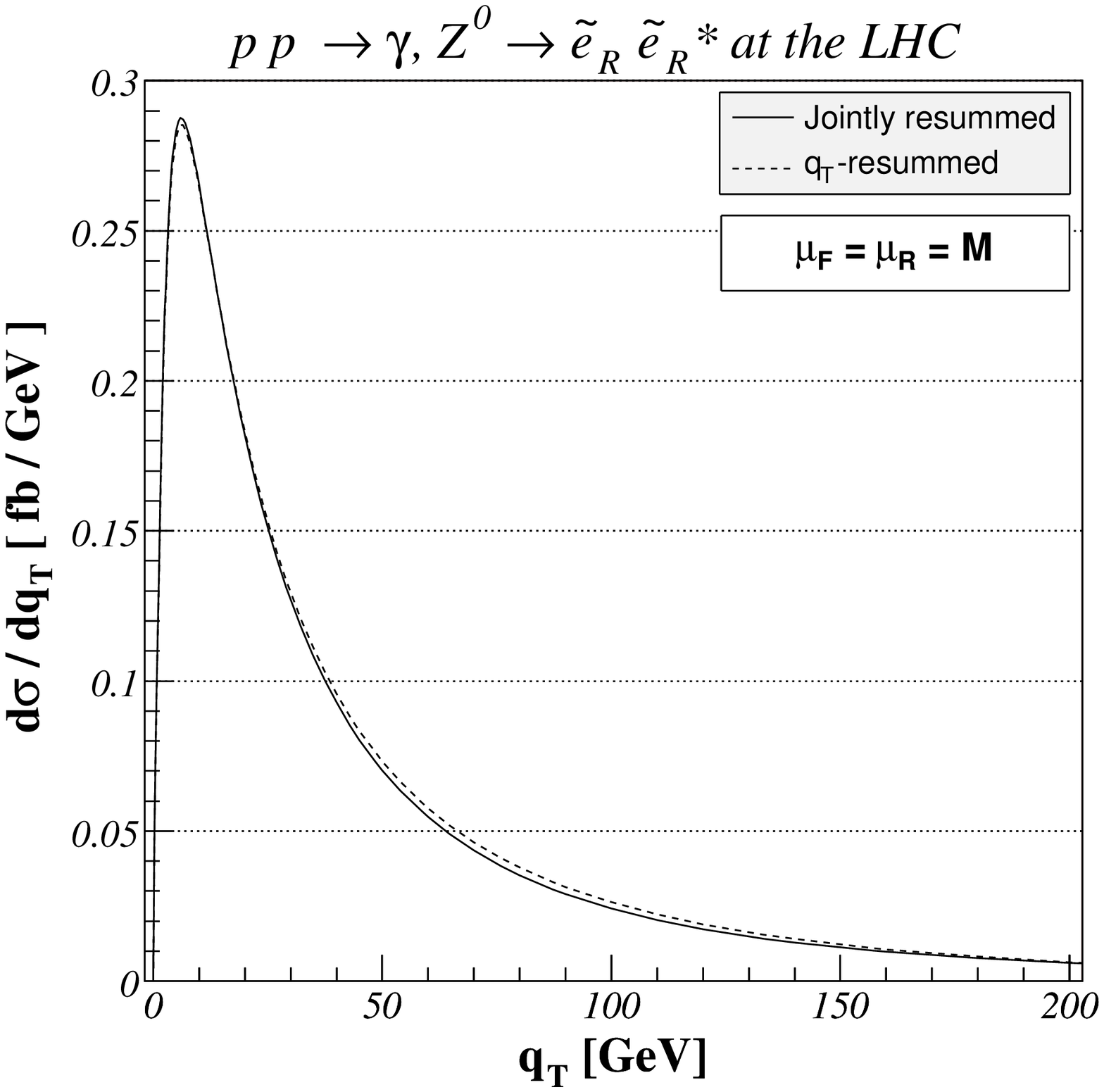}
\caption{\label{fig:44}Transverse-momentum distribution for
selectron pair production at the LHC ion the framework of joint
and $q_T$ resummation.}
\end{figure}

In Fig.\ \ref{fig:43}, we integrate the previous equations with
respect to $M^2$, taking as lower limit the energy threshold for
$\tilde{e}_{R} \tilde{e}_R^\ast$ production and as upper limit the
hadronic energy, $\sqrt s_h$ = 14 TeV. We plot the LO result
(dashed line) and the total NLL+LO matched result (solid line)
with their uncertainty band relative to scale variation (yellow
band for the LO result and green band for the NLL+LO result). The
effect of resummation is clearly visible at small and intermediate
values of $q_T$, the resummation-improved result reaching a value
that is even 40\% higher than the pure
fixed-order result at $q_T=80$ GeV .\\

The asymptotic expansion of the resummation formula at LO (dotted
line) is in good agreement with LO at small values of $q_T$, from
which we can conclude that the cross section is clearly dominated
by the logarithms that we are resumming in this kinematical
region. In the intermediate-$q_T$ region, we can see that the
expansion is slightly smaller than the fixed-order calculation.
Since this effect was not present in $q_T$-resummation (see Fig.\
\ref{fig:8} in Sec.\ \ref{sec:qtnum}), we deduce that it is purely
related to the threshold-enhanced contributions important in the
large-$M$ region. It is also presented in Fig.\ \ref{fig:44},
where we directly compare jointly- and $q_T$-matched results, the
main difference between the two approaches relying indeed in this
intermediate-$q_T$ region where the jointly-resummed cross section
is 5\%-10\% lower than the
$q_T$-resummed one for 50 GeV $< q_T < $ 100 GeV.\\

In Fig.\ \ref{fig:43}, we estimate the scale dependence by an
independent variation of the factorization and renormalization
scales between $M/2$ and $2M$, and is clearly improved using
resummation rather than pure fixed-order calculations. In the
small and intermediate $q_T$-region the effect of scale variation
is 10\% for the LO result, while it is always less than 5\% for
the NLL+LO curve.\\

We also study the dependence of the total NLL+LO matched result on
non-perturbative effects. We show the quantity $\Delta$ defined in
Eq.\ (\ref{eq:delta}), \bea \Delta = \frac{d\sigma^{\rm
(res.+NP)}(\mu_R=\mu_F=M)-d\sigma^{(\rm
res.)}(\mu_R=\mu_F=M)}{d\sigma^{(\rm res.)} (\mu_R=\mu_F=M)},\eea
as a function of the transverse momentum of the slepton pair. The
parameter $\Delta$ gives thus an estimate of the contributions
from the different NP parameterizations (LY-G, BLNY, KN) that we
included in the resummed formula, which can be found in Eqs.\
(\ref{eq:NP1}), (\ref{eq:NP2}) and (\ref{eq:NP3}). They are under
good control, since they are always less than 5\% for $q_{T}>$ 5
GeV and thus considerably smaller than the resummation effects.\\

\begin{figure}
\centering
\includegraphics[width=.7\columnwidth]{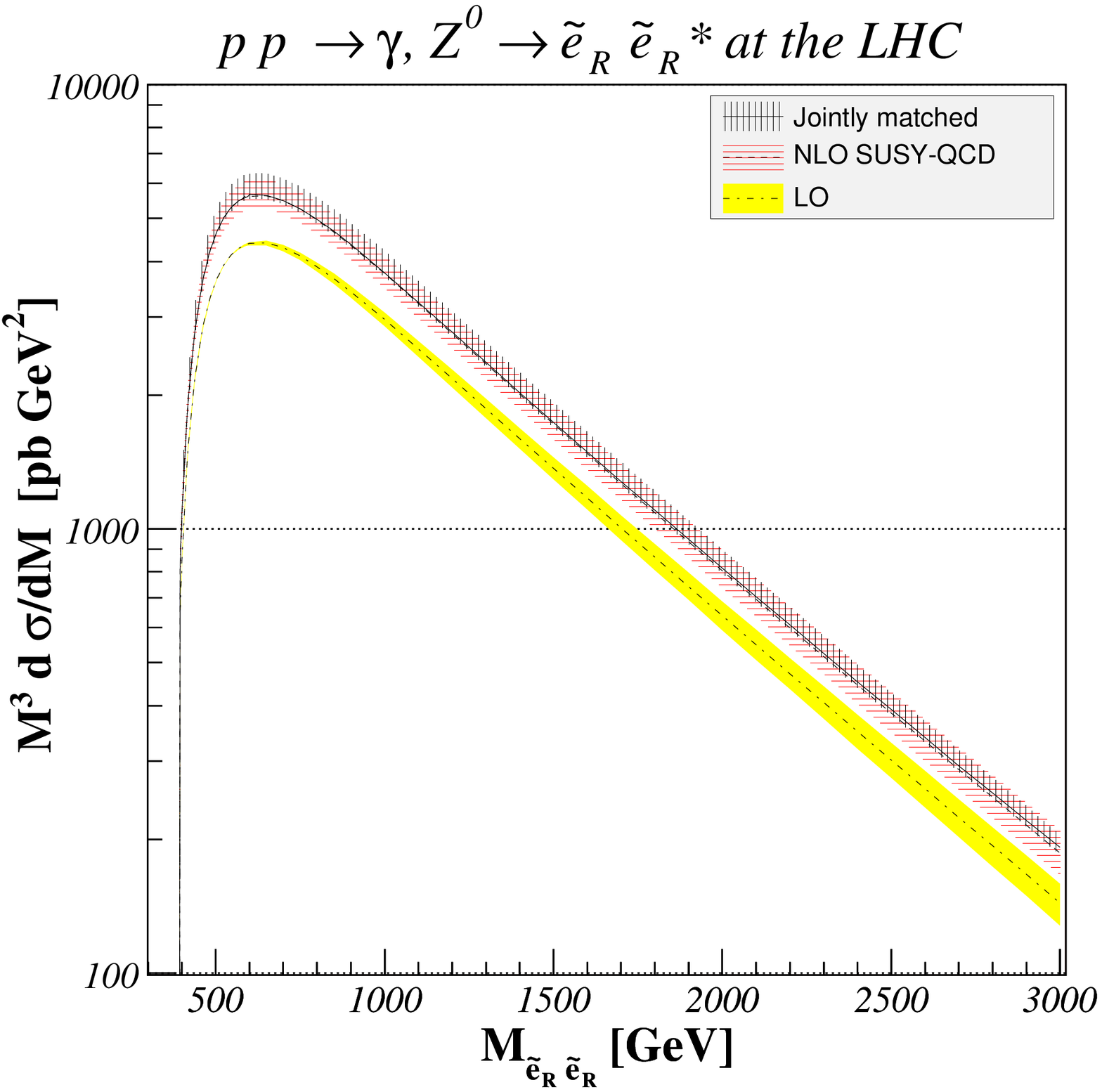}
\caption{\label{fig:45}Invariant-mass distribution
$M^3\,\td\sigma/\td M$ of $\tilde{e}_R$ pairs at the LHC. We show
the total NLL+NLO matched and the fixed-order NLO SUSY-QCD and LO
QCD results, including the respective scale uncertainties as
vertically hatched, horizontally hatched and shaded bands.}
\includegraphics[width=.7\columnwidth]{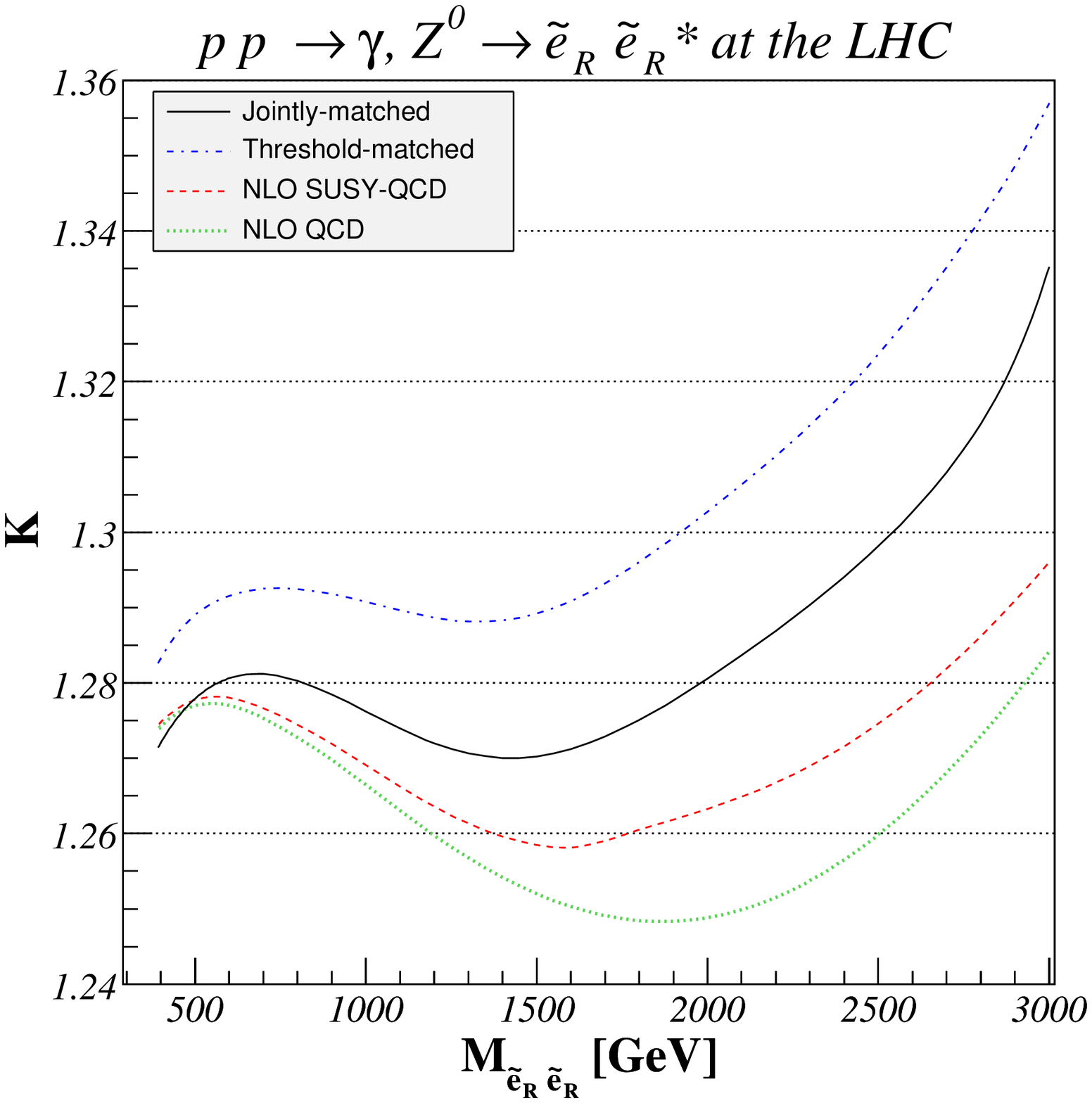}
\caption{\label{fig:46}$K$-factors as defined in Eq.\
(\ref{eq:Kbis}) for $\tilde{e}_R$ pair production at the LHC. We
show the total NLL+NLO jointly- and threshold-matched result, as
well as the fixed-order NLO SUSY-QCD and QCD results.}
\end{figure}

The invariant-mass distribution $M^3\td\sigma/\td M$ for
$\tilde{e}_R$-pair production at the LHC is obtained by
integrating the equations given above with respect to the
transverse-momentum $q_T$, and is shown in Fig.\ \ref{fig:45}. As
in the previous section, the differential cross section
$\td\sigma/\td M$ has been multiplied by a factor $M^3$ in order to
remove the leading mass dependence of propagator and phase space
factors. We can again see the $P$-wave behaviour relative to the
production of scalar particles, since the invariant-mass
distribution rises above the threshold at $\sqrt{s}=2
m_{\tilde{e}_R}$ with the third power of the slepton velocity and
peaks at about 200 GeV above threshold (both for $M^3\td\sigma/\td
M$ and the not shown $\td\sigma/\td M$ differential distribution),
before falling off steeply due to the $s$-channel propagator and
the decreasing parton luminosity. Let us note that we use the MRST
LO 2001 \cite{Martin:2002dr} set of parton
distribution functions for the LO predictions.\\

In the large-$M$ region, the resummed cross section is 30\% higher
than the leading-order cross section, but this represents only a
3\% increase with respect to the NLO SUSY-QCD result. In the
small-$M$ region, much further then from the hadronic threshold,
resummation effects are rather limited, inducing a modification of
the NLO results smaller than 1\%.\\

The shaded, horizontally and vertically shaded bands in Fig.\
\ref{fig:45} represent the theoretical uncertainties for the LO,
NLO SUSY-QCD, and the jointly-matched predictions. At LO, the
dependence coming only from the factorization scale increases with
the momentum-fraction $x$ of the partons in the proton (i.e.\ with
$M$), being thus larger in the right part of the figure, but this
dependence is largely reduced at NLO due to the factorization of
initial-state singularities in the PDFs. However, we add the
dependence due to the renormalization scale in the coupling
$\alpha_s(\mu_R)$, leading to a total variation of about 7\%-11\%.
After resummation, the total scale uncertainty is finally reduced
to only 7\%-8\% for the matched result, the reduction being of
course more important in the large-$M$ region, where the
resummation effects are more
important.\\

In Fig.\ \ref{fig:46}, we show the cross section correction
factors \bea \label{eq:Kbis} K^i = \frac{{\rm d}\sigma^i / {\rm
d}M}{{\rm d}\sigma^{\rm LO} / {\rm d}M} \eea as a function of the
invariant-mass $M$. $i$ labels the corrections induced by NLO QCD,
NLO SUSY-QCD, joint- and threshold-resummation, these two last
calculations being matched with the NLO SUSY-QCD result. \\

At small invariant mass $M$, the resummation is less important,
since we are quite far from the hadronic threshold, as shown in
the left part of the plot. At larger $M$, the logarithms become
important and lead to a larger increase of the resummed
$K$-factors over the fixed-order one. We also show the difference
between threshold and joint resummations, which is only about one
or two percents. It is due to the choice of the Sudakov form
factor $\mathcal G$ and of the $\mathcal H$-function, which
reproduces correctly transverse-momentum resummation in the limit
of $b\to\infty$, $N$ being fixed, but which presents some
differences with the pure threshold limit $b\to 0$ and
$N\to\infty$, as it was the case for joint resummation for Higgs
and electroweak boson production \cite{Kulesza:2002rh,
Kulesza:2003wn}. However, this effect is under good control, since
it is much smaller than the theoretical scale uncertainty of about
7\%.

\newpage $~$\\ \newpage

\chapter{Squark and gaugino production and decays at hadron
colliders}\label{ch:NMFV}

In many models, the non-coloured charginos and neutralinos belong
to the class of the lightest supersymmetric particles, their low
masses counterbalancing the small weak cross sections for direct
production. Theoretically, full NLO SUSY-QCD calculations have
been performed \cite{Beenakker:1999xh}, while experimentally,
their subsequent decay to gold-plated trilepton signatures
\cite{Baer:1986vf, Nath:1987sw} have been exploited in several CDF
and D0 analyses at the Tevatron \cite{Abachi:1995ek, Abe:1998qm,
Barger:1998hp, Matchev:1999nb, Baer:1999bq} and will be
investigated at the LHC \cite{Hinchliffe:1996iu, Abdullin:1998pm}.
Due to their strong coupling, squarks should be abundantly
produced at hadron colliders, and hadroproduction cross sections
and decay has therefore been studied in detail at NLO SUSY-QCD
\cite{Beenakker:1996ch}. The production of top
\cite{Beenakker:1997ut} and bottom \cite{Berger:2000mp} squarks
with large helicity mixing has received particular attention, and
both QCD one-loop and electroweak tree-level contributions have
been calculated for non-diagonal, diagonal and mixed top and
bottom squark pair production \cite{Bozzi:2005sy}. Very recently,
flavour violation has been considered in the context of collider
searches \cite{Bozzi:2007me}. Concerning the associated production
of squarks and gauginos, NLO SUSY-QCD cross sections in cMFV SUSY
have been calculated some times ago \cite{Berger:1999mc} and
generalized to NMFV scenarios \cite{Bozzi:2007me}.\\

In the following, we use for the sake of simplicity the generic
notation \bea \left\{\mathcal{C}^1_{a b c}, \mathcal{C}^2_{a b c}
\right\} = \left\{L_{a b c},R_{a b c} \right\} \eea for the
couplings defined in Sec.\ \ref{sec:coupling}, while the
propagators appearing as mass-subtracted Mandelstam variables read
\bea \begin{array}{l c l c l c l c} s_w &=& s-m_W^2&,~& s_z &=&
s-m_Z^2 &,\\  t_{\tilde{\chi}^0_k} &=& t -
m_{\tilde{\chi}^0_k}^2&,~&  u_{\tilde{\chi}^0_k} &=&
u-m_{\tilde{\chi}^0_k}^2&, \\ t_{\tilde{\chi}_j} &=&
t-m_{\tilde{\chi}^\pm_j}^2 &,~& u_{\tilde{\chi}_j} &=&
u-m_{\tilde{\chi}^\pm_j}^2 &, \\ t_{\tilde{g}} &=& t -
m^2_{\tilde{g}} &,~& u_{\tilde{g}} &=& u - m^2_{\tilde{g}} &,
\\ t_{\tilde{q}_i} &=& t-m_{\tilde{q}_i}^2 &,~&
u_{\tilde{q}_i} &=& u-m_{\tilde{q}_i}^2 &. \end{array}&& \eea

\section{NMFV squark-antisquark pair hadroproduction}

\subsection{Analytical results}\label{sec:sqSQ1}

\begin{figure}
 \centering
 \includegraphics[width=0.75\columnwidth]{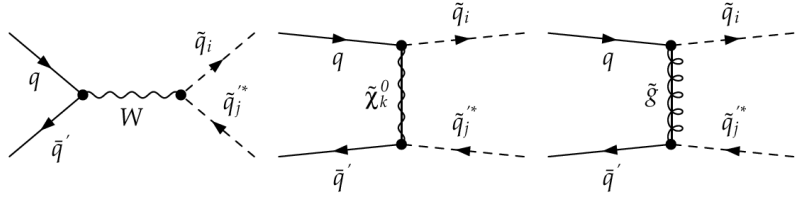}
 \caption{\label{fig:26}Tree-level Feynman diagrams for the production of
          charged squark-antisquark pairs in quark-antiquark collisions.}
\end{figure}

In NMFV SUSY, the production of charged squark-antisquark pairs
\bea q(h_a, p_a)\, \bar{q}^\prime(h_b, p_b) \to \tilde{u}_i(p_1)\,
\tilde{d}^\ast_j(p_2), \eea where $i,j=1,...,6$ label up- and
down-type squark mass eigenstates, $h_{a,b}$ helicities, and
$p_{a,b,1,2}$ four-momenta, proceeds from an equally charged
quark-antiquark initial state through the tree-level Feynman
diagrams shown in Fig.\ \ref{fig:26}. The corresponding cross
section can be written in a compact way as \cite{Bozzi:2007me}
\bea \frac{{\rm d} \hat{\sigma}^{q\bar{q}'}_{h_a, h_b}}{dt} &=&
(1-h_a) (1+h_b) \Bigg[ \frac{\mathcal{W}}{s_w^2} + \bigg(
\sum_{k,l=1,...,4}
\frac{\mathcal{N}^{kl}_{11}}{t_{\tilde{\chi}^0_k}\,
t_{\tilde{\chi}^0_l}}\bigg) +
\frac{\mathcal{G}_{11}}{t_{\tilde{g}}^2} + \bigg( \sum_{k=1,...,4}
\frac{\mathcal{[NW]}^k}{t_{\tilde{\chi}^0_k}\, s_w} \bigg)\nn\\
&+& \frac{\mathcal{[GW]}}{t_{\tilde{g}}\, s_w} \Bigg] + (1-h_a)
(1-h_b) \Bigg[ \bigg( \sum_{k,l=1,...,4}
\frac{\mathcal{N}^{kl}_{12}}{t_{\tilde{\chi}^0_k}\,
t_{\tilde{\chi}^0_l}}\bigg) +
\frac{\mathcal{G}_{12}}{t_{\tilde{g}}^2}\Bigg] \nonumber\\ &+&
(1+h_a) (1+h_b) \Bigg[ \bigg( \sum_{k,l=1,...,4}
\frac{\mathcal{N}^{kl}_{21}}{t_{\tilde{\chi}^0_k}\,
t_{\tilde{\chi}^0_l}}\bigg) +
\frac{\mathcal{G}_{21}}{t_{\tilde{g}}^2}\Bigg] \nonumber\\ &+&
(1+h_a)(1-h_b) \Bigg[ \bigg( \sum_{k,l=1,...,4}
\frac{\mathcal{N}^{kl}_{22}}{t_{\tilde{\chi}^0_k}\,
t_{\tilde{\chi}^0_l}}\bigg) +
\frac{\mathcal{G}_{22}}{t_{\tilde{g}}^2}\Bigg] \label{eq:sqsq1}
\eea thanks to the form factors \bea \mathcal{W} &=& \frac{\pi\,
\alpha^2}{16\, x_W^2\, (1-x_W)^2\,  s^2} \left| L^\ast_{q q^\prime
W}\, L_{\tilde{u}_i \tilde{d}_j W}\right|^2 \left( u\, t -
m^2_{\tilde{u}_i}\,  m^2_{\tilde{d}_j}\right),~\nonumber\\
\mathcal{N}_{mn}^{kl} &=&  \frac{\pi\, \alpha^2}{x_W^2\, (1 -
x_W)^2\, s^2}  \mathcal{C}^n_{\tilde{d}_j q^\prime
\tilde{\chi}_k^0}\,  \mathcal{C}^{m\ast}_{\tilde{u}_i q
\tilde{\chi}_k^0}\,  \mathcal{C}^{n\ast}_{\tilde{d}_j
\tilde{\chi}_l^0}\,  \mathcal{C}^m_{\tilde{u}_i q
\tilde{\chi}_l^0}\, \nn\\ &\times& \Bigg[ \left( u\, t -
m^2_{\tilde{u}_i}\, m^2_{\tilde{d}_j}\right) \delta_{mn} + \left(
m_{\tilde{\chi}^0_k}\, m_{\tilde{\chi}^0_l}\, s
\right) \left(1-\delta_{mn} \right) \Bigg],~\nonumber\\
\mathcal{G}_{mn} &=& \frac{2\, \pi\, \alpha_s^2}{9\, s^2} \left|
\mathcal{C}^{n\ast}_{\tilde{d}_j q^\prime \tilde{g}}\,
\mathcal{C}^m_{\tilde{u}_i q \tilde{g}}\right|^2 \Bigg[ \left( u\,
t - m^2_{\tilde{u}_i}\, m^2_{\tilde{d}_j}\right) \delta_{mn} +
\left(m_{\tilde{g}}^2\, s \right) \left(1-\delta_{mn} \right)
\Bigg],~\nonumber \\ \mathcal{[NW]}^k &=& \frac{\pi\,
\alpha^2}{6\, x_W^2\, (1\!-\!x_W)^2\, s^2}\, {\rm Re}\! \left[\!
L^\ast_{q q^{\prime} W} L_{\tilde{u}_i \tilde{d}_j W}
L_{\tilde{u}_i q \tilde{\chi}_k^0} L^\ast_{\tilde{q}_j q^\prime
\tilde{\chi}_k^0} \right]\! \left(\! u\, t \!-\!
m^2_{\tilde{u}_i}\, m^2_{\tilde{d}_j}\!\right),\!\nonumber\\
\mathcal{[GW]} &=& \frac{4\, \pi\, \alpha_s\, \alpha}{18\, x_W\,
(1 \!-\! x_W)\, s^2}\, {\rm Re}\! \left[\! L^\ast_{\tilde{u}_i q
\tilde{g}} L_{\tilde{d}_j q^\prime \tilde{g}} L^\ast_{q q^\prime
W} L_{\tilde{u}_i \tilde{d}_j W} \right]\! \left(\! u\, t \!-\!
m^2_{\tilde{u}_i}\, m^2_{\tilde{d}_j}\!\right),\! \eea which
combine coupling constants and Dirac traces of the squared and
interference diagrams. In cMFV SUSY, superpartners of heavy
flavours can only be produced through the purely left-handed
$s$-channel $W$-exchange, since the $t$-channel diagrams are
suppressed by the small bottom and negligible top quark densities
in the proton, and one recovers the result in Ref.\
\cite{Bozzi:2005sy}. In NMFV, $t$-channel exchanges can, however,
contribute to heavy-flavour final state production from
light-flavour initial states and even
become dominant, due to the strong gluino coupling.\\

\begin{figure}
 \centering
 \includegraphics[width=\columnwidth]{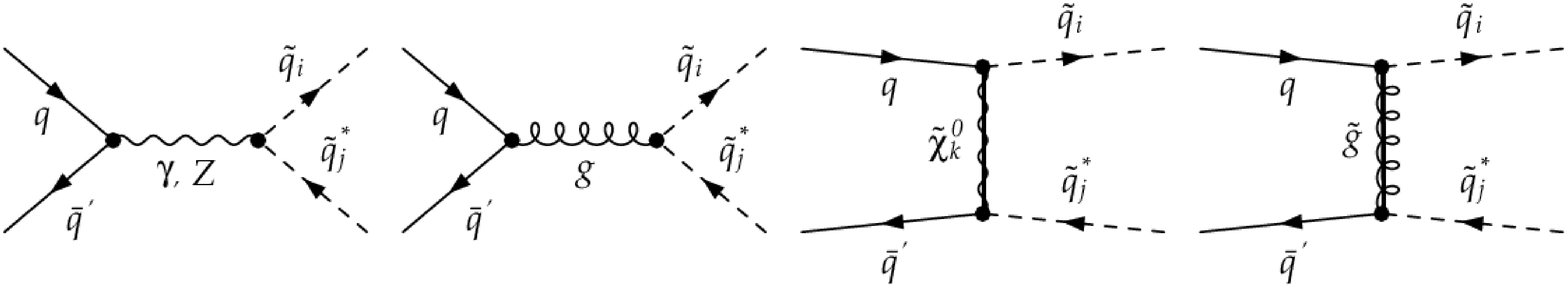}
 \includegraphics[width=\columnwidth]{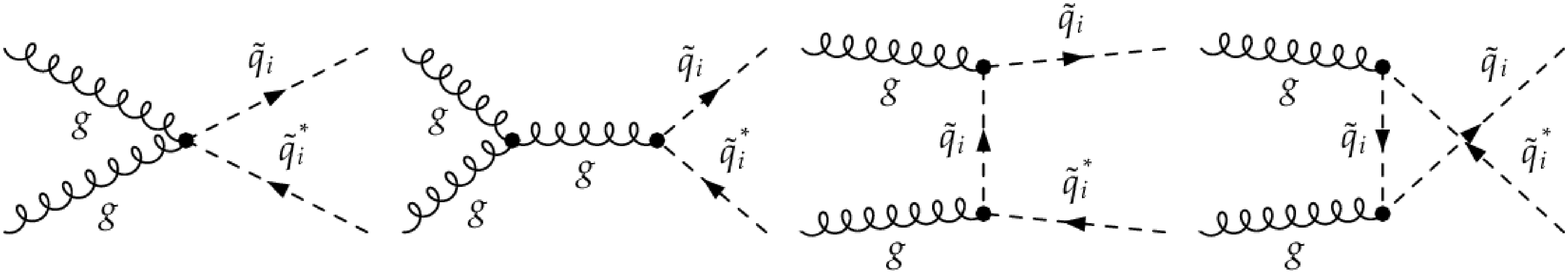}
 \caption{\label{fig:27}Tree-level Feynman diagrams for the production of
          neutral squark-antisquark pairs in quark-antiquark (top) and
          gluon-gluon collisions (bottom).}
\end{figure}
Neutral squark-antisquark pair production in NMFV proceeds either
from equally neutral quark-antiquark initial states \bea q(h_a,
p_a)\, \bar{q}^\prime(h_b, p_b) \to \tilde{q}_i(p_1)\,
\tilde{q}^\ast_j(p_2) \eea through the five different
gauge-boson/gaugino exchanges shown in Fig.\ \ref{fig:27} (top) or
from gluon-gluon initial states \bea g(h_a,p_a)\, g(h_b, p_b) \to
\tilde{q}_i(p_1)\, \tilde{q}^\ast_i(p_2) \eea through the purely
strong couplings shown in Fig.\ \ref{fig:27} (bottom). The
differential cross section for quark-antiquark scattering
\cite{Bozzi:2007me} \bea \frac{{\rm d}
\hat{\sigma}^{q\bar{q}'}_{h_a, h_b}}{dt} &=& (1-h_a) (1+h_b)
\Bigg[ \frac{\mathcal{Y}}{s^2} + \frac{\mathcal{Z}_1}{s_z^2} +
\frac{\mathcal{G}}{s^2} +
\frac{\tilde{\mathcal{G}}_{11}}{t_{\tilde{g}}^2} +
\frac{\mathcal{[YZ]}_1}{s\, s_z} +
\frac{[\tilde{\mathcal{G}}\mathcal{Y}]_1}{t_{\tilde{g}}\, s} +
\frac{[\tilde{\mathcal{G}}\mathcal{Z}]_1}{t_{\tilde{g}}\, s_z}
\nonumber\\ &+&
\frac{[\tilde{\mathcal{G}}\mathcal{G}]_1}{t_{\tilde{g}}\, s} +
\sum_{k,l=1,...,4}
\bigg(\frac{\mathcal{N}^{kl}_{11}}{t_{\tilde{\chi}^0_k}\,
t_{\tilde{\chi}^0_l}}\bigg) + \sum_{k=1,...,4} \bigg(
\frac{\mathcal{[NY]}^k_1}{t_{\tilde{\chi}^0_k}\, s} +
\frac{\mathcal{[NZ]}^k_1}{t_{\tilde{\chi}^0_k}\, s_z} +
\frac{\mathcal{[NG]}^k_1}{t_{\tilde{\chi}^0_k}\, s} \bigg)
\Bigg]\nonumber \\ &+& (1+h_a) (1-h_b) \Bigg[
\frac{\mathcal{Y}}{s^2} + \frac{\mathcal{Z}_2}{s_z^2} +
\frac{\mathcal{G}}{s^2} +
\frac{\tilde{\mathcal{G}}_{22}}{t_{\tilde{g}}^2} +
\frac{\mathcal{[YZ]}_2}{s\, s_z} +
\frac{[\tilde{\mathcal{G}}\mathcal{Y}]_2}{t_{\tilde{g}}\, s} +
\frac{[\tilde{\mathcal{G}}\mathcal{Z}]_2}{t_{\tilde{g}}\, s_z}
\nonumber\\ &+&
\frac{[\tilde{\mathcal{G}}\mathcal{G}]_2}{t_{\tilde{g}}\, s} +
\sum_{k,l=1,...,4}
\bigg(\frac{\mathcal{N}^{kl}_{22}}{t_{\tilde{\chi}^0_k}\,
t_{\tilde{\chi}^0_l}}\bigg) + \sum_{k=1,...,4} \bigg(
\frac{\mathcal{[NY]}^k_2}{t_{\tilde{\chi}^0_k}\, s} +
\frac{\mathcal{[NZ]}^k_2}{t_{\tilde{\chi}^0_k}\, s_z} +
\frac{\mathcal{[NG]}^k_2}{t_{\tilde{\chi}^0_k}\, s} \bigg)
\Bigg]\nonumber\eea\bea &+& (1-h_a) (1-h_b) \Bigg[
\frac{\tilde{\mathcal{G}}_{12}}{t_{\tilde{g}}^2} +
\sum_{k,l=1,...,4}
\bigg(\frac{\mathcal{N}^{kl}_{12}}{t_{\tilde{\chi}^0_k}\,
t_{\tilde{\chi}^0_l}}\bigg) \Bigg]\hspace{2.5cm} \nn \\ &+&
(1+h_a) (1+h_b) \Bigg[
\frac{\tilde{\mathcal{G}}_{21}}{t_{\tilde{g}}^2} +
\sum_{k,l=1,...,4}
\bigg(\frac{\mathcal{N}^{kl}_{21}}{t_{\tilde{\chi}^0_k}\,
t_{\tilde{\chi}^0_l}}\bigg) \Bigg] \label{eq:sqsq2}\eea involves
many different form factors,\bea \mathcal{Y} &=& \frac{\pi\,
\alpha^2\, e_q^2\, e_{\tilde{q}}^2\, \delta_{ij}\,
\delta_{qq^\prime}}{s^2} \left( u\, t - m^2_{\tilde{q}_i}\,
m^2_{\tilde{q}_j^\prime}\right),~ \nonumber \\ \mathcal{Z}_m &=&
\frac{\pi\, \alpha^2}{16\, s^2\, x_W^2 (1-x_W)^2} \left|
L_{\tilde{q}_i \tilde{q}_j Z} +R_{\tilde{q}_i \tilde{q}_j Z}
\right|^2\, \left( C^m_{q q^\prime Z}\right)^2 \left( u\, t -
m^2_{\tilde{q}_i}\, m^2_{\tilde{q}_j^\prime} \right) ,~\nonumber
\\ \mathcal{G} &=& \frac{2\, \pi\, \alpha_s^2\, \delta_{ij}\,
\delta_{qq^\prime}}{9\, s^2} \left( u\, t - m^2_{\tilde{q}_i}\,
m^2_{\tilde{q}_j^\prime}\right) ,~ \nonumber\\
\mathcal{N}_{mn}^{kl} &=& \frac{\pi\, \alpha^2}{x_W^2\, (1 -
x_W)^2\, s^2}\, \mathcal{C}^{m\ast}_{\tilde{q}_i q
\tilde{\chi}_k^0}\, \mathcal{C}^m_{\tilde{q}_i q
\tilde{\chi}_l^0}\, \mathcal{C}^n_{\tilde{q}_j q^\prime
\tilde{\chi}_k^0} \mathcal{C}^{n\ast}_{\tilde{q}_j q^\prime
\tilde{\chi}_l^0}\nn\\ &\times& \Bigg[  \left( u\, t -
m^2_{\tilde{q}_i}\, m^2_{\tilde{q}_j}\right)\, \delta_{mn} +
\left( m_{\tilde{\chi}^0_k}\, m_{\tilde{\chi}^0_l}\,  s
\right)\, \left( 1-\delta_{mn} \right) \Bigg] ,~ \nonumber \\
\tilde{\mathcal{G}}_{mn} &=& \frac{2\, \pi\, \alpha_s^2}{9\,
s^2}\, \left| \mathcal{C}^m_{\tilde{q}_i q \tilde{g}}\,
\mathcal{C}^{n\ast}_{\tilde{q}_j q^\prime \tilde{g}}\right|^2
\Bigg[ \left( u\, t - m^2_{\tilde{q}_i}\,
m^2_{\tilde{q}_j}\right)\, \delta_{mn} + \left( m_{\tilde{g}}^2\,
s \right)\, \left(1-\delta_{mn} \right) \Bigg],~ \nonumber\\
\mathcal{[YZ]}_m &=&  \frac{\pi\, \alpha^2\, e_q\, e_{\tilde{q}}\,
\delta_{ij}\, \delta_{qq^\prime}}{2\, s^2\, x_W (1-x_W)}\, {\rm
Re} \left[ L_{\tilde{q}_i \tilde{q}_j Z} + R_{\tilde{q}_i
\tilde{q}_j Z} \right]\, C^m_{q q^\prime Z} \left( u\, t -
m^2_{\tilde{q}_i}\,
m^2_{\tilde{q}_j^\prime}\right) ,~ \nonumber\\
\mathcal{[NY]}_m^k &=& \frac{2\, \pi\, \alpha^2\, e_q\,
e_{\tilde{q}}\, \delta_{ij}\, \delta_{qq^\prime}}{3\, x_W\, (1 -
x_W)\, s^2}\, {\rm Re} \left[\mathcal{C}^m_{\tilde{q}_i q
\tilde{\chi}_k^0}\, \mathcal{C}^{m\ast}_{\tilde{q}_j q^\prime
\tilde{\chi}_k^0} \right] \left( u\, t - m^2_{\tilde{q}_i}\,
m^2_{\tilde{q}_j}\right),~ \nonumber
\\ \mathcal{[NZ]}_m^k &=& \frac{\pi\, \alpha^2}{6\, x_W^2\, (1 -
x_W)^2\, s^2}\, {\rm Re} \left[\mathcal{C}^m_{\tilde{q}_i q
\tilde{\chi}_k^0}\, \mathcal{C}^{m\ast}_{\tilde{q}_j q^\prime
\tilde{\chi}_k^0} \left( L_{\tilde{q}_i \tilde{q}_j Z} +
R_{\tilde{q}_i \tilde{q}_j Z} \right)\right]\, \mathcal{C}^m_{q
q^\prime Z}\,\nn\\&\times& \left( u\, t -
m^2_{\tilde{q}_i}\, m^2_{\tilde{q}_j}\right),~ \nonumber\\
\mathcal{[NG]}_m^k &=& \frac{8\, \pi\, \alpha\, \alpha_s\,
\delta_{ij}\, \delta_{qq^\prime}}{9\, x_W\, (1 - x_W)\, s^2}\,
{\rm Re} \left[\mathcal{C}^m_{\tilde{q}_i q \tilde{\chi}_k^0}\,
\mathcal{C}^{m\ast}_{\tilde{q}_j q^\prime \tilde{\chi}_k^0}
\right] \left( u\, t -
m^2_{\tilde{q}_i}\, m^2_{\tilde{q}_j}\right),~ \nonumber\\
\big[\tilde{\mathcal{G}} \mathcal{G}\big]_m &=& - \frac{4\, \pi\,
\alpha_s^2\, \delta_{ij}\, \delta_{qq^\prime}}{27\, s^2}\, {\rm
Re} \left[\mathcal{C}^{m\ast}_{\tilde{q}_i q \tilde{g}}\,
\mathcal{C}^m_{\tilde{q}_j q^\prime \tilde{g}} \right] \left( u\,
t - m^2_{\tilde{q}_i}\,
m^2_{\tilde{q}_j}\right),~ \nonumber\\
\mathcal{[\tilde{\mathcal{G}} Z]}_m &=& \frac{2\, \pi\, \alpha\,
\alpha_s}{9\, x_W(1 \!-\! x_W)\, s^2}\, {\rm Re}
\left[\mathcal{C}^{m\ast}_{\tilde{q}_i q \tilde{g}}
\mathcal{C}^m_{\tilde{q}_j q^\prime \tilde{g}} \left(\!
L_{\tilde{q}_i \tilde{q}_j Z} \!+\! R_{\tilde{q}_i \tilde{q}_j Z}
\right) \!\right] \mathcal{C}^m_{q q^\prime Z}\, \left(\! u\, t
\!-\! m^2_{\tilde{q}_i}\, m^2_{\tilde{q}_j}\!\right),\nn\\
\mathcal{[\tilde{\mathcal{G}} Y]}_m &=& \frac{8\, \pi\, \alpha\,
\alpha_s\, e_q\, e_{\tilde{q}}\, \delta_{ij}\,
\delta_{qq^\prime}}{9\, s^2}\, {\rm Re}
\left[\mathcal{C}^{m\ast}_{\tilde{q}_i q \tilde{g}}\,
\mathcal{C}^m_{\tilde{q}_j q^\prime \tilde{g}} \right] \left( u\,
t - m^2_{\tilde{q}_i}\, m^2_{\tilde{q}_j}\right),~ \eea since only
very few interferences (those between strong and electroweak
channels of the same propagator type) are eliminated due to colour
conservation. On the other hand, the gluon-initiated cross section
\bea \frac{{\rm d} \hat{\sigma}^{gg}_{h_a, h_b}}{dt} \!=\!
\frac{\pi\alpha_s^2}{128 s^2}\! \left[\! 24 \left(\!
1\!-\!2\frac{t_{\tilde{q}_i} u_{\tilde{q}_i}}{s^2}\!\right) \!-\!
\frac{8}{3}\!\right]\!\! \left[\! (1\!-\!h_a h_b)\!-\!2 \frac{s
m_{\tilde{q}_i}^2}{t_{\tilde{q}_i} u_{\tilde{q}_i}} \!\left(\!
(1\!-\!h_a h_b) \!-\! \frac{s m_{\tilde{q}_i}^2}{t_{\tilde{q}_i}
u_{\tilde{q}_i}}\! \right)\!\right] \label{eq:sqsq3}~\eea involves
only the strong coupling constant and is thus quite compact. In
the case of cMFV, our results agree with those in Ref.\
\cite{Bozzi:2005sy} for diagonal and non diagonal squark
helicities in the final state. Diagonal production of identical
squark-antisquark mass eigenstates is, of course, dominated by the
strong quark-antiquark and gluon-gluon channels. Their relative
importance depends on the partonic luminosity and thus on the type
and energy of the hadron collider under consideration.
Non-diagonal production of squarks of different helicity or
flavour involves only electroweak and gluino-mediated
quark-antiquark scattering, and the relative importance of these
processes depends largely on the gluino mass.

\subsection{Numerical results}\label{sec:sqsq}

\begin{figure}
 \centering
 \includegraphics[height=0.32\textheight]{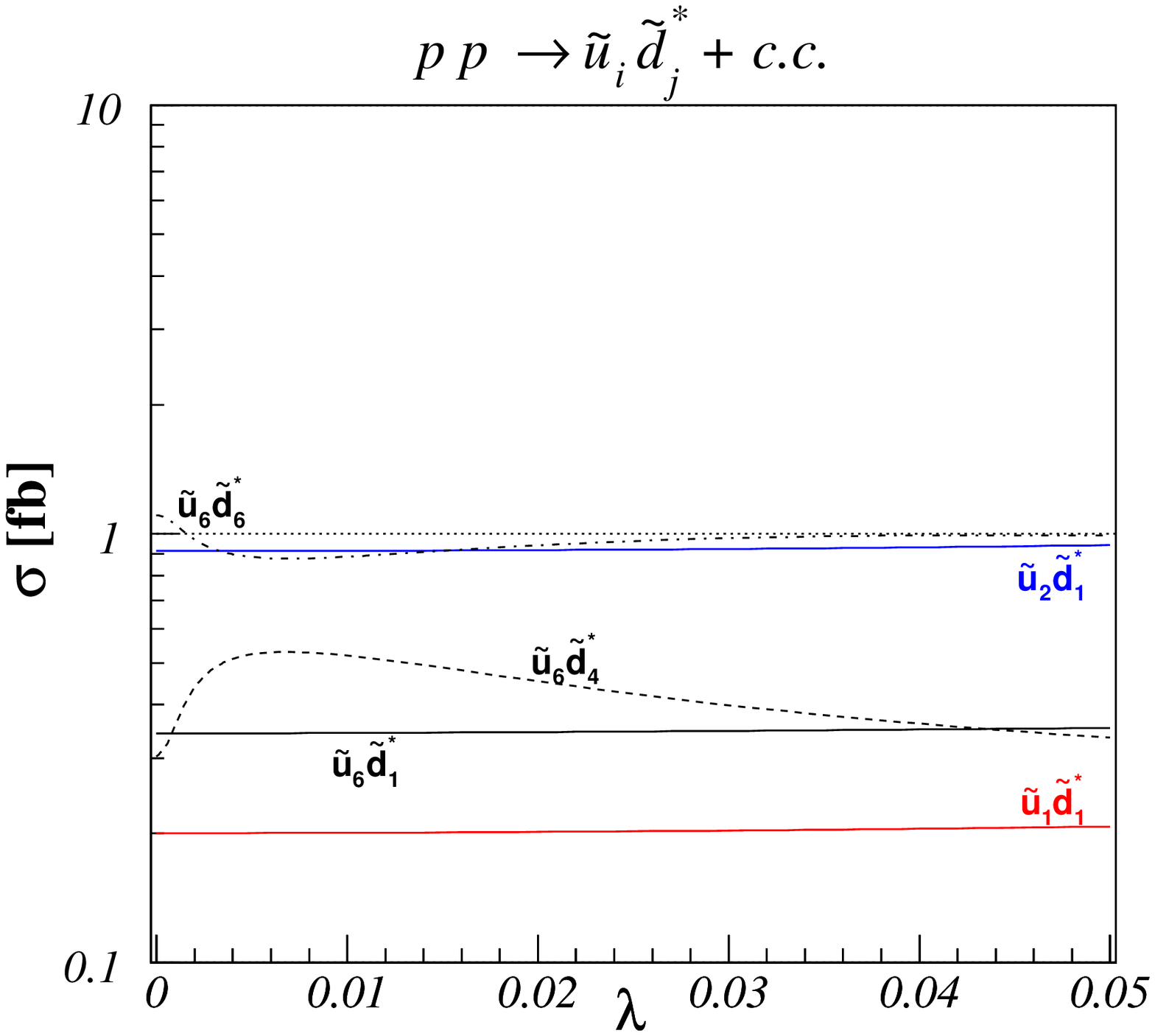}
 \includegraphics[height=0.32\textheight]{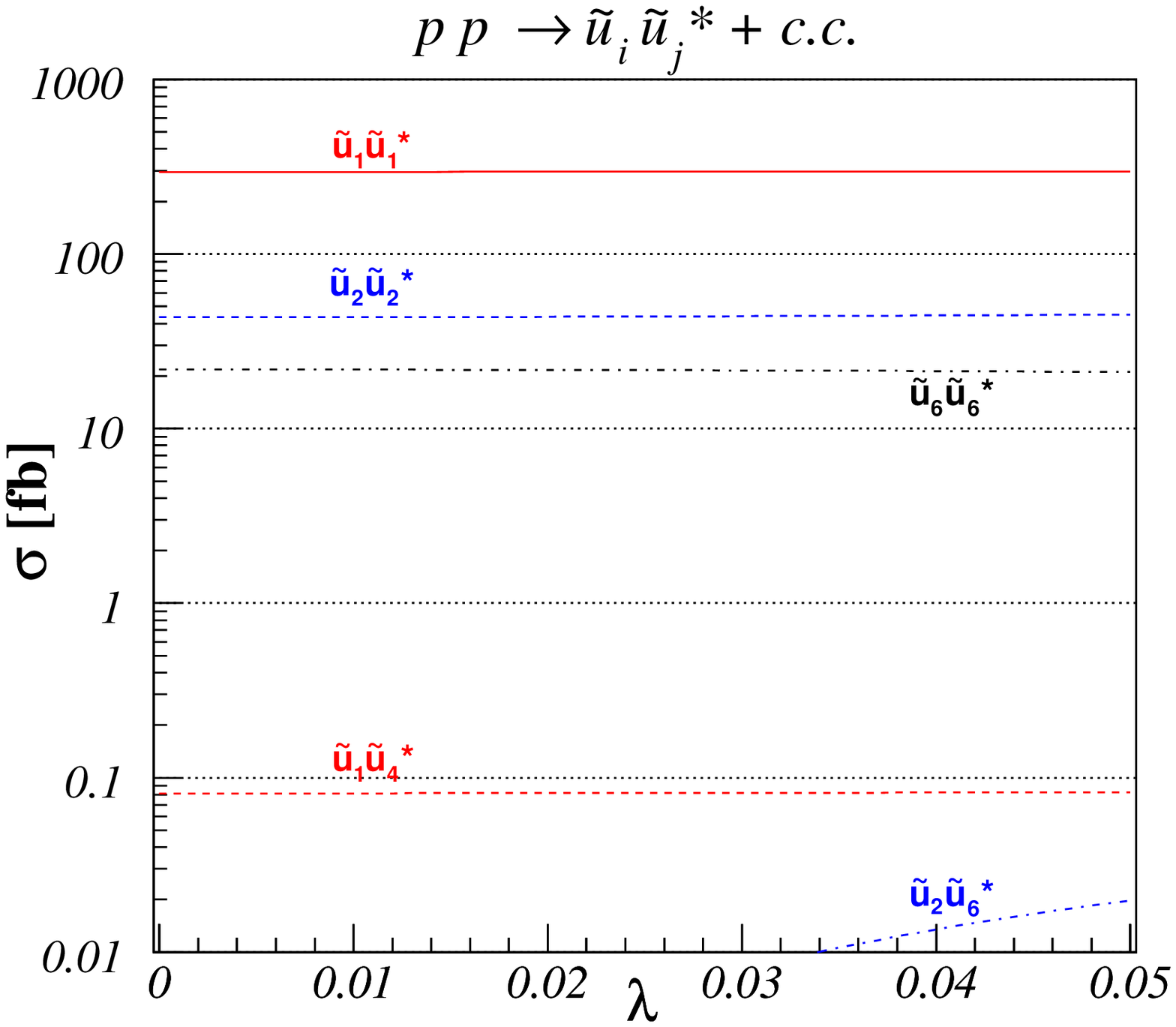}
 \includegraphics[height=0.32\textheight]{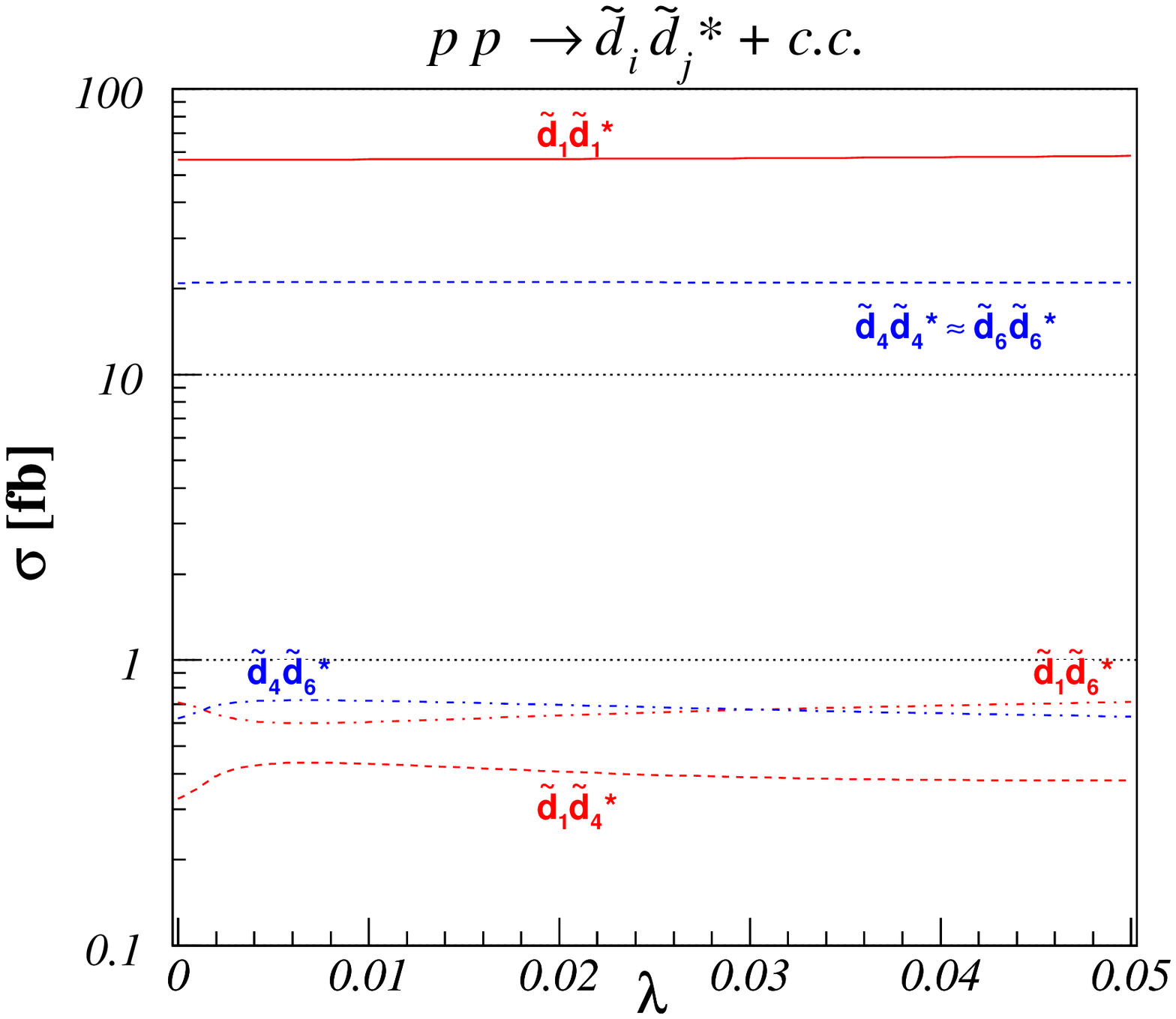}
 \caption{\label{fig:30a}Cross sections for charged squark-antisquark
 (top) and neutral up-type (centre) and down-type (bottom)
 squark-antisquark pair production at the LHC in our benchmark
 scenario A.}
\end{figure}
\begin{figure}
 \centering
 \includegraphics[height=0.32\textheight]{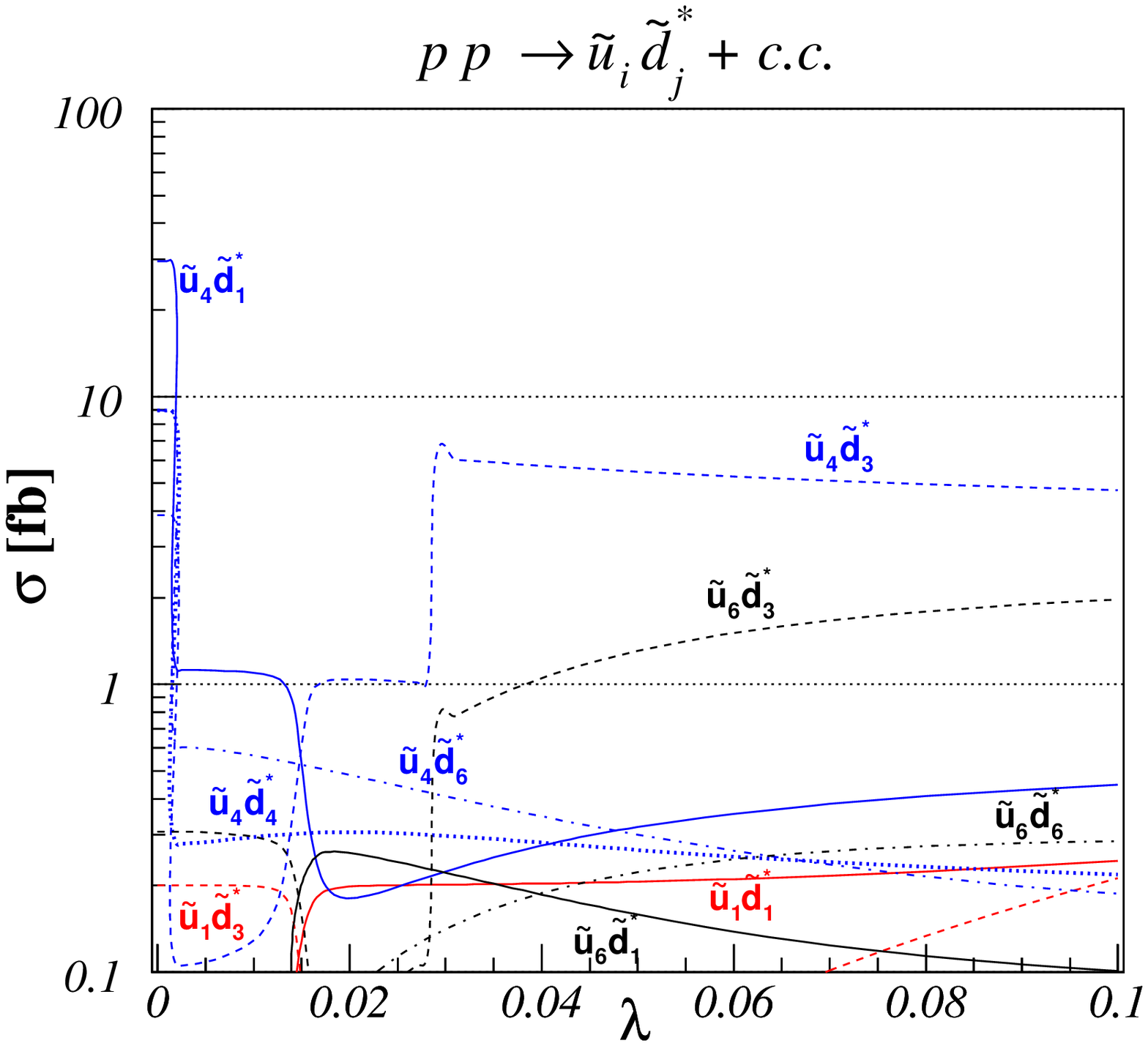}
 \includegraphics[height=0.32\textheight]{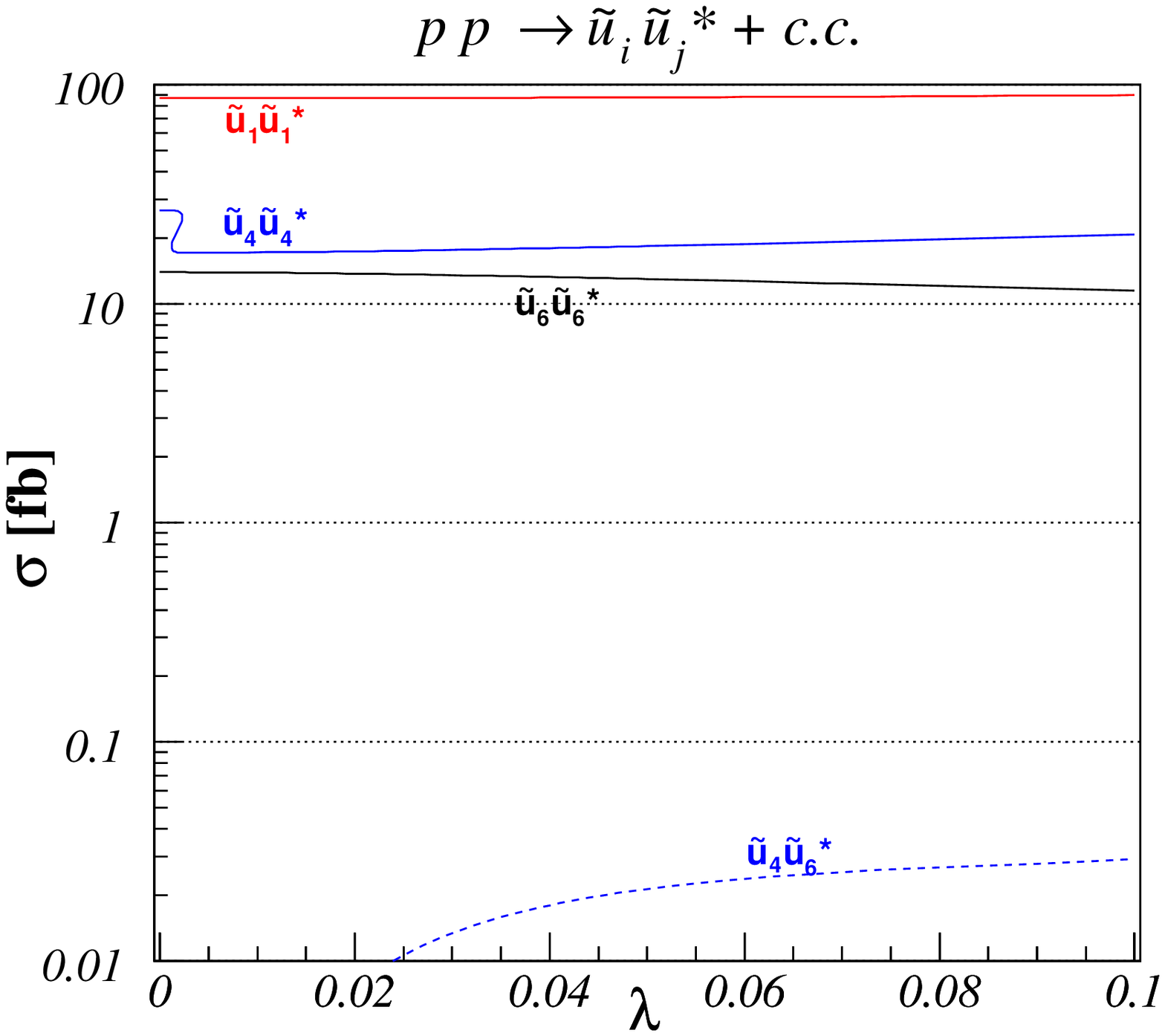}
 \includegraphics[height=0.32\textheight]{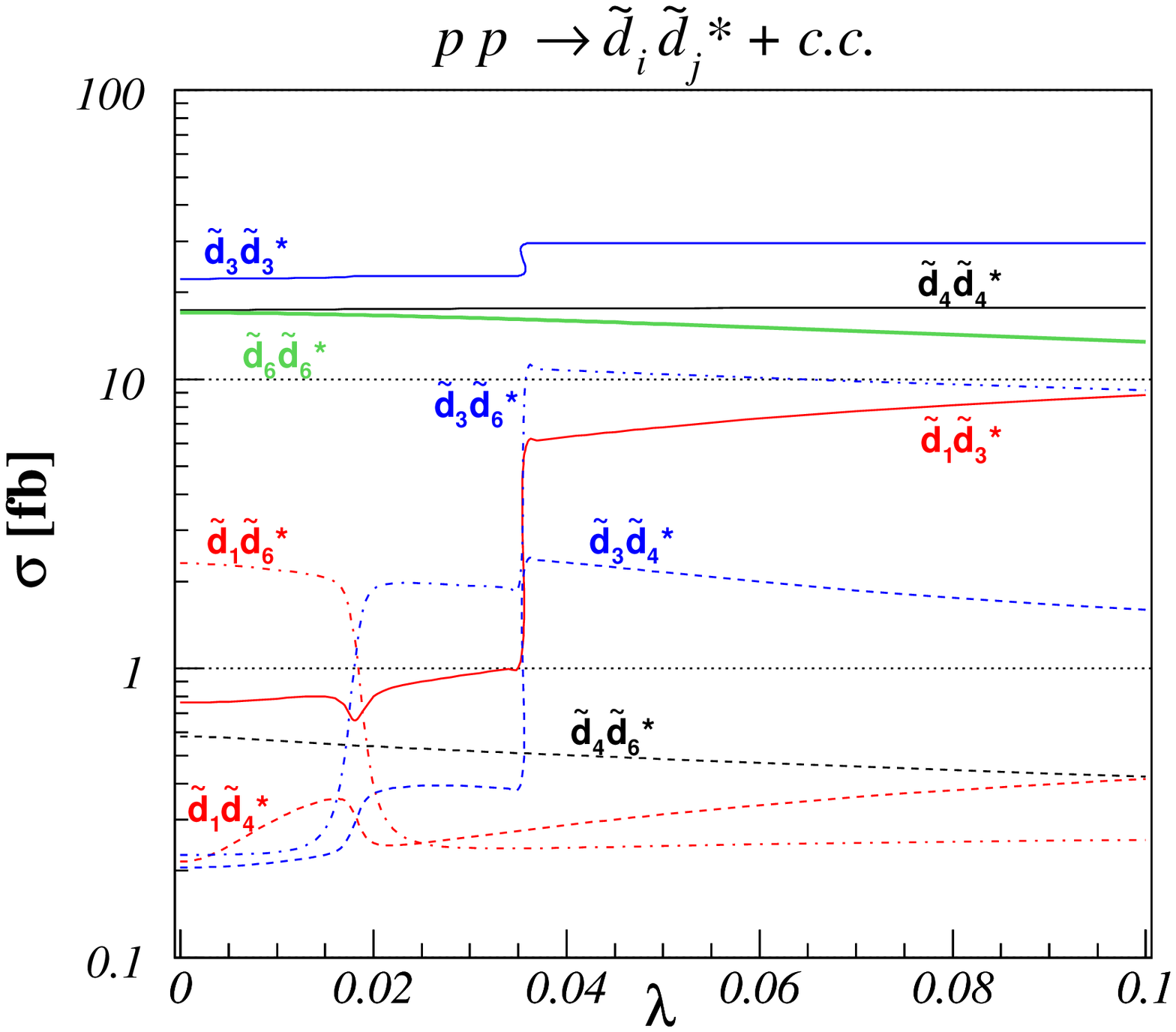}
 \caption{\label{fig:30b}Same as Fig.\ \ref{fig:30a} for our benchmark
          scenario B.}
\end{figure}
\begin{figure}
 \centering
 \includegraphics[height=0.32\textheight]{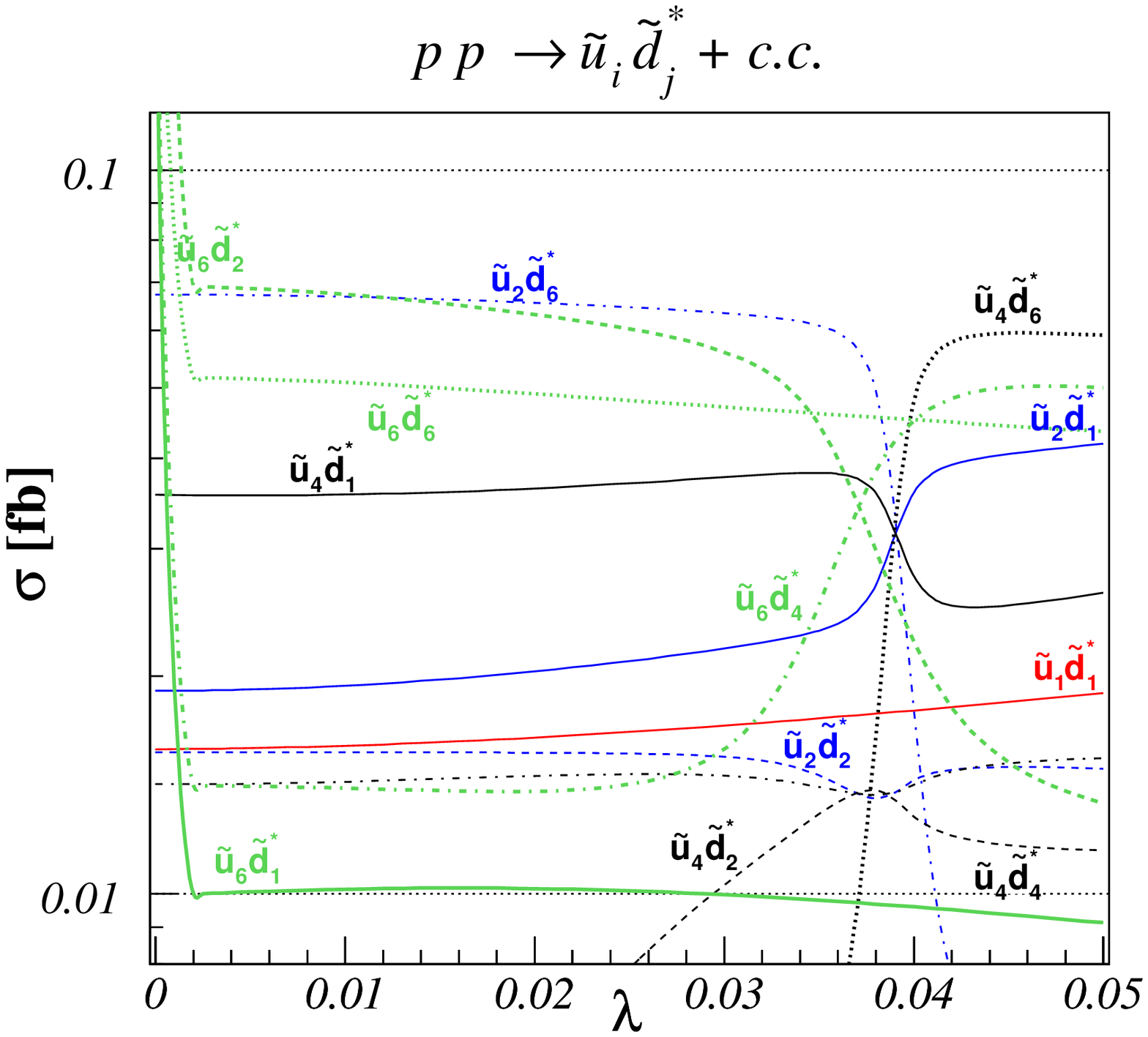}
 \includegraphics[height=0.32\textheight]{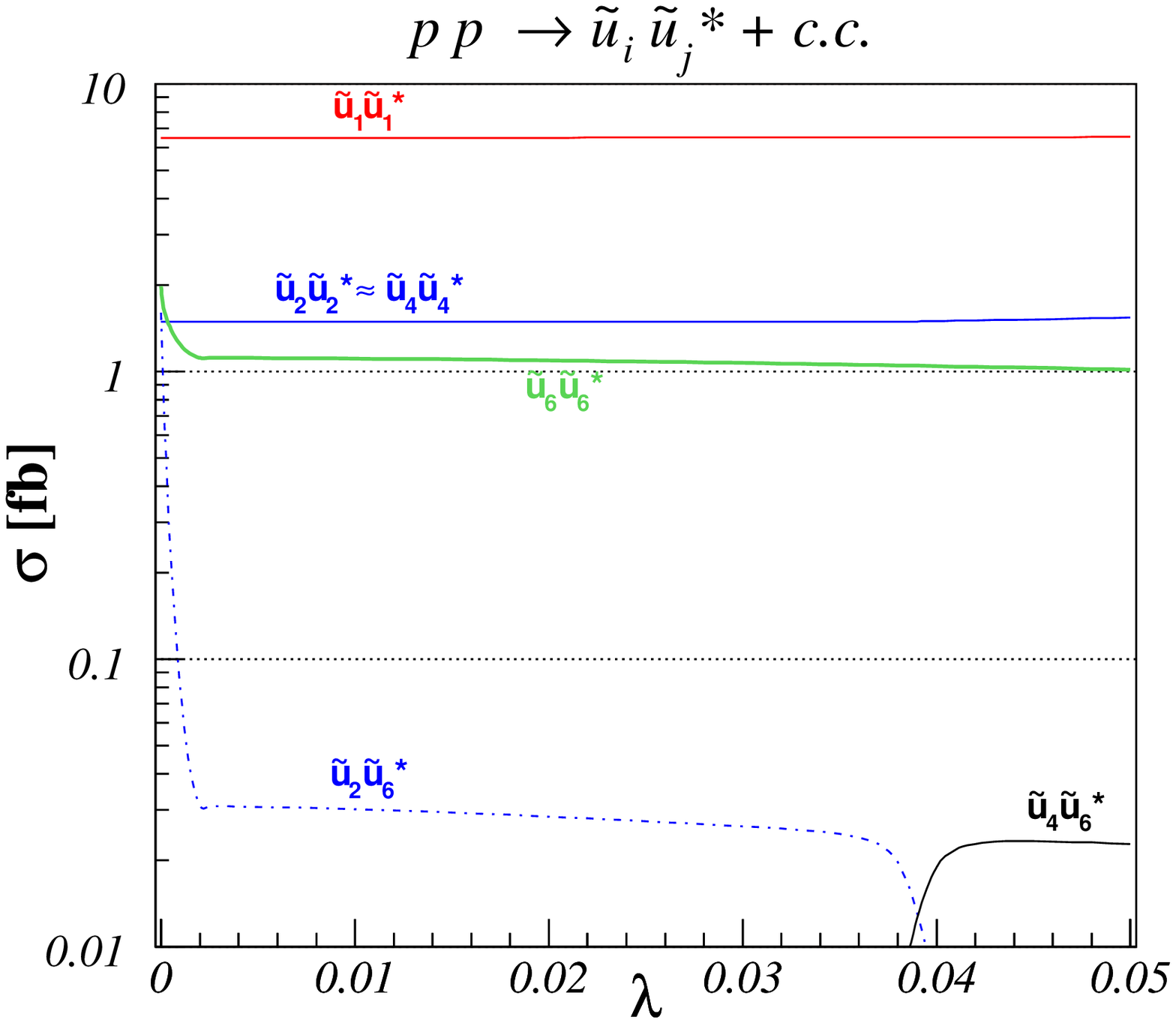}
 \includegraphics[height=0.32\textheight]{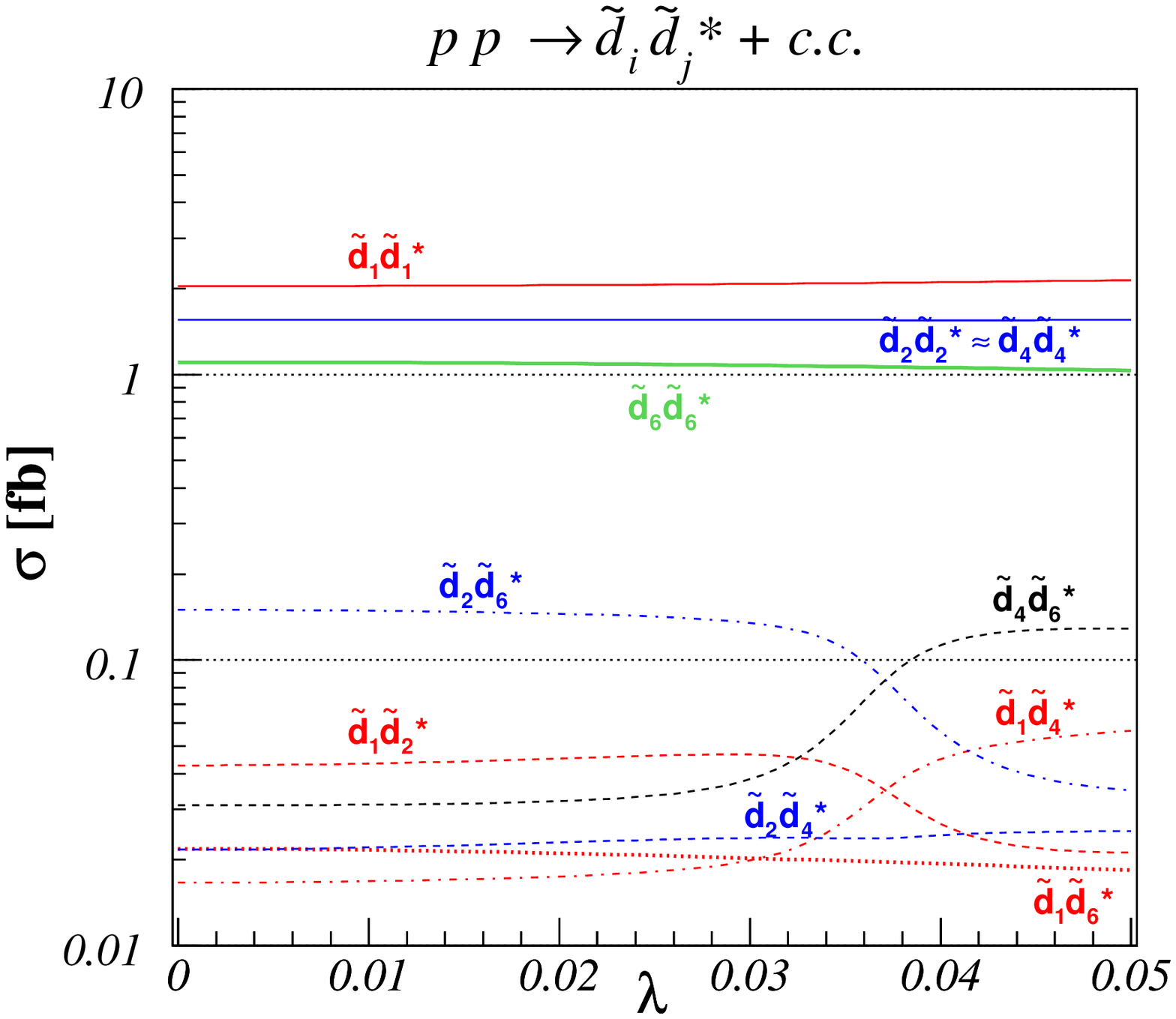}
 \caption{\label{fig:30c}Same as Fig.\ \ref{fig:30a} for our benchmark
          scenario C.}
\end{figure}
\begin{figure}
 \centering
 \includegraphics[height=0.32\textheight]{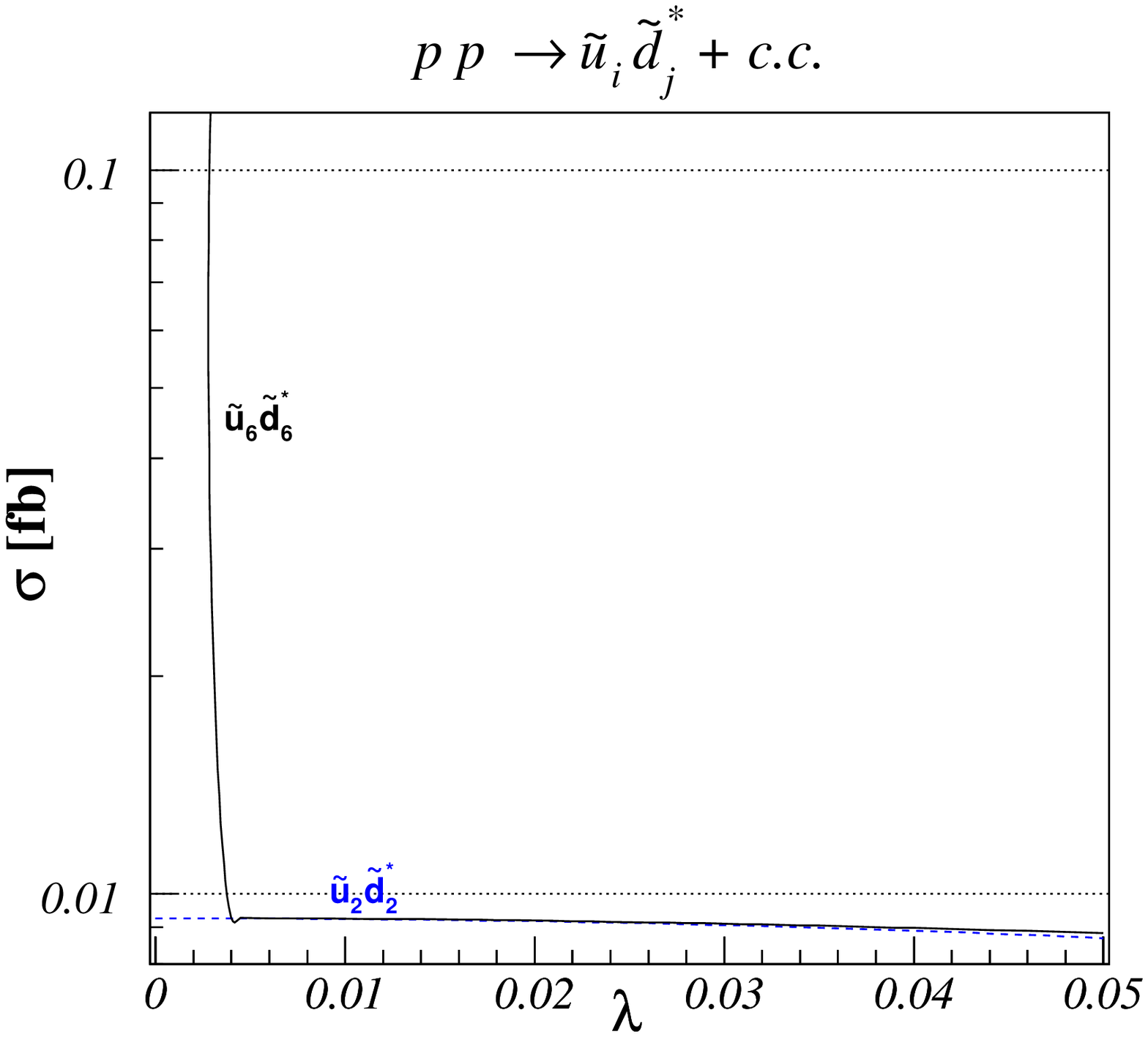}
 \includegraphics[height=0.32\textheight]{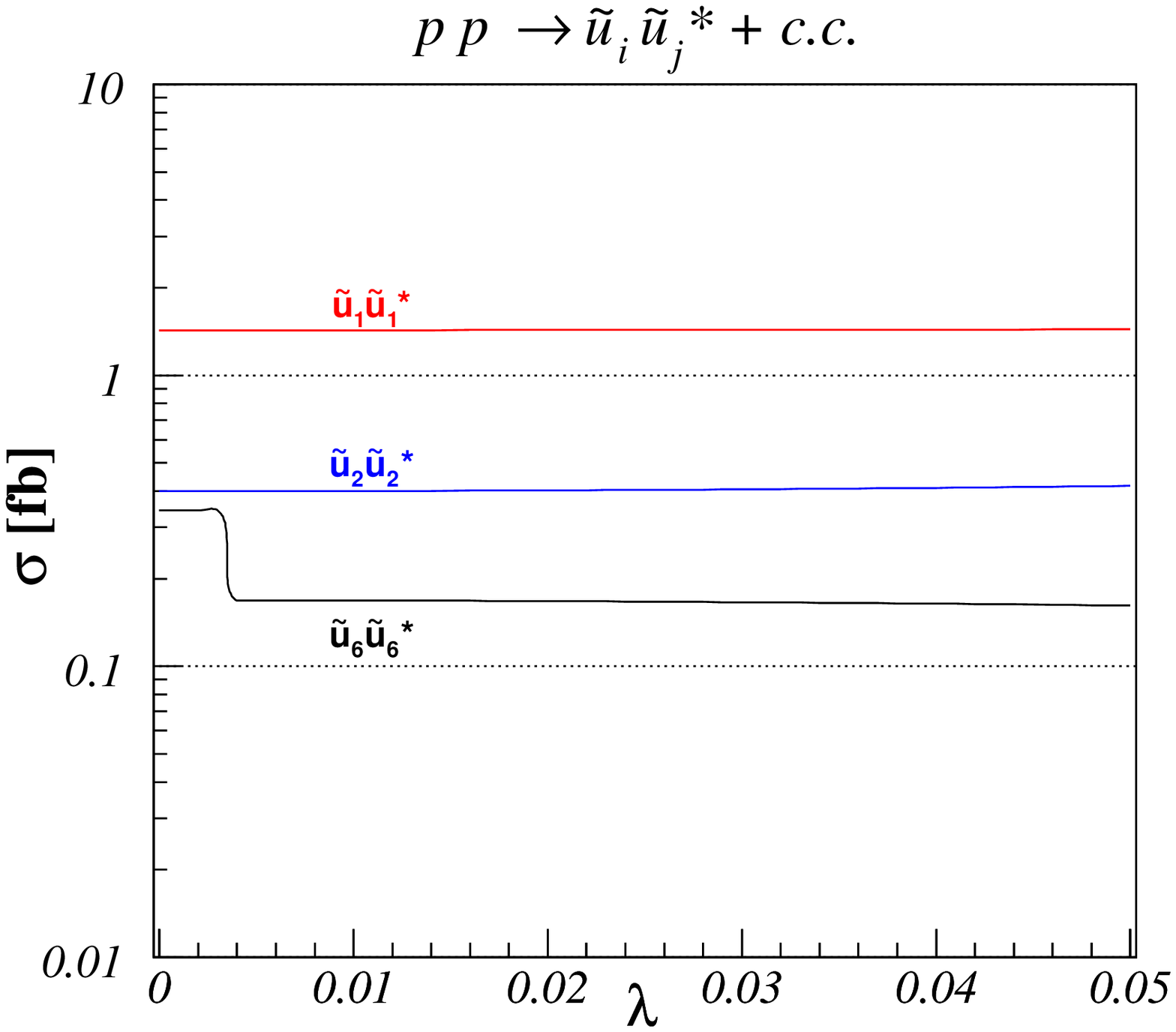}
 \includegraphics[height=0.32\textheight]{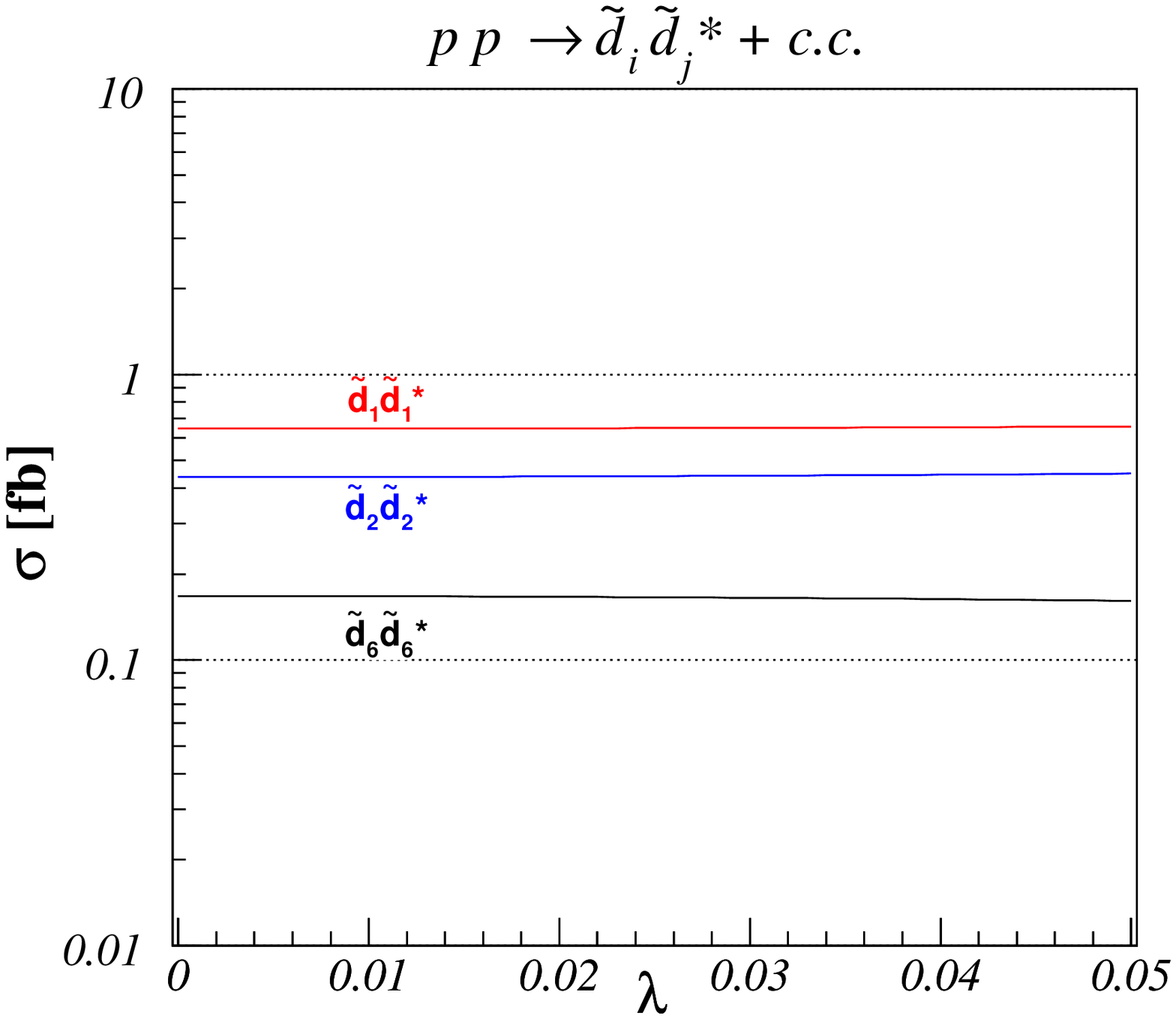}
 \caption{\label{fig:30d}Same as Fig.\ \ref{fig:30a} for our benchmark
          scenario D.}
\end{figure}

We present numerical predictions for the production cross sections
of squark-antisquark pairs in NMFV SUSY at the CERN LHC. Again, we
use the QCD factorization theorem to compute the total unpolarized
hadronic cross sections \bea \sigma &~=&
\int_{4m^2/s_h}^1\!\td\tau\!\!
\int_{-1/2\ln\tau}^{1/2\ln\tau}\!\!\td y \int_{t_{\min}}^{t_{\max}}
\td t \ f_{a/A}(x_a,M_a^2) \ f_{b/B}(x_b,M_b^2) \
{\td\hat{\sigma}\over\td t},~ \eea where the relevant partonic cross
section d$\hat{\sigma}$/d$t$ is convolved with universal parton
densities. For consistency with our leading order QCD calculation
in the collinear approximation, where all squared quark masses,
the top excepted, are much lower than the partonic centre-of-mass
energy, we employ the LO set of the latest CTEQ6 global parton
density fit \cite{Pumplin:2002vw}, which includes $n_f=5$
``light'' quark flavours and the gluon, but no top-quark density.
For gluon initial states and gluon or gluino exchanges, the strong
coupling constant $\alpha_s(\mu_R)$ is calculated with the
corresponding LO value of $\Lambda_{\rm LO}^{n_f=5}=165$ MeV. We
identify the renormalization scale $\mu_R$ with the factorization
scales $M_a=M_b$ and set the scales to the average mass of the
final state SUSY particles $i$ and $j$, $m=(m_i+m_j)/2$.\\

We take current masses and widths of the electroweak gauge bosons,
electroweak mixing angle, electromagnetic fine structure constant
and CKM matrix elements, while the soft SUSY-breaking masses at
the electroweak scale are computed thanks  the computer program
SPheno 2.2.3 \cite{Porod:2003um}. Then we introduce NMFV as in
Sec.\ \ref{sec:scan} through the parameter $\lambda$ defined in
Eq.\ (\ref{eq:lambda}). The physical masses of the SUSY particles
and the mixing angles, taking into account NMFV, are obtained with
the computer program FeynHiggs 2.5.1 \cite{Heinemeyer:1998yj}.\\

The numerical cross sections for charged squark-antisquark and
neutral up- and down-type squark-antisquark pair production are
shown in Figs.\ \ref{fig:30a}, \ref{fig:30b}, \ref{fig:30c} and
\ref{fig:30d} for the benchmark scenarios A, B, C and D described
in Sec.\ \ref{sec:scan}, respectively. The magnitudes of the cross
sections vary from the barely visible level of $10^{-2}$ fb for
weak production of heavy final states to large cross sections of
$10^2$ to $10^3$ fb for the strong production of diagonal
squark-antisquark pairs. Unfortunately, these processes, whose
cross sections are largest, are practically insensitive to the
flavour violation parameter $\lambda$, as the strong gauge
interaction is insensitive to quark flavours.\\

Some of the subleading, non-diagonal cross sections show, however,
sharp transitions, in particular down-type squark-antisquark
production at the benchmark point B (bottom panel on Fig.\
\ref{fig:30b}), but also other squark-antisquark production
processes. At $\lambda=0.02$, the cross sections for $\tilde{d}_1
\tilde{d}_6^\ast$ and $\tilde{d}_3\tilde{d}_6^\ast$ switch places.
The concerned down-type squark mass differences are rather small
(see lower right panel of Fig.\ \ref{fig:05}), and this flip is
then mainly due to the interplay between the different strange and
bottom quark densities in the proton and the level reordering
phenomenon discussed in Sec.\ \ref{sec:bench}, since at this value
of $\lambda$, the flavour contents of $\tilde{d}_1$ and
$\tilde{d}_3$ are exchanged (see Fig.\ \ref{fig:010p}).\\

At $\lambda=0.035$, the flavour content of the eigenstate
$\tilde{d}_3$ switches from $\tilde{s}_R$ to $\tilde{d}_R$,
leading to a sharp increase of the cross sections for $\tilde{d}_3
\tilde{d}_6^\ast$ and $\tilde{d}_1 \tilde{d}_3^\ast$, since these
final states can then be produced from down-type valence quarks.
For $\lambda > 0.035$, the cross section of the latter process,
$\tilde{d}_1 \tilde{d}_3^\ast$ production, increases with the
strange squark content of $\tilde{d}_1$ (see Fig.\
\ref{fig:010p}). Let us note that all of these non-diagonal squark
pair production cross sections are mainly dominated by the
exchange of strongly coupled gluinos despite of their large mass.

\subsection{cMFV limit}

We investigate in this subsection the production of diagonal (same
mass-eigenstates), non-diagonal (different mass eigenstates) and
mixed (different flavours) third generation squark-antisquark
pairs within the limit of the constrained Minimal Flavour
Violation, where the squark mixing is parameterized only by the
two mixing angles $\theta_{\tilde{t}}$ and $\theta_{\tilde{b}}$.
The cMFV limit of Eq.\ (\ref{eq:sqsq2}) in the case of diagonal
squark-antisquark pair production, keeping only strong
contributions and showing explicitly the mixing angle dependence,
is given by \cite{Bozzi:2005sy} \bea
{\td\hat{\sigma}_{h_a,h_b}^{q\bar{q}}\over\td t} &=&
{4\pi\alpha_s^2\over 9s^2}\,\le1-h_ah_b\re\,{tu-\ms^4\over s^2}
\nonumber\\ &-& {4\pi\alpha_s^2\over27s^2}\, \le1\pm(h_b-h_a)
\cos2\theta_{\tilde{q}} - h_ah_b\re\, {tu-\ms^4\over s
t_{\tilde{g}}}\,\delta_{qb}\,\delta_{\tilde{q}\tilde{b}}
\nonumber\\ &+& { \pi\alpha_s^2\over18s^2}\ \le {(1\!+\!h_a
h_b)\,(1\!-\!\cos4\theta_{\tilde{q}})\,m_{\tilde{g}}^2 s\!+\!
(1\!-\!h_a h_b)\,(3\!+\!\cos4\theta_{\tilde{q}})\, (t
u\!-\!\ms^4)\over t_{\tilde{g}}^2} \rp \nonumber\\ & & \lp \pm\,4
\cos2\theta_{\tilde{q}}\,(h_b-h_a){tu-\ms^4\over
t_{\tilde{g}}^2}\re \delta_{qb}\,
\delta_{\tilde{q}\tilde{b}},\hspace{4cm} \eea where the upper sign
holds for $\tilde{b}_1$ and the lower sign for $\tilde{b}_2$
production. Stops are produced only through $s$-channel gluon
exchange due to negligible top quark PDFs in the proton. Even for
sbottom production, $t$-channel gluino contributions are
suppressed by small bottom PDFs and the heavy gluino propagator.
The cross section for the gluon-fusion initiated subprocess in
Eq.\ (\ref{eq:sqsq3}) is left unchanged, since it is mixing
independent. In the case of no squark mixing, our results agree
with the double-polarized and unpolarized cross
sections in Ref.\ \cite{Gehrmann:2004xu}.\\

For non-diagonal production, $s$-channel strong diagrams are not
present. However, we take into account the PDF-suppressed gluino
$t$-channel and the quark-induced tree-level electroweak diagrams,
proceeding either through an $s$-channel $Z$-boson exchange or a
$t$-channel neutralino exchange, \bea
{\td\hat{\sigma}_{h_a,h_b}^{q\bar{q}}\over\td t} &=&
{\pi\alpha^2\over s^2} {\sin^2(2\hs)\le(1-h_a)(1+h_b)L_{q q
Z}^2+(1+h_a)(1-h_b)R_{q q Z}^2\re\over  32\, x_W^2\,
(1-x_W)^2\,s_z^2} \nn \\ &&\times \le tu-m_{\tilde{q}_1}^2
m_{\tilde{q}_2}^2\re\nonumber \\ &-& {\pi\alpha\alpha_s\over
s^2}{\sin^2(2\hs)\le(1-h_a)(1+h_b)L_{q q Z}-(1+h_a)(1-h_b)R_{q q
Z}\re\over 9\, x_W\, (1-x_W)\,s_z\,\tg}\
\delta_{qb}\,\delta_{\tilde{q}\tilde{b}} \nn \\ &&\times
\le tu-m_{\tilde{q}_1}^2 m_{\tilde{q}_2}^2\re\nonumber \\
&+& \ {\pi\alpha_s^2\over9s^2}\ \le {\le(1+h_ah_b)(3+\cos4\hs) -
4(h_a+h_b)\cos2\hs\re\mg^2 s\over\tg^2} \rp \nonumber \\& & \lp +
{2\sin^22\hs(1-h_ah_b) (tu-m_{\tilde{q}_1}^2
m_{\tilde{q}_2}^2)\over \tg^2}\re
\delta_{qb}\,\delta_{\tilde{q}\tilde{b}} \nonumber \\ &-&
{\pi\alpha^2\over s^2} \!\sum_i\! {\sin^2(2\hs) \le(1\!-\!h_a)
(1\!+\!h_b) L_{q q Z} L_{\tilde{q} q \tilde{\chi}_i^0}^2 \!-\!
(1\!+\!h_a)(1\!-\!h_b) R_{q q Z} R_{\tilde{q} q
\tilde{\chi}_i^0}^2\re\over 12\, x_W^2 (1\!-\!x_W)^2\,
s_z\,\ti}\nn\\&&\times \le tu-m_{\tilde{q}_1}^2
m_{\tilde{q}_2}^2\re \delta_{qb}\,
\delta_{\tilde{q}\tilde{b}}\nonumber \\ &+& {\pi\alpha^2\over
s^2}\sum_{i,j} \le {\delta_{qb}\, \delta_{\tilde{q}\tilde{b}}
\over (1+\delta_{ij})\, x_W^2\, (1-x_W)^2\, \ti\, \tj} \re \Bigg[
\Big\{ (1 + h_ah_b) (3 + \cos4\hs) \nn\\ && - 4(h_a +
h_b)\cos2\hs\Big\} L_{\tilde{q} q \tilde{\chi}_i^0} L_{\tilde{q} q
\tilde{\chi}_j^0} R_{\tilde{q} q \tilde{\chi}_i^0} R_{\tilde{q} q
\tilde{\chi}_j^0}\, \mi \mj s + (tu-m_{\tilde{q}_1}^2
m_{\tilde{q}_2}^2) \nn \\ &&\times\! \sin^22\hs
\Big\{\!(\!1\!-\!h_a\!) (\!1\!+\!h_b\!) L_{\tilde{q} q
\tilde{\chi}_i^0}^2 L_{\tilde{q} q \tilde{\chi}_j^0}^2\! +\!
(1\!+\!h_a) (1\!-\!h_b) R_{\tilde{q} q \tilde{\chi}_i^0}^2
R_{\tilde{q} q \tilde{\chi}_j^0}^2\!\Big\}\! \Bigg],\! \nn \eea
\vspace{-.8cm}\bea ~ \eea  where we have summed over
$\tilde{q}_1\tilde{q}_2^\ast+\tilde{q}_2 \tilde{q}_1^\ast$ final
states. The coupling strengths \bea L_{\tilde{q} q
\tilde{\chi}^0_i} &=& \bigg[ (e_q - T^3_q)\, s_W N_{i1} + T^3_q\,
c_W N_{i2} \bigg],~\\ -R_{\tilde{q} q \tilde{\chi}_i^0}^\ast &=&
e_q\, s_W\, N_{i1},~\eea correspond the the
quark-squark-neutralino interactions in the limit of non-mixing
left- and right-handed (s)quarks, provided that we neglect the
relatively small Yukawa couplings, which is the case here since
top (s)quarks do not take part in the
considered subprocess.\\

Finally, mixed squark-pair production, i.e.\ a pair of one bottom
plus one top squark, proceeds at tree-level only through an
$s$-channel exchange of a charged $W^\pm$-boson, since the
$t$-channel exchanges of Fig.\ \ref{fig:26} are PDF-suppressed due
to the negligible top density inside the proton. The
$\tilde{t}_1\tilde{b}_1$ production cross section is obtained from
Eq.\ (\ref{eq:sqsq1}) by keeping only the $W$ contribution,
\bea{\td\hat{\sigma}_{h_a,h_b}^{q\bar{q}'}\over\td t} &=&
{\pi\alpha^2\over s^2}\ |V_{qq'}|^2 |V_{t b}|^2
\le{tu-m_{\tilde{t}_1}^2 m_{\tilde{b}_1}^2\over s^2}\re
{\cos^2\hp\,\cos^2\hm\,(1\!-\!h_a)(1\!+\!h_b)\over 4 x_W^2
(1-m_W^2/s)^2},~~~~~~\label{eq:sqsq1cmfv} \eea where we show
explicitly the dependence on the CKM matrix elements. For the
mixed production of the heavier squark mass eigenstates, the
corresponding index 1 has to be replaced by 2 and
the squared cosine of the mixing angle by the squared sine.\\

\begin{figure}
 \centering
 \includegraphics[width=0.7\columnwidth]{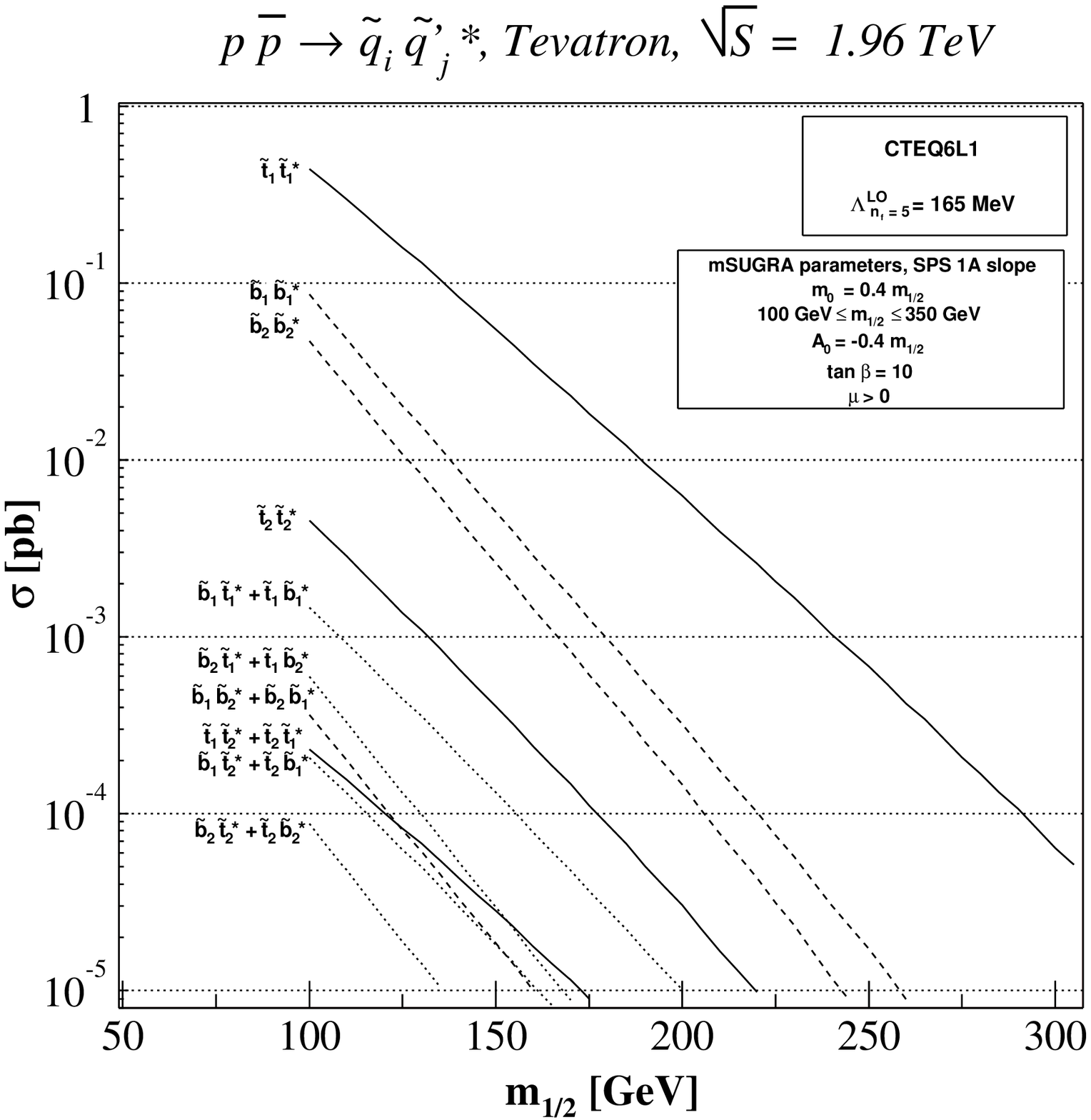}
 \includegraphics[width=0.7\columnwidth]{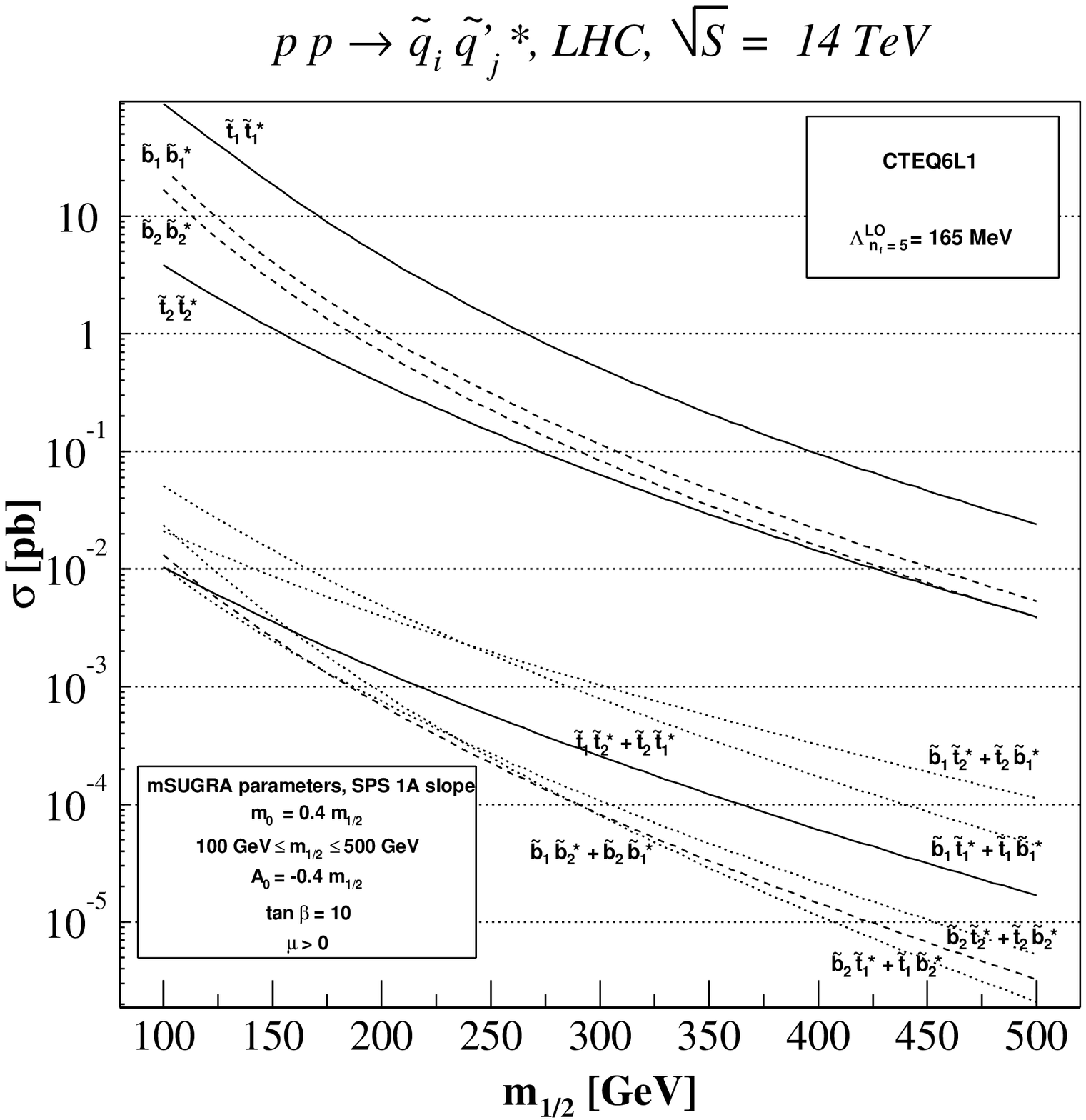}
 \caption{\label{fig:31}Production cross sections for top (full), bottom
 (dashed), and mixed top and bottom squarks (dotted) at the Tevatron
 (top figure) and at the LHC (bottom figure) as a function of the common fermion
 mass $m_{1/2}$ in the mSUGRA model line SPS 1a \cite{Allanach:2002nj}.}
\end{figure}

In the following, we use older values of the SM parameters
\cite{Eidelman:2004wy}, and the physical masses of the squarks and
mixing matrices are calculated using the computer program SuSpect
Version 2.3 \cite{Djouadi:2002ze}. As a first benchmark, we choose
the model line SPS 1a \cite{Allanach:2002nj} described in Sec.\
\ref{sec:thresh} and show in the top panel of the Fig.\
\ref{fig:31} that cMFV non-diagonal and mixed third generation
squark production will be difficult to discover at the Tevatron
due to its limited centre-of-mass energy. As one can see, only
diagonal production of the lighter top squark mass eigenstate will
be visible in the full region of the mSUGRA parameter space shown
here with the expected final integrated luminosity of 8.9
fb$^{-1}$. For diagonal sbottom production, the accessible
parameter space is already reduced to $m_{1/2}\leq 225$ GeV.
Non-diagonal and mixed squark production could only be discovered,
if the common fermion mass $m_{1/2}$ is not much larger than 100
GeV. As expected for a $p\bar{p}$ collider, the cross sections are
very much dominated by $q\bar{q}$ annihilation, even for the
diagonal channels, and $gg$ initial states contribute
at most 15\% in the case of diagonal light stop production.\\

The LHC with its much larger centre-of-mass energy of and design
luminosity of 300 fb$^{-1}$ will, in contrast, have no problem in
producing all combinations of squarks in sufficient numbers. The
hierarchy between the strong diagonal production channels of
${\cal O}(\alpha_s^2)$ and the electroweak non-diagonal and mixed
channels of ${\cal O}(\alpha^2)$ is, however, clearly visible in
the bottom panel of Fig.\ \ref{fig:31}, the latter being about two
orders of magnitude smaller. The LHC being a $pp$ collider and the
average $x$-value in the PDFs being considerably smaller, the
diagonal channels are enhanced by the high $gg$ luminosity, which
dominates their cross sections by up to 93 \%. Among the
electroweak ${\cal O}(\alpha^2)$ processes, mixed production of
top and bottom squarks is favoured over non-diagonal top or bottom
squark production by the possibility of two light masses and a
positive charge in the final state, which is more easily produced
by the charged $pp$ initial state.\\

\begin{figure}
 \centering
 \includegraphics[width=0.7\columnwidth]{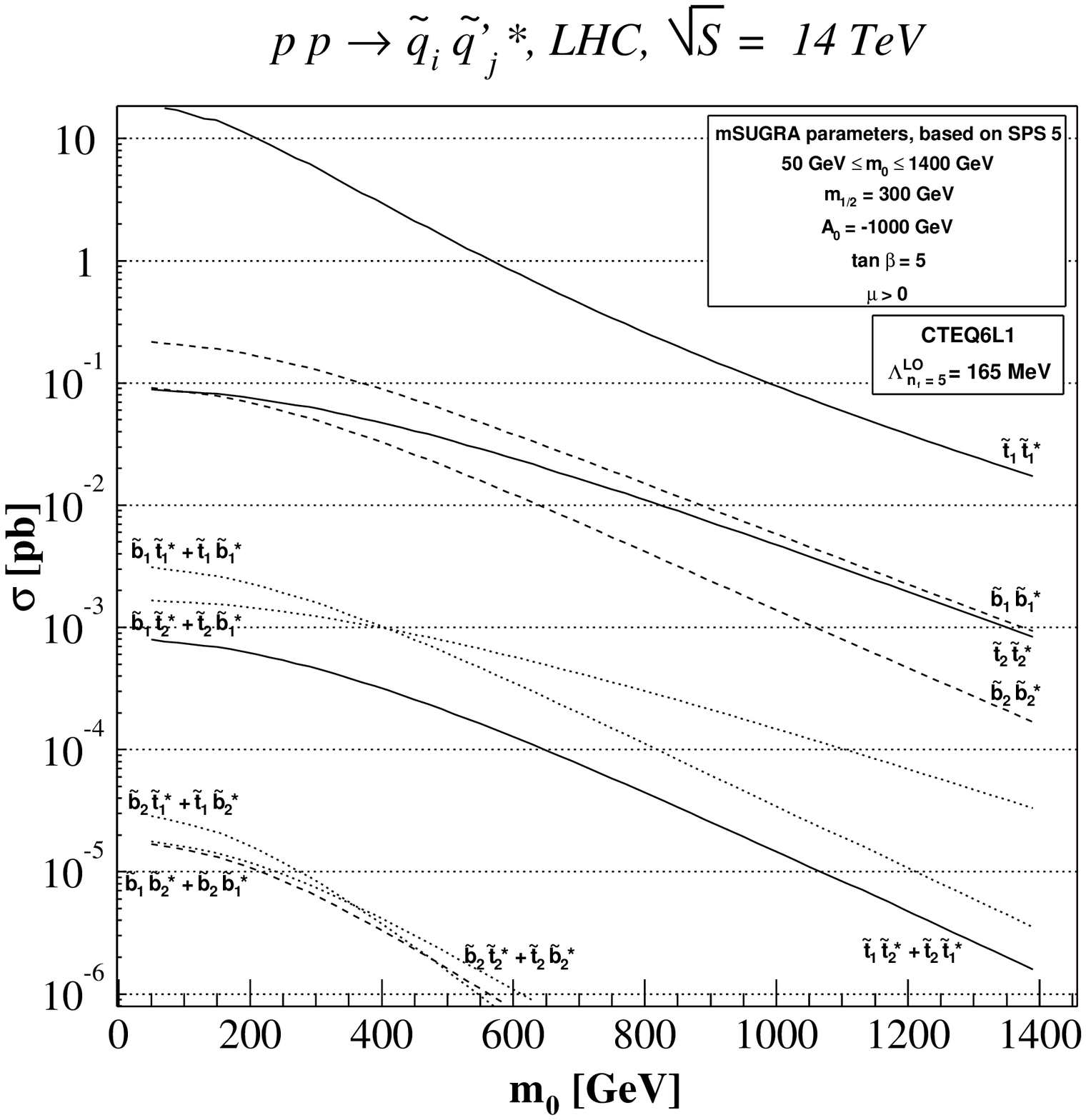}
 \caption{\label{fig:32a}Production cross sections for top (full), bottom
 (dashed), and mixed top and bottom squarks (dotted) at the LHC in the
 mSUGRA model SPS 5 as a function of the scalar mass $m_0$.
 \cite{Allanach:2002nj}.}
\end{figure}

We show in Figs.\ \ref{fig:32a} numerical results for the LHC,
varying $m_0$ independently to test the sensitivity of the cross
section on the squark masses and mixing, starting from the
benchmark point SPS 5 \cite{Allanach:2002nj}, which is a mSUGRA
scenario with low $\tan\beta=5$, large $m_{1/2}=350$ GeV, and
large negative $A_0=-1000$ GeV, leading to heavier sbottoms of 566
and 655 GeV, a heavy $\tilde{t}_2$ of 651 GeV, but also a light
$\tilde{t}_1$ of 259 GeV. We see a clear hierarchy between the
dominant pair production of the lighter stop, strong pair
production of the heavier stop and sbottoms, charged and neutral
electroweak production of final states involving at least one
light squark, and finally charged and neutral electroweak
production of the heavier squarks, which may only be visible up to
$m_0\leq 600$ GeV. The more pronounced hierarchy can be explained
by the considerable squark mass differences in SPS 5, which lead
to additional phase space suppression for the heavier squarks.

\subsection{QCD one-loop contributions for non-diagonal squark
pair production}

\begin{figure}
 \centering
 \includegraphics[width=.75\columnwidth]{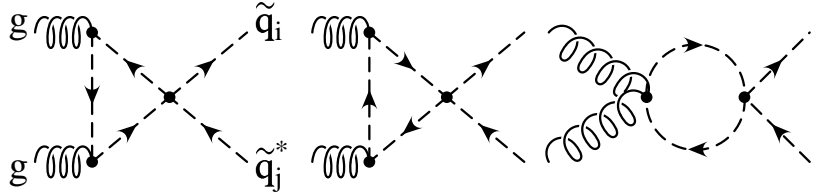}
 \caption{\label{fig:29}Subdominant loop-level QCD Feynman diagrams for
 non-diagonal ($i\neq j$) squark production at hadron colliders.}
\end{figure}

Within QCD and cMFV SUSY, we have also investigated the
possibility to produce non-diagonal squark pairs by the
rescattering of diagonal squark pairs through four-squark vertices
in the final state, in the limit of a decoupled gluino. The
corresponding Feynman diagrams, shown in Fig.\ \ref{fig:29}, have
one-loop topology and are therefore suppressed by additional
squark propagators and/or annihilations within the squark loop.
The squared helicity amplitude for the production of non-diagonal
stop pairs in gluon-gluon collisions is given by
\cite{Bozzi:2005sy} \bea \frac{d\hat{\sigma}_{h_a,h_b}^{gg}}{dt}
&=& (1\!+\!h_a h_b) \le \frac{37 \alpha_s^4 \sin^2{\left(4
\theta_{\tilde{t}}\right)}}{27648 \pi s^{4}} |\Delta
\ln_{\tilde{t}}|^2\rp \!+\! \sum_{\tilde{q} \neq \tilde{t}}
\frac{5 \alpha_{s}^{4} \cos^2{\left(2 \theta_{\tilde{q}}\right)}
\sin^2{\left(2 \theta_{\tilde{t}}\right)}}{3072 \pi s^{4}} |\Delta
\ln_{\tilde{q}}|^2\nonumber\\ &+& \sum_{\tilde{q} \neq \tilde{t}}
\frac{5 \alpha_{s}^{4} \cos{\left(2 \theta_{\tilde{q}}\right)}
\cos{\left(2 \theta_{\tilde{t}}\right)} \sin^2{\left(2
\theta_{\tilde{t}}\right)}}{2304 \pi s^{4}} {\rm Re}\left(\Delta
\ln_{\tilde{q}} \Delta \ln_{\tilde{t}}\right)\nonumber\\
&+& \!\!\!\lp \!\!\sum_{\tilde{q},\tilde{q}'\neq \tilde{t};
\tilde{q}\neq \tilde{q}'} \frac{5 \alpha_{s}^{4} \cos{\left(2
\theta_{\tilde{q}}\right)} \cos{\left(2
\theta_{\tilde{q}'}\right)} \sin^2{\left(2
\theta_{t}\right)}}{1536 \pi s^{4}} {\rm Re}\left(\Delta
\ln_{\tilde{q}} \Delta \ln_{\tilde{q'}}\right)\re, \eea where \bea
\Delta \ln_{\tilde{q}} = m_{\tilde{q}_1}^2
\ln^2{\left(-x_{\tilde{q}_1}\right)} - m_{\tilde{q}_2}^2
\ln^2{\left(-x_{\tilde{q}_2}\right)},{\rm ~~~~with~~}
x_{\tilde{q}_{i}} = \frac{1 - \beta_{\tilde{q}_{i}}}{1 +
\beta_{\tilde{q_{i}}}},~\eea and
$\beta_{\tilde{q}_i}=\sqrt{1-4m_{\tilde{q}_i}^2}$ is the velocity
of the $i^{\rm th}$ squark mass eigenstate. In the limit of
degenerate light squarks, only top and bottom squark loops survive
loop annihilations, and the squared helicity amplitude simplifies
to \bea \frac{d\hat{\sigma}_{h_a,h_b}^{gg}}{dt} &=& (1+h_a h_b)
\le \frac{37 \alpha_{s}^{4} \sin^2{\left(4
\theta_{\tilde{t}}\right)}}{27648 \pi s^{4}} |\Delta
\ln_{\tilde{t}}|^2\rp + \frac{5 \alpha_{s}^{4} \cos^2{\left(2
\theta_{\tilde{b}}\right)} \sin^2{\left(2
\theta_{\tilde{t}}\right)}}{3072 \pi s^{4}} |\Delta
\ln_{\tilde{b}}|^2\nonumber\\ &+& \lp \frac{5 \alpha_{s}^{4}
\cos{\left(2 \theta_{\tilde{b}}\right)} \cos{\left(2
\theta_{\tilde{t}}\right)} \sin^2{\left(2
\theta_{\tilde{t}}\right)}}{2304 \pi s^{4}} {\rm Re}\left(\Delta
\ln_{\tilde{b}} \Delta \ln_{\tilde{t}}\right)\re.
\label{eq:qcdboucle}\eea These expressions have been summed over
$\tilde{t}_1 \tilde{t}_2^\ast + \tilde{t}_2\tilde{t}_1^\ast$ final
states and generalize the corresponding result in Ref.\
\cite{Beenakker:1997ut}, where only top squark loops were taken
into account. For non-diagonal sbottom production, top and bottom
squark indices have to be exchanged in the equations above.\\

\begin{figure}
 \centering
 \includegraphics[width=.7\columnwidth]{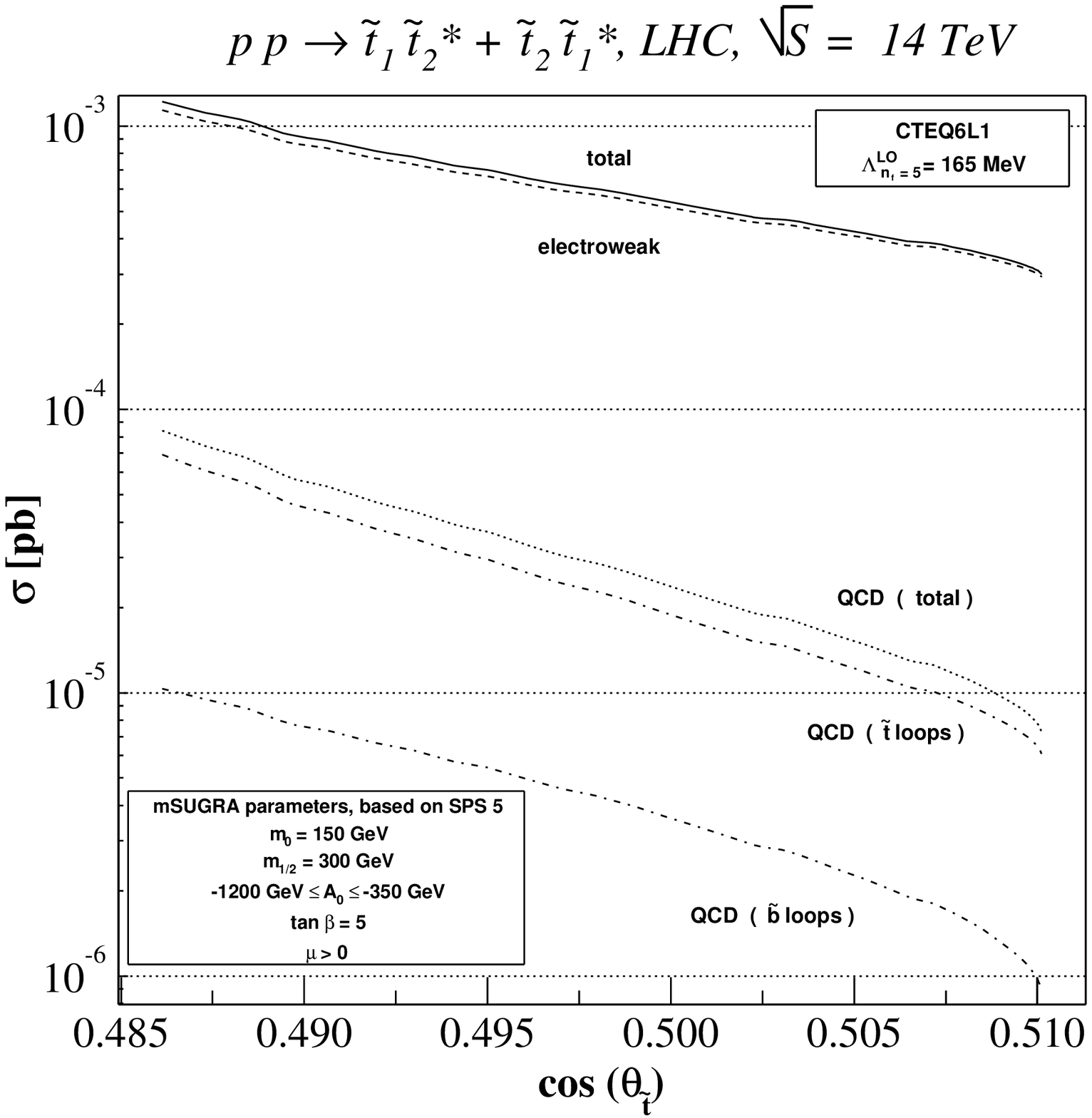}
 \caption{\label{fig:33}Contributions of tree-level electroweak and
 loop-level QCD processes to non-diagonal stop production at the LHC in
 the mSUGRA model SPS 5 \cite{Allanach:2002nj} with a light top squark as a
 function of the cosine of the top squark mixing angle, together with their
 sums.} \vspace{.1cm}
 \includegraphics[width=.7\columnwidth]{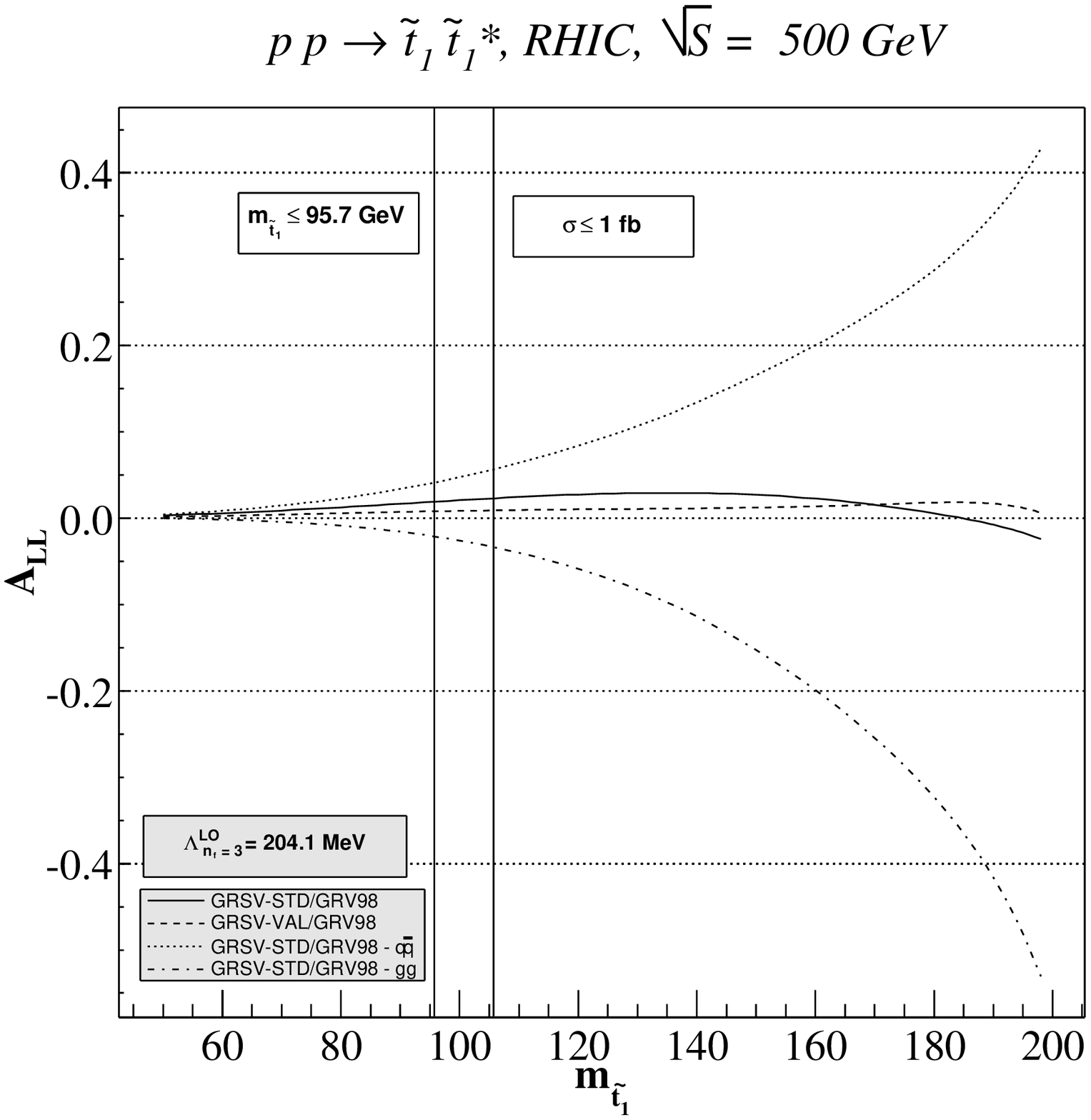}
 \caption{\label{fig:34}Contributions of the $q\bar{q}$ (dotted) and $gg$
 (dot-dashed) initial states to the longitudinal double-spin asymmetry
 $A_{LL}$ at RHIC, together with their sum (full), for stop pair production
 as a function of the light top squark mass. The total asymmetry using
 GRSV-VAL (dashed) instead of GRSV-STD (full) parton densities
 \cite{Gluck:2000dy} is also shown.}
\end{figure}

For the non-diagonal third generation squark production
$\tilde{t}_1\tilde{t}_2^\ast+ \tilde{t}_2\tilde{t}_1^\ast$ and
$\tilde{b}_1\tilde{b}_2^\ast+ \tilde{b}_2\tilde{b}_1^\ast$, the
only contributing diagrams of Fig.\ \ref{fig:27} is the
electroweak $Z$-boson exchange, since the QCD $gg$-initiated
processes and the gluon $s$-channel are only open for diagonal
squark pair production, and the neutralino and gluino diagrams are
suppressed by the negligible top and small bottom densities in the
proton. Indeed, for the mSUGRA scenario SPS 5, their contributions
are found to be six to eight orders of magnitude smaller than the
$Z$-contribution. Within the SPS 5 scenario and at the LHC, Fig.\
\ref{fig:33} shows that for non-diagonal stop production, the QCD
loop-contributions from Fig.\ \ref{fig:29} are smaller than the
electroweak tree-level ones by about one order of magnitude,
whereas we naively expect the ${\cal O}(\alpha^2)$ and ${\cal
O}(\alpha_s^4)$ diagrams to contribute with roughly equal
strength. As already mentioned, this can be easily explained by
the presence of additional heavy squark propagators in the loop
diagrams. Here, we consider not only loops involving top, but also
bottom squarks, which do not cancel in Eq.\ (\ref{eq:qcdboucle}),
if the masses of the two bottom squarks are unequal. However, the
non-diagonal elements in the squark mass matrices are proportional
to the relevant SM quark mass and $m_b\ll m_t$, so that mixing
effects are less important for sbottoms than for stops.
Consequently, sbottom loops contribute about one order of
magnitude less than stop loops, as can also be seen in Fig.\
\ref{fig:33}. SUSY-QCD loop diagrams involving gluino exchanges
have not been calculated here, as they are of ${\cal
O}(\alpha_s^4)$ and require in addition the presence of heavy top
quark and gluino propagators in the loop. In the SPS 5 scenario,
we have indeed a heavy gluino of mass $m_{\tilde{g}}=725$ GeV. The
dependence on the cosine of the stop mixing angle in the
electroweak cross section of Eq.\ (\ref{eq:sqsq1cmfv}) and in the
the sbottom loop contribution in Eq.\ (\ref{eq:qcdboucle}) through
$\sin^2(2\theta_{\tilde{t}})=4\cos^2\theta_{\tilde{t}}~(1-
\cos^2\theta_{\tilde{t}})$ is clearly visible in Fig.\
\ref{fig:33}. In contrast, the stop loop contribution in Eq.\
(\ref{eq:qcdboucle}) has a steeper dependence through
$\sin^2(4\theta_{\tilde{t}})=16\cos^2\theta_{\tilde{t}}~
(1-\cos^2\theta_{\tilde{t}})~(1-2\cos^2\theta_{\tilde{t}})^2$,
which is also visible in the figure.

\subsection{Double-spin asymmetries}

As a possible application of the polarization dependence of our
analytical results, we show in Fig.\ \ref{fig:34} the double-spin
asymmetry of diagonal light stop production at RHIC (within cMFV
SUSY). As this polarized $pp$ collider has only a rather small
centre-of-mass energy of $\sqrt{s_h}=500$ GeV, the observable stop
mass range is obviously very limited. Already for
$m_{\tilde{t}_1}>106$ GeV, the unpolarized cross section drops
below 1 fb, while stop masses below 96 GeV are most likely already
excluded \cite{Yao:2006px}. This leaves only a very small mass
window of 10 GeV for possible observations. In Fig.\ \ref{fig:34}
one can clearly see the rise of the asymmetry for $q\bar{q}$ and
$gg$ initial states, as the stop mass and the correlated $x$-value
in the PDFs grows. However, as the two asymmetries are
approximately of equal size, but opposite sign, the total
observable asymmetry rests below the 5\% level in the entire mass
range shown. This is true for both choices of polarized parton
densities, GRSV 2000 standard (STD) as well as valence (VAL)
\cite{Gluck:2000dy}. For consistency, the unpolarized cross
sections have been calculated in this case using the GRV 98 parton
density set \cite{Gluck:1998xa}, as in Sec.\ \ref{sec:slLO}.

\section{NMFV squark pair production}

\subsection{Analytical results}
\begin{figure}[!h]
 \centering
 \includegraphics[width=0.75\columnwidth]{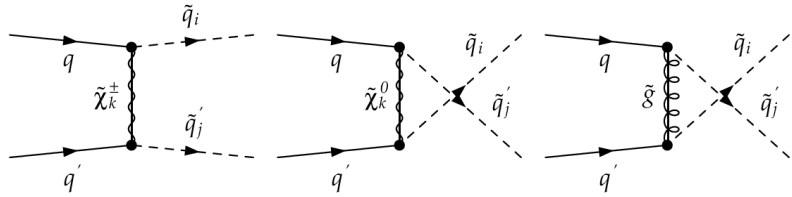}
 \caption{\label{fig:35}Tree-level Feynman diagrams for the production of
          one down-type squark ($\tilde{q}_i$) and one up-type squark
          ($\tilde{q}_j^\prime$) in the collision of an up-type quark ($q$)
          and a down-type quark ($q^\prime$).}
\end{figure}

While squark-antisquark pairs are readily produced in $p\bar{p}$
collisions, e.g.\ at the Tevatron, from valence quarks and
antiquarks, $pp$ colliders have a larger quark-quark luminosity
and will thus more easily lead to squark pair production. The
potentially flavour-violating production of one down-type and one
up-type squark \bea q(h_a,p_a)\,q'(h_b,p_b) \to
\tilde{d}_i(p_1)\,\tilde{u}_j(p_2), \eea in the collision of an
up-type quark $q$ and a down-type quark $q'$ proceeds through the
$t$-channel chargino or $u$-channel neutralino and gluino
exchanges shown in Fig.\ \ref{fig:35}. The corresponding cross
section within NMFV SUSY \cite{Bozzi:2007me} \bea
\frac{d\hat{\sigma}^{qq'}_{h_a, h_b}}{dt} &=& (1\! - \!h_a) (1\! -
\!h_b) \Bigg[\bigg( \sum_{k=1,2 \atop l=1,2}\!
\frac{\mathcal{C}^{kl}_{11}}{t_{\tilde{\chi}_k}\,
t_{\tilde{\chi}_l}}\bigg) \! + \! \bigg( \sum_{k=1,\ldots,4\atop
l=1,\ldots,4}\! \frac{\mathcal{N}_{11}^{kl}}
{u_{\tilde{\chi}_k^0}\, u_{\tilde{\chi}_l^0}}\bigg) \! + \!
\frac{\mathcal{G}_{11}}{u_{\tilde{g}}^2}\nn \\ &\! + \!& \bigg(
\sum_{k=1,2\atop l=1,\ldots,4}\!
\frac{\mathcal{[CN]}^{kl}_{11}}{t_{\tilde{\chi}_k}\,
u_{\tilde{\chi}_l^0}}\bigg) + \bigg( \sum_{k=1,2}\!
\frac{\mathcal{[CG]}^k_{11}}{t_{\tilde{\chi}_k}\,
u_{\tilde{g}}}\bigg) \Bigg] \! + \! (1\! + \!h_a) (1\! + \!h_b)
\Bigg[ \bigg( \sum_{k=1,2 \atop l=1,2} \!
\frac{\mathcal{C}^{kl}_{22}}{t_{\tilde{\chi}_k}\,
t_{\tilde{\chi}_l}}\bigg) \nn \\ &\! + \!&
\bigg(\sum_{k=1,\ldots,4 \atop l=1,\ldots,4} \!
\frac{\mathcal{N}_{22}^{kl}}{u_{\tilde{\chi}_k^0}\,
u_{\tilde{\chi}_l^0}}\bigg) \! + \!
\frac{\mathcal{G}_{22}}{u_{\tilde{g}}^2} + \bigg( \sum_{k=1,2\atop
l=1,\ldots,4}\!
\frac{\mathcal{[CN]}^{kl}_{22}}{t_{\tilde{\chi}_k}\,
u_{\tilde{\chi}_l^0}}\bigg) \! + \! \bigg( \sum_{k=1,2}\!
\frac{\mathcal{[CG]}^k_{22}}{t_{\tilde{\chi}_k}\,
u_{\tilde{g}}}\bigg) \Bigg]\nn \\ &\! + \!& (1\! - \!h_a) (1\! +
\!h_b) \Bigg[\bigg( \sum_{k=1,2 \atop l=1,2} \!
\frac{\mathcal{C}^{kl}_{12}}{t_{\tilde{\chi}_k}\,
t_{\tilde{\chi}_l}}\bigg) \! + \! \bigg(\sum_{k=1,\ldots,4 \atop
l=1,\ldots,4}\!
\frac{\mathcal{N}_{12}^{kl}}{u_{\tilde{\chi}_k^0}\,
u_{\tilde{\chi}_l^0}}\bigg) \! + \!
\frac{\mathcal{G}_{12}}{u_{\tilde{g}}^2}\nn \\ &\! + \! & \bigg(
\sum_{k=1,2\atop l=1,\ldots,4}\!
\frac{\mathcal{[CN]}^{kl}_{12}}{t_{\tilde{\chi}_k}\,
u_{\tilde{\chi}_l^0}}\bigg) \! + \! \bigg( \sum_{k=1,2}\!
\frac{\mathcal{[CG]}^k_{12}}{t_{\tilde{\chi}_k}\,
u_{\tilde{g}}}\bigg)\Bigg]  \! + \! (1\! + \!h_a) (1\! - \!h_b)
\Bigg[\bigg( \sum_{k=1,2 \atop l=1,2} \!
\frac{\mathcal{C}^{kl}_{21}}{t_{\tilde{\chi}_k}\,
t_{\tilde{\chi}_l}}\bigg)\nn \\ &\! + \!&
\bigg(\sum_{k=1,\ldots,4\atop l=1,\ldots,4}\!
\frac{\mathcal{N}_{21}^{kl}}{u_{\tilde{\chi}_k^0}\,
u_{\tilde{\chi}_l^0}}\bigg) \! + \!
\frac{\mathcal{G}_{21}}{u_{\tilde{g}}^2} \! + \! \bigg(
\sum_{k=1,2\atop l=1,\ldots,4}\!
\frac{\mathcal{[CN]}^{kl}_{21}}{t_{\tilde{\chi}_k}\,
u_{\tilde{\chi}_l^0}}\bigg) \! + \! \bigg( \sum_{k=1,2}\!
\frac{\mathcal{[CG]}^k_{21}}{t_{\tilde{\chi}_k}\,
u_{\tilde{g}}}\bigg) \Bigg]~~ \eea involves the form factors \bea
\mathcal{C}_{mn}^{kl} &=& \frac{\pi\, \alpha^2}{4\, x_W^2\,
s^2}\mathcal{C}^n_{\tilde{u}_j q^\prime \tilde{\chi}_k^\pm}\,
\mathcal{C}^{m\ast}_{\tilde{d}_i q \tilde{\chi}_k^\pm}\,
\mathcal{C}^{n\ast}_{\tilde{u}_j q^\prime \tilde{\chi}_l^\pm}\,
\mathcal{C}^m_{\tilde{d}_i q \tilde{\chi}_l^\pm}\, \Bigg[ \left(
u\, t - m^2_{\tilde{d}_i}\, m^2_{\tilde{u}_j}\right)
\left(1-\delta_{mn} \right)\nn\\ && + m_{\tilde{\chi}^\pm_k}\,
\,m_{\tilde{\chi}^\pm_l}\, s\, \delta_{mn} \Bigg],~ \nonumber\\
\mathcal{N}_{mn}^{kl} &=& \frac{\pi\, \alpha^2}{x_W^2 (1 \!-\!
x_W)^2 s^2}\mathcal{C}^{m\ast}_{\tilde{u}_j q \tilde{\chi}_k^0}
\mathcal{C}^{n\ast}_{\tilde{d}_i q' \tilde{\chi}_k^0}
\mathcal{C}^m_{\tilde{u}_j q \tilde{\chi}_l^0}
\mathcal{C}^n_{\tilde{d}_i q' \tilde{\chi}_l^0} \Bigg[ \left( u\,
t \!-\! m^2_{\tilde{d}_i}\,
m^2_{\tilde{u}_j}\right) \left( 1\!-\!\delta_{mn} \right)\nn \\
&& + m_{\tilde{\chi}^0_k}\, m_{\tilde{\chi}^0_l}\, s\, \delta_{mn}
\Bigg],~ \nonumber\\ \mathcal{G}_{mn} &=& \frac{2\, \pi\,
\alpha_s^2 }{9\, s^2} \left| \mathcal{C}^m_{\tilde{u}_j q
\tilde{g}}\, \mathcal{C}^n_{\tilde{d}_i q' \tilde{g}}\right|^2
\Bigg[ \left( u\, t - m^2_{\tilde{d}_i}\, m^2_{\tilde{u}_j}\right)
\left( 1-\delta_{mn} \right) +
m_{\tilde{g}}^2\, s\,\delta_{mn} \Bigg],~ \nonumber \\
\mathcal{[CN]}^{kl}_{mn} &=& \frac{\pi\, \alpha^2}{3\, x_W^2\, (1
- x_W)\, s^2} {\rm Re} \left[ \mathcal{C}^n_{\tilde{u}_j q^\prime
\tilde{\chi}_k^\pm}\, \mathcal{C}^{m\ast}_{\tilde{d}_i q
\tilde{\chi}_k^\pm}\, \mathcal{C}^m_{\tilde{u}_j q
\tilde{\chi}_l^0}\, \mathcal{C}^n_{\tilde{d}_i q'
\tilde{\chi}_l^0} \right]\nn
\\ &&\times \Bigg[ \left( u\, t - m^2_{\tilde{d}_i}\,
m^2_{\tilde{u}_j}\right) \left( \delta_{mn}-1 \right) +
m_{\tilde{\chi}^\pm_k}\, m_{\tilde{\chi}^0_l}\, s\, \delta_{mn}
\Bigg],~~  \nonumber \\ \mathcal{[CG]}^k_{mn} &=& \frac{4\, \pi\,
\alpha\, \alpha_s}{9\, s^2\, x_W} {\rm Re} \left[
\mathcal{C}^n_{\tilde{u}_j q^\prime \tilde{\chi}_k^\pm}\,
\mathcal{C}^{m\ast}_{\tilde{d}_i q \tilde{\chi}_k^\pm}\,
\mathcal{C}^{m\ast}_{\tilde{u}_j q \tilde{g}}\,
\mathcal{C}^{n\ast}_{\tilde{d}_i q' \tilde{g}} \right] \Bigg[
\left( u\, t - m^2_{\tilde{d}_i}\,
m^2_{\tilde{u}_j}\right) \left( \delta_{mn} - 1 \right)\nn \\
&& + m_{\tilde{\chi}^\pm_k}\, m_{\tilde{g}}\, s\, \delta_{mn}
\Bigg], \eea where the neutralino-gluino interference term is
absent due to colour conservation. The cross section for the
charge-conjugate production of antisquarks from antiquarks can be
obtained from the equations above by replacing
$h_{a,b}\to-h_{a,b}$. Heavy-flavour final states are completely
absent in cMFV due to the negligible top quark and small bottom
quark densities in the proton and can thus only be obtained in
NMFV.\\

\begin{figure}
 \centering
 \includegraphics[width=\columnwidth]{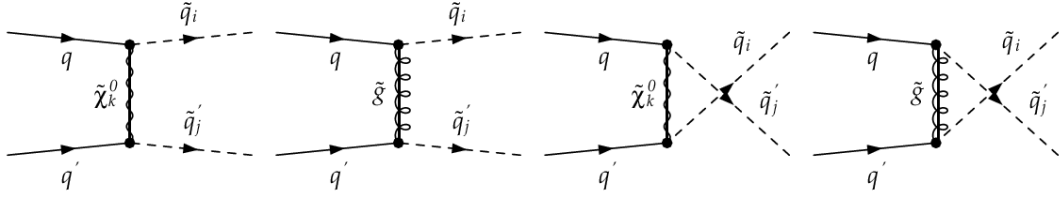}
 \caption{\label{fig:36}Tree-level Feynman diagrams for the production of
          two up-type or down-type squarks.}
\end{figure}

The Feynman diagrams for pair production of two up- or down-type
squarks with non-minimal flavour violation \bea q(h_a,p_a)\,
q'(h_b,p_b) &\to&\tilde{q}_i(p_1)\, \tilde{q}_j(p_2) \eea are
shown in Fig.\ \ref{fig:36}. In NMFV SUSY, neutralino and gluino
exchanges can lead to identical squark flavours for different
quark initial states, so that both $t$- and $u$-channels
contribute and may interfere. The cross section
\cite{Bozzi:2007me} \bea \frac{d\hat{\sigma}^{qq'}_{h_a, h_b}}{dt}
&=& (1-h_a) (1-h_b) \Bigg[\bigg( \sum_{k=1,\ldots,4\atop
l=1,\ldots,4} \frac{[\mathcal{NT}]_{11}^{kl}}
{t_{\tilde{\chi}_k^0}\, t_{\tilde{\chi}_l^0}} +
\frac{[\mathcal{NU}]_{11}^{kl}} {u_{\tilde{\chi}_k^0}\,
u_{\tilde{\chi}_l^0}} + \frac{[\mathcal{NTU}]_{11}^{kl}}
{t_{\tilde{\chi}_k^0}\, u_{\tilde{\chi}_l^0}} \bigg) +
\frac{[\mathcal{GT}]_{11}}{t_{\tilde{g}}^2}\nn \\ &+&
\frac{[\mathcal{GU}]_{11}}{u_{\tilde{g}}^2}+
\frac{[\mathcal{GTU]}_{11}}{u_{\tilde{g}}t_{\tilde{g}}}+ \bigg(
\sum_{k=1,\ldots,4} \frac{[\mathcal{NGA}]_{11}^{k}}
{t_{\tilde{\chi}_k^0}\, u_{\tilde{g}}} +
\frac{[\mathcal{NGB}]_{11}^{k}} {u_{\tilde{\chi}_k^0}\,
t_{\tilde{g}}} \bigg)\Bigg]\frac{1}{1+\delta_{ij}} \nonumber
\\ &+& (1+h_a) (1+h_b) \Bigg[\bigg( \sum_{k=1,\ldots,4\atop l=1,\ldots,4}
\frac{[\mathcal{NT}]_{22}^{kl}} {t_{\tilde{\chi}_k^0}\,
t_{\tilde{\chi}_l^0}} + \frac{[\mathcal{NU}]_{22}^{kl}}
{u_{\tilde{\chi}_k^0}\, u_{\tilde{\chi}_l^0}} +
\frac{[\mathcal{NTU}]_{22}^{kl}} {t_{\tilde{\chi}_k^0}\,
u_{\tilde{\chi}_l^0}}\bigg) +
\frac{[\mathcal{GT]}_{22}}{t_{\tilde{g}}^2} \nn
\\&+& \frac{[\mathcal{GU]}_{22}}{u_{\tilde{g}}^2}+
\frac{[\mathcal{GTU]}_{22}}{u_{\tilde{g}}t_{\tilde{g}}}+ \bigg(
\sum_{k=1,\ldots,4} \frac{[\mathcal{NGA}]_{22}^{k}}
{t_{\tilde{\chi}_k^0}\, u_{\tilde{g}}} +
\frac{[\mathcal{NGB}]_{22}^{k}} {u_{\tilde{\chi}_k^0}\,
t_{\tilde{g}}} \bigg)\Bigg]\frac{1}{1+\delta_{ij}} \nonumber\\
&+& (1-h_a) (1+h_b) \Bigg[\bigg( \sum_{k=1,\ldots,4\atop
l=1,\ldots,4} \frac{[\mathcal{NT}]_{12}^{kl}}
{t_{\tilde{\chi}_k^0}\, t_{\tilde{\chi}_l^0}} +
\frac{[\mathcal{NU}]_{12}^{kl}} {u_{\tilde{\chi}_k^0}\,
u_{\tilde{\chi}_l^0}} + \frac{[\mathcal{NTU}]_{12}^{kl}}
{t_{\tilde{\chi}_k^0}\, u_{\tilde{\chi}_l^0}} \bigg) +
\frac{[\mathcal{GT]}_{12}}{t_{\tilde{g}}^2} \nn \\ &+&
\frac{[\mathcal{GU]}_{12}}{u_{\tilde{g}}^2}+
\frac{[\mathcal{GTU]}_{12}}{u_{\tilde{g}}t_{\tilde{g}}} + \bigg(
\sum_{k=1,\ldots,4} \frac{[\mathcal{NGA}]_{12}^{k}}
{t_{\tilde{\chi}_k^0}\, u_{\tilde{g}}} +
\frac{[\mathcal{NGB}]_{12}^{k}} {u_{\tilde{\chi}_k^0}\,
t_{\tilde{g}}} \bigg)\Bigg]\frac{1}{1+\delta_{ij}} \nonumber\\
&+& (1+h_a) (1-h_b) \Bigg[\bigg( \sum_{k=1,\ldots,4\atop
l=1,\ldots,4} \frac{[\mathcal{NT}]_{21}^{kl}}
{t_{\tilde{\chi}_k^0}\, t_{\tilde{\chi}_l^0}} +
\frac{[\mathcal{NU}]_{21}^{kl}} {u_{\tilde{\chi}_k^0}\,
u_{\tilde{\chi}_l^0}} + \frac{[\mathcal{NTU}]_{21}^{kl}}
{t_{\tilde{\chi}_k^0}\, u_{\tilde{\chi}_l^0}} \bigg) +
\frac{[\mathcal{GT]}_{21}}{t_{\tilde{g}}^2} \nn \\ &+&
\frac{[\mathcal{GU]}_{21}}{u_{\tilde{g}}^2} +
\frac{[\mathcal{GTU]}_{21}}{u_{\tilde{g}}t_{\tilde{g}}}+ \bigg(
\sum_{k=1,\ldots,4} \frac{[\mathcal{NGA}]_{21}^{k}}
{t_{\tilde{\chi}_k^0}\, u_{\tilde{g}}} +
\frac{[\mathcal{NGB}]_{21}^{k}} {u_{\tilde{\chi}_k^0}\,
t_{\tilde{g}}} \bigg)\Bigg] \frac{1}{1+\delta_{ij}} \eea depends
therefore on the form factors \bea [\mathcal{NT}]_{mn}^{kl} &=&
\frac{\pi\, \alpha^2}{x_W^2\, (1 - x_W)^2\,
s^2}\mathcal{C}^{n\ast}_{\tilde{q}_j q^\prime \tilde{\chi}_k^0}\,
\mathcal{C}^{m\ast}_{\tilde{q}_i q \tilde{\chi}_k^0}\,
\mathcal{C}^n_{\tilde{q}_j q^\prime \tilde{\chi}_l^0}\,
\mathcal{C}^m_{\tilde{q}_i q \tilde{\chi}_l^0}\, \Bigg[ \left( u\,
t - m^2_{\tilde{q}_i}\, m^2_{\tilde{q}_j}\right)\nn\\ && \times
\left( 1-\delta_{mn} \right) + m_{\tilde{\chi}^0_k}\,
m_{\tilde{\chi}^0_l}\, s\, \delta_{mn} \Bigg],~\nn\\
\large[\mathcal{NU}\large]_{mn}^{kl} &=& \frac{\pi\,
\alpha^2}{x_W^2\, (1 - x_W)^2\,
s^2}\mathcal{C}^{n\ast}_{\tilde{q}_i q^\prime\tilde{\chi}_k^0}\,
\mathcal{C}^{m\ast}_{\tilde{q}_j q \tilde{\chi}_k^0}\,
\mathcal{C}^n_{\tilde{q}_i q^\prime \tilde{\chi}_l^0}\,
\mathcal{C}^m_{\tilde{q}_j q \tilde{\chi}_l^0}\, \Bigg[ \left( u\,
t - m^2_{\tilde{q}_i}\, m^2_{\tilde{q}_j}\right) \nn\\ && \times
\left( 1-\delta_{mn} \right) + m_{\tilde{\chi}^0_k}\,
m_{\tilde{\chi}^0_l}\, s\, \delta_{mn} \Bigg],~ \nn \\
\large[\mathcal{NTU}\large]_{mn}^{kl} &=& \frac{2\,\pi\,
\alpha^2}{3\, x_W^2\, (1 - x_W)^2\, s^2} {\rm Re} \left[
\mathcal{C}^{m\ast}_{\tilde{q}_i q \tilde{\chi}_k^0}\,
\mathcal{C}^{n\ast}_{\tilde{q}_j q^\prime \tilde{\chi}_k^0}\,
\mathcal{C}^n_{\tilde{q}_i q^\prime \tilde{\chi}_l^0}\,
\mathcal{C}^m_{\tilde{q}_j q \tilde{\chi}_l^0}\right]\nn\\
&&\times \Bigg[ \left( u\, t - m^2_{\tilde{q}_i}\,
m^2_{\tilde{q}_j}\right) \left( 1-\delta_{mn} \right) +
m_{\tilde{\chi}^0_k}\,
m_{\tilde{\chi}^0_l}\, s\, \delta_{mn} \Bigg],~ \nonumber\\
\large[\mathcal{GT}\large]_{mn} &=& \frac{2\, \pi\,
\alpha_s^2}{9\, s^2} \left| \mathcal{C}^n_{\tilde{q}_j q^\prime
\tilde{g}}\, \mathcal{C}^m_{\tilde{q}_i q \tilde{g}}\right|^2
\Bigg[ \left( u\, t - m^2_{\tilde{q}_i}\, m^2_{\tilde{q}_j}\right)
\left( 1-\delta_{mn} \right) +
m_{\tilde{g}}^2\, s\,\delta_{mn} \Bigg],~ \nonumber \\
\large[\mathcal{GU}\large]_{mn} &=& \frac{2\, \pi\, \alpha_s^2
}{9\, s^2} \left|\mathcal{C}^m_{\tilde{q}_i q^\prime \tilde{g}}\,
\mathcal{C}^n_{\tilde{q}_j q \tilde{g}}\right|^2 \Bigg[ \left( u\,
t - m^2_{\tilde{q}_i}\,m^2_{\tilde{q}_j}\right) \left(
1-\delta_{mn} \right) + m_{\tilde{g}}^2\,s\,\delta_{mn} \Bigg],~
\nonumber \\ \large[\mathcal{GTU}\large]_{mn} &=& \frac{-4\, \pi\,
\alpha_s^2 }{27\, s^2} {\rm Re} \left[ \mathcal{C}^m_{\tilde{q}_i
q \tilde{g}}\, \mathcal{C}^n_{\tilde{q}_j q^\prime
\tilde{g}}\mathcal{C}^{m \ast}_{\tilde{q}_i q^\prime \tilde{g}}\,
\mathcal{C}^{n \ast}_{\tilde{q}_j q \tilde{g}} \right] \Bigg[
\left( u\, t - m^2_{\tilde{q}_i}\, m^2_{\tilde{q}_j}\right)
\left(1-\delta_{mn} \right)\nn\\ && + m_{\tilde{g}}^2\,
s\,\delta_{mn} \Bigg],~ \nonumber \\
\large[\mathcal{NGA}\large]_{mn}^k &=& \frac{8\, \pi\, \alpha
\alpha_s}{9\, s^2\, x_W\, (1 - x_W)} {\rm Re} \left[
\mathcal{C}^{n\ast}_{\tilde{q}_j q^\prime \tilde{\chi}_k^0}\,
\mathcal{C}^{m\ast}_{\tilde{q}_i q \tilde{\chi}_k^0}\,
\mathcal{C}^{m \ast}_{\tilde{q}_i q^\prime \tilde{g}}\,
\mathcal{C}^{n \ast}_{\tilde{q}_j q \tilde{g}} \right] \Bigg[
\left( u\, t - m^2_{\tilde{q}_i}\, m^2_{\tilde{q}_j}\right)\nn \\
&& \times \left( 1-\delta_{mn} \right) + m_{\tilde{\chi}^0_k}\,
m_{\tilde{g}}\, s\,\delta_{mn}  \Bigg],~ \nonumber \\
\large[\mathcal{NGB}\large]_{mn}^k &=& \frac{8\, \pi\, \alpha
\alpha_s}{9\, s^2\, x_W\, (1 - x_W)} {\rm Re} \left[
\mathcal{C}^{n\ast}_{\tilde{q}_i q^\prime \tilde{\chi}_k^0}\,
\mathcal{C}^{m\ast}_{\tilde{q}_j q
\tilde{\chi}_k^0}\,\mathcal{C}^{n \ast}_{\tilde{q}_j q^\prime
\tilde{g}}\, \mathcal{C}^{m\ast}_{\tilde{q}_i q \tilde{g}} \right]
\Bigg[ \left( u\, t - m^2_{\tilde{q}_i}\, m^2_{\tilde{q}_j}\right)
\nn\\ && \times \left( 1-\delta_{mn} \right) +
m_{\tilde{\chi}^0_k}\, m_{\tilde{g}}\, s\,\delta_{mn} \Bigg]. \eea
Gluinos will dominate over neutralino exchanges due to their
strong coupling, and the two will only interfere in the mixed $t$-
and $u$-channels due to colour conservation. At the LHC, up-type
squark pair production should dominate over mixed up-/down-type
squark production and down-type squark pair production, since the
proton contains two valence up-quarks and only one valence
down-quark. As before, the charge-conjugate production of
antisquark pairs is obtained by making the replacement
$h_{a,b}\to-h_{a,b}$. If we neglect electroweak contributions as
well as squark flavour and helicity mixing and sum over left- and
right-handed squark states, our results agree with those of Ref.\
\cite{Dawson:1983fw}.

\subsection{Numerical results}

As in Sec.\ \ref{sec:sqsq}, we employ the LO set of the latest
CTEQ6 global parton density fit \cite{Pumplin:2002vw}, with five
active flavours and the gluon, and the strong coupling constant is
calculated with the value of $\Lambda_{\rm LO}^{n_f=5}=165$ MeV.
The renormalization and factorization scales are set to the
average mass of the final state SUSY particles, and the SUSY
masses and mixings are again computed with the help of SPheno and
FeynHiggs.\\

\begin{figure}
 \centering
 \includegraphics[height=0.32\textheight]{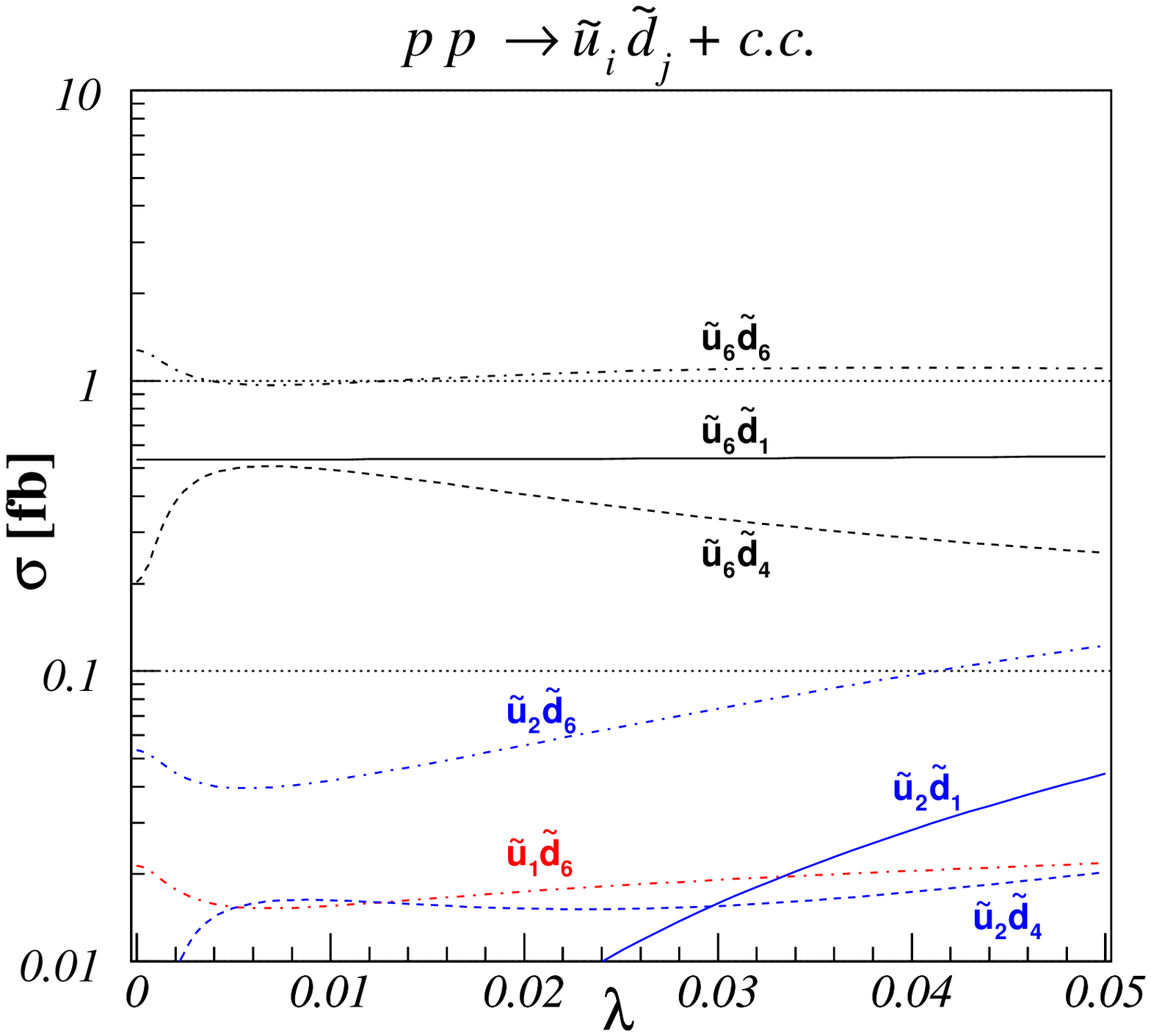}
 \includegraphics[height=0.32\textheight]{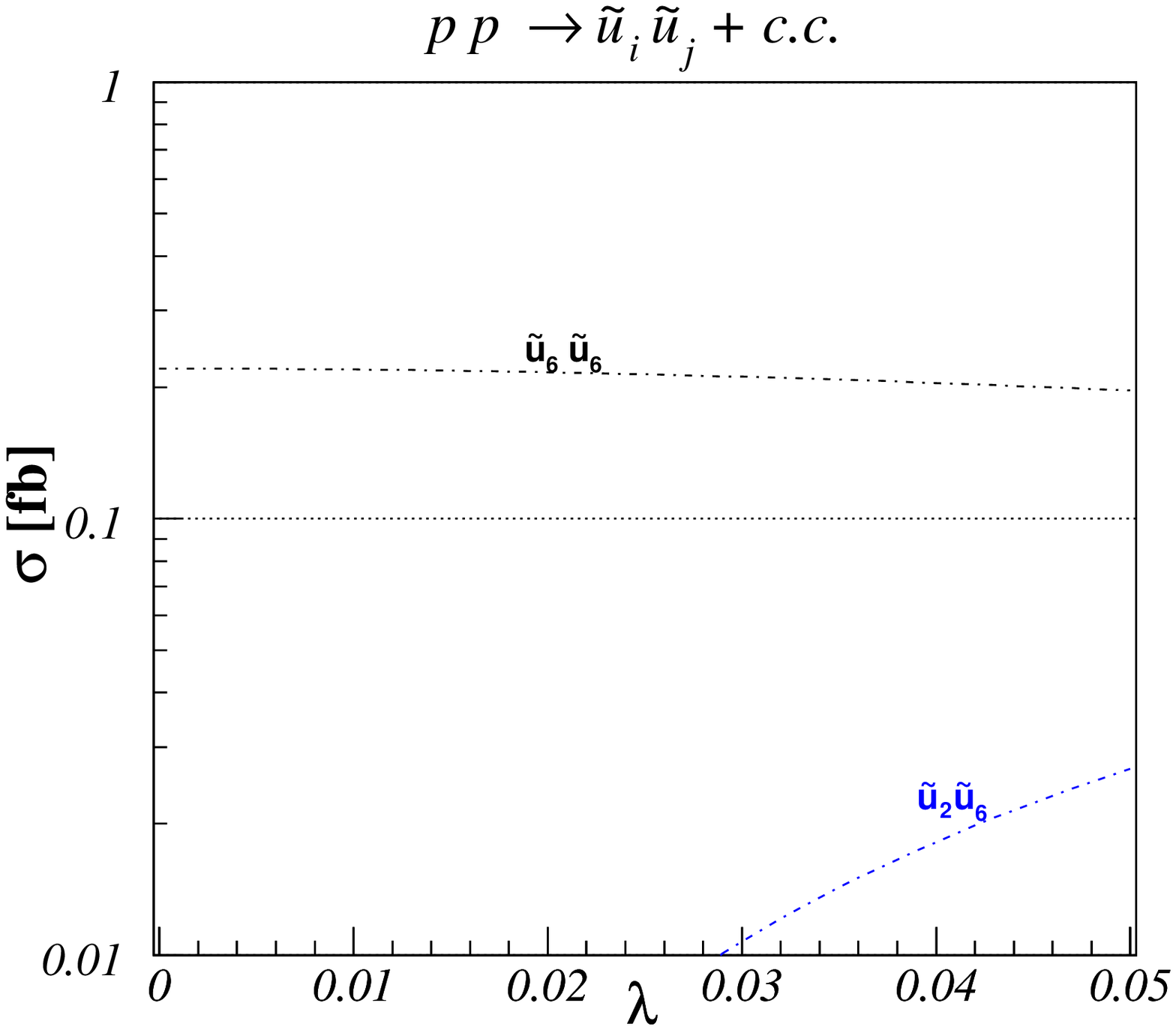}
 \includegraphics[height=0.32\textheight]{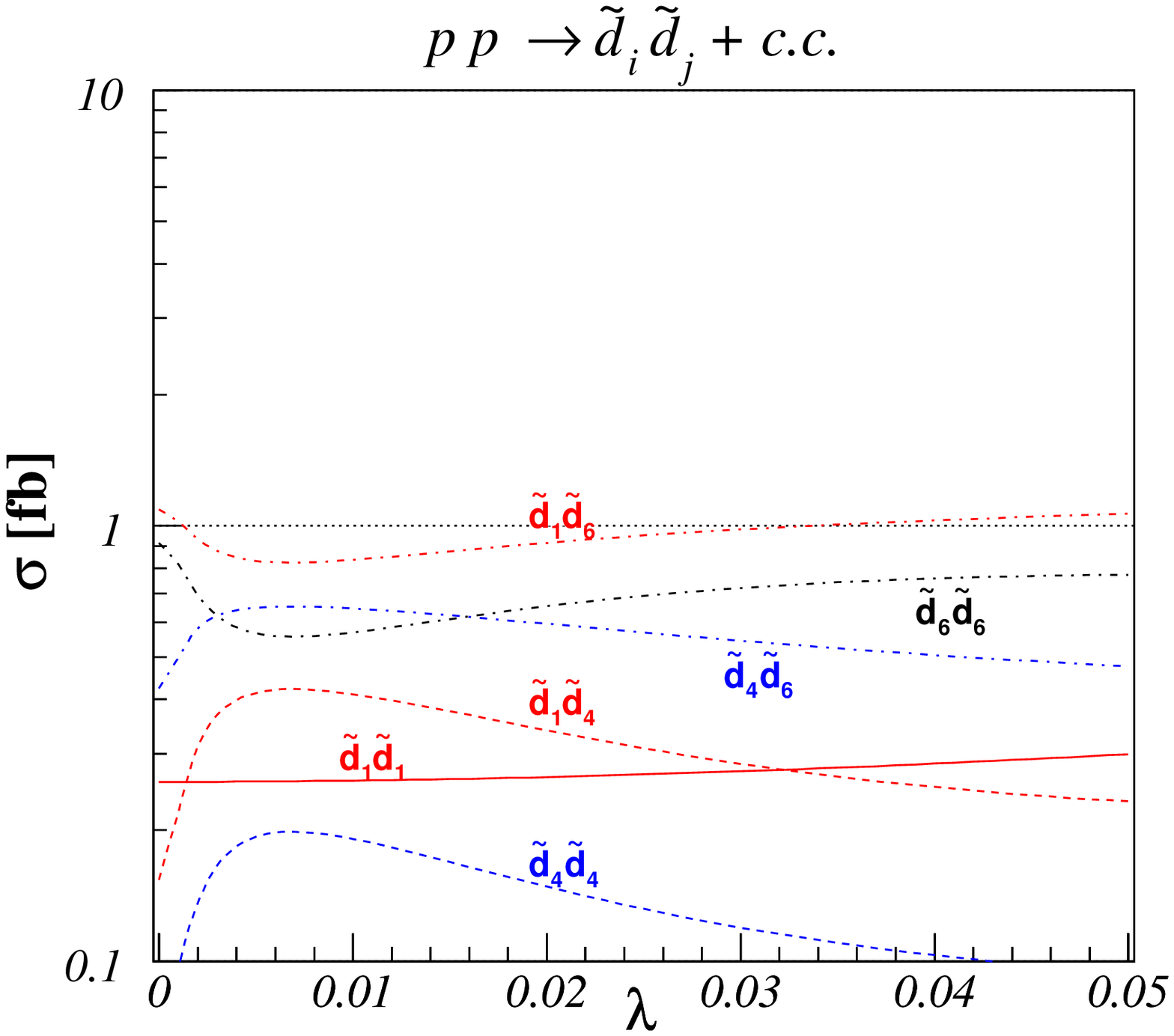}
 \caption{\label{fig:37a}Cross sections for mixed (top),
           up-type (centre) and down-type (bottom) squark-squark pair production
           at the LHC in our benchmark scenario A.}
\end{figure}
\begin{figure}
 \centering
 \includegraphics[height=0.32\textheight]{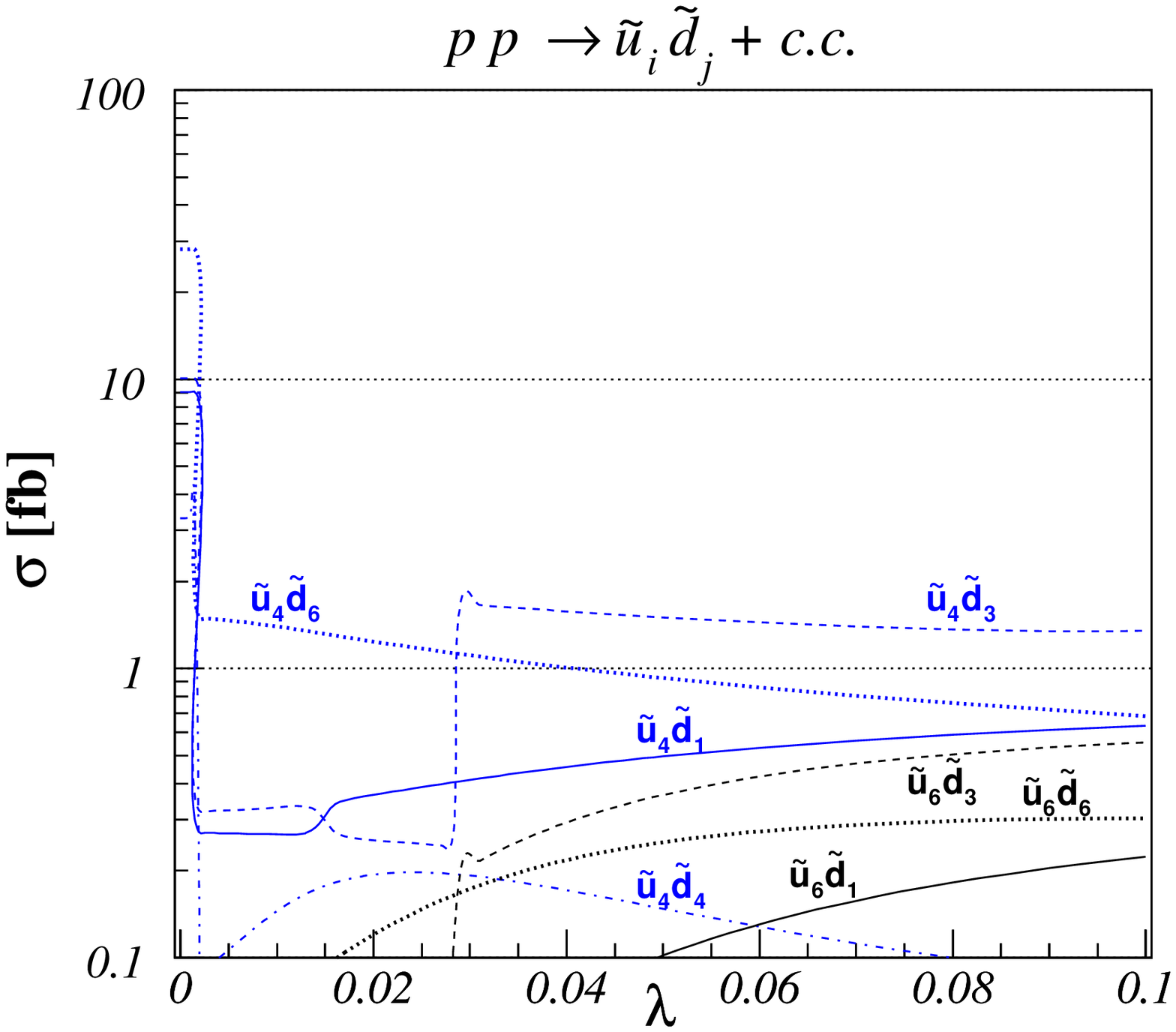}
 \includegraphics[height=0.32\textheight]{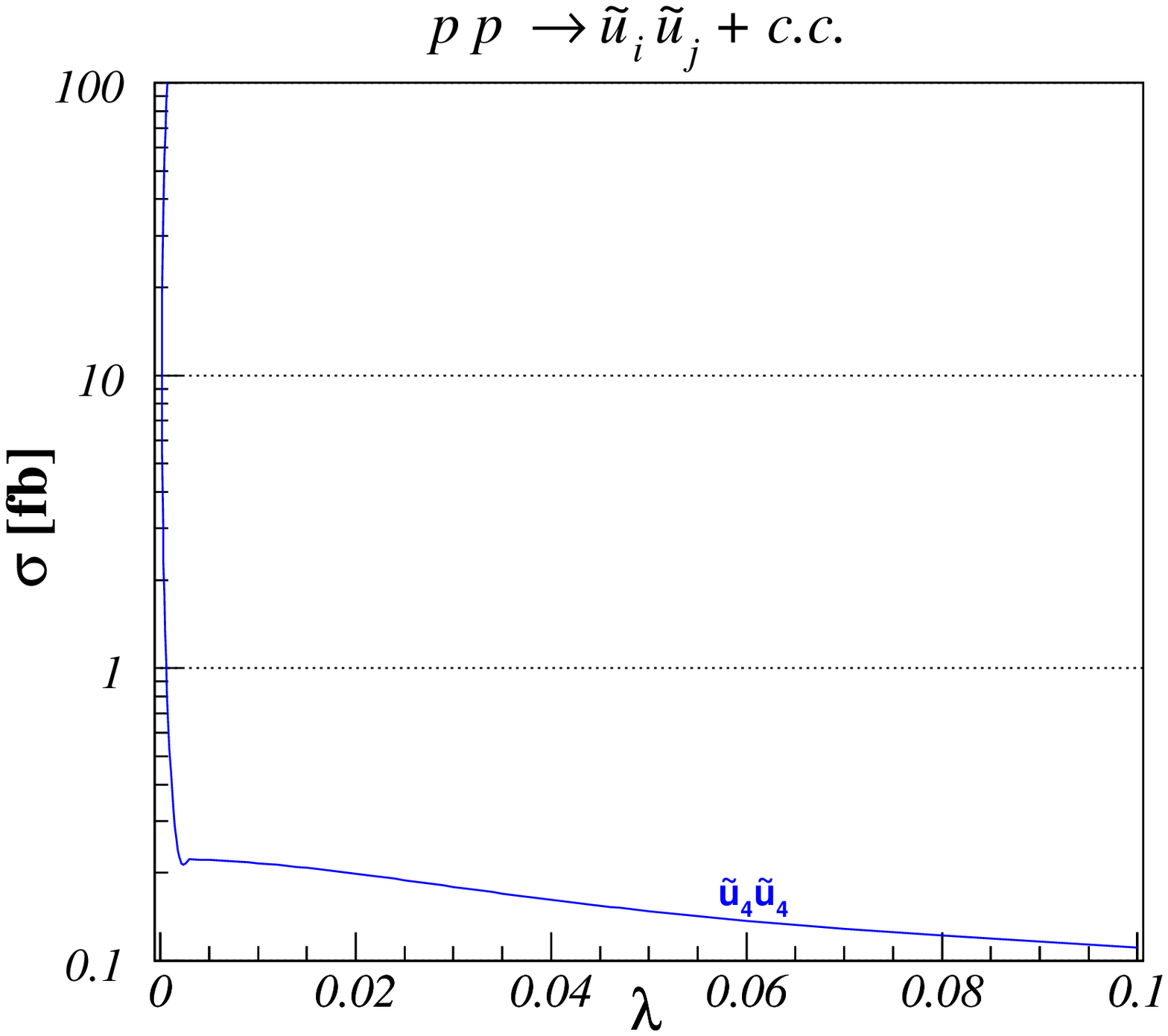}
 \includegraphics[height=0.32\textheight]{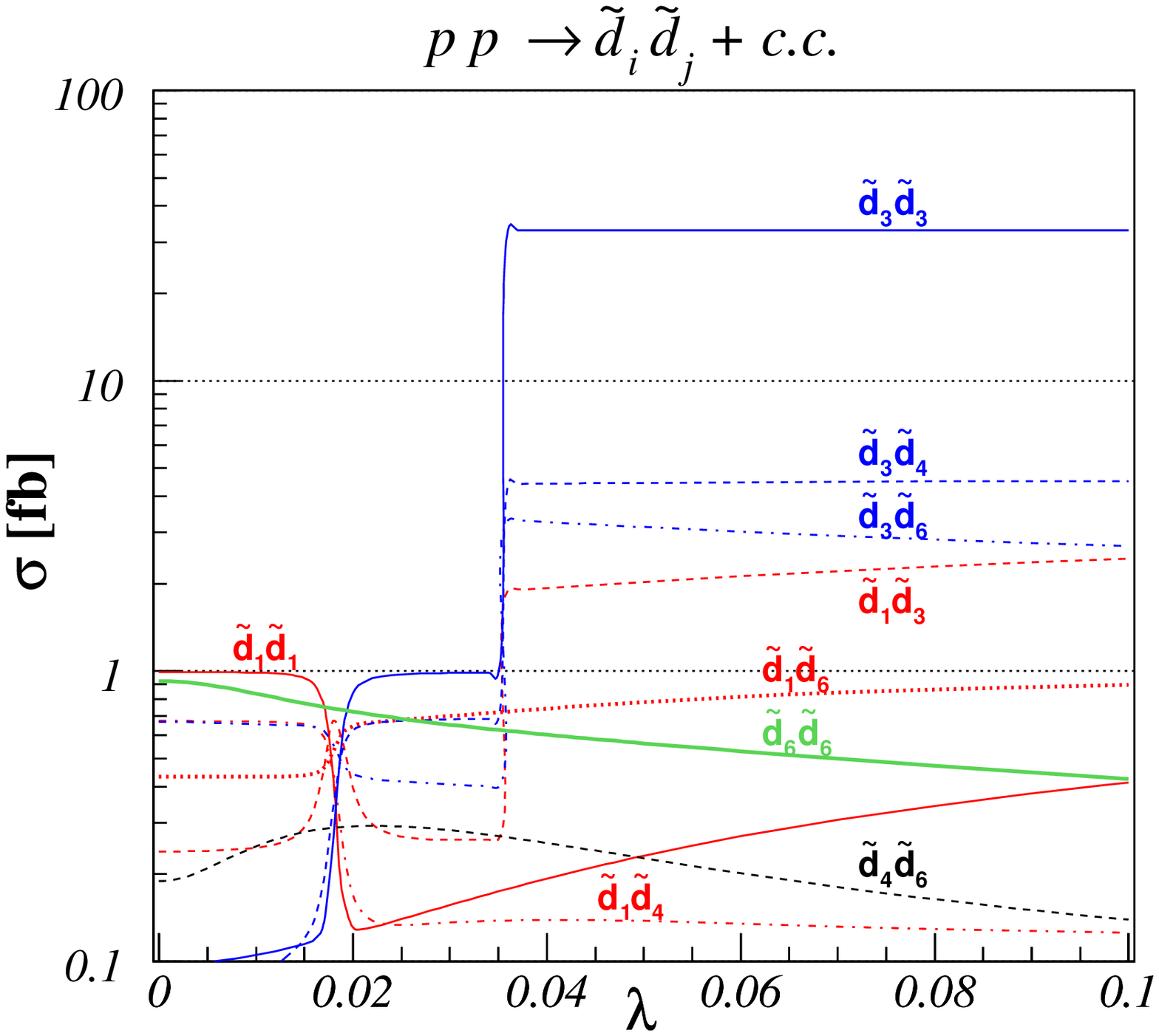}
 \caption{\label{fig:37b}Same as Fig.\ \ref{fig:37a} for our benchmark
          scenario B.}
\end{figure}
\begin{figure}
 \centering
 \includegraphics[height=0.32\textheight]{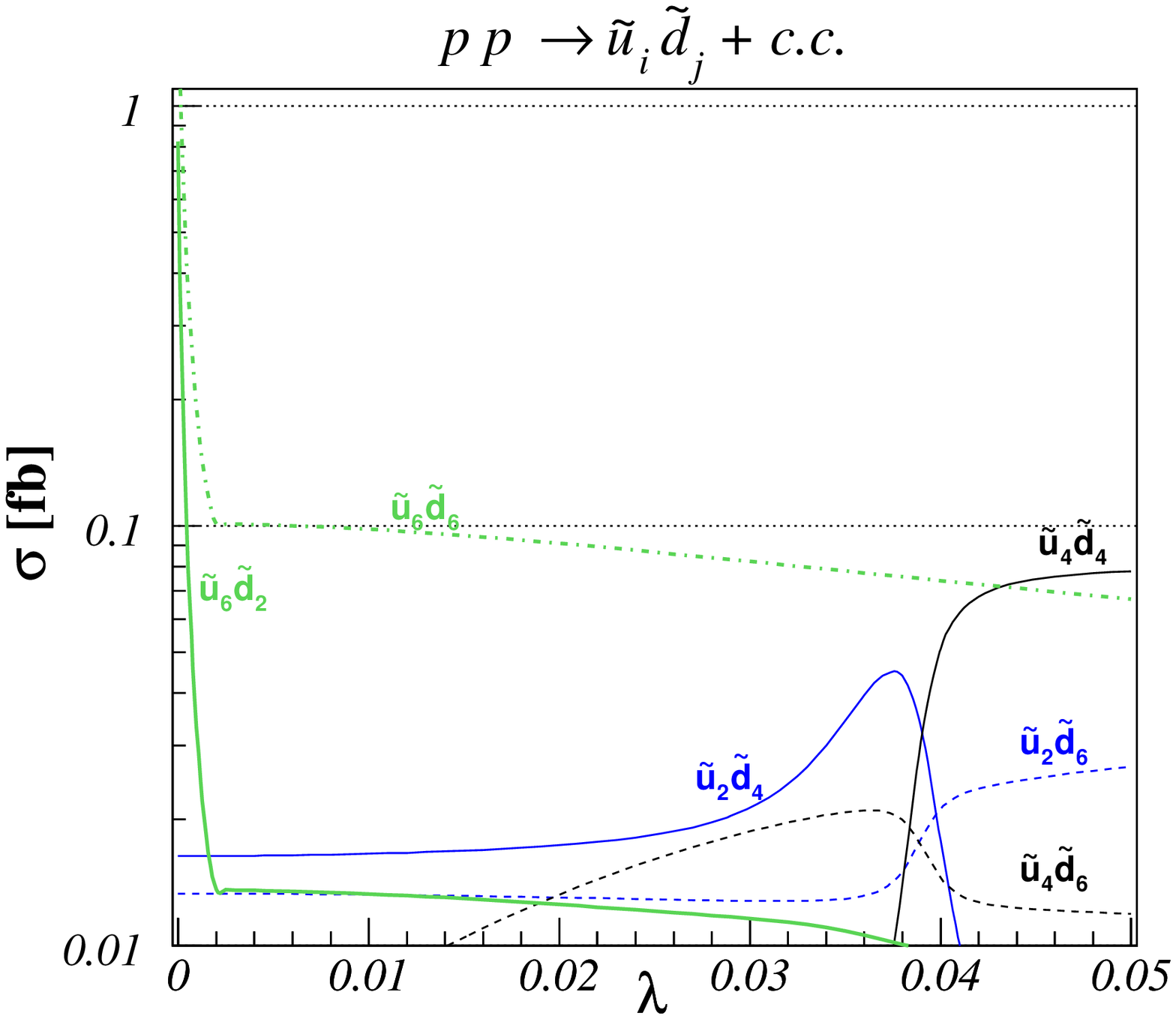}
 \includegraphics[height=0.32\textheight]{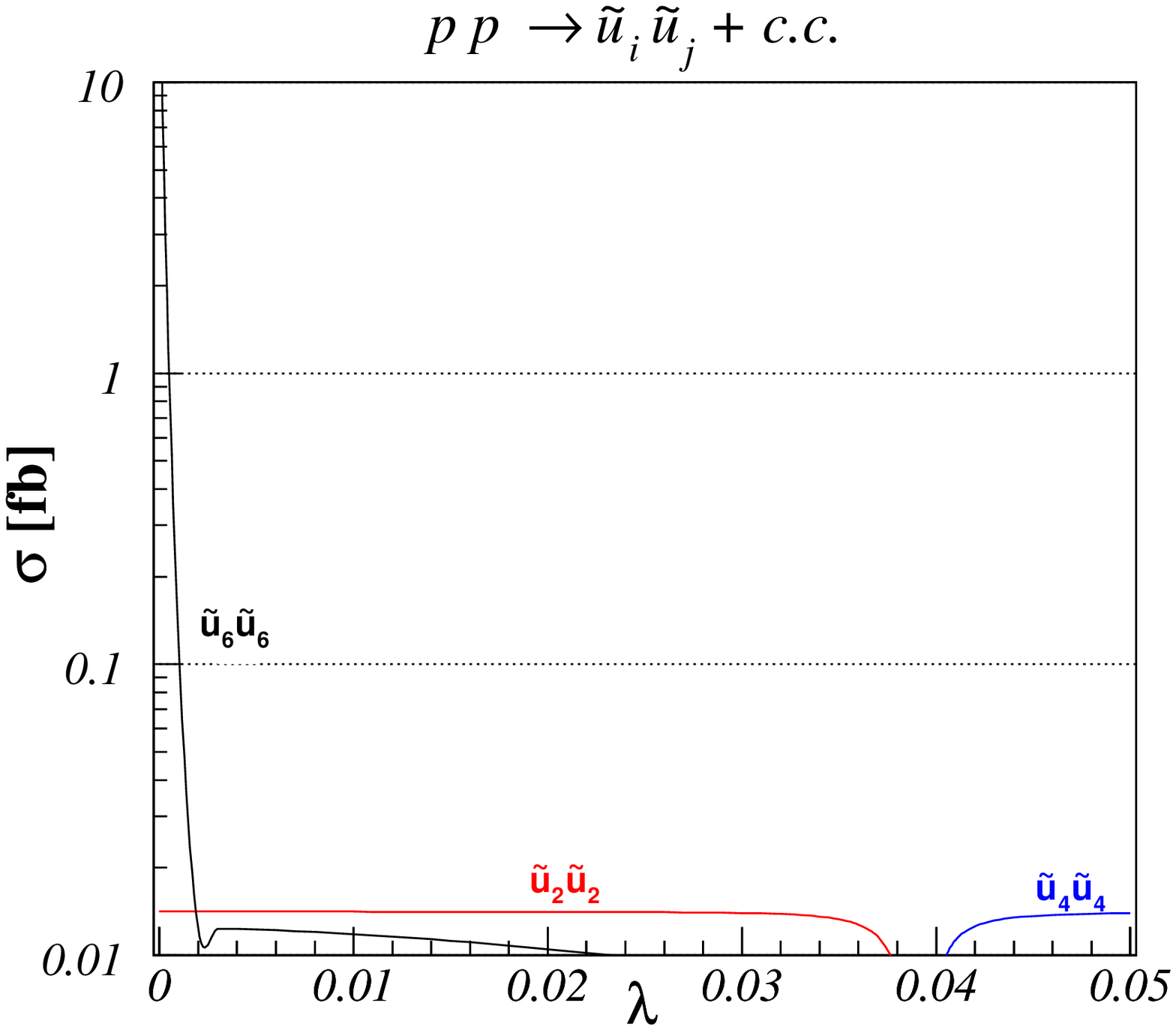}
 \includegraphics[height=0.32\textheight]{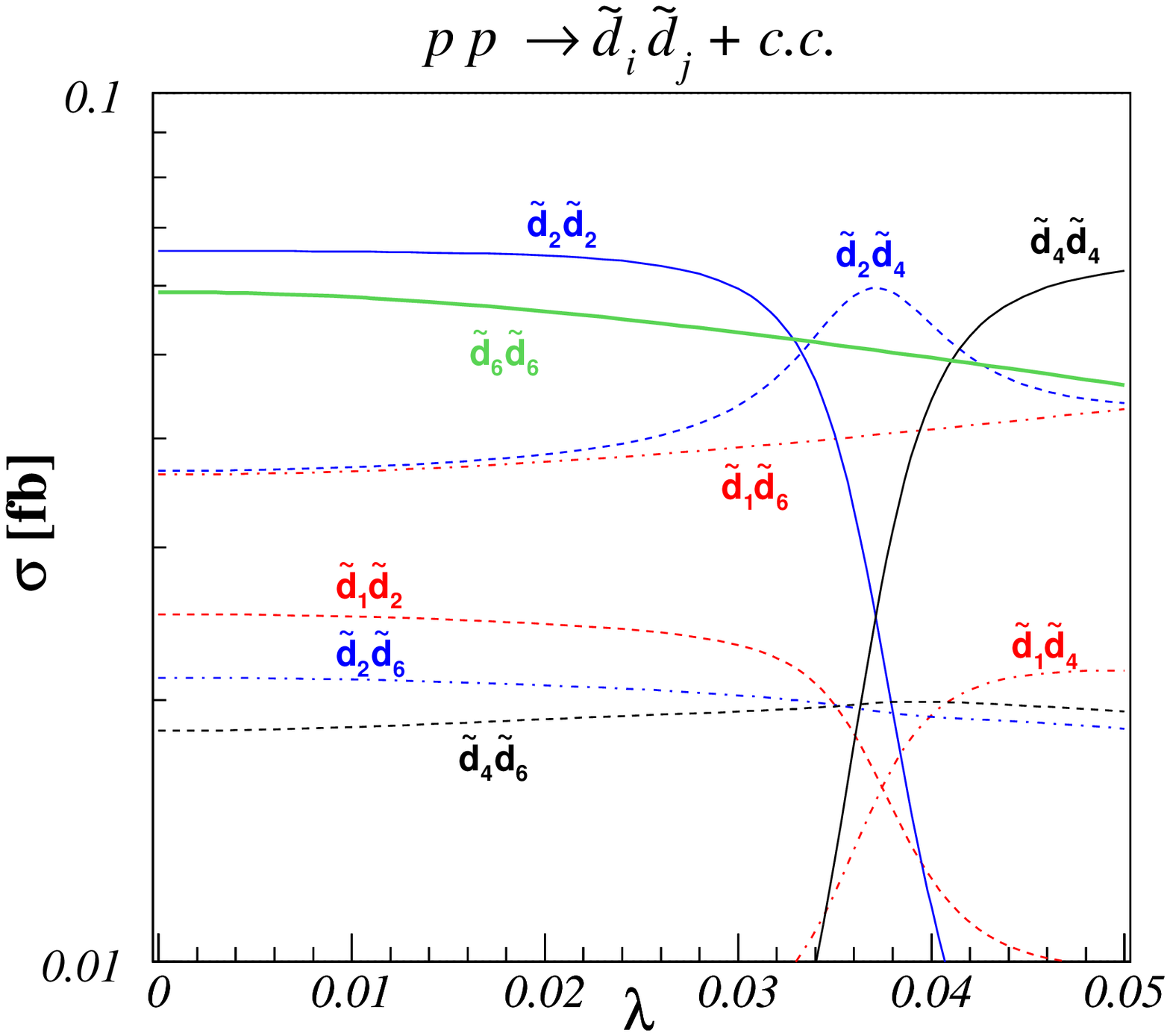}
 \caption{\label{fig:37c}Same as Fig.\ \ref{fig:37a} for our benchmark
          scenario C.}
\end{figure}
\begin{figure}
 \centering
 \includegraphics[height=0.32\textheight]{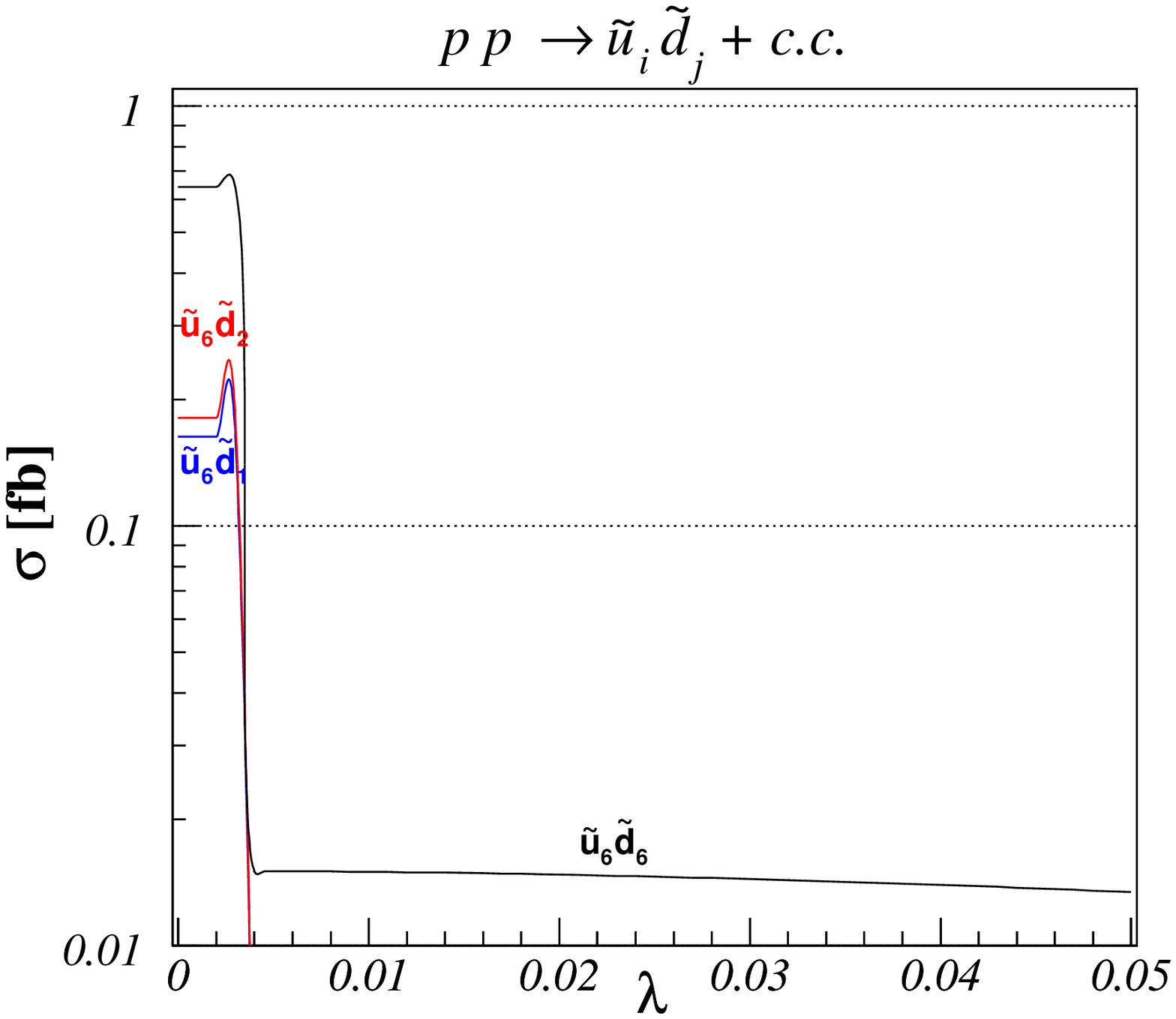}
 \includegraphics[height=0.32\textheight]{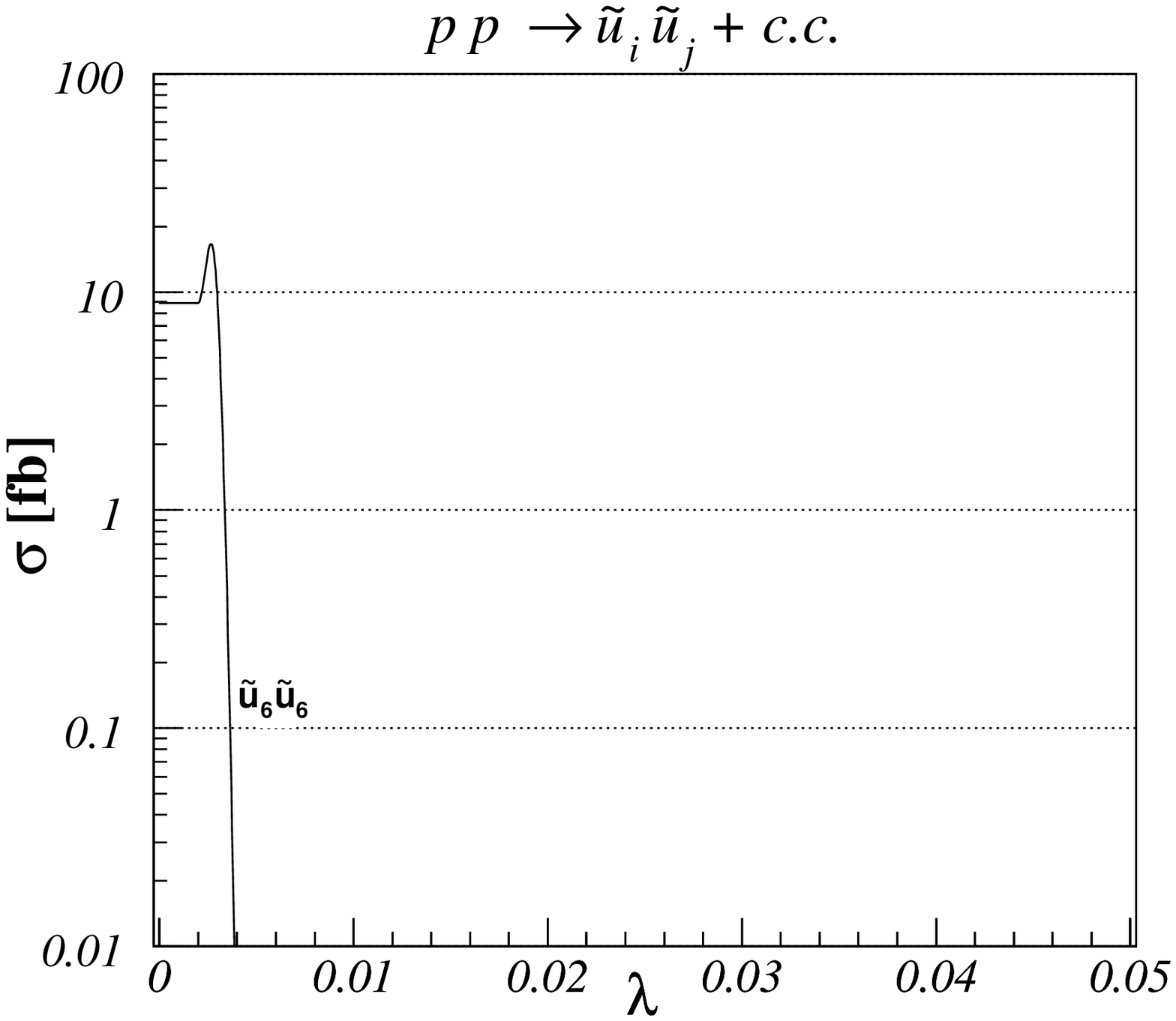}
 \includegraphics[height=0.32\textheight]{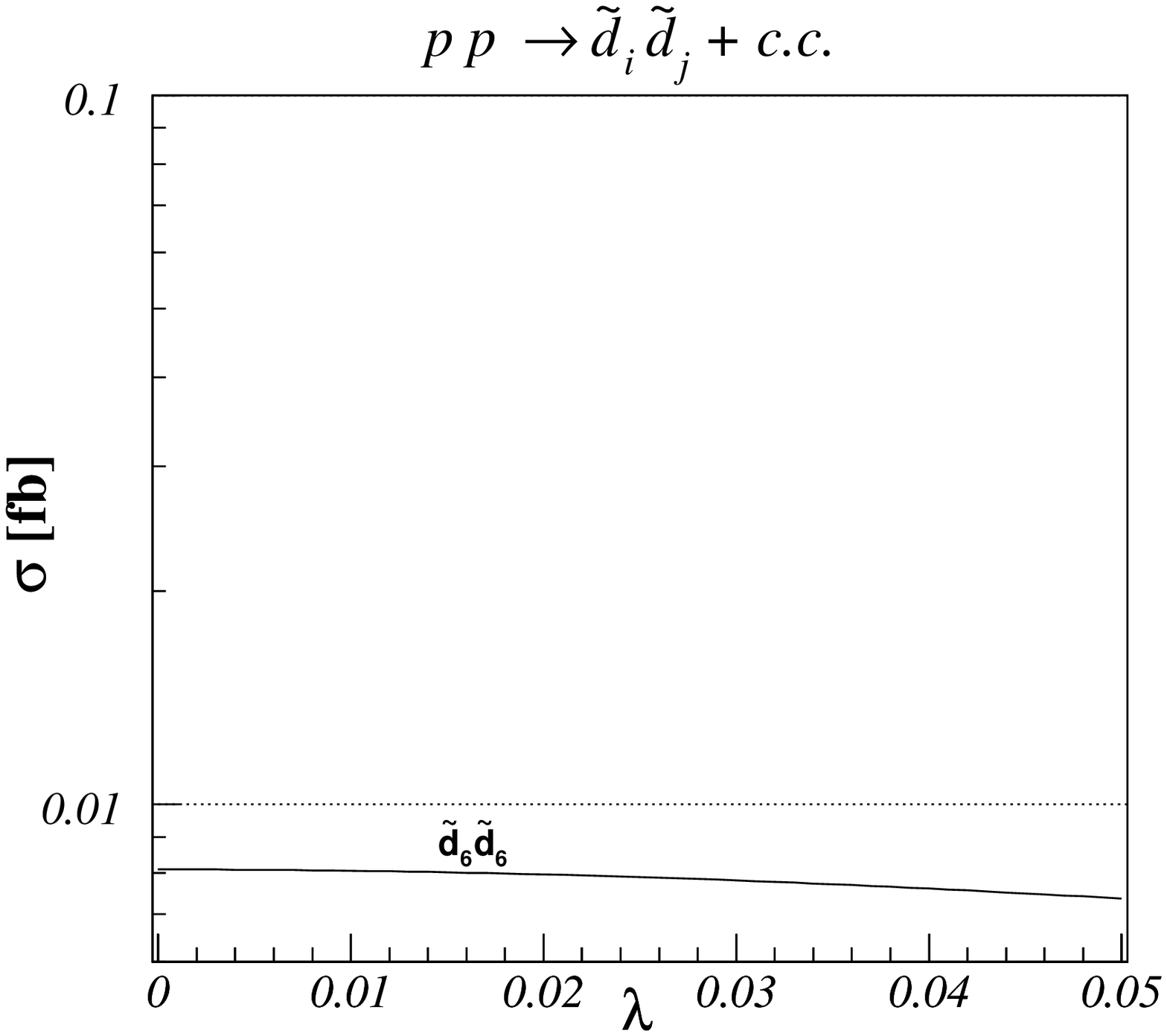}
 \caption{\label{fig:37d}Same as Fig.\ \ref{fig:37a} for our benchmark
          scenario D.}
\end{figure}

The numerical cross sections for mixed, up- and down-type
squark-squark pair production are shown in Figs.\ \ref{fig:37a},
\ref{fig:37b}, \ref{fig:37c} and \ref{fig:37d} for the benchmark
scenarios A, B, C and D described in Sec.\ \ref{sec:scan},
respectively, where we show their dependence on the flavour
violating parameter $\lambda$ (see Eq.\ (\ref{eq:lambda})). While
the diagonal channels are practically insensitive to flavour
violation, some of the non-diagonal cross sections show sharp
transitions, as for example at the benchmark point C (Fig.\
\ref{fig:37c}) where they occur between the
$\tilde{u}_2/\tilde{u}_4$ and $\tilde{d}_2/\tilde{d}_4$ states,
which are pure charm/strange squarks below/above $\lambda=0.035$.
As a side-remark we note that an interesting perspective might be
the exploitation of these $t$-channel contributions to second- and
third-generation squark production for the determination of
heavy-quark densities in the proton. This requires, of course,
efficient experimental techniques for heavy-flavour tagging.

\section{NMFV associated production of squarks and gauginos}

\subsection{Analytical results}

\begin{figure}
 \centering
 \includegraphics[width=0.5\columnwidth]{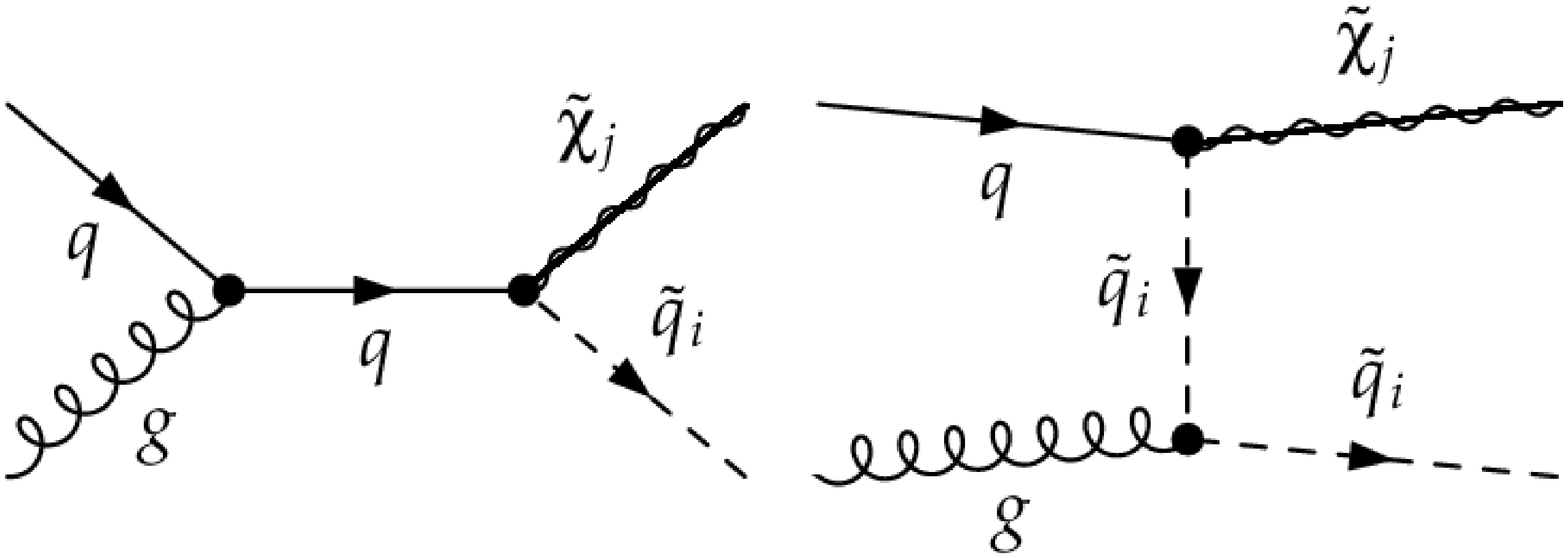}
 \caption{\label{fig:38}Tree-level Feynman diagrams for the associated
          production of squarks and gauginos.}
\end{figure}

\begin{figure}
 \centering
 \includegraphics[width=0.49\columnwidth]{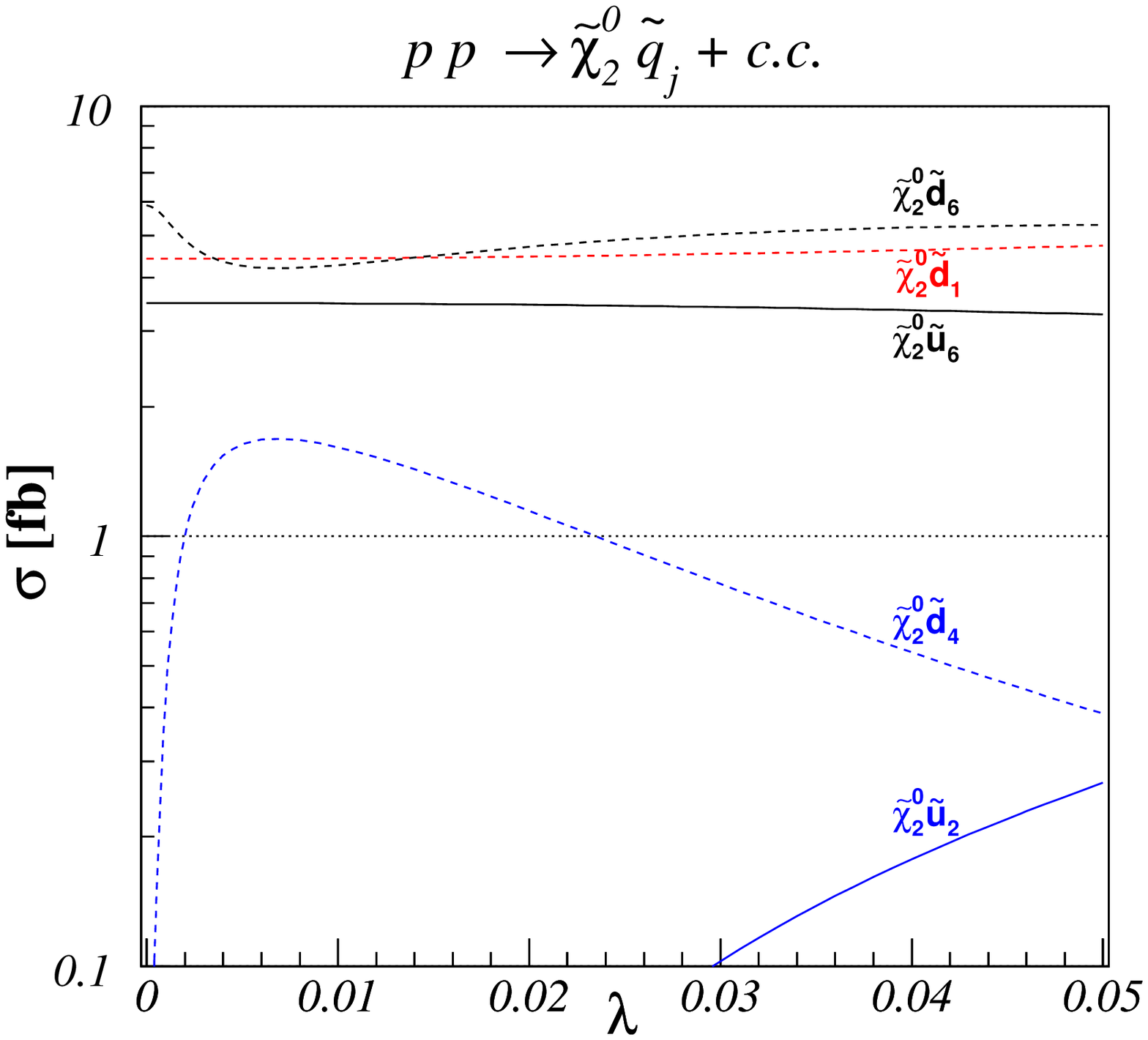}
 \includegraphics[width=0.49\columnwidth]{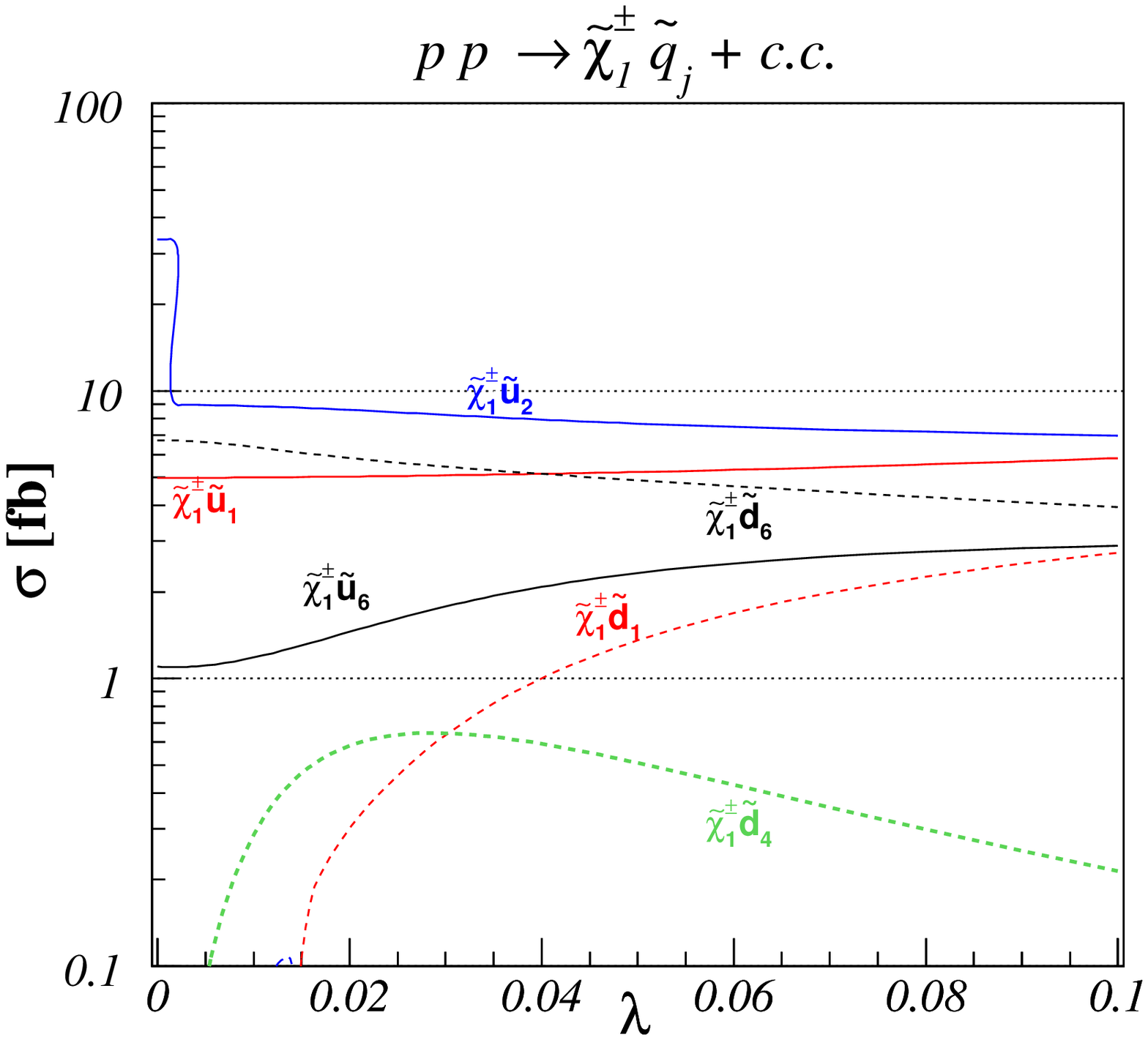}
 \caption{\label{fig:39a}Cross sections for
          associated production of squarks with charginos (left) and
          neutralinos (right) at the LHC in our benchmark
          scenario A.}\vspace{4mm}
 \includegraphics[width=0.49\columnwidth]{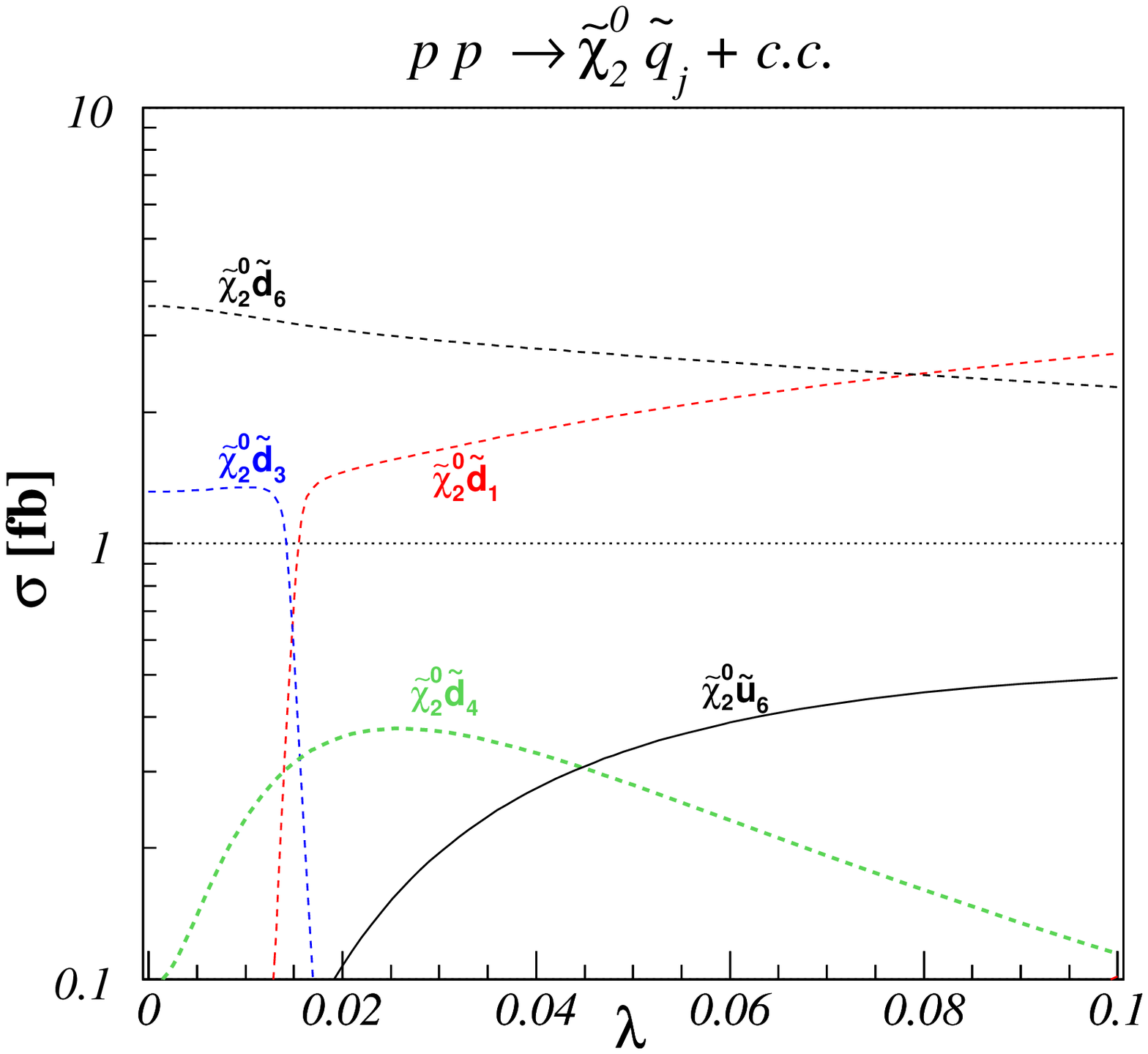}
 \includegraphics[width=0.49\columnwidth]{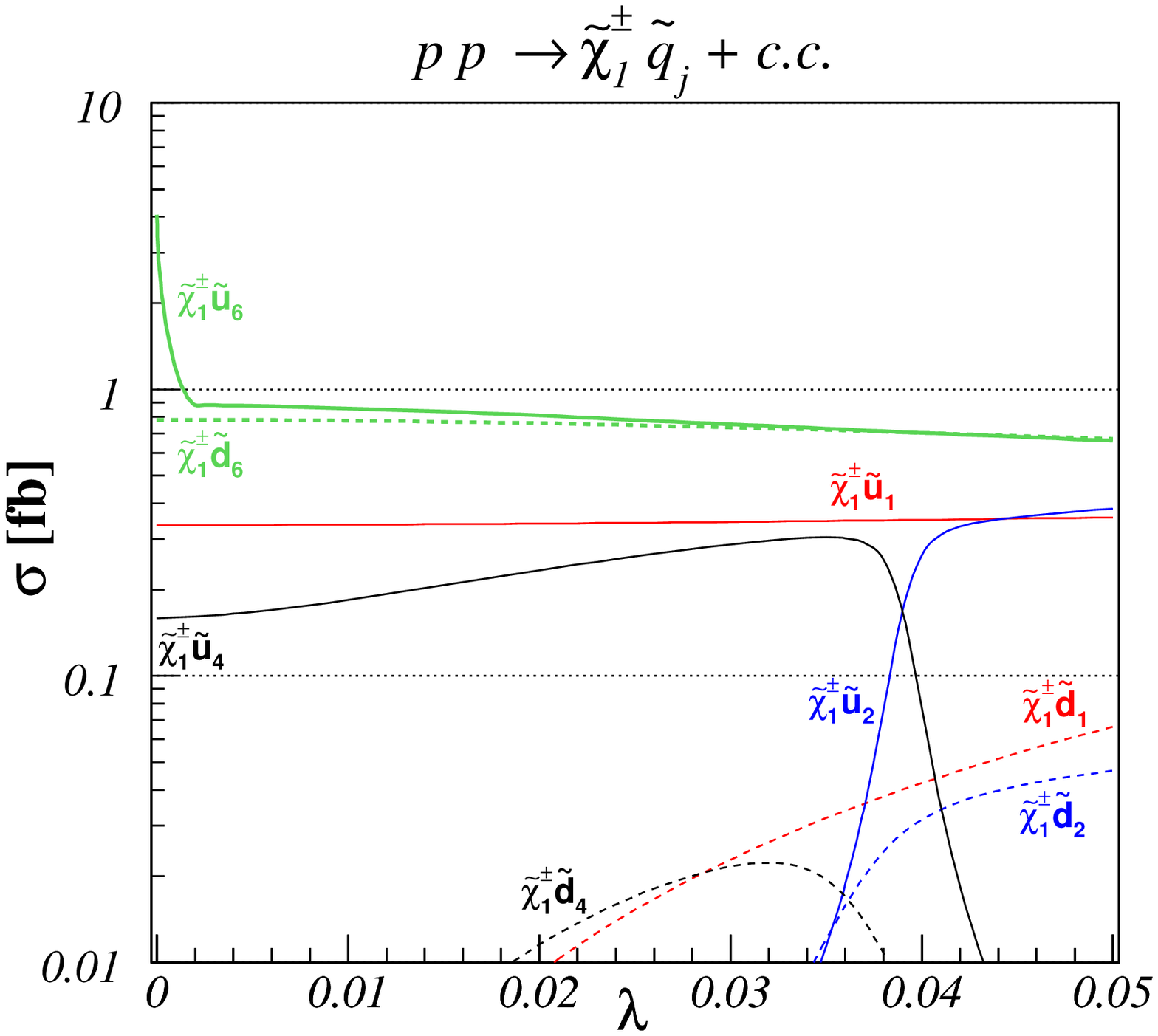}
 \caption{\label{fig:39b}Same as Fig.\ \ref{fig:39a} for our benchmark
          scenario B.}
\end{figure}
\begin{figure}
 \centering
 \includegraphics[width=0.49\columnwidth]{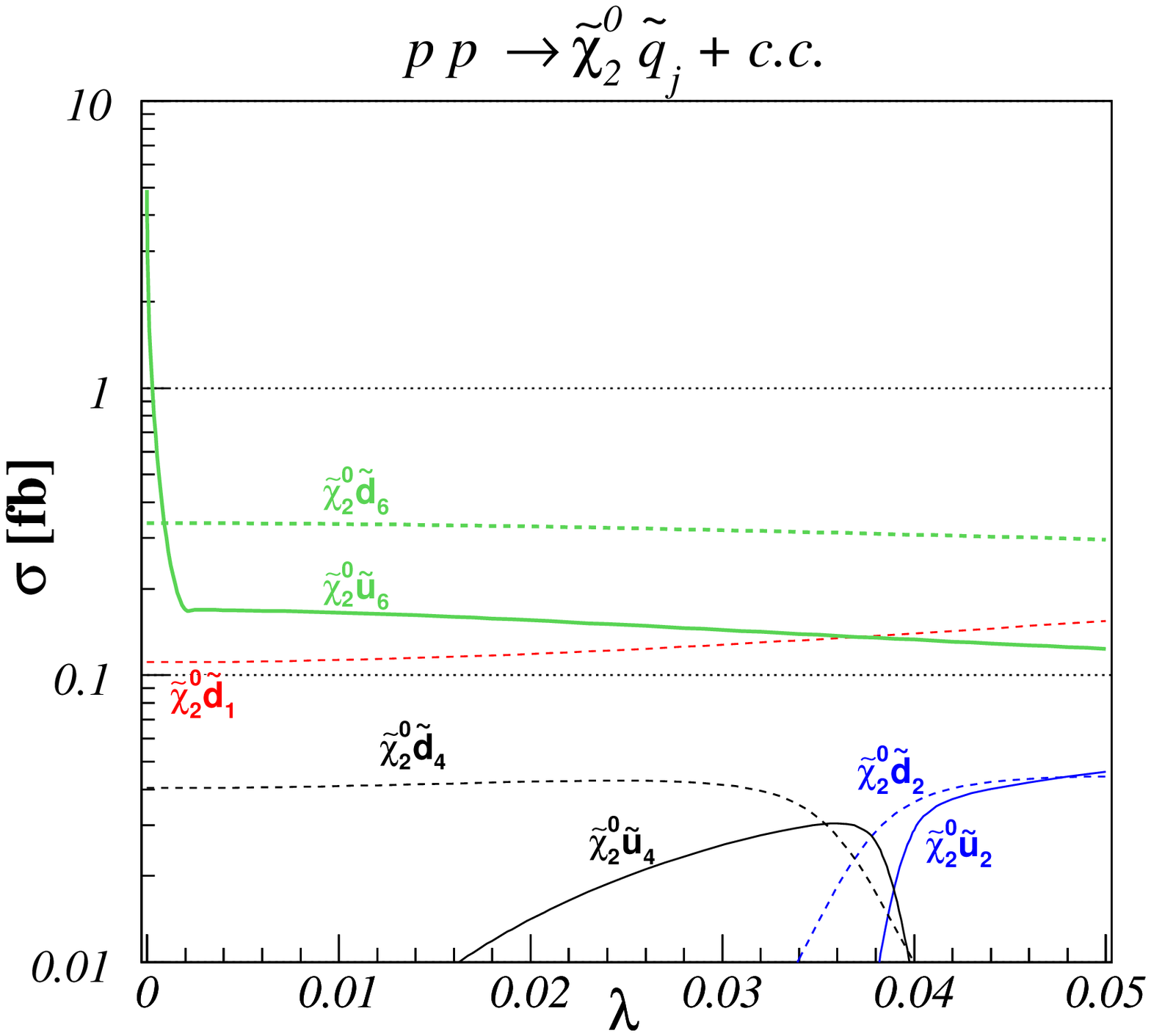}
 \includegraphics[width=0.49\columnwidth]{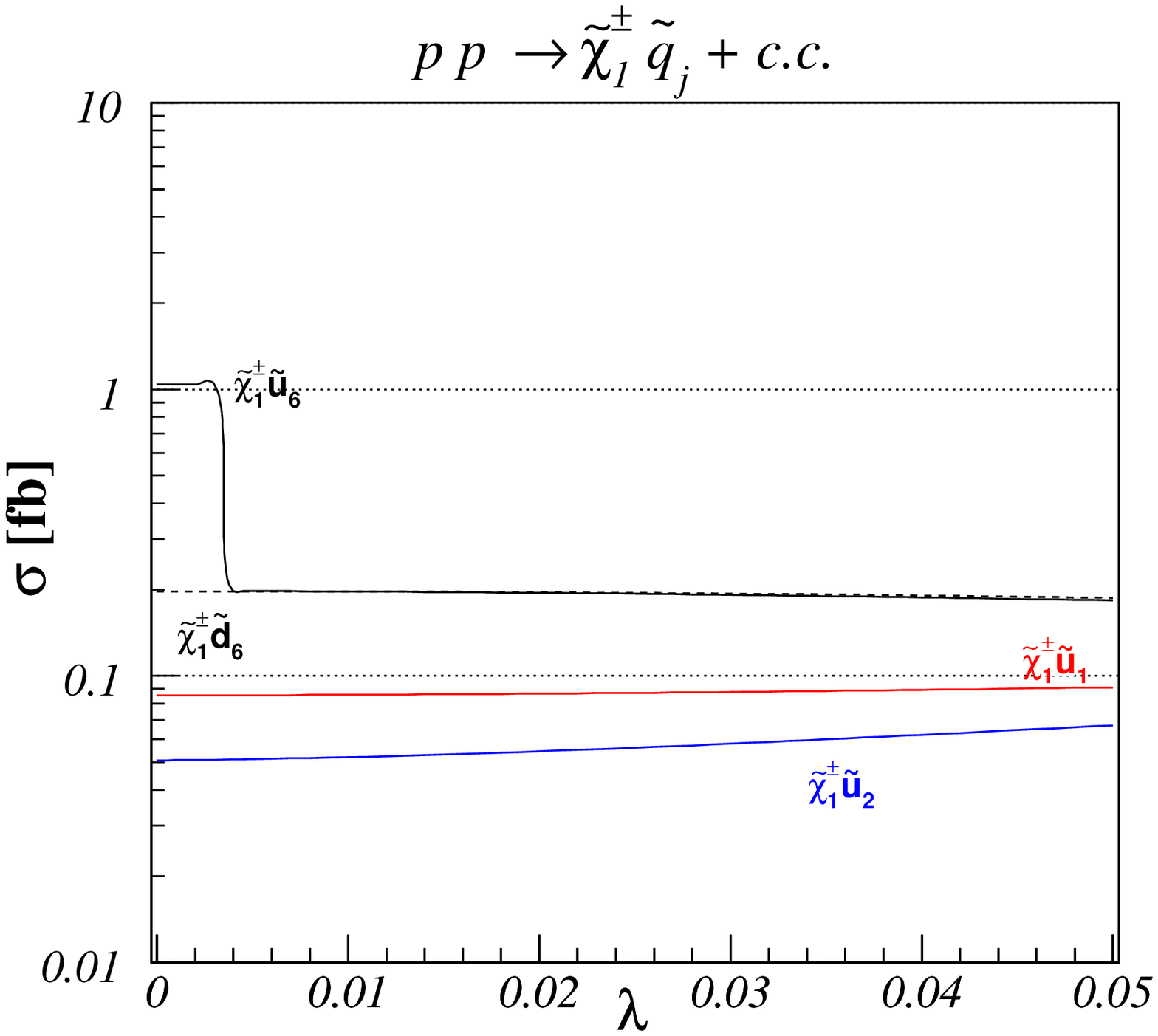}
 \caption{\label{fig:39c}Same as Fig.\ \ref{fig:39a} for our benchmark
          scenario C.}\vspace{4mm}
 \includegraphics[width=0.49\columnwidth]{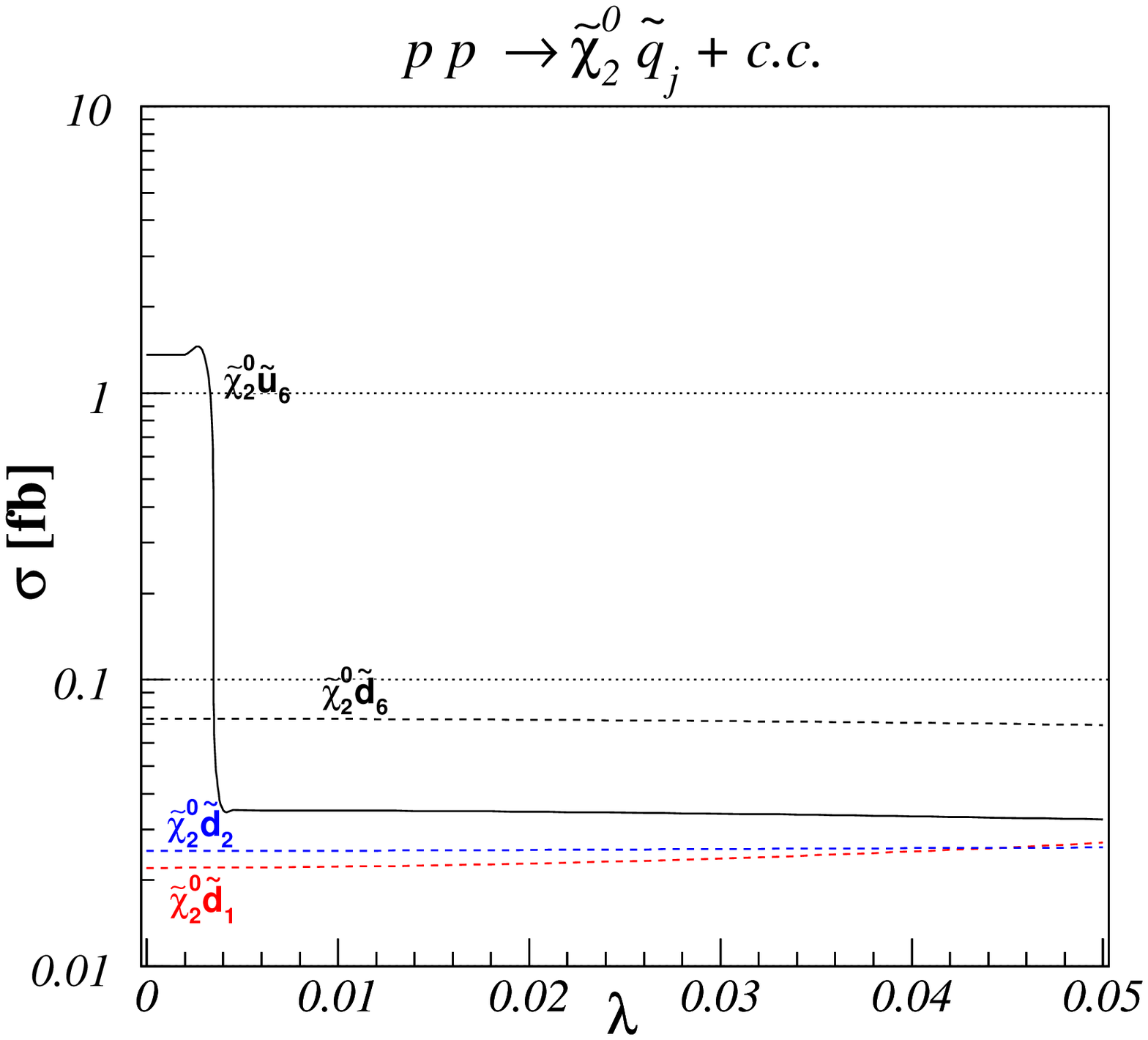}
 \includegraphics[width=0.49\columnwidth]{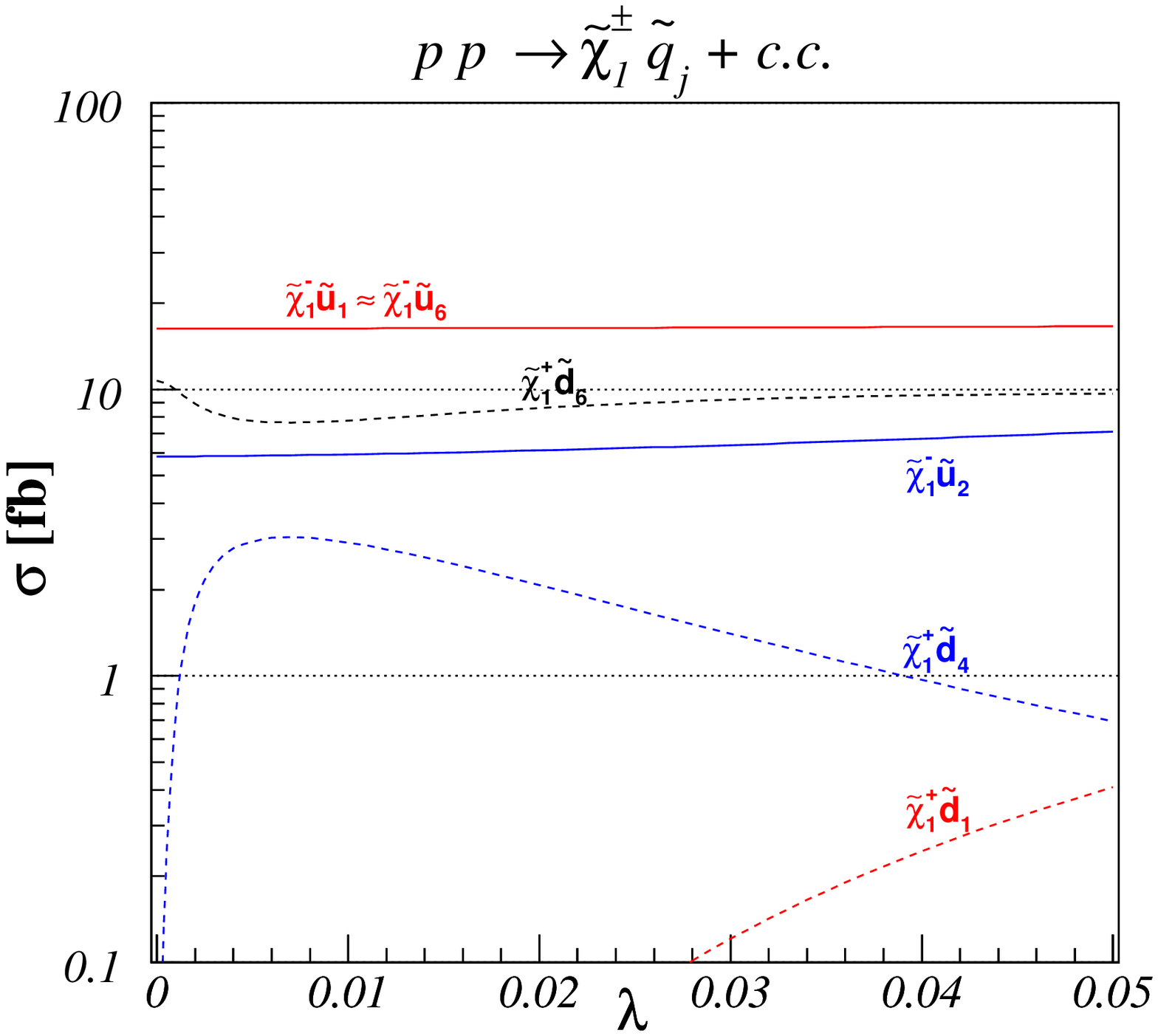}
 \caption{\label{fig:39d}Same as Fig.\ \ref{fig:39a} for our benchmark
          scenario D.}
\end{figure}

The associated production of squarks and neutralinos or charginos
\bea q(h_a,p_a)\, g(h_b,p_b) \to\tilde{\chi}_j(p_1)\,
\tilde{q}_i(p_2) \eea is a semi-weak process that originates from
quark-gluon initial states and has both an $s$-channel quark and a
$t$-channel squark contribution. They involve both a
quark-squark-gaugino vertex that can in general be flavour
violating. The corresponding Feynman diagrams can be seen in Fig.\
\ref{fig:38}. The squark-gaugino cross section is given by
\cite{Bozzi:2007me} \bea \frac{d \hat{\sigma}^{qg}_{h_a, h_b}}{dt}
&=& \frac{\pi\, \alpha\, \alpha_s}{n_{\tilde\chi}\, s^2} \Bigg\{
\frac{-u_{\tilde{\chi}_j}}{s} \bigg[(1-h_a)(1-h_b) \left|
L_{\tilde{q}_i q \tilde{\chi}_j}\right|^2 + (1+h_a)(1+h_b) \left|
R_{\tilde{q}_i q \tilde{\chi}_j}\right|^2\bigg] \nonumber
\\ &+& \frac{t_{\tilde{\chi}_j}\left( t + m^2_{\tilde{q}_i}
\right)}{t_{\tilde{q}_i}^2} \bigg[(1-h_a)\left| L_{\tilde{q}_i q
\tilde{\chi}_j}\right|^2 + (1+h_a) \left| R_{\tilde{q}_i q
\tilde{\chi}_j}\right|^2\bigg] \nonumber
\\ & +& \frac{2\,(u t \!-\! m^2_{\tilde{q}_i}\,
m_{\tilde{\chi}_j}^2)}{s\, t_{\tilde{q}_i}} \bigg[(1\!-\!h_a) (1
\!-\! h_b) \left| L_{\tilde{q}_i q \tilde{\chi}_j}\right|^2 \!+\!
(1 \!+\! h_a)(1 \!+\! h_b) \left| R_{\tilde{q}_i q
\tilde{\chi}_j}\right|^2\bigg] \nonumber
\\  &+& \frac{t_{\tilde{\chi}_j} (t_{\tilde{\chi}_j} -
u_{\tilde{q}_i}) }{s\, t_{\tilde{q}_i}} \bigg[(1-h_a)\left|
L_{\tilde{q}_i q \tilde{\chi}_j}\right|^2 + (1+h_a) \left|
R_{\tilde{q}_i q \tilde{\chi}_j}\right|^2\bigg] \Bigg\}, \eea
where $n_{\tilde\chi}=6x_W(1-x_W)$ for neutralinos and
$n_{\tilde\chi}=12 x_W$ for charginos. Note that the $t$-channel
diagram involves the coupling of the gluon to scalars and does
thus not depend on its helicity $h_b$. The cross section of the
charge-conjugate process can be obtained by taking $h_a\to-h_a$.
For non-mixing squarks and gauginos, we agree again with the
results of Ref.\ \cite{Dawson:1983fw}.

\subsection{Numerical results}

Again, we employ the LO set of the latest CTEQ6 global parton
density fit \cite{Pumplin:2002vw}, with five active flavours and
the gluon, and the strong coupling constant is calculated with the
value of $\Lambda_{\rm LO}^{n_f=5}=165$ MeV. The renormalization
and factorization scales are set to the average mass of the final
state SUSY particles, and the SUSY masses and mixings are computed
with the help of SPheno and FeynHiggs.\\

The numerical cross sections for associated production of squarks
with charginos and neutralinos production as a function of the
flavour violating parameter $\lambda$ are shown in Figs.\
\ref{fig:39a}, \ref{fig:39b}, \ref{fig:39c} and \ref{fig:39d} for
the benchmark scenarios A, B, C and D described in Sec.\
\ref{sec:scan}, respectively. The cross sections for the
semi-strong production of very light gaugino and not too heavy
squark vary from $10^{-1}$ fb to $10^2$ fb and are quite sensitive
to the flavour violation parameter $\lambda$ thanks to the
quark-squark-gaugino vertex in diagrams of Fig.\ \ref{fig:38}.
Smooth transitions are observed for the associated production of
third-generation squarks with charginos and neutralinos, and in
particular for the scenarios A and B.\\

For benchmark point A (Fig.\ \ref{fig:39a}), the cross section for
$\tilde{d}_4$ production decreases with its strange squark content
(see Fig.\ \ref{fig:09p}), while the bottom squark content
increases at the same time. For benchmark point B (Fig.\
\ref{fig:39b}), the same (opposite) happens for $\tilde{d}_6$
($\tilde{d}_1$), while the cross sections for $\tilde{u}_6$
increase/decrease with its charm/top squark content. Even in
constrained Minimal Flavour Violation, the associated production
of stops and charginos is a particularly useful channel for SUSY
particle spectroscopy, as can be seen from the fact that cross
sections vary over several orders of magnitude among our four
benchmark points (see also Ref.\ \cite{Beccaria:2006wz}).

\section{NMFV gaugino pair production}

\subsection{Analytical results}

\begin{figure}
 \centering
 \includegraphics[width=0.75\columnwidth]{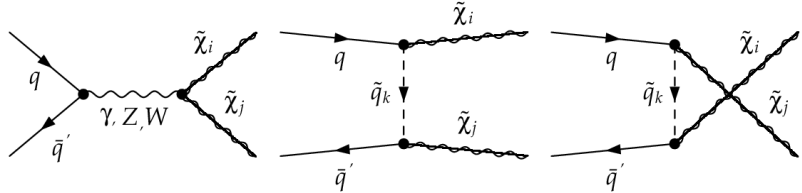}
 \caption{\label{fig:40}Tree-level Feynman diagrams for the production
          of gaugino pairs.}
\end{figure}

We consider the purely electroweak production of gaugino pairs
\bea  q(h_a,p_a)\, \bar{q}^\prime(h_b,p_b) \to
\tilde{\chi}_i(p_1)\, \tilde{\chi}_j(p_2) \eea from
quark-antiquark initial states, where flavour violation can occur
via the quark-squark-gaugino vertices in the $t$- and $u$-channels
(see Fig.\ \ref{fig:40}). However, if it were not for different
parton density weights, summation over complete squark multiplet
exchanges would make these channels insensitive to the exchanged
squark flavour. Furthermore there are no final state squarks that
could be experimentally tagged. The cross section can be expressed
generically as \cite{Bozzi:2007me} \bea \frac{d
\hat{\sigma}^{q\bar{q}'}_{h_a, h_b}}{dt} &=& \frac{\pi \alpha^2}{3
s^2}(1-h_a) (1+h_b) \Big[ \left| Q^u_{LL} \right|^2
u_{\tilde{\chi}_i} u_{\tilde{\chi}_j} + \left| Q_{LL}^t \right|^2
t_{\tilde{\chi}_i} t_{\tilde{\chi}_j} + 2 {\rm
Re} [Q_{LL}^{u\ast} Q_{LL}^t] \nn \\
&\times & m_{\tilde{\chi}_{i}} m_{\tilde{\chi}_{j}} s \Big] +
\frac{\pi \alpha^2}{3 s^2}(1+h_a) (1-h_b) \Big[ \left| Q_{RR}^u
\right|^2 u_{\tilde{\chi}_i} u_{\tilde{\chi}_j}+ \left| Q_{RR}^t
\right|^2 t_{\tilde{\chi}_i} t_{\tilde{\chi}_j} \nn \\ & +&  2
{\rm Re} [Q_{RR}^{u\ast} Q_{RR}^t] m_{\tilde{\chi}_{i}}
m_{\tilde{\chi}_{j}} s \Big] + \frac{\pi \alpha^2}{3 s^2}(1+h_a)
(1+h_b) \Big[ \left| Q_{RL}^u \right|^2 u_{\tilde{\chi}_i}
u_{\tilde{\chi}_j} \nn \\ & + &  \left| Q_{RL}^t \right|^2
t_{\tilde{\chi}_i} t_{\tilde{\chi}_j} \!+\! {\rm Re}
[Q_{RL}^{u\ast} Q_{RL}^t] (u t \!-\! m^2_{\tilde{\chi}_{i}}
m^2_{\tilde{\chi}_{j}})
\Big] \!+\! \frac{\pi \alpha^2}{3 s^2}(1\!-\!h_a) (1\!-\!h_b) \nn \\
& \times & \Big[ \left| Q_{LR}^u \right|^2 u_{\tilde{\chi}_i}
u_{\tilde{\chi}_j} \!+\! \left| Q_{LR}^t \right|^2
t_{\tilde{\chi}_i} t_{\tilde{\chi}_j} \!+\! {\rm Re}
[Q_{LR}^{u\ast} Q_{LR}^t] (u t \!-\! m^2_{\tilde{\chi}_{i}}
m^2_{\tilde{\chi}_{j}}) \Big],~~~~~\label{eq:xsecgg} \eea i.e.\ in
terms of generalized charges. For
$\tilde{\chi}_i^-\tilde{\chi}_j^+$-production, these charges are
given by \bea  Q_{LL}^{u-+} &=& \Bigg(\frac{e_q \delta_{ij}
\delta_{qq^\prime}}{s} - \frac{L_{q q^\prime Z} O^{\prime
R\ast}_{ij}}{2 \, x_W \, (1-x_W) \,s_z} + \sum_{k=1}^6
\frac{L_{\tilde{d}_k q^\prime \tilde{\chi}_i^\pm}
L^\ast_{\tilde{d}_k q \tilde{\chi}_j^\pm}}{2\, x_W\,
u_{\tilde{d}_k}}\Bigg) ,~\nonumber\\ Q_{LL}^{t-+} &=&
\Bigg(\frac{e_q \delta_{ij} \delta_{qq^\prime}}{s} - \frac{L_{q
q^\prime Z} O^{\prime L\ast}_{ij}}{2 \,x_W \,(1-x_W) \,s_z} -
\sum_{k=1}^6 \frac{L^\ast_{\tilde{u}_k q^\prime
\tilde{\chi}_j^\pm} L_{\tilde{u}_k q \tilde{\chi}_i^\pm}}{2
\,x_W\, t_{\tilde{u}_k}}\Bigg),~ \nonumber\\ Q_{RR}^{u-+} &=&
\Bigg(\frac{e_q \delta_{ij} \delta_{qq^\prime}}{s} - \frac{R_{q
q^\prime Z} O^{\prime L\ast}_{ij}}{2 \, x_W \, (1-x_W) \,s_z}+
\sum_{k=1}^6 \frac{R_{\tilde{d}_k q^\prime \tilde{\chi}_i^\pm}
R^\ast_{\tilde{d}_k q \tilde{\chi}_j^\pm}}{2\,  x_W\,
u_{\tilde{d}_k}}\Bigg),~ \nonumber\\ Q_{RR}^{t-+} &=&
\Bigg(\frac{e_q \delta_{ij} \delta_{qq^\prime}}{s} - \frac{R_{q
q^\prime Z} O^{\prime R\ast}_{ij}}{2 \,x_W \,(1-x_W) \,s_z} -
\sum_{k=1}^6 \frac{R^\ast_{\tilde{u}_k q^\prime
\tilde{\chi}_j^\pm} R_{\tilde{u}_k q \tilde{\chi}_i^\pm}}{2
\,x_W\, t_{\tilde{u}_k}}\Bigg),~ \nonumber\\ Q_{LR}^{u-+} &=&
\sum_{k=1}^6 \frac{R_{\tilde{d}_k q^\prime \tilde{\chi}_i^\pm}
L^\ast_{\tilde{d}_k q \tilde{\chi}_j^\pm}}{2\,  x_W\,
u_{\tilde{d}_k}} ,~ \nonumber\\ Q_{LR}^{t-+} &=&  \sum_{k=1}^6
\frac{R^\ast_{\tilde{u}_k q^\prime  \tilde{\chi}_j^\pm}
L_{\tilde{u}_k q \tilde{\chi}_i^\pm}}{2  \,x_W\,
t_{\tilde{u}_k}},~ \nonumber\\ Q_{RL}^{u-+} &=&  \sum_{k=1}^6
\frac{L_{\tilde{d}_k q^\prime  \tilde{\chi}_i^\pm}
R^\ast_{\tilde{d}_k q \tilde{\chi}_j^\pm}}{2\,x_W\,
u_{\tilde{d}_k}} ,~ \nonumber\\ Q_{RL}^{t-+} &=&\sum_{k=1}^6
\frac{L^\ast_{\tilde{u}_k q^\prime \tilde{\chi}_j^\pm}
R_{\tilde{u}_k q \tilde{\chi}_i^\pm}}{2 \,x_W \, t_{\tilde{u}_k}}.
\eea Note that there is no interference between $t$- and
$u$-channel diagrams due to (electromagnetic) charge conservation.
The cross section for chargino pair production in
$e^+e^-$-collisions can be deduced by setting $e_q \to e_l = -1$,
$L_{q q^\prime Z}\to L_{e e Z} = (2\,T^{3}_l - 2\,e_l\,x_W)$ and
$R_{q q^\prime Z}\to R_{e e Z} = - 2\,e_l\, x_W$. Neglecting all
Yukawa couplings, we can then reproduce the
calculations of Ref.\ \cite{Choi:1998ei}.\\

The charges of the chargino-neutralino associated production are
given by \bea Q_{LL}^{u+0} &=&  \frac{1}{\sqrt{2\,(1-x_W)}\, x_W}
\left[ \frac{O^{L\ast}_{ji}  L^\ast_{qq^\prime W}}{\sqrt{2}\,s_w}
+ \sum_{k=1}^6  \frac{L_{\tilde{u}_k q^\prime
\tilde{\chi}_i^\pm}^\ast  L_{\tilde{u}_k q
\tilde{\chi}_j^0}^\ast}{u_{\tilde{u}_k}} \right],~\nonumber \\
Q_{LL}^{t+0} &=& \frac{1}{\sqrt{2\,(1-x_W)}\, x_W} \left[
\frac{O^{R\ast}_{ji} L^\ast_{qq^\prime W}}{\sqrt{2}\,s_w} -
\sum_{k=1}^6 \frac{L_{\tilde{d}_k q \tilde{\chi}_i^\pm}^\ast
L_{\tilde{d}_k  q^\prime \tilde{\chi}_j^0}}{t_{\tilde{d}_k}}
\right],~\nonumber \\ Q_{RR}^{u+0} &=&
\frac{1}{\sqrt{2\,(1-x_W)}\, x_W} \sum_{k=1}^6
\frac{R_{\tilde{u}_k q^\prime \tilde{\chi}_i^\pm}^\ast
R_{\tilde{u}_k q
\tilde{\chi}_j^0}^\ast}{u_{\tilde{u}_k}} ,~  \nonumber \\
Q_{RR}^{t+0} &=& \frac{-1}{\sqrt{2\,(1-x_W)}\, x_W}  \sum_{k=1}^6
\frac{R_{\tilde{d}_k q \tilde{\chi}_i^\pm}^\ast R_{\tilde{d}_k
q^\prime \tilde{\chi}_j^0}}{t_{\tilde{d}_k}} ,~  \nonumber\\
Q_{LR}^{u+0} &=& \frac{1}{\sqrt{2\,(1-x_W)}\, x_W}  \sum_{k=1}^6
\frac{R_{\tilde{u}_k q^\prime  \tilde{\chi}_i^\pm}^\ast
L_{\tilde{u}_k q
\tilde{\chi}_j^0}^\ast}{u_{\tilde{u}_k}} ,~ \nonumber \\
Q_{LR}^{t+0} &=& \frac{1}{\sqrt{2\,(1-x_W)}\, x_W} \sum_{k=1}^6
\frac{L_{\tilde{d}_k q \tilde{\chi}_i^\pm}^\ast R_{\tilde{d}_k
q^\prime
\tilde{\chi}_j^0}}{t_{\tilde{d}_k}} ,~ \nonumber \\
 Q_{RL}^{u+0} &=& \frac{1}{\sqrt{2\,(1-x_W)}\, x_W} \sum_{k=1}^6
 \frac{L_{\tilde{u}_k q^\prime \tilde{\chi}_i^\pm}^\ast
 R_{\tilde{u}_k q \tilde{\chi}_j^0}^\ast}{u_{\tilde{u}_k}} ,~
 \nonumber \\ Q_{RL}^{t+0} &=& \frac{1}{\sqrt{2\,(1-x_W)}\, x_W}
 \sum_{k=1}^6 \frac{R_{\tilde{d}_k q \tilde{\chi}_i^\pm}^\ast
 L_{\tilde{d}_k q^\prime \tilde{\chi}_j^0}}{t_{\tilde{d}_k}}.
\eea The charge-conjugate process is again obtained by making the
replacement $h_{a,b}\to -h_{a,b}$ in Eq.\ (\ref{eq:xsecgg}). In
the case of non-mixing squarks with neglected Yukawa couplings, we
agree with the results of Ref.\ \cite{Beenakker:1999xh}, provided
we correct a sign in their Eq.\ (2) as described in Ref.\
\cite{Spira:2000vf}.\\

Finally, the charges for the neutralino pair production are given
by \bea  Q_{LL}^{u00} &=&  \frac{1}{x_W\,(1-x_W)\,\sqrt{1+
\delta_{ij}}} \left[ \frac{L_{q  q^\prime Z} O^{\prime\prime
L}_{ij} }{2 s_z} + \sum_{k=1}^6  \frac{L_{\tilde{Q}_k q^\prime
\tilde{\chi}_i^0} L_{\tilde{Q}_k q
\tilde{\chi}_j^0}^\ast}{u_{\tilde{Q}_k}} \right] ,~ \nonumber
\\ Q_{LL}^{t00} &=& \frac{1}{x_W\,(1-x_W)\,\sqrt{1+ \delta_{ij}}}
\left[ \frac{L_{q q^\prime Z} O^{\prime\prime R}_{ij} }{2 s_z} -
\sum_{k=1}^6 \frac{L_{\tilde{Q}_k q \tilde{\chi}_i^0}^\ast
L_{\tilde{Q}_k q^\prime \tilde{\chi}_j^0}}{t_{\tilde{Q}_k}}
\right] ,~ \nonumber
\\ Q_{RR}^{u00} &=&  \frac{1}{x_W\,(1-x_W)\,\sqrt{1+ \delta_{ij}}} \left[
\frac{R_{q q^\prime Z} O^{\prime\prime R}_{ij} }{2 s_z} +
\sum_{k=1}^6  \frac{R_{\tilde{Q}_k q^\prime \tilde{\chi}_i^0}
R_{\tilde{Q}_k q \tilde{\chi}_j^0}^\ast}{u_{\tilde{Q}_k}} \right]
,~ \nonumber \\ Q_{RR}^{t00} &=& \frac{1}{x_W\,(1-x_W)\,\sqrt{1+
\delta_{ij}}} \left[ \frac{R_{q q^\prime Z} O^{\prime\prime
L}_{ij} }{2 s_z} - \sum_{k=1}^6 \frac{R_{\tilde{Q}_k q
\tilde{\chi}_i^0}^\ast R_{\tilde{Q}_k q^\prime
\tilde{\chi}_j^0}}{t_{\tilde{Q}_k}} \right] ,~\nonumber\\
Q_{LR}^{u00} &=& \frac{1}{x_W\,(1-x_W)\,\sqrt{1+ \delta_{ij}}}
\sum_{k=1}^6 \frac{R_{\tilde{Q}_k q^\prime \tilde{\chi}_i^0}
L_{\tilde{Q}_k q
\tilde{\chi}_j^0}^\ast}{u_{\tilde{Q}_k}} ,~\nonumber \\
Q_{LR}^{t00} &=& \frac{1}{x_W\,(1-x_W)\,\sqrt{1+ \delta_{ij}}}
\sum_{k=1}^6 \frac{L_{\tilde{Q}_k q \tilde{\chi}_i^0}^\ast
R_{\tilde{Q}_k q^\prime \tilde{\chi}_j^0}}{t_{\tilde{Q}_k}} ,~
\nonumber\eea\bea Q_{RL}^{u00} &=& \frac{1}{x_W\,(1-x_W)\,\sqrt{1+
\delta_{ij}}} \sum_{k=1}^6 \frac{L_{\tilde{Q}_k q^\prime
\tilde{\chi}_i^0} R_{\tilde{Q}_k q
\tilde{\chi}_j^0}^\ast}{u_{\tilde{Q}_k}} ,~ \nonumber \\
Q_{RL}^{t00} &=& \frac{1}{x_W\,(1-x_W)\,\sqrt{1+ \delta_{ij}}}
\sum_{k=1}^6 \frac{R_{\tilde{Q}_k q \tilde{\chi}_i^0}^\ast
L_{\tilde{Q}_k q^\prime \tilde{\chi}_j^0}}{t_{\tilde{Q}_k}}, \eea
which agrees with the results of Ref.\ \cite{Gounaris:2004fm} in
the case of non-mixing squarks.

\subsection{Numerical results}

\begin{figure}
 \centering
 \includegraphics[width=0.49\columnwidth]{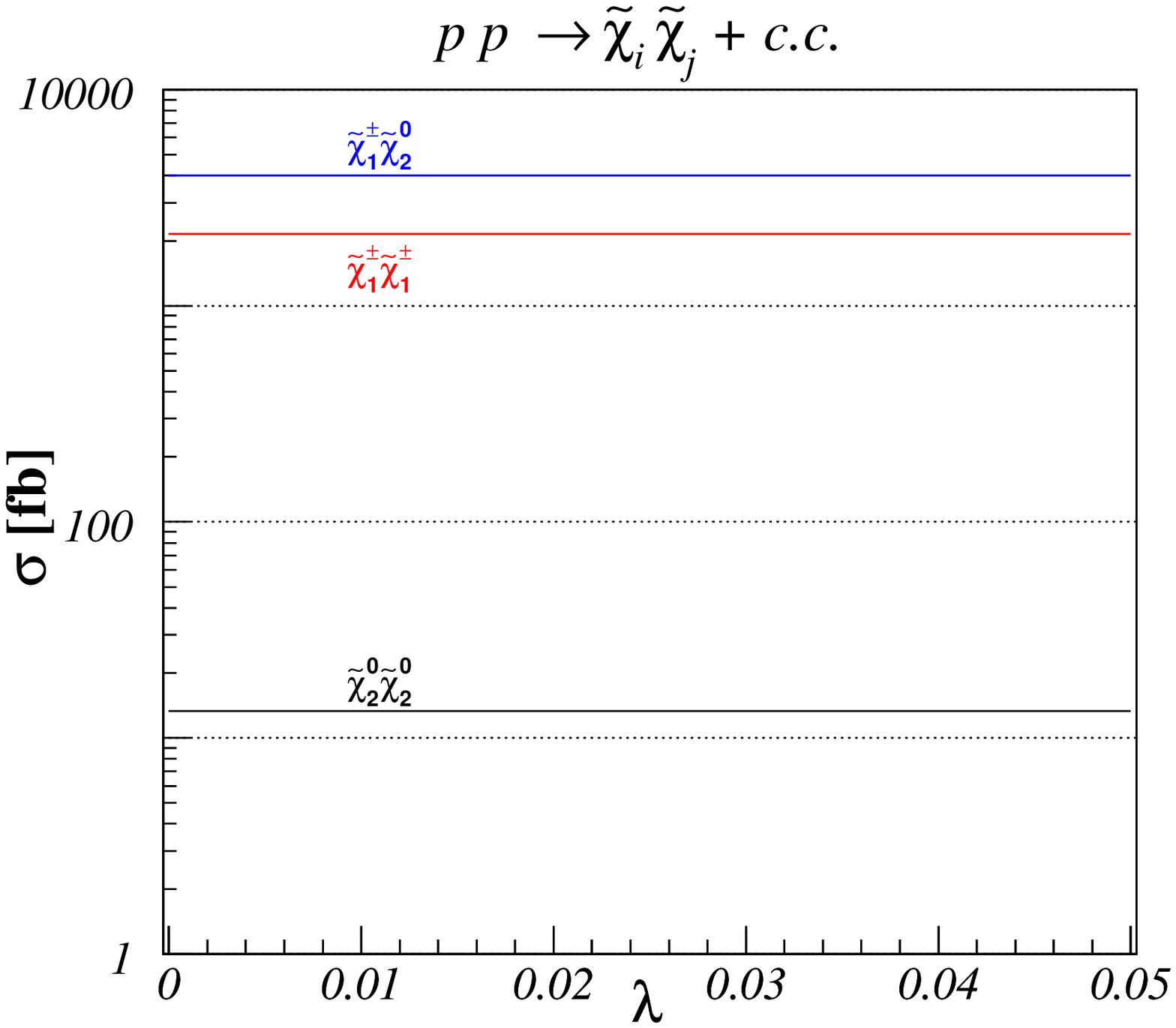}
 \includegraphics[width=0.49\columnwidth]{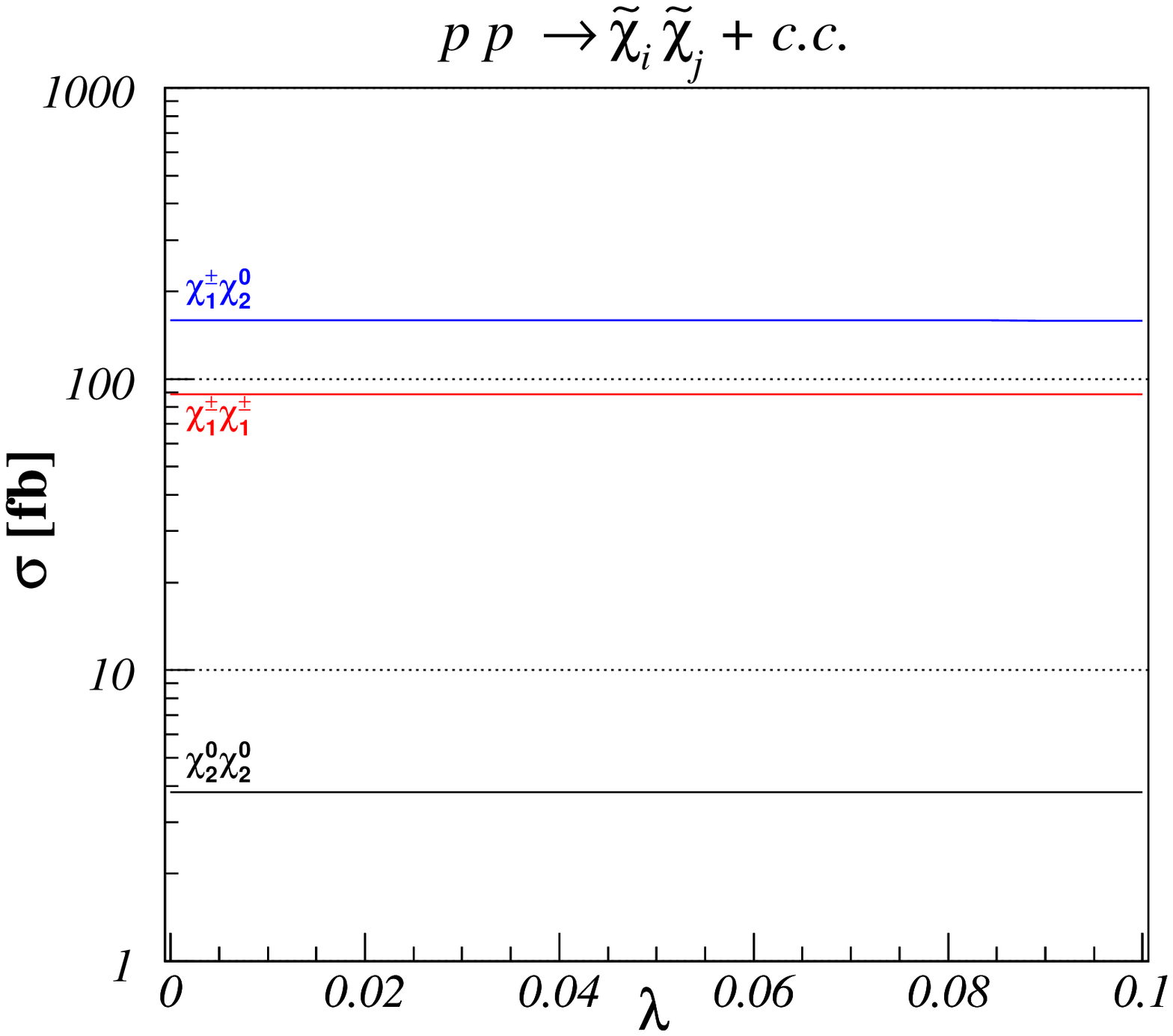}
 \caption{\label{fig:41a}Cross sections for
          gaugino pair production at the LHC in our benchmark
          scenario A (left) and B (right).}
\end{figure}
\begin{figure}
 \centering
 \includegraphics[width=0.49\columnwidth]{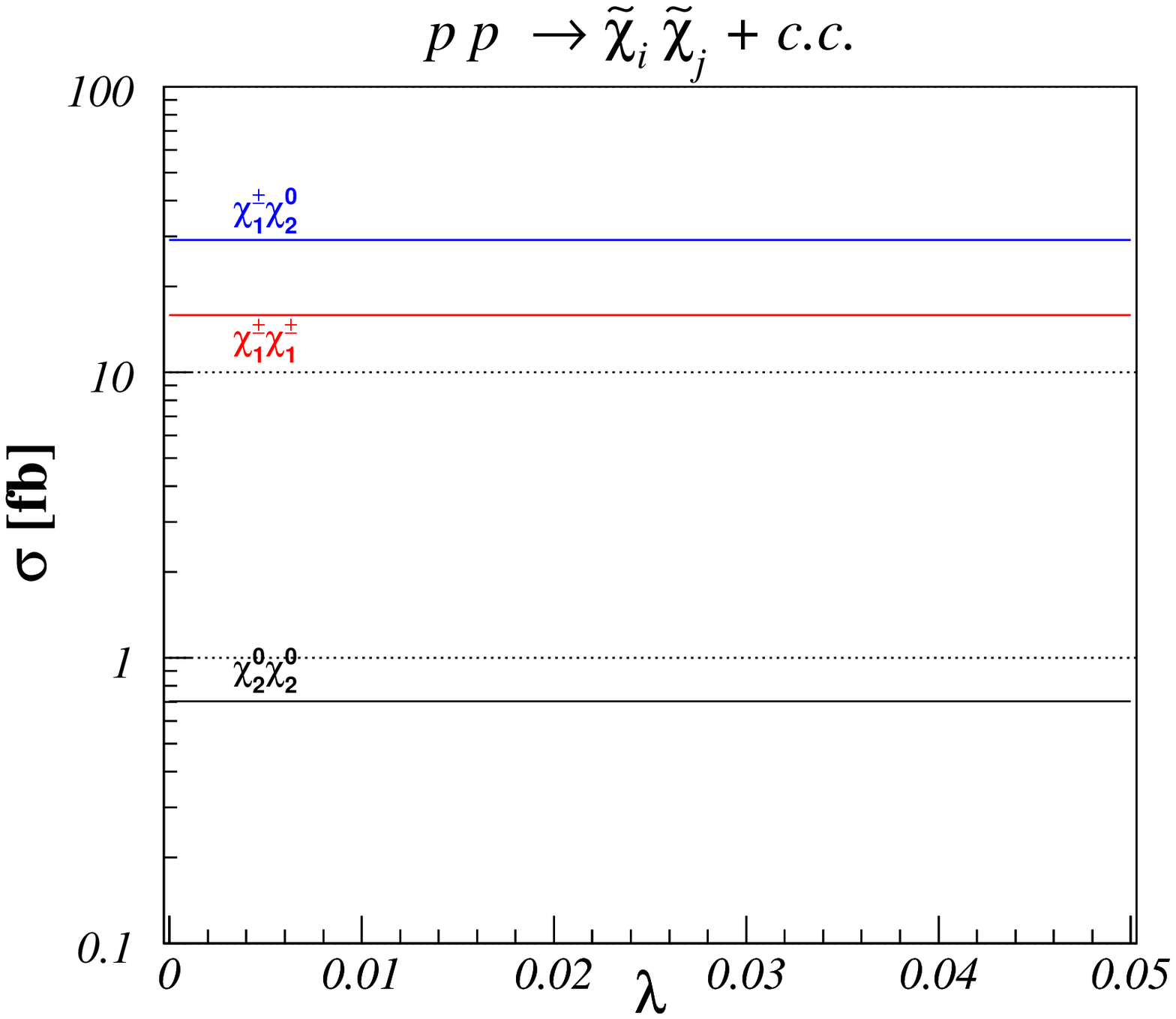}
 \includegraphics[width=0.49\columnwidth]{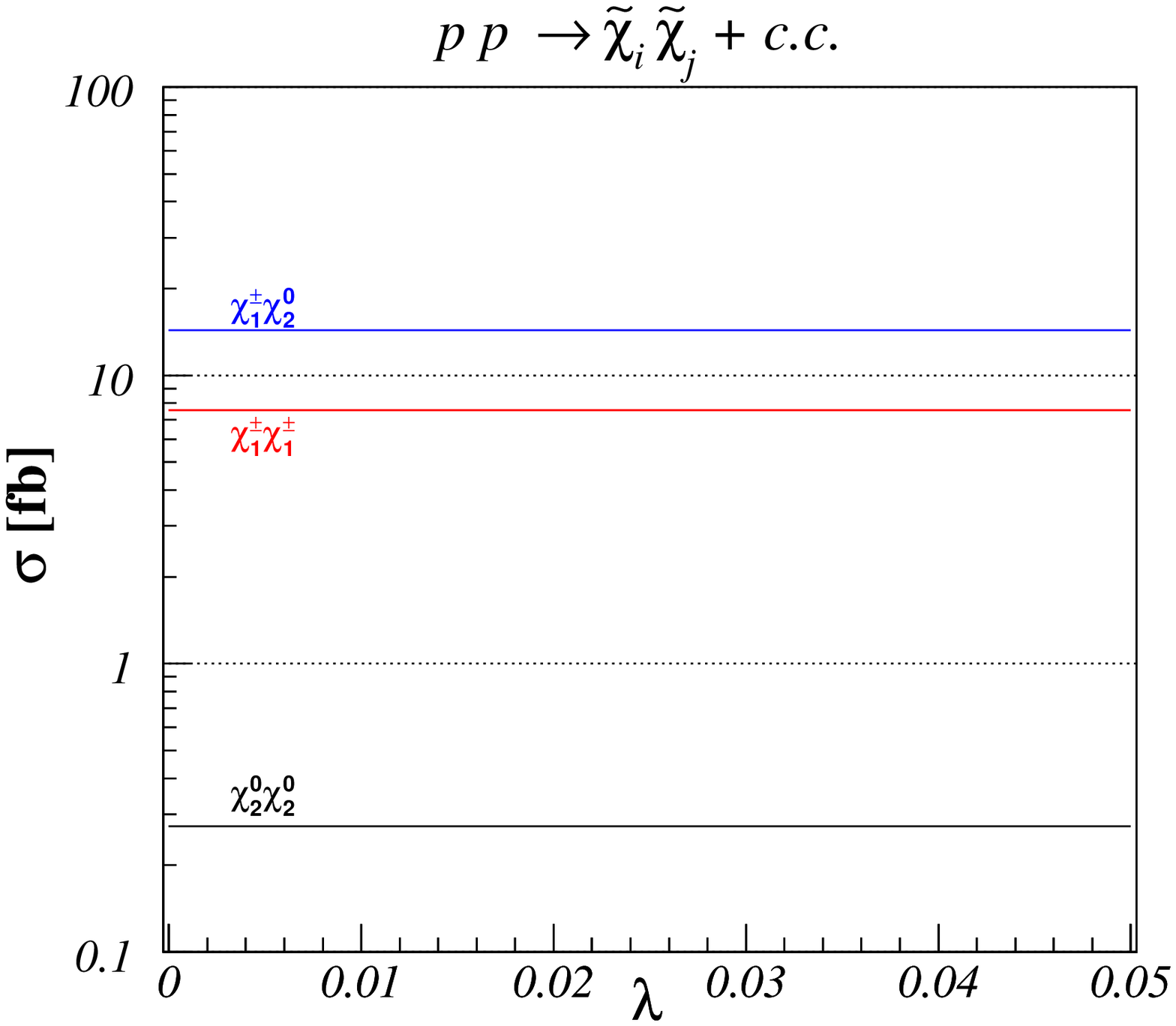}
 \caption{\label{fig:41b}Same as Fig.\ \ref{fig:41a} for our benchmark
          scenario C (left) and D (right).}
\end{figure}

As in the previous sections, the SUSY masses and mixings are
computed with the help of SPheno and FeynHiggs, and the LO set of
the CTEQ6 parton density are used. The numerical cross sections
for gaugino pair production are shown in Figs.\ \ref{fig:41a} and
\ref{fig:41b} for our benchmark scenario A, B, C, and D. Their
cross sections are rather large, but practically insensitive to
the flavour violation parameter $\lambda$, since they are summed
over exchanged squark flavours.

\section{Impact of flavour violation on squark and gaugino production}

In the previous sections, we have performed an extensive analysis
of squark and gaugino hadroproduction and decays in non-minimal
flavour violating supersymmetry. Within the super-CKM basis, we
have taken into account the possible misalignment of quark and
squark rotations and computed all squared helicity amplitudes for
the production and the decay widths of squarks and gauginos in
compact analytic form, verifying that our results agree with the
literature in the case of non-mixing squarks whenever possible. In
order to obtain numerical predictions for hadron colliders, we
have implemented all our results in a flexible computer program,
allowing us to discuss in detail the dependence of the cross
section on flavour violation. An illustrative summary of flavour
violating hadroproduction cross section contributions for
third-generation squarks and/or gauginos is presented in Tab.
\ref{tab:6}, together with the competing flavour-diagonal
contributions, which are the only contributions in cMFV SUSY.

\begin{table}\centering
  \caption{\label{tab:6}Dominant $s$-, $t$-, and $u$-channel contributions to
          the flavour violating hadroproduction of third-generation squarks
          and/or gauginos and the competing dominant flavour-diagonal
          contributions.}\vspace{.2cm}
\begin{tabular}{c|ccc}
  \underline{Exchange} & $s$ & $t$ & $u$ \\
  Final State & & & \\
  \hline
  $\tilde{t}\tilde{b}^*$ & $W$ & NMFV-$\tilde{g}$ & - \\
  $\tilde{b}\tilde{s}^*$ & NMFV-$Z$ & NMFV-$\tilde{g}$ & - \\
  $\tilde{t}\tilde{c}^*$ & NMFV-$Z$ & NMFV-$\tilde{g}$ & - \\
  \hline
  $\tilde{t}\tilde{b}$ & - & - & NMFV-$\tilde{g}$ \\
  $\tilde{b}\tilde{b}$ & - & $\tilde{g}$ & $\tilde{g}$ \\
  $\tilde{t}\tilde{t}$ & - & NMFV-$\tilde{g}$ & NMFV-$\tilde{g}$ \\
  \hline
  $\tilde{\chi}^0\tilde{b}$ & $b$ & $\tilde{b}$ & - \\
  $\tilde{\chi}^\pm\tilde{b}$ & NMFV-$c$ & NMFV-$\tilde{b}$ & - \\
  $\tilde{\chi}^0\tilde{t}$ & NMFV-$c$ & NMFV-$\tilde{t}$ & - \\
  $\tilde{\chi}^\pm\tilde{t}$ & $b$ & $\tilde{t}$ & - \\
  \hline
  $\tilde{\chi}\tilde{\chi}$ & $\gamma,Z,W$ & $\tilde{q}$ & $\tilde{q}$ \\
 \end{tabular}
\end{table}

\section{NMFV decays of squarks, gluino and gauginos}

\subsection{Squark decays}

\begin{figure}
 \centering
 \includegraphics[width=0.75\columnwidth]{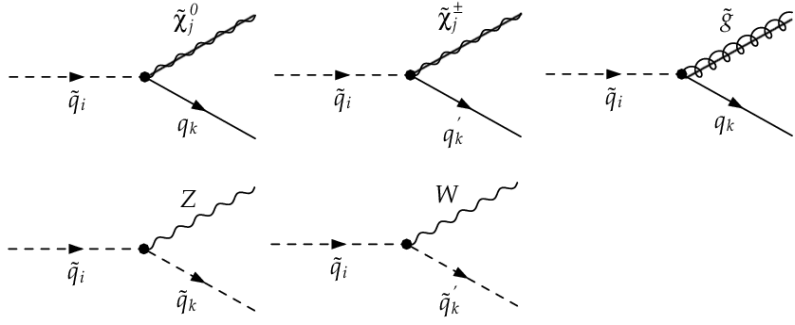}
 \caption{\label{fig:42a}Tree-level Feynman diagrams for squark decays
          into gauginos and quarks (top) and into electroweak gauge bosons
          and lighter squarks (bottom).}
\end{figure}

We turn now from SUSY particle production to decay processes and
show in Fig.\ \ref{fig:42a} the possible decays of squarks into
gauginos and quarks (top) as well as into electroweak gauge bosons
and lighter squarks (bottom). Both processes can in general induce
flavour violation. The decay widths of the former are given by
\bea \Gamma_{\tilde{q}_i \to \tilde{\chi}^0_j q_k} &=&
\frac{\alpha}{2\, m^3_{\tilde{q}_i}\, x_W\, (1-x_W)}
\Bigg(\bigg(m^2_{\tilde{q}_i} - m^2_{\tilde{\chi}^0_j} -
m^2_{q_k}\bigg) \bigg(\left| L_{\tilde{q}_i q_k \tilde{\chi}_j^0}
\right|^2 + \left| R_{\tilde{q}_i q_k \tilde{\chi}_j^0}
\right|^2\bigg) \nonumber \\ &-& 4\, m_{\tilde{\chi}^0_j}\,
m_{q_k}\, {\rm Re}\left[L_{\tilde{q}_i q_k \tilde{\chi}_j^0}
R^\ast_{\tilde{q}_i q_k \tilde{\chi}_j^0}\right]\Bigg)\,
\lambda^{1/2}(m^2_{\tilde{q}_i}, m^2_{\tilde{\chi}^0_j},
m^2_{q_k}),~ \label{eq:sqneq}\\ \Gamma_{\tilde{q}_i \to
\tilde{\chi}^\pm_j q^\prime_k} &=& \frac{\alpha}{4\,
m^3_{\tilde{q}_i}\, x_W} \Bigg(\bigg(m^2_{\tilde{q}_i} -
m^2_{\tilde{\chi}^\pm_j} - m^2_{q^\prime_k}\bigg) \bigg(\left|
L_{\tilde{q}_i q^\prime_k \tilde{\chi}_j^\pm} \right|^2 + \left|
R_{\tilde{q}_i q^\prime_k \tilde{\chi}_j^\pm}
\right|^2\bigg)\nonumber \\ &-& 4\, m_{\tilde{\chi}^\pm_j}\,
m_{q^\prime_k}\, {\rm Re}\left[L_{\tilde{q}_i q^\prime_k
\tilde{\chi}_j^\pm} R^\ast_{\tilde{q}_i q^\prime_k
\tilde{\chi}_j^\pm}\right]\Bigg)\,
\lambda^{1/2}(m^2_{\tilde{q}_i}, m^2_{\tilde{\chi}^\pm_j},
m^2_{q^\prime_k}) ,~\label{eq:sqchq}\\
\Gamma_{\tilde{q}_i \to \tilde{g} q_k} &=& \frac{2\, \alpha_s}{3\,
m^3_{\tilde{q}_i}\, x_W} \Bigg(\bigg(m^2_{\tilde{q}_i} -
m^2_{\tilde{g}} - m^2_{q_k}\bigg) \bigg(\left| L_{\tilde{q}_i q_k
\tilde{g}} \right|^2 + \left| R_{\tilde{q}_i q_k \tilde{g}}
\right|^2\bigg)\nn \\&-& 4\, m_{\tilde{g}}\, m_{q_k}\, {\rm
Re}\left[L_{\tilde{q}_i q_k \tilde{g}} R^\ast_{\tilde{q}_i q_k
\tilde{g}}\right]\Bigg)\, \lambda^{1/2}(m^2_{\tilde{q}_i},
m^2_{\tilde{g}}, m^2_{q_k}), \eea while those of the latter are
given by \bea \Gamma_{\tilde{q}_i \to Z \tilde{q}_k} &=&
\frac{\alpha}{16\, m^3_{\tilde{q}_i}\, m^2_Z x_W (1-x_W)} \left|
L_{\tilde{q}_i \tilde{q}_k Z} \!+\! R_{\tilde{q}_i \tilde{q}_k
Z}\right|^2\, \lambda^{3/2}(m^2_{\tilde{q}_i}, m^2_Z,
m^2_{\tilde{q}_k}),~~~~~~\\ \Gamma_{\tilde{q}_i \to W^\pm
\tilde{q}^\prime_k} &=& \frac{\alpha}{16\, m^3_{\tilde{q}_i}\,
m^2_W\, x_W\, (1-x_W)} \left| L_{\tilde{q}_i \tilde{q}^\prime_k W}
\right|^2\, \lambda^{3/2}(m^2_{\tilde{q}_i}, m^2_W,
m^2_{\tilde{q}^\prime_k}). \eea The usual K\"allen function is
\bea \lambda(x,y,z) = x^2 + y^2 + z^2 - 2 (x\,y + y\,z + z\,x).
\eea In MFV, our results agree with those of Ref.\
\cite{Bartl:1994bu}.

\subsection{Gluino decays}

\begin{figure}
 \centering
 \includegraphics[width=0.25\columnwidth]{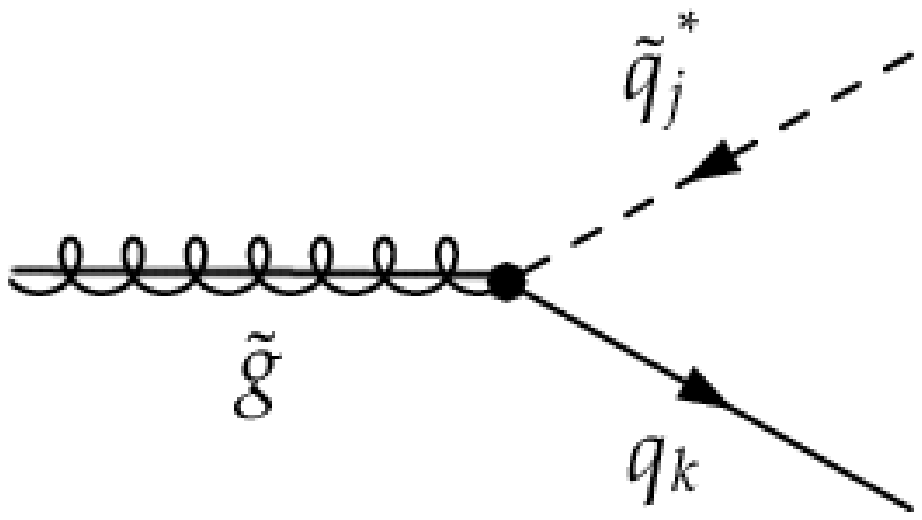}
 \caption{\label{fig:42b}Tree-level Feynman diagram for gluino decays
          into squarks and quarks.}
 \includegraphics[width=0.75\columnwidth]{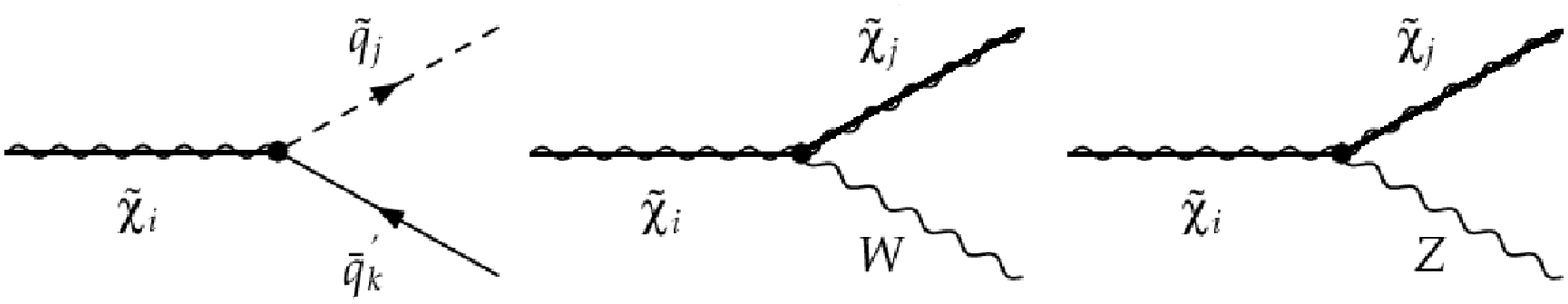}
 \caption{\label{fig:42c}Tree-level Feynman diagrams for gaugino decays
          into squarks and quarks (left) and into lighter gauginos and
          electroweak gauge bosons (centre and right).}
\end{figure}

Heavy gluinos can decay strongly into squarks and quarks as shown
in Fig.\ \ref{fig:42b}. The corresponding decay width \bea
\Gamma_{\tilde{g} \to \tilde{q}^\ast_j  q_k} &=&
\frac{\alpha_s}{8\, m^3_{\tilde{g}}} \Bigg(\bigg(m^2_{\tilde{g}} -
m^2_{\tilde{q}_j} + m^2_{q_k}\bigg) \bigg(\left| L_{\tilde{q}_j
q_k \tilde{g}} \right|^2 + \left| R_{\tilde{q}_j q_k \tilde{g}}
\right|^2\bigg)\nn\\ &+& 4\, m_{\tilde{g}}\, m_{q_k}\, {\rm
Re}\left[L_{\tilde{q}_j q_k \tilde{g}} R^\ast_{\tilde{q}_j q_k
\tilde{g}}\right]\Bigg)\, \lambda^{1/2}(m^2_{\tilde{g}},
m^2_{\tilde{q}_j}, m^2_{q_k}) \eea can in general also induce
flavour violation. In MFV, our result agrees again with the one of
Ref.\ \cite{Bartl:1994bu}.

\subsection{Gaugino decays}

Heavier gauginos can decay into squarks and quarks as shown in
Fig.\ \ref{fig:42c} (left) or into lighter gauginos and
electroweak gauge bosons (Fig.\ \ref{fig:42c} centre and right).
The analytical decay widths are \bea \Gamma_{\tilde{\chi}_i^\pm
\to \tilde{q}_j \bar{q}^\prime_k} &=& \frac{3\, \alpha}{8\,
m^3_{\tilde{\chi}_i^\pm}\, x_W} \Bigg(\bigg(
m^2_{\tilde{\chi}_i^\pm} - m^2_{\tilde{q}_j} +
m^2_{q^\prime_k}\bigg) \bigg(\left| L_{\tilde{q}_j q^\prime_k
\tilde{\chi}_i^\pm} \right|^2 + \left| R_{\tilde{q}_j q^\prime_k
\tilde{\chi}_i^\pm} \right|^2\bigg) \nonumber \\&+& 4\,
m_{\tilde{\chi}_i^\pm}\, m_{q^\prime_k}\, {\rm
Re}\left[L_{\tilde{q}_j q^\prime_k \tilde{\chi}_i^\pm}
R^\ast_{\tilde{q}_j q^\prime_k \tilde{\chi}_i^\pm}\right]\Bigg) \,
\lambda^{1/2}(m^2_{\tilde{\chi}_i^\pm}, m^2_{\tilde{q}_j},
m^2_{q^\prime_k}) \eea and \bea \Gamma_{\tilde{\chi}_i^\pm \to
\tilde{\chi}_j^0 W^\pm} &=& \frac{\alpha}{8\,
m^3_{\tilde{\chi}_i^\pm} m_W^2 x_W} \Bigg(\bigg(
m^4_{\tilde{\chi}_i^\pm} + m^4_{\tilde{\chi}_j^0} - 2\, m^4_W +
m^2_{\tilde{\chi}_i^\pm}\, m^2_W + m^2_{\tilde{\chi}_j^0}\, m^2_W
\nn\\&-& 2\, m^2_{\tilde{\chi}_i^\pm}\, m^2_{\tilde{\chi}_j^0}
\bigg)\, \bigg(\left| O^L_{ij} \right|^2 + \left| O^R_{ij}
\right|^2\bigg) - 12\, m_{\tilde{\chi}_i^\pm}\, m_W^2\,
m_{\tilde{\chi}_j^0}\, {\rm Re}\left[O^L_{ij}
O^{R\ast}_{ij}\right]\Bigg)\nn\\&\times&
\lambda^{1/2}(m^2_{\tilde{\chi}_i^\pm}, m^2_{\tilde{\chi}_j^0},
m^2_W) ,~\eea\bea \Gamma_{\tilde{\chi}_i^\pm \to
\tilde{\chi}_j^\pm Z} &=& \frac{\alpha}{8\,
m^3_{\tilde{\chi}_i^\pm} m_Z^2 x_W (1\!-\!x_W)} \Bigg(\bigg(
m^4_{\tilde{\chi}_i^\pm} \!+\! m^4_{\tilde{\chi}_j^\pm} \!-\! 2\,
m^4_Z \!+\! m^2_{\tilde{\chi}_i^\pm}\, m^2_Z  \!+\!
m^2_{\tilde{\chi}_j^\pm}\, m^2_Z\nn\\ &-& 2\,
m^2_{\tilde{\chi}_i^\pm}\, m^2_{\tilde{\chi}_j^\pm} \bigg)\,
\bigg(\left| O^{\prime L}_{ij} \right|^2 + \left| O^{\prime
R}_{ij} \right|^2\bigg) - 12\, m_{\tilde{\chi}_i^\pm}\, m_Z^2\,
m_{\tilde{\chi}_j^\pm}\, {\rm
Re}\left[O^{\prime L}_{ij} O^{\prime R\ast}_{ij}\right]\Bigg)\nn\\
&\times& \lambda^{1/2}(m^2_{\tilde{\chi}_i^\pm},
m^2_{\tilde{\chi}_j^\pm}, m^2_Z) \eea for charginos and \bea
\Gamma_{\tilde{\chi}_i^0 \to \tilde{q}_j \bar{q}_k} &=& \frac{3\,
\alpha}{4\, m^3_{\tilde{\chi}_i^0}\, x_W\, (1-x_W)} \Bigg(\bigg(
m^2_{\tilde{\chi}_i^0} - m^2_{\tilde{q}_j} + m^2_{q_k}\bigg)
\bigg(\left| L_{\tilde{q}_j q_k \tilde{\chi}_i^0} \right|^2 +
\left| R_{\tilde{q}_j q_k \tilde{\chi}_i^0} \right|^2\bigg)
\nonumber \\&+& 4\, m_{\tilde{\chi}_i^0}\, m_{q_k}\, {\rm
Re}\left[L_{\tilde{q}_j q_k \tilde{\chi}_i^0} R^\ast_{\tilde{q}_j
q_k \tilde{\chi}_i^0}\right]\Bigg) \,
\lambda^{1/2}(m^2_{\tilde{\chi}_i^0}, m^2_{\tilde{q}_j},
m^2_{q_k}) \eea and \bea \Gamma_{\tilde{\chi}_i^0 \to
\tilde{\chi}_j^\pm W^\mp} &=& \frac{\alpha}{8\,
m^3_{\tilde{\chi}_i^0}\, m_W^2\, x_W} \Bigg(\bigg(
m^4_{\tilde{\chi}_i^0} + m^4_{\tilde{\chi}_j^\pm} - 2\, m^4_W +
m^2_{\tilde{\chi}_i^0}\, m^2_W + m^2_{\tilde{\chi}_j^\pm}\,
m^2_W\nn\\ &-& 2\, m^2_{\tilde{\chi}_i^0}\,
m^2_{\tilde{\chi}_j^\pm} \bigg)\, \bigg(\left| O^L_{ij} \right|^2
+ \left| O^R_{ij} \right|^2\bigg) - 12\, m_{\tilde{\chi}_i^0}\,
m_W^2\, m_{\tilde{\chi}_j^\pm}\, {\rm Re}\left[O^L_{ij}
O^{R\ast}_{ij}\right]\Bigg) \nn \\ &\times&
\lambda^{1/2}(m^2_{\tilde{\chi}_i^0},
m^2_{\tilde{\chi}_j^\pm}, m^2_W) ,~ \\
\Gamma_{\tilde{\chi}_i^0 \to \tilde{\chi}_j^0 Z} &=&
\frac{\alpha}{8\, m^3_{\tilde{\chi}_i^0}\, m_Z^2\, x_W\, (1-x_W)}
\Bigg(\bigg( m^4_{\tilde{\chi}_i^0} + m^4_{\tilde{\chi}_j^0} - 2\,
m^4_Z + m^2_{\tilde{\chi}_i^0}\, m^2_Z + m^2_{\tilde{\chi}_j^0}\,
m^2_Z\nn \\ &-& 2\, m^2_{\tilde{\chi}_i^0}\,
m^2_{\tilde{\chi}_j^0} \bigg)\, \bigg(\left| O^{\prime L}_{ij}
\right|^2 + \left| O^{\prime R}_{ij} \right|^2\bigg) - 12\,
m_{\tilde{\chi}_i^0}\, m_Z^2\, m_{\tilde{\chi}_j^0}\, {\rm
Re}\left[O^{\prime\prime L}_{ij} O^{\prime\prime
R\ast}_{ij}\right]\Bigg)\nn\\ &\times&
\lambda^{1/2}(m^2_{\tilde{\chi}_i^0}, m^2_{\tilde{\chi}_j^0},
m^2_Z) \eea for neutralinos, respectively. Chargino decays into a
slepton and a neutrino (lepton and sneutrino) can be deduced from
the previous equations by taking the proper limits, i.e.\ by
removing colour factors and down-type masses in the coupling
definitions. Our results agree then with those of Ref.\
\cite{Obara:2005qi} in the limit of non-mixing sneutrinos. Note
that the same simplifications also permit a verification of our
results for squark decays into a gaugino and a quark in Eqs.\
(\ref{eq:sqneq}) and (\ref{eq:sqchq}) when compared to their
leptonic counterparts in Ref.\ \cite{Obara:2005qi}.

\newpage $~$\\ \newpage

\chapter{Conclusion and outlook} \label{ch:concl}

The Standard Model of particle physics provides a successful
description of presently known phenomena, except for neutrino
physics. However, despite of its success, a set of conceptual
problems do not have a solution within the framework of the SM,
such as the origin of mass, gauge coupling unification, or the
hierarchy problem. Several attempts have been made to solve these
problems, leading to various theories beyond the SM, even if there
is still no experimental evidence of their existence. Each of
these theories predicts new particles, with masses lying in the
TeV-range, i.e.\ the discovery reach of present and future hadron
colliders, the Tevatron and the LHC, which will then be able to
put constraints on these models, or conclude about their
(non-)existence.\\

In this thesis, we have considered the production of sleptons,
squarks and gauginos of the Minimal Supersymmetric Standard Model.
Cross sections for SUSY particles production at hadron colliders
have been extensively studied in the past at leading order and
also at next-to-leading order of perturbative QCD, since they are
expected to receive important contributions from radiative
corrections. These corrections include large logarithms, which
have to be resummed in order to get reliable perturbative results.
We have thus performed a first and extensive study of the
resummation effects for SUSY particle pair production at hadron
colliders, focusing on Drell-Yan like slepton-pair and
slepton-sneutrino associated production in minimal supergravity
and gauge-mediated SUSY-breaking scenarios. We have presented
accurate transverse-momentum and invariant-mass distributions, as
well as total cross sections, resumming soft-gluon emission
contributions to all orders in the strong coupling.\\

Monte Carlo event generators are also commonly used, especially by
experimentalists, to calculate observables depending on the
soft-gluon emission from the initial state partons in hadronic
collisions. This allows to compare experimental data to
theoretical predictions and to simulate experimental signatures,
when there are no experimental data yet. These programs usually
implement the hard scattering process at the leading order,
matching it with parton showering, the latter parameterizing the
initial- and final-state radiation. Let us note that recently,
event generators using next-to-leading order calculations have
also been developed. It is expected that both Monte Carlo parton
showering and resummation calculations should accurately describe
the effects of soft-gluon emission from the incoming partons. A
comparison between the predictions given by the two approaches
would certainly be useful in testing their reliability. Finally,
other SUSY particle production processes, such as gaugino- or
squark pair hadroproduction should also be considered. \\

The quest for supersymmetric particles at hadron colliders will
nevertheless rely on our ability of predicting both the SUSY
signal and the SM processes being the backgrounds for these
searches. In this work, we have focused on presenting accurate
predictions for a slepton-pair signal, i.e.\ two highly energetic
lepton and a large amount of missing energy. We have not
considered the background consisting mainly in lepton pairs coming
from $WW$ and $t\bar t$ decays. A detailed phenomenological study,
including all background contributions, remains to be performed,
in order to propose proper experimental cuts to enhance the signal
over the background ratio. Furthermore, the sensitivity of these
cuts to a complete experimental environment should also be
investigated.\\

In non-minimal supersymmetric models, novel effects of flavour
violation may occur. In this case, the flavour structure in the
squark sector cannot be directly deduced from the trilinear Yukawa
couplings of the fermion and Higgs supermultiplets. We have
performed a precise numerical analysis of the experimentally
allowed parameter space, considering minimal supergravity
scenarios with non-minimal flavour violation, looking for regions
allowed by low-energy, electroweak precision, and cosmological
data. Leading order cross sections for the production of squarks
and gauginos at hadron colliders have been implemented in a
flexible computer program, allowing us to study in detail the
dependence of these cross sections on flavour violation.\\

A full experimental study including heavy-flavour tagging
efficiencies, detector resolutions, and background processes
would, of course, be very interesting in order to establish the
experimental significance of NMFV. While the implementation of our
analytical results in a general-purpose Monte Carlo generator
should now be straight-forward, such a detailed experimental study
is beyond the scope of this work. Other possible extensions of our
work include the investigation of other SUSY-breaking mechanisms
or the computation of the radiative corrections for all of these
processes within the NMFV framework.\\

\end{document}